\documentstyle[12pt,twoside,graphicx]{book}

\newcommand{\be}{\begin{equation}}
\newcommand{\ee}{\end{equation}}
\newcommand{\ba}{\begin{array}}
\newcommand{\ea}{\end{array}}
\newcommand{\bea}{\begin{eqnarray}}
\newcommand{\eea}{\end{eqnarray}}
\newcommand{\bi}{\begin{itemize}}
\newcommand{\ei}{\end{itemize}}

\def\vec#1{\mathchoice{\mbox{\boldmath$\displaystyle\bf#1$}}
{\mbox{\boldmath$\textstyle\bf#1$}}
{\mbox{\boldmath$\scriptstyle\bf#1$}}
{\mbox{\boldmath$\scriptscriptstyle\bf#1$}}}

\def\bbbz{{\mathchoice {\hbox{$\sf\textstyle Z\kern-0.4em Z$}}
{\hbox{$\sf\textstyle Z\kern-0.4em Z$}}
{\hbox{$\sf\scriptstyle Z\kern-0.3em Z$}}
{\hbox{$\sf\scriptscriptstyle Z\kern-0.2em Z$}}}}

\chardef\ii="10

\def\bbbc{{\mathchoice {\setbox0=\hbox{\rm C}\hbox{\hbox
to0pt{\kern0.4\wd0\vrule height0.9\ht0\hss}\box0}}
{\setbox0=\hbox{$\textstyle\hbox{\rm C}$}\hbox{\hbox
to0pt{\kern0.4\wd0\vrule height0.9\ht0\hss}\box0}}
{\setbox0=\hbox{$\scriptstyle\hbox{\rm C}$}\hbox{\hbox
to0pt{\kern0.4\wd0\vrule height0.9\ht0\hss}\box0}}
{\setbox0=\hbox{$\scriptscriptstyle\hbox{\rm C}$}\hbox{\hbox
to0pt{\kern0.4\wd0\vrule height0.9\ht0\hss}\box0}}}}

\newcommand{\fcaption}[1]{
        \refstepcounter{figure}
        \setbox\@tempboxa = \hbox{\footnotesize Fig.~\thefigure. #1}
        \ifdim \wd\@tempboxa > 5in
           {\begin{center}
        \parbox{5in}{\footnotesize\smalllineskip Fig.~\thefigure. #1}
            \end{center}}
        \else
             {\begin{center}
             {\footnotesize Fig.~\thefigure. #1}
              \end{center}}
        \fi}
\newcommand{\tcaption}[1]{
        \refstepcounter{table}
        \setbox\@tempboxa = \hbox{\footnotesize Table~\thetable. #1}
        \ifdim \wd\@tempboxa > 5in
           {\begin{center}
        \parbox{5in}{\footnotesize\smalllineskip Table~\thetable. #1}
            \end{center}}
        \else
             {\begin{center}
             {\footnotesize Table~\thetable. #1}
              \end{center}}
        \fi}
\begin{document}
\begin{titlepage}

\begin{center}
\vspace*{2.0cm}

\vspace{1.5cm} {\LARGE{\it\bf Quantum Systems with Hidden
Symmetry.

\vspace{0.5cm} Interbasis Expansions}}

\vspace{2.0cm}

{\Large\bf L.G. Mardoyan, G.S. Pogosyan, A.N. Sissakian

\vspace{0.5cm}

and

\vspace{0.5cm}

V.M. Ter-Antonyan}

\vspace{5cm}
\vfill {\bf DUBNA, YEREVAN}

\vfill {\bf 2023}
\end{center}

\end{titlepage}

\tableofcontents


\newpage
\addcontentsline{toc}{chapter}{Preface to the English edition}
\begin{center}
 {\bf Preface to the English edition}
\end{center}

This monograph is the English version of the book "Quantum systems with hidden symmetry.
Interbasis expansions" published in 2006 by the publishing house FIZMATLIT (Moscow) in Russian.
When compiling this version of the book, typos and inaccuracies noted since the release of the
Russian edition have been corrected.

We are grateful to Alexander Gusev for the help he provided in compiling the layout of the book.

\vspace{0.3cm}
October 2023 \hfill  L.G. Mardoyan\footnote{Joint Institute for Nuclear Research, Dubna, Russia}$^,$\footnote{Yerevan State University, Yerevan, Armenia \\ e-mail: mardoyan@theor.jinr.ru, pogosyan@ysu.am}

\hfill G.S. Pogosyan$^{\rm2}$
\addcontentsline{toc}{chapter}{Preface to the Russian edition}
\vspace{0.8cm}
\begin{center}
{\bf Preface to the Russian edition}
\end{center}

The purpose of this book is to bring together methods and results of research related to the problem
of interbasis expansions in the theory of the simplest systems with hidden symmetry, such as
two- and three-dimensional hydrogen atoms, circular and multidimensional oscillators. The monograph
consists of five chapters and two appendices. It is based on the work of the authors
\cite{ARUT1,ARUT2,ARUT3,DMPST,DPST,T-5,T-7,POG-10,T-9,T-10,T-11,T-1,T-2,T-4,MPS1,MPSTA2,MPSTA1,
T-3,MPS,MPSTA4,MPTA,MPSTA5,MPSTA11,MPSTA12,MPSTA13,PSMT,POGOSYAN1,POGOSYAN2,POGOSYAN3,POGOSYAN4,
POGOSYAN5,POGOSYAN6,TERANT}, carried out at the Laboratory of Theoretical Physics of the Joint Institute 
of Nuclear Research and at the Department of Theoretical Physics of Yerevan State University. 
Due to space limitations, the monograph did not include a one-dimensional hydrogen atom, 
an anharmonic oscillator, systems with hidden symmetry in spaces of constant
curvature, and dyone-oscillator duality. In the future we intend to dedicate a separate monograph on these issues.

To our deep regret, our friend and co-author, Professor V.M.Ter-Antonyan, is not with us today.
A talented theoretical physicist and a brilliant expert in quantum mechanics. His early death
did not allow him to take part in the final phase of the preparation of this book;
we dedicate our work to his blessed memory.

In conclusion, we consider it our pleasant duty to thank R.M.Avakyan, S.I.Vinitsky, V.G.Kadyshevsky,
I.V.Komarov, A.V.Matvienko, R.M.Mir-Kasimov, L.I.Ponomarev and E.V. Chubaryan for discussing many
of the issues raised in the book.

We express special gratitude to our co-authors E.M.Akopyan, P.Winternitz, B.Wolf, L.S. Davtyan, H.Groshe,
A.A.Izmestyev, V.Kallis, E.Kalnins, Kh.Karayan, M.Kibler, I.Lukach, V.Miller, A.Nersessian and I.Patera
for long-term and fruitful scientific cooperation.

We are also grateful to A.R.Balabekyan and K.E.Muradyan for the help they provided in preparing the book manuscript.

\vspace{0.3cm}
2006     \hfill L.G.Mardoyan

\hfill G.S.Pogosyan

\hfill A.N.Sissakian

\newpage

\addcontentsline{toc}{chapter}{Introduction}

\chapter*{Introduction}
\pagestyle{myheadings}
\markboth{INTRODUCTION}{INTRODUCTION}

The given monograph, as the heading suggests, is devoted to the
investigation of interbasis expansions in quantum systems with
hidden symmetry. The interest in the problem has not been waning
for already half a century, and the flow of literature where this
or that aspect of the question is being touched upon is presently
as stable as ten years ago. The ideas and methods characteristic
of this area of theoretical Physics initially emerged during the
study of particles behaviour in quantum and oscillator fields and
were further developed in the quantum theory of field \cite{SHIR}, in
the field theory with fundamental length \cite{KADYSH}, in Physics of
elementary particles \cite{ELCHAS}, theory of lasers \cite{DICKE}, in the model of
shells \cite{ELLIOT1}, collective model of nuclei \cite{ELLIOT2}, in the modern
theory of (super)strings \cite{WITTEN}, in different physical systems with
magnetic monopole \cite{GIBBONS,ZWANZIG}, in the supersymmetric quantum
mechanics \cite{KHARE,LUC-POG-1} and quantum gravitation  \cite{FUJIK}.

Usually symmetry transformations suggest transmissions and
rotations, i.e. transformations which do not change the operator
of kinetic energy. The symmetries themselves are meanwhile
identical to those inherent to the interaction potential. A
conceptually new step was the discovery of the so-called hidden
and dynamic symmetries which meant more than the invariance of the
interaction potential in relation to this or that geometrical
transformation. The group of hidden symmetry affects both the
potential and kinetic parts of the Hamiltonian and represents a
principally new type of symmetries inherent to the limited class
of potentials.

The theory of systems with hidden symmetry has undergone several
stages of development. It has emerged from the Kepler's problem
where still Laplace \cite{LAPLACE} introduced the additional vector preserving
quantity in the orbit plane and directed toward the big axis of
the ellipsis and modulo equal to the excentriciteth. The Laplacian
integral of motion was later rediscovered by Runge \cite{RUNGE} and Lentz
subsequently \cite{LENZ}, and his generalization for the case of quantum
mechanics was elegantly applied by Pauli \cite{PAULI} for the calculation
of the discrete spectrum of the hydrogen atom \cite{LL}. Long before
Pauli Bertrand \cite{BERTRAN} made a fundamental contribution to the theory
of systems with hidden symmetry and proved that among all the
central-symmetric fields it is only in the

Coulomb and oscillator fields where all the finite trajectories of
particles motion are closed. Almost a century later Bertrand's
theorem brought about not a less significant fact in quantum
mechanics, i.e. an accidental degeneracy of the energetic spectrum
of the hydrogen atom and isotropic oscillator, and the explanation
of this fact led Fock \cite{FOCK1,FOCK2} to the concept of the hidden symmetry in
quantum-mechanical systems. Fock showed that the non-relativistic
Coulomb problem in the discrete spectrum at the transition to the
impulse space and stereographic projection comes to the integral
equation invariant to the group of four-dimensional rotations $O(4)$.
In Schr\"{o}dinger's starting equation this symmetry is not
seen, thus it was called a hidden symmetry.  $O(4)$-symmetry and
the $O(4)\subset O(3)$ condition accounted for the fact that in the
Coulomb field together with the angular momentum there should
exist another vector integral of motion. Bargman \cite{BARGMAN} established
the connection between the approaches of Pauli and Fock and showed
that the properly normalized integrals of the Laplacian motion and
the angular momentum obey the commutational relations
characteristic of the generators of $O(4)$ group.

The discovery of the hidden system of the hydrogen atom and the
success of symmetric schemes in physics of elementary particles
stimulated the development of group-theoretical approach to the
problems connected with the Coulomb field. The spread of the Fock
method to the continuous spectrum \cite{BANDER,PER-POP,POPOV} showed that one
deals with the three-dimensional hyperboloid instead of the
three-dimensional sphere, and with the infinite-dimensional
unitary representations of the Lorentz $O(3,1)$ group instead of the finite
representations of the $O(4)$ group. The case of the zero energy
required a special investigation \cite{CHEN} and indicated that here we deal
with the non-relativistic Galileo group or with the group of
motions of three-dimensional space $E(3)$ isomorphic to it. Also
the multidimensional Coulomb problem was investigated. In Levi's
work \cite{LEVY} the discrete spectrum and the wave functions of
multidimensional hydrogen atom were found, in Gy\"orgyi and Revai's
work \cite{revai} the result was purely algebraically corroborated  and
in Kuznetzov's work \cite{KUZNECOV} the multidimensional Coulomb wave
functions in the continuous spectrum were calculated. The
symmetric aspects of the multidimensional $n\geq 2$ Coulomb problem
were considered in Alliluev's work \cite{ALLILUEV} and the group of the dynamic
symmetry of one-dimensional problem was found in the works \cite{DPST1,LMPST1}.

The initial thoughts of Fock and Bargmann developed also in the
direction of the so-called groups of non-invariance \cite{MUKUNDA}. As
different from the dynamic group, where the Hamiltonian acts as
Casimir operator, part of the elements of the non-invariance group do
not commutate with the problem's Hamiltonian, but rather act as
raising and decreasing operators on the wave functions of levels.
A noncompact group, containing a dynamic group as a subgroup, is
usually selected as an non-invariance group. Information of the
invariance group, as stated by Gell-Mann \cite{GELMAN} and Dothan
\cite{DOTAN}, permits to predict the whole spectrum of its excited states
according to several known levels of the system. The further
development of this direction is presented in the monograph \cite{MANKO1}.
The description of the hydrogen atom in the framework of the
ladder representation of the group $U(2,2)$ can be found in the
review work \cite{TODOR}. For completeness sake let us bring here also
different works devoted to the symmetric aspects of the Coulomb
field \cite{BACRY,BARUT1,ENGLEF,FRONC,GYORGYI,HAN,HYLLERAAS, LAPORTE1,LAPORTE2,
MANKO2,MARIWALLA,MUSTO,PODOLSKY,PRAT1,PRAT2,MUKUNDA1}.

The accidental degeneration of energy levels of $n$-dimensional
isotropic oscillator was first introduced in terms of the hidden
symmetry $U(n)$ in the works of Jauch \cite{JAUCH1} and Jauch and Hill \cite{JAUCH2}.
The analogous result was later obtained by Baker \cite{BAKER}. In Alliluev's
work \cite{ALLILUEV} the connection between the quantization of the circle
oscillator and the quantization of a certain operator obeying the
commutation relations characteristic of the angular momentum
operator was established. In Demkov's works \cite{DEMKOV1} the symmetry
group of the isotropic oscillator accounting for the additional
degeneration of energy levels was considered. The symmetry group
of the anisotropic oscillator was found in the works \cite{DEMKOV2,DULOCK,ILKAL,MAIELLA,
VENDRAMIN}.

Apart from hydrogen atom and isotropic oscillator, hidden symmetry
is also characteristic of non-relativistic spinless charged
particle moving in homogeneous magnetic field \cite{Lippmann}; neutron in the
field of linear conductor \cite{PRONKO} and of some other potentials
(mainly singular generalizations of the Coulomb or oscillator
potentials) \cite{DRAGAN,EVANS,FMSUW,IPSW1,KMJP2,KMJP3,KKMP1,MSVW,FSUW}
\footnote{In Evans' paper \cite{EVANS}, generalized systems with hidden symmetry are called
superintegrable systems.}. Part of it, such as
Harthman potential \cite{HARTMAN}, found application in molecular physics.

It is noteworthy that the connection between the two groups of
hidden symmetry $SO(n+1)$ and $SU(n)$, in its turn, leads to the
concept of Coulomb-oscillator or, to be more specific,
dyon-oscillator duality in quantum mechanics. For instance,
$(n+1)$-dimensional radial Schr\"{o}dinger's equation with the Coulomb
potential is identical to the $2n$-dimensional oscillator equation
accurate within duality transformation \cite{TERANT}. According to Hurwitz
theorem \cite{HUR} the complete conformity (not only for radial parts)
is possible only for special dimensions, namely that of $(2,2),
(3,4)$ and $(5,8)$. The dual transformation in these cases is named
after Levi-Civita \cite{LC}, Kustaanheimo and Steifel \cite{KS} and Hurwitz
\cite{DMPST}. For the case of spaces of constant curvature these
transformations were generalized in the works \cite{KMP7,KMP9,NERPOG}.

Similar to the Euclidean space, investigation of systems with
hidden symmetry in spaces of constant curvature both in the
positive and in the negative has a rich history. Such system was
first investigated in Schr\"{o}dinger's works \cite{SCHR}, who via
factorization method found the energy spectrum for the potential
analogous to the Coulomb's potential on the three-dimensional $S_{3}$
sphere and showed that complete degeneracy by orbital and
azimuthal quantum numbers takes place as in the case of flat
space. Not a long after Stevenson \cite{STEV} came to the same result by
Schr\"{o}dinger's equation direct solution and found the
nonstandardized wave functions of the hydrogen atom. Analogous
result was obtained in the works of Infeld \cite{INF} and Infeld and
Schild \cite{IS} for the case of a space of negative curvature. Later
in Higg's \cite{HIG}, Leemon's \cite{LEEMON}, Kurochkin and Otchik's \cite{KO}, and
Kurochkin, Otchik and Bogush's \cite{BKO} works it was shown that the
presence of degeneracy for the energetic spectrum of Coulomb's
problem and harmonic oscillator on the sphere on Lobachevsky
three-dimensional space is accounted for by the presence of
additional motion integrals, namely that of Runge-Lentz vector
analog (for Coulomb's potential) and Demkov's tensor (for the
oscillator). However it turned out that the commutational
relations for the components of these operators along with the
components of the angular momentum are of nonlinear character and
generate algebraic structure which can be considered as an
expansion of Lee's algebra  (Poisson's algebra in classical
mechanics), namely quadratic algebra\footnote{For the first time, such algebras were introduced by
Sklyanin and found application in many areas of physics: in solving the classical Yang-Baxter equations
\cite{SKLYAN1,SKLYANIN2}, in statistics \cite{ESSLER}, in the case of exactly solvable classical problems
\cite{GRAN1}, for describing some two-dimensional \cite{DAS1,VILE} and three-dimensional
\cite{DAS2,GRAN3,GRAN1} superintegrable systems, as well as in calculating interbasis transitions in the
Coulomb and oscillatory problems on a three-dimensional sphere and a hyperboloid \cite{GZLa,GZLb}.}.
As a result the notion of
the group of hidden symmetry does not make sense for the Coulomb's
and oscillator problems on the spaces of constant curvature and
one can talk only about the algebra of hidden symmetry. Despite
the above-mentioned peculiarities, the existing commutational
relations appear to be enough for the derivation of the energetic
spectrum  formula and the multiplicity of the hydrogen atom
degeneracy and the isotropic oscillator on the $S_{3}$ sphere
\cite{HIG,KO}. The Coulomb's and oscillator problems on the spheres and
hyperboloids were further developed in the works
\cite{BIJ1,BIJ2, BOR,GRO1,GRO2,GROP5,GROP4,GROP1,HPSV1,IK,KATA,OR,VMPSS}
and were applied for the construction of a multiparticle wave function \cite{BB},
for the solution of the problem of two Coulomb centres at large intercentre
separation \cite{BOT}, the description of the quarkonium spectrum \cite{IZ1} and the excitons in
quantum dots \cite{GRITKUR}.

Quantum systems with hidden symmetry possess a significant
peculiarity, i.e. the variables in Scr\"{o}dinger's equation,
describing such systems, are separated in some coordinate
systems\footnote{The question of the separation of variables in the Schr\"{o}dinger
equation in three-dimensional Euclidean space was solved by Eisenhart \cite{EISENHART}.
A similar problem for the Helmholtz equation in two- and three-dimensional spaces of constant curvature
was completely solved in the well-known paper by Olevskii \cite{OLEV}.
A graphical procedure for constructing arbitrary orthogonal coordinate systems in an $n$-dimensional
Euclidean space and on an $n$-dimensional sphere is given in the article by Kalnins and Miller
\cite{KMJ} and well described in the monograph by Kalnins \cite{KAL1}.
The case of subgroup coordinate systems on a sphere and a hyperboloid of arbitrary dimension is analyzed in
detail in the works of Vilenkin \cite{VILEN2} and Vilenkin, Kuznetsov and Smorodinsky \cite{VKS}.},
and the corresponding solutions form complete bases for the rest
of quantum numbers at the given energy. This peculiarity is one of
the manifestations of the accidental degeneracy. A good example of
such character is the isotropic oscillator  separated in the eight
coordinate systems, namely in the Cartesian, spherical,
cylindrical, circular elliptic, sphero-conal, prolate and oblate
spheroidal and ellipsoid, and the hydrogen atom permitting of the
separation of variables in the four coordiante systems, namely
spherical, sphero-conal, parabolic and prolate spheroidal. From
the official point of view the importance of variables separation
in several coordinate systems is obvious. In the spectroscopy of
hydrogen-like systems the spherical coordinate system is used, in
the investigation of Stark effect the parabolic system is used,
and in the problem of electron motion in the field with two fixed
Coulomb centres the prolate spheroidal coordinate system is used.
The choice of a specific basis is predetermined by convenience,
thus the necessity for the expansion from one basis to the other
arises. An example of such an interbasis transition is the
expansion of a plane wave over spherical in scattering theory. At
present the theory of interbasis expansion forms an independent
direction in the theory of systems with hidden symmetry, and most
of its aspects are well represented in successful sources
\cite{BANDER,BAZ, KMW,KSMOR1,MANKO1,MIL1,POPOV}. Some interconverting transformations are
applied in the three-body problem \cite{FAIFMAN}, low-energy nuclear physics
\cite{SMIRNOV} and in calculating different integrals in quantum molecular
theory \cite{BELL}.

Apparently the first interbasis expansion in the theory of the
hydrogen atom is the Stone's \cite{STONE} result which in the momentum
representation got the expansion of the parabolic basis over the
spherical one. Later, based on work by Fock \cite{FOCK1} and Bargmann \cite{BARGMAN},
Park \cite{PARK} found an expansion of the coordinate parabolic basis of
the hydrogen atom over the spherical one. Then Tarter \cite{TARTER}
repeated the result of Park by a pure analytical approach, i.e.
without group-theoretical methods. Perelomov and Popov \cite{PER-POP2}
obtained an expansion of the Rutherford wavefunction over the
spherical basis and Majumdar and Basu \cite{BESU} found an expansion of
the general, not necessarily scattering, parabolic basis over the
spherical one. Coulson and Robertson \cite{COULSON1} investigated the hydrogen
atom in spheroidal coordinates, while Coulson and Joseph \cite{COJO} led the
problem of the expansion of the spheroidal basis over the
spherical to the solution of the algebraic system of homogeneous
equations. In the work of Johanson and Lippmann \cite{Lippmann} the quantum
analysis of the motion of non-relativistic particle in homogeneous
magnetic field is given and two bases describing the Landau levels
\cite{LANDAU-30,LL}, i.e. the Cartesian and the cylindrical, are found. The
problem of interbasis expansions is well-formulated, rather than
solved in the work \cite{Lippmann}. The interbasis expansions in the
isotropic oscillator for flat space have been discussed in the
works of Pluhar and Tolar \cite{TOLAR}, Chacon and de Llano \cite{CHACON} and
later in the works \cite{MPSTA8,MPSTA9,MPSTA10,POGOSYAN6}. The case of three-dimensional
sphere is considered in details in the work \cite{HPSV1}.

Conventionally three methods for the solution of interbasis
expansions can be singled out: group-theoretical, analytical and
the asymptotic method. The notion of group-theoretical approach is
based on reducing the problem investigated to the problem soluble
in the framework of group theory. Park's work \cite{PARK} (as well as
\cite{GZLa,GZLb,HKPS1,KPS}) can serve as an illustration of this method
where it is proved that the problem of the expansion of the
hydrogen atom parabolic basis over the spherical is equivalent to
the problem of angular momentum addition in quantum mechanics. The
group-theoretical approach is elegant, but its application
presupposes high mastery of the information of the hidden symmetry
and is not supported by common recipe. In other words, every
system requires a specific approach. The analytical approach, the
example of which is Tarter's work \cite{TARTER}, is the direct calculation
of overlap integral between different bases. As a rule, these
integrals are quite complicated and the integration is possible
only after the expansion in series of integrands. The most
difficult part of this programme is the further contraction of the
resulting series into the given expressions for various
polynomials or functions. Actually, the last stage of the
described calculations can be realized in exceptional cases.

The third approach is suggested in our works \cite{T-5,T-1,T-2,T-4,T-3}.
The essence of asymptotic method is the observation that most
bases are a great deal simplified in this or that limit by one or
several variables. Thus, it is convenient to pass to interbasis
expansions in the necessary limit and only afterwards to integrate
by the rest variables. This method simplifies the problem of
integrating to a great extent. The accumulation points are
selected with the aim of preserving the possibility to use any
condition of orthogonality which can express the expansion
coefficient by the overlap integral taken by free variables. The
last condition is the most essential. Usually it is the matter of
tending to infinity (discrete spectrum) or to the zero (continuous
spectrum), i.e. transition to large or little distances from the
source. The method of asymptotic is especially effective for the
investigation of complex systems, such as multi-dimensional
isotropic oscillator.

In a broad sense, the problem of interbasis expansions can be
formulated in two ways. In the first formulation it is supposed
the bases are known (i.e. the appropriate Schr\"{o}dinger equation
has been solved) and the expansion coefficients are sought. The
asymptotic method is applied for this approach. In the second
formulation the apparent type of the expanded basis is unknown,
but given a system of mutually commuting operators, for which this
is an eigenbasis, is given. In this way the solution is sought in
matrix formulation and the result is obtained in the form of
recurrent relations, from which the coefficients of sought
expansions are to be defined. The described method is mostly
effective when the expanded basis is not expressed by
hypergeometrical functions (such as the elliptic basis of
two-dimensional hydrogen atom), and the possible expansion over
the other basis is defined by one summation (i.e. the two bases
differ only by one quantum number). In a more complicated variant,
when the expansion of one basis over the other is possible only in
the form of multiple sums by several quantum numbers and the
corresponding coefficients are not factorized, one deals with
multiple recurrent relations, which lead in their turn to the
characteristic equations for cubic (or of higher dimension)
matrices.

The given monograph is devoted to the solution of the problem of
interbasis relations (and in some cases the bases are sought) in
the following quantum systems with hidden symmetry: two-and
three-dimensional hydrogen atom, circular oscillator,
non-relativistic charged spinless particle, moving in a homogeneous
magnetic field, multidimensional isotropic oscillator,
four-dimensional oscillator and the MIC-Kepler problem. Of special
consideration in the book are also bases obtained by the division
of variables in the coordinates containing a dimensional $R$
parameter (elliptical bases of the two-dimensional hydrogen atom
and circular oscillator, spheroidal bases of hydrogen atom and
isotropic oscillator). In more common double-centered problems the
parameter $R$ has the sense of distance between the centres. In the
limits $R \rightarrow 0$ and $R\rightarrow \infty$ such coordinates are degenerated into more
simple and common. Bases, satisfying the "accordance principle"
and requiring complex bases to transfer into simple at $R \rightarrow 0$ and
$R\rightarrow \infty$, are considered in the book.

\newpage
\chapter{Two dimensional Hydrogen atom}
\markboth{CHAPTER 1. TWO DIMENSIONAL HYDROGEN ATOM}{}
It is customary to call the electron-proton system with the
interaction potential \cite{MAC,SHIBUYA,Zaslow} the two-dimensional
hydrogen atom
\bea U (x,y) = -\frac{\alpha}{\sqrt{x^2+y^2}} \eea
where $\alpha = eZ$. From a physical point of view this type of a
system is most likely idealization reflecting the behavior of a
hydrogen atom in the fields forcing electron and proton to be in the
same plane. It is not improbable that this situation stems from the
separation of variables that is applied when solving any real
problem.

The two-dimensional hydrogen atom is a convenient object for
elucidating a lot of questions arising in investigation of more
complex systems. This concerns, first of all, hidden symmetry
inherent in hydrogen-like atoms but having a geometrically obvious
interpretation only in the case of the two-dimensional hydrogen atom
\cite{SHIBUYA}. Comparing energy spectra and probability
distribution densities of the two- and three-dimensional hydrogen
atom one can easily determine the factor of dimension for
hydrogen-like systems and its effect on physically observable
regularities.

The two-dimensional hydrogen atom, being a system with hidden
symmetry, can be simultaneously studied in several systems of
coordinates, namely, polar \cite{Zaslow}, two mutually perpendicular
parabolic \cite{MAC} and elliptic \cite{T-4,MPS} ones. In the
absence of external fields these bases are mathematically equivalent
though it is obvious that a polar basis is more preferable in
describing the spectroscopy of the two-dimensional hydrogen atom;
whereas the parabolic one, in analyzing an analog of the Rutherford
scattering. The true purpose of the parabolic and elliptic bases
(which are not obvious from a point of view of geometric symmetry of
the Coulomb field) is the description of the behavior of the
two-dimensional hydrogen atom in the homogeneous electric field and
in the field of the other proton, i.e., with the Stark effect (the
variables can be separated in one of the two parabolic systems of
coordinates depending on the direction of the field) and with a
plane problem of two fixed Coulomb centers.

In the framework of perturbation theory with degeneration these
bases acquire the status of correct unperturbed wave functions to be
used for construction of necessary matrix elements, energy
corrections, etc. The above-listed bases exhaust all the bases that
can be obtained within the theory of separation of variables
\cite{MORS-FESHBACH}. Such a limited number of bases makes the
analysis of these bases and establishment of expansions between them
attractive.

The present chapter starts with the analysis of the fundamental
bases of the two-dimensional hydrogen atom, i.e., the bases that are
eigenfunctions of the Hamiltonian and one of the generators of the
hidden symmetry group $SO(3)$. In terms of the procedure of
separation of variables the case in point is the solutions of the
Schr\"odinger equation in the polar and two mutually perpendicular
parabolic systems of coordinates. In the second section, the
relations connecting the fundamental bases of the two-dimensional
hydrogen atom-the parabolic basis with the polar one, are derived.
The third section is devoted to the solution of the Schr\"odinger
equation for the two-dimensional hydrogen atom in the elliptic
coordinates. The fourth section is the calculation of the
coefficients entering into mutual expansions between elliptic,
polar, and parabolic bases with given parity. In the fifth section,
the trinomial recurrence relations for the coefficients connecting
the elliptic basis with the polar and parabolic ones are introduced.
In the sixth section, the obtained recurrence relations are used to
calculate elliptic corrections to the polar and parabolic bases. In
the seventh section the fundamental bases for the two-dimensional
hydrogen atom in the continuous spectrum are constructed. The rest
sections of this chapter are devoted to the problem of interbases
relations in the continuous spectrum: section 8 is the expansion of
the Rutherford function over partial waves, section 9 deals with the
relation between the polar and parabolic bases, section 10 is
devoted to a transition from the region of the continuous spectrum
to that of the discrete one, section 11 deals with the relation
between two parabolic bases, and finally, section 12 is the
expansion of the elliptic basis over the polar one.

\section{Fundamental Bases of the Two-Dimensional Hydrogen Atom}
\markboth{CHAPTER 1. TWO DIMENSIONAL HYDROGEN ATOM}{1.1. FUNDAMENTAL BASES}

\textbf{1.1.1.}     In the polar system of coordinates
\bea x=r\cos\varphi, \qquad y=r\sin\varphi, \qquad r > 0, \,\,\,\,
\varphi \in [0, 2\pi), \eea
the Schr\"odinger equation describing the two-dimensional hydrogen
atom \footnote{In this chapter we everywhere use the Coulomb
system of units $(\hbar=m=e=1)$}
\bea \label{hyd-pol1} \hat H \Psi=
\left\{-\frac{1}{2}\left(\frac{\partial^2}{\partial x^2} +
\frac{\partial^2}{\partial y^2}\right) -
\frac{1}{\sqrt{x^2+y^2}}\right\}\Psi = E\psi \eea
has the form
\bea \label{hyd-pol2} \frac{\partial^2 \Psi}{\partial r^2} +
\frac{1}{r} \frac{\partial \Psi}{\partial r} + \frac{1}{r^2}
\frac{\partial^2 \Psi}{\partial \varphi^2} + 2\left(E +
\frac{1}{r}\right) \Psi =  0. \eea
Equation (\ref{hyd-pol1}) is separated if one chooses the wave
function as a product $\Psi (r,\varphi) =  R(r) \Phi(\varphi)$ where
the angular function is $\Phi(\varphi) = e^{im\varphi}/\sqrt{2\pi}$,
\, $m = 0, \pm 1,\pm 2, ....$ and the radial function $R(r)$ obeys
the equation
\bea \label{hyd-pol3} \frac{d^2 R(r)}{d r^2} + \frac{1}{r} \frac{d
R(r)}{d r} + 2\left(E + \frac{1}{r} - \frac{m^2}{2r^2}\right) R(r) =
0. \eea
In this section, we consider only the states of the discrete
spectrum when $E<0$. Let us pass in (\ref{hyd-pol2}) to new
variables $\rho = 2 \sqrt{-2E} r = 2\omega r$ and make a
substitution $R(r) = \rho^{|m|} e^{-\rho/2} W(\rho)$. As a result,
for the function $W(\rho)$ we arrive at the equation for the
degenerate hypergeometric function \cite{LL}
\bea \label{hyd-pol4} \rho \frac{d^2 W(\rho)}{d \rho^2} +
\biggl[2|m|+1 - \rho\biggr] \frac{d W(\rho)}{d \rho} - \left(|m| +
\frac{1}{2} - \frac{1}{\omega}\right) W(\rho) =  0 \eea
the solution of which determines the discrete spectrum under the
condition $|m|+ 1/2 - 1/\omega = - n_{r}$ with $n_{r}$ being an
integer nonnegative number, i.e.,
\bea \label{hyd-pol5} E_n = -\frac{1}{2(n+\frac12)^2}, \eea
where $n=n_r+|m|$  and, consequently, runs integer nonnegative
values. At given $n$ the quantum numbers $n_r$ and $|m|$ correspond
to different states, i.e., energy levels are degenerate. One can
easily see that in this case the degeneracy multiplicity of the
$n$-th energy level equals $2n+1$. To each $E$ there correspond the
normalized wave functions
\bea \label{hyd-pol6} \Psi_{n m} (r, \varphi) &=&
\frac{\omega^3}{\pi} \, \sqrt{\frac{(n+|m|)!}{(n-|m|)!}} \,
\frac{(2\omega r)^{|m|}}{(2|m|)!} \, e^{- \omega r} \, F(- n+|m|; \,
2|m|+ 1; \, 2\omega r) \, e^{im\varphi}= \nonumber
\\[3mm]
&=& \frac{\omega^3}{\pi} \, \sqrt{\frac{(n-|m|)!}{(n+|m|)!}}(2\omega
r)^{|m|} \, e^{- \omega r} \, L_{n-|m|}^{2|m|} (2\omega r) \,
e^{im\varphi}, \eea
where $L_{n}^{\alpha} (r)$ are the Laguerre polynomials \cite{BE2}
and the quantum number $m$ at given $n$  is limited to values $0,
\pm 1, \pm 2, .... \pm n$. It should be noted that a hypergeometric
series is determined as
\bea \label{hypergeo} F\left(\alpha; \, \gamma; \, x\right) \, =
\sum_{s=0}^{\infty}\,
\frac{\left(\alpha\right)_s\,x^s}{s!\,\left(\gamma\right)_s}, \eea
where $\left(\gamma\right)_n=\Gamma\left(\gamma +n
\right)/\Gamma\left(\gamma\right)$ is the Pochhammer symbol.

\textbf{1.1.2.} In the parabolic coordinates
\bea \label{hyd-pol7} \mu^2 = \sqrt{x^2+y^2} + x, \qquad \nu^2 =
\sqrt{x^2+y^2} - x, \eea
the Schr\"odinger equation (\ref{hyd-pol1}) has the form
\bea \label{hyd-pol9} \left\{ - \frac{1}{2(\mu^2+\nu^2)}
\left(\frac{\partial^2}{\partial \mu^2}+ \frac{\partial^2
\Psi}{\partial \nu^2}\right) - \frac{2}{\mu^2+\nu^2} \right\}\Psi =
E \Psi. \eea
Choosing the wave function $\Psi$ as a product $\Psi(\mu, \nu) =
\psi_1 (\mu) \psi_2 (\nu)$, after separation of variables we arrive
at two ordinary differential equations
\bea \label{hyd-pol10} \frac{1}{2}\frac{d^2 \psi_1}{d \mu^2} +
(\beta_1 + E \mu^2) \psi_1 &=& 0,
\\[2mm]
\label{hyd-pol11} \frac{1}{2}\frac{d^2 \psi_2}{d \nu^2} + (\beta_2 +
E \nu^2) \psi_2 &=& 0 \eea
in which parabolic constants $\beta_1$, $\beta_2$ are related by
$\beta_1 + \beta_2 = 2$.

The separated equations (\ref{hyd-pol10}) and (\ref{hyd-pol11}) are
seen to formally coincide with the Schr\"odinger equation for the
one-dimensional harmonic oscillator where the role of potential
energy is played by the functions $V(\mu) = - E \mu^2$ and $V(\nu) =
- E \nu^2$; and of eigenvalues, by the parabolic separation
constants $\beta_1$ and $\beta_2$. Hence, the repulsive oscillator
corresponds to the continuous spectrum states $E>0$ and $E=0$ is a
free particle motion. At $E<0$ equations (\ref{hyd-pol10}) and
(\ref{hyd-pol11}) describe an ordinary linear oscillator whose
solutions determine the discrete spectrum under the conditions
\bea \label{hyd-pol12} \beta_1 =  \sqrt{-2E}\,  (n_1 + 1/2), \qquad
\beta_2 =  \sqrt{-2E}\, (n_2 + 1/2), \eea
where $n_1$ and $n_2$ are integer nonnegative numbers that are
expressed in terms of the Hermite polynomials multiplied by the
Gaussian exponent
\bea \label{hyd-pol13} \psi_1 (\mu) = C_1 \, e^{ -
\frac{1}{2}\sqrt{-2E}\mu^2} H_{n_1} [(-2E)^{1/4}\, \mu], \quad
\psi_2 (\nu) = C_2 \, e^{ - \frac{1}{2}\sqrt{-2E}\nu^2} H_{n_2}
[(-2E)^{1/4}\, \nu], \eea
It should be noted that the Hermite polynomials are related with
degenerate hypergeometric functions by
\bea
\label{hermit}
H_n (x) = \Biggl\{\begin{array}{l}
\left(-1\right)^{\frac{n}{2}}\,
\frac{n!}{\left(\frac{n}{2}\right)!}\,
F\left(-\frac{n}{2};\,\frac{1}{2};\,x^2\right)\,, \qquad  n \quad {\rm is \,\, even}, \\
[5mm]  \left(-1\right)^{\frac{n-1}{2}}\,
\frac{n!}{\left(\frac{n-1}{2}\right)!}\,2x\,
F\left(-\frac{n-1}{2};\,\frac{3}{2};\,x^2\right)\,, \qquad  n \quad
{\rm is \,\, odd}.
\end{array} \eea
Excluding from relations (\ref{hyd-pol12}) the constants $\beta_1$
and $\beta_2$ we arrive at the following expression for energy
levels:
\bea \label{hyd-pol14} E = -
\frac{1}{2\left(\frac{n_1+n_2}{2}+\frac{1}{2}\right)^2}, \qquad n_1,
n_2 = 0,1,2,..... \eea
which contains extra states of the discrete spectrum in the case of
hydrogen atom and coincides with an analogous formula derived in
separating variables in the polar system of coordinates provided
that $n_1+n_2$ is always even. This fact has a simple explanation.
From the definition of the parabolic system of coordinates
\bea \label{hyd-pol7-0} x =  \frac{1}{2} (\mu^2 - \nu^2), \qquad y =
\mu \nu \eea
one can see that to the pairs $(\mu, \nu)$ and $(- \mu, - \nu)$
there formally corresponds the same point on the plane $(x,y)$ and,
consequently, the wave function $\Psi(\mu, \nu)$ should be even with
respect to the change $(\mu, \nu) \rightarrow (- \mu, - \nu)$. This
means that the functions $\psi_1 (\mu)$ and $\psi_2 (\nu)$ should
have the same parity. Taking into account the properties of the
Hermite polynomials under inversion $x\to -x$: $H_{n} (-x) = (-1)^n
H_{n} (x)$ we get that
\bea \label{hyd-pol15} \Psi_{n_1 n_2} (- \mu, - \nu) = \psi_{n_1} (-
\mu) \psi_{n_2} (-\nu) = (-1)^{n_1+n_2} \psi_{n_1} (\mu) \psi_{n_2}
(\nu) \eea
and, consequently, only the states for which the sum $n_1+n_2$ is
even belong to a set of solutions of the two-dimensional hydrogen
atom in the parabolic system of coordinates. Now if a new quantum
number $n = \frac12(n_1+n_2)$, $n=0,1,2...$ is introduced, then
(\ref{hyd-pol13}) totally coincides with (\ref{hyd-pol5}) derived
earlier for the energy spectrum of the two-dimensional hydrogen atom
and the degeneracy multiplicity, as it should be, equals $2n+1$.

Transformation between the Cartesian and parabolic coordinates
$(\mu, \nu)$ will be unambiguous if one assumes that the latter vary
in the interval $0 \leq \mu < \infty$ and $-\infty < \nu < \infty$.
Thus, these parabolic wave functions $\Psi_{n_1 n_2}$ normalized to
the whole plane $(x, y)$ should satisfy the condition (1.1.20)
\bea \label{hyd-pol16} \int_0^{\infty} \int_{-\infty}^{\infty} \,
\Psi_{n_1 n_2}(\mu, \nu) \, \Psi_{n_1' n_2'}^{*}(\mu, \nu) \,
(\mu^2+\nu^2) d\mu d\nu = \delta_{n_1 n_1'} \, \delta_{n_2 n_2'}
\eea
which, after simple calculations, results in
\bea \label{hyd-pol17} \Psi_{n_1 n_2} (\mu, \nu) =
\left(\frac{\omega^3}{\pi}\right)^\frac{1}{2} \, \frac{e^{ -
\frac{1}{2}\omega(\mu^2+\nu^2)}} {\sqrt{2^{n_1+n_2}(n_1)!(n_2)!}} \,
H_{n_1} (\sqrt{\omega}\, \mu)\, H_{n_2} (\sqrt{\omega}\, \nu), \eea
where $\omega = \sqrt{-2E_n}= 1/(n+1/2)$, and the quantum numbers
$n_1$ and $n_2$ have the same parity.

\vspace{0.5cm} \noindent \textbf{1.1.3.} \, In a similar manner, one can
solve the problem of eigenfunctions and eigenvalues for the
two-dimensional hydrogen atom in the second parabolic system of
coordinates
\bea
\label{hyd-pol7-1}
x =
\bar \mu \bar \nu
\qquad
y =
\frac{1}{2} ({\bar \mu}^2 - {\bar \nu}^2).
\eea
It follows from (\ref{hyd-pol7-1}) that parabolic coordinates of the first and
second type are related by transformation of rotation by an angle
$\frac{\pi}{2}$:
\bea
\label{hyd-pol7-2}
\bar \mu = \frac{\mu+\nu}{\sqrt 2},
\qquad
\bar \nu = \frac{\mu-\nu}{\sqrt 2}\, .
\eea
It is obvious that by virtue of invariance of equation (\ref{hyd-pol9}) with
respect to arbitrary rotations in the plane $(\mu,\nu)$ it follows that the
second parabolic basis can be derived from the first one by a simple change of
variables $\mu \to \bar \mu$, $\nu \to \bar \nu$ and quantum numbers $n_1 \to
\bar n_1$, $n_2 \to \bar n_2$:
\bea
\label{hyd-pol017}
\Psi_{\bar n_1 \bar n_2} (\bar \mu, \bar \nu) =
\left(\frac{\omega^3}{\pi}\right)^\frac{1}{2} \,
\frac{e^{ - \frac{1}{2}\omega(\bar \mu^2+\bar \nu^2)}}
{\sqrt{2^{\bar n_1+\bar n_2}(\bar n_1)!(\bar n_2)!}}
\,
H_{\bar n_1} (\sqrt{\omega}\, \bar \mu)\,
H_{\bar n_2} (\sqrt{\omega}\, \bar \nu)\, ,
\eea
where again the parabolic quantum numbers $\bar n_1$ and $\bar n_2$ have the
same parity.

\vspace{0.5cm} \noindent \textbf{1.1.4.} \, "Accidental degeneracy" in the two-dimensional
hydrogen atom indicates the existence of higher symmetry than purely geometric
symmetry of the Schr\"{o}dinger equation. To determine this symmetry, we use
the method of separation of variables. Excluding from
equations(\ref{hyd-pol10}) and (\ref{hyd-pol11}) the energy parameter
$\omega^2$ and passing instead of $\beta_1$ and $\beta_2$ to the new separation
constant
\bea
\label{symdym1-01}
\beta = \frac{1}{2}(\beta_1- \beta_2) = \sqrt{-\frac {E}{2}} \,
(n_1-n_2)
=
\sqrt{-2E} \, p, \qquad\qquad
-n \leq p \leq n\, ,
\eea
we arrive at the equation ${\cal P} \Psi_{np} (\mu, \nu) = p \,
\Psi_{np} (\mu, \nu)$ where the operator ${\cal P}$ is determined by
expression
\bea
\label{symdym1-1}
2 \sqrt{-2E} \, {\cal P} = \frac{1}{\mu^2+\nu^2}
\left[\mu^2\frac{\partial^2}{\partial \nu^2}-
\nu^2\frac{\partial^2}{\partial \mu^2} +
2(\mu^2-\nu^2)\right]\,.
\eea
Now the parabolic wave function is determined by a different pair of numbers
$n= (n_1+n_2)/2$ and $p= (n_1-n_2)/2$ instead of $(n_1, n_2)$. Proceeding in
this way one can show that the second parabolic basis satisfies the equation
${\cal K} \Psi_{nk} (\bar \mu, \bar \nu) = k \, \Psi_{nk} (\bar \mu, \bar \nu)$
where the operator ${\cal K}$ is obtained from(\ref{symdym1-1}) by the change
$(\mu, \nu)\to (\bar \mu, \bar \nu)$, and the quantum number $k = (\bar n_1 -
\bar n_2)/2$. It is finally obvious that the polar basis satisfies the equation
$L \Psi_{nm}(r,\varphi) = m \Psi_{nm}(r,\varphi)$ where the operator $L$
coincides with the $l_z$ component of the angular momentum operator and equals
$L=-i\partial/\partial \varphi$. Rewriting ${\cal P}$ and  ${\cal K}$ in the
Cartesian system of coordinates
\bea
\label{symdym1-2}
{\cal P}
&=&
\frac{1}{\sqrt{-2E}}
\left(\frac{x}{\sqrt{x^2+y^2}}
+ x \frac{\partial^2}{\partial y^2}
-
y \frac{\partial}{\partial x} \frac{\partial}{\partial y}
- \frac{1}{2} \frac{\partial}{\partial x}\right)\,,
\\[2mm]
\label{symdym1-3}
{\cal K}
&=&
\frac{1}{\sqrt{-2E}}
\left(\frac{y}{\sqrt{x^2+y^2}}
+ y \frac{\partial^2}{\partial x^2}
-
x \frac{\partial}{\partial y} \frac{\partial}{\partial x}
- \frac{1}{2} \frac{\partial}{\partial y}\right)\,,
\eea
we can see that these operators are two-dimensional analogs of the Runge-Lentz
vector (see Chapter 3) and together with the operator
\bea
\label{symdym1-4}
L = i\left(y\frac{\partial}{\partial x}-
 x \frac{\partial}{\partial y}\right)\, ,
\eea
determine the Fock symmetry of the two-dimensional hydrogen atom. It can
immediately be verified that the operators ${\cal P}$, ${\cal K}$ and $L$ (on
the wave functions of the two-dimensional hydrogen atom with fixed energy
$E<0$) satisfy the commutation relations
\bea
\label{symdym1-5}
\{{\cal P}, {\cal K} \} = i L,
\qquad
\{L, {\cal P} \} = i {\cal K},
\qquad
\{{\cal K}, L \} = i {\cal P}\, ,
\eea
and generate a linear algebra isomorphic to the algebra $so(3)$. The latter is
called the invariance algebra for the two-dimensional hydrogen atom, as each of
the operators ${\cal P}$, ${\cal K}$ and $L$ commutes with the Hamiltonian
$\cal H$. As is well known, from the algebra $so(3)$ one can pass to the group
$SO(3)$ as well as to the group $SU(2)$. On the other hand, degeneracy
multiplicity of the $n$th energy level equals $2n+1$, $n=0,1,2..$ and coincides
with the dimension of representations of the group $SO(3)$ \footnote{A detailed
description of all the problems having to do to the two-dimensional hydrogen
atom symmetry can be found in \cite{MAC} and a brilliant book by Englefield
\cite{ENGLEF}.}. Thus, accidental degeneracy of the two-dimensional Coulomb problem
can also be formulated in terms of the symmetry group $SO(3)$ also called the
group of dynamic or hidden symmetry. The knowledge of the group of dynamic
symmetry gives a possibility to reproduce in a purely algebraic way the formula
for the energy spectrum of the two-dimensional hydrogen atom. Indeed, let us
construct the operator ${\cal L}^2 = {\cal P}^2 + {\cal K}^2 + L^2$ that (being
a scalar) is expressed in terms of the Hamiltonian
\bea
\label{symdym1-6}
{\cal L}^2 = - \frac{1}{2E}
\left(1+\frac{1}{2}\cal H \right)\,,
\eea
and, on the other hand, being the Casimir operator for the group $SO(3)$, it
satisfies the equation ${\cal L}^2 \Psi_{nm} = n(n+1)\Psi$. Hence, elementary
calculations lead to the sought formula (\ref{hyd-pol5}).

Thus, within the method of separation of variables, eigenvalues for the
generators of the group of hidden symmetry acquire the meaning of separation
constants and eigenfunctions, which are common for the Hamiltonian and each of
the generators, are solutions or as we call them {\it fundamental bases} of the
Schr\"{o}dinger equation in different systems of coordinates. Information of
the fundamental bases and the group of hidden symmetry is collected in Table
\ref{t11}


\begin{table}[t]
\caption{Fundamental bases and generators of the group of hidden
symmetry $SO(3)$}\label{t11}
\noindent
\begin{tabular}{|l|l|l|l|}
\hline
&&&\\
$x{=}r\cos\varphi$&$0{\leq} r {<}\infty$&${\cal H}\Psi_{nm} {=} E_n\Psi_{nm}$&
$\Psi_{nm} = \left(\frac{\omega^3}{\pi}\right)^\frac{1}{2}\,
\frac{(2\omega r)^{|m|}\,e^{-\omega r}\, e^{im\varphi}}
{\sqrt{(n-m)![(n+m)!]^{-1}}}\,
L_{n-|m|}^{2|m|}(2\omega r)$\\
&&&\\
$y{=}r\sin\varphi$& $0{\leq} \varphi {<} 2\pi$
&$L\Psi_{nm} {=} m \Psi_{nm}$&
$L= -i \frac{\partial}{\partial\varphi}; \,\,
-n\leq m\leq n, \,\, \omega=\sqrt{-2E_n}$\\
&&&\\
\hline
&&&\\
$x{=}\frac{\mu^2{-}\nu^2}{2}$&$0{\leq} \mu {<} \infty$&
${\cal H}\Psi_{np} {=} E_n \Psi_{np}$&
$\Psi_{np} = \left(\frac{\omega^3}{\pi}\right)^\frac{1}{2}
\frac{H_{n+p}(\sqrt\omega \mu)
H_{n-p}(\sqrt\omega \nu)}{\sqrt{2^{2n}(n+p)!(n-p)!}}
e^{-\frac{\omega}{2}(\mu^2+\nu^2)}$\\
&&&\\
$y{=} \mu \nu$ & ${-}\infty{<}\nu{<}\infty$& ${\cal P} \Psi_{np} {=} p \Psi_{np}$&
${\cal P} = \frac{1}{2\omega}\frac{1}{\mu^2+\nu^2}
\left[\mu^2\frac{\partial^2}{\partial \nu^2}-
\nu^2\frac{\partial^2}{\partial \mu^2}+2(\mu^2-\nu^2)\right]$\\
&&&\\
\hline
&&&\\
$x{=}\bar \mu \bar \nu$&  $0{\leq} \bar \mu{<}\infty$& ${\cal H} \Psi_{nk} {=} E_n \Psi_{nk}$
& $\Psi_{nk}$ and ${\cal K}$ are obtained from $\Psi_{np}$ and ${\cal P}$\\
&&
& by the change $\mu \rightarrow \bar\mu$,\, $\nu \rightarrow \bar\nu$, \, $p\rightarrow k$,
\\
$y{=}\frac{{\bar \mu}^2{-}{\bar \nu}^2}{2}$& ${-}\infty{<}\bar v{<}\infty$& ${\cal K}
\Psi_{nk} {=} k \Psi_{nk}$
&   $p$ and $k \in (-n,n)$
\\
&&&\\
\hline
\end{tabular}
\end{table}

\section{Relation between the fundamental bases of the two-dimensional hydrogen atom}
\markboth{CHAPTER 1. TWO DIMENSIONAL HYDROGEN ATOM}{1.2. RELATION BETWEEN THE FUNDAMENTAL BASES}

The above-obtained fundamental bases are alternative descriptions of the
two-dimensional hydrogen atom, and an arbitrary state with a given energy can
be represented as expansion over any of these bases.In a particular case, the
expanded state can also be fundamental. In this section, we deal with
expansions of this particular type. They can be assigned a geometric meaning by
taking in the space of the group of hidden symmetry $SO(3)$ the Cartesian
system of coordinates each axis of which is put in correspondence with its own
fundamental basis (see Fig. \ref{f11}). It follows from the commutation relations
between $L, {\cal P}, {\cal K}$ and the meaning of the fundamental basis that
expansions between these bases may be interpreted as rotations by a right angle
in the relevant coordinate planes. As rotations are plane, the expansion
coefficients should coincide up to phase multipliers with the Wigner $d$
-function of the right angle. Unfortunately, the form of phase multipliers of
$d$-functions cannot be obtained from simple geometric reasonings. Before
proceeding to a direct calculation justifying the above-described picture let
us tabulate the final results in Table \ref{t12}.

\begin{figure}[t]
\unitlength=1mm
\special{em:linewidth 0.4pt}
\linethickness{0.4pt}
\begin{picture}(115.00,84.00)
\put(58.00,30.00){\vector(0,1){54.00}}
\put(58.00,30.00){\vector(1,0){53.00}}
\put(58.00,30.00){\vector(-1,-1){23.00}}
\put(35.00, 2.00){\makebox(0,0)[cc]{${\cal P}, \Psi_{np}$}}
\put(67.00,81.00){\makebox(0,0)[cc]{$L,\Psi_{nm}$}}
\put(115.00,22.00){\makebox(0,0)[cc]{${\cal K},\Psi_{nk}$}}
\end{picture}
\caption{}
\label{f11}
\end{figure}

 The first column and the upper row of this table
represent the fundamental bases of the two-dimensional hydrogen atom and the
remaining cells represent the coefficients of interbasis expansions. The case
in point is the expansions of the bases of the left column over the bases of
the upper row. These expansions imply the summation over quantum numbers
$k,p,m$, respectively.

\begin{table}[t]
\caption{Coefficients of the fundamental interbasis expansions}\label{t12}
\begin{center}
\begin{tabular}{|c|c|c|c|}\hline
&&&\cr
& $\Psi_{nm}(r,\varphi)$ & $\Psi_{np}(\mu,\nu)$
& $\Psi_{nk}(\bar \mu, \bar \mu)$
\cr
&&&\cr
\hline
&&&\cr
$\Psi_{nm}(r,\varphi)$ & 1 & $(i)^{n-p}d_{pm}^n\left(\frac{\pi}{2}\right)$&
$(i)^{n+k+m}d_{km}^n \left(\frac{\pi}{2}\right)$
\cr
&&&\cr
\hline
&&&\cr
$\Psi_{np}(\mu,\nu)$ & $(-i)^{n-p}d_{pm}^n \left(\frac{\pi}{2}\right)$
& 1 & $d_{kp}^n
\left(\frac{\pi}{2}\right)$
\cr
&&&\cr
\hline
&&&\cr
$\Psi_{nk}(\bar \mu, \bar \nu)$ & $(-i)^{n+k+m}d_{km}^n
\left(\frac{\pi}{2}\right)$ & $d_{kp}^n\left(\frac{\pi}{2}\right)$ & 1
\cr
&&&\cr
\hline
\end{tabular}
\end{center}
\end{table}

Let us start with the expansion of the first parabolic basis
(\ref{hyd-pol17}) over the polar one (\ref{hyd-pol6}):
\bea
\label{inter1.1}
\Psi_{np} (\mu, \nu) = \sum_{m=-n}^n
\,
V_{n pm} \,
\Psi_{nm}(r,\varphi)\, .
\eea
One can easily be convinced that the overlapping integral between these bases
is very complicated to calculate. The following method is more effective ({\it
the asymptotic method}). Let us pass in equation (\ref{inter1.1}) to the limit
of large $r$ (respectively, $\mu$ and $\nu$) and use the asymptotic form of the
degenerate hypergeometric function
\bea
F (-N; \alpha; x) \stackrel{x \to \infty}{\longrightarrow}
(-1)^N \,
\frac{\Gamma(\alpha)}{\Gamma(\alpha+N)}\, x^N.
\eea
In this limit the bases are considerably simplified and contain $r$ in the same
power. Therefore, dependences on $r$ are reduced in both the parts of the
expansion (\ref{inter1.1}) and the orthonormalization condition of the
functions $e^{im\varphi}$ leads to the result
\bea
\label{inter1.2}
V_{n p m} \,
=
(-1)^{n-|m|}  \,
\frac{2^n}{\pi}
\sqrt{\frac{(n+|m|)!(n-|m|)!}{(n+p)!(n-p)!}}
\int_0^{\pi}
(\cos\varphi)^{n+p}\,
(\sin\varphi)^{n-p}\,
e^{-2im \varphi}\, d \varphi \,.
\eea
Calculation of the integral over $d\varphi$ is elementary. Expanding
$(\cos\varphi)^{n+p}$ by the binomial formula
\bea
(\cos\varphi)^{n+p} = \frac{1}{2^{n+p}} \,
\sum^{n+p}_{s=0}\, e^{i\varphi(n+p-2s)} \,
\left(\matrix{n+p
\cr
s
\cr}\right)\,,
\eea
where the brackets denote the binomial coefficients, and taking into account
that according to \cite{BE1} at $n
> - 1$
\bea
\int\limits_0^{\pi}
(\sin\varphi)^{n}\, e^{iN \varphi}\, d \varphi
=
\frac{\pi}{2^n} \,
\frac{e^{i\frac{\pi}{2}N} \,\Gamma(n+1)}
{\Gamma\left(\frac{n+N}{2}+1\right)
\Gamma\left(\frac{n-N}{2}+1\right)}\, ,
\eea
we arrive at the result
\bea
\label{inter1.3}
\frac{(-1)^m}{\pi}
\int\limits_0^{\pi}
(\cos\varphi)^{n+p}\,
(\sin\varphi)^{n-p}\,
e^{- 2im \varphi}\, d \varphi
=
\frac{i^{n+p}}{2^{2n}}
\sum_{s=0}^{n+p}\,
(-1)^{s}
\left(\matrix{n+p
\cr
s
\cr}\right)
\left(\matrix{
n-p
\cr
m-p+s
\cr}\right)\,.
\eea
Using the representation for a small Wigner $d$-function of the argument
$\pi/2$ \cite{VAR}, known from the angular momentum theory,
\bea
\label{inter1.4}
d_{M,M'}^J\left(\frac{\pi}{2}\right)
=\frac{(-1)^{M-M'}}{2^J}
\sqrt{\frac{(J+M)!(J-M)!}{(J+M')!(J-M')!}}
\sum(-1)^k
\left(\matrix{
J+M'\,\,
\cr
k
\,\,
\cr}\right)
\left(\matrix{
J-M'\,\,
\cr
k+M-M'
\,\,
\cr}\right)
\nonumber
\eea
and comparing it with the right-hand side of (\ref{inter1.3}) we find the
coefficients $V_{n p}^{nm}$. Finally, taking into account the symmetry relation
$d_{M,M'}^J (\beta) = (-1)^{M-M'}d_{M',M}^J(\beta)$ we arrive at the conclusion
that expansion (\ref{inter1.1}) has the form
\bea
\label{2.2}
\Psi_{n p}(\mu, \nu)=
(-i)^{n-p}
\sum_{m = -n}^n
d_{p,m}^n
\left(\frac{\pi}{2}\right)
\Psi_{n m}(r, \varphi).
\eea
The Wigner functions $d_{pm}^n \left(\frac{\pi}{2}\right)$ are tabulated
completely and, therefore, there is no problem in using expansion
(\ref{inter1.1}) at certain values of the quantum numbers $n$ and $p$.

Let us now pass to the expansion of the second parabolic basis over the first
one
\bea
\Psi_{n k}(\bar \mu, \bar \nu)=
\sum_{p=-n}^n \,
T_{nkp} \,
\Psi_{n p}(\mu, \nu).
\eea
The coefficients of this expansion can be calculated in a similar way. However,
the following approach is obviously more appropriate here.In the fourth
chapter, it will be shown that under the rotation of the system of coordinates
by an arbitrary angle $\alpha$ the product of Hermite polynomials is
transformed according to the rule
\bea
\frac{H_{n+k}(x\cos\alpha-y\sin\alpha)H_{n-k}(x\sin\alpha
+y\cos\alpha)}{\sqrt{(n+k)!(n-k)!}}=
\sum_{p=-n}^n d_{k,p}^n (2\alpha)
\frac{H_{n+p}(x)H_{n-p}(y)}{\sqrt{(n+p)!(n-p)!}}.
\eea
Using this rule we immediately have
\bea
\label{2.3}
\Psi_{n k}(\bar \mu, \bar \nu) = \sum_{p=-n}^n
d_{k,p}^n
\left(\frac{\pi}{2}\right)
\Psi_{n p}(\mu, \nu).
\eea
Expansion of the second parabolic basis over r the polar one can be derived
with the help of the addition theorem \cite{VAR}
\bea
\sum_{M''=-J}^J d_{M,M''}^J
\left(\frac{\pi}{2}\right) d_{M'',M'}^J
\left(\frac{\pi}{2}\right)
e^{-i\frac{\pi}{2}M''}=
(-i)^{M+M'}
d_{M,M'}^J\left(\frac{\pi}{2}\right),
\eea
and it has the form
\bea
\label{2.4}
\Psi_{n k}(\bar \mu, \bar \nu)=
(-i)^{n+k}
\sum_{m=-n}^n (-i)^m
d_{k, m}^n \left(\frac{\pi}{2}\right)
\Psi_{n m}(r, \varphi).
\eea
So the geometric approach really justifies a simple geometric picture, we have
started out presentation. The expansion coefficients inverse to expansions
(\ref{2.2})-(\ref{2.3}) coincide with those shown in Table \ref{t12}. This fact is a
consequence of orthonormalization of the expansion coefficients.

\section{Explicitly factorized elliptic basis of the two-dimen\-si\-onal hydrogen atom}
\markboth{CHAPTER 1. TWO DIMENSIONAL HYDROGEN ATOM}{1.3. EXPLICITLY FACTORIZED ELLIPTIC BASIS }

The position of an arbitrary point ($x, y$) on the upper or lower half-plane
can be described by defining two quantities $r_1$ and $r_2$ determining its
distance from the points $(0,0)$ è $(R,0)$ (see Fig. \ref{f12}). An approach like this
is convenient in two-center problems when at the points $(0,0)$ and $(R,0)$
there are sources creating a field in which a motion of a particle with the
coordinates($x, y$)is studied.

Elliptic coordinates as $R \to 0$ and $R \to \infty$ turn into polar and
parabolic ones if the positions of the Coulomb center and a charged particle in
the limiting transition process remain fixed. Hence,the "principle of
correspondence" follows which states that all formulae obtained in the elliptic
coordinates should turn into the corresponding polar and parabolic analogs in
the given limits.Though this principle is natural,the realization of the
limiting transition seems to be nontrivial as, on the one hand, the elliptic
Coulomb functions are expressed in terms of the quantities obeying three-term
recurrence relations and, on the other hand, the polar and parabolic wave
functions are constructed on the basis of two-term recurrence relation.
Therefore, before proceeding to limiting transitions in the wave functions,
matrix elements, etc. one should find out the way three-term recurrence
relations turn into two-term ones within these limits.

\begin{figure}[t]
\begin{center}
\unitlength=1.00mm
\special{em:linewidth 0.4pt}
\linethickness{0.4pt}
\begin{picture}(103.00,65.00)
\put(15.00,8.00){\line(3,2){51.00}}
\put(66.00,42.00){\line(1,-2){16.99}}
\put(66.00,42.00){\circle{2.00}}
\put(66.00,49.00){\makebox(0,0)[cc]{$(x,y)$}}
\put(35.00,31.00){\makebox(0,0)[cc]{$r_1$}}
\put(79.00,32.00){\makebox(0,0)[cc]{$r_2$}}
\put(19.00,5.00){\makebox(0,0)[cc]{$0$}}
\put(83.00,4.00){\makebox(0,0)[cc]{$R$}}
\put(15.00,3.00){\vector(0,1){62.00}}
\put(21.00,61.00){\makebox(0,0)[cc]{$Y$}}
\put(10.00,8.00){\vector(1,0){93.00}}
\put(99.00,14.00){\makebox(0,0)[cc]{$X$}}
\end{picture}
\end{center}
\caption{}
\label{f12}
\end{figure}

\noindent The elliptic coordinates $\xi$ and $\eta$ are determined
in terms of $r_1$ and $r_2$ as follows:
\bea
\label{EL-COOR-1}
\cosh \xi = (r_1+r_2)/R,
\qquad
\cos \eta = (r_1-r_2)/R,
\eea
and change in the limits $0 \leq \xi < \infty$, $0 \leq \eta < 2\pi$. The
elliptic coordinates ($\xi, \eta$) and ($\xi, 2\pi-\eta$)correspond to the
points symmetric with respect to the axis of abscissas. The lines $\xi=$const
and  $\eta=$const make the sum and difference of distances $r_1$ and $r_2$
constant and, therefore, are confocal ellipses and parabolas with the focuses
at the points $(0,0)$ and $(R,0)$. The elliptic coordinates are orthogonal. The
Cartesian coordinates are expressed in terms of  $\xi$ and $\eta$ as follows:
\bea
\label{EL-COOR-2}
x = \frac{R}{2}(\cosh \xi \cos\eta +1),
\qquad
y = \frac{R}{2} \sinh \xi \sin \eta.
\eea
The Laplacian and the two-dimensional volume element in the elliptic
coordinates have the form
\bea
\label{EL-COOR-3}
\Delta &=& \frac{\partial^2}{\partial x^2} +
\frac{\partial^2}{\partial y^2}
= \frac{4}{R^2(\cosh^2\xi - \cos^2\eta)}
\left(\frac{\partial^2}{\partial \xi^2} +
\frac{\partial^2}{\partial \eta^2}\right),
\\[2mm]
d V &=& dx dy = \frac{R^2}{4} (\cosh^2\xi - \cos^2\eta) d\xi d\eta.
\label{EL-COOR-0-3}
\eea
It follows from the definition of the elliptic system of coordinates
(\ref{EL-COOR-2}) that as $R \to 0$ and $R \to \infty$
\bea
\begin{array}{ll}
\label{EL-COOR-4}
\cosh\xi \stackrel{R\to 0}{\longrightarrow} \frac{2r}{R},
&
\cos\eta \stackrel{R\to 0}{\longrightarrow} \cos\phi,
\\[3mm]
\cosh\xi \stackrel{R\to \infty}{\longrightarrow} 1 + \frac{\nu^2}{R},
&
\cos\eta \stackrel{R\to \infty}{\longrightarrow} -1 + \frac{\mu^2}{R}.
\label{EL-COOR-5}
\end{array}
\eea
Limiting transitions in(\ref{EL-COOR-4}) and (\ref{EL-COOR-5}) are carried out
at a fixed position of the point ($x,y$) and a fixed origin of the coordinates.
From the elliptic system of coordinates one can also get the Cartesian one if
the origin of the coordinates is transferred to the point $(R/2,0)$ and then
the left and right focuses are moved away to $-\infty$ and $+\infty$,
respectively (see Chapter 3).

\vspace{0.5cm} \subsection {Separation of variables} \vspace{0.2cm}

In the elliptic coordinates the two-dimensional hydrogen atom is described by
the Schr\"{o}dinger equation
\bea
\label{EL-COOR-6}
\frac{\partial^2\Psi}{\partial \xi^2} +
\frac{\partial^2\Psi}{\partial \eta^2} -
\left[\frac{\omega^2 R^2}{4}(\cosh^2\xi - \cos^2\eta)
+ R (\cosh\xi - \cos\eta)\right]\Psi = 0.
\eea
Here $\omega = \sqrt{-2E}$ and the Coulomb system of units is accepted in which
$m=\hbar=e=1$. One can see from the structure of equation(\ref{EL-COOR-6}) that
the variables in it are separated . The substitution

\bea
\label{EL-COOR-7}
\Psi (\xi, \eta) = X(\xi) Y(\eta)
\eea
transforms the equation in partial derivatives (\ref{EL-COOR-6}) into two
ordinary differential equations
\bea
\label{EL-COOR-8}
\frac{d^2 Y(\eta)}{d\eta^2} -
\left(R\cos\eta + \frac{\omega^2 R^2}{4}\cos^2\eta
\right) Y(\eta) = + Q Y(\eta),
\\[2mm]
\label{EL-COOR-9}
\frac{d^2 X(\xi)}{d\xi^2} +
\left(R\cosh\xi + \frac{\omega^2 R^2}{4}\cosh^2\xi
\right) X(\xi) = - Q X(\xi)
\eea
in which $Q$ is the separation constant in the elliptic coordinates. These
equations can be written down in a unified form
\bea
\label{EL-COOR-10}
\frac{d^2 Z(\zeta)}{d\zeta^2} -
\left(R\cos\zeta + \frac{\omega^2 R^2}{4}\cos^2\zeta
\right) Z(\zeta) = + \lambda Z(\zeta)
\eea
where $\zeta$ is complex. One can easily verify that at $\zeta = \eta \in [0,
2\pi]$ and $\zeta = \xi \in [0, \infty)$ equation (\ref{EL-COOR-10}) results in
(\ref{EL-COOR-8}) and (\ref{EL-COOR-9}), respectively .  Thus, in the complex
plane $\zeta$ the domain containing the half-plane ${\cal R}e \zeta = 0$,
${\cal I}m\zeta \geq 0$ and the interval ${\cal I}m\zeta = 0$, $0 \leq {\cal
R}e \zeta < 2\pi$ is physical. The functions $X(\xi)$ and $Y(\eta)$ are
expressed in terms of $Z(\zeta)$ as follows:
$$
X(\xi) = Z(i\xi), \qquad Y(\eta) = Z(\eta).
$$
The invariance of the Cartesian coordinates $x$ and $y$ with respect to each of
the transformations
$$
\xi \rightarrow \xi + 2\pi i, \qquad  \eta \rightarrow \eta + 2\pi
$$
results in the unambiguity condition
$$
X(\xi) = X(\xi+2i\pi), \qquad Y(\eta)=Y(\eta+2\pi),
$$
that is equal to the complex periodicity condition
\bea
\label{EL-COOR-11}
Z(\zeta) = Z(\zeta+2\pi).
\eea
In what follows we will be interested in solutions of  (\ref{EL-COOR-10}) that
are finite at all the points of the physical domain and satisfy the periodicity
condition (\ref{EL-COOR-11}).

\vspace{0.5cm} \subsection {Coulomb elliptic wave functions}
\vspace{0.2cm}

Equation (\ref{EL-COOR-10}) belongs to the class of ordinary differential
equations with periodic coefficients. After the substitution
\bea
\label{EL-COOR-12}
Z(\zeta) = e^{-\frac{\omega R}{2}\cos\zeta} W(\zeta)
\eea
and a subsequent change $\zeta = 2u$ it leads to the Ince equation \cite{ARSS}
\bea
\label{EL-COOR-13}
\frac{d^2 W}{d u^2} +  \alpha \sin 2u \frac{d W}{d u} +
(g -  p \alpha \cos 2u) W = 0
\eea
in which the parameters $\alpha$, $g$ and $p$ equal
$$
\alpha = 2\omega R,
\qquad
g = \omega^2 R^2 - 4Q,
\qquad
p = -1 + 2/\omega.
$$
The most natural way of solving equations of the type(\ref{EL-COOR-13}) is the
expansion of the function $W(u)$ in the Fourier series. However, with an
approach like this there is no appropriate criterion for investigation of
convergence of series obtained. In this connection let us make one more change
$t = \cos u$ and transform equation (\ref{EL-COOR-13}) to the algebraic form
\bea
\label{EL-COOR-14}
(1-t^2) \frac{d^2 W}{d t^2} -  t(2\alpha+1-2\alpha t^2)
\frac{d W}{d t} + (g + p\alpha -2 p \alpha t^2) W = 0.
\eea
Equation (\ref{EL-COOR-14}) has two singularities $t= \pm 1$ at finite $t$ and,
therefore, its solution has to be sought in the form
\bea
\label{EL-COOR-014}
W (t) =  (1-t^2)^{\gamma}\, F(t).
\eea
Studying the behavior of solution (\ref{EL-COOR-014}) near $t^2=1$ we obtain
for $\gamma$ two possible values - $0$ and $1/2$. Thus, there are two types of
solutions
$$
W^{(+)} (\xi) =
\sum_{s=0}^{\infty} a_{s} \left(\cos \frac{\xi}{2}\right)^{s},
\qquad
W^{(-)}(\xi) =
\sin\frac{\xi}{2} \, \sum_{s=0}^{\infty} b_{s} \left(\cos\frac{\xi}{2}\right)^{s}.
$$
The first solution is even and the second is odd in $\xi$. The periodicity
condition (\ref{EL-COOR-11}) requires that summation over $s$ in $W^{(+)}
(\xi)$ and $W^{(-)} (\xi)$ should be carried out over even or odd values,
respectively:
\bea
\label{EL-COOR-15}
W^{(+)}(\xi)
&=& \sum_{s=0}^{\infty} a_{2s} \left(\cos\frac{\xi}{2}\right)^{2s},
\\[2mm]
\label{EL-COOR-16}
W^{(-)}(\xi)
&=& \sin\frac{\xi}{2}  \,
\sum_{s=0}^{\infty} b_{2s} \left(\cos\frac{\xi}{2}\right)^{2s+1}.
\eea
Substitution of series (\ref{EL-COOR-15}) and (\ref{EL-COOR-16}) in equation
(\ref{EL-COOR-14}) leads to the recurrence relations
\bea
\label{EL-COOR-19}
(s+1/2)(s+1)\, a_{2s+2} &+& \left[\frac{g +  \alpha p}{4}
- s(s+\alpha)\right]\, a_{2s} +
\alpha\left(s - 1 - \frac{p}{2}\right)\, a_{2s-2} = 0,
\\[3mm]
\label{EL-COOR-20}
(s+3/2)(s+1)\, b_{2s+3} &+& \left[\frac{g +  \alpha p}{4}
- s(s+\alpha) - \left(2s+1+\frac{\alpha}{2}\right)\right]\, b_{2s+1}+
\nonumber
\\[3mm]
&+&
\alpha\left(s - \frac{p}{2}\right)\, b_{2s-1} = 0.
\eea

Let us study the convergence of series (\ref{EL-COOR-15}) and
(\ref{EL-COOR-16})by the method suggested in\cite{COULSON1}. Let at large
values of $s$
\bea
\label{EL-COOR-21}
\frac{a_{2s+2}}{a_{2s}} \sim c_0 + \frac{c_1}{s} + ....
\eea
Then at the same values of $s$
\bea
\label{EL-COOR-22}
\frac{a_{2s}}{a_{2s-2}} \sim c_0 + \frac{c_1}{s-1} + ....
= c_0 + \frac{c_1}{s} + O(\frac{1}{s^2}).
\eea
Let us divide the recurrence relation (\ref{EL-COOR-19}) into $a_{2s-2}$ and
use (\ref{EL-COOR-21}) and (\ref{EL-COOR-22}). Restricting ourselves to the
first two principal terms with respect to $s$ we derive the equations
\bea
\label{EL-COOR-23}
c_0 (c_0 -1) = 0,
\qquad
c_0 (2c_1 + \frac{3}{2}c_0) - (c_1+ \alpha c_0) +\alpha = 0
\eea
the roots of which are
\bea
\label{EL-COOR-24}
a) \,\,\,\,\, c_0 = 0, \quad  c_1 = \alpha,
\qquad\qquad
b) \,\,\,\,\, c_0 = 1, \quad  c_1 = - 3/2.
\eea

In the case "a"
\bea
\label{EL-COOR-25}
Z(\zeta) \sim e^{\frac{\omega R}{2}\cos\zeta}
\eea
and as  $\zeta \to i\infty$ this solution will not be finite.

In the case "b"
\bea
\label{EL-COOR-26}
Z(\zeta) > \exp\{1/2 +1/2 (1 - \omega R)\cos\zeta\}
\eea
and, consequently, this solution as $\zeta \to i\infty$ does not tend to zero
at all $\omega R$. Analogous analysis of the recurrence relation
(\ref{EL-COOR-20}) also leads to two cases: a) $c_0=0, c_1=\alpha$; \, b)
$c_0=1, c_1 =-1/2$, of which the latter is the new one. In this case
\bea
\label{EL-COOR-27}
W (\zeta) \sim \sum_{s} \frac{\left(\cos\frac{\zeta}{2}\right)^s}
{s^{\frac12}}
\eea
i.e., $W(\zeta)$ diverges at $\zeta=0$ and $\zeta = 2\pi$. It follows from the
above analysis that series (\ref{EL-COOR-15}) and (\ref{EL-COOR-16}) will be
finite in the whole physical domain only if they cut off.

Let $a_{2n}$ and $a_{2n'+1}$ be the last nonzero coefficients in
(\ref{EL-COOR-15}) and (\ref{EL-COOR-16}). Then it follows from the recurrence
formulae (\ref{EL-COOR-19}) and (\ref{EL-COOR-20}),if one puts in them $s=n+1$
and $s=n'+1$, respectively, that $n=n'$ and the energy spectrum has the form
$$
E_n = - \frac{1}{2(n+1/2)^2},
\qquad n=0,1,2,...
$$
i.e., coincides with the well-known result.

Let us write down the recurrence relations(\ref{EL-COOR-19}) and
(\ref{EL-COOR-20}) so that the cutoff conditions of series (\ref{EL-COOR-15})
and (\ref{EL-COOR-16}) should be fulfilled automatically:
\bea
\label{EL-COOR-29}
\alpha_{2s} \, a_{2s+2} +
\beta_{2s} \, a_{2s} +
R \gamma_{2s} \, a_{2s-2} = 0,
\eea
\bea
\label{EL-COOR-30}
\alpha_{2s} = (s+1/2)(s+1),
\quad
\gamma_{2s} = 2 \omega (s-n-1),
\nonumber
\eea
\bea
\label{EL-COOR-31}
\beta_{2s} = -s^2 +  \frac{2(n-2s)}{2n+1} R - \lambda(R),
\eea
\bea
\label{EL-COOR-32}
\alpha_{2s+1} \, b_{2s+3} +
\beta_{2s+1} \, b_{2s+1} +
R \gamma_{2s+1} \, b_{2s-1} = 0,
\eea
\bea
\alpha_{2s+1} = (s+3/2)(s+1),
\quad
\gamma_{2s+1} = 2 \omega (s-n),
\nonumber
\eea
\bea
\label{EL-COOR-33}
\beta_{2s+1} = - (s+1)^2 +  \frac{2(n-1-2s)}{2n+1} R - \lambda(R),
\eea
where the following notation is introduced:
\bea
\label{EL-COOR-029}
\lambda(R) = Q(R) - \frac{\omega^2 R^2}{4}.
\eea
The condition $a_0=a_1=1$, $a_{-1} = a_{-2}=0$ should be added to the above
formulae. Each of the recurrence relations (\ref{EL-COOR-29}) and
(\ref{EL-COOR-32}) is a system of homogeneous equations compatible only under
the condition of equality to zero of the relevant determinants
\bea
\label{EL-COOR-34}
D_{2n}^{(+)} (\lambda) =
\left|
\begin{array}{ccccccc}
\beta_0&\alpha_0&&\cdot&& \\
R \gamma_2&\beta_2&\alpha_2&&\\
\cdot&\cdot&\cdot&\cdot&&\cdot&\cdot\\
&&&\cdot&R\gamma_{2n-2}&\beta_{2n-2}&\alpha_{2n-2}\\
&&&\cdot&&R\gamma_{2n}&\beta_{2n}
\end{array}
\right| = 0,
\eea
\bea
\label{EL-COOR-0-35}
D_{2n+1}^{(-)} (\lambda) =
\left|
\begin{array}{ccccccc}
\beta_1&\alpha_1&&\cdot&& \\
R \gamma_3&\beta_3&\alpha_3&&\\
\cdot&\cdot&\cdot&\cdot&&\cdot&\cdot\\
&&&\cdot&R\gamma_{2n-3}&\beta_{2n-3}&\alpha_{2n-3}\\
&&&\cdot&&R\gamma_{2n-1}&\beta_{2n-1}
\end{array}
\right| = 0.
\eea
The roots of equations (\ref{EL-COOR-34}) and (\ref{EL-COOR-0-35}) give
eigenvalues for the separation constant $\lambda_{n}^{(\pm)}(R)$ . As regards
the determinants of this type, it is known \cite{COULSON1} that they have real
and different roots. It means that eigenvalues for the elliptic separation
constant are real and can be numbered by the entire index $q$, namely,
$\lambda_{n}^{(\pm)}(R)\to \lambda_{nq}^{(\pm)}(R)$, where $0\leq q \leq n$ and
$1\leq q' \leq n$ for $\lambda_{nq}^{(+)}(R)$ and $\lambda_{nq'}^{(-)}(R)$,
respectively. The multiplicity degeneracy of the  $n$th level equals $2n+1$, as
it must be.

It is the practice to express polynomial solutions of equation
(\ref{EL-COOR-13}) in terms of $C_{n}^{q}(u,R)$ and $S_{n}^{q}(u,R)$ and the
functions (\ref{EL-COOR-12}) in terms of $hc_{n}^{q}(u,R)$ and
$hs_{n}^{q}(u,R)$. In this notation physically admissible solutions of
\ref{EL-COOR-13}) have the form
\bea
\label{EL-COOR-35}
C^{q}_{n}(u, R)
&=& \sum_{s=0}^{n} a_{2s}^{n, q}(R) (\cos u)^{2s},
\\[2mm]
\label{EL-COOR-36}
S^{q}_{n}(u, R)
&=& \sin u \,
\sum_{s=0}^{n} b_{2s+1}^{n, q}(R) \, (\cos u)^{2s+1},
\eea
and their respective solutions of equation (\ref{EL-COOR-8}) are
\bea
\label{EL-COOR-37}
hc_{n}^{q}(u,R)  &=&
e^{-\frac{\omega R}{2}\cos 2u}\,
C^{q}_{n}(u, R),
\\[2mm]
\label{EL-COOR-38}
hs_{n}^{q}(u,R)  &=&
e^{-\frac{\omega R}{2}\cos 2u}\,
S^{q}_{n}(u, R).
\eea
It is known that under a definite numeration the index $q$ gives the number of
zeroes of the polynomials $C_{n}^{q}(u,R)$ and $S_{n}^{q}(u,R)$ and,
consequently, eigenvalues of the elliptic separation constant can be regulated
in number of units of the wave functions (\ref{EL-COOR-37})
and(\ref{EL-COOR-38}).

At $\zeta =i\xi$ solutions (\ref{EL-COOR-37}) and (\ref{EL-COOR-38}) give
radial and at $\zeta = \eta$ angular wave functions. As radial wave functions,
it is convenient to deal not with $hc_{n}^{q}(i\frac{\xi}{2},R)$ and
$hs_{n}^{q}(i\frac{\xi}{2},R)$, but with solutions $hc_{n}^{q'}(i\frac{\xi}{2}
+\frac{\pi}{2};- R)$ and  $hs_{n}^{q'}(i\frac{\xi}{2} +\frac{\pi}{2};- R)$,
which,as one can easily see,differ from the former ones only by the factor
\cite{ARSS}. Indeed, let us rewrite equation (\ref{EL-COOR-14}) in the form
$$
\frac{d^2 W^{(\pm)}}{d u^2} +  \alpha \sin 2u
\frac{d W^{(\pm)}}{d u} +
(-4\lambda^{(\pm)}_n (R) -  p \alpha \cos 2u) W^{(\pm)} = 0
$$
and make the change $u \to u + \frac{\pi}{2}$ and $R\to- R$
$$
\frac{d^2 W^{(\pm)}(u, -R)}{d u^2} +  \alpha \sin 2u
\frac{d W^{(\pm)}(u, -R)}{d u} +
(-4\lambda^{(\pm)}_n (-R) -  p \alpha \cos 2u) W^{(\pm)}(u, -R) = 0.
$$
Comparing the last two equations we can see that the sets of eigenvalues for
the constants $\lambda^{(\pm)}_n (R)$ and $\lambda^{(\pm)}_n (-R)$ coincide and
only their numeration changes. Hence it follows that for any $q$ there can be
found such $q'$ that $\lambda^{(\pm)}_{nq} (R) = \lambda^{(\pm)}_{nq'} (-R)$
and,
consequently,
\bea C^{q}_{n}(i\frac{\xi}{2}, R) &=&  a \, C^{q'}_{n}(i\frac{\xi}{2}+
\frac{\pi}{2}, -R),
\\[2mm]
S^{q}_{n}(i\frac{\xi}{2}, R)
&=&  b \,
S^{q'}_{n}(i\frac{\xi}{2}+ \frac{\pi}{2}, -R)
\eea
with $|a|=|b|=1$. The coefficients $a_{2s}(-R)$ and $b_{2s+1}(-R)$ are
calculated from the recurrence relations differing from
(\ref{EL-COOR-29})-(\ref{EL-COOR-33}) by the change $R \to - R$.

Thus, the elliptic wave functions of the two-dimensional hydrogen atom are
divided into two subbases
\bea
\label{EL-COOR-41}
\Psi_{n q_1 q_2}^{(+)} (\xi,\eta; R)
&=&
C_{n q_1 q_2}^{(+)} (R)
\,
hc_{n}^{q_1}\left(\frac{\eta}{2}; R\right)
\,\,
hc_{n}^{q_2}\left(i\frac{\xi}{2} +\frac{\pi}{2}; \, - R\right),
\\[2mm]
\label{EL-COOR-42}
\Psi_{n q_1 q_2}^{(-)} (\xi,\eta; R)
&=&
C_{n q_1 q_2}^{(-)} (R)
\,
hc_{n}^{q_1}\left(\frac{\eta}{2}; R\right)
\,\,
hs_{n}^{q_2}\left(i\frac{\xi}{2} +\frac{\pi}{2}; \, - R\right)
\eea
the first of which is even and the second is odd with respect to the change
$\eta \to - \eta$. The quantum number $q_1$ numbers in ascending order the
values for the separation constant $\lambda_{nq_1}^{(\pm)}$ and determines the
number of zeroes of the functions $hc_{n}^{q_1}\left(\frac{\eta}{2}; R\right)$
and $hs_{n}^{q_1}\left(\frac{\eta}{2}; R\right)$,the so-called angular Coulomb
elliptic functions , with even in $\eta$ sub-basis $0\leq q_1 \leq n$ and in
odd $1\leq q_1 \leq n$. The general number of zeroes of the elliptic wave
function (\ref{EL-COOR-7}) at a given discrete value of $E_n$ equals $n$. The
number $q_1$ at fixed $n$ determines $q_2=n-q_1$ which gives the number of
zeroes for the radial Coulomb elliptic function
$hs_{n}^{q_2}\left(i\frac{\xi}{2} + \frac{\pi}{2}; \, - R\right)$ and
$hs_{n}^{q_2}\left(i\frac{\xi}{2} +\frac{\pi}{2}; \, - R\right)$, with
\bea
\label{EL-COOR-39}
\lambda_{n q_1}^{(\pm)} (R) &=&  \lambda_{n q_2}^{(\pm)} (-R).
\eea
It should be emphasized that our functions $C_n^q$ and $S_n^q$ differ from the
generally accepted Eins polynomials \cite{ARSS} by normalization
(\ref{EL-COOR-43}) and the requirement that the elliptic
subbases(\ref{EL-COOR-41}) and (\ref{EL-COOR-42}) should turn into the
corresponding polar and parabolic  subbases of the two-dimensional hydrogen
atom as  $R\to 0$ and $R\to \infty$.

Tables \ref{t13} and \ref{t14} give explicit expressions of the elliptic wave functions for
the lowest quantum numbers.

\begin{table}[t]
\caption{ Elliptic basis of the two-dimensional hydrogen atom
$\Psi_{nq_1 q_2}^{(+)}(\xi,\eta;R)$ is the product of two functions, each
depending only on $\xi$ and $\eta$,respectively, and being even with respect to
its argument.}
\label{t13}

\begin{center}
\begin{tabular}{|c|c|c|}\hline
&&\\
$N$&$\Psi_{nq_1 q_2}^{(+)}(\xi,\eta;R)$&$\lambda_{q_1}^{(+)}$
\\
&&\\
\hline
&&\\
$0$& $\sqrt{\frac{8}{\pi}}e^{-R(\cosh\xi+\cos\eta)}$&
$\lambda^{(+)}=0$\\
&&\\
\hline
&&\\
$1$&$\sqrt{\frac{8(\lambda^{(+)}+1)}
{27\pi(2\lambda^{(+)}+1)}}\,
e^{- \frac{R}{3}(\cosh\xi+\cos\eta)}$&
$\lambda^{(+)}(\lambda^{(+)}+1)$
\\
&&\\
&$\left[1-\left(\lambda^{(+)}+\frac{2R}{3}\right)(\cosh\xi-1)\right]\,
\left[1-\left(\lambda^{(+)} + \frac{2R}{3}\right)
(\cosh\xi-1)\right]$&
$=\frac{4}{9}R^2$
\\
&&\\
\hline
&&\\
$2$&
$\frac{4}{5\sqrt{10\pi}}
\left\{1+\frac{25}{12}\left(\frac{\lambda^{(+)}}{R}\right)^2+
\frac{1}{3}\left(\frac{\lambda^{(+)}}{\lambda^{(+)}+4}\right)^2
\right\}^{-\frac{1}{2}}$
$e^{- \frac{R}{5}(\cosh\xi+\cos\eta)}$&
$\lambda^{(+)}(\lambda^{(+)}+1)$
\\
&&\\
&$\left[1-\left(\lambda^{(+)} + \frac{4R}{5}\right)
(\cosh\xi-1)+\frac{\lambda^{(+)}}{6}
\left(\lambda^{(+)} + 1 + \frac{4R}{5}\right)(\cosh\xi-1)^2
\right]$
&$\times(\lambda^{(+)}+4)$
\\
&&\\
&$\left[1+\left(\lambda^{(+)}-\frac{4R}{5}\right)(1+\cos\eta)+
\frac{\lambda^{(+)}}{6}\left(\lambda^{(+)}+1
- \frac{4R}{5}\right)(1+\cos\eta)^2\right]$
&$=\frac{16R^2}{25}(\lambda^{(+)}+3)$\\
&&\\
\hline
\end{tabular}
\end{center}

\end{table}

\begin{table}[h!t]
\caption
{Elliptic basis of the two-dimensional hydrogen atom.
$\Psi_{nq_1 q_2}^{(-)}(\xi,\eta;R)$ is the product of two functions, each
depending only on $\xi$ and $\eta$, respectively,and being odd with respect to
its argument.}\label{t14}
\begin{center}
\begin{tabular}{|c|c|c|}\hline
&&\\
$N$&$\Psi_{nq_1 q_2}^{(-)} (\xi, \eta; R)$&$\lambda_{q_1}^{(-)}$
\\
&&\\
\hline
&&\\
$1$& $-i\frac{4}{9}\sqrt{\frac{2}{3\pi}}\cosh\xi \,
\sin\eta \,
e^{- \frac{R}{3}(\cosh\xi+\cos\eta)}$&
$\lambda^{(-)} = -1$
\\
&&\\
\hline
&&\\
$2$&$-i\frac{24}{25}\frac{R}{\sqrt{30\pi}}
\sqrt{\frac{\lambda^{(-)}+4}
{2\lambda^{(-)'}+5}} \, \cosh\xi \, \sin\eta  \,
e^{- \frac{R}{5}(\cosh\xi+\cos\eta)}$&
$(\lambda^{(-)}+1)$
\\
&&\\
&$\times\left\{1-\frac{1}{3}
\left(\lambda^{(-)}+1+\frac{2R}{5}\right)(\cosh\xi-1)
\right\}$&
$\times (\lambda^{(-)}+4)$
\\
&&\\
&$\times \left\{1+\frac{1}{3}\left(\lambda^{(-)}+
1-\frac{2R}{5}\right)(1+\cos\eta)\right\}$
&$=\frac{4}{25}R^2$
\\
&&\\
\hline
&&\\
$3$&
$-i\frac{8R}{49}
\sqrt{\frac{3}{7\pi}}\left\{1-\frac{49}{10}\left(\frac{\lambda^{(-)}+1}
{R}\right)+\frac{3}{5}\left(\frac{\lambda^{(-)'}+1}
{\lambda^{(-)}+9}\right)^2\right\}^{-\frac{1}{2}}
\sinh\xi \, \sin\eta \,$&
$(\lambda^{(-)}+1)$
\\
&&\\
&$\times \, e^{- \frac{R}{7}(\cosh\xi+\cos\eta)} \,
\biggl[1-\frac{1}{3}\left(\lambda^{(-)}+1+\frac{4R}{7}\right)
(\cosh\xi-1)$
&$\times (\lambda^{(-)}+4)$
\\
&&\\
&$+ \frac{\lambda^{(-)'}+1}{30}
\left(\lambda^{(-)}+4+\frac{4R}{7}\right)
(\cosh\xi-1)\biggr]
\biggl[1+\frac{1}{3}\left(\lambda^{(-)}+1-\frac{4R}{7}\right)
$
&$\times (\lambda^{(-)}+9)$
\\
&&\\
&$\times (1+\cos\eta) + \frac{\lambda^{(-)}+1}{30}
\left(\lambda^{(-)} + 4 - \frac{4R}{7}\right)(1+\cos\eta)^2 \biggr]$
&$ = \frac{16R^2}{49}(\lambda^{(-)}+6)$
\\
&&\\
\hline
\end{tabular}
\end{center}
\end{table}

\subsection {Orthogonality}

Let us choose normalization constants $C_{n q_1 q_2}^{(\pm)} (R)$ so that the
following condition be fulfilled:
\bea
\label{EL-COOR-43}
\frac{R^2}{4} \,
\int_{0}^{\infty} \int_{0}^{2\pi}\,
\psi_{n q_1 q_2}^{(\pm)*} (\xi,\eta; R) \,
\psi_{n q_1 q_2}^{(\pm)} (\xi,\eta; R) \,
(\cosh^2\xi - \cos^2\eta)\, d\xi\, d\eta
= 1.
\eea
It follows from the periodicity condition(\ref{EL-COOR-11}) and oddness of
functions (\ref{EL-COOR-38}) that bases (\ref{EL-COOR-41}) and
(\ref{EL-COOR-42}) are mutually orthogonal at all $n$, $q_1$, $q_2$ and $q_1'$,
$q_2'$. This in its turn means that each of these bases in itself is not
complete.

The elements of each of the bases (\ref{EL-COOR-41}) and (\ref{EL-COOR-42}),
being eigenfunctions of the Hamiltonian, are orthogonal in $n$, i.e., at
$n\not= n'$
\bea
\label{EL-COOR-44}
\int \, \psi_{n' q_1 q_2}^{(\pm)*}
(\xi,\eta; R) \, \psi_{n q_1 q_2}^{(\pm)} (\xi,\eta; R) \, d V = 0.
\eea
There are two more orthogonality conditions that are consequences of the
degeneracy of the energy spectrum in quantum numbers $q$. Indeed, equations
(\ref{EL-COOR-8}) and (\ref{EL-COOR-9}) at $q_1 \not= q_2$ and $q_1' \not=
q_2'$ with the use of the trick well known in the Sturm-Liouville problem
\cite{MORS-FESHBACH} result in

\bea
\int_{0}^{\infty} \,
hc_{n}^{q_1}\left(i\frac{\xi}{2} +\frac{\pi}{2}; \, - R\right)
hc_{n}^{q_2 *}\left(i\frac{\xi}{2} +\frac{\pi}{2}; \, - R\right)
d \xi = 0,
\\[2mm]
\int_{0}^{\infty} \,
hs_{n}^{q_1}\left(i\frac{\xi}{2} +\frac{\pi}{2}; \, - R\right)
hs_{n}^{q_2 *}\left(i\frac{\xi}{2} +\frac{\pi}{2}; \, - R\right)
d \xi = 0,
\\[2mm]
\int_{0}^{2\pi} \,
hc_{n}^{q_1*}\left(\frac{\eta}{2}; \, R\right)
hc_{n}^{q_2}\left(\frac{\eta}{2}; \,  R\right)
d \eta =
\int_{0}^{2\pi} \,
hs_{n}^{q_1'*}\left(\frac{\eta}{2}; \, R\right)
hs_{n}^{q_2'}\left(\frac{\eta}{2}; \,  R\right)
d \eta = 0.
\eea
Using these formulae one can easily show that at $q \not= q'$
\bea
\label{EL-COOR-45}
\int \, \psi_{n q_1 q_2}^{(\pm)*}
(\xi,\eta; R) \, \psi_{n q_1' q_2'}^{(\pm)} (\xi,\eta; R) \, d V = 0.
\eea

\subsection {Limit $R \to 0$}

Let us show that as $R \to 0$ the elliptic basis turns into the polar one.
First, consider the recurrence relation (\ref{EL-COOR-29}). As $R \to 0$ in
(\ref{EL-COOR-34}) one can reject the elements $R\gamma_{2s}$, whereupon the
determinant turns into the product of diagonal elements. Denoting the number of
the diagonal element equal to zero by $m$ we get $\beta_{2m} = 0$, i.e.,
$$
\lambda_{n q_1}^{(+)} (0) =  \lambda_{n q_2}^{(+)} (0) = -m^2.
$$
Hence it follows that in the limit $R \to 0$ the quantum numbers $q_1$ and
$q_2$ tend, respectively, to the values $m$ and $n-m$, and
$$
\beta_{2s} (\pm R) \rightarrow \tilde\beta_{2s} =  (m-s)(m+s).
$$

Writing down now the recurrence relations (\ref{EL-COOR-29}) successively for
$s = 0,1,2,...,$ $m-1$ and keeping in mind that $a_{-2}=0$ one can easily be
convinced that as $R \to 0$ (\ref{EL-COOR-29}) transforms to
\bea
\label{EL-COOR-46}
\alpha_{2s} a_{2s+2} (\pm R) + \tilde\beta_{2s} a_{2s}(\pm R) = 0
\eea
if $0 \leq s \leq m-1$.

In a similar way, starting from above, i.e., from the condition $a_{2n+2}=0$ we
arrive at the conclusion that as $R \to 0$ and $m+1 \leq s \leq n$
(\ref{EL-COOR-29}) is substituted by
\bea
\label{EL-COOR-47}
\tilde\beta_{2s} a_{2s} (\pm R) \pm R \gamma_{2s} a_{2s-2}(\pm R) = 0
\eea
At $s=m$, expanding $\lambda(R)$ in the Taylor series we obtain
$$
\tilde\beta_{2m} \stackrel{R\to 0}\longrightarrow
\pm \varepsilon_{2m} R + O(R^2),
\qquad
\varepsilon  =  \frac{n-2m}{n+1/2} -
\left(\frac{d \lambda_{n,m}^{(+)}}{d R}\right)_{R=0},
$$
and, therefore, (\ref{EL-COOR-29}) at $s=m$ becomes
\bea
\label{EL-COOR-48}
\alpha_{2m} a_{2m+1} (\pm R) \pm R \varepsilon_{2m} a_{2m} (\pm R)
\pm R \gamma_{2m} a_{2m-2} (\pm R) = 0.
\eea
According to (\ref{EL-COOR-46}) and (\ref{EL-COOR-47}),
$$
a_{2m+2} (\pm R) = \mp \frac{R \gamma_{2m+2}}{\tilde\beta_{2m+2}} \,
a_{2m} (\pm R),
\qquad
a_{2m-2} (\pm R)=  \frac{\alpha_{2m-2}}{\tilde\beta_{2m-2}} \,
a_{2m} (\pm R),
$$
and substituting these formulae in (\ref{EL-COOR-48}) we have
$$
\left(\frac{d \lambda_{n,m}^{(+)}}{d R}\right)_{R=0} = 0.
$$
The formula derived is a restriction under which the cutoff conditions at
$s=-1$ and $s=n+1$ are consistent. Analogous analysis of the recurrence
formulae (\ref{EL-COOR-32})leads to the following recurrence relations:
$$
\alpha_{2s+1} b_{2s+3} (\pm R) +
\tilde\beta_{2s+1} b_{2s+1}(\pm R) = 0,
\qquad
0 \leq s \leq m-2,
$$
$$
\tilde\beta_{2s+1} b_{2s+1} (\pm R) \pm
R \gamma_{2s+1} b_{2s-1}(\pm R) = 0,
\qquad
m \leq s \leq n-1,
$$
in which $\tilde\beta_{2s+1} = (m+s+1)(m-s-1)$, and to the condition
$$
\left(\frac{d \lambda_{n,m-1}^{(-)}}{d R}\right)_{R=0} = 0.
$$
Now using the two-term recurrence relations (\ref{EL-COOR-46}) and
(\ref{EL-COOR-47}) and keeping in mind that $a_0=1$ we get
\bea
\label{EL-COOR-49}
a_{2s}(\pm R) \stackrel{R\to 0}\longrightarrow (-1)^s
\frac{\tilde\beta_{0} \tilde\beta_{2}\cdot \tilde\beta_{2s-2}}
{\alpha_{0}\alpha_{2}\cdot\alpha_{2s-2}} =
\frac{(m)_s (-m)_s}{(1/2)_s s!}
\eea
at $1 \leq s \leq m$ and
\bea
\label{EL-COOR-50}
a_{2s+2m}(\pm R)
&\stackrel{R\to 0}\longrightarrow & (-1)^m \,
(\pm 2\omega R)^s  \,
\frac{(-n+m)_s}{(2m+1)_s} \, \frac{a_{2m}(\pm R)}{s!}
\nonumber
\\[2mm]
&=&
(\pm 2\omega R)^s  \,
\frac{(-n+m)_s }{(2m+1)_s} \, \frac{2^{2m-1}}{s!}
\eea
at $1 \leq s \leq n-m$.

Let us now consider the limit $R \to 0$ in the wave functions. According to
formulae (\ref{EL-COOR-4}),
\bea
\label{EL-COOR-51}
C_{n}^{q} \left(\frac{\eta}{2}; R\right)
&\stackrel{R\to 0}\longrightarrow&
\sum_{s=0}^{n} a_{2s}(R) \left(\cos\frac{\varphi}{2}\right)^{2s},
\\[2mm]
\label{EL-COOR-52}
C_{n}^{q} \left(i\frac{\xi}{2}+ \frac{\pi}{2}; -R\right)
&\stackrel{R\to 0}\longrightarrow&
\sum_{s=0}^{n} \frac{a_{2s}(-R)}{R^s}\, (-r)^s.
\eea
Substituting in here (\ref{EL-COOR-49}) and (\ref{EL-COOR-50}) we notice that
polynomials take the form
\bea
\label{EL-COOR-53}
C_{n}^{q} \left(\frac{\eta}{2}; R\right)
&\stackrel{R\to 0}\longrightarrow&
{_2F_1}\left(-m, \, m; \, \frac12; \, \cos^2\frac{\varphi}{2}\right)
= (-1)^m \, \cos m\varphi,
\\[2mm]
\label{EL-COOR-54}
C_{n}^{q} \left(i\frac{\xi}{2}+ \frac{\pi}{2}; -R\right)
&\stackrel{R\to 0}\longrightarrow&
2^{2m-1} \, \left(\frac{r}{R}\right)^m \,
{_1F_1}(-n+m;\, 2m+1; \, 2\omega r).
\eea
The described scheme of reasoning can also be applied to the functions
$S_{n}^{q'} \left(\frac{\eta}{2}; R\right)$ and \\ $S_{n}^{q'}
\left(i\frac{\xi}{2}+ \frac{\pi}{2}; -R\right)$. Here is the result
\bea
\label{EL-COOR-55}
S_{n}^{q'} \left(\frac{\eta}{2}; R\right)
&\stackrel{R\to 0}\longrightarrow&
\frac{\sin\varphi}{2}
{_2F_1}\left(1-m, \, 1+m; \, \frac32; \, \cos^2\frac{\varphi}{2}\right)=
\nonumber
\\[2mm]
&=&
(-1)^{m-1} \, \frac{\sin m\varphi}{2m},
\\[2mm]
\label{EL-COOR-56}
S_{n}^{q'} \left(i\frac{\xi}{2}+ \frac{\pi}{2}; -R\right)
&\stackrel{R\to 0}\longrightarrow&
(-i) \frac{4^{m-1}}{m} \, \left(\frac{r}{R}\right)^m \,
{_1F_1}(-n+m;\, 2m+1; \, 2\omega r).
\eea
In what follows we will show that
\bea
\label{EL-COOR-57}
|C_{n q_1 q_2}^{(+)} (R)|
&\stackrel{R\to 0}\longrightarrow&
\sqrt{\frac{8\omega^3}{\pi}} \,
\left(\frac{\omega R}{2}\right)^m \,
\sqrt{\frac{(n+m)!}{(n-m)!}} \,
\frac{1}{(2m)!},
\\[2mm]
\label{EL-COOR-58}
|C_{n q_1' q_2'}^{(-)} (R)|
&\stackrel{R\to 0}\longrightarrow&
\sqrt{\frac{8\omega^3}{\pi}} \,
\left(\frac{\omega R}{2}\right)^m \,
\frac{4m^2}{(2m)!}
\sqrt{\frac{(n+m)!}{(n-m)!}}.
\eea
Using these limiting expressions and formulae (\ref{EL-COOR-53}) -
(\ref{EL-COOR-56}) one can easily be convinced that
\bea
\label{EL-COOR-59}
\Psi_{n q_1 q_2}^{(\pm)} (\xi,\eta; R)
\stackrel{R\to 0}\longrightarrow
\sqrt{2}\,
\Psi_{n m}^{(\pm)} (r, \varphi),
\eea
where the signs $(+)$ and $(-)$ on the right correspond to the polar wave
functions with definite parity with respect to inversion $\varphi \to -
\varphi$:
\bea
\label{EL-COOR-60}
\Psi_{n m}^{(\pm)} (r, \varphi) =
R_{nm}(r) \, \frac{1}{\sqrt{2\pi}} \,
\left
\{\matrix{\cos m\varphi,
\qquad 0 \leq m \leq n,
\cr
\cr
i\sin m\varphi, \qquad 1 \leq m \leq n,
\cr}
\right.,
\eea
and the radial functions $R_{nm}(r)$ are determined by the expression
\bea
\label{EL-COOR-060}
R_{nm}(r) =
\frac{\sqrt{2\omega^3}}{(2|m|)!}\,
\sqrt{\frac{(n+|m|)!}{(n-|m|)!}}\,
\, (2\omega r)^{|m|}\,
e^{- \omega r} \,\,
F(- n+|m|; \, 2|m|+ 1; \, 2\omega r).
\eea

\subsection{Limit  $R \to \infty$}

Now let us study the parabolic limit. As $R \to \infty$ in  (\ref{EL-COOR-34})
one can neglect the finite terms and reduce the determinant to the product of
diagonal terms. Denoting by $n_1$ the number of the diagonal element equal to
zero we have $\beta_{n_1}=0$ and, therefore,
\bea
\label{EL-COOR-P1}
\lambda_{n q_1}^{(+)} (R)
\stackrel{R\to \infty}\longrightarrow
\lambda_{n n_1}^{(+)} (R)
=
\frac{R(n-n_1)}{n+1/2} + \frac{R^2}{(2n+1)^2},
\eea
where $n_1 = 0,2,...$. Using an analogous line of reasoning relative to the
recurrence relation for $a_{2s}(-R)$ we get
\bea
\label{EL-COOR-P2}
\lambda_{n q_2}^{(+)} (-R)
\stackrel{R\to \infty}\longrightarrow
\lambda_{n n_2}^{(+)} (-R)
=
- \frac{R(n-n_1)}{n+1/2} + \frac{R^2}{(2n+1)^2},
\eea
with $n_2 = 0,2,....$. It follows from the condition (\ref{EL-COOR-39}) that
$(n_1+n_2)/2 = n$, i.e., $n_1$ and $n_2$ are parabolic quantum numbers.

It is seen from (\ref{EL-COOR-P1}) and (\ref{EL-COOR-P2}), and
(\ref{EL-COOR-30}) that at $n_1 \not=2s$ and $n_2 \not=2s$
\bea
\label{EL-COOR-P4}
\beta_{2s} (R)
\stackrel{R\to \infty}\longrightarrow
R \,\beta_{2s}^{(1)},
\qquad
\beta_{2s} (-R)
\stackrel{R\to \infty}\longrightarrow
R \,\beta_{2s}^{(2)},
\nonumber
\eea
where the quantities $\beta_{2s}^{(1)}$ and $\beta_{2s}^{(2)}$ are already
independent of $R$ and have the form
\bea
\label{EL-COOR-P5}
\beta_{2s}^{(1)} = (n_1-2s)/(n+1/2),
\\[2mm]
\label{EL-COOR-P6}
\beta_{2s}^{(2)} = (2s-n_2)/(n+1/2).
\eea
These formulae together with the cutoff condition $a_{-2}=0$ show that in the
limit $R\to \infty$ the three-term recurrence relations (\ref{EL-COOR-29}) turn
into the following two-term ones :
\bea
\label{EL-COOR-P7}
\alpha_{2s} a_{2s+2} (R) + R \beta_{2s}^{(1)} a_{2s} (R)= 0,
\qquad
0 \leq s \leq \frac{n_1}{2} - 1,
\\[2mm]
\label{EL-COOR-P8}
\alpha_{2s} a_{2s+2} (-R) + R \beta_{2s}^{(2)} a_{2s} (-R)= 0,
\qquad
0 \leq s \leq \frac{n_2}{2} - 1.
\eea
In a similar way, using the cutoff conditions $a_{2n+2}=0$ we have
\bea
\label{EL-COOR-P9}
\beta_{2s}^{(1)} a_{2s} (R) + \gamma_{2s} a_{2s-2} (R)= 0,
\qquad
\frac{n_1}{2} + 1 \leq s \leq n,
\\[2mm]
\label{EL-COOR-P10}
\beta_{2s}^{(2)} a_{2s} (-R) - \gamma_{2s} a_{2s-2} (-R)= 0,
\qquad
\frac{n_2}{2} + 1 \leq s \leq n.
\eea

Now let us consider the case $n_1 = 2s$ and $n_2 = 2s$. Then instead of
(\ref{EL-COOR-P1}) and (\ref{EL-COOR-P2}) one should write down
\bea
\label{EL-COOR-P11}
\lambda_{n q_1}^{(+)} (R)
&\stackrel{R\to \infty}\longrightarrow&
\lambda_{0}^{(+)} + \frac{R(n-n_1)}{n+1/2} + \frac{R^2}{(2n+1)^2},
\nonumber
\\[2mm]
\lambda_{n q_2}^{(+)} (- R)
&\stackrel{R\to \infty}\longrightarrow&
\lambda_{0}^{(+)} - \frac{R(n-n_2)}{n+1/2} + \frac{R^2}{(2n+1)^2},
\nonumber
\eea
from where $\beta_{n_1} = - n_1^2/4 -  \lambda_{0}^{(+)}$, $\beta_{n_2} = -
n_2^2/4 - \lambda_{0}^{(+)}$. The constant $\lambda_{0}^{(+)}$ can be derived
from (\ref{EL-COOR-29}) at  $s=n_1/2$ if one substitutes in it the expressions

\bea
\label{EL-COOR-P12}
a_{n_1-2} = - \frac{1}{R} \frac{\alpha_{n_1-2}}{\beta_{n_1-2}^{(1)}}\,
a_{n_1},
\qquad
a_{n_1+2} = - \frac{\gamma_{n_1+2}}{\beta_{n_1+2}^{(1)}}\,
a_{n_1},
\nonumber
\eea
that are a consequence of (\ref{EL-COOR-P7}) and (\ref{EL-COOR-P9}). The result
is:
\bea
\label{EL-COOR-P13}
\lambda_{0}^{(+)}  =
\frac{n_1^2}{2} - n (n_1 + 1/2).
\nonumber
\eea

In a similar way, we study the limit $R\to \infty$ in the recurrence relations
(\ref{EL-COOR-32}). We have
\bea
\label{EL-COOR-P14}
\lambda_{n q_1}^{(-)} (R)
&\stackrel{R\to \infty}\longrightarrow&
\frac{R(n-n_1)}{n+1/2} + \frac{R^2}{(2n+1)^2},
\\[2mm]
\label{EL-COOR-P15}
\lambda_{n q_2}^{(-)} (-R)
&\stackrel{R\to \infty}\longrightarrow&
- \frac{R(n-n_1)}{n+1/2} + \frac{R^2}{(2n+1)^2},
\eea
with $n_1 = 0,2,.....$, $n_2 = 0,2,....$.  According to (\ref{EL-COOR-39}),
$(n_1+n_2)/2= n$. From (\ref{EL-COOR-P14}), (\ref{EL-COOR-P15}) è
(\ref{EL-COOR-31}) it follows that $n_1 \not= 2s+1$, $n_2 \not= 2s+1$
\bea
\label{EL-COOR-P16}
\beta_{2s+1} (R) \stackrel{R\to \infty}\longrightarrow
R \,\beta_{2s+1}^{(1)},
\qquad
\beta_{2s+1} (-R) \stackrel{R\to \infty}\longrightarrow
R \,\beta_{2s+1}^{(2)},
\nonumber
\eea
where
\bea
\label{EL-COOR-P17}
\beta_{2s+1}^{(1)} = (n_1-2s-1)/(n+1/2),
\qquad
\beta_{2s}^{(2)} = (n_2-2s-1)/(n+1/2).
\eea
These formulae together with the cutoff condition $b_{-1} = 0$ give
\bea
\label{EL-COOR-P18}
\alpha_{2s+1} b_{2s+3} (R) + R \beta_{2s+1}^{(1)} b_{2s+1} (R)= 0,
\qquad
0 \leq s \leq \frac{n_1-3}{2},
\\[2mm]
\label{EL-COOR-P19}
\alpha_{2s+1} b_{2s+3} (-R) + R \beta_{2s+1}^{(2)} b_{2s+1} (-R)= 0,
\qquad
0 \leq s \leq \frac{n_2-3}{2},
\eea
and with the cutoff condition $a_{2n+1}=0$:
\bea
\label{EL-COOR-P20}
\beta_{2s+1}^{(1)} b_{2s+1} (R) + \gamma_{2s+1} b_{2s-1} (R) = 0,
\qquad
\frac{n_1+1}{2} \leq s \leq n-1,
\\[2mm]
\label{EL-COOR-P21}
\beta_{2s+1}^{(2)} b_{2s+1} (-R) - \gamma_{2s+1} b_{2s-1} (-R) = 0,
\qquad
\frac{n_2+1}{2} \leq s \leq n-1.
\eea
Let us consider the case when $n_1 = 2s+1$ and $n_2 = 2s+1$. As $R\to\infty$ we
have
\bea
\label{EL-COOR-P22}
\lambda_{n q_1}^{(-)} (R)
&\stackrel{R\to \infty}\longrightarrow&
\lambda_{0}^{(-)} + \frac{R(n-n_1)}{n+1/2} + \frac{R^2}{(2n+1)^2},
\nonumber
\\[2mm]
\lambda_{n q_2}^{(-)} (- R)
&\stackrel{R\to \infty}\longrightarrow&
\lambda_{0}^{(-)} - \frac{R(n-n_2)}{n+1/2} + \frac{R^2}{(2n+1)^2},
\nonumber
\eea
whence
$$
\beta_{n_1} = - \frac{(n_1-1)^2}{4} -  n_1 - \lambda_{0}^{(-)},
\qquad
\beta_{n_2} = - \frac{(n_2-1)^2}{4} -  n_2 -  \lambda_{0}^{(-)}.
$$
The constant $\lambda_{0}^{(-)}$ is determined from (\ref{EL-COOR-31}) at  $s =
(n_1-1)/2$ and equals
\bea
\label{EL-COOR-P23}
\lambda_{0}^{(-)}  =  \frac{(n_1)^2}{2} - n (n_1 + 1/2).
\nonumber
\eea

Let us proceed to study the limit
 $R\to \infty$ in the wave functions. The functions $C_{n}^{q}
\left(\frac{\eta}{2}; R\right)$ and $C_{n}^{q} \left(i\frac{\xi}{2}+
\frac{\pi}{2}; -R\right)$, according to (\ref{EL-COOR-5}), behave as follows:
\bea
\label{EL-COOR-P24}
C_{n}^{q} \left(\frac{\eta}{2}; R\right)
&\stackrel{R\to \infty}\longrightarrow&
\sum_{s=0}^{n} \frac{a_{2s}(R)}{R^s}
\left(\frac{\mu^2}{2}\right)^{s},
\\[2mm]
\label{EL-COOR-P024}
C_{n}^{q} \left(i\frac{\xi}{2}+ \frac{\pi}{2}; -R\right)
&\stackrel{R\to \infty}\longrightarrow&
\sum_{s=0}^{n} \frac{a_{2s}(-R)}{R^s}
\left(-\frac{\nu^2}{2}\right)^{s}.
\eea
It follows from the two-term recurrence relations (\ref{EL-COOR-P7}) -
(\ref{EL-COOR-P10}) that
\bea
\label{EL-COOR-P25}
a_{2s} (R)
&\stackrel{R\to \infty}\longrightarrow&
(-R)^s \, \frac{\beta_0^{(1)}\cdot\beta_2^{(1)}
\cdots \beta_{2s-2}^{(1)}}
{\alpha_{0}\cdot\alpha_{2} \cdots \alpha_{2s-2}},
\qquad
1 \leq s \leq \frac{n_1}{2},
\\[2mm]
\label{EL-COOR-P26}
a_{2s} (-R)
&\stackrel{R\to \infty}\longrightarrow&
(-R)^s \, \frac{\beta_0^{(2)}\cdot\beta_2^{(2)} \cdots
\beta_{2s-2}^{(2)}}
{\alpha_{0}\cdot\alpha_{2} \cdots \alpha_{2s-2}},
\qquad
1 \leq s \leq \frac{n_2}{2},
\\[2mm]
\label{EL-COOR-P27}
a_{2s} (R)
&\stackrel{R\to \infty}\longrightarrow&
(-1)^{s-\frac{n_1}{2}} \, \frac{\gamma_{n_1+2}\cdot\gamma_{n_1+4}
\cdots\gamma_{2s}}
{\beta_{n_1+2}^{(1)}\cdot\beta_{n_1+4}^{(1)} \cdots
\beta_{2s}^{(1)}} a_{n_1}
\qquad
\frac{n_1}{2} + 1 \leq s \leq n,
\\[2mm]
\label{EL-COOR-P28}
a_{2s} (-R)
&\stackrel{R\to \infty}\longrightarrow&
\frac{\gamma_{n_1+2}\cdot\gamma_{n_1+4}
\cdots
\gamma_{2s}}
{\beta_{n_1+2}^{(2)}\cdot\beta_{n_1+4}^{(2)} \cdots
\beta_{2s}^{(2)}} a_{n_2}
\qquad
\frac{n_2}{2}+ 1 \leq s \leq n,
\eea
and, therefore, the functions $C_{n}^{q} \left(\frac{\eta}{2}; R\right)$ and
$C_{n}^{q} \left(i\frac{\xi}{2}+ \frac{\pi}{2}; -R\right)$ as $R\to\infty$ turn
into polynomials of power
 $n_1/2$ and $n_2/2$. According to
(\ref{EL-COOR-P25}), (\ref{EL-COOR-P26}), (\ref{EL-COOR-30}) and
(\ref{EL-COOR-P5}), (\ref{EL-COOR-P6}),
\bea
\label{EL-COOR-P29}
a_{2s} (R)
&\stackrel{R\to \infty}\longrightarrow&
\frac{(-n_1/2)_s}{(1/2)_s} \, \frac{(2\omega R)^s}{s!},
\qquad
1 \leq s \leq \frac{n_1}{2},
\nonumber
\\[2mm]
\label{EL-COOR-P30}
a_{2s} (-R)
&\stackrel{R\to \infty}\longrightarrow&
\frac{(-n_2/2)_s}{(1/2)_s} \, \frac{(-2\omega R)^s}{s!},
\qquad
1 \leq s \leq \frac{n_2}{2},
\eea
and, consequently,
\bea
\label{EL-COOR-P31}
C_{n}^{q} \left(\frac{\eta}{2}; R\right)
&\stackrel{R\to \infty}\longrightarrow&
F \left(- \frac{n_1}{2}; \, \frac12; \, \omega \mu^2\right),
\\[2mm]
\label{EL-COOR-P32}
C_{n}^{q} \left(i\frac{\xi}{2}+ \frac{\pi}{2}; -R\right)
&\stackrel{R\to \infty}\longrightarrow&
F \left(- \frac{n_2}{2}; \, \frac12; \, \omega \nu^2\right).
\eea
The functions $S_{n}^{q} \left(\frac{\eta}{2}; R\right)$ and $S_{n}^{q}
\left(i\frac{\xi}{2}+ \frac{\pi}{2}; -R\right)$ as $R\to\infty$ are
\bea
\label{EL-COOR-P33}
S_{n}^{q} \left(\frac{\eta}{2}; R\right)
&\stackrel{R\to \infty}\longrightarrow&
\frac{\mu}{\sqrt{2R}}\,
\sum_{s=0}^{n-1} \frac{b_{2s+1}(R)}{R^s}
\left(\frac{\mu^2}{2}\right)^{s},
\\[2mm]
\label{EL-COOR-P033}
S_{n}^{q} \left(i\frac{\xi}{2}+ \frac{\pi}{2}; -R\right)
&\stackrel{R\to \infty}\longrightarrow&
\frac{i\nu}{\sqrt{2R}}\,
\sum_{s=0}^{n-1} \frac{b_{2s+1}(-R)}{R^s}
\left(-\frac{\nu^2}{2}\right)^{s}.
\eea
It follows from the two-term recurrence relations (\ref{EL-COOR-P18}) -
(\ref{EL-COOR-P21}) that
\bea
\label{EL-COOR-P34}
b_{2s+1} (R)
&\stackrel{R\to \infty}\longrightarrow&
(-R)^s \, \frac{\beta_1^{(1)}\cdot\beta_3^{(1)}
\cdots \beta_{2s-1}^{(1)}}
{\alpha_{1}\cdot\alpha_{3} \cdots \alpha_{2s-1}},
\qquad
1 \leq s \leq \frac{n_1-1}{2},
\\[2mm]
\label{EL-COOR-P35}
b_{2s+1} (-R)
&\stackrel{R\to \infty}\longrightarrow&
(-R)^s \, \frac{\beta_1^{(2)}\cdot\beta_3^{(2)} \cdots
\beta_{2s-1}^{(2)}}
{\alpha_{1}\cdot\alpha_{3} \cdots \alpha_{2s-1}},
\qquad
1 \leq s \leq \frac{n_2-1}{2},
\\[2mm]
\label{EL-COOR-P36}
b_{2s+1} (R)
&\stackrel{R\to \infty}\longrightarrow&
(-1)^{s-\frac{n_1-1}{2}} \, \frac{\gamma_{n_1+1}\cdot\gamma_{n_1+3}
\cdots\gamma_{2s+1}}
{\beta_{n_1+1}^{(1)}\cdot\beta_{n_1+3}^{(1)} \cdots
\beta_{2s+1}^{(1)}} \, a_{n_1},
\quad
\frac{n_1+3}{2} \leq s \leq n-1,
\\[2mm]
\label{EL-COOR-P37}
b_{2s+1} (-R)
&\stackrel{R\to \infty}\longrightarrow&
\frac{\gamma_{n_1+1}\cdot\gamma_{n_1+3}
\cdots
\gamma_{2s+1}}
{\beta_{n_1+1}^{(2)}\cdot\beta_{n_1+3}^{(2)} \cdots
\beta_{2s+1}^{(2)}} \, a_{n_2},
\qquad
\frac{n_2+3}{2} \leq s \leq n-1,
\eea
and,therefore, the functions  $S_{n}^{q} \left(\frac{\eta}{2}; R\right)$ and
$S_{n}^{q} \left(i\frac{\xi}{2}+ \frac{\pi}{2}; -R\right)$ turn into
polynomials of power $(n_1-1)/2$ and $(n_2-1)/2$. According to these formulae,
and (\ref{EL-COOR-P17}) and (\ref{EL-COOR-32}) we have
\bea
\label{EL-COOR-P38}
b_{2s+1} (R)
&\stackrel{R\to \infty}\longrightarrow&
\frac{(-(n_1-1)/2)_s}{(3/2)_s} \, \frac{(2\omega R)^s}{s!},
\qquad
1 \leq s \leq \frac{n_1-1}{2},
\nonumber
\\[2mm]
\label{EL-COOR-P39}
b_{2s+1} (-R)
&\stackrel{R\to \infty}\longrightarrow&
\frac{(-(n_2-1)/2)_s}{(3/2)_s} \, \frac{(-2\omega R)^s}{s!},
\qquad
1 \leq s \leq \frac{n_2-1}{2},
\eea
and hence
\bea
\label{EL-COOR-P40}
S_{n}^{q} \left(\frac{\eta}{2}; R\right)
&\stackrel{R\to \infty}\longrightarrow&
\frac{\mu}{\sqrt{2R}}\,
F \left(- \frac{n_1-1}{2}; \, \frac32; \, \omega \mu^2\right),
\\[2mm]
\label{EL-COOR-P41}
S_{n}^{q} \left(i\frac{\xi}{2}+ \frac{\pi}{2}; -R\right)
&\stackrel{R\to \infty}\longrightarrow&
- \frac{i\nu}{\sqrt{2R}}\,
F \left(- \frac{n_2-1}{2}; \, \frac32; \, \omega \nu^2\right).
\eea
It will be shown below that
\bea
\label{EL-COOR-P46}
|C_{n q_1 q_2}^{(+)} (R)|
&\longrightarrow&
\sqrt{\frac{\omega^3}{\pi}} \,
\frac{\sqrt{(n_1)!(n_2)!}}
{2^n \, \left(\frac{n_1}{2}\right)!\left(\frac{n_2}{2}\right)!},
\\[2mm]
\label{EL-COOR-P47}
|C_{n q_1 q_2}^{(-)} (R)|
&\longrightarrow&
\sqrt{\frac{\omega^3}{\pi}} \,
\frac{8\omega R}{2^n} \,
\frac{\sqrt{(n_1)!(n_2)!}}
{\left(\frac{n_1-1}{2}\right)!\left(\frac{n_2-1}{2}\right)!}.
\eea
With this in hand we have
\bea
\label{EL-COOR-P48}
\Psi_{n q_1 q_2}^{(\pm)} (\xi,\eta; R)
\longrightarrow
\Psi_{n_1 n_2}^{(\pm)} (\mu, \nu),
\eea
where the parabolic wave function is determined according to formula
(\ref{hyd-pol17}), and the quantum numbers $n_1$ and $n_2$ are even for the
function $\Psi_{n_1 n_2}^{(+)}$ and odd for $\Psi_{n_1 n_2}^{(-)}$.

\subsection{Normalization constants}

Now we proceed to discuss the problem of determining the elliptic normalization
constants $C_{n q_1 q_2}^{(+)}$ and $C_{n q_1' q_2'}^{(-)}$. It follows from
the normalization condition and the explicit form of the functions $\psi_{n q_1
q_2}^{(\pm)} (\xi,\eta; R)$ è $\psi_{n q_1' q_2'}^{(\pm)} (\xi,\eta; R)$ that
\bea
\label{EL-COOR-45-0}
2R^2\, |C_{n q_1 q_2}^{(+)} (R)|^2
&\times&
\sum^{n}_{s,s',t,t'=0} (-1)^{t+t'} \, a_{2s}(R) \, a_{2s'}(R) \,
a_{2t}(-R) \, a_{2t'}(-R) \,\times
\nonumber
\\[2mm]
&\times&
\{ I_{s,s'}^{1/2} J_{t,t'}^{-1/2} +
I_{s,s'}^{-1/2} J_{t,t'}^{1/2} \}
= 1,
\\[3mm]
\label{EL-COOR-45-1}
2R^2\, |C_{n q_1' q_2'}^{(-)} (R)|^2
&\times&
\sum^{n-1}_{s,s',t,t'=0} (-1)^{t+t'} \, a_{2s+1}(R) \, a_{2s'+1}(R) \,
a_{2t+1}(-R) \, a_{2t'+1}(-R) \,\times
\nonumber
\\[2mm]
&\times&
\{{\tilde I}_{s,s'}^{1/2} {\tilde J}_{t,t'}^{-1/2} +
{\tilde I}_{s,s'}^{-1/2} {\tilde J}_{t,t'}^{1/2} \}
= 1,
\eea
where summation over all four indices is in the limits $(0, n)$, and
$I_{ss'}^{|m|}$ and ${\cal I}_{tt'}^{|m|}$ are expressed in terms of degenerate
hypergeometric functions of first and second kind
\begin{eqnarray}
F\left(a; b; z\right) &=& \frac{\Gamma(b)}{\Gamma(a)\Gamma(b-a)}\,
\int\limits_0^1\,e^{zt}t^{a-1}(1-t)^{b-a-1}dt,
\label{sfh.3.24}
\\ [3mm]
\psi\left(a; b; z\right) &=& \frac{1}{\Gamma(a)}\,
\int\limits_0^\infty\,e^{-zt}t^{a-1}(1+t)^{b-a-1}dt,
\label{sfh.3.25}
\end{eqnarray}
as follows:
\bea
I_{s,s'}^{\pm 1/2}
&=&
\frac{\Gamma(1\pm 1/2)\Gamma(s+s'+1\pm 1/2)}
{\Gamma(s+s'+2\pm 1)} \,
F (s+s'+1\pm 1/2; \, s+s'+2\pm 1; \, -2\omega R),
\nonumber
\\[2mm]
J_{t,t'}^{\pm 1/2}
&=&
\Gamma(t+t'+1\pm 1/2)
\psi (t+t'+1\pm 1/2; \, t+t'+2\pm 1; \, -2\omega R),
\nonumber
\\[2mm]
{\tilde I}_{s,s'}^{\pm 1/2}
&=&
\frac{\Gamma(2\pm 1/2)\Gamma(s+s'+2\pm 1/2)}
{\Gamma(s+s'+4\pm 1)} \,
F (s+s'+2\pm 1/2; \, s+s'+4\pm 1; \, -2\omega R),
\nonumber
\\[2mm]
{\tilde J}_{t,t'}^{\pm 1/2}
&=&
\Gamma(t+t'+2\pm 1/2)
\Psi (t+t'+2\pm 1/2; \, t+t'+2\pm 1; \, -2\omega R).
\nonumber
\eea
Using the asymptotics
\bea
\label{EL-COOR-45-3}
F (a; \, c; \, x)
&\stackrel{x\to 0}\longrightarrow& 1,
\qquad
\psi (a; \, n+1; \, x)
\stackrel{x\to 0}\longrightarrow
\frac{\Gamma(n)}{\Gamma(a)} \,
\left(\frac{1}{x}\right)^n,
\nonumber
\\[2mm]
F (a; \, c; \, x)
&\stackrel{x\to \infty}\longrightarrow&
\frac{\Gamma(c)}{\Gamma(c-a)} \,
(-x)^{-a},
\qquad
\psi (a; \, n+1; \, x) \stackrel{x\to \infty}\longrightarrow
\left(\frac{1}{x}\right)^a,
\nonumber
\eea
after some tedious calculations one can derive formulae (\ref{EL-COOR-P46}) and
(\ref{EL-COOR-47}) if one assumes that the phases in the normalization
constants $C_{n q_1 q_2}^{(\pm)}$ behave as $\Phi_{0}^{(\pm)} \to \pi m$ and
$\Phi_{\infty}^{(\pm)} \to \frac{\pi}{2}(n_1+n_2-1)$ if $n_1$ and $n_2$ are
odd, and $\Phi_{\infty}^{(\pm)} \to \frac{\pi}{2}(n_1+n_2)$ if $n_1$ and $n_2$
are even.

\section{Interbasis expansions in the two-dimensional hydrogen atom}
\markboth{CHAPTER 1. TWO DIMENSIONAL HYDROGEN ATOM}{1.4. INTERBASIS EXPANSIONS}
This section is devoted to expansions between the elliptic, polar, and
parabolic sub-bases of the two-dimensional hydrogen atom with given parity.

\subsection{Expansion of the parabolic basis over the polar one with
given parity} Let us divide the parabolic and polar bases into sub-bases with
definite parity with respect to transformation $y \to - y$, i.e., $\varphi \to
- \varphi$ and $\nu \to - \nu$. Analytically, both the parabolic sub-bases are
determined by the same formula and the difference is only in the values run by
the indices. The expansion of parabolic sub-bases over the polar ones under
such a choice has the form
\bea
\Psi_{np}^{(\pm)}(\mu,\nu) =
\sum_{m=0;1}^{n} \,
V_{npm}^{(\pm)} \,
\Psi_{nm}^{(\pm)}(r,\varphi),
\label{d1}
\eea
where the polar sub-bases $\Psi_{nm}^{(\pm)}(r,\varphi)$ are determined by
formula (\ref{EL-COOR-60}). The summation over $m$ is carried out within the
limits $0\leq m \leq n$ and $1\leq m \leq n$ for even and odd sub-bases,
respectively. The coefficient $V_{npm}^{(+)}$ at $m=0$ is calculated by the
substitution $r=0$ and is given by expression
\bea
V_{np, 0}^{(+)}
=\frac{(-1)^n}{2^n}\frac{\sqrt{(n+p)!(n-p)!}}
{\left(\frac{n+p}{2}\right)!\left(\frac{n-p}{2}\right)!}.
\label{d2}
\eea
To calculate the rest of the coefficients in even expansion, one should take
the limit $\varphi\rightarrow \pi$ and use the orthogonality of radial wave
functions specific to the two-dimensional hydrogen atom (see Appendix A)
\bea
\int_{0}^{\infty} R_{nm}(r)R_{nm'}(r)\frac{dr}{r}
=
\frac{\omega^3}{m}\delta_{mm'}.
\label{d3}
\eea
This allows one to write down the expansion coefficients as integrals of two
hypergeometric series. The use of the formula
\begin{eqnarray}
\label{d4}
\int_0^\infty\, e^{-z}z^{m-1}F(a; b; z)F(c; d; z)dz
&=&
\frac{\Gamma(m)\Gamma(d)\Gamma(d-c-m)}{\Gamma(d-c)\Gamma(d-m)}\times
\nonumber
\\[3mm]
&\times&
{_3F}_2 \left\{
\begin{array}{l}
a, m, 1+m-d \cr \\
b,  1 +m-c-d
\end{array}
\Biggr| 1 \right\},
\end{eqnarray}
the validity of which is justified by the expansion of one of the integrand
degenerate hypergeometric functions in powers of $z$ and further integration,
results in
\bea
V_{npm}^{(+)}
=
\frac{(-1)^{n-m}n!}{2^{n-1}\left(\frac{n+p}{2}\right)!
\left(\frac{n-p}{2}\right)!}
\sqrt{\frac{(n+p)!(n-p)!}{(n+m)!(n-m)!}} {_3F_2}
\left\{\matrix{-\frac{n-p}{2},\,\,m,\,\, -m \cr \cr \frac{1}{2},\,\,-n
\cr}\Biggr|1\right\}.
\label{d5}
\eea
The coefficients of odd expansion are calculated in a similar way. The only
difference is the inclusion of the boundary condition
\begin{eqnarray*}
\lim_{\varphi\to\pi}\frac{\sin m\varphi}{\sin \varphi}
= (-1)^{m+1}\, \pi.
\end{eqnarray*}
The final result is
\bea V_{npm}^{(-)} &=& \frac{i(-1)^{n-m}m(n-1)!}{2^{n-2}
\left(\frac{n+p-1}{2}\right)!\left(\frac{n-p-1}{2}\right)!}
\sqrt{\frac{(n+p)!(n-p)!}{(n+m)!(n-m)!}} \times\nonumber
\\[2mm]
&\times&
{_3F_2} \left\{\matrix{ -\frac{n-p-1}{2},\,\,m+1,\,\, 1-m \cr \cr
\frac{3}{2},\,\,1-n\cr}\Biggr|1\right\}.
\label{d6}
\eea
Formulae (\ref{d2}), (\ref{d5}) and (\ref{d6}) can be written down in a unique
way
\bea
V_{npm}^{(\pm)}
&=&
\frac{(-1)^{n-m}n!}{2^{n-1}
\Gamma\left(\frac{n+p}{2}+1\right)
\Gamma\left(\frac{n-p}{2}+1\right)}
\sqrt{\frac{(n+p)!(n-p)!}{(n+m)!(n-m)!}}\times
\nonumber
\\[2mm]
&\times& \left(1-\frac{1}{2}\delta_{m,0}\right) {_3F_2}
\left\{\matrix{ -\frac{n-p}{2},\,\,m,\,\, -m \cr \cr
\frac{1}{2},\,\,-n \cr}\Biggr|1\right\}
\label{d7}
\eea
by using the identity \cite{BA} four times,
\bea
{_3F_2}\left\{\matrix{ a,\,\,a',\,\, -n\cr \cr c',\,\,1-n-c
\cr}\Biggr|1\right\} =
\frac{\Gamma(c+a+n)\Gamma(c)}{\Gamma(c+a)\Gamma(c+n)}
{_3F_2} \left\{\matrix{ a,\,\,c'-a',\,\, -n \cr \cr c',\,\,c+a
\cr}\Biggr|1\right\},
\label{d8}
\eea
thus resulting in the relation
\begin{eqnarray*}
{_3F_2} \left\{\matrix{ -\frac{n-p-1}{2},\,\,1+m,\,\, 1-m \cr \cr
\frac{3}{2},\,\,1-n \cr}\Biggr|1\right\} =
\frac{-in\Gamma\left(\frac{n+p+1}{2}\right)
\Gamma\left(\frac{n-p+1}{2}\right)}
{2m\Gamma\left(\frac{n+p}{2}+1\right)\Gamma\left(\frac{n-p}{2}-1\right)}
{_3F_2} \left\{\matrix{ -\frac{n-p}{2},\,\,m,\,\, -m \cr \cr
\frac{1}{2},\,\,-n \cr}\Biggr|1\right\}.
\end{eqnarray*}
The results (\ref{2.2}) and (\ref{d7}) tell about the relation between the
Wigner $d$-function of the argument  $\pi/2$ the hypergeometric functions
${_3F_2}(\dots|1)$. The relation between these objects in particular cases was
also discussed in \cite{SMOR-SHEL}.

\subsection{Expansion of the elliptic basis over the polar
one }  Let us write down the sought expansion
\bea
\Psi_{nq}^{(\pm)}(\xi,\eta;R)= \sum_{m=0;1}^n \,
W_{nqm}^{(\pm)} \,
\Psi_{nm}^{(\pm)}(r,\varphi).
\label{d9}
\eea
Here, as in the previous case, the summation is in the limits $0\leq m \leq n$
and $1\leq m \leq n$ for even and odd sub-bases, respectively. Taking the limit
$r \to 0$ we get
\bea
W_{nq0}^{(+)} \,
=
\left(\frac{\pi}{\omega^3}\right)^{\frac{1}{2}}
C_{nq}^{(+)}(R).
\label{d10}
\eea
The remaining coefficients are calculated by analogy with what has been done in
section 1 for the case $m\ne 0$. This leads to the formulae
\bea
W_{nqm}^{(+)}
&=&
(-1)^{m}\sqrt{2\pi}\frac{m}{\omega^3}
C_{nq}^{(+)}(R)e^{\frac{\omega R}{2}}\times
\nonumber
\\[2mm]
&\times&
\int_{0}^{\infty}hc_n^q
\left(\frac{i\xi}{2}+\frac{\pi}{2},-R\right)
R_{nm}\left(\frac{R}{2}(\cosh\xi-1)\right)
\frac{\sinh\xi}{\cosh\xi-1}d\xi,
\label{d11}
\\[2mm]
W_{nqm}^{(-)}
&=&i (-1)^{m}\sqrt{\frac{\pi}{2}}
C_{nq}^{(-)}(R)e^{\frac{\omega R}{2}}\times
\nonumber
\\[2mm]
&\times& \int_{0}^{\infty}hs_n^q
\left(\frac{i\xi}{2}+\frac{\pi}{2},-R\right)
R_{nm}\left(\frac{R}{2}(\cosh\xi-1)\right) d\xi.
\label{d12}
\eea
Using formulae (\ref{EL-COOR-35}) - (\ref{EL-COOR-38}), (\ref{EL-COOR-060}) and
integrating over $\xi$ we have
\bea
W_{nqm}^{(+)}
&=&
\left(\frac{\pi}{\omega^3}\right)^{\frac{1}{2}}
\frac{(-1)^m2n!C_{nq}^{(+)}(R)}{\sqrt{(n+m)!(n-m)!}}
\sum_{s=0}^{m}\frac{a_{2s}(-R)}{(-2\omega R)^s}
\frac{(-m)_s(m)_s}{(-n)_s},
\nonumber
\\[3mm]
W_{nqm}^{(-)}
&=&
\left(\frac{\pi}{\omega^3}\right)^{\frac{1}{2}}
\frac{(-1)^{m-1}m(n-1)!C_{nq}^{(-)}(R)}{\sqrt{(n+m)!(n-m)!}}
\sum_{s=0}^{m-1}\frac{b_{2s+1}(-R)}{(-2\omega R)^{s+1}}
\frac{(1-m)_s(m+1)_s}{(1-n)_s}.
\nonumber
\eea
\subsection{Expansion of the elliptic basis over the
parabolic one} Let us write down the expansion of the elliptic sub-bases over
the parabolic ones
\bea
\Psi_{nq}^{(\pm)}(\xi,\eta;R) =
\sum_{p=-n}^{n}\,
U_{nqp}^{(\pm)}\,
\Psi_{np}^{(\pm)}(\mu,\nu).
\label{d15}
\eea
Substituting the parabolic coordinates in the right-hand side of this equality
by the elliptic ones
\begin{eqnarray*}
\mu^2=\frac{R}{2}(\cosh\xi+1)(1+\cos\eta), \qquad
\nu^2=\frac{R}{2}(\cosh\xi-1)(1-\cos\eta),
\end{eqnarray*}
taking $\eta$ equal to $\pi$, and using orthonormalization of the Hermite
polynomials we have
\bea
U_{nqp}^{(+)}\,
&=&
(-1)^{\frac{n+p}{2}}\frac{2^{p}}{\omega}
\frac{\left(\frac{n+p}{2}\right)!C_{nq}^{(+)}(R)}{\sqrt{(n+p)!(n-p)!}}
\int_0^{\infty}\exp\left\{-\frac{\omega R}{2}(\cosh\xi-2)\right\}
\sqrt{R(\cosh\xi+1)}\times
\nonumber
\\[2mm]
&\times&
hc_n^q\left(i\frac{\xi}{2}+\frac{\pi}{2};-R\right)\,
H_{n-p}\left(\sqrt{\omega R(\cosh\xi-1)}\right)
d\xi,
\label{d16}
\\[3mm]
U_{nqp}^{(-)}\,
&=&
(-1)^{\frac{n+p-1}{2}}\frac{2^{p-1}}{\omega^{\frac{3}{2}}}
\frac{\left(\frac{n+p-1}{2}\right)!
C_{nq}^{(-)}(R)}{\sqrt{(n+p)!(n-p)!}}
\int_0^{\infty} \,
\exp\left\{-\frac{\omega R}{2}(\cosh\xi-2)\right\}\times
\nonumber
\\[2mm]
&\times& hs_{n}^{n-q}\left(i\frac{\xi}{2}+\frac{\pi}{2};-R\right)\,
H_{n-p}\left(\sqrt{\omega R(\cosh\xi-1)}\right) \,
d\xi.
\label{d17}
\eea
Let us express the Hermite polynomials in these formulae in terms of the
degenerate hypergeometric functions, according to (\ref{hermit}), and use
formulae (\ref{EL-COOR-35})--(\ref{EL-COOR-38}). This allows us to carry out
integration and obtain
\bea
U_{nqp}^{(+)}\,
&=&
(-1)^{\frac{n+p}{2}}
\frac{2^n}{\omega^{\frac{3}{2}}}
\frac{\left(\frac{n+p}{2}\right)!C_{nq}^{(+)}(R)}{\sqrt{(n+p)!(n-p)!}}
\sum_{s=0}^{n}\frac{a_{2s}(-R)}{(-2\omega R)^s}
\frac{\Gamma(s+1)\Gamma\left(s+\frac{1}{2}\right)}{
\Gamma\left(s+1-\frac{n-p}{2}\right)},
\nonumber
\\[3mm]
U_{nqp}^{(-)}\,
&=&
(-1)^{\frac{n+p-1}{2}}
\frac{2^{n-2}}{\omega^{\frac{3}{2}}R}
\frac{\left(\frac{n+p-1}{2}\right)!
C_{nq}^{(-)}(R)}{\sqrt{(n+p)!(n-p)!}}
\sum_{s=0}^{n-1}\frac{a_{2s+1}(-R)}{(-2\omega R)^s}
\frac{\Gamma(s+1)\Gamma\left(s+\frac{3}{2}\right)}{
\Gamma\left(s+1-\frac{n-p-1}{2}\right)}.
\nonumber
\eea

\subsection{Limiting transitions $R\to 0$ and $R\to
\infty$} Let us trace the limiting transitions $R\to 0$ and $R\to \infty$ in
formulae (\ref{d11}), (\ref{d12}), (\ref{d16}), and (\ref{d17}). Making use of
the behavior of the functions $hc_n^q$, $hs_{n}^q$ and normalization
coefficients $C_{nq}^{(+)}(R)$ and $C_{nq}^{(-)}(R)$ as $R\to 0$ we immediately
have
\begin{eqnarray*}
W_{nq0}^{(+)}\,
\stackrel{R\to 0}\longrightarrow
2\sqrt 2 \delta_{q0},
\qquad
W_{nqm}^{(\pm)}\,
\stackrel{R\to 0}\longrightarrow
2\sqrt 2 \delta_{qm}.
\end{eqnarray*}
Now let us substitute these limiting expressions in (\ref{d11}), (\ref{d12})
and take into account (\ref{d4}). Then after using formula (\ref{d8}) twice
\begin{eqnarray*}
{_3F_2}
\left\{\matrix{-\frac{n-p}{2},\,\,m+\frac{1}{2},\,\,
-m+\frac{1}{2} \cr \cr
\frac{1}{2},\,\,\ \frac{1}{2}-n \cr}\Biggr|1\right\}
=
\frac{\Gamma\left(\frac{n+p+1}{2}\right)n!}
{\Gamma\left(n+\frac{1}{2}\right)\Gamma\left(\frac{n+p}{2}\right)}
{_3F_2}
\left\{\matrix{-\frac{n-p}{2},\,\,m,\,\,-m \cr \cr
\frac{1}{2},\,\,\-n \cr}\Biggr|1\right\},
\end{eqnarray*}
\begin{eqnarray*}
{_3F_2}
\left\{\matrix{-\frac{n-p-1}{2},\,\,m+\frac{1}{2},\,\,
\frac{1}{2}-m\cr \cr
\frac{3}{2},\,\,\frac{1}{2}-n \cr}\Biggr|1\right\}=
\frac{\Gamma\left(\frac{n+p+2}{2}\right)(n-1)!}
{\Gamma\left(n+\frac{1}{2}\right)\Gamma\left(\frac{n+p-1}{2}\right)}
{_3F_2}
\left\{\matrix{-\frac{n-p-1}{2},\,\,m+1,\,\,1-m \cr \cr
\frac{3}{2},\,\,1-n \cr}\Biggr|1\right\}
\end{eqnarray*}
we come to the conclusion that at $m\ne 0$
\begin{eqnarray*}
U_{nqp}^{(\pm)}\,
\stackrel{R\to 0}\longrightarrow
\pm\frac{1}{\sqrt 2}
\langle n,p|n,m\rangle.
\end{eqnarray*}
The factor $\frac{1}{\sqrt 2}$ arises due to the normalization condition
accepted by us
\begin{eqnarray*}
\int |\psi_{nq}^{(\pm)}(\xi,\eta;R)|^2 dv=1,
\qquad
\int |\psi_{nm}^{(\pm)}(r,\varphi)|^2 dv=\frac{1}{2}.
\end{eqnarray*}
As $R\to \infty$ with the use of formulae (\ref{EL-COOR-P32}),
(\ref{EL-COOR-P41}), (\ref{EL-COOR-P46}), and (\ref{EL-COOR-P47}) we derive

\begin{eqnarray*}
W_{nqm}^{(\pm)}\,
\stackrel{R\to \infty}\longrightarrow
V_{npm}^{(\pm)}, \qquad
W_{nqp}^{(\pm)}
\stackrel{R\to \infty}\longrightarrow
\delta_{p,2q-n}.
\end{eqnarray*}

\section{Recurrence relations and the elliptic basis of the two-dimensional hydrogen atom}
\markboth{CHAPTER 1. TWO DIMENSIONAL HYDROGEN ATOM}{1.5. RECURRENCE RELATIONS AND THE ELLIPTIC BASIS}
We have already dwelled upon two possible statements of the problem of
interbasis expansions (depending on whether the known explicit form of the
expanded basis is presumed or not ). In the second and fourth sections we dealt
with the first statement of the problem. Here we will construct the elliptic
basis. Moreover, the case in point is not just the solution in one or another
way of the Schr\"{o}dinger equation in the elliptic coordinates but the
construction of such superpositions as polar or first parabolic bases (at given
energy) which could be eigenfunctions of the elliptic integral of motion
$\Lambda$ and correspond to eigenvalues $\lambda_q^{(\pm)}$. The idea of using
additional integrals of motion to determine interbasis expansions was borrowed
from \cite{COULSON1}, where it was put forward and applied to the problem of
finding spheroidal wave functions of the hydrogen atom.

\subsection{Elliptic integral of motion}

After excluding the energy parameter $\omega^2$ from equations
(\ref{EL-COOR-8}) and (\ref{EL-COOR-8}) we arrive at the operator
\bea
\label{3.1}
{\cal Q} =
\frac{1}{\cosh^2\xi-\cos^2\eta}
\left(\cos^2\eta\frac{\partial^2}{\partial\xi^2}+
ch^2\xi\frac{\partial^2}{\partial\eta^2}\right) -
\frac{R \cosh\xi\cos\eta}{\cosh\xi+\cos\eta}.
\eea
The constant $Q$ is the eigenvalue of this operator and the elliptic
sub-bases (\ref{EL-COOR-41}) and (\ref{EL-COOR-42}) are the
eigenfunctions. As within the limits $R \rightarrow 0$ and $R
\rightarrow \infty$  the elliptic system of coordinates transforms
to the polar and first parabolic one, it is clear {\it a priori}
that the operator ${\cal Q}$ should be a linear combination of
operators $L$ and ${\cal P}$ which in these limits turns into $L^2$
and ${\cal P}$, respectively. The choice of ${L}^2$ instead of $L$
is due to the fact that as $R \rightarrow 0$, i.e., $\eta
\rightarrow \phi$, the parity with respect to the inversion
$\eta\rightarrow -\eta$ is conserved. To determine the weight
factors a free constant in the above-mentioned linear combination we
turn in expression (\ref{3.1}) to the Cartesian coordinate and
compare the obtained result with the form of the operators $L^2,
{\cal P}$ and ${\cal H}$ written in terms of the Cartesian
coordinates as well. After some tedious calculations we get
\bea
\label{3.01}
{\cal Q} = - L^2 - \omega R {\cal P} -  \frac{R^2}{2} {\cal H}.
\eea
The quantity $\lambda(R)$ introduced in (\ref{EL-COOR-029}) is the eigenvalue
of the operator
\bea
\label{3.0-01}
\Lambda =
{\cal Q} +  \frac{R^2}{2} {\cal H} =
- L^2 - \omega R {\cal P}.
\eea

 \subsection{Elliptic basis}

In this subsection we will introduce the recurrence relations determining the
coefficients in the expansion of the elliptic basis of the two-dimensional
hydrogen atom over the polar and parabolic ones. First, let us consider the
superposition of polar bases
\bea
\label{4.1}
\Psi_{nq}^{(\pm)}(\xi, \eta; R) = \sum_{m = -n}^n  \,
W_{nqm}^{(\pm)}(R) \, \Psi_{nm}(r, \varphi)
\eea
(in contrast with (\ref{d9}) in (\ref{4.1}) on the right there is a polar basis
having no definite parity). By definition, these superpositions are the
eigenfunctions of the operator $\hat\Lambda$ corresponding to eigenvalues
$\lambda_q^{(\pm)}$
\bea
\label{4.2}
\hat\Lambda \Psi_{nq}^{(\pm)}=
\lambda_q^{(\pm)} \Psi_{nq}^{(\pm)}.
\eea
As $\eta \rightarrow - \eta$ the polar angle $\varphi$ changes the sign and,
therefore, taking account of the symmetry properties of the elliptic bases
$\Psi_{nq}^{(+)}$ and $\Psi_{nq}^{(-)}$ with respect to the inversion $\eta
\rightarrow - \eta$ we get from(\ref{4.1}) the following condition
\bea
\label{4.3}
W_{nq,-m}^{(\pm)}(R) = \pm W_{nqm}^{(\pm)}(R).
\eea
Substitute (\ref{4.1}) and (\ref{4.2}),  multiply the resultant equation by
$\Psi_{nm'}^*(r,\phi)$, and integrate it over the two-dimensional volume. Then
the orthonormalization of the polar basis leads to two systems of homogeneous
equations which should be satisfied by the expansion coefficients  (\ref{4.1})
\bea
\label{4.4}
\sum_{m=-n}^n
\left({\cal P}_{mm'}+
\frac{\lambda_q^{(\pm)}+{m'}^2}{\omega R}\delta_{mm'}
\right)\, W_{nqm}^{(\pm)} (R)=0.
\eea
Here ${\cal P}_{mm'}$ implies the matrix element of the generator ${\cal P}$ with
respect to polar bases, i.e.,
\bea
\label{4.5}
{\cal P}_{mm'}=
\int \Psi_{nm'}^{*}(r, \varphi)\, {\cal P} \, \Psi_{nm}(r, \varphi) dv.
\eea
This matrix element can be calculated by two methods depending on whether the
operator ${\cal P}$ is written in the polar coordinates or the polar bases are
expanded over the parabolic ones. To have a clear idea of computational
possibilities of the procedure of interbasis expansions, we will consider both
the methods. In the first method, we write the operator ${\cal P}$ in the polar
coordinates
\bea
{\cal P} =
\frac{1}{\omega}
\left\{\cos\varphi\left(1+\frac{1}{r}\frac{\partial^2}{\partial\varphi^2}+
\frac{1}{2}\frac{\partial}{\partial r}\right)+
\sin\varphi\left(\frac{\partial}{\partial r}-\frac{1}{2r}\right)
\frac{\partial}{\partial\phi}\right\}.
\nonumber
\eea
Upon substitution of this operator in the expression for the matrix element
${\cal P}_{m'm}$ and trivial integration over the polar angle we arrive at the formula
\bea
{\cal P}_{m'm}=
\frac{1}{2\omega}
(\delta_{m',m+1}t_{m+1,m}+t_{m,m-1}\delta_{m',m-1})
\nonumber
\eea
in which the quantity $t_{m,m-1}$ is derived from $t_{m+1,m}$ by changing $m
\rightarrow m-1$, and $t_{m+1,m}$ has the form
\bea
t_{m+1,m}=
\int_{0}^\infty
rR_{n,m+1}(r)
\left[1+\left(m+\frac{1}{2}\right)
\left(\frac{d}{dr}-\frac{m}{r}\right)\right]
R_{nm}(r)dr.
\nonumber
\eea
A direct calculation of this integral is time consuming and tedious. It is
easier to use the following way. Let us write down two radial Schr\"{o}dinger
equations, one with the quantum number $m$ and the other with $m'$, multiply
the first by $R_{nm'}$ and the second by $R_{nm}$, subtract one from the other,
and integrate over $dr$. After the above manipulations one can easily establish
the identity
\bea
\int_0^\infty
rR_{nm'}(r)
\left[\frac{d}{dr}-
\frac{{m'}^2-m^2-1}{2r}\right]
R_{nm}(r)dr =0
\nonumber
\eea
which implies that the contribution to the integral $t_{m+1,m}$ comes only from
the first term in square brackets so that
\bea
t_{m+1,m}=
\int_0^\infty
r\, R_{n m+1}(r)\, R_{nm}dr=
-\omega\sqrt{(n-m)(n+m+1)}
\nonumber
\eea
and for the matrix element ${\cal P}_{m'm}$ we have
\bea
\label{4.6}
{\cal P}_{m'm}=
-\frac{1}{2}\sqrt{(n-m)(n+m+1)}\delta_{m,m'-1}-
\frac{1}{2}\sqrt{(n+m)(n-m+1)}\delta_{m,m'+1}.
\eea
Now following the recipe of the second method, namely, substituting in
(\ref{4.5}) the expansion of the polar basis over the first parabolic one one
can immediately arrive at
\bea
{\cal P}_{m'm}=
\sum_{p=-n}^n \, p\, d_{pm}^n \left(\frac{\pi}{2}\right)\,
d_{pm'}^n\left(\frac{\pi}{2}\right).
\nonumber
\eea
This formula with the use of the recurrence relation known from the angular
momentum theory \cite{VAR}
\bea
-Md_{MM'}^J\left(\frac{\pi}{2}\right)&=&
\frac{1}{2}\sqrt{(J+M')(J-M'+1)}
d_{M,M'-1}^J\left(\frac{\pi}{2}\right)+
\nonumber
\\[3mm]
&+&
\frac{1}{2}\sqrt{(J-M')(J+M'+1)}
d_{M,M'+1}^J \left(\frac{\pi}{2}\right)
\label{4.7}
\eea
 and the orthonormalization condition
\bea
\sum_{M''=-J}^J d_{M''M}^J (\beta)
d_{M''M'}^J(\beta)=\delta_{MM'}
\nonumber
\eea
leads to the above result (\ref{4.6}). From the above calculations one can
easily evaluate the efficiency of using interbasis expansions for calculation
of matrix elements. Formula(\ref{4.6}) and the system of equations (\ref{4.1})
result in the three-term recurrence relations:
\bea
\label{4.8}
\frac{\lambda_q^{(\pm)}(R)+m^2}{\omega R}\, W_{nqm}^{(\pm)}(R)+
&=&
\frac{1}{2}\sqrt{(n-m)(n+m+1)}\, W_{nqm+1}^{(\pm)}(R)
\nonumber
\\ [2mm]
&+&
\frac{1}{2} \sqrt{(n+m)(n-m+1)} \, W_{nqm-1}^{(\pm)}(R).
\eea
This result is a starting point in the programme of calculating the expansion
coefficients (\ref{4.1}). By virtue of the condition (\ref{4.3})  only $n+1$
equations $(0\leq m\leq n)$ are independent at the sign $<<+>>$; and  $n$
equations $(1\leq m\leq n)$, at the sign $<<->>$.  Setting the corresponding
determinants equal to zero we derive equations used to determine eigenvalues of
the separation constants $\lambda_{q}^{(\pm)}(R)$. Upon determination of the
eigenvalues for the elliptic separation constants $\lambda_q^{(\pm)}(R)$
recurrence relations (\ref{4.8}) are solved together with the normalization
conditions
\bea
\label{4.9}
\sum_{m=-n}^n
{W_{nqm}^{(\pm)}}^*(R)
W_{nqm}^{(\pm)}(R)=1.
\eea
The above-described method can word for word be repeated for the expansion of
the elliptic basis over the parabolic one:
\bea
\label{4.10}
\Psi_{nq}^{(\pm)}(\xi,\eta;R)=
\sum_{p=-n}^n
U_{nqp}^{(\pm)}(R)\Psi_{np}(\mu,\nu).
\eea
Here, like in (\ref{4.1}), the left-hand part is assumed $\it a priori$ to be
unknown and the constructive idea implies that $\Psi_{nqm}^{(\pm)}$ satisfies
equation (\ref{4.2}). Using this line of reasoning we arrive at the three-term
recurrence relations
\bea
\label{4.11}
&&\left(\lambda_q^{(\pm)}+\omega pR+\frac{1}{2}(n^2+n-p^2)\right)
U_{nqm}^{(\pm)}(R)=
\nonumber
\\[2mm]
&=&\frac{1}{4}\sqrt{(n-p)(n-p-1)(n+p+1)(n+p+2)}
U_{nqm+2}^{(\pm)}+
\nonumber
\\[2mm]
&+&\frac{1}{4}\sqrt{(n+p)(n+p-1)(n-p+1)(n-p+2)}
U_{nqm-2}^{(\pm)}(R),
\eea
that have to be solved together with the normalization condition
\bea
\label{4.12}
\sum_{p=-n}^n
U_{nqp}^{*(\pm)}(R)
U_{nqp}^{(\pm)} (R)=1.
\eea
It is obvious that the elliptic sub-bases obtained from (\ref{4.11}) and
(\ref{4.12}) may differ by a phase factor from the sub-bases constructed on the
basis of relations (\ref{4.8}) and (\ref{4.9}). The reason for such a
difference lies in the necessity of taking square roots in the process of
calculation; in this case, there is no general way of choosing phases of the
roots which wood guarantee a full coincidence of the results following from
(\ref{4.8}), (\ref{4.9}) and (\ref{4.11}), (\ref{4.12}). To get around these
tedious details and, at the same time, to achieve consistency of the phases,
one should merely substitute in (\ref{4.1}) the expansion of the polar basis
over the first parabolic one and having compared the obtained result with
(\ref{4.10}) write down
\bea
\label{4.13}
U_{nqp}^{(\pm)}(R)=
(i)^{n-p}
\sum_{m=-n}^n d_{pm}^n
\left(\frac{\pi}{2}\right)
W_{nqm}^{(\pm)}(R).
\eea
At known $W_{nqm}^{(\pm)}(R)$ formula (\ref{4.13}) is more convenient for
calculation of the coefficients $U_{nqp}^{(\pm)}(R)$ than the recurrence
relation (\ref{4.10}).

\section{Elliptic corrections to the polar and parabolic bases of the two-dimensional
hydrogen atom}
\markboth{CHAPTER 1. TWO DIMENSIONAL HYDROGEN ATOM}{1.6. ELLIPTIC CORRECTIONS TO THE POLAR AND PARABOLIC BASES}

At $R\rightarrow 0$ and $R\rightarrow \infty$ the operator $\Lambda$ tends to
$(- L^2)$ and $(-\omega R {\cal P})$, respectively, i.e., the elliptic
sub-bases turn into the polar and parabolic ones with given parity and the
quantum number $q$ retains the meaning of the quantity giving the number of
zeroes of the function depending on the variable $\eta$. The behavior of the
elliptic separation constants $\lambda_q^{(\pm)}(R)$ as $R\rightarrow 0$ and
$R\rightarrow\infty$ is determined by the formulae
\bea
\label{5.01}
\lambda_q^{(-)}(R)
&\stackrel{R\to 0}\longrightarrow& -q^2,
\qquad 0\leq q\leq n,
\\[2mm]
\label{5.02}
\lambda_q^{(-)}(R)
&\stackrel{R\to 0}\longrightarrow&
-q^2,  \qquad
1\leq q\leq n,
\\[2mm]
\label{5.03}
\lambda_q^{(+)}(R)
&\stackrel{R\to \infty}\longrightarrow&
-\omega R (2q-n),
\qquad 0\leq q\leq n,
\\[2mm]
\label{5.04}
\lambda_q^{(-)}(R)
&\stackrel{R\to \infty}\longrightarrow&
-\omega R (2q-n-1),
\qquad 1\leq q\leq n.
\eea
Let us elucidate how the coefficients $W_{nqm}^{(\pm)}(R)$ behave within the
limits considered. It follows from  (\ref{4.1}) that
\bea
\label{5.1}
W_{nqm}^{(\pm)}(R)=
\int \Psi_{nm}^*(r,\varphi)
\Psi_{nq}^{(\pm)}(\xi,\eta;R)dv
\eea
and, therefore, the dependence of the coefficients $W_{nqm}^{(\pm)}(R)$ on $R$
is determined by quantum numbers $n$ and $q$ only. This means that as
$R\rightarrow 0$ and  $R\rightarrow\infty$ all the coefficients
$W_{nqm}^{(\pm)}(R)$ in the three-term recurrence relation (\ref{4.8}) have the
same order in $R$. Now multiply (\ref{4.8}) by $\omega R$ and turn  $R$ to
zero. It follows from (\ref{5.01}), (\ref{5.02}) that within this limit the
expansion coefficients (\ref{4.1}) are equal to zero at $m\not=q$. Nonzero
values of  $W_{nqm}^{(\pm)}(R)$ are determined from the normalization condition
(\ref{4.9}) and equal
\bea
W_{nqm}^{(\pm)}(0)=\frac{1}{2}\delta_{qm},
\qquad
1\leq m\leq n,
\qquad
W_{nq0}^{(\pm)}(0)=\delta_{q0}.
\eea
In the limit $R\rightarrow \infty$, as is seen from (\ref{5.03}) and
(\ref{5.04}), the three-term recurrence relations (\ref{4.8}) take the form
\bea
-(2q-n)W_{nqm}^{(+)}(\infty)
&=&
\frac{1}{2}\sqrt{(n-m)(n+m+1)}W_{nqm+1}^{(+)}(\infty)+
\nonumber
\\[2mm]
&+&\frac{1}{2}\sqrt{(n+m)(n-m+1)}W_{nqm-1}^{(+)}(\infty),
\\[2mm]
-(2q-n-1)W_{nqm}^{(-)}(\infty)
&=&
\frac{1}{2}\sqrt{(n-m)(n+m+1)}W_{nqm+1}^{(-)}(\infty)+
\nonumber
\\[2mm]
&+&\frac{1}{2}\sqrt{(n+m)(n-m+1)}W_{nqm-1}^{(-)}(\infty).
\eea
A comparison of these formulae with the recurrence relations (\ref{4.7})
testifies to that the limiting values for the coefficients
$W_{nqm}^{(\pm)}(\infty)$ with an accuracy up to a phase factor, which depends
only on the quantum numbers $n$ and $q$, coincide with the Wigner $d$ -
function of the right angle. To provide a transition in expansion (\ref{4.1})
as $R\rightarrow \infty$ suffice it to choose these phase factors as follows:
\bea
W_{nqm}^{(+)}(\infty)=(-i)^{n-(2q-n)}
d_{2q-n,m}^n
\left(\frac{\pi}{2}\right),
\\[2mm]
W_{nqm}^{(-)}(\infty)=(-i)^{n-(2q-n-1)}
d_{2q-n-1,m}^n
\left(\frac{\pi}{2}\right).
\eea
Then using formula (\ref{4.13})one can easily be convinced that within the
limits $R\rightarrow 0$ and $R\rightarrow\infty$ expansion (\ref{4.10}) turns
into the expansion of the polar basis over the first parabolic basis and into
identical transformation, respectively.

Let us now find corrections to the polar and parabolic bases at small and large
$R$.

At $\omega R\ll 1$ the operator $\omega R {\cal P}$ is considered perturbed. In
this case, the polar wave functions with definite parity will be unperturbed
wave functions. Solutions to the equations
\bea
\label{ee1}
(- L^2 - \omega R {\cal P})
\Psi_{nq}^{(\pm)}(\xi,\eta;R)=
\lambda_{nq}^{(\pm)}(R)
\Psi_{nq}^{(\pm)}(\xi,\eta;R)
\eea
should be sought in the form
\bea
\label{ee2}
\Psi_{nq}^{(\pm)}(\xi,\eta;R)=
\sum_m
T_{nqm}^{(\pm)}(R)
\Psi_{nm}^{(\pm)}(r,\varphi).
\eea
Here summation over $m$ is in the limits $0\leq m\leq n$ and $1\leq m\leq n$
for even and odd solutions, respectively. Beyond the scope of perturbation
theory, expansion (\ref{ee2})  has the meaning of expansion between two
equivalent solutions of the same Schr\"{o}dinger equation.

According to the formulae of the stationary perturbation theory \cite{LL},
\bea
\lambda_{nq}^{(\pm)}(R)
&=&\lambda_{nq}^{(\pm)}(0)-
\omega R {\cal P}_{qq}^{(\pm)}+ \omega^2R^2\, {\sum_{m}}^{'} \,
\frac{|{\cal P}_{mq}^{(\pm)}|^2}{m^2-q^2},
\nonumber
\\[2mm]
\Psi_{nq}^{(\pm)}(\xi,\eta;R)
&=&
\Psi_{nq}^{(\pm)}(r,\varphi)+
\omega R \, {\sum_{m}}^{'}\, \frac{{\cal P}_{mq}^{(\pm)}}{q^2-m^2}\,
\Psi_{nm}^{(\pm)}(r,\varphi)
\nonumber
\eea
where
\bea
\label{ee3}
{\cal P}_{mq}^{(\pm)}
= \int \Psi_{nm}^{*(\pm)}(r,\varphi) {\cal P}
\Psi_{nq}^{(\pm)}(r,\varphi)\, dv
= \frac{1}{4}
\left[{\cal P}_{mq}+{\cal P}_{-m,-q}\pm {\cal P}_{m,-q}\pm {\cal P}_{-m,q}\right].
\eea
We have already calculated the matrix element ${\cal P}_{mq}^{(\pm)}$. Therefore, with
(\ref{4.6}) taken into account we have
\bea
\label{ee6}
{\cal P}_{mq}^{(\pm)}
&=&
-\frac{1}{4}
\sqrt{(n-q)(n+q+1)}\delta_{m,q+1}-
\frac{1}{4}\sqrt{(n+q)(n-q+1)}\delta_{m,q-1}-
\nonumber
\\[2mm]
&-&
\frac{1+(-1)^{t}}{8}\sqrt{(n+q)(n-q+1)}
\delta_{m,1-q},
\eea
where $t$ is the parity of states with respect to which the matrix elements
(\ref{ee3}) are taken. It follows from (\ref{ee6}) that ${\cal P}_{qq}^{(\pm)}=0$. As
$\lambda_{nq}^{(\pm)}(0) = -q^2$, then at $\omega R\ll 1$ we have
\bea
\lambda_{nq}^{(\pm)}(R)
&=&
-q^2-\frac{\omega^2R^2}{8}
\frac{n^2+n+q^2}{4q^2-1}+
\nonumber
\\[2mm]
&+&
\frac{1+(-1)^{t}}{2}
\frac{\omega^2R^2}{8}n(n+1)
\left(\delta_{q0}-\frac{3}{2}\delta_{qn}\right),
\nonumber
\\[2mm]
\Psi_{nq}^{(\pm)}(\xi,\eta;R)
&=&
\Psi_{nq}^{(\pm)}(r,\varphi)+
\frac{\omega R}{4}
\Biggl\{\frac{(n-q)(n+q+1)}{2q+1}\Psi_{n,q+1}^{(\pm)}(r,\varphi)-
\nonumber
\\[2mm]
&-&\frac{\sqrt{(n+q)(n-q+1)}}{2q-1}
\Psi_{n,q-1}^{(\pm)}(r,\varphi)-
\nonumber
\\[2mm]
&-&
\frac{1+(-1)^{t}}{2}\frac{\sqrt{(n+q)(n-q+1)}}{2q-1}
\Psi_{n,1-q}^{(\pm)}(r,\varphi)\Biggr\}.
\nonumber
\eea

Now let us consider the case  when $\omega R\gg 1$. Having divided both the
sides of equation (\ref{ee1}) by $\omega R$ and considering the operator
$\left(-\frac{L^2}{\omega R}\right)$ to be perturbation and the parabolic wave
functions with definite parity to be proper wave functions of an unperturbed
system, instead of (\ref{ee2}) we get
\bea
\Psi_{nq}^{(\pm)}(\xi,\eta;R) = \sum_{p}
U_{nqp}^{(\pm)}(R)\, \Psi_{np}^{(\pm)}(\mu,\nu).
\nonumber
\eea
At even and odd wave functions summation is over values of $p$ having the same
parity as $n$ or $n-1$, respectively. It follows from (\ref{5.03}),
(\ref{5.04}) and formulae of perturbation theory that at $\omega R\gg 1$ with
an accuracy of first order quantities
\bea
\lambda_{nq}^{(\pm)}(R)
&=&
-\omega Rq^{(\pm)}-(\hat L^2)_{q^{\pm},q^{\pm}},
\nonumber
\\[2mm]
U_{nqp}^{(\pm)}(R)
&=&
\delta_{p,q^{\pm}}-\frac{1}{\omega R}
\frac{(\hat L^2)_{q^{\pm},p}}{q^{\pm}-p}
\nonumber
\eea
where $q^{+} = 2q-n$, $q^{-} = 2q-n-1$ and in the second term of the lower
relation  the condition $q^{\pm}\ne p$ holds. The matrix elements of the
operator $L^2$ of the parabolic bases with definite parity are calculated on
the same principle, we used in deriving formula (\ref{ee6}). This scheme of
calculations results in the formulae
\bea
(L^2)_{kp}
&=&
-\frac{1}{4}\sqrt{(n+k)(n+k-1)(n-k+1)(n-k+2)}
\delta_{p,k-2}-
\nonumber
\\[2mm]
&-&
\frac{1}{4}\sqrt{(n+k+1)(n+k+2)(n-k)(n-k-1)}
\delta_{p,k+2}+\frac{1}{2}(n^2+n-k^2)\delta_{p,k}.
\nonumber
\eea
Using this result one can easily show that at $\omega R\gg 1$
\bea
\lambda_{nq}^{(\pm)}(R)
&=&
-\omega Rq^{\pm}-\frac{n^2+n-(q^{\pm})^2}{2},
\nonumber
\\[2mm]
\Psi_{nq}^{(\pm)}(\xi,\eta;R)
&=&\Psi_{nq^{\pm}}^{(\pm)}(\mu,\nu)+
\nonumber
\\[2mm]
&+&
\frac{1}{8\omega R}
\Biggl[\sqrt{(n-q^{\pm})(n-q^{\pm}-1)(n+q^{\pm}+1)(n+q^{\pm}+2)}
\Psi_{n,q^{\pm}+2}(\mu, \nu)+
\nonumber
\\[2mm]
&+&
\sqrt{(n+q^{\pm})(n+q-1)(n-q^{\pm}+1)(n-q^{\pm}+2)}
\Psi_{n,q^{\pm}-2}(\mu, \nu)\Biggr].
\nonumber
\eea
With the matrix elements ${\cal P}_{mq}^{(\pm)}$ and  $(L^2)_{q^{\pm}p}$ one can also
calculate the next terms of the series of perturbation theory.

\section{Fundamental bases of the two-dimensional hydrogen atom
in the continuous spectrum}
\markboth{CHAPTER 1. TWO DIMENSIONAL HYDROGEN ATOM}{1.7. FUNDAMENTAL BASES IN THE CONTINUOUS SPECTRUM}

\subsection{Polar basis}
In the polar coordinates, after separation of variables the
Schr\"{o}dinger equation can easily be put in a radial equation in which for
the continuous spectrum, i.e. $E>0$, it is easier to pass to the new variable
$\rho = - 2ik r$, where $k=\sqrt{2E}$. As a result, the radial equation itself
takes the form
\bea
\label{cont-rad1}
\frac{d^2 R}{d \rho^2} + \frac{1}{\rho} \frac{d R}{d \rho}
- \left(\frac{m^2}{\rho^2} + \frac{1}{4} - \frac{i}{k \rho}\right)
R = 0.
\eea
Studying the behavior of the wave function as $\rho \to 0$ and $\rho \to
\infty$ one can easily see that the solution of equation (\ref{cont-rad1})
should be sought for in the form
\bea
\label{cont-rad2}
R (\rho) = e^{-\rho/2} \, \rho^{|m|} \, W(\rho).
\eea
Then the function $W(\rho)$ obeys the equation for the degenerate
hypergeometric function
\bea
\label{cont-rad-03}
\rho \frac{d^2 W}{d \rho^2} + [(2|m|+1) - \rho]
\frac{d W}{d \rho} - \left(|m| + \frac{1}{2} - \frac{i}{k}\right)
W = 0,
\eea
and the radial wave function should be determined the solution of this equation
regular at zero
\bea
\label{cont-rad3}
R_{r m} (r, \varphi)
=
C_{km} \,
\frac{(-2ikr)^{|m|}}{(2|m|)!} \,
e^{ikr} \,
F\left(\frac{1}{2}+|m|-\frac{i}{k};\, 2|m|+1; \, -2ikr \right).
\eea
The normalization factor $C_{km}$ can easily be determined by passing to large
$r$ and using the asymptotic expansion
\bea
\label{cont-rad4}
F(\alpha; \, \gamma; \, z) \approx
\frac{\Gamma(\gamma)}{\Gamma(\gamma-\alpha)} (-z)^{-\alpha}
+
\frac{\Gamma(\gamma)}{\Gamma(\alpha)} e^{z} z^{\alpha-\gamma}.
\eea
Substituting this asymptotic expansion in expression for $R_{km}$, after simple
transformations we cone to the formula
\bea
\label{cont-rad4-1}
R_{km} (r) &\approx&
\frac{C_{km}\, e^{-\pi/2k}}
{|\Gamma(|m|+ \frac12 + \frac{i}{k})|} \,
\frac{(2|m|)!}{\sqrt{2k}} \times\\
\nonumber
\\
&\times&
\frac{2}{\sqrt{r}} \,
\cos\left[kr + \frac{1}{k}\ln 2kr - \frac{\pi}{2} (|m|+1)
+ \delta_{|m|}(k)\right]
\nonumber
\eea
in which the quantity $\delta_{|m|}(k)$ is determined as an argument of the
$\Gamma$ function
\bea
\label{cont-rad5}
\delta_{|m|} (k) =
\arg \Gamma\left(|m|+ \frac12 - \frac{i}{k}\right).
\eea
Now it is obvious that the wave function $\Psi_{km}(r, \varphi) = R_{km} (r)\,
e^{im\varphi}/\sqrt{2\pi}$ normalized by the condition
\bea
\label{2.7}
\int_0^{\infty}
\int_0^{2\pi}
\Psi_{km}(r, \varphi)
\Psi_{k'm'}^*(r, \varphi)r dr d\varphi
= 2\pi \delta (k-k') \delta_{mm'},
\eea
should be determined by the expression
\bea
\label{2.6}
\Psi_{km}(r, \varphi)
&=&
C_{km} \,
\frac{(2kr)^{|m|}}{(2|m|)!} \,
e^{ikr} \,
F\left(\frac{1}{2}+|m|-\frac{i}{k}; 2|m|+1; -2ikr\right)
\frac{e^{im\varphi}}{\sqrt{2\pi}},
\eea
where
\bea
\label{2.8}
C_{km}=
e^{\frac{\pi}{2k}} \,
\sqrt{2k} \,
\left|\Gamma\left(\frac{1}{2}+|m|-\frac{i}{k}\right)\right|.
\eea

\subsection{Parabolic basis} In the parabolic coordinates $\mu$ and $\nu$
the method of separation of variables leads to the equations
\bea
\frac{d^2\psi_1(\tilde \mu)}{d{\tilde \mu}^2}+
\left(\frac{{\tilde  \mu}^2}{4}+
\frac{1+\beta}{k}\right)
\psi_1(\tilde  \mu)
&=& 0,
\\ [2mm]
\frac{d^2 \psi_2 (\tilde  \nu)}{d{\tilde  \nu} ^2}
+\left( \frac{\tilde \nu^2}{4}+\frac{1-\beta}{k}\right)
\psi_2(\tilde \nu)
&=& 0,
\eea
in which $\tilde  \mu = \sqrt{2k} \mu, \tilde  \nu = \sqrt{2k} \nu$, $\beta$ is
the separation constant and $\Psi(\tilde  \mu, \tilde  \nu) = \psi_1(\tilde
\mu) \psi_2 (\tilde \nu)$. As is known \cite{BE2}, the real solutions of the
equation
\bea
\frac{d^2\omega}{d^2z}+
\left(\frac{z^2}{4}-\varrho\right)\omega=0
\eea
on the real axis can be chosen as follows:
\bea
\omega_1(z)
&=&
\qquad
\frac{\Gamma\left(\frac{3}{4}-\frac{i\varrho}{2}\right)}
{\sqrt{\pi} \,  2^{\frac{3}{4}+\frac{i\varrho}{2}}}
\left\{D_{i\varrho-\frac{1}{2}}(e^{i\frac{\pi}{4}}z)+
D_{i\varrho-\frac{1}{2}}(-e^{i\frac{\pi}{4}}z)\right\},
\\[2mm]
\omega_2(z)
&=&
- \frac{ e^{-\frac{i\pi}{4}}
\Gamma \left(\frac{1}{4}-\frac{i\varrho}{2}\right)}
{{\sqrt \pi} \, 2^{\frac{1}{4}+\frac{i\varrho}{2}}}
\left\{D_{i\varrho-\frac{1}{2}}(e^{i\frac{\pi}{4}}z)-
D_{i\varrho-\frac{1}{2}}(-e^{i\frac{\pi}{4}}z)\right\},
\eea
where $D_\nu(z)$ is the parabolic cylinder function. Taking into account that,
according to \cite{BE2},
\bea
D_\nu(z)=2^{\frac{\nu}{2}}e^{-\frac{z^2}{4}}
\left\{\frac{\Gamma\left(\frac{1}{2}\right)}
{\Gamma\left(\frac{1-\nu}{2}\right)} \,
F\left(\frac{\nu}{2}; \, \frac{1}{2};\, \frac{z^2}{2}\right)+
\frac{z}{\sqrt2}\frac{\Gamma\left(-\frac{1}{2}\right)}
{\Gamma\left(-\frac{\nu}{2}\right)} \,
F\left(\frac{1-\nu}{2}; \, \frac{3}{2}; \, \frac{z^2}{2}\right)
\right\},
\eea
we have
\bea
\omega_1(z)
=
e^{-\frac{iz^2}{4}} \,
F\left(\frac{1}{4}-\frac{i\varrho}{2};\,
\frac{3}{2};\, \frac{iz^2}{2}\right),
\qquad
\omega_2(z)
=
e^{-\frac{iz^2}{4}}\, z  \,
F\left(\frac{3}{4}-\frac{i\varrho}{2};\,
\frac{3}{2};\, \frac{iz^2}{2}\right).
\nonumber
\eea
As in the case with negative energies, the parabolic wave function $\Psi(\mu,
\nu)$ should be even in the change $(\mu, \nu) \rightarrow(- \mu, - \nu)$ and,
consequently, the functions $\psi_1(\tilde \mu)$ and $\psi_2(\tilde \nu)$
should have the same parity. Thus, in the parabolic system of coordinates there
can be two sets of wave functions, namely,

\bea
\label{2.10}
\psi_{k\beta}^{(+)}( u,  v)
&=&
C_{k\beta}^{(+)}e^{ik\frac{ u^2+ v^2}{2}}
F\left(\frac{1}{4}-\frac{i}{2k}-\frac{i\beta}{2k};
\frac{1}{2}; -ik u^2\right)\times
\nonumber
\\[2mm]
&\times&
F\left(\frac{1}{4}-\frac{i}{2k}+\frac{i\beta}{2k};\frac{1}{2};
-ik v^2\right),
\\[2mm]
\label{2.11}
\psi_{k\beta}^{(-)} ( u,  v)
&=&
C_{k\beta}^{(-)}( u v)e^{ik\frac{ u^2+ v^2}{2}}
F\left(\frac{3}{4}-\frac{i}{2k}-\frac{i\beta}{2k};
\frac{3}{2}; -ik u^2\right)\times
\nonumber
\\[2mm]
&\times&
F\left(\frac{3}{4}-\frac{i}{2k}+\frac{i\beta}{2k};
\frac{3}{2};-ik v^2\right).
\eea
As the functions $\psi_{k\beta}^{(+)}(\mu, \nu)$ and $\psi_{k\beta}^{(-)}(\mu,
\nu)$ have different parity with respect to $\nu$, then
\bea
\label{2.12}
\int_0^{\infty}\int_{-\infty}^{\infty}
\Psi_{k \beta}^{(\mp)}(\mu, \nu)
\Psi_{k' \beta'}^{(\pm)^*}(\mu, \nu)
(\mu^2+ \nu^2) du dv = 0,
\eea
and, consequently, none of the sets is complete. We assume that the wave
functions \ref{2.10}) and (\ref{2.11}) satisfy the following orthonormalization
conditions:
\bea
\label{2.13a}
\int_0^{\infty}\int_{-\infty}^{\infty}
\Psi_{k \beta}^{(\pm)}(\mu, \nu)
\Psi_{k' \beta'}^{(\pm)^*}(\mu, \nu)
(\mu^2+ \nu^2) du dv  =
2\pi \, \delta(k-k') \, \delta(\beta-\beta').
\eea
The constants $C_{k\beta}^{(\pm)}$ ensuring these conditions will be calculated
later in this chapter .

\section{Expansion of the Rutherford wave function of the two-dimensional
hydrogen atom over partial waves}
\markboth{CHAPTER 1. TWO DIMENSIONAL HYDROGEN ATOM}{1.8. EXPANSION OF THE RUTHERFORD WAVE FUNCTION}
The wave function describing the Rutherford scattering at $\nu \rightarrow \pm
\infty$ and all $\mu$ should be like
$$
\Psi (\mu, \nu) \rightarrow e^{ik \frac{\mu^2 - \nu^2}{2}}.
$$
It is obvious that functions like this can be constructed only from the
solutions (\ref{2.10}), provided that $\beta = - 1 - \frac{ik}{2}$. Indeed,
using the Kummer transformation \cite{BE1}
$$
F(\alpha; \, \gamma;\, z)  =  e^{-z} \,
F(\gamma-\alpha; \, \gamma;\, -z),
$$
one can easily show that
\bea
\label{2.14}
\Psi_k (\mu, \nu) = C_k \,  e^{ik \frac{\mu^2-\nu^2}{2}} \,
F\left(\frac{i}{k}; \, \frac{1}{2}; \, ik \nu^2 \right).
\eea
Taking the limit $\nu \rightarrow \infty$ and using the asymptotic expansion
(\ref{cont-rad4}) it is easy to establish that at large distances from the
center the wave function (\ref{2.14}) with an accuracy up to logarithmic
distortions typical for the Coulomb field represents a superposition of an
incident plane and divergent circular waves. These logarithmic distortions
apart, and requiring that the amplitude of an incident plane wave should be
equal to unity, for the normalization constant $C_k$ and the scattering
amplitude $f(\varphi)$, we get
\bea
\label{ras-rad4}
C_k
&=&
\frac{e^\frac{\pi}{2k}}{\sqrt\pi}  \,
\Gamma\left(\frac{1}{2}-\frac{i}{k}\right),
\\[2mm]
\label{ras-rad5}
f(\varphi)
&=&
\frac{e^\frac{i\pi}{4}}{\sqrt{2k}}  \,
\frac{\Gamma\left(\frac{1}{2}-\frac{i}{k}\right)}
{\Gamma\left(\frac{i}{k}\right)}\,
\frac{\exp\left(\frac{2i}{k}\ln\sin\frac{\varphi}{2}\right)}
{\sin\frac{\varphi}{2}}.
\eea
It can be seen from the formula (\ref{ras-rad5}) that the scattering amplitude has
variable $k$ both poles and zeros, and the poles of the amplitude
correspond to the poles of $\Gamma$ - the function in the numerator, and to the zeros
- poles of $\Gamma$ - the function in the denominator. As it should
be the poles of the scattering amplitude are located in the region of negative
energies and in their totality determine the energy spectrum of the two-dimensional
a hydrogen atom. The situation with zeros is more curious. It is easy to verify that
amplitude zeros (\ref{ras-rad5}) correspond to the energy spectrum
three-dimensional hydrogen atom. This behavior of the zeros of the scattering amplitude
is not typical for a three-dimensional problem and therefore it is natural to assume that
that it is a consequence of the dependence of the scattering amplitude
(\ref{ras-rad5}) and wave function (\ref{2.14}) of the factor,
defining the dimension of the space in which the process takes place
scattering.

The anomalous behavior of the amplitude (\ref{ras-rad5}) is also reflected in the scattering cross section.
Indeed, passing to ordinary units, it is easy to show that the two-dimensional analogue of the formula
for the Rutherford scattering cross section has the form
\bea
\label{ras-rad6}
d \sigma = \frac{\alpha}{2M v^2}  \,
\tanh \left(\frac{\alpha \pi}{\hbar v}\right) \,
\frac{d \varphi}{\sin^2\frac{\varphi}{2}}.
\eea
This formula is remarkable in the following respect. First, it includes Planck's constant and, consequently,
the cross section of two-dimensional Rutherford scattering does not coincide with its classical limit.
Secondly, it depends on the energy and the fine structure constant in a non-power-law manner.
The only generality of the expression (\ref{ras-rad6}) with the Rutherford formula is that both results
predict that the cross section is independent of the sign of the charge of the scattered particles.

Consider now the expansion of the Rutherford wave function (\ref{2.14}) in the polar basis (\ref{2.6})
\bea
\label{2.21}
\Psi_k(\mu, \nu)=
\sum_{m=-\infty}^\infty \,
W_{km} \, \Psi_{km}(r, \varphi)
\eea
Multiplying (\ref{2.21}) by $e^{-im\varphi}$ and integrating over $d\varphi$,
after simple transformations we get
\bea
\label{2.22}
&&W_{km}
\sqrt{2k}\, \frac{(2kr)^{|m|}}{(2|m|)!} \,
F\left(\frac{1}{2}+|m|-\frac{i}{k};
2|m|+1; -2ikr\right)=
\nonumber
\\[2mm]
&=& \frac{ \, \Gamma \left(\frac{1}{2}-\frac{i}{k}\right)}
{|\Gamma\left(\frac{1}{2}+|m|-\frac{i}{k}\right)|} \,
\frac{1}{2\pi} \,
\int_0^{2\pi} \, e^{-im\varphi}  \,
F\left[\frac{1}{2}-\frac{i}{k};\frac{1}{2};
-ikr(1-\cos\varphi)\right]\, d\varphi.
\eea
Consider the last integral. Expanding the degenerate hypergeometric
function in series, we have:
\bea
{\cal I}_m
&=&
\frac{1}{2\pi}\int_0^{2\pi}
F\left[\frac{1}{2}-\frac{i}{k};\, \frac{1}{2};\,
-ikr(1-\cos\varphi)\right] \, e^{-im\varphi} d\varphi=
\nonumber
\\[2mm]
&=&
\frac{1}{2\pi} \,
\sum_{n=0}^\infty
\frac{(-ikr)^n}{n!}
\frac{\left(\frac{1}{2}-\frac{i}{k}\right)_n}
{\left(\frac{1}{2}\right)_n} \,
\int_0^{2\pi}(1-\cos\varphi)^n e^{-im\varphi}d\varphi.
\eea
It is easy to show that
\bea
\int_0^{2\pi} \, (1-\cos\varphi)^n\, e^{-im\varphi} \,
d\varphi = (-1)^m \frac{\sqrt\pi \, 2^n \, n!
\Gamma\left(n+\frac{1}{2}\right)}
{\Gamma(n-m+1)\Gamma(n+m+1)}.
\eea
As can be seen, this integral is nonzero only under the condition $n \geq |m|$, so that
\bea
{\cal I}_m
=
\frac{(-1)^m}{\Gamma\left(\frac{1}{2}-\frac{i}{k}\right)} \,
\sum_{n=|m|}^\infty
\,
\frac{\Gamma\left(\frac{1}{2}-\frac{i}{k}+n\right) \, (-2ikr)^n}
{\Gamma(n-m+1)\Gamma(n+m+1)}.
\eea
Replacing $n$ with $n+|m|$, after some simple transformations we get
\bea
{\cal I}_m =
\frac{\Gamma\left(\frac{1}{2}-\frac{i}{k}+|m|\right)}
{\Gamma\left(\frac{1}{2}-\frac{i}{k}\right)}
\frac{(2ikr)^{|m|}}{(2|m|)!}
F\left(\frac{1}{2}+|m|-\frac{i}{k};2|m|+1;-2ikr\right).
\eea
Returning now to (\ref{2.22}) we arrive at the following expression for the coefficients $W_{km}$
\bea
\label{2.23}
W_{km}=
\frac{(i)^{|m|}}{\sqrt k} \,
\exp\left\{\arg\Gamma\left(\frac{1}{2}+|m|-\frac{i}{k}\right)\right\}.
\eea
It follows from (\ref{cont-rad5}) that $W_{km}$ is expressed in terms of the Coulomb scattering
phase and, therefore, the transformation itself (\ref{2.21}) can be rewritten as
\bea
\label{2.24}
\Psi_k (\mu, \nu)=
\sum_{m = - \infty}^{\infty} \, (i)^{|m|} \,
\frac{e^{i\delta_{|m|}}}{\sqrt k} \, \Psi_{km}(r, \varphi),
\eea
which is the expansion of the Rutherford wave function in terms of partial waves.

The Coulomb scattering phase in (\ref{2.24}) can be easily expressed in terms of
analytic continuation of the Wigner $d$-function included in the expansion (\ref{2.2})
to the domain of complex values of the moment.
Indeed, using the formula
\bea
d_{m, -J}^J\left(\frac{\pi}{2}\right)=
\sqrt{\frac{(2J)!}{(J+m)!(J-m)!}}
\frac{(-1)^{J+m}}{2^J}
\eea
and substituting $J = (i/k - 1/2)$ into it, after simple calculations
we come to the conclusion that
\bea
e^{i\delta_m}=
\frac{\pi^{\frac{1}{4}}}{(-i)^m(-1)^{\frac{i}{k}-\frac{1}{2}}}
\sqrt{\frac{\Gamma\left(\frac{1}{2}-\frac{i}{k}\right)}
{\Gamma \left(\frac{i}{k}\right)}}
d_{m, \frac{1}{2}-\frac{i}{k}}^{\frac{i}{k}-\frac{1}{2}}
\left(\frac{\pi}{2}\right).
\eea
It follows from the obtained result that the decomposition (\ref{2.24})
is an analytic continuation of the transformation (\ref{2.2}) from $E<0$ to $E>0$ with further
selection of solutions corresponding to Rutherford scattering. As a result, the symmetry of the
two-dimensional Coulomb scattering problem hidden in the transformation (\ref{2.24}) becomes
explicit\footnote{Note that, according to \cite{PER-POP2}, in a similar three-dimensional problem
there is also agreement between transformations in the discrete spectrum and in the theory
scattering (see Chapter 3).}.

\section{Relation between parabolic and polar bases
two-dimensional hydrogen atom in the continuous spectrum}
\markboth{CHAPTER 1. TWO DIMENSIONAL HYDROGEN ATOM}{1.9. RELATION BETWEEN PARABOLIC AND POLAR BASES}
We now study the general problem of expansions relating bases (\ref{2.6}),
(\ref{2.10}) and (\ref{2.11}) to each other. Let us write the expansion of
the parabolic bases (\ref{2.10}) and (\ref{2.11}) in the polar basis (\ref{2.6}):
\bea
\label{2.25}
\Psi_{k\beta}^{(+)}(\mu, \nu)
&=&
\sum_{m=-\infty}^\infty W_{k\beta m}^{(+)}\,
\Psi_{km}(r, \varphi),
\\[3mm]
\label{2.26}
\Psi_{k\beta}^{(-)}(\mu, \nu)
&=&
\sum_{m=-\infty}^\infty W_{k\beta m}^{(-)}\,
\Psi_{km}(r, \varphi).
\eea
The parabolic and polar wave functions included in the expansions (\ref{2.25}) and
(\ref{2.26}) are infinite series of hypergeometric type. The behavior of each of
them at large values of the variable is determined by the formula (\ref{cont-rad4})
and has the form of oscillations, as it happens in the case of the Coulomb radial
wave function (\ref{cont-rad4-1}). Hence it is clear that the methods that we use
so effectively in calculating the coefficients of interbasis expansions in the discrete spectrum
are absolutely inapplicable in the case of a continuous spectrum. To calculate the
coefficients of interbasis expansions in the continuous spectrum - $W_{k\beta m}^{(\pm)}$,
we use another method, proposed for the first time in \cite{POGOSYAN2} and based on the
simple behavior of wave functions near the center, i.e. for $r\sim 0$.

\subsection{Calculation of transition matrices}

Let's pass in the left parts of the relations (\ref{2.25}) and (\ref{2.26})
from parabolic to polar coordinates, then multiply on both sides (\ref{2.25}) and
(\ref{2.26}) by $e^{-im\varphi}$ and we integrate within $(0, 2\pi)$. Then
\bea
\label{2.27}
W_{k\beta m}^{(+)}
&\cdot&
F\left(\frac{1}{2}+|m|-\frac{i}{k}; \, 2|m|+1; \, -2ikr\right)=
\nonumber
\\[2mm]
&=&
\frac{C_{k\beta}^{(+)}}{C_{km}} \,
\frac{(2|m|)!}{2^{|m|}}
\,
\sum_{s=0}^\infty
\sum_{t=0}^\infty
\frac{(v)_s}{\left(\frac{1}{2}\right)_s}\,
\frac{(w)_t}{\left(\frac{1}{2}\right)_t}\,
\frac{(-ikr)^{s+t-|m|}}{s!t!}\,
A_{s t}^m,
\\[2mm]
\label{2.28}
W_{k\beta m}^{(-)}
&\cdot&
F\left(\frac{1}{2}+|m| -\frac{i}{k}; \, 2|m|+1; \, -2ikr\right)=
\nonumber
\\[2mm]
&=&
\frac{C_{k\beta}^{(-)}}{C_{km}} \,
\frac{2i\, (2|m|)!}{2^{|m|}} \,
\sum_{s=0}^\infty
\sum_{t=0}^\infty
\frac{\left(\frac{1}{2}+v\right)_s}{\left(\frac{3}{2}\right)_s}\,
\frac{\left(\frac{1}{2}+w\right)_t}{\left(\frac{3}{2}\right)_t}\,
\frac{(-ikr)^{s+t-|m|}}{s!t!} \,  B_{st}^m.
\eea
These formulas use the following notation
$v = \frac{1}{4}-\frac{i}{2k}(1+\beta)$,
$w=\frac{1}{4}-\frac{i}{2k}(1-\beta)$ and
\bea
A_{st}^m
&=&
\frac{1}{\sqrt{2\pi}}
\int_0^{2\pi} (1+\cos\varphi)^s
(1-\cos\varphi)^t
e^{-im\varphi} d\varphi,
\\[2mm]
B_{st}^m
&=&
\frac{1}{\sqrt{2\pi}}
\int_0^{2\pi}(1+\cos\varphi)^s
(1-\cos\varphi)^t\sin\varphi
e^{-im\varphi}d\varphi.
\eea
It is obvious that $A_{st}^m$ and $B_{st}^m$ are nonzero only under the conditions $s+t\geq |m|$
and $s+t+1\geq|m|$ and therefore, all terms of the series contain $r$ to a non-negative degree.
Now we can pass to the limit $r\rightarrow 0$ in (\ref{2.27}) and (\ref{2.28}).
As a result, only those members of the series survive for which $s+t=|m|$ $s+t=|m|-1$ and we
arrive at the following expressions
\bea
W_{k\beta m}^{(+)}
&=&
\frac{C_{k\beta}^{(+)}}{C_{km}}
\frac{(2|m|)!}{2^{|m|}} \,
\sum_{s=0}^{|m|}
\frac{(v)_s}{\left(\frac{1}{2}\right)_s}
\frac{(w)_{|m|-s}}{\left(\frac{1}{2}\right)_{|m|-s}}
\frac{A_{s,|m|-s}^m}{s!(|m|-s)!},
\\[2mm]
W_{k\beta m}^{(-)}
&=&
\frac{C_{k\beta}^{(-)}}{C_{km}}
\frac{2i\, (2|m|)!}{2^{|m|}} \,
\sum_{s=0}^{|m|-1}
\frac{\left(\frac{1}{2}+v\right)_s}{\left(\frac{3}{2}\right)_s}
\frac{\left(\frac{1}{2}+w\right)_{|m|-1-s}}
{\left(\frac{3}{2}\right)_{|m|-1-s}}
\frac{B_{s,|m|-1-s}^m}{s!(|m|-1-s)!}.
\eea
Taking into account that
\bea
A_{s,|m|-s}^m =
(-1)^{|m|-s}\frac{\sqrt{2\pi}}{2^{|m|}},
\qquad
B_{s,|m|-1-s}^m = i \, {\rm{\mbox sgn}} (m) \, A_{s,|m|-s}^m
\eea
(sgn is a sign function) and a formula (\ref{d8}) connecting between
are two generalized hypergeometric functions $_3F_2$ of unit
argument, after simple calculations we have
\bea
\label{2.31a}
W_{k\beta m}^{(+)}
&=&
(-1)^{|m|}\,
\sqrt{2\pi}\,
\frac{C_{k\beta}^{(+)}}{C_{km}}\,
\frac{\Gamma(w+v+|m|)}{\Gamma(w+v)}\,
{_3F_2}
\left\{\matrix{
v,\,\, |m|,\,\,
-|m|\,\,
\cr
\cr
\frac{1}{2},\,\, v+w
\,\,
\cr}\Bigg|1\right\},
\\[2mm]
\label{2.31b}
W_{k\beta m}^{(-)}
&=&
(-1)^{|m|-1}m\sqrt{2\pi} \,
\frac{C_{k\beta}^{(-)}}{C_{km}} \,
\frac{\Gamma(w+v+|m|)}{\Gamma(w+v-1)}\,\times
\nonumber
\\[2mm]
&\times&
{_3F_2}
\left\{\matrix{
\frac{1}{2}+v,\,\,1+ |m|,\,\,
1-|m|\,\,
\cr
\cr
\frac{3}{2},\,\, v+w+1
\,\,
\cr}\Bigg|1\right\}.
\eea
The obtained formulas completely solve the problem of transition from the
Coulomb parabolic basis of the two-dimensional hydrogen atom to the polar one in
continuous spectrum.

Let us now write out the integral representations for the coefficients
interbasis expansions $W_{k\beta m}^{(\pm)}$.
It can be verified by direct calculation that for $Re(a+\alpha +1)>1$,
$Re(a+\beta+1)>1$ and $N=0,1,...$ the formula
\bea
{_3F_2}\left\{\matrix{
- N,\,\, N+2\alpha+2\beta+1,\,\,
-a+\alpha\,\,
\cr
\cr
\alpha+\beta+1,\,\, \alpha+\beta-2a
\,\,
\cr}\Bigg|1\right\} =
\frac{\Gamma(2a-N-\alpha-\beta+1)\Gamma(\alpha+\beta+1)}
{\Gamma(2a-\alpha-\beta+1)\Gamma(\alpha+\beta+N+1)}\times
\nonumber
\\[3mm]
\times
\left(\frac{1}{2}\right)^{2a+\alpha+\beta+1}
\frac{N! \, \Gamma(2a+N+\alpha+\beta)}
{\Gamma(a+\alpha+1) \Gamma(a+\beta+1)}
\int_{-1}^1
(1-x)^{a+\beta}(1+x)^{a+\alpha}P_N^{(\alpha+\beta, \alpha+\beta)}(x)dx,
\nonumber
\eea
where $P_N^{(\gamma,\gamma)}(x)$ is a Jacobi polynomial. Taking advantage of this
identity and noting that according to \cite{SEGO}
\bea
P_{|m|}^{(-\frac{1}{2},- \frac{1}{2})}(\cos\varphi)
=
\frac{(2|m|-1)!!}{(2|m|)!!}\, \cos|m|\varphi,
\qquad
P_{|m|-1}^{(\frac{1}{2},\frac{1}{2})}(\cos\varphi)
=
2\frac{(2|m|-1)!!}{(2|m|)!!} \,
\frac{\sin|m|\varphi}{\sin\varphi},
\nonumber
\eea
we arrive at the required integral representations
\bea
\label{2.32a}
W_{k\beta m}^{(+)}
&=&
2^{v+w}\sqrt{2\pi}\frac{C_{k\beta}^{(+)}}{C_{km}}
\frac{\Gamma (1+|m|-v-w)}{\Gamma\left(\frac{1}{2}-v\right)
\Gamma\left(\frac{1}{2}-w\right)}\times
\nonumber
\\[2mm]
&\times&
\int_0^\pi(1-\cos\varphi)^{-v}(1+\cos\varphi)^{-w} \cos m\varphi \,
d\varphi,
\\[2mm]
\label{2.32b}
W_{k\beta m}^{(-)}
&=&
2^{v+w-1}\sqrt{2\pi}\frac{C_{k\beta}^{(-)}}{C_{km}}
\frac{\Gamma (1+|m|-v-w)}{\Gamma\left(1-v\right)
\Gamma\left(1-w\right)}\times
\nonumber
\\[2mm]
&\times&
\int_0^\pi(1-\cos\varphi)^{-v}(1+\cos\varphi)^{-w}
\sin m\varphi \, d\varphi.
\eea
Integral representations (\ref{2.32a}) and (\ref{2.32b}) will be needed
us in the next section when calculating the parabolic normalization
constants $C_{k\beta}^{(\pm)}$ and constructing an inverse expansion
polar basis on parabolic.

\subsection{Transition matrix properties and normalization constants}

In this section, some general relations are obtained, which
obey the transition matrices $W_{k\beta m}^{(\pm)}$, computed
constants $C_{k\beta}^{(\pm)}$ providing the conditions
orthonormality (\ref{2.13a}), and the inverse transformations
expansions (\ref{2.25}) and (\ref{2.26}).

{\bf 1.} Substituting the expansions (\ref{2.25}) and (\ref{2.26})
into the conditions (\ref{2.13a}) and (\ref{2.12}), we get
\bea
\label{2.34-1}
\sum_{m=-\infty}^{\infty}
W_{k\beta m}^{(\pm)}\, W_{k{\beta}' m}^{(\mp)^*}
&=& 0,
\\[2mm]
\label{2.34}
\sum_{m=-\infty}^{\infty}
W_{k\beta m}^{(\pm)}\, W_{k{\beta}' m}^{(\pm)^*}
&=&
\delta (\beta-\beta').
\eea
The relation (\ref{2.34}) allows us to calculate the normalization constants
$C_{k\beta}^{(\pm)}$. From (\ref{2.34}) and integral representations
(\ref{2.32a}) and (\ref{2.32b}) it follows that
\bea
\label{2.35a}
&&\delta(\beta- \beta')
=
\frac{2\pi}{k}
\frac{e^{-\frac{\pi}{k}} C_{k\beta'}^{(+)} C_{k\beta'}^{(+)^*}}
{\Gamma\left(\frac{1}{4}+\frac{i}{2k}+\frac{i\beta}{2k}\right)
\Gamma\left(\frac{1}{4}+\frac{i}{2k}-\frac{i\beta}{2k}\right)
\Gamma\left(\frac{1}{4}-\frac{i}{2k}-\frac{i\beta'}{2k}\right)
\Gamma\left(\frac{1}{4}-\frac{i}{2k}+\frac{i\beta'}{2k}\right)}\times
\nonumber
\\[2mm]
&\times&
\int_0^\pi \int_0^\pi \,
\frac{A^{(+)}(\varphi, \varphi') \, d\varphi \, d\varphi'}
{(1-\cos\varphi)^{v(\beta)}
(1+\cos\varphi)^{w(\beta)}
(1-\cos\varphi')^{v^*(\beta')}
(1+\cos\varphi')^{w^*(\beta')}},
\\[2mm]
\label{2.35b}
&&\delta(\beta- \beta')
=
\frac{\pi}{2k}
\frac{e^{-\frac{\pi}{k}} C_{k\beta'}^{(-)} C_{k\beta'}^{(-)}}
{\Gamma\left (\frac{3}{4}+\frac{i}{2k}+\frac{i\beta}{2k}\right)
\Gamma\left( \frac{3}{4}+\frac{i}{2k}-\frac{i\beta}{2k}\right)
\Gamma\left(\frac{3}{4}-\frac{i}{2k}-\frac{i\beta'}{2k}\right)
\Gamma\left(\frac{3}{4}-\frac{i}{2k}+\frac{i\beta'}{2k}\right)}\times
\nonumber
\\[2mm]
&\times&
\int_0^\pi
\int_0^\pi \,
\frac{A^{(-)}(\varphi, \varphi') \, d\varphi \, d\varphi'}
{(1-\cos\varphi)^{v(\beta)}
(1+\cos\varphi)^{w(\beta)}
(1-\cos\varphi')^{v^*(\beta')}
(1+\cos\varphi')^{w^*(\beta')}},
\eea
where
\bea
A^{(+)}(\varphi, \varphi')
&=&
\sum_{m=-\infty}^{\infty}
\cos m \varphi \cos m \varphi'
= \pi\delta(\varphi-\varphi'),
\nonumber
\\[2mm]
A^{(-)}(\varphi, \varphi')
&=&
\sum_{m=-\infty}^\infty
\sin m\varphi \sin m \varphi'
= \pi\delta(\varphi-\varphi').
\nonumber
\eea
The last two formulas take into account that $-\pi\leq\varphi-\varphi'\leq\pi$.
Using now the expressions for $A^{(\pm)}(\varphi, \varphi')$
from (\ref{2.35a}) and (\ref{2.35b}) we have:
\bea
\label{2.36a}
\delta(\beta-\beta')=
\frac{2\pi^2e^{-\frac{\pi}{k}} C_{k\beta'}^{(+)} C_{k\beta'}^{(+)^*}
I(\beta, \beta')}
{k\Gamma\left (\frac{1}{4}+\frac{i}{2k}+\frac{i\beta}{2k}\right)
\Gamma\left( \frac{1}{4}+\frac{i}{2k}-\frac{i\beta}{2k}\right)
\Gamma\left(\frac{1}{4}-\frac{i}{2k}-\frac{i\beta'}{2k}\right)
\Gamma\left(\frac{1}{4}-\frac{i}{2k}+\frac{i\beta'}{2k}\right)},
\\[2mm]
\label{2.36b}
\delta(\beta-\beta')=
\frac{\pi^2e^{-\frac{\pi}{k}} C_{k\beta'}^{(-)}
C_{k\beta'}^{(-)} I(\beta,\beta')}
{2k\Gamma\left (\frac{3}{4}+\frac{i}{2k}+\frac{i\beta}{2k}\right)
\Gamma\left( \frac{3}{4}+\frac{i}{2k}-\frac{i\beta}{2k}\right)
\Gamma\left(\frac{3}{4}-\frac{i}{2k}-\frac{i\beta'}{2k}\right)
\Gamma\left(\frac{3}{4}-\frac{i}{2k}+\frac{i\beta'}{2k}\right)},
\eea
where
\bea
\label{2.36-1}
I(\beta, \beta')=
\int_0^\pi
\frac{d\varphi}{\sin\varphi}
\left(\frac{1-\cos\varphi}
{1+\cos\varphi}\right)^{\frac{i}{k}(\beta-\beta')}.
\eea
Let us change the variable in the last integral according to
$\cos\varphi = \tanh\mu$ and use the fact that
$$
\frac{1-\cos\varphi}{1+\cos\varphi} = e^{-2\mu}.
$$
Then
\bea
\label{2.37}
I(\beta, \beta')=
\int_{-\infty}^\infty
e^{-\frac{i\mu}{k}(\beta-\beta')} d\mu =
2\pi k \delta (\beta-\beta').
\eea
From (\ref{2.36a}), (\ref{2.36b}), and (\ref{2.37}) it follows that
with a real choice of normalization constants $C_{k\beta}^{(\pm)}$
we have
\bea
\label{2.38a}
C_{k\beta}^{(+)}
&=&
\frac{e^{\frac{\pi}{2k}}}{\sqrt{4\pi^3}}
\left|\Gamma\left(\frac{1}{4}+\frac{i}{2k}+\frac{i\beta}{2k}\right)
\Gamma\left(\frac{1}{4}+\frac{i}{2k}-\frac{i\beta}{2k}\right)\right|,
\\[2mm]
\label{2.38b}
C_{k\beta}^{(-)}
&=&
\frac{e^{\frac{\pi}{2k}}}{\sqrt{\pi^3}}
\left|\Gamma\left(\frac{3}{4}+\frac{i}{2k}+\frac{i\beta}{2k}\right)
\Gamma\left(\frac{3}{4}+\frac{i}{2k}-\frac{i\beta}{2k}\right)\right|.
\eea

{\bf 2.} We now calculate the following integrals:
\bea
V_{m m'}^{(\pm)}=
\int_{-\infty}^\infty
W_{k\beta m}^{(\pm)}W_{k\beta' m}^{(\pm)^*}d\beta.
\nonumber
\eea
Using integral representations (\ref{2.32a}) and
(\ref{2.32b}), we get
\bea
\label{2.39a}
V_{m m'}^{(+)}
&=&
\frac{(i)^{|m'|-|m|}}{2k\pi^2}
\frac{\Gamma \left(\frac{1}{2}+|m|+\frac{i}{k}\right)
\Gamma\left(\frac{1}{2}+|m'|-\frac{i}{k}\right)}
{\left|\Gamma\left(\frac{1}{2}+|m|+\frac{i}{k}\right)
\Gamma\left(\frac{1}{2}+|m'|+\frac{i}{k}\right)\right|}\times
\nonumber
\\[2mm]
&\times&
\int_0^\pi d\varphi
\int_0^\pi d\varphi'
\frac{\cos m\varphi}{(\sin\varphi)^{\frac{1}{2}-\frac{i}{k}}}
\frac{\cos m'\varphi'}{(\sin\varphi')^{\frac{1}{2}+\frac{i}{k}}}
G(\varphi, \varphi),
\\[3mm]
\label{2.39b}
V_{m m'}^{(-)}
&=&
\frac{(i)^{|m'|-|m|}}{2k\pi^2}
\frac{\Gamma \left(\frac{1}{2}+|m|+\frac{i}{k}\right)
\Gamma\left(\frac{1}{2}+|m'|-\frac{i}{k}\right)}
{\left|\Gamma\left(\frac{1}{2}+|m|+\frac{i}{k}\right)
\Gamma\left(\frac{1}{2}+|m'|+\frac{i}{k}\right)\right|}\times
\nonumber
\\[2mm]
&\times&
\int_0^\pi d\varphi
\int_0^\pi d\varphi'
\frac{\sin m\varphi}{(\sin\varphi)^{\frac{1}{2}-\frac{i}{k}}}
\frac{\sin m'\varphi'}{(\sin\varphi')^{\frac{1}{2}+\frac{i}{k}}}
G(\varphi, \varphi),
\eea
where
\bea
G(\varphi, \varphi') = \int_{-\infty}^{\infty}
\left\{\frac{(1-\cos\varphi)(1+\cos\varphi')}
{(1+\cos\varphi)(1-\cos\varphi')}\right\}^
{\frac{i\beta}{2k}}d \beta=
2\pi k \delta(\nu-\nu')
\nonumber
\eea
and $\cos\varphi=th\nu, \cos\varphi'=th\nu'$. Substituting the resulting
expression for $G(\varphi,\varphi')$ in (\ref{2.39a}) and (\ref{2.39b}),
given the definition of Chebyshev polynomials \cite{BE2}
\bea
T_n(\cos\varphi)=\cos n\varphi,
\qquad
U_n(\cos\varphi)=
\frac{\sin(n+1)\varphi}{\sin\varphi}
\nonumber
\eea
and orthonormality conditions
\bea
\int_{-1}^1
\frac{dx}{\sqrt{1-x^2}}T_{|m|}(x)T_{|m'|}(x)
&=&
\frac{\pi}{2}(\delta_{m, m'}+\delta_{m, -m'}),
\nonumber
\\[2mm]
\int_{-1}^1
\sqrt{1-x^2}dx
U_{|m|-1}(x)U_{|m'|-1}(x)
&=&
\frac{\pi}{2}(\delta_{m, m'}-\delta_{m,-m'}),
\nonumber
\eea
we have
\bea
\label{2.40}
\int_{-\infty}^\infty
W_{k\beta m}^{(\pm)}W_{k\beta m'}^{(\pm)^*}d\beta
= \frac{1}{2}(\delta_{m m'}\pm\delta_{m,-m'}).
\eea
This formula allows you to wrap transformations (\ref{2.25})
and (\ref{2.26}) and obtain the following expansion of the polar basis in terms of
parabolic
\bea
\label{2.41}
\Psi_{km}(r,\varphi) =
\int_{-\infty}^\infty \, \left\{
W_{k\beta m}^{(+)^*}\, \Psi_{k\beta}^{(+)}(\mu, \nu)
+ W_{k\beta m}^{(-)^*}\, \Psi_{k\beta}^{(-)}(\mu, \nu)
\right\}d\beta.
\eea
Using the reality of parabolic normalizations
constants $C_{k\beta}$, the explicit form of the phase factor in
polar normalization constant $C_{km}$ and the relation
(\ref{d8}), after some calculations we arrive at the following
properties of transition matrices:
\bea
(W_{k\beta m}^{(\pm)})^*=\pm W_{k\beta m}^{(\pm)}.
\nonumber
\eea

\section{Transition from continuous to discrete spectrum}
\markboth{CHAPTER 1. TWO DIMENSIONAL HYDROGEN ATOM}{1.10. TRANSITION FROM CONTINUOUS TO DISCRETE SPECTRUM}
Let's see how from the transformations (\ref{2.25}),
(\ref{2.26}) and (\ref{2.41}) belonging to the region $E>0$, one can
get transformations valid for $E<0$ (\ref{2.2}) and
its inverse transformation of the polar wave function to
parabolic. This will allow us to simultaneously receive more
useful forms of writing transition matrices in the discrete spectrum.

{\bf 1.10.1.}
Let's go to (\ref{2.25}) and (\ref{2.26}) to the region of negative
energies, i.e. Let's make the substitutions:
\begin{eqnarray*}
k\rightarrow i|k|=i\left(n+\frac{1}{2}\right)^{-1}=i\lambda
\end{eqnarray*}
- in the polar wave function,
\begin{eqnarray*}
\frac{1}{4}-\frac{(1+\beta)}{2|k|}= -\frac{n_1}{2},
\quad
n_1=0,2, 4,\dots;
\qquad
\frac{1}{4}-\frac{(1-\beta)}{2|k|} = -\frac{n_2}{2},
\quad n_2=0, 2, 4, \dots
\end{eqnarray*}
- in the parabolic wave function $\Psi_{k\beta}^{(+)}(\mu,\nu)$;
\begin{eqnarray*}
\frac{3}{4}-\frac{(1+\beta)}{2|k|}= -\frac{n_1}{2}
+\frac{1}{2}, \quad n_1=1, 3, 5,\dots;
\qquad
\frac{1}{4}-\frac{(1-\beta)}{2|k|}= -\frac{n_2}{2}+\frac{1}{2},
\quad n_2=1, 3, 5,\dots
\end{eqnarray*}
- in the parabolic wave function $\Psi_{k\beta}^{(-)}(\mu,\nu)$.

Here, all wave functions refer to the same energy level and
so $n= (n_1+n_2)/2$. Then given the formula (\ref{hermit}),
we have
\bea
\Psi_{n_1 n_2}^{(+)} (\mu, \nu)
&=& \sum_{m=-n}^n {\tilde W}_{n_1 n_2 m}^{(+)}
\Psi_{nm}(r,\varphi),
\label{2.43a}
\\[2mm]
\Psi_{n_1 n_2}^{(-)} (\mu, \nu)
&=& \sum_{m=-n}^n {\tilde W}_{n_1 n_2 m}^{(-)}
\Psi_{nm}(r, \varphi),
\label{2,43b}
\eea
where $\Psi_{nm}(r,\varphi)$ is the polar wave function of the discrete
spectrum (\ref{hyd-pol6}), $\Psi_{n_1 n_2}^{(+)}$ and $\Psi_{n_1 n_2}^{(-)}$
coincides with (\ref{hyd-pol17}) for $n_1$ and $n_2$ even and $n_1$ and $n_2$
odd, respectively, and
\bea
{\tilde W}_{n_1 n_2 m}^{(+)}
&=&
C_{n_1 n_2}^{nm} \, \frac{(n)!}{\left(\frac{n_1}{2}\right)!
\left(\frac{n_2}{2}\right)!}
{_3F_2} \left\{\matrix{ -\frac{n_1}{2},\,\, |m|,\,\, -|m|\,\, \cr
\cr \frac{1}{2},\,\, -n \,\, \cr}\Biggr|1\right\}, \label{2.44a}
\\[3mm]
\label{2.44b}
{\tilde W}_{n_1 n_2 m}^{(-)}
&=& i C_{n_1 n_2}^{nm}\,
\frac{2m(n-1)!}{\left(\frac{n_1-1}{2}\right)!
\left(\frac{n_2-1}{2}\right)!} {_3F_2} \left\{\matrix{
-\frac{1-n_1}{2},\,\,1+ |m|,\,\, 1-|m|\,\, \cr \cr
\frac{3}{2},\,\,1-n \,\, \cr}\Biggr|1\right\},
\eea
where
\begin{eqnarray*}
C_{n_1 n_2}^{nm} = \frac{(-1)^n}{2^n}\,
\sqrt{\frac{(n_1)!(n_2)!}{(n+|m|)!(n-|m|)!}}.
\end{eqnarray*}
The formulas (\ref{2.44a}) and (\ref{2.44b}) can be written in a single
form. Indeed, according to (\ref{d8})
\begin{eqnarray*}
{_3F_2} \left\{\matrix{ -\frac{n_1}{2},\,\, |m|,\,\, -|m|\,\,
\cr \cr \frac{1}{2},\,\, -n \,\, \cr}\Biggr|1\right\}
&=&
\frac{(-1)^{\frac{n_1}{2}}\Gamma\left(\frac{1}{2}\right)
\Gamma\left(|m|+\frac{1}{2}\right)\Gamma(|m|+1)
\Gamma\left(\frac{n_2}{2}+1\right) }
{\Gamma\left(n+1\right)\Gamma\left(\frac{1}{2}+\frac{n_1}{2}\right)
\Gamma\left(1+|m|-\frac{n_1}{2}\right)
\Gamma\left(\frac{1}{2}+|m|-\frac{n_1}{2}\right)}\times
\\[2mm]
&\times& {_3F_2} \left\{\matrix{
\frac{1-n_1}{2},\,\,-\frac{n_1}{2},\,\, -n+|m| \cr \cr
\frac{1}{2}+|m|-\frac{n_1}{2},\,\, 1+|m|-\frac{n_1}{2} \,\,
\cr}\Biggr|1\right\},
\end{eqnarray*}
\begin{eqnarray*}
{_3F_2} \left\{\matrix{ \frac{1-n_1}{2},\,\,1+ |m|,\,\, 1-|m| \cr
\cr \frac{3}{2},\,\,1 -n \,\, \cr}\Biggr|1\right\}
&=&
\frac{(-1)^{\frac{n_1-1}{2}}\Gamma\left(|m|+\frac{1}{2}\right)
\Gamma(|m|)\Gamma\left(\frac{3}{2}\right)
\Gamma\left(1+\frac{n_2}{2}\right)}
{\Gamma(n)\Gamma\left(1+\frac{n_1}{2}\right)
\Gamma\left(1+|m|-\frac{n_1}{2}\right)
\Gamma\left(\frac{1}{2}+|m|-\frac{n_1}{2}\right)}\times
\\[2mm]
&\times& {_3F_2} \left\{\matrix{ \frac{1-n_1}{2},\,\,
-\frac{n_1}{2},\,\, -n+|m| \cr \cr \frac{1}{2}+|m|-\frac{n_1}{2}
\,\, 1+|m|-\frac{n_1}{2},\,\,\cr}\Biggr|1\right\},
\end{eqnarray*}
where the first relation refers to the case when $n_1$ and $n_2$
are even, and the second - when $|m|\geq1$ and $n_1$, $n_2$ are odd.
Given the last two formulas, and the formulas (\ref{2.44a}) and (\ref{2.44b})
we obtain the expansion of the parabolic wave function of the discrete
spectrum (\ref{hyd-pol17}) by polar basis (\ref{hyd-pol6})
\bea
\Psi_{n_1 n_2}(\mu, \nu)
=
\sum_{m=-n}^n {\tilde W}_{n_1 n_2 m}\, \Psi_{nm}(r,\varphi).
\label{2.45} \eea
Here
\bea
&&{\tilde W}_{n_1 n_2 m}
= \frac{(-1)^{n+\frac{n_1}{2}}}{2^n}
\left[\frac{1+(-1)^{n_1}}{2}+\frac{1-(-1)^{n_1}}{2}{\rm sgn}(m)\right]
\frac{(2|m|)!}{\Gamma(2|m|-n_1+1)}\times
\nonumber
\\[2mm]
&\times&
\sqrt{\frac{(n_1)!}{(n_2)!(n+|m|)!(n-|m|)!}}\,
{_3F_2}\left\{\matrix{ \frac{1-n_1}{2},\,\, - \frac{n_1}{2},\,\, -n+|m|
\cr \cr \frac{1}{2}+|m|-\frac{n_1}{2},\,\, 1+|m|-\frac{n_1}{2}
\,\, \cr}\Biggr|1\right\}.
\label{2.46}
\eea
The transition matrix (\ref{2.46}) can be represented as coefficients
Clebsch-Gordan group $SU(2)$, if we use the formula expressing
these coefficients in terms of the generalized hypergeometric function ${_3F}_2$
from argument $1$ \cite{VAR}:
\begin{eqnarray}
C_{a \alpha; b \beta}^{c\gamma}
&=&
\frac{\delta_{\gamma, \alpha+\beta}\Delta(abc)}
{(a+b-c)!(c-b+\alpha)!(c-a+\gamma)!}
\left[\frac{(2c+1)(a+\alpha)!(b-\beta)!(c+\gamma)!(c-\gamma)!}
{(a-\alpha)!(b+\beta)!} \right]^{1/2}\times
\nonumber
\\[2mm]
\label{ecsf.1.13}
&\times&
{_3F}_2 \left\{\begin{array}{l}
-a-b+c, -a+\alpha, -b-\beta \\
-a+c-\beta+1,  -b+c+\alpha+1  \\
\end{array}
\biggr| 1 \right\}.
\end{eqnarray}
The $\Delta$ symbol included in this formula has the form
\begin{eqnarray}
\label{ecsf.1.14}
\Delta(abc) = \sqrt{\frac{(a+b-c)!(a-b-c)!(b-a+c)!}
{(a+b+c+1)!}}\,.
\end{eqnarray}
Comparing (\ref{2.46}) with the formula (\ref{ecsf.1.13}) it is easy to see that
that the hypergeometric functions coincide if the equalities
\begin{eqnarray*}
c=|m|-\frac{1}{2},
\quad
a=b= \frac{n}{2} - \frac{1}{4},
\quad
\gamma=-\frac{1}{2},
\quad
\alpha=-\frac{1}{4}+\frac{n_2-n_1}{4}
\quad
\beta=-\frac{1}{4}-\frac{n_2-n_1}{4}.
\end{eqnarray*}
Substituting the values of the parameters obtained above
$c,a,\gamma,\alpha,\beta$ into the representation (\ref{ecsf.1.13}), we find that
\bea
{\tilde W}_{n_1 n_2 m}&=&
\left[\frac{1+(-1)^{n_1}}{2}+
\frac{1-(-1)^{n_1}}{2} {\rm sgn}(m)\right]\times \nonumber \\
\label{2.47} \\
&\times& \frac{(-1)^{n+\frac{n_1}{2}}}{\sqrt 2} 
C_{\frac{n}{2}-\frac{1}{4},
\frac{n_2-n_1}{4}-\frac{1}{4};
\frac{n}{2} - \frac{1}{4},
\frac{n_1-n_2}{4}-\frac{1}{4}}^{|m|-\frac{1}{2},-\frac{1}{2}}.
\nonumber
\eea
We got an interesting result. Transition matrix (\ref{2.47})
is expressed in terms of the Clebsch-Gordan coefficients of the group $SU(2)$,
if formally they are extended to quarter-integer values of the moment.
$3j$ - a symbol with quarter-integer values also occurs in other
\cite{SMOR-SHEL} problems (see also Chapter 5).

Let us now show that the transition matrix taken in this form can
be expressed in terms of $d$ - Wigner functions of the argument
$\frac{\pi}{2}$. Considering that according to \cite{VAR}
\begin{eqnarray}
C_{a \alpha; b \beta}^{c \gamma}
&=&
\frac{(-1)^{a-c+\beta}}{2^{J+1}}
\left[\frac{(2c+1)(c+\gamma)!(J+1)!(J-2c)!}
{(a-\alpha)!(a+\alpha)!(b-\beta)!
(b+\beta)!(c-\gamma)!(J-2a)!(J-2b)!}\right]^{1/2}
\nonumber
\\[2mm]
\label{KG.1.1}
&\times& \int\limits_{-1}^{1}\,(1-x)^{a-\alpha}\,(1+x)^{b-\beta}\,
\frac{d^{c-\gamma}}{\left(d x\right)^{c-\gamma}}\left[ \left(1
-x\right)^{J-2a}\,\left(1 +x\right)^{J-2b}\right]dx,
\end{eqnarray}
($J=a+b+c$) and formula
\begin{eqnarray*}
T_{|m|}(\cos \theta)=
\frac{(-1)^{|m|}\sqrt\pi}{2^{|m|}\Gamma\left(|m|+\frac{1}{2}\right)}
(1-x^2)^{\frac{1}{2}}
\frac{d^{|m|}}{dx^{|m|}}(1-x^2)^{|m|-\frac{1}{2}} = \cos |m|\theta,
\end{eqnarray*}
(\ref{2.47}) can be written as
\begin{eqnarray*}
{\tilde W}_{n_1 n_2 m}
&=&
(-1)^{n+n_1-|m|}\,
\left[\frac{1+(-1)^{n_1}}{2}+\frac{1-(-1)^{n_1}}{2}\,
{\rm sgn}(m)\right] \,
\sqrt{\frac{2^{n_1+n_2}(n+|m|)!(n-|m|)!}{n_1! n_2!}}
\\[3mm]
&\times&
\int_{0}^{\pi}\, (\cos\varphi)^{n_1}\,
(\sin\varphi)^{n_2}\,
e^{-2i|m|\varphi}\, d\varphi.
\end{eqnarray*}
Comparing the last result with the integral representation
(\ref{inter1.2}) we have
\begin{eqnarray*}
{\tilde W}_{n_1 n_2 m}= \frac{(-1)^{n+n_1-|m|}}{\pi}
\left\{\frac{1+(-1)^{n_1}}{2}+\frac{1-(-1)^{n_1}}{2}sgn(m)\right\}
d_{|m|, \frac{n_1-n_2}{2}}^{\frac{n_1+n_2}{2}}
\left(\frac{\pi}{2}\right).
\end{eqnarray*}
Finally, using the $d$-property of the Wigner function \cite{VAR}
\bea
d_{M,M'}^J\left(\frac{\pi}{2}\right)=
(-1)^{J-M'}d_{-M,M'}^J\left(\frac{\pi}{2}\right)
\eea
we conclude that the transition matrix ${\tilde W}_{n_1 n_2 m}$ is exactly
coincides with the matrix that implements the decomposition (\ref{inter1.1}).
So, consistency of results (\ref{inter1.1}),
(\ref{2.25}) and (\ref{2.26}) is proved.

{\bf 1.10.2.}
Let us now turn to the discrete spectrum in the expansion
(\ref{2.41}). According to (\ref{2.41})
\bea
\overline{\Psi}_{km}(r, \varphi)
&=& \frac{1}{\Gamma\left(\frac{1}{2}+|m|-\frac{i}{k}\right)}
\, \frac{1}{2\pi i}
\,
\int_{\frac{1}{4}-\frac{i}{2k}\infty}^{\frac{1}{4}+\frac{i}{2k}\infty}
\biggl[L_k^{(+)}(z)\, \overline{W}_{kzm}^{(+)} \,
\overline{\Psi}_{kz}^{(+)}(\mu, \nu)
\nonumber
\\ [2mm]
&+&
L_k^{(-)}(z)\, \overline{W}_{kzm}^{(-)} \,
\overline{\Psi}_{kz}^{(-)}(\mu,\nu)\biggr]\, dz,
\label{2.48}
\eea
where $z=\frac{1}{4}-\frac{i(1+\beta)}{2k}$ and
\begin{eqnarray*} L_k^{(+)}(z) &=&
\Gamma(z)\Gamma\left(\frac{1}{2}-z\right)\Gamma\left(
\frac{1}{2}-\frac{i}{k}-z\right)\Gamma\left(\frac{i}{k}+z \right),
\\[2mm]
L_k^{(-)}(z) &=&
\Gamma(1-z)\Gamma\left(\frac{1}{2}+z\right)\Gamma\left(
1-\frac{i}{k}-z\right)\Gamma\left(\frac{1}{2}+\frac{i}{k}+z
\right),
\\[2mm]
\overline{W}_{kzm}^{(+)} &=& \frac{(-1)^{|m|}}{2\pi^2}
\frac{\sqrt{2\pi}}{\Gamma\left(\frac{1}{2}+\frac{i}{k}\right)}
{_3F_2} \left\{\matrix{ \frac{1}{2}-z,\,\, |m|,\,\, -|m| \cr \cr
\frac{1}{2},\,\, \frac{1}{2}+\frac{i}{k} \cr}\Biggr|1\right\},
\\[2mm]
\overline{W}_{kzm}^{(-)} &=& \frac{(-1)^{|m|-1}}{\pi^2}
\frac{im\sqrt{2\pi}}{\Gamma\left(\frac{3}{2}+\frac{i}{k}\right)}
{_3F_2} \left\{\matrix{ 1-z,\,\,1+ |m|,\,\, 1-|m| \cr \cr
\frac{3}{2},\,\, \frac{3}{2}+\frac{i}{k} \cr}\Biggr|1\right\},
\\[2mm]
{\overline\Psi}_{kz}^{(+)}(\mu,\nu) &=&
e^{ik\frac{\mu^2+\nu^2}{2}} F\left(z; \frac{1}{2}; -ik\mu^2\right)
F\left(\frac{1}{2}-\frac{i}{k}-z; \frac{1}{2}; -ik\nu^2\right),
\\[2mm]
\overline{\Psi}_{kz}^{(+)}(\mu, \nu) &=& (-4ik\mu \nu)
e^{ik\frac{\mu^2+\nu^2}{2}} F\left(\frac{1}{2}+z; \frac{3}{2};
-ik\mu^2\right) F\left(1-\frac{i}{k}-z; \frac{3}{2};
-ik\nu^2\right),
\\[2mm]
\overline{\Psi}_{km}(r,\varphi) &=& \frac{(-2ikr)^{|m|}}{(2|m|)!}
e^{ikr} F\left(\frac{1}{2}+|m|-\frac{i}{k};2|m|+1; -2ikr\right)
\frac{e^{im\varphi}}{\sqrt{2\pi}}. \end{eqnarray*}
The singularities of the integrand in (\ref{2.48}) with respect to $z$ coincide
with poles of the function $L_k^{(\pm)}(z)$ and are shown in Figures \ref{f13} and \ref{f14}.

\begin{figure}[t]
\unitlength=1.00mm \special{em:linewidth 0.4pt}
\linethickness{0.4pt}
\begin{picture}(141.00,110.00)
\put(19.00,50.00){\vector(1,0){121.00}}
\put(75.00,00.00){\vector(0,1){110.00}}
\put(92.00,10.00){\vector(0,1){90.00}}
\put(75.00,50.00){\circle*{2.00}}
\put(60.00,50.00){\circle*{2.00}}
\put(45.00,50.00){\circle*{2.00}}
\put(30.00,50.00){\circle*{2.00}}
\put(30.00,50.00){\circle{4.00}}
\put(45.00,50.00){\circle{4.00}}
\put(60.00,50.00){\circle{4.00}}
\put(75.00,30.00){\circle*{2.00}}
\put(122.00,30.00){\circle*{2.00}}
\put(110.00,30.00){\circle*{2.00}}
\put(98.00,30.00){\circle*{2.00}}
\put(98.00,55.00){\makebox(0,0)[cc]{$\frac{1}{2}$}}
\put(110.00,55.00){\makebox(0,0)[cc]{$\frac{3}{2}$}}
\put(122.00,55.00){\makebox(0,0)[cc]{$\frac{5}{2}$}}
\put(98.00,95.00){\makebox(0,0)[cc]{$C$}}
\put(68.00,104.00){\makebox(0,0)[cc]{$\Im m Z$}}
\put(141.00,54.00){\makebox(0,0)[cc]{$\Re e Z$}}
\put(30.00,55.00){\makebox(0,0)[cc]{$-3$}}
\put(45.00,55.00){\makebox(0,0)[cc]{$-2$}}
\put(60.00,55.00){\makebox(0,0)[cc]{$-1$}}
\put(71.00,55.00){\makebox(0,0)[cc]{$0$}}
\put(79.00,27.00){\makebox(0,0)[cc]{$\frac{i}{k}$}}
\put(125.00,52.00){\vector(1,1){5.00}}
\put(130.00,57.00){\vector(0,0){0.00}}
\put(133.00,59.00){\makebox(0,0)[cc]{$\Gamma\left(\frac{1}{2}-z\right)$}}
\put(27.00,52.00){\vector(-1,1){5.00}}
\put(18.00,60.00){\makebox(0,0)[cc]{$\Gamma(z)$}}
\put(125.00,29.00){\vector(1,-1){4.00}}
\put(129.00,25.00){\vector(0,0){0.00}}
\put(131.00,22.00){\makebox(0,0)[cc]{$\Gamma\left(\frac{1}{2}-\frac{i}{k}-z\right)$}}
\put(28.00,29.00){\vector(-1,-1){4.00}}
\put(21.00,22.00){\makebox(0,0)[cc]{$\Gamma\left(\frac{i}{k}+z\right)$}}
\put(122.00,50.00){\makebox(0,0)[cc]{$\bigotimes$}}
\put(110.00,50.00){\makebox(0,0)[cc]{$\bigotimes$}}
\put(98.00,50.00){\makebox(0,0)[cc]{$\bigotimes$}}
\put(60.00,30.00){\makebox(0,0)[cc]{$\times$}}
\put(45.00,30.00){\makebox(0,0)[cc]{$\times$}}
\put(30.00,30.00){\makebox(0,0)[cc]{$\times$}}
\put(75.00,50.00){\circle{4.00}}
\end{picture}
\caption{Poles of the function $L_{k}^{(+)}(Z)$.}
\label{f13}
\end{figure}

\begin{figure}[t]
\unitlength=1mm \special{em:linewidth 0.4pt} \linethickness{0.4pt}
\begin{picture}(143.00,115.00)
\put(20.00,60.00){\vector(1,0){120.00}}
\put(80.00,00.00){\vector(0,1){115.00}}
\put(95.00,10.00){\vector(0,1){95.00}}
\put(69.00,60.00){\circle*{2.00}}
\put(55.00,60.00){\circle*{2.00}}
\put(40.00,60.00){\circle*{2.00}}
\put(40.00,60.00){\circle{4.00}}
\put(55.00,60.00){\circle{4.00}}
\put(69.00,60.00){\circle{4.00}}
\put(105.00,60.00){\makebox(0,0)[cc]{$\times$}}
\put(115.00,60.00){\makebox(0,0)[cc]{$\times$}}
\put(125.00,60.00){\makebox(0,0)[cc]{$\times$}}
\put(125.00,40.00){\circle*{2.00}}
\put(115.00,40.00){\circle*{2.00}}
\put(105.00,40.00){\circle*{2.00}}
\put(83.00,40.00){\line(-1,0){6.00}}
\put(77.00,40.00){\line(0,1){0.00}}
\put(69.00,40.00){\makebox(0,0)[cc]{$\bigotimes$}}
\put(55.00,40.00){\makebox(0,0)[cc]{$\bigotimes$}}
\put(40.00,40.00){\makebox(0,0)[cc]{$\bigotimes$}}
\put(84.00,42.00){\makebox(0,0)[cc]{$-\frac{i}{k}$}}
\put(105.00,66.00){\makebox(0,0)[cc]{$1$}}
\put(115.00,66.00){\makebox(0,0)[cc]{$2$}}
\put(125.00,66.00){\makebox(0,0)[cc]{$3$}}
\put(143.00,57.00){\makebox(0,0)[cc]{$\Re e Z$}}
\put(99.00,100.00){\makebox(0,0)[cc]{$C$}}
\put(85.00,112.00){\makebox(0,0)[cc]{$\Im mZ$}}
\put(36.00,63.00){\vector(-1,1){3.00}}
\put(30.00,68.00){\makebox(0,0)[cc]{$\Gamma\left(\frac{1}{2}+z\right)$}}
\put(40.00,66.00){\makebox(0,0)[cc]{$-\frac{5}{2}$}}
\put(55.00,66.00){\makebox(0,0)[cc]{$-\frac{3}{2}$}}
\put(69.00,66.00){\makebox(0,0)[cc]{$-\frac{1}{2}$}}
\put(38.00,39.00){\vector(-1,-1){4.00}}
\put(31.00,31.00){\makebox(0,0)[cc]{$\Gamma\left(\frac{1}{2}+\frac{i}{k}+z\right)$}}
\put(128.00,37.00){\vector(1,-1){4.00}}
\put(136.00,30.00){\makebox(0,0)[cc]{$\Gamma\left(1-\frac{i}{k}-z\right)$}}
\put(128.00,62.00){\vector(1,1){6.00}}
\put(137.00,69.00){\makebox(0,0)[cc]{$\Gamma(1-z)$}}
\end{picture}
\caption{Poles of the function $L_k^{(-)}(Z)$.}
\label{f14}
\end{figure}
Let us consider the integrals
\begin{eqnarray*}
I^{(\pm)}(k,m,R)=
\frac{1}{\Gamma\left(\frac{1}{2}+|m|-\frac{i}{k}\right)}
\frac{1}{2\pi i}
\int_{\frac{1}{4}-\frac{iR}{2k}}^{\frac{1}{4}+\frac{iR}{2k}}
f^{(\pm)}(k,m,z)dz,
\end{eqnarray*}
in which
\begin{eqnarray*}
f^{(\pm)}(k,m,z)= L_k^{(\pm)}(z)\overline{W}_{kzm}^{(\pm)}
\overline{\Psi}_{kz}^{(\pm)}(\mu,\nu).
\end{eqnarray*}
We perform analytic continuation with respect to $k$ to the domain $k \rightarrow
i\lambda+\eta =i\left(n+\frac{1}{2}\right)^{-1}+\varepsilon$ where
$\lambda>0$, $\varepsilon $ is small and positive. In this case, the contour
integration will approach the real axis, and the poles will occupy
provisions presented in Figures \ref{f15} and \ref{f16}.

\begin{figure}[t]\unitlength=1.00mm \special{em:linewidth 0.4pt}
\linethickness{0.4pt}
\begin{picture}(145.00,120.00)
\put(10.00,60.00){\vector(1,0){135.00}}
\put(80.00,00.00){\vector(0,1){120.00}}
\put(80.00,60.00){\circle*{2.00}}
\put(70.00,60.00){\circle*{2.00}}
\put(60.00,60.00){\circle*{0.00}}
\put(60.00,60.00){\circle*{2.00}}
\put(50.00,60.00){\circle*{2.00}}
\put(32.00,60.00){\circle*{2.00}}
\put(15.00,60.00){\circle*{2.00}}
\put(15.00,60.00){\circle{4.00}}
\put(32.00,60.00){\circle{4.00}}
\put(50.00,60.00){\circle{4.00}}
\put(60.00,60.00){\circle{4.00}}
\put(70.00,60.00){\circle{4.00}}
\put(80.00,60.00){\circle{4.00}}
\put(23.00,60.00){\makebox(0,0)[cc]{$\times$}}
\put(32.00,40.00){\circle*{2.00}}
\put(50.00,40.00){\circle*{2.00}}
\put(60.00,40.00){\circle*{2.00}}
\put(70.00,40.00){\circle*{2.00}}
\put(80.00,40.00){\circle*{2.00}}
\put(90.00,40.00){\circle*{2.00}}
\put(101.00,40.00){\circle*{2.00}}
\put(110.00,40.00){\circle*{2.00}}
\put(31.00,38.00){\vector(-1,-1){4.00}}
\put(25.00,31.00){\makebox(0,0)[cc]{$\Gamma(-n-z+i\varepsilon)$}}
\put(115.00,40.00){\circle*{0.00}}
\put(117.00,40.00){\circle*{0.00}}
\put(119.00,40.00){\circle*{0.00}}
\put(121.00,40.00){\circle*{0.00}}
\put(123.00,40.00){\circle*{0.00}}
\put(125.00,40.00){\circle*{0.00}}
\put(96.00,60.00){\makebox(0,0)[cc]{$\bigotimes$}}
\put(106.00,60.00){\makebox(0,0)[cc]{$\bigotimes$}}
\put(108.00,62.00){\vector(1,1){4.00}}
\put(115.00,67.00){\makebox(0,0)[cc]{$\Gamma\left(\frac{1}{2}-z\right)$}}
\put(106.00,67.00){\makebox(0,0)[cc]{$\frac{3}{2}$}}
\put(96.00,67.00){\makebox(0,0)[cc]{$\frac{1}{2}$}}
\put(77.00,65.00){\makebox(0,0)[cc]{$0$}}
\put(70.00,65.00){\makebox(0,0)[cc]{$-1$}}
\put(60.00,65.00){\makebox(0,0)[cc]{$-2$}}
\put(50.00,65.00){\makebox(0,0)[cc]{$-3$}}
\put(47.00,65.00){\circle*{0.00}}
\put(45.00,65.00){\circle*{0.00}}
\put(42.00,65.00){\circle*{0.00}}
\put(40.00,65.00){\circle*{0.00}}
\put(38.00,65.00){\circle*{0.00}}
\put(32.00,65.00){\makebox(0,0)[cc]{$-n$}}
\put(15.00,65.00){\makebox(0,0)[cc]{$-n-1$}}
\put(23.00,54.00){\makebox(0,0)[cc]{$-n-\frac{1}{2}$}}
\put(23.00,63.00){\vector(0,1){7.00}}
\put(23.00,73.00){\makebox(0,0)[cc]{$\Gamma\left(-\frac{1}{2}-n-i\varepsilon+z\right)$}}
\put(58.00,63.00){\vector(-1,4){1.67}}
\put(55.00,73.00){\makebox(0,0)[cc]{$\Gamma(z)$}}
\put(145.00,54.00){\makebox(0,0)[cc]{$\Re e Z$}}
\put(99.33,73.00){\line(-1,0){0.33}}
\put(137.00,80.00){\makebox(0,0)[cc]{$C$}}
\bezier{48}(94.00,71.00)(95.00,76.00)(102.00,75.00)
\put(85.00,117.00){\makebox(0,0)[cc]{$\Im m Z$}}
\put(76.00,45.00){\line(-1,0){28.00}}
\put(12.00,45.00){\vector(1,0){36.00}}
\put(102.00,75.00){\vector(1,0){41.00}}
\put(94.00,71.00){\line(-1,-2){12.00}}
\bezier{28}(82.00,47.00)(81.00,45.00)(76.00,45.00)
\end{picture}
\caption{The poles of the function $L_k^{(\pm)}(Z)$ and the integration contour for
$k \to i\lambda+\varepsilon$.}
\label{f15}
\end{figure}

\begin{figure}[t]
\unitlength=1.00mm \special{em:linewidth 0.4pt}
\linethickness{0.4pt}
\begin{picture}(146.00,140.00)
\put(10.00,80.00){\vector(1,0){135.00}}
\put(80.00,10.00){\vector(0,1){130.00}}
\put(70.00,80.00){\circle*{2.00}}
\put(60.00,80.00){\circle*{2.00}}
\put(40.00,80.00){\circle*{2.00}}
\put(20.00,80.00){\circle*{2.00}} \put(20.00,80.00){\circle{4.00}}
\put(40.00,80.00){\circle{4.00}} \put(60.00,80.00){\circle{4.00}}
\put(70.00,80.00){\circle{4.00}}
\put(30.00,80.00){\makebox(0,0)[cc]{$\bigotimes$}}
\put(43.00,77.00){\circle*{0.00}}
\put(45.00,77.00){\circle*{0.00}}
\put(47.00,77.00){\circle*{0.00}}
\put(47.00,77.00){\circle*{0.00}}
\put(49.00,77.00){\circle*{0.00}}
\put(51.00,77.00){\circle*{0.00}}
\put(53.00,77.00){\circle*{0.00}}
\put(55.00,77.00){\circle*{0.00}}
\put(57.00,77.00){\circle*{0.00}}
\put(95.00,80.00){\makebox(0,0)[cc]{$\times$}}
\put(105.00,80.00){\makebox(0,0)[cc]{$\times$}}
\put(115.00,80.00){\makebox(0,0)[cc]{$\times$}}
\put(125.00,80.00){\makebox(0,0)[cc]{$\times$}}
\put(95.00,86.00){\makebox(0,0)[cc]{$1$}}
\put(105.00,86.00){\makebox(0,0)[cc]{$2$}}
\put(115.00,86.00){\makebox(0,0)[cc]{$3$}}
\put(125.00,86.00){\makebox(0,0)[cc]{$4$}}
\put(128.00,82.00){\circle*{0.00}}
\put(130.00,82.00){\circle*{0.00}}
\put(132.00,82.00){\circle*{0.00}}
\put(134.00,82.00){\circle*{0.00}}
\put(134.00,82.00){\circle*{0.00}}
\put(125.00,65.00){\circle*{2.00}}
\put(115.00,65.00){\circle*{2.00}}
\put(105.00,65.00){\circle*{2.00}}
\put(95.00,65.00){\circle*{2.00}}
\put(70.00,65.00){\circle*{2.00}}
\put(60.00,65.00){\circle*{2.00}}
\put(48.00,65.00){\circle*{2.00}}
\put(40.00,85.00){\makebox(0,0)[cc]{$-n-\frac{1}{2}$}}
\put(60.00,85.00){\makebox(0,0)[cc]{$-\frac{3}{2}$}}
\put(70.00,85.00){\makebox(0,0)[cc]{$-\frac{1}{2}$}}
\put(86.00,136.00){\makebox(0,0)[cc]{$\Im m Z$}}
\put(146.00,74.00){\makebox(0,0)[cc]{$\Re e Z$}}
\put(48.00,60.00){\makebox(0,0)[cc]{$-n+\frac{1}{2}$}}
\put(129.00,65.00){\circle*{0.00}}
\put(131.00,65.00){\circle*{0.00}}
\put(133.00,65.00){\circle*{0.00}}
\put(135.00,65.00){\circle*{0.00}}
\put(137.00,65.00){\circle*{0.00}}
\put(60.00,62.00){\vector(0,-1){8.00}}
\put(60.00,50.00){\makebox(0,0)[cc]{$\Gamma\left(\frac{1}{2}-n+i\varepsilon-z\right)$}}
\put(30.00,84.00){\vector(0,1){11.00}}
\put(30.00,100.00){\makebox(0,0)[cc]{$\Gamma(n+1-i\varepsilon+z)$}}
\put(58.00,83.00){\vector(-1,4){3.00}}
\put(55.00,99.00){\makebox(0,0)[cc]{$\Gamma\left(\frac{1}{2}+z\right)$}}
\put(127.00,84.00){\vector(1,1){3.00}}
\put(133.00,89.00){\makebox(0,0)[cc]{$\Gamma(1-z)$}}
\put(11.00,71.00){\vector(1,0){37.00}}
\put(49.00,71.00){\line(1,0){29.00}}
\put(93.00,92.00){\vector(1,0){49.00}}
\put(138.00,96.00){\makebox(0,0)[cc]{$C$}}
\put(89.00,88.00){\line(-2,-5){6.00}}
\bezier{24}(83.00,73.00)(81.00,71.00)(78.00,71.00)
\bezier{28}(89.00,88.00)(91.00,92.00)(94.00,92.00)
\end{picture}

\caption{The poles of the function $L_k^{(-)}(Z)$ and the integration contour for
$k\to i\lambda+\varepsilon$.}
\label{f16}
\end{figure}

We close the contour in the upper half-plane $z$ and write
\begin{eqnarray*}
I^{(+)}(i\lambda + \varepsilon, m, R) &=&
\frac{1}{\Gamma(\varepsilon-n+|m|)} \Biggl[\,\sum_{t=0,2}^{2n}Res
f^{(+)}\left(k,m;-\frac{t}{2}\right)+ \sum_{t=2n+2}^{t_R^{(1)}}Res
f^{(+)}\left(k,m;-\frac{t}{2}\right)+
\\ [2mm]&+& \sum_{t=1,3}^{t_R^{(2)}}Res
f^{(+)}\left(k,m;-n-\frac{t}{2}\right)- \frac{1}{2\pi
i}\int_{\gamma_R} f^{(+)}(k,m;z)dz\Biggr], \\ [3mm]
I^{(-)}(i\lambda+\varepsilon,m, R) &=& \frac{1}{\Gamma
(\varepsilon -n+|m|)} \Biggl[\,\sum_{t=1,3}^{2n-1}Res
f^{(-)}\left(k,m;-\frac{t}{2}\right)+ \sum_{t=2n+1}^{t_R^{(3)}}Res
f^{(-)}\left(k,m;-\frac{t}{2}\right)+
\\ [2mm] &+&\sum_{t=2,4}^{t_R^{(4)}}Res
f^{(-)}\left(k,m;-n-\frac{t}{2}\right)- \frac{1}{2\pi
i}\int_{\gamma_R} f^{(-)}(k,m;z)dz\Biggr].
\end{eqnarray*}

\noindent
Here $\gamma_R$ is a semicircle of radius $R$, $T_R^{(i)}$ are the extreme
the left poles of each type that are inside
contour $C$. For $\varepsilon \rightarrow 0$ the contributions of the last three
terms in the previous formulas are finite, and the first sums tend to
infinity, so the dependence on $R$ drops out and one can
write
\bea
I^{(+)}(i\lambda +\varepsilon,m)
&=&
\frac{1}{\Gamma(\varepsilon -n+|m|)}
\sum_{t=0,2}^{2n}(-1)^{\frac{t}{2}}
\frac{\Gamma\left(\frac{1}{2}+\frac{t}{2}\right)
\Gamma\left(n+\frac{1}{2}-
\frac{t}{2}\right)}{\Gamma\left(1+\frac{t}{2}\right)}\times
\nonumber
\\[2mm]
&\times& \Gamma\left(\varepsilon -n+\frac{t}{2}\right)
\overline{W}_{i\lambda+\varepsilon,-\frac{t}{2},m}^{(+)}
\overline{\Psi}_{i\lambda+\varepsilon,-\frac{t}{2}}^{(+)}(\mu,\nu),
\label{2.49a}
\\[2mm]
I^{(-)}(i\lambda +\varepsilon,m)
&=&
\frac{1}{\Gamma(\varepsilon -n+|m|)}
\sum_{t=1,3}^{2n-1}(-1)^{\frac{t-1}{2}}
\frac{\Gamma\left(1+\frac{t}{2}\right)
\Gamma\left(n+1-\frac{t}{2}\right)}
{\Gamma\left(\frac{1}{2}+t\right)}\times
\nonumber
\\[2mm]
&\times& \Gamma\left(\varepsilon-n+\frac{t}{2}+\frac{1}{2}\right)
\overline{W}_{i\lambda+\varepsilon,-\frac{t}{2},m}^{(-)}
\overline{\Psi}_{i\lambda+\varepsilon,-\frac{t}{2}}^{(-)}(\mu,\nu).
\label{2.49b}
\eea
Considering now the property of gamma functions
\begin{eqnarray*}
\frac{\Gamma(z)}{\Gamma(z-n)}=(-1)^n
\frac{\Gamma(n+1-z)}{\Gamma(1-z)}
\end{eqnarray*}
and directing $\varepsilon \rightarrow 0$ to (\ref{2.49a})
and (\ref{2.49b}) instead of (\ref{2.48}) we have
\bea \overline{\Psi}_{i\lambda, m}(r,\varphi)
&=& (-1)^{n-|m|}
(n-|m|)! \Biggl\{\sum_{n_1=0,2}^{2n}
\frac{\Gamma\left(\frac{1}{2}+
\frac{n_1}{2}\right)\Gamma\left(\frac{1}{2}
+\frac{n_2}{2}\right)} {\left(\frac{n_1}{2}\right)!
\left(\frac{n_2}{2}\right)!} \,
\overline{W}_{i\lambda,-\frac{n_1}{2},m}^{(+)}
\overline{\Psi}_{i\lambda,-\frac{n_1}{2}}^{(+)}(\mu, \nu)+
\nonumber\\[2mm]
&+& \sum_{n_1=1,3}^{2n-1}(-1)^{N_1}
\frac{\Gamma\left(1+\frac{n_1}{2}\right)
\Gamma\left(1+\frac{n_2}{2}\right)}
{\left(\frac{n_1-1}{2}\right)!\left(\frac{n_2-1}{2}\right)!}
\overline{W}_{i\lambda,-\frac{n_1}{2},m}^{(-)}
\overline{\Psi}_{i\lambda,-\frac{n_1}{2}}^{(-)}(\mu,\nu)\Biggr\},
\label{2.50}
\eea
where $n=\frac{n_1+n_2}{2}$. Applicable to coefficients
$\overline{W}_{i\lambda,-\frac{N_1}{2},m}^{(\pm)}$ transformation
(\ref{d8}), we pass from degenerate hypergeometric functions
to the Hermite polynomials (\ref{hermit}), and take into account that
\begin{eqnarray*}
\overline{\Psi}_{i\lambda,m}(r,\varphi)=
\left(\frac{1}{2\lambda^3}\right)^{\frac{1}{2}}
\sqrt{\frac{(n-|m|)!}{(n+|m|)!}}\Psi_{nm}(r,\varphi),
\end{eqnarray*}
where $\Psi_{nm}(r,\varphi)$ is the polar wave function of the discrete
spectrum (\ref{hyd-pol6}). Then instead of (\ref{2.50}) we get
\begin{eqnarray*}
\Psi_{nm}(r,\varphi)= \sum_{n_1=0,2}^{2n} \,
\widetilde{W}_{n_1 n_2m}^{(+)^*} \, \Psi_{n_1 n_2}^{(+)}(\mu,\nu)+
\sum_{n_1=1,3}^{2n-1}\, \widetilde{W}_{n_1 n_2m}^{(-)^*}\,
\Psi_{n_1 n_2}^{(-)}(\mu,\nu).
\end{eqnarray*}
The formulas (\ref{2.44a}), (\ref{2.44b}), (\ref{2.46}) allow the latter
write the result in a more compact form:
\begin{eqnarray*}
\Psi_{nm}(r,\varphi) = \sum_{n_1=0,1}^{2n}\,
\widetilde{W}_{n_1 n_2m}^* \, \Psi_{n_1 n_2}(\mu,\nu).
\end{eqnarray*}
It was proved above that this formula is equivalent to the decomposition
polar basis over a parabolic one in the discrete spectrum.
So, interbasis transitions in discrete and continuous spectra
agreed.

\section{Transformations connecting the parabolic bases of the two-dimensional
hydrogen atom in continuous spectrum}
\markboth{CHAPTER 1. TWO DIMENSIONAL HYDROGEN ATOM}{1.11. CONNECTING THE PARABOLIC BASES IN CONTINUOUS SPECTRUM}

In the previous section, we showed that the completeness property in
continuous spectrum has a system of wave functions, including
parabolic bases of both parities.
It follows from what has been said that the expansions of interest to us can be
are written as follows:
\begin{eqnarray}
\Psi_{k\overline{\beta}}^{(+)}\left(\overline{\mu}, \overline{\nu}\right)
&=& \int_{-\infty}^{\infty}\left[P_{k\beta\overline{\beta}}^{(++)}
\Psi_{k\beta}^{(+)}\left(\mu, \nu\right)
+ P_{k\beta\overline{\beta}}^{(+-)}
\Psi_{k\beta}^{(-)}\left(\mu, \nu\right)\right]d\beta,
\label{par-par-5}
\\ [3mm]
\Psi_{k\overline{\beta}}^{(-)}
\left(\overline{\mu}, \overline{\nu}\right)
&=&
\int_{-\infty}^{\infty}\left[P_{k\beta\overline{\beta}}^{(-+)}
\Psi_{k\beta}^{(+)}\left(\mu, \nu\right)
+ P_{k\beta\overline{\beta}}^{(--)}
\Psi_{k\beta}^{(-)}\left(\mu, \nu\right)\right]d\beta,
\label{par-par-6}
\end{eqnarray}
where the second parabolic basis is $\Psi_{k\overline{\beta}}^{(+)}$,
$\Psi_{k\overline{\beta}}^{(-)}$ are obtained from (\ref{2.10}) and
(\ref{2.11}) by replacing $\beta\to \overline{\beta}$,
$(\mu, \nu) \to (\overline{\mu}, \overline{\nu})$.
It can be seen from the formulas (\ref{2.10}) and (\ref{2.11}) that the interbasic
overlap integrals are difficult to compute directly. That's why we
we use an indirect method based on the above
expansions of parabolic subbases in polar ones (\ref{2.25}):
\begin{eqnarray}
\Psi_{k\beta}^{(\pm)}\left(\mu, \nu\right)
&=& \sum_{m=-\infty}^{\infty}W_{k\beta m}^{(\pm)}\,
\Psi_{km}\left(r, \varphi\right),
\label{par-par-7}
\\ [3mm]
\Psi_{k\overline{\beta}}^{(\pm)}\left(\overline{\mu},
\overline{\nu}\right)
&=& \sum_{m=-\infty}^{\infty}e^{im\pi/2}
W_{k\overline{\beta} m}^{(\pm)}\,
\Psi_{km}\left(r, \varphi\right).
\label{par-par-8}
\end{eqnarray}
Note that the decomposition (\ref{par-par-8}) is obtained from
(\ref{par-par-7}), given that the replacement
$(\mu, \nu) \leftrightarrow (\overline{\mu}, \overline{\nu})$
is equivalent to the transformation $x \leftrightarrow y$. From
(\ref{par-par-5})-(\ref{par-par-8}) it is easy to show that the coefficients
$P$ are related to the coefficients $W$ by the following relations:
\begin{eqnarray}
P_{k\beta\overline{\beta}}^{(r,t)}
= \sum_{m=-\infty}^{\infty}i^m
W_{k\overline{\beta} m}^{(r)}\,
W_{k\overline{\beta},- m}^{(t)*},
\label{par-par-9}
\end{eqnarray}
where the indices $(r)$ and $(t)$ take the values $(+)$ and $(-)$.
The formula (\ref{par-par-9}) forms the basis for further calculations.
Substituting the integral representations of the coefficients
$W_{k\sigma m}^{(\pm)}$ (\ref{2.32a}) and (\ref{2.32b})
to (\ref{par-par-9}) and using the identity
\begin{eqnarray*}
\sum_{m=\infty}^\infty\,e^{im\varphi} =
\sum_{m=\infty}^\infty\,\delta\left(\varphi - 2\pi m\right),
\end{eqnarray*}
after quite a lot of calculations, we get the following
integral representations for the coefficients $P$:
\begin{eqnarray*}
P_{k\beta\overline{\beta}}^{(++)} &=& \frac{\pi}{2}
A_{k\beta\overline{\beta}}^{(++)}\left(
B_{k\beta\overline{\beta}} + B_{k\beta,{-\overline{\beta}}}^* +
B_{k,-\beta{\overline{\beta}}}^* -B_{k,-\beta{-\overline{\beta}}}^*
\right), \\ [2mm]
P_{k\beta\overline{\beta}}^{(+-)} &=&  \frac{i\pi}{2}
A_{k\beta\overline{\beta}}^{(+-)}\left(
B_{k\beta\overline{\beta}} - B_{k\beta,{-\overline{\beta}}}^* +
B_{k,-\beta{\overline{\beta}}}^* -B_{k,-\beta{-\overline{\beta}}}^*
\right), \\ [2mm]
P_{k\beta\overline{\beta}}^{(-+)} &=& - \frac{i\pi}{2}
A_{k\beta\overline{\beta}}^{(-+)}\left(
B_{k\beta\overline{\beta}} + B_{k\beta,{-\overline{\beta}}}^* -
B_{k,-\beta{\overline{\beta}}}^* -B_{k,-\beta{-\overline{\beta}}}^*
\right), \\ [2mm]
P_{k\beta\overline{\beta}}^{(--)} &=& \frac{\pi}{2}
A_{k\beta\overline{\beta}}^{(--)}\left(
B_{k\beta\overline{\beta}} - B_{k\beta,{-\overline{\beta}}}^* -
B_{k,-\beta{\overline{\beta}}}^* + B_{k,-\beta{-\overline{\beta}}}^*
\right).
\end{eqnarray*}
Here the quantity $B_{k\beta\overline{\beta}}$ is given by the integral
\begin{eqnarray*}
B_{k\beta\overline{\beta}} = \int_0^{\pi/2}
\left(1-\sin\varphi\right)^{-a_{-\overline{\beta}}}
\left(1+\sin\varphi\right)^{a_{\overline{\beta}}}
\left(1-\cos\varphi\right)^{-a_{-\beta}}
\left(1+\cos\varphi\right)^{-a_{\beta}}
d\varphi
\end{eqnarray*}
and the constants $A$ are defined by the expressions
\begin{eqnarray*}
A_{k\beta\overline{\beta}}^{(++)} &=& \frac{\pi}{2k}
\frac{C_{k\overline{\beta}}^{(+)}C_{k\beta}^{(+)}e^{-\pi/k}}
{\Gamma\left(a_{\overline{\beta}}^*\right)
\Gamma\left(a_{-\overline{\beta}}^*\right)
\Gamma\left(a_{\beta}\right)
\Gamma\left(a_{-\beta}\right)}, \\ [2mm]
A_{k\beta\overline{\beta}}^{(+-)} &=&  \frac{\pi}{k}
\frac{C_{k\overline{\beta}}^{(+)}C_{k\beta}^{(-)}e^{-\pi/k}}
{\Gamma\left(a_{\overline{\beta}}^*\right)
\Gamma\left(a_{-\overline{\beta}}^*\right)
\Gamma\left(\frac{1}{2}+a_{\beta}\right)
\Gamma\left(\frac{1}{2}+a_{-\beta}\right)}, \\ [2mm]
A_{k\beta\overline{\beta}}^{(-+)} &=& - \frac{\pi}{k}
\frac{C_{k\overline{\beta}}^{(-)}C_{k\beta}^{(+)}e^{-\pi/k}}
{\Gamma\left(\frac{1}{2}+a_{\overline{\beta}}^*\right)
\Gamma\left(\frac{1}{2}+a_{-\overline{\beta}}^*\right)
\Gamma\left(a_{\beta}\right)
\Gamma\left(a_{-\beta}\right)}, \\ [2mm]
A_{k\beta\overline{\beta}}^{(--)} &=& \frac{\pi}{2k}
\frac{C_{k\overline{\beta}}^{(-)}C_{k\beta}^{(-)}e^{-\pi/k}}
{\Gamma\left(\frac{1}{2}+a_{\overline{\beta}}^*\right)
\Gamma\left(\frac{1}{2}+a_{-\overline{\beta}}^*\right)
\Gamma\left(\frac{1}{2}+a_{\beta}\right)
\Gamma\left(\frac{1}{2}+a_{-\beta}\right)}.
\end{eqnarray*}
The quantity $a_\sigma$ is defined by the formula
\begin{eqnarray*}
a_\sigma = \frac{1}{4} - \frac{i}{2k}(1+\sigma).
\label{par-par-4}
\end{eqnarray*}
Finally, making the change $\sin\varphi = \tanh u$ in
$B_{k,\beta{\overline{\beta}}}$ and $B_{k,-\beta \overline{\beta}}$ and
$\cos\varphi = \tanh u$ in and $B_{k,\beta{-\overline{\beta}}}^* $ and
$B_{k,-\beta{-\overline{\beta}}}^* $ and introducing the function
\begin{eqnarray*}
g_{\beta}(u) = \left(\cosh u-1\right)^{-a_{\beta}^*}
\left(\cosh u + 1\right)^{-a_{\beta}^*}
\end{eqnarray*}
we arrive at the following integral representations:
\begin{eqnarray}
P_{k\beta\overline{\beta}}^{(++)}
&=& \pi A_{k\beta\overline{\beta}}^{(++)} \int_0^\infty
\cos\left(\frac{\overline{\beta}u}{k}\right)
\left[g_{\beta}(u) + g_{-\beta}(u)\right]du, \label{par-par-10}  \\ [2mm]
P_{k\beta\overline{\beta}}^{(+-)}
&=& -\pi A_{k\beta\overline{\beta}}^{(+-)} \int_0^\infty
\sin\left(\frac{\overline{\beta}u}{k}\right)
\left[g_{\beta}(u) + g_{-\beta}(u)\right]du, \label{par-par-11}  \\ [2mm]
P_{k\beta\overline{\beta}}^{(-+)}
&=& -i\pi A_{k\beta\overline{\beta}}^{(-+)} \int_0^\infty
\cos\left(\frac{\overline{\beta}u}{k}\right)
\left[g_{\beta}(u) - g_{-\beta}(u)\right]du, \label{par-par-12} \\ [2mm]
P_{k\beta\overline{\beta}}^{(--)}
&=& i\pi A_{k\beta\overline{\beta}}^{(--)} \int_0^\infty
\sin\left(\frac{\overline{\beta}u}{k}\right)
\left[g_{\beta}(u) - g_{-\beta}(u)\right]du.
\label{par-par-13}
\end{eqnarray}
The resulting formulas allow one to directly check the series
general relations to which the coefficients must obey
$P_{k\beta\overline{\beta}}$. With their help, you can
analytic continuation of interbasic transformations
(\ref{par-par-5}) and (\ref{par-par-6}) into the region of discrete
spectrum and restore the results obtained in the above. We are not
Let's focus on these questions and move on to
calculation of integrals (\ref{par-par-10}) - (\ref{par-par-13}).

To find the explicit form of the coefficients $P_{k\beta\overline{\beta}}$
we use the well-known formulas \cite{BE1}
\begin{eqnarray}
\cos\left(\frac{\overline{\beta}u}{k}\right)
&=&
\cosh\left(\frac{u}{2}\right)^{-2i\overline{\beta}/k}\,
F\left(\frac{i\overline{\beta}}{k}, \frac{1}{2}+
\frac{i\overline{\beta}}{k}; \frac{1}{2}; \tanh^2\frac{1}{2}u
\right), \label{par-par-14}  \\ [2mm]
\sin\left(\frac{\overline{\beta}u}{k}\right)
&=& \frac{2\overline{\beta}}{k}\tanh\left(\frac{u}{2}\right)
\cosh\left(\frac{u}{2}\right)^{-2i\overline{\beta}/k} \times \nonumber \\
\\
&\times& F\left(1+\frac{i\overline{\beta}}{k}, \frac{1}{2}+
\frac{i\overline{\beta}}{k}; \frac{3}{2}; \tanh^2\frac{1}{2}u
\right). \nonumber \label{par-par-15}
\end{eqnarray}
Substituting (\ref{par-par-5}) and (\ref{par-par-6}) into integral
views (\ref{par-par-10}) - (\ref{par-par-13}), decomposing
hypergeometric functions $F$ into a series and taking into account the relation
\cite{BE1}
\begin{eqnarray*}
\int_0^\infty \left(\sinh u\right)^\alpha\,
\left(\cosh u\right)^{-\gamma} du =
\frac{\Gamma\left(\frac{1+ \alpha}{2}\right)
\Gamma\left(\frac{\gamma - \alpha}{2}\right)}
{2\Gamma\left(\frac{1+\gamma}{2}\right)}
\end{eqnarray*}
valid for $\Re e\alpha > -1,\, \Re e (\alpha - \gamma) > 0$,
we arrive at the following expressions for $P_{k\beta\overline{\beta}}$:
\begin{eqnarray}
P_{k\beta\overline{\beta}}^{(++)} &=&
\frac{\sqrt {\pi^3}}{2^pk}e^{-\pi/k}\,C_{k\overline{\beta}}^{(+)}
C_{k\beta}^{(+)} \frac{\Gamma\left(1 - a_{\overline{\beta}}\right)}
{\Gamma\left(\frac{1}{2} - a_{\overline{\beta}}\right)}
\Biggl[\frac{1}{\Gamma\left(a_{\beta}\right)
\Gamma\left(1 - p- a_{\beta}\right)}\times
\nonumber \\ [2mm]
&\times&
{_3F}_2 \left\{
\begin{array}{l}
-p,  \frac{1}{2}-p, a_{-\beta} \cr \\
\frac{1}{2},  1 - p- a_{\beta}
\end{array}
\Biggr| 1 \right\}+ \left(\beta \to -\beta\right)\Biggr],
\label{par-par-16}  \\ [2mm]
P_{k\beta\overline{\beta}}^{(+-)} &=&
-\frac{ip\sqrt {\pi^3}}{2^pk}e^{-\pi/k}\,C_{k\overline{\beta}}^{(+)}
C_{k\beta}^{(-)} \frac{\Gamma\left(1 - a_{\overline{\beta}}\right)}
{\Gamma\left(\frac{1}{2} - a_{\overline{\beta}}\right)}
\Biggl[\frac{1}{\Gamma\left(\frac{1}{2}+a_{\beta}\right)
\Gamma\left(\frac{3}{2} - p- a_{-\beta}\right)}\times
\nonumber \\ [2mm]
&\times&
{_3F}_2 \left\{
\begin{array}{l}
\frac{1}{2}-p,  1-p, \frac{1}{2}+a_{\beta} \cr \\
\frac{3}{2},  \frac{3}{2} - p- a_{-\beta}
\end{array}
\Biggr| 1 \right\}+ \left(\beta \to -\beta\right)\Biggr],
\label{par-par-17}  \\ [2mm]
P_{k\beta\overline{\beta}}^{(-+)} &=& -
\frac{i\sqrt {\pi^3}}{2^{p+1}k}e^{-\pi/k}\,C_{k\overline{\beta}}^{(-)}
C_{k\beta}^{(+)} \frac{\Gamma\left(\frac{1}{2} -
a_{-\overline{\beta}}\right)}
{\Gamma\left(1 - a_{\overline{\beta}}\right)}
\Biggl[\frac{1}{\Gamma\left(a_{\beta}\right)
\Gamma\left(1 - p- a_{\beta}\right)}\times
\nonumber \\ [2mm]
&\times&
{_3F}_2 \left\{
\begin{array}{l}
-p,  \frac{1}{2}-p, a_{-\beta} \cr \\
\frac{1}{2},  1 - p- a_{\beta}
\end{array}
\Biggr| 1 \right\}+ \left(\beta \to -\beta\right)\Biggr],
\label{par-par-18} \\ [2mm]
P_{k\beta\overline{\beta}}^{(--)} &=&
-\frac{p\sqrt {\pi^3}}{2^{p+1}k}e^{-\pi/k}\,C_{k\overline{\beta}}^{(-)}
C_{k\beta}^{(-)} \frac{\Gamma\left(\frac{1}{2} -
a_{-\overline{\beta}}\right)}
{\Gamma\left(1 - a_{\overline{\beta}}\right)}
\Biggl[\frac{1}{\Gamma\left(\frac{1}{2}+a_{\beta}\right)
\Gamma\left(\frac{3}{2} - p- a_{-\beta}\right)}\times
\nonumber \\ [2mm]
&\times&
{_3F}_2 \left\{
\begin{array}{l}
\frac{1}{2}-p,  1-p, \frac{1}{2}+a_{\beta} \cr \\
\frac{3}{2},  \frac{3}{2} - p- a_{-\beta}
\end{array}
\Biggr| 1 \right\}+ \left(\beta \to -\beta\right)\Biggr],
\label{par-par-19}
\end{eqnarray}
where the notation $p=-i\overline{\beta}/k$ is introduced.

Now, substituting into (\ref{par-par-9}) the explicit expressions of the coefficients
$W_{k\beta m}^{(\pm)}$ from (\ref{2.31a}) and (\ref{2.31b})
after comparing with formulas (\ref{par-par-16}) - (\ref{par-par-19})
it is easy to derive the following mathematical relations:
\begin{eqnarray*}
\sum_{m}i^m
{_3F}_2 \left\{
\begin{array}{l}
-|m|,  |m|, a_{\overline{\beta}} \cr \\
\frac{1}{2},  \frac{1}{2} - \frac{i}{k}
\end{array}
\Biggr| 1 \right\}
{_3F}_2 \left\{
\begin{array}{l}
-|m|, |m|, \frac{1}{2}-a_{\beta} \cr \\
\frac{1}{2},  \frac{1}{2} + \frac{i}{k}
\end{array}
\Biggr| 1 \right\} =
\frac{\Gamma\left(1 - a_{-\overline{\beta}}\right)
\sqrt {\pi^3}}{2^p\Gamma\left(\frac{1}{2} - a_{\overline{\beta}}\right)
\cosh\frac{\pi}{k}}\times \\ [2mm]
\times
\left[\frac{1}{\Gamma\left(a_{\beta}\right)
\Gamma\left(1 - p- a_{\beta}\right)}
{_3F}_2 \left\{
\begin{array}{l}
-p,  \frac{1}{2}-p, a_{-\beta} \cr \\
\frac{1}{2},  1 - p- a_{\beta}
\end{array}
\Biggr| 1 \right\}+ \left(\beta \to -\beta\right)\right],
\end{eqnarray*}
\begin{eqnarray*}
\sum_{m}i^m m
{_3F}_2 \left\{
\begin{array}{l}
-|m|,  |m|, a_{\overline{\beta}} \cr \\
\frac{1}{2},  \frac{1}{2} - \frac{i}{k}
\end{array}
\Biggr| 1 \right\}
{_3F}_2 \left\{
\begin{array}{l}
1-|m|, 1+|m|, 1-a_{\beta} \cr \\
\frac{3}{2},  \frac{3}{2} + \frac{i}{k}
\end{array}
\Biggr| 1 \right\} =
\frac{ip\left(\frac{1}{2} + \frac{i}{k}\right)\sqrt {\pi^3}}
{2^p\cosh\frac{\pi}{k}}\times
\\ [2mm]
\times
\frac{\Gamma\left(1 - a_{-\overline{\beta}}\right)}
{\Gamma\left(\frac{1}{2} - a_{\overline{\beta}}\right)}
\left[\frac{1}{\Gamma\left(\frac{1}{2}+a_{\beta}\right)
\Gamma\left(\frac{3}{2} - p- a_{\beta}\right)}
{_3F}_2 \left\{
\begin{array}{l}
1-p,  \frac{1}{2}-p, \frac{1}{2}+ a_{-\beta} \cr \\
\frac{3}{2},  \frac{3}{2} - p- a_{\beta}
\end{array}
\Biggr| 1 \right\}+ \left(\beta \to -\beta\right)\right],
\end{eqnarray*}
\begin{eqnarray*}
\sum_{m}i^m m^2
{_3F}_2 \left\{
\begin{array}{l}
1-|m|, 1+|m|, \frac{1}{2}+a_{\overline{\beta}} \cr \\
\frac{3}{2},  \frac{3}{2} - \frac{i}{k}
\end{array}
\Biggr| 1 \right\}
{_3F}_2 \left\{
\begin{array}{l}
1-|m|, 1+|m|, 1-a_{\beta} \cr \\
\frac{3}{2},  \frac{3}{2} + \frac{i}{k}
\end{array}
\Biggr| 1 \right\}=
\\ [2mm]
= -
\frac{p\left(\frac{1}{4} + \frac{1}{k^2}\right)}
{2^{p+1}\cosh\frac{\pi}{k}}
\frac{\sqrt {\pi^3}\Gamma\left(\frac{1}{2} - a_{-\overline{\beta}}\right)}
{\Gamma\left(1 - a_{\overline{\beta}}\right)}
\Biggl[\frac{1}{\Gamma\left(\frac{1}{2}+a_{\beta}\right)
\Gamma\left(\frac{3}{2} - p- a_{\beta}\right)}\times
\\ [2mm]
\times
{_3F}_2 \left\{
\begin{array}{l}
\frac{1}{2}-p,  1-p, \frac{1}{2}+a_{-\beta} \cr \\
\frac{3}{2},  \frac{3}{2} - p- a_{\beta}
\end{array}
\Biggr| 1 \right\}+ \left(\beta \to -\beta\right)\Biggr].
\end{eqnarray*}
Note that in the particular case $\overline{\beta} = 0$ the coefficients
$P$ are noticeably simplified:
\begin{eqnarray*}
P_{k\beta 0}^{(+-)} = P_{k\beta 0}^{(--)} =0, \qquad
P_{k\beta 0}^{(++)} = \frac{e^{-\pi/k}}{k}C_{k\beta}^{(+)}
\cosh\frac{\pi\beta}{2k},    \qquad
P_{k\beta 0}^{(-+)} = -\frac{e^{-\pi/k}}{k}C_{k\beta}^{(+)}
\sinh\frac{\pi\beta}{2k}.
\end{eqnarray*}
Since the coefficients $W$ from (\ref{par-par-9}) are real,
follows the symmetry property
$P_{k\beta \overline{\beta}} = P_{k\overline{\beta} \beta}$.
Therefore, similar expressions are valid for the particular case
$\beta = 0$. Finally, for $\beta = \overline{\beta} =0$ we have
\begin{eqnarray*}
P_{k00}^{(+-)} = P_{k00}^{(--)} = P_{k00}^{(-+)} = 0, \qquad
P_{k\beta 0}^{(++)} = \frac{\left|\Gamma\left(
\frac{1}{4}+\frac{i}{2k}\right)\right|^2}
{2k\sqrt{\pi^3}}.
\end{eqnarray*}

\section{Algebraization of elliptic wave functions of two-dimensional
hydrogen atom in continuous spectrum}
\markboth{CHAPTER 1. TWO DIMENSIONAL HYDROGEN ATOM}{1.12.  ELLIPTIC WAVE FUNCTIONS IN CONTINUOUS SPECTRUM}

In the fifth section of this chapter, we showed that elliptic
Coulomb wave functions are also eigenfunctions
elliptic integral of motion $\Lambda$, which for the case
continuous spectrum can be written in the following form:
\bea
\label{CONT-ELLIP-1}
\Lambda = {L}^2 +  k R\, {\hat{\cal P}}.
\eea
Here $k = \sqrt{2E}$, and the operator ${\hat{\cal P}}$
is determined by the formula
\bea
\label{CONT-ELLIP-2}
{\hat{\cal P}} =
\frac{1}{2k}\frac{1}{\mu^2+\nu^2}
\left\{\mu^2\frac{\partial^2}{\partial \nu^2} -
\nu^2\frac{\partial^2}{\partial \mu^2} +
2(\mu^2+\nu^2)\right\}.
\eea
Let us denote the elliptic Coulomb wave functions with given
parity with respect to the transformation $\eta \rightarrow - \eta$
through $\Psi_{k \lambda}^{(\pm)}$, where $\lambda$ are eigenvalues
operator values (\ref{CONT-ELLIP-1}), which we will also provide
indexes $\ll\pm\gg$:
\bea
\label{CONT-ELLIP-3}
\Lambda \, \Psi_{k\lambda}^{(\pm)}(\xi, \eta) =
\lambda^{(\pm)} \, \Psi_{k\lambda}^{(\pm)}(\xi, \eta).
\eea
Let us expand the elliptic Coulomb wave functions in terms of the polar
basis of the two-dimensional hydrogen atom in the continuous spectrum
\bea
\label{CONT-ELLIP-4}
\Psi_{k \lambda}^{(\pm)} (\xi, \eta)=
\sum_{m=0}^{\infty}
W_{k \lambda m}^{(\pm)} \,
\Psi_{km}^{(\pm)} (r, \varphi).
\eea
Here the polar basis $\Psi_{km}^{(\pm)}$ is defined by the formula
(\ref{EL-COOR-60}).
From (\ref{CONT-ELLIP-3}) and (\ref{CONT-ELLIP-4}) in the standard way
we obtain equations that determine the form of the coefficients $W^{(\pm)}$ and
eigenvalues of the elliptic quantum number $\lambda$:
\bea
\label{CONT-ELLIP-5}
\sum_{m=0}^{\infty}
\left[\left(m^2-\lambda^{(+)}\right)
\pi\delta(k'-k)\delta_{mm'}(1+\delta_{m0})+
kR {\cal P}_{k'm';km}^{(+)}\right]\, W_{k\lambda m}^{(+)}=0,
\\[2mm]
\label{CONT-ELLIP-6}
\sum_{m=0}^{\infty}
\left[\left(m^2-\lambda^{(-)}\right)
\pi\delta(k'-k)\delta_{mm'}+
kR {\cal P}_{k'm';km}^{(-)}\right]\, W_{k\lambda m}^{(-)}=0,
\eea
in which
\bea
\label{CONT-ELLIP-7}
{\cal P}_{k'm';km}^{(\pm)} =  \int \Psi_{k'm'}^{*(\pm)}
{\hat{\cal P}} \Psi_{km}^{(\pm)}\, dV.
\eea
Let's calculate the matrix elements (\ref{CONT-ELLIP-7}).
Going to (\ref{CONT-ELLIP-2}) to polar coordinates, after
integration over $d\varphi$ we obtain
\bea
{\cal P}_{k'm';km}^{(+)}
&=&
\frac{1}{4k}
\left[(\delta_{m,m'+1}+\delta_{m,1-m'})E_{k'm';km}+
\delta_{m,m'-1}D_{k'm';km}\right],
\nonumber
\\[2mm]
{\cal P}_{k'm';km}^{(-)}
&=&
\frac{1}{4k}
\left[\delta_{m,m'+1}+E_{k'm';km}+\delta_{m,m'-1}D_{k'm';km}\right],
\nonumber
\eea
where
$$
E_{k'm';km} =  \int_{0}^{\infty}
R_{k'm'}^{*}(r)
\hat A_m R_{km}(r)rdr,
\qquad
D_{k'm';km} = \int_{0}^{\infty}
R_{k'm'}^{*}(r)
\hat A_{-m} R_{km}(r)rdr
$$
and the operator $\hat A_m$ has the form
\bea
\hat A_m = 1-\left(m-\frac{1}{2}\right)
\left(\frac{d}{dr}+\frac{m}{r}\right).
\nonumber
\eea
Further acting by the operator $\hat A_m$ on the radial wave function
$R_{km}$ and using the relations \cite{BE1}
\bea
\frac{d}{dx}F(a,c;x)=\frac{c-1}{x}\{F(a,c-1;x)-F(a,c;x)\},
\nonumber
\\[2mm]
\gamma(\gamma-1)F(a,\gamma-1;x)
-\gamma(\gamma-1+x)F(a,\gamma;x)+(\gamma-a)x
F(a,\gamma+1;x)=0,
\nonumber
\\[2mm]
\beta F(a,\beta;x)-\beta F(a-1,\beta;x)-xF(a,\beta+1;x)=0,
\nonumber
\eea
in which
$$
a = m+\frac{1}{2}-\frac{i}{k},
\qquad
c=2m+1,
\qquad
x=-2ikr
\nonumber
\\[2mm]
\gamma=c-1,
\qquad
\beta=c-2
$$
it can be proved that the operator $\hat A_m$ behaves like a lowering
operator:
\bea
\hat A_{m}\, R_{km}
= - k\, \sqrt{\left(m-\frac{1}{2}\right)^2+\frac{1}{k^2}}\, R_{k,m-1}.
\nonumber
\eea
From this it is also obvious that
\bea
\hat A_{-m}\, R_{km}
=-k\sqrt{\left(m+\frac{1}{2}\right)^2+\frac{1}{k^2}}\, R_{k,m+1}.
\nonumber
\eea
The obtained formulas together with the condition (\ref{2.7}) lead to the conclusion that
\bea
{\cal P}_{k'm';km}^{(+)}
&=&
-\frac{\pi}{2}\delta(k'-k)
\Biggl[\sqrt{\left(m-\frac{1}{2}\right)^2+\frac{1}{k^2}}
\left(\delta_{m,m'+1}+\delta_{m,1-m'}\right)+
\nonumber
\\[2mm]
\label{CONT-ELLIP-8}
&+&\sqrt{\left(m+\frac{1}{2}\right)^2+\frac{1}{k^2}}
\delta_{m,m'+1}\Biggr],
\\[2mm]
{\cal P}_{k'm';km}^{(-)} &=&
-\frac{\pi}{2}\delta(k'-k)
\Biggl[\sqrt{\left(m-\frac{1}{2}\right)^2+\frac{1}{k^2}}
\delta_{m,m'+1}+
\nonumber
\\[2mm]
\label{CONT-ELLIP-9}
&+&
\sqrt{\left(m+\frac{1}{2}\right)^2 +\frac{1}{k^2}}\delta_{m,m'-1}\Biggr].
\eea
Substituting (\ref{CONT-ELLIP-8}) and (\ref{CONT-ELLIP-9}) into
(\ref{CONT-ELLIP-5}) and (\ref{CONT-ELLIP-6}), we get three-term
recurrent relations
\bea
\frac{1}{2}
\sqrt{\left(m+\frac{1}{2}\right)^2+\frac{1}{k^2}}
W_{k\lambda,m+1}^{(+)}
&+&
\frac{1}{2}\sqrt{\left(m-\frac{1}{2}\right)^2+\frac{1}{k^2}}
\left[W_{k\lambda,m-1}^{(+)}+W_{k\lambda,1-m}^{(+)}\right]+
\nonumber
\\[2mm]
&+&
\frac{\lambda^{(+)}-m^2}{kR}\left[W_{k\lambda m}^{(+)}
+ W_{k\lambda,-m}\right]=0,
\nonumber
\\[2mm]
\frac{1}{2}
\sqrt{\left(m+\frac{1}{2}\right)^2+\frac{1}{k^2}}
W_{k\lambda,m+1}^{(-)}
&+&
\frac{\lambda^{(-)}-m^2}{kR}W_{k\lambda m}^{(-)}+
+\frac{1}{2}\sqrt{\left(m-\frac{1}{2}\right)^2+\frac{1}{k^2}}
W_{k\lambda,m-1}^{(-)}=0.
\nonumber
\eea
It follows from the decomposition (\ref{CONT-ELLIP-4}) that these equations must
be decided subject to additional conditions
$$
W_{k\lambda,-1}^{(+)} = W_{k\lambda,0}^{(-)} = 0.
$$
Thus, we have proved that the problem of the elliptic basis of a two-dimensional
hydrogen atom in the continuous spectrum can be reduced to the problem of studying
two three-term recurrence relations. This algebraization is convenient
to construct a perturbation theory for $R\ll 1$, i.e. to calculate
elliptic corrections to the polar basis.

\newpage

\chapter{ Hydrogen Atom}
\markboth{CHAPTER 2.  HYDROGEN ATOM}{}

In this chapter we will consider the following issues: hidden symmetry of the
hydrogen atom; expansion of spherical and parabolic bases of the hydrogen atom
in the discrete and continuous spectra; solution of the Schr\"{o}dinger
equation for the hydrogen atom in the prolate spheroidal coordinates and the
behavior of the prolate spheroidal basis as $R\to 0$ and $R\to\infty$ ($R$ is
the dimensional parameter entering into the definition of the prolate
spheroidal coordinates); expansion of the prolate spheroidal basis of the
hydrogen atom over the spherical and parabolic ones; spheroidal corrections to
the spherical and parabolic bases. In the presentation of this material we
follow papers \cite{ARUT1,T-1,MPSTA1,POGOSYAN1}.
\section{Hidden symmetry of the hydrogen atom}
\markboth{CHAPTER 2.  HYDROGEN ATOM}{2.1. HIDDEN SYMMETRY OF THE HYDROGEN ATOM}

The Hamiltonian of the hydrogen atom (everywhere in this chapter we use the
Coulomb units $\hbar=e=M=1$)
\begin{eqnarray}
{\hat H} = \frac{1}{2}{\hat {\bf p}}^2 - \frac{1}{r},
\qquad r= \sqrt{x_1^2+x_2^2+x_3^2},
\label{symh.1.1}
\end{eqnarray}
where ${\hat {\bf p}}^2 = {\hat p}_i {\hat p}_i$ and ${\hat p}_i= - \imath
\partial/\partial x_i$ $(i=1,2,3)$. Owing to that the potential
$-1/r$ is spherically symmetric the Hamiltonian (\ref{symh.1.1})
commutes with the orbital moment operator ${\hat{\bf l}}= {\bf
r}\times {\hat {\bf p}}$ whose components play the role of integrals
of motion. We use instead of the orbital moment vector ${\hat{\bf
l}}$ the second rank antisymmetric tensor (dual to the axial vector
${\hat{\bf l}}$)
\begin{eqnarray}
{\hat l}_{ij} = \epsilon_{ijk}{\hat l}_{k}
\equiv x_i {\hat p}_j - x_j {\hat p}_i,
\label{symh.1.2}
\end{eqnarray}
With the fundamental commutation relations
\begin{eqnarray}
\left[{\hat p}_i, f(x_1,x_2,x_3)\right]= -
i\frac{\partial f}{\partial x_i},
\qquad
i = 1,2,3,
\label{symh.1.5}
\end{eqnarray}
one can easily establish the commutation law
\begin{eqnarray}
\left[{\hat l}_{ik}, x_j\right] =
\imath \delta_{ij}x_k - \imath \delta_{kj}x_i,
\qquad
\left[{\hat l}_{ik}, {\hat p}_j\right] =
\imath \delta_{ij}{\hat p}_k -
\imath \delta_{kj}{\hat p}_i,
\label{symh.1.6}
\\[3mm]
\left[{\hat l}_{ij}, {\hat l}_{mn}\right] = \imath
\delta_{im}{\hat l}_{jn} - \imath \delta_{in}{\hat l}_{jm}
- \imath \delta_{jm}{\hat l}_{in}
+ \imath \delta_{jn}{\hat l}_{im}.
\label{symh.1.7}
\end{eqnarray}

The last relation indicates that the operators ${\hat l}_{ij}$ form an algebra
isomorphic to the algebra $so(3)$, thus confirming that the group of geometric
symmetry of the hydrogen atom is the group of three-dimensional rotations -
$SO(3)$. A direct calculation can show that the Runge-Lenz vector \cite{LL},
\begin{eqnarray}
{\hat {\bf A}} = \frac{1}{2}\left({\hat{\bf l}}\times {\hat{\bf p}}-
{\hat{\bf p}}\times {\hat{\bf l}}\right) +\frac{{\bf r}}{r}
\label{symh.1.3}
\end{eqnarray}
which, with account taken of (\ref{symh.1.2}), can be rewritten in the form
\begin{eqnarray}
{\hat A}_j = \frac{1}{2}\left({\hat l}_{ij}{\hat p}_i+
{\hat p}_i{\hat l}_{ij}\right) +\frac{x_j}{r},
\qquad j=1,2,3 \,,
\label{symh.1.4}
\end{eqnarray}
also commutes with Hamiltonian (\ref{symh.1.1}). Indeed, using auxiliary
commutation relations
\begin{eqnarray*}
\left[{\hat p}_i, \frac{x_j}{r}\right]
= -\frac{\imath}{r}\delta_{jn} + \imath
\frac{x_ix_j}{r^3},
\qquad
\left[{\hat H}, {\hat p}_i\right] = \imath
\frac{x_ix_j}{r^3},\qquad \left[{\hat H}, \frac{x_j}{r}\right]
=\frac{x_j}{r^3}-\frac{\imath}{r}{\hat p}_j
+ \imath \frac{x_ix_j}{r^3}{\hat p}_i,
\end{eqnarray*}
one can prove that $\left[{\hat H}, {\hat A}_j\right]=0$, $(j=1,2,3)$ and,
consequently, the components of the vector ${\hat {\bf A}}$ are also the
integrals of motion. The conserved quantities ${\hat A}_j$, which are not
related with purely geometric invariance of Hamiltonian (\ref{symh.1.1}), are
often called the integrals of motion.

Let us construct the algebra of the three-dimensional hydrogen atom symmetry.
Using formulae (\ref{symh.1.6}) and (\ref{symh.1.7}) we get
\begin{eqnarray}
\left[{\hat l}_{ij}, {\hat A}_j\right] =
\imath \delta_{ik}{\hat A}_j - \imath \delta_{kj}{\hat A}_i.
\label{symh.1.8}
\end{eqnarray}
Taking into account the last formula and the commutation relations
\begin{eqnarray*}
\left[{\hat A}_i, {\hat p}_j\right]
&=& -\delta_{ij}\left({\hat {\bf p}}^2-
\frac{1}{r}\right) + \imath {\hat p}_i{\hat p}_j -
\imath \frac{x_ix_j}{r^3},
\\[3mm]
\left[{\hat A}_i, \frac{x_j}{r}\right]
&=& \frac{\imath}{r} {\hat l}_{ij} + \frac{\imath}{r}
\left(x_i{\hat p}_j+x_j{\hat p}_i\right)-2\frac{x_ix_j}{r^3}
- \frac{x_ix_j}{r^3}x_k{\hat p}_k
- \frac{\imath}{r}\delta_{ij}x_k{\hat p}_k
\end{eqnarray*}
after simple calculations we get
\begin{eqnarray}
\left[{\hat A}_i, {\hat A}_j\right] = - 2\imath {\hat H}{\hat l}_{ij},
\label{symh.1.08}
\end{eqnarray}
The obtained commutation relations (\ref{symh.1.7}), (\ref{symh.1.8}) and
(\ref{symh.1.08}) generate the infinite dimensional linear Lie algebra. To
close the algebra, we use the fact that at fixed value of negative energy (the
state of the discrete spectrum) in (\ref{symh.1.08}) the Hamiltonian can be
changed by the constant $E$ and assume ${\hat A'}_i = {\hat A}_i/\sqrt{-2E}$.
For the operator ${\hat A'}$ the commutation rules (\ref{symh.1.8}) and
(\ref{symh.1.08}) take the form
\begin{eqnarray}
\left[{\hat l}_{ij}, {\hat A'}_j\right] =
\imath \delta_{ik}{\hat A'}_j - \imath \delta_{kj}{\hat A'}_i,
\qquad
\left[{\hat A'}_i, {\hat A'}_j\right]=\imath {\hat l}_{ij}.
\label{symh.1.9}
\end{eqnarray}
Now we introduce the new operators ${\hat D}_{\mu\nu}$ $(\mu, \nu =1,2,3,4)$:
${\hat D}_{ij}\equiv {\hat l}_{ij}$ $(i,j =1,2,3)$ and ${\hat D}_{4i}\equiv
{\hat A'}_{i}$. Then instead of the commutation relations (\ref{symh.1.7}) and
(\ref{symh.1.9}) we get
\begin{eqnarray}
\left[{\hat D}_{\mu \nu}, {\hat D}_{\lambda \rho}\right]=
\imath \delta_{\mu\lambda}{\hat D}_{\nu\rho}-
\imath \delta_{\mu\rho}{\hat D}_{\nu\lambda}-
\imath \delta_{\nu\lambda}{\hat D}_{\mu\rho}
+ \imath \delta_{\nu\rho}{\hat D}_{\mu\lambda},
\label{symh.1.10}
\end{eqnarray}
i.e., the components of the operator ${\hat D}_{\mu\nu}$ coincide with the
generators of the group $SO(4)$. Thus, it is proved that in the case of the
discrete spectrum the group $SO(4)$ is the group of "hidden" or dynamic
symmetry of the hydrogen atom \cite{FOCK1,FOCK2,LL}. In so doing one can prove
that the Lorentz group $SO(3,1)$ is the group of "hidden" symmetry of the
hydrogen atom describing the continuous spectrum. Indeed, passing from the
operators ${\hat A}_i$ to ${\hat A''}_i = {\hat A}_i/\sqrt{2E}$ we get instead
of (\ref{symh.1.9}) the commutation rules
\begin{eqnarray}
\left[{\hat l}_{ij}, {\hat A''}_j\right] = \imath \delta_{ik}{\hat
A''}_j - \imath \delta_{kj}{\hat A''}_i, \qquad \left[{\hat
A''}_i, {\hat A''}_j\right]= - \imath {\hat l}_{ij},
\label{symh.1.09}
\end{eqnarray}
which together with (\ref{symh.1.7}) form the Lie algebra $so(3,1)$. In the
space of wave functions with zero energy, as it follows from (\ref{symh.1.08}),
the Runge-Lenz vector components commute with each other. The algebra of
operators ${\hat l}_{ij}, {\hat A}_j$ contains the Abelian subalgebra and is
isomorphic to the three-dimensional Euclidean algebra $e(3)$\footnote{This kind
of limiting transition from one Lie algebra to another which is not isomorphic
to the initial one is called the Wigner-In\"{o}n\"{u} contractions \cite{IW}
(see also \cite{IPSW4,IPSW2,POGSIS}).}.

The explicit form of the "hidden" symmetry of the hydrogen atom can be seen if
one passes in the Schr\"{o}dinger equation from the coordinate representation
to the momentum one. After stereographic projection it takes the form of the
equation describing a free motion on the three-dimensional sphere (discrete
spectrum) or three-dimensional hyperboloid (continuous spectrum) whose isometry
groups are the $SO(4)$ and $SO(3,1)$ groups, respectively \cite{BANDER}. The
hidden symmetry is responsible for the separation of variables in the
Schr\"{o}dinger equation for the hydrogen atom in several orthogonal systems of
coordinates: spherical, parabolic, spheroidal, and sphero-conical
\footnote{Separation of variables in the sphero-conical system of coordinates
is specified by the spherical symmetry of the hydrogen atom Hamiltonian.}. If
solutions in the spherical and parabolic system were known even at the dawn of
the development of quantum mechanics, a spheroidal analysis of the hydrogen
atom was carried out at the end of the 1950s by Coulson and Robertson
\cite{COULSON1} and later by Coulson and Joseph \cite{COJO}.

Information on the group of hidden symmetry allows one to calculate in a purely
algebraic way eigenvalues of the hydrogen atom energy. From \cite{BARUT-RON} it
is known that two Casimir operators for the group $SO(4)$ have the form
\begin{eqnarray*}
{\hat C}_2 =\frac{1}{2}{\hat D}_{\mu \nu}{\hat D}_{\mu \nu}, \qquad
{\hat C}_3 = \epsilon_{\mu \nu \lambda \rho}{\hat D}_{\mu \nu}
{\hat D}_{\lambda \rho},
\end{eqnarray*}
or taking into account the definitions of the operators ${\hat D}_{\mu \nu}$
\begin{eqnarray}
{\hat C}_2 = \frac{1}{2}{\hat l}_{ij}^2
- \frac{1}{2E}{\hat A}_{i}^2,
\qquad
{\hat C}_3 = \frac{1}{\sqrt{-2E}}\,
\epsilon_{ijk}{\hat l}_{ij}{\hat A}_{k}.
\label{symh.1.11}
\end{eqnarray}
Commutation rules(\ref{symh.1.6}) and the equalities ${\hat l}_{ij}^2 = 2{\hat
l}_{ij}x_i{\hat p}_j$ and ${\hat l}_{ij}^2{\hat {\bf p}}^2 = 2{\hat
l}_{ij}{\hat l}_{ik}{\hat p}_j{\hat p}_k$ allow one to write down ${\hat
A}_{i}^2$ in a simple form
\begin{eqnarray*}
{\hat A}_{i}^2 = 2E{\hat l}_{ij}^2 + 2E +1.
\end{eqnarray*}
whence we have
\begin{eqnarray}
{\hat C}_2 = - \frac{1}{2E} - 1.
\label{symh.1.12}
\end{eqnarray}
Using the explicit form of the operators ${\hat D}_{\mu \nu}$ and the following
simple identities $\epsilon_{ijk}x_i{\hat l}_{jk} = 2\epsilon_{ijk}x_ix_j {\hat
p}_{k} \equiv 0$ and
 $\epsilon_{ijk}{\hat l}_{ij}{\hat p}_{k} =
2\epsilon_{ijk}x_i{\hat p}_{j}{\hat p}_{k} \equiv 0$, after short calculations
we get
\begin{eqnarray*}
{\hat C}_{3} = \frac{4}{\sqrt{-2E}}
\left[-\left({\hat {\bf p}}^2-\frac{1}{r}\right)\epsilon_{ijk}
x_i{\hat l}_{jk} +\left(x_m{\hat p}_m - \imath\right)\epsilon_{ijk}
{\hat l}_{jk}{\hat p}_k\right] = 0.
\end{eqnarray*}
On the other hand, according to \cite{BARUT-RON,PER-POP1}, eigenvalues of the
Casimir operators of the group $SO(4)$ can be represented as
\begin{eqnarray}
C_2 = \mu_1(\mu_1+2) + \mu_2, \qquad C_3 = 8(\mu_1 +1)\mu_2,
\label{symh.1.13}
\end{eqnarray}
where $\mu_1$ and $\mu_2$ are nonnegative integers with $\mu_1 \geq \mu_2 \geq
0$. It follows from the condition $C_3=0$ that $\mu_2 = 0$ and
\begin{eqnarray*}
- \frac{1}{2E} -1 = \mu_1(\mu_1 +2),
\end{eqnarray*}
whence we find an expression for the spectrum of energies
\begin{eqnarray}
E_n  = - \frac{1}{2n^2},
\label{symh.1.14}
\end{eqnarray}
where $n= \mu_1 + 1$ is the principal quantum number taking integer positive
values.

\section{Solution to the Park--Tarter problem}
\markboth{CHAPTER 2.  HYDROGEN ATOM}{2.2. SOLUTION TO THE PARK--TARTER PROBLEM}
In this section, a simple method is suggested to calculate the coefficients of
the unitary transformation connecting spherical and parabolic wave functions of
the hydrogen atom in the discrete spectrum.

Let us write down the unitary transformation we are interested in as
\begin{eqnarray}
\Psi_{n_1n_2m}(\mu, \nu, \varphi) =
\sum_{l=|m|}^{n-1}
\,
W^{nl}_{n_1n_2m}
\,
\Psi_{nlm}(r, \theta, \varphi),
\label{PT.1.1}
\end{eqnarray}
where the principal quantum number $n=n_1+n_2+|m|+1$, and the spherical and
parabolic wave functions $\Psi_{nlm}(r,\theta,\varphi)$ and
$\Psi_{n_1n_2m}(\mu, \nu, \varphi)$ correspond to the same energy level and are
chosen in the following way \cite{LL}:
\begin{eqnarray}
\Psi_{nlm}(r, \theta, \varphi)
&=& R_{nl}(r)\,Y_{lm}(\theta, \varphi),
\label{sh.1.1}
\\[3mm]
\Psi_{n_1n_2m}(\mu, \nu, \varphi)
&=& \frac{1}{n^2}
f_{n_1 m}(\mu)\,f_{n_2 m}(\nu) \,
\frac{e^{im\varphi}}{\sqrt{\pi}},
\label{ph.1.1}
\end{eqnarray}
where
\begin{eqnarray}
R_{nl}(r)
&=& \frac{2}{n^2(2l+1)!}
\sqrt{\frac{(n+l)!}{(n-l-1)!}}\,\left(\frac{2r}{n}\right)^l
e^{-\frac{r}{n}}\, F\left(-n+l+1, 2l+2,\frac{2r}{n}\right),
\label{rh.1.1}
\\[3mm]
Y_{lm}(\theta, \varphi)
&=& \frac{(-1)^{\frac{m-|m|}{2}}}{2^l l!}
\sqrt{\frac{(2l+1)(l+|m|)!}{4\pi (l-|m|)!}}\frac{e^{im\varphi}}
{\left(\sin\theta\right)^{|m|}}
\,\frac{d^{l-|m|}}{\left(d\cos\theta\right)^{l-|m|}}
\left(\cos^2\theta -1\right)^l.
\label{spf.1.1}
\end{eqnarray}
and
\begin{eqnarray}
f_{p q}(x) = \frac{1}{|q|!} \sqrt{\frac{(p+|q|)!}{p!}}
e^{-x/2n}\left(\frac{x}{n}\right)^{|q|/2} F\left(-p;|q|+1;
\frac{x}{n}\right).
\label{ph.1.2}
\end{eqnarray}
The problem of calculation of interbasis coefficients $W^{nl}_{n_1n_2m}$ can be
essentially simplified if one uses the method of asymptotics. Indeed, passing
in the left-hand side of relation (\ref{PT.1.1}) from the parabolic coordinates
to the spherical ones, according to the formulae
\begin{eqnarray}
\mu = r(1+\cos\theta), \qquad \nu = r(1-\cos\theta),
\label{sp-pr.1.1}
\end{eqnarray}
and taking the limit $r \to \infty$, one can easily show that the dependence on
$r$ in (\ref{PT.1.1}) disappears . Multiplying both sides of the asymptotic
equality thus obtained by $Y_{lm}^*(\theta, \varphi)$ and integrating over the
solid angle we get
\begin{eqnarray}
W_{n_1n_2m}^{nl}
&=& \frac{(-1)^{\frac{m-|m|}{2}}}{2^{n+l} l!}
\sqrt{\frac{(2l+1)(l+|m|)!(n+l)!(n-l-1)!}{(l-|m|)!(n_1)!(n_2)!
(n_1+|m|)!(n_2+|m|)!}}\times
\nonumber\\[3mm]
\label{h.1.1}
&\times&
\int\limits_{-1}^{1}\,(1-x)^{n_2}\,(1+x)^{n_1}\,
\frac{d^{l-|m|}}{\left(d x\right)^{l-|m|}}
\left(1 -x^2\right)^l\, dx,
\end{eqnarray}
where $x=\cos\theta$. Now comparing \ref{h.1.1}) with the well-known integral
representation for the Clebsch-Gordan coefficients of the group $SU(2)$
(\ref{KG.1.1}) we immediately arrive at the sought result
\begin{eqnarray}
W_{n_1n_2m}^{nl} = (-1)^{l+n_2+\frac{m-|m|}{2}}
\,\,C_{\frac{n-1}{2},\,\frac{n_1-n_2+|m|}{2};\,\,\,
\frac{n-1}{2},\,\frac{n_2-n_1+|m|}{2}}^{l,\,|m|}\,.
\label{h.1.2}
\end{eqnarray}
Thus, transformation (\ref{PT.1.1}) and its inverse transformation can be
written as
\begin{eqnarray}
\Psi_{n_1n_2m}
&=& (-1)^{n_2+\frac{m-|m|}{2}}
\sum_{l=|m|}^{n-1}\,(-1)^{l}\,
C_{\frac{n-1}{2},\,\frac{n_1-n_2+|m|}{2};\,\,\,
\frac{n-1}{2},\,\frac{n_2-n_1+|m|}{2}}^{l,\,|m|}\,
\Psi_{nlm},
\label{PT.1.2}
\\[3mm]
\Psi_{nlm}
&=& (-1)^{l+\frac{m-|m|}{2}}
\sum_{n_2=0}^{n-|m|-1}\,(-1)^{n_2}\,
C_{\frac{n-1}{2},\,\frac{n-1}{2}-n_2;\,\,\,
\frac{n-1}{2},\,n_2+|m|-\frac{n-1}{2}}^{l,\,|m|}\,
\Psi_{n_1n_2m}.
\label{PT.1.3}
\end{eqnarray}
These formulae are in agreement with those derived by Park \cite{PARK}.

Tarter's result \cite{TARTER} for the expansion coefficients (\ref{PT.1.1}) can
be reproduced from relation (\ref{h.1.2}) if one uses the following formula
connecting the Clebsch--Gordan coefficients with the generalized hypergeometric
function ${_3F}_2$ of the unit argument
\begin{eqnarray}
C_{a \alpha ; b \beta}^{c\gamma}
&=& \left[
\frac{(2c+1)(b-a+c)!(a+\alpha)!(b+\beta)!(c+\gamma)!}
{(b-\beta)!(c-\gamma) ! (a+b-c)!(a-b+c)!(a+b+c+1)!} \right] ^{1/2}\times
\nonumber \\[2mm]
\label{KG.1.3}
&\times&
\delta_{\gamma,\alpha+\beta}\frac{(-1)^{a-\alpha}}{\sqrt{(a-\alpha)!}}
\frac{(a+b-\gamma)!}{(b-a+\gamma)!} {_3F}_2 \left\{
\begin{array}{l}
-a+\alpha, c+\gamma+1, -c+\gamma \\
\gamma-a-b,  b-a+\gamma+1  \\
\end{array}
\Biggr| 1 \right\}.
\end{eqnarray}
Now comparing (\ref{h.1.2}) with the formula (\ref{KG.1.3}) we obtain
ratio
\begin{eqnarray}
W_{n_1n_2m}^{nl} &=& (-1)^{l+\frac{m-|m|}{2}}\,
\sqrt{\frac{(2l+1)(n_1+|m|)!(n_2+|m|)!(l+|m|)!}
{(n_1)!(n_2)!(l-|m|)!(n-l -1)!(n+l)!}}\times
\nonumber
\\[3mm]
\label{h.1.3}
&\times&  \frac{(n -|m|-1)!}{|m|!}\,
{_3F}_2 \left\{
\begin{array}{l}
-n_2,  -l +|m|, l+|m|+1
\\ [2mm] |m|+1,   -n+|m|+1
\end{array}
\biggr| 1 \right\}.
\end{eqnarray}
The transformation coefficients $W_{n_1n_2m}^{nl}$ in (\ref{h.1.3}) coincide to
the phase factor $(-1)^{\frac{m+|m|}{2}}$ with Tarter's matrix \cite{TARTER}.
The existing distinction is due to the choice of a phase factor in the
spherical function $Y_{lm}$.

\section{Prolate spheroidal basis of the hydrogen atom}
\markboth{CHAPTER 2.  HYDROGEN ATOM}{2.3. PROLATE SPHEROIDAL BASIS OF THE HYDROGEN ATOM }
Spheroidal coordinates are a natural tool for investigation of many problems of
mathematical physics \cite{KOMPOCLA}. In quantum mechanics these coordinates
are used to describe the behavior of a charged particle in the field of two
Coulomb centers. The distance $R$ between the centers is taken as a dimensional
parameter specifying spheroidal coordinates and has a dynamic meaning, i.e.,
enters into the energy--spectrum expression. If the charge of one of the
centers is put as zero, one arrives at a one-center problem and the parameter
$R$ becomes purely kinematic. This simplifies the problem considerably. At the
same time the mathematical structure of the spheroidal equations remains the
same as the energy enters into both the radial and angular equations.
Consequently, the spheroidal analysis of a hydrogen atom becomes the first step
in the investigation of the two-center Coulomb problem.

\subsection{Prolate spheroidal coordinates}

Now we present the information necessary for further description. The prolate
spheroidal coordinates $\xi, \eta, \varphi$ are determined in the following
way:
\begin{eqnarray}
x = \frac{R}{2}\sqrt{(\xi^2-1)(1-\eta^2)}\cos \varphi, \quad y =
\frac{R}{2}\sqrt{(\xi^2-1)(1-\eta^2)}\sin \varphi,  \quad z =
\frac{R}{2}(\xi \eta +1), \label{sfh.3.1}
\end{eqnarray}
where $\xi \in [1, \infty)$, $\eta \in [-1, 1]$, $\varphi \in [0, 2\pi)$, $R
\in [0, \infty)$. The differential elements of length, volume, and the Laplace
operator have the form
\begin{eqnarray}
dl^2 &=& \frac{R^2}{4}\left[(\xi^2 - \eta^2)\left(
\frac{d\xi^2}{\xi^2-1}+\frac{d\eta^2}{1-\eta^2}\right)+
(\xi^2-1)(1-\eta^2)d\varphi^2\right],
\label{sfh.3.2}
\\[2mm]
dV &=& \frac{R^3}{8}(\xi^2 - \eta^2)d\xi d\eta d \varphi,
\label{sfh.3.3}
\\
\Delta &=& \frac{4}{R^2(\xi^2 {-} \eta^2)}\left[
\frac{\partial}{\partial \xi}(\xi^2-1)\frac{\partial}{\partial
\xi} + \frac{\partial}{\partial
\eta}(1-\eta^2)\frac{\partial}{\partial \eta} \right] +
\frac{4}{R^2(\xi^2{-}1)(1{-}\eta^2)} \frac{\partial^2}{\partial
\varphi^2}.
\label{sfh.3.4}
\end{eqnarray}
The parameter $R$ is an interfocus distance, and as $R \to 0$ and $R \to
\infty$ the prolate spheroidal coordinates turn into spherical and parabolic
coordinates, respectively. Indeed, the terms of an infinite set of
three-dimensional spheroidal coordinates (\ref{sfh.3.1}) differ from each other
by a value of the parameter $R$. At arbitrary $R$ the prolate spheroidal
coordinates are related with spherical and parabolic coordinates as follows:
\begin{eqnarray}&
\xi = \frac{r}{R} + \sqrt{1-\frac{2r}{R}\cos\theta +
\frac{r^2}{R^2}}, \qquad   \eta = \frac{r}{R} -
\sqrt{1-\frac{2r}{R}\cos\theta + \frac{r^2}{R^2}}, \label{sfh.3.5}
\\&
\xi = \frac{\mu + \nu}{R} + \sqrt{1-\frac{\mu - \nu}{R} +
\frac{(\mu + \nu)^2}{R^2}}, \quad    \eta = \frac{\mu + \nu}{R} -
\sqrt{1-\frac{\mu - \nu}{R} + \frac{(\mu + \nu)^2}{R^2}}.
\label{sfh.3.6}
\end{eqnarray}
Formulae (\ref{sfh.3.5}) and (\ref{sfh.3.6}) make it possible to study the
behavior of the spheroidal coordinates at small and large $R$. One can easily
show that as $R \to 0$
\begin{eqnarray}
\xi \to \frac{2r}{R}, \qquad \eta \to \cos\theta, \label{sfh.3.7}
\end{eqnarray}
and as $R \to \infty$
\begin{eqnarray}
\xi \to \frac{\nu}{R}+1, \qquad \eta \to \frac{\mu}{R}-1.
\label{sfh.3.8}
\end{eqnarray}
In both the limits the point $(x, y, z)$ and the coordinates of the Coulomb
center are thought to be fixed.

\subsection{Explicitly factorized prolate spheroidal basis
of the hydrogen atom }

In this subsection, we give a prolate spheroidal basis of the hydrogen atom
turning at small and large $R$ into the spherical and parabolic bases of the
hydrogen atom, respectively. This fact distinguishes our solution from that
obtained earlier in the well-known paper \cite{COULSON1}. For details
concerning the spheroidal basis of the hydrogen atom and the relevant
references see the monograph \cite{KOMPOCLA}.

Separation of variables
\begin{eqnarray}
\Psi_{nqm}\left(\xi, \eta, \varphi; R\right)
= C_{nqm}\left(R\right)\,
\Pi_{nqm}\left(\xi; R\right)\,
\Xi_{nqm}\left(\eta; R\right)\,
\frac{e^{im\varphi}}{\sqrt{2\pi}}
\label{sfh.3.9}
\end{eqnarray}
in the prolate spheroidal coordinates $\xi,\eta,\varphi$ results in the
equations
\begin{eqnarray}
\left[\frac{d}{d\xi}(\xi^2-1)\frac{d}{d\xi}
-\frac{m^2}{\xi^2-1}+\frac{ER^2}{2}(\xi^2-1)
+ R\xi\right] \Pi\left(\xi; R\right)
&=&A\Pi\left(\xi; R\right),
\label{sfh.3.12}
\\ [3mm]
\left[\frac{d}{d\eta}(1-\eta^2)\frac{d}{d\eta}-\frac{m^2}{1-\eta^2}+
\frac{ER^2}{2}(1-\eta^2)
-R\eta\right] \Xi\left(\eta; R\right)
&=&-A\Xi\left(\eta; R\right).
\label{sfh.3.13}
\end{eqnarray}
Here we used the following notation: $C_{nqm}\left(R\right)$ is the
normalization factor, the functions $\Pi_{nqm}\left(\xi; R\right)$ and
$\Xi_{nqm}\left(\eta; R\right)$ are commonly called the oblate radial and
angular spheroidal Coulomb functions, $A(R)$ is the spheroidal separation
constant, $n$ is the principal quantum number, $m$ is the azimuth quantum
number, and $q$ is the quantum number varying in the limits $1\leq q \leq
n-|m|$ and denoting in ascending order $n-|m|$ values of the separation
constant $A(R)$. At $q\not= q'$ the following orthogonality conditions are
valid:
\begin{eqnarray}
\label{SPHERIDAL-ORT1}
\int\limits_{1}^{\infty}\,
\Pi_{nq'm}^{*}\left(\xi; R\right)
\Pi_{nqm}\left(\xi; R\right)
\, d\xi
&=& 0,
\\[2mm]
\label{SPHERIDAL-ORT2}
\int\limits_{-1}^{1}\,
\Xi_{nq'm}^{*}\left(\eta; R\right)
\Xi_{nqm}\left(\eta; R\right)\,
d\eta
&=&0.
\end{eqnarray}
Let us single out singularities in (\ref{sfh.3.12}) and (\ref{sfh.3.13})
\begin{eqnarray}
\label{SPHERIDAL-1.1}
\Pi_{nqm}\left(\xi; R\right)
&=&
e^{-\xi R/2n} \, (\xi^2-1)^{\frac{|m|}{2}} \,
f_{nqm}\left(\xi; R\right),
\\[2mm]
\label{SPHERIDAL-1.2}
\Xi_{nqm}\left(\eta; R\right)
&=&
e^{-\eta R/2n} \, (1-\eta^2)^{\frac{|m|}{2}} \,
g_{nqm}\left(\eta; R\right).
\end{eqnarray}
Representing the functions $f_{nqm}\left(\xi; R\right)$ and $g_{nqm}\left(\eta;
R\right)$ in power series $(\xi-1)$ and $(1+\eta)$,that these series cut off
and obtain the energy spectrum of the hydrogen atom. As a result, we arrive at
the following polynomials
\begin{eqnarray}
\label{sfh.3.14}
f_{nqm}\left(\xi; R\right)
&=& \sum_{s=0}^{n-|m|-1}\, a_s(R) \left(\xi -1\right)^s,
\\ [3mm]
\label{sfh.3.15}
g_{nqm}\left(\eta; R\right)
&=& \sum_{s=0}^{n-|m|-1}\, b_s(R) \left(1+\eta\right)^s,
\end{eqnarray}
where the expansion coefficients $a_s$ and $b_s$ obey the trinomial recurrence
relations :
\begin{eqnarray}
\alpha_s a_{s+1} + \beta_sa_s + R\gamma_s a_{s-1}&=&0,
\label{sfh.3.16}
\\ [3mm]
-\alpha_s b_{s+1} + {\tilde \beta}_sb_s + R\gamma_s b_{s-1}&=&0,
\label{sfh.3.17}
\end{eqnarray}
in which
\begin{eqnarray}
\label{sfh.3.18}
\alpha_s
&=& 2(s+1)(s+|m|+1),
\\[3mm]
\label{sfh.3.19}
\beta_s
&=& \left(s+|m|\right)\left(s+|m|+1\right) +
\frac{R}{n}\left(n-|m|-2s-1\right)-\frac{R^2}{4n^2}+ A_q(R),
\\[3mm]
\label{sfh.3.20}
{\tilde \beta}_s
&=& \left(s+|m|\right)\left(s+|m|+1\right)
- \frac{R}{n}\left(n-|m|-2s-1\right)
- \frac{R^2}{4n^2}+A_q(R),
\\ [3mm]
\label{sfh.3.21}
\gamma_s &=& \frac{1}{n}\left(n-|m|-s\right).
\end{eqnarray}
The "cut-off" conditions: $a_{-1} = a_{n-|m|}\equiv 0$, $b_{-1}=b_{n-|m|}\equiv
0$ и $a_0=b_0=1$. hold as well. The normalization constant $C_{nqm}(R)$ is
calculated from the condition
\begin{eqnarray}
\label{sfh.3.22}
\frac{R^3}{8}\int\limits_{1}^{\infty}\,\int\limits_{-1}^{1}\,
\int\limits_{0}^{2\pi}\,
\left|\psi_{nqm}\left(\xi, \eta, \varphi; R\right)\right|^2
\left(\xi^2-\eta^2\right)\,d\xi d\eta d \varphi = 1
\end{eqnarray}
and it equals
\begin{eqnarray}
\label{sfh.3.23}
\left|C_{nqm}(R)\right|^2 = \frac{1}{2^{4|m|+1}R^3}
\left[\sum_{ss'tt'}\,2^{s+s'+t+t'}\,a_sa_{s'}b_tb_{t'}
\left(I_{ss'}^{|m|}{\cal I}_{tt'}^{|m|+1}+
I_{ss'}^{|m|+1}{\cal I}_{tt'}^{|m|}\right)\right]^{-1},
\end{eqnarray}
where the summation over all four indices is carried out in the limits $(0,
n-|m|-1)$, and $I_{ss'}^{|m|}$ and ${\cal I}_{tt'}^{|m|}$ are expressed in
terms of the degenerate hypergeometric functions as follows (see formulae
(\ref{sfh.3.24}) and (\ref{sfh.3.25}) from Chapter 1):
\begin{eqnarray}
\label{sfh.3.26}
I_{ss'}^{|m|} &=& \left(|m|+s+s'\right)!
\psi\left(|m|+s+s'+1; 2|m|+s+s'+2; \frac{2R}{n}\right),
\\[3mm]
\label{sfh.3.27}
{\cal I}_{tt'}^{|m|} &=& \frac{|m|!\left(|m|+t+t'\right)!}
{\left(2|m|+t+t'+1\right)!}
F\left(|m|+t+t'+1; 2|m|+t+t'+2; -\frac{2R}{n}\right).
\end{eqnarray}

\section{Limiting transitions in recurrence relations}
\markboth{CHAPTER 2.  HYDROGEN ATOM}{2.4. LIMITING TRANSITIONS IN RECURRENCE RELATIONS}
The recurrence relation (\ref{sfh.3.16}) is a set of linear homogeneous
equations $a_s$ and, therefore, the corresponding determinant should vanish.
Hence one can determine eigenvalues of the separation constant $A(R)$. We shall
prove below that in the limit $R\to 0$ and $R\to \infty$ (\ref{sfh.3.16}) and
(\ref{sfh.3.17}) turn into binomial recurrence relations.

\subsection{Limit $R\to \infty$.} In the limit $R\to
\infty$ one can reject the coefficients $\alpha_s$  in the above determinant
"at the background" of infinitely large coefficients $\beta_s$ and $R\gamma_s$
and write down the determinant as a product of all the coefficients $\beta_s$.
Then the condition of equality of the determinant to zero requires that one of
the factors $\beta_s$ vanish. Denoting the number $s$ corresponding to this
factor by $n_2$, we have
\begin{eqnarray}
\label{sfh.4.1}
A(R)\stackrel{R\to \infty}{\longrightarrow}
- \frac{R}{n} \left(n-|m|-2n_2-1\right) + \frac{R^2}{4n^2}.
\end{eqnarray}
Using this line of reasonings with reference to (\ref{sfh.3.17}) one can show
that there exists such $n_1$ for which $\tilde{\beta}_{n_1}=0$ and
\begin{eqnarray}
\label{sfh.4.2}
A(R) \stackrel{R\to \infty}{\longrightarrow}
\frac{R}{n}\left(n-|m|-2n_1-1\right)+\frac{R^2}{4n^2}.
\end{eqnarray}
Formulae (\ref{sfh.4.1}) and (\ref{sfh.4.2}) are compatible if
$n=n_1+n_2+|m|+1$, i.e., if $n_1$ and $n_2$ are parabolic quantum numbers.

It is seen from relations (\ref{sfh.3.19}), (\ref{sfh.3.20}), (\ref{sfh.4.1})
and (\ref{sfh.4.2}) that at $s\neq n_1$, $s\neq n_2$
\begin{eqnarray}
\label{sfh.4.3}
\beta_s \stackrel{R\to \infty}{\longrightarrow}
R\beta_s^{(1)},
\qquad
\tilde{\beta}_s \stackrel{R\to \infty}{\longrightarrow}
R\beta_s^{(2)},
\end{eqnarray}
where the quantities $\beta_s^{(1)}$ and $\beta_s^{(2)}$ are independent of $R$
and have the form
\begin{eqnarray}
\label{sfh.4.4}
\beta_s^{(1)} = 2\frac{n_2-s}{n}, \qquad
\beta_s^{(2)} = -2\frac{n_1-s}{n}.
\end{eqnarray}
These formulae and the cut-off conditions $a_1=b_1\equiv 0$ show that in the
limit $R\to \infty$ the trinomial recurrence relations (\ref{sfh.3.16}) and
(\ref{sfh.3.17}) transform into the following binomial ones:
\begin{eqnarray}
\label{sfh.4.5}
\alpha_s a_{s+1} + R\beta_s^{(1)}a_s
&=& 0,
\\[3mm]
\label{sfh.4.6}
- \alpha_sb_{s+1} +R\beta_s^{(2)}b_s
&=& 0,
\end{eqnarray}
if $0\leq s \leq n_1-1$ and $0\leq s \leq n_2-1$. From (\ref{sfh.4.3}) and
(\ref{sfh.4.4}), and the conditions $a_{n-|m|}=b_{n-|m|}\equiv 0$ one can prove
in a similar way the relations
\begin{eqnarray}
\beta_s^{(1)}a_s + \gamma_s a_{s-1}
&=& 0,
\label{sfh.4.7}
\\ [3mm]
\beta_s^{(2)}b_s + \gamma_s b_{s-1}
&=& 0,
\label{sfh.4.8}
\end{eqnarray}
if $n_1+1 \leq s \leq n-|m|-1$ и $n_2+1 \leq s \leq n -|m|-1$.

Now let us consider the case when $s=n_1$ and $s=n_2$. According to
(\ref{sfh.4.1}) and (\ref{sfh.4.2}) as $R\to \infty$ the expansion of the
function $A(R)$ includes only the terms $R^{-k}$ with $k \leq 2$ and,
therefore,
\begin{eqnarray}
A(R)\stackrel{R\to \infty}{\longrightarrow}
A_0 -
\frac{R}{n}\left(n-|m|-2n_2-1\right)+\frac{R^2}{4n^2},
\label{sfh.4.9}
\\ [3mm]
A(R)\stackrel{R\to\infty}{\longrightarrow}
A_0 +
\frac{R}{n}\left(n-|m|-2n_1-1\right)+\frac{R^2}{4n^2},
\label{sfh.4.10}
\end{eqnarray}
(one should obey the condition $n=n_1+n_2+|m|+1$). Substituting (\ref{sfh.4.9})
and (\ref{sfh.4.10}) into (\ref{sfh.3.19}) and (\ref{sfh.3.20}), we have
\begin{eqnarray*}
\beta_{n_2} =  A_0 +
\left(n_2+|m|\right)\left(n_2+|m|+1\right), \qquad
{\tilde\beta}_{n_2} =  A_0 +
\left(n_1+|m|\right)\left(n_1+|m|+1\right).
\label{sfh.4.11}
\end{eqnarray*}
The constant $A_0$ can be determined from the recurrence relation
(\ref{sfh.3.16}) at $s=n_2$ if the following expressions are substituted in it:
\begin{eqnarray*}
a_{n_2-1}=-\frac{1}{R}\frac{\alpha_{n_2-1}}{\beta_{n_2-1}^{(1)}}a_{n_2},
\qquad
a_{n_2+1} = -\frac{\gamma_{n_2+1}}{\beta_{n_2-1}^{(1)}}.
\label{sfh.4.13*}
\end{eqnarray*}
The result is:
\begin{eqnarray}
A_0 = 2n_2^2 -2n_2\left(n-|m|-1\right) -
\left(|m|+1\right)\left(n-1\right).
\label{sfh.4.14}
\end{eqnarray}
It should be noted that the last formula was derived in \cite{COULSON1} from
the analysis of the recurrence relations derived upon representing the
functions $f(\xi; R)$ and  $g(\eta; R)$ as series in Laguerre polynomials of
the variables $\xi$ and $\eta$.

\subsection{Limit $R\to 0$.} This limit can be
investigated in a similar way. The corresponding binomial recurrence relations
have the form
\begin{eqnarray}
\alpha_s a_{s+1} +\overline{\beta}_s a_s
&=& 0,
\label{sfh.4.15}
\\ [3mm]
- \alpha_s b_{s+1} +\overline{\beta}_s b_s
&=& 0,
\label{sfh.4.16}
\end{eqnarray}
if $0\leq s \leq l-|m|-1$ and
\begin{eqnarray}
\overline{\beta}_s a_s + R \gamma_s a_{s-1}
&=& 0,
\label{sfh.4.17}
\\[3mm]
\overline{\beta}_s b_s + R \gamma_s b_{s-1}
&=& 0,
\label{sfh.4.18}
\end{eqnarray}
and if $l-|m|+1 \leq s \leq n-|m|-1$. In these relations
\begin{eqnarray}
\overline{\beta}_s = A(0) + (s+|m|)(s+|m|+1).
\label{sfh.4.19}
\end{eqnarray}
The case $s = l-|m|$ stands out. At small $R$ we have
\begin{eqnarray}
A(R)\stackrel{R\to 0}{\longrightarrow} A(0) +
A'(0)\, R + O\left(R^2\right).
\label{sfh.4.20}
\end{eqnarray}
It follows from the Schr\"{o}dinger equation that $A(0) = -l(l+1)$. Taking this
into account as well as (\ref{sfh.3.19}) and (\ref{sfh.3.20}) we get
\begin{eqnarray}
\beta_{l-|m|} \stackrel{R\to 0}{\longrightarrow}
\varepsilon_{l-|m|}R + O\left(R^2\right),
\label{sfh.4.21}
\\[3mm]
{\tilde \beta}_{l-|m|} \stackrel{R\to 0}{\longrightarrow}
\overline{\varepsilon}_{l-|m|}R+O\left(R^2\right),
\label{sfh.4.22}
\end{eqnarray}
where
\begin{eqnarray*}
\varepsilon_{l-|m|}
= A'(0) + \frac{n+|m|-2l-1}{n},
\qquad
\overline{\varepsilon}_{l-|m|}
= A'(0) - \frac{n+|m|-2l-1}{n}.
\end{eqnarray*}
These formulae are used to establish the form of recurrence relations
(\ref{sfh.3.16}) and (\ref{sfh.3.17}) at $s=l-|m|$
\begin{eqnarray}
\alpha_{l-|m|} a_{l-|m|+1} + R\varepsilon_{l-|m|}a_ {l-|m|}+
R\gamma_{l-|m|} a_{l-|m|-1}&=&0,
\label{sfh.4.25}
\\ [3mm]
-\alpha_{l-|m|} b_{l-|m|+1} + R\overline{\varepsilon}_{l-|m|}b_{l-|m|}
+ R\gamma_{l-|m|} b_{l-|m|-1}&=&0.
\label{sfh.4.26}
\end{eqnarray}
According to (\ref{sfh.4.15}) - (\ref{sfh.4.18}), the quantities $a_{l-|m|+1}$,
$a_{l-|m|-1}$, $b_{l-|m|+1}$ and $b_{l-|m|-1}$ are expressed via $a_{l-|m|}$
and $b_{l-|m|}$, respectively. Therefore, (\ref{sfh.4.25}) and (\ref{sfh.4.26})
should lead to constrains under which the cut-off conditions at $s=-1$ and
$s=n-|m|$ are consistent. Using (\ref{sfh.4.21}) and (\ref{sfh.4.22}) one can
easily show that the condition
\begin{eqnarray}
A'(0) = 0.
\label{ZERO}
\end{eqnarray}
is just such a constraint. In what follows, this condition will be derived on
other grounds and then discussed in a proper way.

\section{Limiting transitions for the wave functions}
\markboth{CHAPTER 2.  HYDROGEN ATOM}{2.5. LIMITING TRANSITIONS FOR THE WAVE FUNCTIONS}
In this section, it will be shown that binomial recurrence relations derived
above generate parabolic and spherical bases of the hydrogen atom.

\subsection{Limit $R\to \infty$.} According to
(\ref{sfh.3.8}), (\ref{sfh.3.14}), and (\ref{sfh.3.15}), the functions
$f\left(\xi; R\right)$ and $g\left(\eta; R\right)$ behave as follows:
\begin{eqnarray}
f_{nqm} \left(\xi; R \right)
&\stackrel{R\to \infty}{\longrightarrow}&
\sum_{s=0}^{n_1+n_2}\,\frac{a_s(R)}{R^s}\,\mu^s,
\label{sfh.4.027}
\\[3mm]
\qquad
g_{nqm}
\left(\eta; R\right)
&\stackrel{R\to \infty}{\longrightarrow}&
\sum_{s=0}^{n_1+n_2}\,\frac{b_s(R)}{R^s}\,\nu^s.
\label{sfh.4.27}
\end{eqnarray}
Let us clarify the behavior of the coefficients $a_s(R)$ and $b_s(R)$ in the
limit $R\to\infty$. It follows from binomial recurrence relations
(\ref{sfh.4.5}) - (\ref{sfh.4.8}) that
\begin{eqnarray}
a_s(R) &\stackrel{R\to \infty}{\longrightarrow}& (-1)^s
\frac{\beta_0^{(1)}\beta_1^{(1)}\dots \beta_{s-1}^{(1)}}
{\alpha_0\alpha_1\dots \alpha_{s-1}}R^s,
\,\,\,\quad\qquad 1\leq s \leq n_2,
\label{sfh.4.28}
\\[3mm]
b_s(R) &\stackrel{R\to \infty}{\longrightarrow}&
\frac{\beta_0^{(2)}\beta_1^{(2)}\dots \beta_{s-1}^{(2)}}
{\alpha_0\alpha_1\dots \alpha_{s-1}}R^s,
\qquad\qquad\qquad 1\leq s \leq n_1,
\label{sfh.4.29}
\\[3mm]
a_s(R) &\stackrel{R\to \infty}{\longrightarrow}& (-1)^{s-n_2}
\frac{\gamma_{n_2+1}\gamma_{n_2+2}\dots \gamma_s}
{\beta_{n_2+1}^{(1)}\beta_{n_2+2}^{(1)}\dots \beta_{s}^{(1)}}a_{n_2},
\qquad n_2+1\leq s \leq n_1+n_2,
\label{sfh.4.30}
\\[3mm]
b_s(R) &\stackrel{R\to \infty}{\longrightarrow}& (-1)^{s-n_1}
\frac{\gamma_{n_1+1}\gamma_{n_1+2}\dots \gamma_s}
{\beta_{n_1+1}^{(2)}\beta_{n_1+2}^{(2)}\dots \beta_{s}^{(2)}}b_{n_1},
\qquad n_1+1\leq s \leq n_1+n_2,
\label{sfh.4.31}
\end{eqnarray}
and, therefore, the functions $f_{nqm}\left(\xi; R\right)$ and
$g_{nqm}\left(\eta; R\right)$ turn into polynomials of powers $n_2$ and $n_1$,
respectively. Substituting in (\ref{sfh.4.027}) and (\ref{sfh.4.28}) the values
of $\alpha_s$, $\beta_s^{(1)}$ and $\beta_s^{(2)}$ from (\ref{sfh.3.18}) and
(\ref{sfh.4.4}) one can easily get the formulae
\begin{eqnarray}
a_s(R) &\stackrel{R\to \infty}{\longrightarrow}&
\frac{(-1)^s}{n^s}\frac{n_2! \,|m|!}{(n_2-s)!\, (s+|m|)!}
\frac{R^s}{s!},
\qquad 1\leq s \leq n_2,
\label{sfh.4.32}
\\[3mm]
b_s(R) &\stackrel{R\to \infty}{\longrightarrow}&
\frac{(-1)^s}{n^s}\frac{n_1! \,|m|!}{(n_1-s)!\, (s+|m|)!}
\frac{R^s}{s!},
\qquad 1\leq s \leq n_1,
\label{sfh.4.33}
\end{eqnarray}
proving the sought behavior of the polynomials $f_{nqm}\left(\xi; R\right)$ and
$g_{nqm}\left(\eta; R\right)$ in the limit $R\to\infty$
\begin{eqnarray}
\lim_{R\to \infty} f_{nqm}\left(\xi; R\right)
&=&
F\left(-n_2; |m|+1; \frac{\mu}{n}\right),
\label{sfh.4.34}
\\[3mm]
\lim_{R\to \infty} g_{nqm}\left(\eta; R\right)
&=&
F\left(-n_1; |m|+1; \frac{\nu}{n}\right).
\label{sfh.4.35}
\end{eqnarray}

\subsection{Limit $R\to 0$.} In the limit $R \to 0$,
according to (\ref{sfh.3.7}), we have
\begin{eqnarray}
\Pi_{nkm}\left(\xi; R\right) &\stackrel{R\to 0}{\longrightarrow}&
e^{-\frac{r}{n}}\left(\frac{2r}{R}\right)^{|m|}\,\sum_{s=0}^{n-|m|-1}\,
\frac{a_s(R)}{R^s} (2r)^{s},
\label{sfh.4.38}
\\[3mm]
\Xi_{nqm}\left(\eta; R\right) &\stackrel{R\to 0}{\longrightarrow}&
\left(\sin\theta\right)^{|m|}\,\sum_{s=0}^{n-|m|-1}\,
b_s (R) \left(1+\cos\theta\right)^s.
\label{sfh.4.39}
\end{eqnarray}
Binomial recurrence relations (\ref{sfh.4.15}) - (\ref{sfh.4.18}) result in
\begin{eqnarray}
a_s(R) &\stackrel{R\to 0}{\longrightarrow}&
(-1)^s\frac{\overline{\beta}_0\overline{\beta}_1
\dots\overline{\beta}_{s-1}}
{\alpha_0\alpha_1\dots \alpha_{s-1}},
\label{sfh.4.40}
\\[3mm]
b_s(R) &\stackrel{R\to 0}{\longrightarrow}&
\frac{\overline{\beta}_0\overline{\beta}_1\dots\overline{\beta}_{s-1}}
{\alpha_0\alpha_1\dots \alpha_{s-1}},
\label{sfh.4.41}
\end{eqnarray}
if $1 \leq s \leq l-|m|$, and
\begin{eqnarray}
a_s(R) &\stackrel{R\to 0}{\longrightarrow}&
(-R)^{s-l+|m|}\frac{\gamma_{l-|m|+1}\gamma_{l-|m|+2}\dots \gamma_s}
{\overline{\beta}_{l-|m|+1}\overline{\beta}_{l-|m|+2}
\dots \overline{\beta}_{s}}a_{l-|m|},
\label{sfh.4.42}
\\[3mm]
b_s(R) &\stackrel{R\to 0}{\longrightarrow}& (-R)^{s-l+|m|}
\frac{\gamma_{l-|m|+1}\gamma_{l-|m|+2}\dots \gamma_s}
{\overline{\beta}_{l-|m|+1}\overline{\beta}_{l-|m|+2}\dots
\overline{\beta}_{s}}b_{l-|m|},
\label{sfh.4.43}
\end{eqnarray}
if $l-|m|+1 \leq s \leq n-|m|-1$. Substituting into (\ref{sfh.4.41}) and
(\ref{sfh.4.42}) the values of $\alpha_s$, $\gamma_s$ and
$\overline{\beta}_{s}$ from (\ref{sfh.3.18}), (\ref{sfh.3.21}), and
(\ref{sfh.4.19}), we get
\begin{eqnarray}
a_{s+l-|m|}(R) &\stackrel{R\to 0}{\longrightarrow}&
\left(\frac{R}{n}\right)^{s}\frac{(-n+l+1)_s}
{s!(2l+2)_s}a_{l-|m|},
\label{sfh.4.44} \\ [3mm]
b_s(R) &\stackrel{R\to 0}{\longrightarrow}&
\frac{(-l+|m|)_s(l+|m|+1)_s}
{2^ss!(|m|+1)_s},
\label{sfh.4.45}
\end{eqnarray}
with $1 \leq s \leq n-l-1$ in the first case and $1 \leq s \leq l-|m|$ in the
second case. These formulae allow one to rewrite the limiting equalities
(\ref{sfh.4.38}) and (\ref{sfh.4.39}) as
\begin{eqnarray}
\Pi_{nkm}\left(\xi; R\right) &\stackrel{R\to 0}{\longrightarrow}&
\frac{|m|!(2l)!}{2^{l-m} l!(l-|m|)!}\,
e^{-\frac{r}{n}}\, \left(\frac{2r}{n}\right)^{l}\,
F\left(-n+l+1; 2l+2; \frac{2r}{n}\right),
\label{sfh.4.46}
\\[3mm]
\Xi_{nqm}\left(\eta; R\right) &\stackrel{R\to 0}{\longrightarrow}&
\left(\sin\theta\right)^{|m|}
\,F\left(-l+|m|, l+|m|+1; |m|+1; \frac{1+\cos\theta}{2}\right).
\label{sfh.4.47}
\end{eqnarray}

\subsection{Limits $R\to 0$ and $R\to\infty$ in the
normalization constant.} Using formulae(\ref{sfh.3.23}) - (\ref{sfh.3.27}) and
the asymptotic relations of the hypergeometric function as $R\to 0$ and
$R\to\infty$ one can show that
\begin{eqnarray}
C_{nqm}(R)
\stackrel{R\to 0}{\longrightarrow}
\frac{e^{i\Phi(0)}}{n^2}\left(\frac{2R}{n}\right)^l
\frac{l!(l+|m|)!}{\left[2^{|m|}|m|!(2l)!\right]^2}
\left[\frac{2(n+l)!(l+|m|)!}{(2l+1)(l-|m|)!(n-l-1)!}\right]^{\frac{1}{2}},
\label{sfh.4.49}
\end{eqnarray}
\begin{eqnarray}
C_{nqm}(R)
\stackrel{R \to \infty}{\longrightarrow}
\frac{e^{i\Phi(\infty)}}{\left(|m|!\right)^2}\,\frac{\sqrt{2}}{n^2}
\left(\frac{R}{2n}\right)^{|m|}\,
\sqrt{\frac{\left(n_1+|m|\right)!\left(n_2+|m|\right)!}
{\left(n_1\right)!\left(n_2\right)!}},
\label{sfh.4.37}
\end{eqnarray}
where $\Phi(R)$ is the phase of the normalization constant $C_{nqm}(R)$ which
can be chosen arbitrary. Under this arbitrariness we put
$$
\Phi (R) = \frac{\pi}{1+R} \, \left(l + \frac{m-|m|}{2}\right)
$$
so that
$$
\Phi (\infty) = 0,
\qquad\qquad
\Phi (0) = \pi \, \left(l + \frac{m-|m|}{2}\right).
$$
Under this choice the prolate spheroidal basis of the hydrogen atom exactly
turns into the spherical and parabolic bases in the limits $R\to 0$ and
$R\to\infty$, respectively.

\section{Expansion over the spherical and parabolic bases of the hydrogen atom}
\markboth{CHAPTER 2.  HYDROGEN ATOM}{2.6. EXPANSION OVER THE SPHERICAL AND PARABOLIC BASES}

In this section, expansions of the spheroidal basis (\ref{sfh.3.9}) over the
other two bases of the hydrogen atom, spherical and parabolic, are obtained.

Let us write down the expansion we are interested in as follows:
\begin{eqnarray}
\Psi_{nqm}\left(\xi, \eta, \varphi; R\right)
&=&
\,\,\,\,\,\,\,\,
\sum_{l=|m|}^{n-1}\, W_{nq}^{lm}(R)\,
\Psi_{nlm}(r, \theta, \varphi),
\label{sfh.5.1}
\\[3mm]
\Psi_{nqm}\left(\xi, \eta, \varphi; R\right)
&=&
\sum_{n_1+n_2=n-|m|+1}\, U_{qm}^{n_1 n_2}(R)\,
\Psi_{n_1n_2m}\left(\mu, \nu, \varphi\right).
\label{sfh.6.1}
\end{eqnarray}
In calculating the coefficients determining interbasis expansions it is
necessary to use at some stage the orthogonality condition. In the previous
section that was the orthogonality of the associated Legendre polynomials. In
what follows, in calculating the coefficients $W_{nq}^{lm}(R)$ and $U_{qm}^{n_1
n_2}(R)$ we will use the orthogonality of the radial part of the hydrogen atom
spherical basis in the quantum number $l$, proved in Appendix A, and the
orthogonality of the function (\ref{ph.1.2}) in the quantum number $n_2$,
respectively.

\subsection{Expansion over the spherical basis.}

In this expansion we perform the following operations: in the wave function
$\Psi_{nlm}(r, \theta, \varphi)$ we pass from the spherical coordinates to the
spheroidal ones
\begin{eqnarray}
r =\frac{R}{2}(\xi + \eta),
\qquad
\cos\theta = \frac{1+\xi\eta}{\xi+\eta},
\label{sfh.5.2}
\end{eqnarray}
let $\eta \to -1$ on both sides of (\ref{sfh.5.1})
\begin{eqnarray}
r \to \frac{R}{2}\left(\xi - 1\right),
\qquad
\cos\theta \to -1,
\label{sfh.5.3}
\end{eqnarray}
and use the orthogonality condition for the radial function (\ref{ort.1.2}).
These operations lead to the formula
\begin{eqnarray}
W_{nq}^{lm}(R)
&=&
(-1)^{l+\frac{m-|m|}{2}}\,
\sqrt{\frac{2l+1}{2}\frac{(l-|m|)!}{(l+|m|)!}} \,
2^{|m|}\,|m|!\, \frac{n^2 R}{2} \,
C_{nqm}(R)\times
\nonumber
\\[3mm]
\label{sfh.5.4}
&\times&
e^{\frac{R}{2n}} \,
\int\limits_{1}^{\infty}\,
\left(\frac{\xi-1}{\xi+1}\right)^{\frac{|m|}{2}}
\, R_{nl}\left[\frac{R}{2}(\xi-1)\right]\,\Pi(\xi; R)\,d\xi.
\end{eqnarray}
Integral in (\ref{sfh.5.4}) can easily be calculated if one passes to the new
variable $t=\xi-1$ and uses formulae (\ref{ort.1.4}), (\ref{ort.1.5}).
Collecting all together we get
\begin{eqnarray}
W_{nq}^{lm}(R)
&=&
(-1)^{l+\frac{m-|m|}{2}}\,
\frac{(2n)^{|m|}\, n^2\, |m|!}{\sqrt{(n+l)!(n-l-1)!}}
\sqrt{\frac{2l+1}{2}\frac{(l-|m|)!}{(l+|m|)!}}\times
\nonumber
\\[3mm]
\label{sfh.5.5}
&\times&
\frac{C_{nqm}(R)}{R^{|m|}} \,
\sum_{s=0}^{l-|m|}\, \frac{a_s(R)}{R^s}\,
\frac{(l+|m|+s)! \, (n-|m|-1-s)! n^s}{(l-|m|-s)!}.
\end{eqnarray}
At $l=|m|$ this formula is considerably simplified:
\begin{eqnarray}
W_{n,q}^{|m|,m}(R)
&=&
(-1)^{\frac{m+|m|}{2}}\,
(2n)^{|m|} \, n^2\, |m|!
\,
\sqrt{\frac{2|m|+1}{2} \frac{(n-|m|-1)!}{(n+|m|)!(2|m|)!}}
\,
\frac{C_{nqm}(R)}{R^{|m|}}.
\nonumber
\end{eqnarray}
In the last expression the dependence of $W_{n,q}^{|m|,m}(R)$ on the
coefficients $a_s(R)$ and $b_s(R)$ is concentrated in the normalization factor
$C_{nqm}(R)$.

\subsection{Parabolic basis expansion.}

Let us substitute into the right-hand side of (\ref{sfh.6.1})
\begin{eqnarray}
\mu = \frac{R}{2}\left(\xi+1\right)\left(1+\eta\right),
\qquad
\nu = \frac{R}{2}\left(\xi-1\right)\left(1-\eta\right),
\label{sfh.6.2}
\end{eqnarray}
assume that $\eta = -1$ in both the parts (\ref{sfh.6.1}), pass to the new
variable $t=R(\xi-1)/n$, and use the orthogonality condition
\begin{eqnarray*}
\int\limits_{0}^{\infty}\,
e^{-t}\,t^{|m|}
F\left(-n_2; |m|+1; t\right)
F\left(-{n'}_2; |m|+1; t\right)dt =
\frac{\left(|m|!\right)^2\,\left(n_2\right)!}{\left(n_2+|m|\right)!}\,
\delta_{n_2{n'}_2}.
\label{sfh.6.3}
\end{eqnarray*}
A chain of such operations results in
\begin{eqnarray}
U_{qm}^{n_1 n_2}(R)
&=&
C_{nqm}(R)\frac{n^2}{\sqrt{2}}
\left(\frac{2n}{R}\right)^{|m|}
\sqrt{\frac{\left(n_1\right)!\left(n_2+|m|\right)!}
{\left(n_2\right)!\left(n_1+|m|\right)!}}\times
\nonumber
\\[3mm]
\label{sfh.6.4}
&\times& \sum_{s=0}^{n-|m|-1}\,\frac{a_s(R)}{R^s}\,
{n}^{s} \,
\int\limits_{0}^{\infty}\, e^{-t}\,t^{|m|+s}
F\left(-n_2; |m|+1; t\right)dt,
\end{eqnarray}
and upon calculating the integral by formulae (\ref{ort.1.4}) and
(\ref{ort.1.5}) we get
\begin{eqnarray}
U_{qm}^{n_1 n_2}(R)
&=& (-1)^{n_2}C_{nqm}(R)
\sqrt{\frac{\left(n_1\right)!}{\left(n_2\right)!}}
\frac{n^2}{\sqrt{2}}\,
\frac{|m|!}{\sqrt{2\left(n_1+|m|\right)!\left(n_2+|m|\right)!}}\times
\nonumber
\\[3mm]
\label{sfh.6.5}
&\times& \left(\frac{2n}{R}\right)^{|m|} \,
\sum_{s=0}^{n_1+n_2}\, \frac{a_s(R)}{R^s}
\,
\frac{s!\left(|m|+s\right)!n^s}{\Gamma\left(s-n_2+1\right)}.
\end{eqnarray}
Formulae (\ref{sfh.5.5}) and (\ref{sfh.6.5}) are rather complicated. The
explicit form of the coefficients $W_{nq}^{lm}(R)$ and $U_{qm}^{n_1 n_2}(R)$
for the lowest quantum numbers will be given below.

\subsection{Limits $R \to 0$ and $R \to \infty$ in the
coefficients $W_{nq}^{lm}(R)$.} With the above complexity of the coefficients
(\ref{sfh.5.5}) and (\ref{sfh.6.5}) taken into account, it is probably
worthwhile to make the limiting transitions $R \to 0$ and $R \to \infty$ in
them. First, this will provide us with a reliable test and, second, will make
it possible to elucidate the transition mechanism of the coefficients
$W_{nq}^{lm}(R)$ and $U_{qm}^{n_1 n_2}(R)$ in the Kronecker symbol and
Clebsch-Gordan coefficients.

As for $R \to 0$ $A_q(R) \to - l'(l'+1)$, then according to (\ref{sfh.4.19}),
${\overline{\beta}}_s = (s+|m|-l')(s+|m|+l'+1)$ and, therefore, for $1\leq s
\leq l'-|m|$
\begin{eqnarray}
a_s (R)
\stackrel{R\to 0}{\longrightarrow}
(-1)^s \, \frac{(-l'+|m|)_s (l'+|m|+1)_s}
{2^s s! (|m|)_s}.
\end{eqnarray}
Suppose that $l'> l$. As for $R \to 0$ the normalization constant $C_{nqm}(R)$
is proportional to $R^{l'}$ and $a_s(R)$ is independent of $R$, the expansion
coefficient $W_{nq}^{lm}(R)$ behaves as $R^{l'-l}$ and tends to zero for all
$l'
> l$. Now let $l' \leq l$. As $R \to 0$, in this case, in the sum
(\ref{sfh.5.5}) only terms with $s> l'-|m|$ may differ from zero and
,consequently,
\begin{eqnarray}
\label{sfh.5.6}
W_{nqm}^l(R)
&\stackrel{R\to 0}{\longrightarrow}&
A_{l'l}^{nm}\,
\sum_{s=0}^{l-l'}\,
\frac{a_{s+l'-|m|}(R)}{R^s}\,
\frac{n^s (l'+l+s)! (n-l'-s-1)!}
{(l-l'+s)!},
\end{eqnarray}
where
\begin{eqnarray}
A_{l'l}^{nm} &=&
\frac{(-1)^{l+l'}}{|m|!\,[(2l')!]^2} \times \nonumber \\
\label{sfh.5.7} \\
&\times& \sqrt{\frac{(2l+1)(l-|m|)! (l'!)^2 [(l'+|m|)!]^3\, 4^{l'-|m|}
(n+l')!}
{(2q+1)(l+|m|)!(l'-|m|)!(n+l)!(n-l-1)!(n-l'-1)!}}.
\nonumber
\end{eqnarray}
Taking into account that, according to (\ref{sfh.4.40}) and (\ref{sfh.4.44}),
\begin{eqnarray}
a_{s+l'-|m|}(R)
&\stackrel{R\to 0}{\longrightarrow}&
\frac{|m|!(2l')!}{2^{l'-|m|}\, l'!(l'+|m|)!}\,
\frac{(-n+l'+1)_s}{s! (2l'+2)_s} \,
\frac{R^s}{n^s},
\end{eqnarray}
let us write formula (\ref{sfh.5.6}) in the form
\begin{eqnarray}
W_{nqm}^l(R) &\stackrel{R\to 0}{\longrightarrow}&
\frac{(l+l')!}{(2l')!} \,
\sqrt{\frac{(2l+1)(n+l')!(l'+|m|)!(l-|m|)!(n-l'-1)!}
{(2l'+1)(n+l)!(l+|m|)!(l'-|m|)!(n-l-1)!}}\times
\nonumber
\\[3mm]
\label{sfh.5.9}
&\times&
\frac{1}{\Gamma(l-l'+1)} \,
{_2F_1}\left(-l+l',\, l+l'+1; \, 2l'+2; \, 1\right).
\end{eqnarray}
The summation of the hypergeometric series by formula (\ref{ort.1.5})
\begin{eqnarray}
{_2F_1}\left(-l+l',\, l+l'+1; \, 2l'+2; \, 1\right)
=
\frac{1}{\Gamma(l'-l+1)}\,
\frac{\Gamma(2l'+2)}{\Gamma(l+l'+2)}
\label{sfh.5.10}
\end{eqnarray}
shows that the coefficients $W_{nq}^{lm}(R)$ as $R\to 0$ equal zero for all
$l'< l$. From the afore-said and formulae (\ref{sfh.5.9}) and (\ref{sfh.5.10})
it immediately follows that
\begin{eqnarray}
W_{nqm}^l(R) \stackrel{R\to 0}{\longrightarrow} \delta_{ql}.
\end{eqnarray}

Let us now consider the limit $R \to \infty$. By virtue of relations
(\ref{sfh.4.28}) and (\ref{sfh.4.28})the behavior of the coefficients $a_s(R)$
at large $R$ is determined as $a_s(R)\sim R^s$ at $1\leq s \leq n_2$ and
$a_s(R)\sim R^{n_2}$ at $n_2+1\leq s \leq n-|m|-1$. First, assume that $n_2
\geq l-|m|$. Then, according to relations (\ref{sfh.4.32}) and
(\ref{sfh.4.37}), we get
\begin{eqnarray}
W_{nqm}^l(R)
&\stackrel{R\to \infty}{\longrightarrow}&
(-1)^{l+\frac{m-|m|}{2}}\frac{(n-|m|-1)!}{|m|!}
\sqrt{\frac{(2l+1)(l+|m|)!(n_1+|m|)!(n_2+|m|)!}{(n_1)!(n_2)!(l-|m|)!
(n+l)!(n-l-1)!}}\times
\nonumber
\\[3mm]
\label{sfh.5.12}
&\times&
{_3F}_2 \left\{
\begin{array}{l}
-n_2, -l+|m|, l+|m|+1 \\
[3mm]
|m|+1, -n+|m|+1\\
\end{array}
\biggr| 1 \right\},
\end{eqnarray}
which completely coincides with the Tarter matrix (\ref{h.1.3}).

At $n_2 < l-|m|$ the upper limit of the summation in \ref{sfh.5.5}) $l-|m|$ can
be replaced by $n_2$. Therefore, using again relations (\ref{sfh.4.32}) and
(\ref{sfh.4.37}) we arrive at the same formula (\ref{sfh.5.5}).

\subsection{Limits $R \to 0$ and $R \to \infty$ in the
coefficients $U_{qm}^{n_1 n_2}(R)$.} As $R \to \infty$, as has already been
mentioned above, there exists such a value of $s$ equal to $n'_2$ for which
$\beta_{n'_2} =0$. According to (\ref{sfh.4.28}) and (\ref{sfh.4.30}), the
ratio $a_s/R^s$  is nonzero in this limit if $0\leq s \leq n'_2$, so that in
the sum (\ref{sfh.6.5}) the maximum value of $s$ can be changed by $n'_2$.
Therefore, with the use of (\ref{sfh.4.28}) and (\ref{sfh.4.37}) we can see
that at $n'_2 < n_2$ the coefficient $U_{qm}^{n_1 n_2}(R)$ vanishes due to the
$\Gamma\left(s-n_2+1\right)$ function in the sum (\ref{sfh.6.5}), and at $n'_2
\geq n_2$ the following limiting equality holds:
\begin{eqnarray*}
U_{qm}^{n_1 n_2}(R) &\rightarrow&
\sqrt{\frac{(n_2)!(n_1)!(n'_2+|m|)!(n'_1+|m|)!}
{(n'_2)!(n'_1)!(n_2+|m|)!(n_1+|m|)!}}
\, \,
\sum_{s=0}^{q-n_2}\,
\frac{(-1)^s}{s! \Gamma(n_2-n'_2-s+1)}.
\end{eqnarray*}
One can easily show that the last expression can take only two values - zero
and unity, depending on whether $n'_2 > n_2$ or $n'_2 = n_2$. Therefore,
\begin{eqnarray*}
U_{qm}^{n_1 n_2}(R) \rightarrow \delta{n'_2 n_2}.
\end{eqnarray*}

Let us pass to the limit $R \to 0$. It follows from (\ref{sfh.4.40}),
(\ref{sfh.4.44}) and (\ref{sfh.4.49}) that in this limit
\begin{eqnarray}
U_{qm}^{n_1 n_2}(R)
&\rightarrow&
(-1)^{n_2+l+\frac{m-|m|}{2}}
\sqrt{\frac{(2l+1)(n_1)!(l+|m|)!(l-|m|)!(n+l)!}{(n_2)!(n_1+|m|)!
(n_2+|m|)!(n-l-1)!}}\times
\nonumber
\\[3mm]
&\times& \frac{l!}{(2l+1)!(l-|m|-n_2)!}
{_3F}_2 \left\{
\begin{array}{l}
-n+l+1, l+1, l-|m|+1 \\
[3mm]
l-|m|-n_2+1, 2l+2\\
\end{array}
\biggr| 1 \right\}.
\nonumber
\end{eqnarray}
Further transformation by formulae (\ref{d8}) and
\begin{eqnarray}
{_3F}_2 \left\{
\begin{array}{l}
c, b, e-a \\
[3mm]
e, 1+b+c-f\\
\end{array}
\biggr| 1 \right\}
&=& \frac{\Gamma(1-f+a)\Gamma(1+b+c-f)}
{\Gamma(1+e-f)\Gamma(1-e-f+a+b+c)}\times
\nonumber
\\[3mm]
\label{3F2.1.2}
&\times&
{_3F}_2 \left\{
\begin{array}{l}
e-a, e-b, e-c \\
[3mm]
e, 1+e-f\\
\end{array}
\biggr| 1 \right\}
\end{eqnarray}
(the latter is taken from \cite{SMOR-SHEP}) leads to the Tarter's result
(\ref{h.1.3}).

\section{Spheroidal corrections to the spherical and parabolic bases}
\markboth{CHAPTER 2.  HYDROGEN ATOM}{2.7. SPHEROIDAL CORRECTIONS}
In this section, we derive trinomial recurrence relations governing the
coefficients $U_{qm}^{n_1n_2}(R)$ and $W_{nq}^{lm}(R)$ and on their basis we
calculate corrections to the spherical and parabolic bases. The initial
formulae for these calculations are the equations governing in the discrete
spectrum the spherical $\Psi_{nlm}(r, \theta, \varphi)$, parabolic
$\Psi_{n_1n_2m}(\mu, \nu, \varphi)$, and spheroidal $\Psi_{nqm}(\xi, \eta,
\varphi)$ bases of the hydrogen atom
\begin{eqnarray}
{\hat H}\Psi_{nlm} = E_n\Psi_{nlm},
\qquad {\hat l}^2\Psi_{nlm} = l(l+1)\Psi_{nlm},
\qquad {\hat l}_z\Psi_{nlm} = m\Psi_{nlm},
\label{sfh.7.1}
\end{eqnarray}
\begin{eqnarray}
{\hat H}\Psi_{n_1n_2m} = E_n \Psi_{n_1n_2m},
\quad
{\hat A}_z \Psi_{n_1n_2m} = \frac{(n_1-n_2)}{n}\Psi_{n_1n_2m},
\quad
{\hat l}_z \Psi_{n_1n_2m} = m \Psi_{n_1n_2m},
\label{sfh.7.2}
\end{eqnarray}
\begin{eqnarray}
{\hat H}\Psi_{nqm} = E_n \Psi_{nqm},
\qquad
{\hat \Lambda}\Psi_{nqm} = {\lambda}_q \Psi_{nqm},
\qquad
{\hat l}_z \Psi_{nqm} = m \Psi_{nqm}.
\label{sfh.7.3}
\end{eqnarray}
Here the operator
\begin{eqnarray}
{\hat \Lambda} = - {\hat l}^2 -  n\, {\hat A}_z,
\label{sfh.7.4}
\end{eqnarray}
where ${\hat l}^2$ is the square of the orbital moment operator and ${\hat
A}_z$ is the $z$ component of the Laplace-Runge-Lenz vector (\ref{symh.1.3}).

\subsection{Recurrence relations for the coefficients
$W_{nq}^{lm}(R)$.} From the definition of the operator ${\hat \Lambda}$ and
prolate spheroidal basis of the hydrogen atom, and the expansion
(\ref{sfh.5.1}) we have the system of equations
\begin{eqnarray}
-\frac{1}{R}\left[{\lambda}_q(R)+l(l+1)\right]\,W_{nq}^{lm}(R) =
\sum_{l'=|m|}^{n-1}W_{nq}^{l'm}(R)\,\left({\hat A}_z\right)_{ll'},
\label{sfh.7.5}
\end{eqnarray}
where
\begin{eqnarray}
\left({\hat A}_z\right)_{ll'} =
\int\,\psi_{nlm}^{*}(r, \theta, \varphi)\,
{\hat A}_z\,\psi_{nl'm}(r, \theta, \varphi)\,dV.
\label{sfh.7.6}
\end{eqnarray}
It is convenient to calculate the matrix elements $\left({\hat
A}_z\right)_{ll'}$ using expansion of the spherical basis of the hydrogen atom
over the parabolic basis (\ref{PT.1.3}). Taking into account the eigenvalue
equations for the operator ${\hat A}_z$ (\ref{sfh.7.2}) and the
orthonormalization of the parabolic basis, instead of relation (\ref{sfh.7.6})
we have
\begin{eqnarray}
\left({\hat A}_z\right)_{ll'}
&=& \frac{(-1)^{l+l'}}{n}\,\sum_{n_2=0}^{n-|m|-1}\, (n-|m|-2n_2-1) \,\,
C_{\frac{n-1}{2},\,\frac{n-1}{2}-n_2;\,\,\,
\frac{n-1}{2},\,n_2+|m|-\frac{n-1}{2}}^{l,\,|m|}\,\times
\nonumber
\\[3mm]
\label{sfh.7.7}
&\times&
C_{\frac{n-1}{2},\,\frac{n-1}{2}-n_2;\,\,\,
\frac{n-1}{2},\,n_2+|m|-\frac{n-1}{2}}^{l',\,|m|}\,.
\nonumber
\end{eqnarray}
Then using the trinomial recurrence relation for the Clebsch-Gordan
coefficients
 $C_{a\alpha;a\beta}^{c,\gamma}$ \cite{VAR}
\begin{eqnarray}
(\beta-\alpha)\,\, C_{a\alpha;a\beta}^{c-1,\gamma}
&=&
\sqrt{\frac{(c-\gamma)(c+\gamma)(2a-c+1)
(2a+c+1)}{(2c-1)(2c+1)}}\,\,
C_{a\alpha;a\beta}^{c\gamma}+
\nonumber
\\ [2mm]
&+&
\sqrt{\frac{(c-\gamma-1)(c+\gamma-1)(2a-c+2)
(2a+c)}{(2c-3)(2c-1)}}\,\,
C_{a\alpha;a\beta}^{c-2,\gamma}
\label{sfh.7.8}
\nonumber
\end{eqnarray}
and the orthogonality relation
\begin{eqnarray}
\sum_{\alpha+\beta=\gamma} C_{a\alpha;b\beta}^{c \gamma}
C_{a\alpha;b\beta}^{c'\gamma'} = \delta_{c'c} \delta_{\gamma'
\gamma},
\label{sfh.7.9}
\nonumber
\end{eqnarray}
for the matrix elements $\left({\hat A}_z\right)_{ll'}$ in the spherical basis
we get the following expression \cite{ENGLEF}:
\begin{eqnarray}
\left({\hat A}_z\right)_{ll'}
&=& -\sqrt{\frac{(l+|m|+1)(l-|m|+1)(n+l+1)(n-l+1)}{n^2(2l+1)(2l+3)}}\,\,
\delta_{l',l+1}-
\nonumber
\\[2mm]
&-&
\sqrt{\frac{(l+|m|)(l-|m|)(n+l)(n-l)}{n^2(2l-1)(2l+1)}}\,\,
\delta_{l',l-1}\,.
\label{sfh.7.10}
\end{eqnarray}
With the last result and equation (\ref{sfh.7.5}) taken into account we arrive
at the sought trinomial recurrence relations
\begin{eqnarray*}
\frac{n}{R}\left[{\lambda}_q(R)+l(l+1)\right]
W_{nqm}^l(R)
=
\sqrt{\frac{(l+|m|)(l-|m|)(n+l)(n-l)}{(2l-1)(2l+1)}}\,\,
W_{nqm}^{l-1}(R)+
\\[2mm]
+
\sqrt{\frac{(l+|m|+1)(l-|m|+1)(n+l+1)(n-l+1)}{(2l+1)(2l+3)}}\,\,
W_{nqm}^{l+1}(R).
\end{eqnarray*}
The trinomial recurrence relations derived allow one to calculate the
eigenvalues $\lambda_q$ and the coefficients $W_{nq}^{lm}(R)$ analytically at
small $n$. For an unambiguous choice of coefficients one should take into
account the normalization condition
\begin{eqnarray}
\sum_{l=|m|}^{n-1}\,\left|W_{nqm}^{l}(R)\right|^2 =1.
\label{sfh.7.13}
\end{eqnarray}

\subsection{Recurrence relations for the coefficients
$U_{qm}^{n_1 n_2}(R)$.} Applying the above-described technique to expansion
(\ref{sfh.6.1}) we arrive at the equations
\begin{eqnarray}
\left[\frac{R}{n}
\left(n-|m|-2n_2-1\right)+\lambda_q\right]\,U_{nqm}^{n_2}(R) =
\sum_{n^{'}_2=0}^{n-|m|-1}\,U_{nqm}^{n^{'}_2}(R)\,
\left({\hat l}^2\right)_{n_2{n'}_2},
\label{sfh.7.14}
\end{eqnarray}
where
\begin{eqnarray}
\left({\hat l}^2\right)_{n_2n_2^{'}} =
\int\,\psi_{n_1n_2m}^{*}(\mu, \nu, \varphi)\,
{\hat l}^2\,\psi_{n_1^{'}n_2^{'}m}(\mu, \nu, \varphi)\,dV.
\label{sfh.7.15}
\end{eqnarray}
If, according to (\ref{PT.1.2}), one expands the parabolic basis of the
hydrogen atom over the spherical one, takes into account the following
trinomial recurrence relation for the Clebsch-Gordan coefficients \cite{KIBLER}
\begin{eqnarray}
[c(c+1) - 2\alpha\beta]\,\,
C^{c,\gamma}_{a \alpha; a \beta}
&=& \sqrt{(a+\alpha)(a-\alpha+1)(a-\beta)(a+\beta+1)}\,\,
C^{c,\gamma}_{a \alpha - 1; a \beta + 1}+
\nonumber
\label{KG.KIB.GR}
\\[2mm]
&+& \sqrt{(a-\alpha)(a+\alpha+1)(a+\beta)(a-\beta+1)}\,\,
C^{c,\gamma}_{a, \alpha + 1; a, \beta - 1},
\nonumber
\end{eqnarray}
and the orthogonality condition \cite{VAR}
\begin{eqnarray}
\sum_{c=|\gamma|}^{a+b} C_{a\alpha ;b\beta }^{c\gamma}
C_{a\alpha';b\beta'}^{c\gamma} = \delta_{\alpha\alpha'}
\delta_{\beta\beta'}, \label{KG.ort.1}
\nonumber
\end{eqnarray}
then for the matrix elements of the operator $\left({\hat
l}^2\right)_{n_2n_2^{'}}$ in the parabolic basis one can prove the formula
\begin{eqnarray}
\left({\hat l}^2\right)_{n_2n_2^{'}}
&=&
\left(n_2+1\right)\left(n-|m|-n_2\right) +
\left(n-n_2\right)\left(n_2+|m|\right)
\delta_{n_2^{'},n_2}-
\nonumber
\\[2mm]
&-& \sqrt{n_2(n-n_2)(n-|m|-n_2)(n_2+|m|)}\,\,
\delta_{n_2^{'},n_2-1}-
\\[2mm]
&-& \sqrt{(n_2+1)(n-n_2-1)(n-|m|-n_2-1)(n_2+|m|+1)}\,\,
\delta_{n_2^{'},n_2+1}.
\nonumber
\label{sfh.7.16}
\end{eqnarray}
After substitution of (\ref{sfh.7.16}) into (\ref{sfh.7.14}) we are led to the
trinomial recurrence relations for $U_{qm}^{n_2}(R) \equiv U_{qm}^{n_1 n_2}(R)$
\begin{eqnarray}
&&\left[\frac{R}{n}\left(n-|m|-2n_2-1\right)+
\lambda_q + C_{n_2}\right]\,U_{qm}^{n_2}(R)=
\nonumber
\\[2mm]
&=&
\sqrt{(n_2+1)(n-n_2-1)(n-|m|-n_2-1)(n_2+|m|+1)}\,\,
U_{qm}^{n_2+1}(R)+
\nonumber
\\[2mm]
&+&
\sqrt{n_2(n-n_2)(n-|m|-n_2)(n_2+|m|)} \,U_{qm}^{n_2-1}(R),
\nonumber
\end{eqnarray}
which have to be solved together with the normalization condition
\begin{eqnarray}
\sum_{l=n_2}^{n-|m|-1}\,\left|U_{qm}^{n_1 n_2}(R)\right|^2 =1.
\label{sfh.7.20}
\end{eqnarray}

\subsection{Spheroidal corrections to the
spherical and parabolic bases of the hydrogen atom.} At small $R$
the second term in the operator $\hat \Lambda$ is perturbation and
the spherical basis is zero approximation. After division by $R$ the
eigenfunction and eigenvalue equation for the operator $\hat
\Lambda$ takes the form
\begin{eqnarray}
\left(-{\hat A}_z -
\frac{1}{R}\,{\hat l}^2 \right)\,\Psi_{nqm}(\xi, \eta, \varphi; R)
= \frac{1}{R}\,\lambda_q\,\Psi_{nqm}(\xi, \eta, \varphi; R).
\label{sfh.7.21}
\end{eqnarray}
As is seen from (\ref{sfh.7.21}), at large $R$ the term ${\hat l}^2/R$ is
perturbation. From the preceding it is seen that the corrections to $\lambda_q$
and $\psi_{nqm}(\xi, \eta, \varphi; R)$ at small and large $R$ can be
calculated in a standard manner. With the help of the well--known formulae from
perturbation theory \cite{LL} and the expressions for the matrix elements of
the operators $\left({\hat l}^2\right)_{n_2n_2^{'}}$ and $\left({\hat
A}_z\right)_{ll'}$, we get:\\
a) the region of small $R$:
\begin{eqnarray}
\lambda_q(R)
&=& - q(q+1) +
\frac{R^2}{2n^2(2q+1)}
\Biggl[\frac{(q+|m|)(q-|m|)(n+q)(n-q)}{q(2q-1)}-
\nonumber
\\[3mm]
&-&
\frac{(q+|m|+1)(q-|m|+1)(n+q+1)(n-q-1)}{(q+1)(2q+3)}\Biggr]\, ,
\label{sfh.7.22}
\\[3mm]
\Psi_{nqm}^{sprd.}
&=& \Psi_{nqm}^{sph.} +  \frac{R}{2n}
\sqrt{\frac{(q+|m|)(q-|m|)(n+q)(n-q)}{q^2(2q-1)(2q+1)}}\,\,
\Psi_{n,q-1,m}^{sph.}+
\nonumber
\\[3mm]
&+&
\frac{R}{2n}
\sqrt{\frac{(q+|m|+1)(q-|m|+1)(n+q+1)(n-q-1)}{(q+1)^2(2q+1)(2q+3)}}
\,\Psi_{n,q+1,m}^{sph.};
\nonumber
\end{eqnarray}
b) the region of large $R$:
\begin{eqnarray*}
\lambda_q(R)
&=&
- (n-q)(q+|m|) - (q+1)(n-|m|-q-1)
- \frac{R}{n}\, \left(n-|m|-2q-1\right),
\\ [3mm]
\Psi_{nqm}^{sphrd.}
&=& \Psi_{nqm}^{пр.} -
\frac{2n}{R}\,
\sqrt{(q+1)(n-|m|-q-1)(n-q-1)(q+|m|+1)}\,
\Psi_{n,q+1,m}^{par.}+
\\ [3mm]
&+&
\sqrt{q(n-|m|-q)(n-q)(q+|m|)}\,
\Psi_{n,q-1,m}^{par.}.
\end{eqnarray*}
The first formula is derived in the second perturbation order; and the rest
three, in the first perturbation order. Higher order corrections are calculated
in a similar way .

We have derived above the condition (\ref{ZERO}) that is a constraint the
fulfillment of which guarantees the consistency of the cutoff at $s=-1$ and
$s=n-|m|$. It follows from theorem (\ref{ort.1.14}) that
\begin{eqnarray}
\left(\frac{\partial {\hat\Lambda}}
{\partial R}\right)_{qq}
= - \left({\hat A}_{z}\right)_{qq}
=
\frac{\partial {\hat\lambda_q}}{\partial R},
\label{sfh.7.11}
\end{eqnarray}
where a mean value is taken from the spheroidal basis of the hydrogen atom.
Taking the limit $R\to\infty$ we pass to the spherical basis and with
(\ref{sfh.7.10}) taken into account we are able to verify that the diagonal
matrix element in (\ref{sfh.7.11}) vanishes. The condition (\ref{ZERO}) also
follows from formula (\ref{sfh.7.22}).

\subsection{Expansion at small quantum numbers.} It has
already been mentioned that in the general case for calculation of eigenvalues
for the separation constant $A(R^2)$ and the prolate spheroidal basis of the
hydrogen atom one needs a computer, though at $0\leq n \leq 3$ it is
unnecessary. Here we give the respective results. Introduce
\begin{eqnarray*}
\widetilde{W}_{nqm}^l(R) = \frac{W_{nqm}^l(R)}{C_{nqm}(R)}.
\end{eqnarray*}
The normalization condition
\begin{eqnarray*}
\sum_{l=|m|}^{n-1}\,\left[W_{nqm}^l(R)\right]^2 = 1
\end{eqnarray*}
results in a convenient formula for calculation of the spheroidal normalization
constant
\begin{eqnarray*}
C_{nqm}(R)=\left\{\sum_{l=|m|}^{n-1}\,
\left[W_{nqm}^l(R)\right]^2\right\}^{-\frac{1}{2}}.
\end{eqnarray*}
Using this formula one can easily show that
\begin{eqnarray*}
\begin{array}{lll}
C_{1q0}(R) = \sqrt{2},& C_{2q1}(R)=\frac{\sqrt{2}}{16}R,&
C_{2q0}(R)=\frac{1}{2\sqrt{2}}
\left(\frac{\lambda+2}{\lambda+1}\right)^{\frac{1}{2}},
\nonumber
\\ [3mm]
C_{3q2}(R) = \frac{\sqrt{2}}{648}R^2, &
C_{3q1}(R) = \frac{R}{27}
\left(\frac{\lambda+6}{\lambda+4}\right)^{\frac{1}{2}},&
C_{3q0}(R)=\sqrt{\frac{2}{27}}\left[1+\frac{27}{8}
\frac{\lambda^2}{R^2}+
\frac{\lambda^2}{2(\lambda + 6)^2}\right]^{-\frac{1}{2}},
\end{array}
\end{eqnarray*}
where $\lambda=\Lambda-R^2/4n^2$. The allowed values for $\lambda$ and the
normalized prolate spheroidal basis are :
\begin{eqnarray*}
\Psi_{1q0} &=& \frac{1}{\sqrt{\pi}}\,e^{-\frac{R}{2}(\xi+\eta)},
\qquad \lambda =0;
\\ [3mm]
\Psi_{2q1} &=& \frac{R}{16\sqrt{\pi}}\,e^{-\frac{R}{4}(\xi+\eta)}
\left[\left(\xi^2-1\right)
\left(1-\eta^2\right)\right]^{\frac{1}{2}}e^{i\varphi},
\qquad \lambda + 2 = 0;
\\ [3mm]
\Psi_{2q0} &=&\frac{1}{4\sqrt{\pi}}\left(\frac{\lambda+2}
{\lambda+1}\right)^{\frac{1}{2}}
\left[1+\frac{\lambda}{2}(1+\xi\eta)-\frac{R}{4}(\xi+\eta)\right]\,
e^{-\frac{R}{4}(\xi+\eta)},
\qquad \lambda(\lambda+2)=\frac{R^2}{4};
\\ [3mm]
\Psi_{3q2} &=& \frac{R^2}{648\sqrt{\pi}}\,e^{-\frac{R}{6}(\xi+\eta)}
\left(\xi^2-1\right)\left(1-\eta^2\right)e^{2i\varphi},
\qquad \lambda+6=0;
\\ [3mm]
\Psi_{3q1} &=& \frac{R}{27}
\left(\frac{\lambda+6}{\lambda+4}\right)^{\frac{1}{2}}\,
e^{-\frac{R}{6}(\xi+\eta)}
\left[\left(\xi^2-1\right)
\left(1-\eta^2\right)\right]^{\frac{1}{2}}\times
\\ [3mm]
&\times& \left[1+\frac{1}{4}(\lambda+2)(1+\xi\eta)-\frac{R}{12}
(\xi+\eta)\right]
\frac{e^{i\varphi}}{\sqrt{2\pi}},
\qquad (\lambda+2)(\lambda+6) = \frac{R^2}{9};
\\ [3mm]
\Psi_{3q0} &=& \frac{1}{\sqrt{27\pi}}\,e^{-\frac{R}{6}(\xi+\eta)}\,
\left[1+\frac{27}{8}\frac{\lambda^2}{R^2}+
\frac{\lambda^2}{2(\lambda + 6)^2}\right]^{-\frac{1}{2}}\,
\Biggl[1-\frac{R}{3}(\xi+\eta)+\frac{\lambda^2}{2}(1+\xi\eta)-
\\ [3mm]
&-& \frac{\lambda R}{24}(\xi+\eta)(1+\xi\eta)+\frac{R^2}{27}
\left(\frac{\lambda+8}{\lambda+6}\right)(\xi+\eta)^2+
\frac{\lambda R^2}{72(\lambda+6)}(1+\xi\eta)^2\Biggr],
\\ [3mm]
&&
\lambda(\lambda+2)(\lambda+6) = \frac{4}{9}R^2\,(\lambda+4).
\end{eqnarray*}

\section{Expansion of the parabolic basis of the hydrogen atom
over the spherical one in the continuous spectrum. Inverse expansion.}
\markboth{CHAPTER 2.  HYDROGEN ATOM}{2.8. EXPANSION OF THE PARABOLIC BASIS OVER THE SPHERICAL ONE}
In \cite{PER-POP2}, a number of problems relating to hidden symmetry in the
Coulomb field were considered, and it was in particular shown that in the case
of scattering, when the Rutherford wave function was used as a parabolic wave
function, interbasis coefficients are expressed in terms of the Clebsch--Gordan
coefficients for the generalized "representations" of the $SU(2)$ group
corresponding to definite complex values of the orbital moment. The problem of
expansion of an arbitrary (i.e., not necessarily the Rutherford one) parabolic
Coulomb wave function of the continuous spectrum over the corresponding
spherical basis was solved in \cite{BESU} by the formula establishing expansion
of the product of two degenerate hypergeometric functions over the Legendre
polynomials and in \cite{POGOSYAN1} by the method of asymptotics ($r\to 0$).
Below, we will dwell upon these papers \cite{POGOSYAN1}.

In the continuous spectrum the orbital moment$l$  runs all integer values
restricted from below by the inequality $l\geq |m|$. In this connection the
expansion of the parabolic basis of the hydrogen atom over the spherical one
has the form
\begin{eqnarray}
\Psi_{k\beta m}(\mu, \nu, \varphi)
= \sum_{i=|m|}^{\infty}\,
W_{k\beta}^{lm}
\Psi_{klm}(r,\theta,\varphi)\,.
\label{3.5}
\end{eqnarray}
The parameter $\beta$ is the eigenvalue of the operator $\frac{1}{2}{A}_z$,
where $A_z$ is the Laplace--Runge--Lenz vector projection (\ref{symh.1.3}) and
has the meaning of the separation constant in the parabolic coordinates
\footnote{In solving the Schr\"{o}dinger equation in the parabolic coordinates
\cite{LL} there arise two constants $\beta_1$ and $\beta_2$ related by
$\beta_1+\beta_2=1$. The parameters $\beta_1$ and $\beta_2$ are expressed in
terms of $\beta$ as follows: $\beta_1=1/2+\beta$, $\beta_2=1/2-\beta$.}. The
quantity $k$ is determined as $k=\sqrt{2E}$ and the wave functions obeying the
normalization conditions
\begin{eqnarray*}
\int\, \Psi_{k\beta m}(\mu,\nu,\varphi)
\Psi_{k'\beta'm'}^*(\mu, \nu, \varphi)dv
&=&
2\pi\delta(k-k')\delta(\beta-\beta')\delta_{mm'}\,,
\\[2mm]
\int\, \Psi_{klm}(r, \theta, \varphi)
\Psi_{k'l'm'}^*(r, \theta, \varphi)dv
&=& 2\pi\delta(k-k')\delta_{ll'}\delta_{mm'}
\end{eqnarray*}
are of the form \cite{FISHER,LL,WENTZER}
\bea
\Psi_{k\beta m}(\mu, \nu, \varphi)
&=& C_{k\beta}^m
f_{k\beta}^m(\mu)f_{k,-\beta}(\nu)
\frac{e^{im\varphi}}{\sqrt{2\pi}}\,,
\label{3.6a}
\\[2mm]
\Psi_{klm}(r,\theta,\varphi)
&=&N_{kl}R_{kl}(r) Y_{lm}(\theta,
\varphi)\,,
\label{3.6b}
\eea
where
\bea
f_{k\beta}(x)
&=& \frac{(-ikx)^{\frac{|m|}{2}}}{(|m|)!}
e^{\frac{ikx}{2}}
{_1F_1}\left(\frac{|m|+1}{2}
- \frac{i}{2k}-\frac{i\beta}{k};|m|+1;-ikx\right)\,,
\label{3.6a1}
\\[2mm]
R_{kl}(r)
&=&
\frac{(-2ikr)^l}{(2l+1)!}e^{ikr}
{_1F_1}\left(l+1-\frac{i}{k}; 2l+2;-2ikr\right)\,,
\label{3.6b1}
\eea
and
\bea
C_{k\beta}^{|m|}
&=&
i^{|m|} \sqrt{\frac{k}{\pi}}e^{\frac{\pi}{2k}}
\left|\Gamma\left(\frac{|m|+1}{2}-\frac{i}{2k}
-\frac{i\beta}{k}\right)
\Gamma\left(\frac{|m|+1}{2}-\frac{i}{2k}
+\frac{i\beta}{k}\right)\right|\,,
\label{3.6a2}
\\[2mm]
N_{kl}
&=&
(i)^l 2ke^{\frac{\pi}{2k}}
\left|\Gamma\left(l+1+\frac{i}{k}\right)\right|\,. \label{3.6b2}
\eea

\subsection{Calculation of interbasis coefficients}

Let us multiply both the sides of transformation (\ref{3.5}) by
$Y_{lm}^*(\theta,\varphi)$, integrate over the solid angle, and pass from the
parabolic coordinates to the spherical ones. Then instead of transformation
(\ref{3.5}) we have
\begin{eqnarray*}
W_{k\beta}^{lm}
&\cdot&{_1F_1}\left(l+1-\frac{i}{k}; 2l+2; -2ikr\right)
=
\frac{(2l+1)!}{2^l(|m|!)^2} \frac{C_{k\beta}^m}{N_{kl}}\times
\nonumber
\\[2mm]
&\times&
\sum_{s=0}^\infty \sum_{t=0}^\infty
\frac{(v)_s}{(|m|+1)_s} \frac{(u)_t}{(|m|+1)_t}
\frac{(-ikr)^{|m|+s+t-l}}{s!t!}Q_{st}^{lm}\,,
\end{eqnarray*}
where
\begin{eqnarray*}
Q_{st}^{lm}=\frac{1}{\sqrt{2\pi}} \int\,
(1-\cos\theta)^{t+\frac{|m|}{2}} (1+\cos\theta)^{s+\frac{|m|}{2}}
Y_{lm}^*(\theta,\varphi)d\Omega\,,
\end{eqnarray*}
and
\begin{eqnarray*}
v=\frac{|m|+1}{2}-\frac{i}{k}\left(\frac{1}{2}+\beta\right)\,,
\qquad
u=\frac{|m|+1}{2}-\frac{i}{k}\left(\frac{1}{2}-\beta\right)\,.
\end{eqnarray*}
The repeated integration by parts convinces us of that  $Q_{st}^{lm}$ is
nonzero only if $s+t+|m|-l\geq 0$ and, therefore, all terms of the series
contain $r$ in the nonnegative degree, so that in the limit, when $r\to 0$, we
get
\begin{eqnarray}
W_{k\beta}^{lm}= \frac{(2l+1)!}{2^l(|m|!)^2}
\frac{C_{k\beta}^m}{N_{kl}} \sum_{s=0}^{l-|m|}
\frac{(v)_s}{(|m|+1)_s} \frac{(u)_{l-|m|-s}}{(|m|+1)_{l-|m|-s}}
\frac{Q_{s,l-|m|-s}^{lm}}{s!(l-|m|-s)!}\,.
\label{3.7}
\end{eqnarray}
The integral $Q_{st}^{lm}$ at $t=l-|m|-s$ turns into the closed expression
\begin{eqnarray*}
Q_{s,l-|m|-s}^{lm}
= (-1)^{l+\frac{m-|m|}{2}-s}
\frac{2^{l+1}l!}{(2l+1)!} \sqrt{\frac{2l+1}{2}(l+|m|)!(l-|m|)!}.
\end{eqnarray*}
Substituting the latter in (\ref{3.7}) and taking into account auxiliary
equalities
\begin{eqnarray*}
(u)_{l-|m|-s}&=&(-1)^s\frac{(u)_{l-|m|}}{(1+|m|-l-u)_s},
\\[2mm]
(|m|+l)_{l-|m|-s}&=&(-1)^s \frac{(u)_{l-|m|}}{(1+|m|-l-u)_s},
\\[2mm]
(l-|m|-s)!&=&(-1)^s\frac{(l-|m|)!}{(-l+|m|)_s},
\end{eqnarray*}
we come to the conclusion that
\begin{eqnarray*}
W_{k\beta}^{lm}
&=&
(-1)^{l+\frac{m-|m|}{2}}
\frac{C_{k\beta}^m}{N_{kl}} \frac{2}{|m|!}
\sqrt{\frac{2l+1}{2}\frac{(l+|m|)!}{(l-|m|)!}}\,
\frac{\Gamma(u+l-|m|)}{\Gamma(u)}\times
\nonumber \\[2mm]
&\times&
{_3F_2} \left\{\matrix{ v,\,\,-l,\,\, -l+|m|,\,\, \cr \cr
|m|+1,\,\, 1+|m|-l-u \,\, \cr} \Biggr |1\right\}.
\end{eqnarray*}
With the use of relation (\ref{d8}) we rewrite the latter result in a more
convenient form
\bea
\label{3.8}
W_{k\beta}^{lm}
&=& (-1)^{l+\frac{m-|m|}{2}}
\frac{C_{k\beta}}{N_{kl}}
\frac{2}{|m|!}
\sqrt{\frac{2l+1}{2}\frac{(l+|m|)!}{(l-|m|)!}}
\frac{\Gamma\left(l+1-\frac{i}{k}\right)}
{\Gamma\left(|m|+1-\frac{i}{k}\right)}\times
\nonumber
\\[2mm]
&\times&
{_3F_2}
\left\{\matrix{
\frac{|m|+1}{2}-\frac{i}{k}\left(\frac{1}{2}+\beta\right),
\,\,l+|m|+1,\,\, -l+|m|,\,\,
\cr
\cr
|m|+1,\,\, 1+|m|-\frac{i}{k}
\,\,
\cr} \Biggr |1\right\}.
\eea
It should be noted that up to now we have not used the explicit form of the
normalization constants $C_{k\beta}$ and $N_{kl}$. In this respect formula
(\ref{3.8}) is valid with any way of normalization of wave functions. In the
accepted normalization
\bea
W_{k\beta}^{lm}
&=&
(-1)^{\frac{m-|m|}{2}}
\frac{\left|\Gamma\left(\frac{|m|+1}{2}-\frac{i}{2k}
-\frac{i\beta}{k}\right)
\Gamma\left(\frac{|m|+1}{2}-\frac{i}{2k}
+\frac{i\beta}{k}\right)\right|}
{\Gamma\left(|m|+1-\frac{i}{k}\right)}\,
\sqrt{\frac{(2l+1)(l+|m|)!}{(l-|m|)!}}\times
\nonumber
\\[3mm]
\label{3.9}
&\times&
\frac{i^{l+|m|}}{\sqrt{2\pi k}} \frac{e^{i\delta_l}}{|m|!}
\,\,
{_3F_2}\left\{\matrix{\frac{|m|+1}{2}-\frac{i}{k}
\left(\frac{1}{2}+\beta\right),\,\,l+|m|+1,\,\,
-l+|m|,\,\, \cr \cr |m|+1,\,\, 1+|m|-\frac{i}{k} \,\, \cr}
\Biggr|1\right\}\,,
\eea
where $\delta_l$ is the Coulomb scattering phase
\begin{eqnarray*}
\delta_l =
\arg\Gamma\left(l+1-\frac{i}{k}\right)\,.
\end{eqnarray*}
It has already been mentioned in the introduction \cite{PARK,LL} that in the
discrete spectrum the problem of finding a transformation between spherical and
parabolic bases reduces to an ordinary procedure of adding two moments and,
therefore, interbasis coefficients coincide with the Clebsch--Gordan
coefficients. The coefficients $W_{k\beta}^{lm}$ can also be expressed in terms
of the Clebsch--Gordan coefficients but continued with respect to their indices
to the complex region. According to formula (\ref{KG.1.3}), relating the
Clebsch--Gordan coefficients with the generalized hypergeometric function
${_3F_2}$ of the unit argument, we have
\begin{eqnarray}
W_{k\beta}^{lm}
&=& (-1)^{v+\frac{m-|m|}{2}}
\frac{C_{k\beta}}{N_{kl}}
\sqrt{\frac{2\Gamma(1-v)\Gamma(1-u)
\Gamma\left(l+1+\frac{i}{k}\right)}
{\Gamma(|m|+1-v)\Gamma(|m|+1-u)\Gamma\left(\frac{i}{k}-l\right)}}\times
\nonumber
\\[3mm]
\label{3.10}
&\times&
C_{\frac{|m|-v-u}{2},\frac{|m|+v-u}{2};
\frac{|m|-v-u}{2},\frac{|m|+u-v}{2}}^{l,|m|}\,.
\end{eqnarray}
Formula (\ref{3.10}) generalizes the result we have obtained in the discrete
spectrum for the continuous spectrum where the group of hidden symmetry is the
Lorentz group $SO(3,1)$ and the moments themselves are complex. Thus, two
equivalent expressions are derived for the coefficients of interbasis
expansions(\ref{3.9}) and  (\ref{3.10}). The first is convenient in concrete
calculations and the second reflects the dynamic symmetry of the Coulomb field
in the continuous spectrum.

In conclusion of this subsection we should like to point out two particular
cases simplifying transformation (\ref{3.5}). These are the case $\beta=0$ when
the hypergeometric function in (\ref{3.9}) can be summed and the Rutherford
scattering. Indeed, assuming in (\ref{3.9}) that $\beta=0$ we have
\bea
W_{k0}^{lm}
&=&
(-1)^{\frac{m-|m|}{2}}\frac{(i)^{l+|m|}}{\sqrt{2k\pi}}
\,
\frac{e^{i\delta_e}}{|m|!}
\,
\sqrt{\frac{(2l+1)(l+|m|)!}{(l-|m|)!}}\times
\nonumber
\\[3mm]
\label{3.22}
&\times&
\frac{\left|\Gamma\left(\frac{|m|+1-\frac{i}{k}}{2}\right)\right|^2}
{\Gamma\left(|m|+1-\frac{i}{k}\right)}\,
{_3F_2}\left\{\matrix{ -l+|m|,\,\,l+|m|+1,\,\,
\frac{|m|+1-\frac{i}{k}}{2},\,\, \cr \cr |m|+1,\,\,
|m|+1-\frac{i}{k} \,\, \cr}\Biggr|1\right\}.
\eea
According to the Watson theorem \cite{BE1, Watson},
\bea
{_3F_2} \left\{\matrix{ -l+|m|,\,\,l+|m|+1,\,\,
\frac{|m|+1-\frac{i}{k}}{2},\,\, \cr \cr |m|+1,\,\,
|m|+1-\frac{i}{k} \,\, \cr}\Biggr|1\right\}
=
\frac{\Gamma\left(\frac{1}{2}\right)\Gamma(|m|+1)
\Gamma\left(1+\frac{|m|+1-\frac{i}{k}}{2}\right)
\Gamma\left(\frac{1}{2}-\frac{|m|+\frac{i}{k}}{2}\right)}
{\Gamma\left(\frac{l+|m|}{2}+1\right)
\Gamma\left(\frac{1+|m|-l}{2}\right)
\Gamma\left(\frac{1-l-\frac{i}{k}}{2}\right)
\Gamma\left(\frac{l-\frac{i}{k}+2}{2}\right)}
\nonumber
\eea
and, therefore, $W_{k0}^{lm}$ differs from zero only at even $(l-|m|)$.
Substituting the latter equality in (\ref{3.22}) and using the formulae
\begin{eqnarray*}
\frac{\Gamma(2z)}{\Gamma(z)}= \frac{2^{2z-1}}{\sqrt\pi}
\Gamma\left(\frac{1}{2} +z\right) \qquad
\Gamma\left(\frac{1}{2}-z\right)\Gamma\left(\frac{1}{2}+z \right)=
\frac{\pi}{\cos\pi z},
\end{eqnarray*}
after long algebraic calculations we get
\begin{eqnarray*}
W_{k0}^{lm}= \frac{(-1)^{\frac{l+m}{2}}}{\sqrt{2k}}
\sqrt{\frac{(2l+1)(l+|m|-1)!!(l-|m|-1)!!}{(l+|m|)!!(l-|m|)!!}}
\frac{1}{\left|\Gamma\left(1+\frac{l+\frac{i}{k}}{2}\right)\right|}\,.
\end{eqnarray*}
At $m=0$ и $\beta=-\frac{1}{2}-\frac{ik}{2}$ formula (\ref{3.9}) naturally
turns into the expansion of the Rutherford wave function over the spherical
basis
\begin{eqnarray}
\label{3.022}
\frac{e^{\frac{\pi}{2k}}}{\sqrt\pi}
\Gamma\left(1-\frac{i}{k}\right) e^{ik\frac{\zeta-\eta}{2}}
{_1F_1}\left(\frac{i}{k};1;ik\eta\right)
= \sum_{l=0}^{\infty}(i)^l \,
\sqrt{2l+1} \frac{e^{i\delta e}}{k}\psi_{kl0}(r,\theta)\,.
\end{eqnarray}
The result analogous to formula (\ref{3.022}) for the Rutherford case was
obtained in \cite{PER-POP}.

\subsection{Orthogonality and inverse transformation}

Let us now consider the expansion of the spherical basis over the parabolic
one. The parameter $\beta$ may also take complex values. So for the Rutherford
wave function $\beta = - \frac{1}{2} - \frac{i}{k}$. Therefore, it is certainly
unclear what region of integration over $\beta$ provides the orthogonality of
the coefficients $W_{k\beta}^{lm}$ in the quantum number $l$.

Let us prove an important property of orthonormalization
\begin{eqnarray}
\label{3.010}
Q^{ll'}=\int\limits_{-\infty}^\infty W_{k\beta}^{lm}
W_{k\beta}^{l'm^*}d\beta
= \delta_{ll'}\,.
\end{eqnarray}
Substitute into (\ref{3.010}) expression (\ref{3.9}) instead of
$W_{k\beta}^{lm}$, and write out the generalized hypergeometric function
${_3F_2}$ as a polynomial and make the change of variables $z=\frac{i\beta}{k}$
\begin{eqnarray*}
Q^{ll'}
&=&
\frac{(i)^{l-l'}}{2\pi k} \frac{1}{(|m|!)^2}
\frac{e^{i(\delta_l-\delta_l')}}
{\left|\Gamma\left(|m|+1-\frac{i}{k}\right)\right|^2}
\sqrt{\frac{(2l+1)(2l'+1)(l+|m|)!(l'+|m|)!}
{(l-|m|)!(l'-|m|)!}}\times
\nonumber
\\[2mm]
&\times& \sum_{s=0}^{l'-|m|} \sum_{t=0}^{l-|m|}
\frac{(l'+|m|+1)_s}{(|m|+1)_s}
\frac{(-l'+|m|)_s}{\left(|m|+1+\frac{i}{k}\right)}
\frac{(l+|m|+1)_t}{(|m|+1)_t}
\frac{(-l'+|m|)_t}{\left(|m|+1-\frac{i}{k}\right)_t}
\frac{B_{st}}{s!t!}\,,
\end{eqnarray*}
where
\begin{eqnarray*}
B_{st}
&=&
(-ik) \int\limits_{-i\infty}^{i\infty}
\Gamma\left(\frac{|m|+1}{2}+\frac{i}{2k}+z+s\right)
\Gamma\left(\frac{|m|+1}{2}-\frac{i}{2k}+t-z \right)\times
\nonumber
\\[2mm]
&\times&
\Gamma\left(\frac{|m|+1}{2}+\frac{i}{2k}-z \right)
\Gamma\left(\frac{|m|+1}{2}-\frac{i}{2k}+z \right)dz\,.
\end{eqnarray*}
According to the Barns lemma \cite{WITWAT},
\begin{eqnarray*}
\frac{1}{2\pi i} \int\limits_{-i\infty}^{i\infty}
\Gamma(\alpha+s)\Gamma(\beta+s)
\Gamma(\gamma-s)\Gamma(\delta-s)ds=
\frac{\Gamma(\alpha+\gamma)\Gamma(\alpha+\delta)\Gamma(\beta+\gamma)
\Gamma(\beta+\delta)}{\Gamma(\alpha+\beta+\gamma+\delta)}\,,
\end{eqnarray*}
if the poles of the expression $\Gamma(\gamma-s)\Gamma(\delta-s)$ are on the
right of the path of integration and the poles of the expression
$\Gamma(\alpha+s)\Gamma(\beta+s)$ are on the left; in this case none of the
poles of the first set coincides with none of the poles of the second set. In
our case, the requirements of this lemma are satisfied and, therefore,
\begin{eqnarray*}
Q^{ll'}
&=&
\frac{(i)^{l-l'}e^{i(\delta_l-\delta_l')}}{(2|m|+1)!}
\sqrt{\frac{(2l+1)(2l'+1)(l+|m|)!(l'+|m|)!}{(l-|m|)!(l'-|m|)!}}\times
\\[2mm]
&\times&
\sum_{s=0}^{l'-|m|}
\frac{(l'+|m|+1)_s(-l'+|m|)_s}{s!(2|m|+2)_S} \sum_{t=0}^{l-|m|}
\frac{(l+|m|+1)_t(-l+|m|)_t(|m|+1+s)_t}{t!(|m|+1)_t(2|m|+2+s)_t}\,.
\end{eqnarray*}
Now let us use the Saalschutz theorem \cite{BE1}
\begin{eqnarray*}
\sum_{p=0}^n \frac{1}{p!}
\frac{(s)_p(b)_p(-n)_p}{(c)_p(1+a+b-c-n)_p}=
\frac{(c-a)_n(c-b)_n}{(c)_n(c-a-b)_b}\,.
\end{eqnarray*}
Substituting in here
\begin{eqnarray*}
a=l+|m|+1, \qquad n=l-|m|, \qquad
b=|m|+1+s, \qquad c=2|m|+2+s,
\end{eqnarray*}
we get
\begin{eqnarray*}
\sum_{t=0}^{l-|m|} \frac{1}{t!}
\frac{(l+|m|+1)_t(-l+|m|)_t(|m|+1+s)_t} {(|m|+1)_t(2|m|+2+s)_t}=
\frac{(-l+|m|+1+s)_{l-|m|}(|m|+1)_{l-|m|}}
{(2|m|+2+s)_{l-|m|}(-l)_{l-|m|}}\,.
\end{eqnarray*}
As a result, $Q^{ll'}$ takes the form
\begin{eqnarray*}
Q^{ll'}
&=& \frac{(-1)^{|m|}(-i)^{l+l'}e^{i(\delta_l-\delta_l')}}
{(l+|m|+1)!}
\sqrt{\frac{(2l+1)(2l'+1)(l+|m|)!(l'+|m|)!}{(l-|m|)!(l'-|m|)!}}\times
\\[2mm]
&\times&
\sum_{s=0}^{l'-|m|}
\frac{(l'+|m|+1)_s(-l'+|m|)_s}{(l+|m|+2)_s\Gamma(s+|m|-l+1)}\,.
\end{eqnarray*}
Since $s_{max}= l'-|m|$, then at $l'<l$ for all $s$ there holds $s+|m|+1-l\leq
0$ and, therefore, $Q^{ll'}=0$. This result also remains valid at $l'>l$, as
the initial expression for $Q^{ll'}$ is symmetric with respect to the change
$l\leftrightarrow l'$. At $l=l'$ in the sum over  $s$ one should take into
account the last term with $s=l'-|m|$. It is easy to verify that $Q^{ll'}=1$.
Thus, formula (\ref{3.010}) is proved.

With the use of formula (\ref{d8}) one can show that the coefficients
$W_{k\beta}^{lm}$ are real, so that (\ref{3.010}) is equivalent to the
condition
\begin{eqnarray}
\label{3.001}
\int\limits_{-\infty}^\infty \, W_{k\beta}^{lm} \, W_{k\beta}^{l'm}
\, d\beta = \delta_{ll'}\,.
\end{eqnarray}
Formulae (\ref{3.001}) and (\ref{3.5}) lead to the expansion of the spherical
basis of the hydrogen atom over the parabolic one
\bea
\Psi_{klm}(r,\theta,\varphi)
= \int\limits_{-\infty}^\infty
W_{k\beta}^{lm}\,
\Psi_{k\beta m}(\xi,\eta,\varphi)\,d\beta\,
\label{3.12} \eea
where integration is over the real basis.

In a particular case $l=|m|$ the coefficients $W_{k\beta}^{l,m}$ are
considerably simplified
\begin{eqnarray*}
W_{k\beta}^{m,|m|}=(-1)^{\frac{m+|m|}{2}}
\frac{C_{k\beta}}{N_{k|m|}} \frac{2}{|m|!}
\sqrt{\frac{(2|m|+1)!}{2}}\,.
\end{eqnarray*}
Substituting the latter formula into(\ref{3.12}), passing to the variables
$q=-ik\xi$ and $p=-ik\eta$, taking account of
\begin{eqnarray*}
|N_{km}|^2
&=& 4k^2\, e^{\frac{\pi}{k}}  \,
\Gamma(|m|+1+2\alpha)\Gamma(|m|+1-2\alpha)\,,
\\[3mm]
|C_{k\beta}|^2
&=&
\frac{k}{\pi} \, e^{\frac{\pi}{k}} \,
\left|\Gamma\left\{\frac{|m|+1}{2}+i(\alpha+\gamma)\right\}\right|^2\,
\left|\Gamma\left\{\frac{|m|+1}{2}+i(\alpha-\gamma)\right\}\right|^2\,,
\end{eqnarray*}
where $\alpha=\frac{1}{2}k$ and $\gamma=\frac{\beta}{k}$, we arrive at the
result
\begin{eqnarray*}
&&\frac{2\pi(|m|!)^2}{(2|m|+1)!}
\Gamma\left(|m|+1+2\alpha\right)\Gamma\left(|m|+1-2\alpha\right)
F\left(|m|+1-2i\alpha;\, 2|m|+2;\, q+p\right)=
\\[3mm]
&=& \int\limits_{-\infty}^{\infty}
\left|\Gamma\left\{\frac{|m|+1}{2}+i(\alpha+\gamma)\right\}\right|^2
\,\left|\Gamma\left\{\frac{|m|+1}{2}+
i(\alpha-\gamma)\right\}\right|^2\times
\\[3mm]
&\times& F\left\{\frac{|m|+1}{2}-i(\alpha+\gamma);|m|+1; q\right\}
F\left\{\frac{|m|+1}{2}-i(\alpha-\gamma);|m|+1;
p\right\}d\gamma\,
\end{eqnarray*}
which was derived by Meixner in \cite{MEIXNER}.

\section{Transition from the continuous spectrum to the discrete one}
\markboth{CHAPTER 2.  HYDROGEN ATOM}{2.9. TRANSITION FROM THE CONTINUOUS SPECTRUM TO THE DISCRETE ONE}
Let us show that formulae (\ref{3.5}) and (\ref{3.12}) turn into the well-known
results for the discrete spectrum (\ref{h.1.1}). Substitute (\ref{3.6a}),
(\ref{3.6b}) and (\ref{3.9}) into expansion (\ref{3.5}) and pass to the limit
$k\rightarrow i|k|=i\sqrt{2E}=\frac{i}{n}$ for the spherical wave function and
$$
\frac{|m|+1}{2}-\frac{i}{k}\left(\frac{1}{2}+\beta\right)
\rightarrow -n_1 \qquad
\frac{|m|+1}{2}-\frac{i}{k}\left(\frac{1}{2}-\beta\right)
\rightarrow -n_2\,
$$
for the parabolic wave function.  Here $n_1$ and $n_2$ are integer nonnegative
numbers and $n=n_1+n_2+|m|+1$. Taking account of the equality
\begin{eqnarray*}
\frac{\Gamma(l+1-n)}{\Gamma(|m|+1-n)}
= (-1)^{l-|m|}\frac{\Gamma(n-|m|)}{\Gamma(n-l)}
\end{eqnarray*}
and passing to the normalization wave functions of the discrete spectrum
(\ref{sh.1.1}) and (\ref{ph.1.1}), we have
\bea
\Psi_{n_1 n_2 m}(\xi,\eta,\varphi) = \sum_{l=|m|}^{n-1}
\tilde W_{n_1n_2}^{lm} \,
\Psi_{nlm}(r,\theta,\varphi)\,,
\label{3.13}
\eea
where the coefficients
\bea
\tilde W_{n_1 n_2}^{lm}
&=&
(-1)^{\frac{m+|m|}{2}}\frac{(n-|m|-1)!}{|m|!}
\sqrt{\frac{(2l+1)(l+|m|)!(n_1+|m|)!(n_2+|m|)!}
{(n+l)!(l-|m|)!(n-l-1)!(n_1)!(n_2)!}}\times
\nonumber \\[3mm]
\label{3.14}
&\times&
{_3F_2} \left\{\matrix{ -l+|m|,\,\,l+|m|+1,\,\,
-n_1,\,\, \cr \cr |m|+1,\,\, |m|+1-n \,\, \cr}
\Biggr|1 \right\}
\eea
with an accuracy up to transformation (\ref{d8}) coincide with the coefficients
$W_{n_1 n_2}^{lm}$ of formula (\ref{h.1.3}). In a similar way one can show that
in the limit of the discrete spectrum the coefficients (\ref{3.10}) take the
form (\ref{h.1.2}).
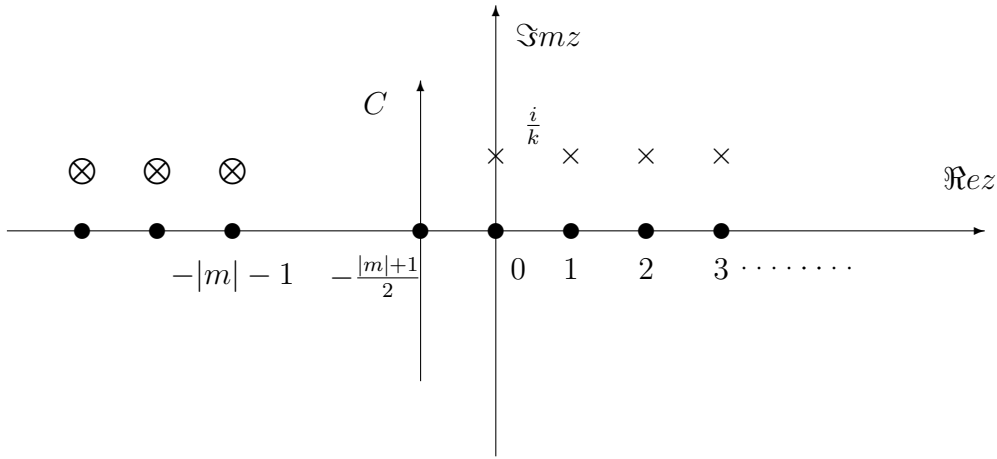
\begin{figure}[t]
\unitlength=1mm \special{em:linewidth 0.4pt} \linethickness{0.4pt}
\begin{picture}(145.00,60.00)(0,30)
\put(15.00,60.00){\vector(1,0){130.00}}
\put(80.00,30.00){\vector(0,1){60.00}}
\put(70.00,40.00){\vector(0,1){40.00}}
\put(64.00,77.00){\makebox(0,0)[cc]{$C$}}
\put(87.00,86.00){\makebox(0,0)[cc]{${\Im}m z$}}
\put(143.00,67.00){\makebox(0,0)[cc]{${\Re}e z$}}
\put(80.00,60.00){\circle*{2.00}}
\put(90.00,60.00){\circle*{2.00}}
\put(100.00,60.00){\circle*{2.00}}
\put(110.00,60.00){\circle*{2.00}}
\put(70.00,60.00){\circle*{2.00}}
\put(45.00,60.00){\circle*{2.00}}
\put(35.00,60.00){\circle*{2.00}}
\put(25.00,60.00){\circle*{2.00}}
\put(80.00,70.00){\makebox(0,0)[cc]{$\times$}}
\put(90.00,70.00){\makebox(0,0)[cc]{$\times$}}
\put(100.00,70.00){\makebox(0,0)[cc]{$\times$}}
\put(110.00,70.00){\makebox(0,0)[cc]{$\times$}}
\put(85.00,74.00){\makebox(0,0)[cc]{$\frac{i}{k}$}}
\put(83.00,55.00){\makebox(0,0)[cc]{$0$}}
\put(90.00,55.00){\makebox(0,0)[cc]{$1$}}
\put(100.00,55.00){\makebox(0,0)[cc]{$2$}}
\put(110.00,55.00){\makebox(0,0)[cc]{$3$}}
\put(113.00,55.00){\circle*{0.00}}
\put(115.00,55.00){\circle*{0.00}}
\put(117.00,55.00){\circle*{0.00}}
\put(119.00,55.00){\circle*{0.00}}
\put(121.00,55.00){\circle*{0.00}}
\put(123.00,55.00){\circle*{0.00}}
\put(123.00,55.00){\circle*{0.00}}
\put(125.00,55.00){\circle*{0.00}}
\put(127.00,55.00){\circle*{0.00}}
\put(121.00,55.00){\circle*{0.00}}
\put(45.00,54.00){\makebox(0,0)[cc]{$-|m|-1$}}
\put(64.00,54.00){\makebox(0,0)[cc]{$-\frac{|m|+1}{2}$}}
\put(25.00,68.00){\makebox(0,0)[cc]{$\bigotimes$}}
\put(35.00,68.00){\makebox(0,0)[cc]{$\bigotimes$}}
\put(45.00,68.00){\makebox(0,0)[cc]{$\bigotimes$}}
\end{picture}

\caption{Poles and the integration contour of the
function $L_{km}(z)$ at $k>0$.}
\label{f21}
\end{figure}
Let us now consider the inverse transformation (\ref{3.12}). Passing from the
integration variable $\beta$ to
\begin{eqnarray*}
z= -\frac{|m|+1}{2}+\frac{i}{k} \left(\frac{1}{2}+\beta\right),
\end{eqnarray*}
we get
\bea
\bar\Psi_{klm}(r,\theta,\varphi)
&=&
\frac{E_{klm}}
{\Gamma\left(|m|+1-\frac{i}{k}\right)}
\,
\frac{1}{2\pi i}
\,
\int\limits_{-\frac{|m|+1}{2}-\frac{i}{k}
\infty}^{-\frac{|m|+1}{2}+\frac{i}{k}\infty} \,
{_3F_2} \left\{\matrix{ -l+|m|,\,\,l+|m|+1,\,\, -z,\,\, \cr \cr
|m|+1,\,\, |m|+1-\frac{i}{k} \,\, \cr} \Biggr|1 \right\}\,\times
\nonumber
\\[3mm]
\label{3.15}
&\times&
\bar\psi_{kzm}(\xi,\eta,\varphi) L_{km}(z)dz\,,
\eea
where
\begin{eqnarray*}
\bar\Psi_{klm}(r,\theta,\varphi)
&=& \frac{(-2ilr)^l}{(2l+1)!}
e^{ikr}\, {_1F_1}\left(l+1-\frac{i}{k};2l+2;-2ikr\right)
Y_{lm}(\theta,\varphi)\,,
\\[3mm]
\bar\Psi_{kzm}(\xi, \eta, \varphi)
&=&
\frac{(-ik\xi)^{\frac{|m|}{2}}}{|m|!}
\frac{(-ik\eta)^{\frac{|m|}{2}}}{|m|!} e^{ik\frac{\xi+\eta}{2}}
\frac{e^{im\varphi}}{\sqrt{2\pi}}\times
\\[3mm]
&\times& {_1F_1}(-z;|m|+1;-ik\xi) \,
{_1F_1}\left(|m|+1-\frac{i}{k}+z;|m|+1; -ik\eta\right),
\\
L_{km}(z)
&=&
\Gamma(-z)\Gamma(|m|+1-z)
\Gamma\left(|m|+1-\frac{i}{k}+z\right)
\Gamma\left(\frac{i}{k}-z\right)\,,
\\[2mm]
E_{klm}
&=&
\frac{(-1)^{\frac{m+|m|}{2}}}
{\Gamma\left(l+1+\frac{i}{k}\right)},
\frac{1}{|m|!} \,
\sqrt{\frac{(2l+1)(l+|m|)!}{(l-|m|)!}}\,.
\end{eqnarray*}

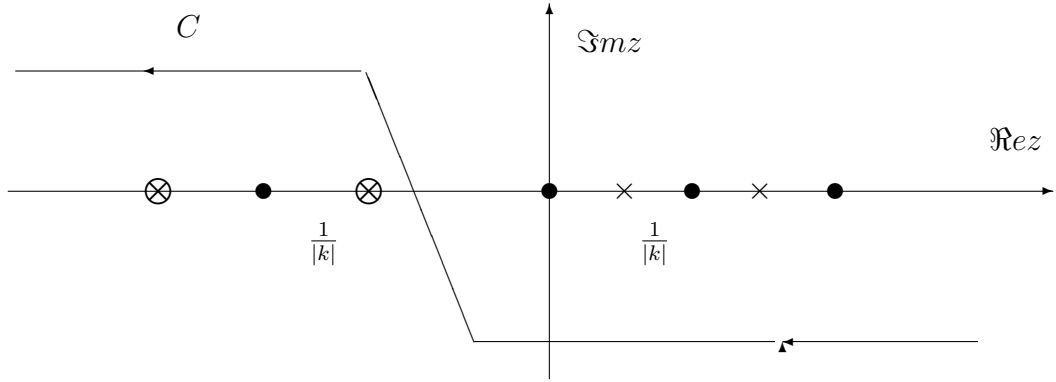
\begin{figure}[t]\unitlength=1.00mm \special{em:linewidth 0.4pt}
\linethickness{0.4pt}

\begin{picture}(149.00,50.00)(0,15)
\put(10.00,40.00){\vector(1,0){139.00}}
\put(82.00,15.00){\vector(0,1){50.00}}
\put(101.00,40.00){\circle*{0.00}}
\put(92.00,40.00){\makebox(0,0)[cc]{$\times$}}
\put(110.00,40.00){\makebox(0,0)[cc]{$\times$}}
\put(58.00,40.00){\makebox(0,0)[cc]{$\bigotimes$}}
\put(30.00,40.00){\makebox(0,0)[cc]{$\bigotimes$}}
\put(90.00,60.00){\makebox(0,0)[cc]{${\Im}m z$}}
\put(144.00,47.00){\makebox(0,0)[cc]{${\Re}e z$}}
\put(139.00,20.00){\vector(-1,0){26.00}}
\put(113.00,20.00){\vector(0,0){0.00}}
\put(28.00,56.00){\line(-1,0){17.00}}
\put(61.00,56.00){\line(0,0){0.00}}
\put(34.00,62.00){\makebox(0,0)[cc]{$C$}}
\put(96.00,33.00){\makebox(0,0)[cc]{$\frac{1}{|k|}$}}
\put(52.00,33.00){\makebox(0,0)[cc]{$\frac{1}{|k|}$}}
\put(112.00,20.00){\line(-1,0){40.00}}
\put(101.00,40.00){\circle*{0.00}}
\put(101.00,40.00){\circle*{0.00}}
\put(120.00,40.00){\circle*{0.00}}
\put(120.00,40.00){\circle*{0.00}}
\put(82.00,40.00){\circle*{0.00}}
\put(82.00,40.00){\circle*{0.00}}
\put(44.00,40.00){\circle*{0.00}}
\put(44.00,40.00){\circle*{0.00}}
\put(72.00,20.00){\line(-2,5){14.33}}
\put(57.00,56.00){\vector(-1,0){29.00}}
\put(44.00,40.00){\circle*{2.00}}
\put(82.00,40.00){\circle*{2.00}}
\put(101.00,40.00){\circle*{2.00}}
\put(120.00,40.00){\circle*{2.00}}
\end{picture}
\caption{Location of the poles of the function $L_{km}(z)$ and the
integration contour in the plane $z$ at $\frac{1}{|k|}<\frac{(|m|+1)}{2}$}
\label{f22}
\end{figure}

Singularities of the integrand in(\ref{3.15}) with respect to the variable z
coincide with the eigenfunctions $L_{km}(z)$ and together with the integration
contour are depicted in Fig. \ref{f21}. The change of the parameter $k$ entails the
change of the position of the poles in the upper half-plane and the change of
the shape of the contour (the latter takes place if the parameter $k$ acquires
the imaginary part).

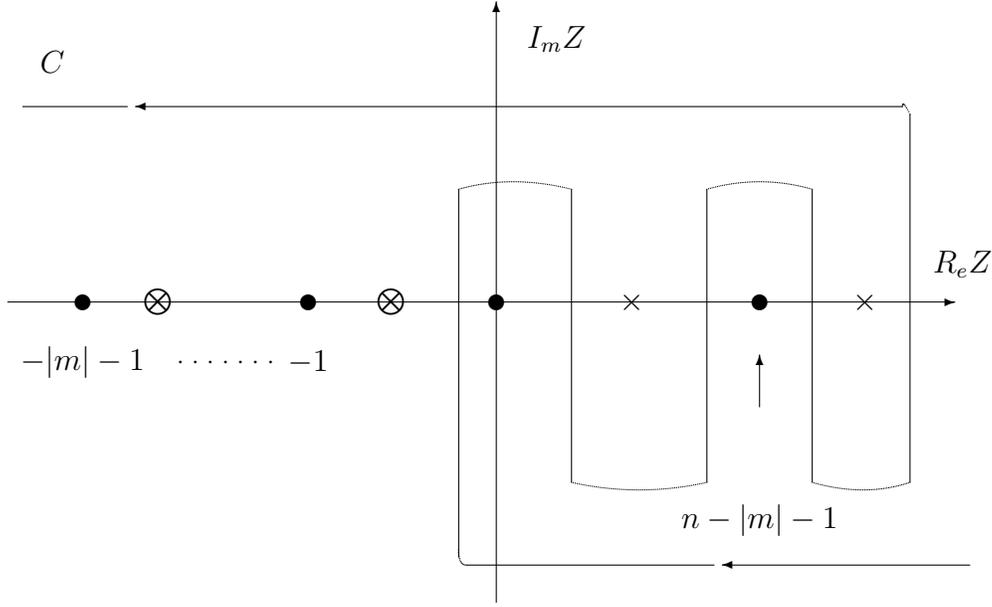
\begin{figure}[t]
\unitlength=1.00mm \special{em:linewidth 0.4pt}
\linethickness{0.4pt}
\begin{picture}(143.00,80.00)
\put(15.00,40.00){\vector(1,0){126.00}}
\put(80.00,00.00){\vector(0,1){80.00}}
\put(80.00,40.00){\circle*{2.00}}
\put(115.00,40.00){\circle*{2.00}}
\put(55.00,40.00){\circle*{2.00}}
\put(25.00,40.00){\circle*{2.00}}
\put(35.00,40.00){\makebox(0,0)[cc]{$\bigotimes$}}
\put(66.00,40.00){\makebox(0,0)[cc]{$\bigotimes$}}
\put(98.00,40.00){\makebox(0,0)[cc]{$\times$}}
\put(129.00,40.00){\makebox(0,0)[cc]{$\times$}}
\put(142.00,45.00){\makebox(0,0)[cc]{$R_eZ$}}
\put(88.00,75.00){\makebox(0,0)[cc]{$I_mZ$}}
\put(55.00,32.00){\makebox(0,0)[cc]{$-1$}}
\put(25.00,32.00){\makebox(0,0)[cc]{$-|m|-1$}}
\put(38.00,32.00){\circle*{0.00}}
\put(40.00,32.00){\circle*{0.00}}
\put(42.00,32.00){\circle*{0.00}}
\put(44.00,32.00){\circle*{0.00}}
\put(46.00,32.00){\circle*{0.00}}
\put(48.00,32.00){\circle*{0.00}}
\put(50.00,32.00){\circle*{0.00}}
\put(143.00,5.00){\vector(-1,0){33.00}}
\put(109.00,5.00){\line(-1,0){33.00}}
\bezier{12}(76.00,5.00)(75.00,5.00)(75.00,7.00)
\put(75.00,6.00){\line(0,1){49.00}}
\put(90.00,55.00){\line(0,-1){39.00}}
\bezier{64}(75.00,55.00)(82.00,57.00)(90.00,55.00)
\put(108.00,55.00){\line(0,-1){39.00}}
\put(122.00,16.00){\line(0,1){39.00}}
\put(135.00,16.00){\line(0,1){49.00}}
\bezier{72}(90.00,16.00)(99.00,14.00)(108.00,16.00)
\bezier{60}(108.00,55.00)(115.00,57.00)(122.00,55.00)
\bezier{12}(135.00,65.00)(134.00,67.00)(134.00,66.00)
\put(134.00,66.00){\vector(-1,0){102.00}}
\put(31.00,66.00){\line(-1,0){14.00}}
\put(21.00,72.00){\makebox(0,0)[cc]{$C$}}
\put(115.00,26.00){\vector(0,1){7.00}}
\put(115.00,11.00){\makebox(0,0)[cc]{$n-|m|-1$}}
\bezier{56}(135.00,16.00)(129.00,14.00)(122.00,16.00)
\end{picture}
\caption{Location of the poles of the function
$L_{mk}(Z)$ and the integration contour at $\frac{1}{|k|}\geq |m|+1$.}
\label{f23}
\end{figure}

Now let us take the limit $k \rightarrow i|k|$. In this case, the upper poles
will be shifted to the real axis, thus moving to the right with respect to the
corresponding lower poles by the same distance $\frac{1}{|k|}$. At small
$\frac{1}{|k|}$ the position of the poles of type $\otimes$ is given in
Fig. \ref{f22}. If $\frac{1}{|k|}>\frac{(|m|+1)}{2}$, then in tending to the real axis
the nearest on the left to the contour pole of type $\otimes$  will have a
tendency to cross the contour, and to avoid a fictitious divergence arisen, it
is necessary to deform the contour itself. Figure \ref{f23} shows the case
corresponding to that when $\frac{1}{|k|}>|m|+1$.

In the discrete spectrum $\frac{1}{|k|}= n\geq |m|+1$, so that in tending to
the real axis $n-|m|-1$ poles of type $\otimes$  approach the poles at the
points $z=1,2,....n-|m|-1$ and clamping the contour prevent us from deforming
it ${}^{1}$.\footnotetext[1] {The so-called pinch of the contour occurs. The
method of identification of singularities of the functions given as integrals
over the known singularities of the integrand was widely used in studying
analogous properties of the Feynman integrals and is expounded in \cite{EDEN}.}
The integral itself becomes divergent in this case. This divergence is
compensated by a relevant divergence in the gamma-function in front of the
integral in (\ref{3.15}). Indeed, assume that $k=\frac{1}{n}+\varepsilon$,
calculate the integral by closing the contour at infinity, and then tend
$\varepsilon$ to zero. The residue at the points $z=|m|+1,|m|+2,.....$ equal
zero, as in the limit $\varepsilon \to 0$ these points correspond to the second
order poles. Simple poles whose residue are finite and do not give a
contribution to(\ref{3.15}) are in the interval $n-|m|\leq z\leq |m|$: as
$\varepsilon \to 0$ the integrand gamma-function tends to infinity. Only simple
poles in the interval $0\leq z\leq n-|m|-1$ remain.

 Taking into account the fact that
$$
{\rm Res}_{z\to - n}\, \Gamma(z) =  \frac{(-1)^n}{n!},
$$
using the equality
\begin{eqnarray*}
\lim\limits_{\varepsilon \to 0}
\frac{\Gamma(-N+i\varepsilon)}{\Gamma(|m|+1-n-i\varepsilon)}=
(-1)^{n-|m|-N-1}\frac{\Gamma(n-|m|)}{\Gamma(N+1)}\,
\end{eqnarray*}
and passing to the wave functions of the discrete spectrum normalized to unity
(\ref{sh.1.1}) and (\ref{ph.1.1}), we have the expansion in which the
coefficients  $W_{n_1 n_2}^{lm}$ are determined by formula (\ref{3.14}). We
come to the conclusion that under an analytic continuation from the region $E >
0$ to the region $E <0$ both direct and inverse transformations  turn into
analogous transformations that take place in the discrete spectrum.

\section{Limiting cases}
\markboth{CHAPTER 2.  HYDROGEN ATOM}{2.10. LIMITING CASES}
In this subsection we consider a limiting transition to a free motion
($\alpha\to0$) and a limiting case of zero energy ($k\to 0$).

\subsection{Transition to a free motion}
Transition to a free motion proceeds in the
following way. Let us restore atomic units (but assume that $\hbar=M=1$),
multiply (\ref{3.5}) by $\sqrt\alpha$, take the limit $\alpha\rightarrow 0$ and
group in a certain way the cofactors in the right-hand side of (\ref{3.5})
\bea
(\sqrt\alpha \Psi_{k\beta m})_{\alpha=0}
= \sum_{l=|m|}^{\infty}
\left(\frac{1}{\sqrt\alpha}W_{k\beta}^{lm}\right)_{\alpha=0}
\,
(\alpha\Psi_{klm})_{\alpha=0}\,.
\label{3.16}
\eea
It is obvious that
\begin{eqnarray*}
(\sqrt\alpha \Psi_{k\beta m})_{\alpha=0}
&\equiv&
\Psi_{k\beta m}^{(0)}(\xi, \eta, \varphi)=
\sqrt{\frac{k}{\pi}} \frac{(-k^2\xi\eta)^{\frac{|m|}{2}}}{(|m|!)^2}
e^{ik\frac{\xi+\eta}{2}}\,
\left|\Gamma\left(\frac{|m|+1}{2}-\frac{i\beta}{k}\right)\right|^2
\,
\frac{e^{im\varphi}}{\sqrt{2\pi}}\times
\\[2mm]
&\times&
F\left(\frac{|m|+1}{2}-\frac{i\beta}{k};|m|+1;-ik\xi\right)
F\left(\frac{|m|+1}{2}+\frac{i\beta}{k};
|m|+1;-ik\eta\right)\,,
\\[3mm]
\left(\frac{1}{\sqrt \alpha}W_{k\beta}^{lm}\right)_{\alpha=0}
&\equiv& S_{k\beta}^{lm} =
(-1)^{\frac{m-|m|}{2}} \frac{(i)^{l+|m|}}{\sqrt{\pi k}}
\sqrt{\frac{2l+1}{2}\frac{(l+|m|)!}{(l-|m|)!}}
\frac{\left|\Gamma\left(\frac{|m|+1}{2}-\frac{i\beta}{k}\right)\right|^2}
{(|m|!)^2}\times
\\[2mm]
&\times& {_3F_2} \left\{\matrix{
\frac{|m|+1}{2}-\frac{i\beta}{k},\,\,-l+|m|,\,\, l+|m|+1,\,\, \cr
\cr |m|+1,\,\, |m|+1 \,\, \cr}\Biggr|1\right\},
\\[3mm]
(\alpha \Psi_{klm})_{\alpha=0}
&\equiv& \Psi_{klm}^{(0)}(r,\theta, \varphi)
=
\frac{(2k)l!}{(2l+1)!} (2kr)^l
e^{ikr} F(l+1;2l+2;-2ikr)Y_{lm}(\theta,\varphi)\,=
\\[3mm]
&=& \sqrt{\frac{2\pi k}{r}}\,
J_{l+\frac{1}{2}}(kr) Y_{lm}(\theta, \varphi),
\end{eqnarray*}
where $\Psi_{k\beta m}^{(0)}(\xi, \eta, \varphi)$ and
$\Psi_{klm}^{(0)}(r,\theta, \varphi)$ are parabolic and spherical wave
functions of a free motion, and $J_{l+\frac{1}{2}}$ is the Bessel spherical
function \cite{BE2}. Thus, it follows from (\ref{3.16}) that
\begin{eqnarray*}
\Psi_{k\beta m}^{(0)}(\xi,\eta,\varphi)=
\sum_{l=|m|}^{\infty}\, S_{k\beta}^{lm} \,
\Psi_{klm}^{(0)}(r,\theta,\varphi)\,.
\end{eqnarray*}
Analogously, one can construct an inverse relation
\begin{eqnarray*}
\Psi_{klm}^{(0)}(r,\theta,\varphi)=
\int_{-\infty}^{\infty}\, S_{k\beta}^{lm}\,
\Psi_{k\beta m}^{(0)}(\xi,\eta,\varphi)d\beta\,.
\end{eqnarray*}
The formulae establishing transitions between parabolic and spherical wave
functions of a free motion were derived in \cite{MILLER-1974} and are in
agreement with our results.

\subsection{Limit $k \to 0$} Let us divide (\ref{3.5}) by $\sqrt k$ and
take the limit $k \to 0$. Then
\bea
\left(\frac{\Psi_k\beta m}{\sqrt k}\right)_{k=0}
=
\sum_{l=|m|}^{\infty}\,
W_{o\beta}^{lm}\,
\left(\frac{\Psi_{klm}}{\sqrt k}\right)_{k=0}\,.
\label{3.17}
\eea
As $k \to 0$ the spherical wave function has the form \cite{LL}
\bea
\frac{\Psi_{klm}}{\sqrt k}|_{k=0} = \sqrt{\frac{4\pi}{r}}
J_{2l+1}(\sqrt{8r}) Y_{lm}(\theta,\varphi)\,.
\label{3.18}
\eea
Let us find the limit of the function $\left(\frac{\Psi_{k\beta m}}{\sqrt
k}\right)$ as $k \to 0$.  With the use of the formula relating the Bessel
functions to the hypergeometric ones ${_0F_1}$ \cite{BE2} we have
\begin{eqnarray*}
{_1F_1}\left\{\frac{|m|+1}{2} -
\frac{i}{k}\left(\frac{1}{2}\pm\beta\right);
|m|+1;-ikx\right\}
&\stackrel{k \to 0} \longrightarrow&
{_0F_1}\left\{|m|+1; -\left(\frac{1}{2}\pm\beta\right)x\right\}=
\\[3mm]
&=&
\frac{|m|!}
{\left[\left(\frac{1}{2}\pm\beta\right)x\right]^{\frac{|m|}{2}}}
J_{|m|}
\left\{2\sqrt{\left(\frac{1}{2}\pm\beta\right)x}\right\}\,.
\end{eqnarray*}
One can easily be convinced that
\begin{eqnarray*}
C_{k\beta} \stackrel{k \to 0}
\longrightarrow i^{|m|}\sqrt{4\pi k}
\left(\frac{\frac{1}{4}-\beta^2}{k^2}\right)^{\frac{|m|}{2}}
\theta\left(\frac{1}{2}-|\beta|\right)\,,
\end{eqnarray*}
i.e. $\left(\frac{\psi_k\beta m}{\sqrt k}\right)$ as $k \to 0$  differs from
zero only under the condition $\frac{1}{2}\leq\beta\leq \frac{1}{2}$.
Therefore, we finally have
\bea
\frac{\psi_{k\beta m}}{\sqrt k}|_{k=0}= \sqrt{4\pi}J_{|m|}
\left\{2\sqrt{\left(\frac{1}{2}+\beta\right)\xi}\right\} J_{|m|}
\left\{2\sqrt{\left(\frac{1}{2}-\beta\right)\eta}
\frac{e^{im\varphi}}{\sqrt{2\pi}}\right\}\,.
\label{3.19}
\eea
Then from the equalities
\begin{eqnarray*}
\frac{\Gamma\left(l+1-\frac{i}{k}\right)}{\Gamma\left(|m|+1-\frac{i}{k}
\right)} \cong \left(-\frac{i}{k}\right)^{l-|m|},
\quad
\frac{C_{k\beta}}{N_{kl}} \cong \frac{(-ik)^{l-|m|}}{\sqrt 2}
\left\{\left(\frac{1}{4}-\beta^2 \right)\right\}^{\frac{|m|}{2}}\,,
\end{eqnarray*}
the fact that within the limit $k\to 0$ the generalized hypergeometric function
turns into the function ${_2F_1}$ and the formulae relating the spherical
function to the hypergeometric one  \cite{VAR}
\begin{eqnarray*}
Y_{lm}(\theta, \varphi)
&=& (-1)^{\frac{m+|m|}{2}}
\sqrt{\frac{2l+1}{2}\frac{(l+|m|)!}{(l-|m|)!}}
\frac{(\sin\theta)^{|m|}}{|m|!2^{|m|}}\times
\\ [3mm]
&\times&
\frac{e^{im\varphi}}{2\pi}\,
{_2F_1}\left(-l+|m|, l+|m|+1; |m|+1; \cos^2\frac{\theta}{2}\right)
\end{eqnarray*}
one can easily obtain that
\bea
W_{0\beta}^{lm}= (-1)^{l-m}\sqrt{4\pi} Y_{lm}(\arccos2\beta, 0)\,.
\label{3.20}
\eea
Substituting (\ref{3.20}), (\ref{3.18}), and (\ref{3.19}) into (\ref{3.17}) we
finally get
\bea
J_{|m|}\left\{\sqrt{8r}\cos\frac{\theta}{2}\cos\frac{\alpha}{2}\right\}
J_{|m|}\left\{\sqrt{8r}\sin\frac{\theta}{2}\sin\frac{\alpha}{2}\right\}
&=&
\frac{4\pi}{\sqrt{r}}\,
\sum_{l=|m|}^{\infty} (-1)^{l-m}\, J_{2l+1}(\sqrt{ 8r})\,\times
\nonumber\\[3mm]
\label{3.21}
&\times&
Y_{lm}(\theta,0) Y_{lm}(\alpha,0)\,,
\eea
with the notation $\alpha=\arccos2\beta$, $\alpha\in (0,\pi)$ is introduced.
For the inverse transformation we have
\begin{eqnarray*}
\sqrt{\frac{2}{r}}J_{2l+1}
(\sqrt{8r})Y_{lm}(\theta,0)
&=& (-1)^{l-m}\, \int\limits_0^{\pi}\,
J_{|m|}\left(\sqrt{8r}\cos\frac{\theta}{2}\cos\frac{\alpha}{2}\right)\times
\\[3mm]
&\times&
J_{|m|}\left(\sqrt{8r}\sin\frac{\theta}{2}\sin\frac{\alpha}{2}\right)
Y_{lm}(\alpha,0)\sin\alpha d\alpha\,.
\end{eqnarray*}
One can easily see that (\ref{3.21})  is a particular case of a more general
relation
\begin{eqnarray*}
&&\frac{z}{2}J_{\mu}(z\cos\varphi\cos\theta)
J_{\nu}(z\sin\varphi\sin\theta)= (\cos\varphi\cos\theta)^\mu
(\sin\varphi\sin\theta)^\nu \times
\\[3mm]
&\times&
\sum_{n=0}^\infty (-1)^n (\mu+\nu+2n+1)J_{\mu+\nu+2n+1}(z)
\frac{\Gamma(\mu+\nu+n+1)\Gamma(n+\nu+1)}{\Gamma(\mu+\nu+1)n!
[\Gamma(\nu+1)]^2}\times
\\[3mm]
&\times&
{_2F_1}(-n,\mu+nu+n+1;\nu+1;{\sin\varphi}^2)
{_2F_1}(-n,\mu+\nu+n+1;\nu+1;{\sin\theta}^2)
\end{eqnarray*}
at $\mu=\nu=|m|$. The last formula was first derived by Bateman in
\cite{BATEMAN}.

Particular relations derived in this section confirm the validity of expansions
(\ref{3.5}) and (\ref{3.12}) and give an idea of their generality.

\newpage

\chapter{Circular Oscillator}
\markboth{CHAPTER 3. CIRCULAR OSCILLATOR}{}
In this chapter we discuss: Expansions in fundamental bases of the circular
oscillator; solutions of the Schr\"{o}dinger equation for the circular
oscillator in the elliptic system of coordinates; limiting transitions $R\to 0$
and $R\to\infty$ in the elliptic basis; expansion of the explicitly factorized
elliptic basis of the circular oscillator over the polar and Cartesian bases;
interbasis expansions in the problem of motion of a nonrelativistic charged
particle in the homogeneous magnetic field. In expounding this chapter we used
the following
papers:\cite{T-2,MPS1,MPSTA4,POGOSYAN3,POGOSYAN4,POGOSYAN5}.

\section{Expansions in fundamental bases}
\markboth{CHAPTER 3. CIRCULAR OSCILLATOR}{3.1. EXPANSIONS IN FUNDAMENTAL BASES}

As is known, as a particle moves in the circular oscillator field, described by
the Hamiltonian\footnote{In this chapter, we use everywhere the system of
units in which $(\hbar=M=\omega=1)$.}
\bea
\label{2D-OSC-1}
{\cal H} = \frac{1}{2}
\left\{{\hat {\bf p}}^2 + r^2\right\},
\qquad\quad
r = \sqrt{(x^2+y^2)}
\eea
in parallel with the orbital moment projection $l_z$, there exists an
additional conserved quantity, two-dimensional symmetric tensor,
\bea
\label{2D-OSC-2}
A_{ij} = \frac{1}{2}\left(p_i p_j + x_i x_j\right),
\qquad\qquad
i,j = 1,2,
\eea
where ${\hat {\bf p}}^2 = {\hat p}_i {\hat p}_i$, and ${\hat p}_i= - \imath
\partial/\partial x_i$\, ($i=1,2$) и $x_1=x$ $x_2=y$. The Hermitian tensor $A_{ij}$
contains three independent components $A_{11}$, $A_{22}$, and $A_{12}$. One can
easily verify that the trace of this tensor coincides with the circular
oscillator Hamiltonian (\ref{2D-OSC-1}), so that only two additional conserved
components remain. Let us construct three conserved operators

\bea
\label{2D-OSC-3}
{\cal P}
&=&
\frac{1}{2}(A_{11} - A_{22}) =
\frac{1}{4}
\left(x^2 - \frac{\partial^2}{\partial x^2} - y^2
+ \frac{\partial^2}{\partial y^2}\right),
\\[3mm]
\label{2D-OSC-03}
{\cal K}
&=&
A_{12} = \frac{1}{2} \left(xy -
\frac{\partial^2}{\partial x y}\right),
\\[3mm]
\label{2D-OSC-003}
L &=& \frac{l_z}{2} = \frac{1}{2i}
\left(x\frac{\partial}{\partial y} -
y\frac{\partial}{\partial x}\right).
\eea
By a direct calculation one can verify that these operators obey the
commutation relations
\bea
\label{2D-OSC-4}
\{{\cal P}, {\cal K} \} = iL,
\qquad
\{L, {\cal P}\} = i{\cal K},
\qquad
\{{\cal K}, L \} = i{\cal P}
\eea
and determine the algebra of symmetry isomorphic to the algebra $so(3)$.

By fundamental bases of the circular oscillator we mean the general
eigenfunctions of the Hamiltonian ${\cal H}$ of the circular oscillator and
each of the operators $({\cal P}, {\cal K}, L)$:
\bea
{\cal H} \Psi_{J+M,J-M}(x,y) &=& (2J+1)\Psi_{J+M,J-M}(x,y),
\quad
{\cal P} \Psi_{J+M,J-M}(x,y) =  M \Psi_{J+M,J-M}(x,y),
\nonumber
\\[3mm]
{\cal H} \Psi_{J+{\bar M},J-{\bar M}}(\bar x, \bar y),
&=&
(2J+1) \Psi_{J+{\bar M},J-{\bar M}}(\bar x, \bar y),
\quad
{\cal K} \Psi_{J+{\bar M},J-{\bar M}}(\bar x, \bar y) = {\bar M}
\Psi_{J+{\bar M},J-{\bar M}}(\bar x, \bar y),
\nonumber
\\[3mm]
{\cal H} \Psi_{2J, 2M'}(r, \varphi)
&=&
(2J+1)\Psi_{2J, 2M'}(r, \varphi),
\qquad\quad
L \Psi_{2J, 2M'}(r, \varphi) = M' \Psi_{2J, 2M'}(r, \varphi).
\nonumber
\eea
These bases are realized by the Schr\"{o}dinger equation solutions in the
polar, Cartesian $(x,y)$, and Cartesian $(\bar x= \frac{1}{\sqrt{2}} (x+y),
\bar y =\frac{1}{\sqrt{2}} (x-y))$ turned by an angle $\pi/4$ systems of
coordinates. Eigenvalues for the operators ${\cal P}$, ${\cal K}$ and $L$ have
the meaning of the corresponding separation constants.

Let us recall the explicit form of fundamental bases
\bea
\Psi_{J+M,J-M}(x,y)
&=&
\frac{(-i)^{J-M}}{2^J \sqrt{\pi}}\,
\frac{e^{-\frac{1}{2}(x^2+y^2)}}{\sqrt{(J+M)!(J-M)!}}
H_{J+M}(x) H_{J-M}(y),
\label{CO.1}
\\ [2mm]
\Psi_{J+\bar M,J-\bar M}(\bar x,\bar y)
&=&
\frac{(-i)^{J-\bar M}}{2^J \sqrt{\pi}}\,
\frac{e^{-\frac{1}{2}(\bar x^2+\bar y^2)}}
{\sqrt{(J+\bar M)!(J-\bar M)!}}
H_{J+\bar M}(x) H_{J-\bar M}(y),
\label{CO.01}
\\ [2mm]
\Psi_{2J, 2|M'|}(r,\varphi)
&=& R_{2J, 2|M'|}(r) \, \frac{e^{2i|M'|\varphi}}{\sqrt{2\pi}},
\label{CO.2}
\eea
where
\bea
R_{2J, 2|M'|}(r)
=
(-1)^{J-|M'|} \,
\sqrt{\frac{2(J+|M'|)!}{(J-|M'|)!}} \,
\frac{(r)^{2|M'|}}{(2|M'|)!} \,
e^{-\frac{r^2}{2}}\,
F\left(-J+|M'|;\, 2|M'|+1; \, r^2\right).
\nonumber
\eea
Phase factors of the wave functions are chosen for simplicity. Quantum numbers
$M$, $\bar M$, and $M'$ vary in the limits : $-J,-J+1, .... J-1, J$. Number $J$
can be both integer and half-integer and it totally determines the energy
spectrum of the circular oscillator: $E=2J+1$. The degeneracy multiplicity of
energy levels of the circular oscillator at fixed value of $J$ equals $2J+1$
and can take both even and odd values (it should be noted that in the case of
the two-dimensional hydrogen atom the degeneracy multiplicity was equal to
$2n+1$ and took only odd values). Hence, it follows that both even and
odd-dimensional irreducible representations of the group $SU(2)$ playing the
role of a group of "hidden" or dynamic symmetry of the circular oscillator are
realized on the fundamental bases.

For each fundamental basis
\bea
{\cal J}^2 \Psi = ({\cal P}^2 + {\cal K}^2 + L^2) \Psi =
\frac{1}{4}({\cal H}^2 -1) \Psi = J(J+1) \Psi.
\eea
Therefore, with the commutation relations (\ref{2D-OSC-4}) taken into account
it is clear that the expansion in the fundamental bases of the circular
oscillator (like for the two-dimensional hydrogen atom ) is realized by the
Wigner $d$-function of the right angle. The result of the relevant calculations
is given in Table \ref{t12}, and the expansions are
\bea
\label{2D-OSC-7}
\Psi_{J+M,J-M}(x,y)
&=&
\sum_{\bar M=-J}^{J}\, d^{J}_{-M, \bar M}\left(\frac{\pi}{2}\right)\,
\Psi_{J+\bar M,J-\bar M}(\bar x,\bar y),
\\[3mm]
\Psi_{J+M,J-M}(x,y)
&=&
\sum_{M'=-J}^{J}\, d^{J}_{M', M}\left(\frac{\pi}{2}\right)\,
\Psi_{2J, 2|M'|}(r,\varphi),
\label{2D-OSC-8}
\\[3mm]
\Psi_{J+\bar M,J-\bar M}(\bar x,\bar y)
&=&
\sum_{M'=-J}^{J}\, d^{J}_{M',- \bar M}\left(\frac{\pi}{2}\right)\,
\Psi_{2J, 2|M'|}(r,\varphi).
\label{2D-OSC-9}
\eea
Summation in these formulae is over the values: $-J, -J+1, ..., J-1, J$.

An interesting result follows from (\ref{2D-OSC-7}). Substituting into expansion
(\ref{2D-OSC-7}) the explicit form of its constituent bases one can easily
derive the formula
\bea
\label{ERMIT-1}
\frac{H_{\mu_1}\left(\frac{x+y}{\sqrt 2}\right)
H_{\mu_2}\left(\frac{x-y}{\sqrt 2}\right)}
{\sqrt{(\mu_1)! (\mu_2)!}}=
\sum_{k=0}^{\mu_1+\mu_2}
d_{\frac{\mu_1+\mu_2}{2}-k, \frac{\mu_1-\mu_2}{2}}^
{\frac{\mu_1+\mu_2}{2}}
\left(\frac{\pi}{2}\right)
\frac{H_{k}(x) H_{\mu_1+\mu_2-k}(y)}
{\sqrt{(k)!(\mu_1+\mu_2-k)!}}.
\eea
At $\mu_2=0$ it turns into the well-known addition theorem for the Hermitian
polynomials cited, e.g., in \cite{KAMPE}
\bea
\label{ERMIT-01}
H_{\mu_1} \left(\frac{x+y}{\sqrt 2}\right)=
\frac{(\mu_1)!}{2^{\frac{\mu_1}{2}}}
\sum_{k=0}^{\mu_1}
\frac{H_k(x) H_{\mu_1-k}(y)}{(k)! (\mu_1-k)!}
\eea
and is consequently its generalization.

Let us now consider the expansion of the Cartesian basis of the circular
oscillator over the polar basis from a different viewpoint. Introduce the polar
and Cartesian bases of the circular oscillator with given parity, i.е., being
eigenfunctions of the sets of operators наборов операторов $\{{\cal H}, {\cal
P}, P_{y}, P_{xy}\}$, $\{{\cal H}, {\cal P}, P_{y}, P_{xy}\}$, respectively,
where by definition
\begin{eqnarray*}
{\hat P}_y\Psi(x,y) &=& \Psi(x,-y), \qquad
{\hat P}_y\Psi(r,\varphi) = \Psi(r,-\varphi), \\ [2mm]
{\hat P}_{xy}\Psi(x,y) &=& \Psi(-x,-y), \qquad
{\hat P}_{xy}\Psi(r,\varphi) = \Psi(r,\varphi+\pi).
\end{eqnarray*}
The Cartesian sub-bases with given parity are determined by the expressions
\begin{eqnarray*}
\Psi_{2k,2n-2k}^{(+)}(x,y)
&=& \overline{H}_{2k}(x)\overline{H}_{2n-2k}(y),
\qquad
\Psi_{2k+1,2n-2k}^{(+)}(x,y) =
\overline{H}_{2k+1}(x)\overline{H}_{2n-2k}(y),
\\ [2mm]
\Psi_{2k,2n-2k+1}^{(-)}(x,y)
&=&
\overline{H}_{2k}(x)\overline{H}_{2n-2k+1}(y),
\qquad
\Psi_{2k+1,2n-2k+1}^{(-)}(x,y) =
\overline{H}_{2k+1}(x)\overline{H}_{2n-2k+1}(y),
\end{eqnarray*}
and the corresponding polar sub-bases are determined by the
formula
\begin{eqnarray*}
\Psi_{2n,2p}^{(+)}(r,\varphi) &=& R_{2n,2p}(r)\frac{\cos2p\varphi}
{\sqrt \pi},
\\ [2mm]
\Psi_{2n+1,2p+1}^{(+)}(r,\varphi) &=& R_{2n+1,2p+1}(r)
\frac{\cos(2p+1)\varphi}{\sqrt \pi},
\\ [2mm]
\Psi_{2n+1,2p+1}^{(-)}(r,\varphi) &=& iR_{2n+1,2p+1}(r)
\frac{\sin(2p+1)\varphi}{\sqrt \pi}, \\ [2mm]
\Psi_{2n+2,2p+2}^{(-)}(r,\varphi) &=& iR_{2n+2,2p+2}(r)
\frac{\sin(2p+2)\varphi}{\sqrt \pi}.
\end{eqnarray*}
Here $n$, $p$ and $k$ are integers with $0\leq p \leq n$ and $0\leq k \leq n$.

Expansion (\ref{2D-OSC-8}) is divided into four expansions
\begin{eqnarray}
\Psi_{2k,2n-2k}^{(+)}(x,y) &=& \sum_{p=0}^n\,\omega^{(+)}_{2p}
\Psi_{2n,2p}^{(+)}(r,\varphi),
\label{CO-INT-1}
\\ [2mm]
\Psi_{2k+1,2n-2k}^{(+)}(x,y) &=& \sum_{p=0}^n\,\omega^{(+)}_{2p+1}
\Psi_{2n+1,2p+1}^{(+)}(r,\varphi),
\label{CO-INT-2}
\\ [2mm]
\Psi_{2k,2n-2k+1}^{(-)}(x,y) &=& \sum_{p=0}^n\,\omega^{(-)}_{2p+1}
\Psi_{2n+1,2p+1}^{(-)}(r,\varphi),
\label{CO-INT-3}
\\ [2mm]
\Psi_{2k+1,2n-2k+1}^{(-)}(x,y) &=& \sum_{p=0}^n\,\omega^{(-)}_{2p+2}
\Psi_{2n+2,2p+2}^{(-)}(r,\varphi).
\label{CO-INT-4}
\end{eqnarray}
Comparing expansions (\ref{CO-INT-1})-(\ref{CO-INT-4}) with (\ref{2D-OSC-8})
one can easily show that
\begin{eqnarray*}
\omega^{(+)}_{2p} &=&
{\sqrt{2}}\left(1-\frac{1}{2}\delta_{p0}\right)
d_{p, 2k-n}^n\left(\frac{\pi}{2}\right), \qquad
\omega^{(+)}_{2p+1} = {\sqrt{2}}
d_{p+\frac{1}{2}, 2k-n+\frac{1}{2}}^{n+\frac{1}{2}}
\left(\frac{\pi}{2}\right),
\\ [2mm]
\omega^{(-)}_{2p+1} &=& {\sqrt{2}}
d_{p+\frac{1}{2}, 2k-n-\frac{1}{2}}^{n+\frac{1}{2}}
\left(\frac{\pi}{2}\right),
\qquad
\omega^{(-)}_{2p+2} = {\sqrt{2}}
d_{p+1, 2k-n}^{n+1}
\left(\frac{\pi}{2}\right).
\end{eqnarray*}
Calculations can be carried out in a different way too. Taking the limit
(\ref{CO-INT-1}) $r \to 0$ we find
\begin{eqnarray*}
\omega^{(+)}_{0} = \frac{(-1)^{n-k}}
{2^{n-1/2}}\frac{\sqrt{(2k)!(2n-2k)!}}
{k!(n-k)!}.
\end{eqnarray*}
In calculating the rest of the coefficients we take into account relation
(\ref{d4}) and the condition of orthogonality over the azimuthal quantum number
$m$ (see Appendix A)
\begin{eqnarray}
\int_0^\infty\, R_{n,m}(r)R_{n, m'}(r)\frac{dr}{r} =
\frac{1}{m}
\delta_{m m'},
\label{CO-INT-5}
\end{eqnarray}
that is valid at $m \not= 0$ and $m' \not= 0$.  As a result we get

\begin{eqnarray}
\omega^{(+)}_{2p} &=&
\frac{(-1)^{n+k-p}}{2^{n-1/2}k!(n-k)!}
\sqrt{\frac{(2k)!(2n-2k)!}{(n+p)!(n-p)!}}
{_3F}_2 \left\{
\begin{array}{l}
-k, -p, p \cr \\
\frac{1}{2},  -n
\end{array}
\Biggr| 1 \right\},
\nonumber
\\ [2mm]
\omega^{(+)}_{2p+1} &=&
\frac{(-1)^{n+k-p}(2p+1)n!}{2^{n}k!(n-k)!}
\sqrt{\frac{(2k+1)!(2n-2k)!}{(n+p+1)!(n-p)!}}
{_3F}_2 \left\{
\begin{array}{l}
-k, -p, p+1 \cr \\
\frac{3}{2},  -n
\end{array}
\Biggr| 1 \right\},\nonumber
\\
\label{OV-CO-INT-1}
\\ 
\omega^{(-)}_{2p+1} &=&
\frac{(-1)^{n+k-p+1}(2p+2)n!}{2^{n+1/2}k!(n-k)!}
\sqrt{\frac{(2k+1)!(2n-2k+1)!}{(n+p+2)!(n-p)!}}
{_3F}_2 \left\{
\begin{array}{l}
-k, -p, p+1 \cr \\
\frac{3}{2},  -n
\end{array}
\Biggr| 1 \right\},
\nonumber
\\ [2mm]
\omega^{(-)}_{2p+2} &=&
\frac{(-1)^{n+k-p}n!}{2^{n}k!(n-k)!}
\sqrt{\frac{(2k)!(2n-2k+1)!}{(n+p+1)!(n-p)!}}
{_3F}_2 \left\{
\begin{array}{l}
-k, -p, p+1 \cr \\
\frac{1}{2},  -n
\end{array}
\Biggr| 1 \right\}.
\nonumber
\end{eqnarray}
In what follows it will be clear that the bases with given $P_x$ and $P_{xy}$
are limiting ($R\to 0$ and $R\to\infty$) for a more general elliptic basis of
the circular oscillator .

\section{Elliptic basis of the circular oscillator}
\markboth{CHAPTER 3. CIRCULAR OSCILLATOR}{3.2. ELLIPTIC BASIS OF THE CIRCULAR OSCILLATOR}

The elliptic coordinates $\xi$ and $\eta$, we will deal with, vary in the
limits $0\leq \xi<\infty, 0\leq \eta\leq 2\pi$ and are connected with the
Cartesian coordinates $x$ and $y$ as follows:
\bea
x = \frac{R}{2}\cosh\xi \cos\eta,
\qquad
y = \frac{R}{2} \sinh\xi \sin\eta.
\label{ECO.1}
\eea
Under this definition, the elliptic system of coordinates (\ref{ECO.1}) differs
from the one we used in Chapter I by the transfer of the origin of coordinates
at the point $(R/2,0)$. As $R\rightarrow 0$ and $R\rightarrow \infty$ the
coordinates thus chosen degenerate into the polar and Cartesian
ones
\bea
\cosh\xi \longrightarrow \frac{2r}{R},
\qquad \cos\eta \longrightarrow \cos\varphi,
\qquad    (R\rightarrow 0), \label{ECO.2a}
\\[2mm]
\sinh\xi \longrightarrow \frac{2y}{R},
\qquad
\cos\eta \longrightarrow \frac{2x}{R},
\qquad (R\rightarrow\infty).
\label{ECO.2b}
\eea
Limiting transitions in (\ref{ECO.2a}) and (\ref{ECO.2b}) are carried out at
the fixed position of the point ($x,y$). The Laplacian and the two-dimensional
volume element for the elliptic system of coordinates (\ref{ECO.1}) coincide
with formulae (\ref{EL-COOR-3}) and (\ref{EL-COOR-0-3}).

Let us denote by $\epsilon$ the circular oscillator energy and write down the
Schr\"{o}dinger equation
\begin{eqnarray*}
\left(\frac{\partial^2}{\partial \xi^2}+\frac{\partial^2}{\partial
\eta^2}\right)\Psi +
\left[\frac{\epsilon R^2}{4}(\cosh 2\xi - \cos 2\eta) -
\frac{R^4}{64}(\cosh^2 2\xi - \cos^2 2\eta)\right]\Psi = 0.
\end{eqnarray*}
This equation, after the substitution
$$
\Psi(\xi,\eta;R^2) = X(\xi; R^2)\, Y(\eta; R^2)
$$
and introduction of the separation constant $A(R^2)$, splits into two ordinary
differential equations
\bea
\left(\frac{d^2}{d\xi^2}+\frac{\epsilon R^2}{4}\cosh 2\xi -
\frac{R^4}{64}\cosh^2 2\xi\right) X(\xi;R^2)
&=& -A(R^2)X(\xi;R^2),
\label{ECO.3a}
\\[2mm]
\left(\frac{d^2}{d\eta^2}+\frac{\epsilon R^2}{4}\cos 2\eta -
\frac{R^4}{64}\cos^2 2\eta\right)Y(\eta;R^2)
&=& + A(R^2)Y(\eta;R^2).
\label{ECO.3b}
\eea
Equations (\ref{ECO.3a}) and (\ref{ECO.3b}) lead to each other under the change
$\eta=i\xi$ and, therefore,
\bea
\Psi(\xi,\eta; R^2) = C(R^2)\, Z(i\xi;R^2)\, Z(\eta;R^2),
\label{ECO.4}
\eea
where $C$ is the normalization constant determined by the condition
\begin{eqnarray}
\int |\psi(\xi,\eta;R^2)|^2 \, dV = 1,
\label{ECO.01-18}
\end{eqnarray}
and $Z(\zeta; R^2)$ is the function being the solution of the equation
\bea
\left(\frac{d^2}{d\zeta^2}-\frac{\epsilon R^2}{4}\cos 2\zeta
+ \frac{R^4}{64}\cos^2 2\zeta\right) Z(\zeta; R^2) =
A(R^2)\, Z(\zeta;R^2),
\label{ECO.5}
\eea
satisfying the periodicity (unambiguity) condition
\bea
Z(\zeta+2\pi; -R^2) = Z(\zeta;R^2)
\label{ECO.6}
\eea
and decreasing as $\zeta \rightarrow i\infty$ (finiteness).

Let us solve equation (\ref{ECO.5}). Introduce the function $W(\zeta; R^2)$,
according to
\begin{eqnarray*}
Z(\zeta;R^2) = e^{-\frac{R^2}{16}\cos 2\zeta}\, W(\zeta;R^2)
\end{eqnarray*}
and transform (\ref{ECO.5}) to the Ince equation \cite{ARSS}
\bea
\frac{d^2W}{d\zeta^2}+\frac{R^2}{4}\sin 2\zeta\frac{dW}{d\zeta}+
\left[\frac{R^4}{64}-A(R^2)-\frac{R^2}{4}(e-1)\cos 2\zeta\right]W=0.
\label{ECO.7}
\eea
Analysis of singular points of equation (\ref{ECO.7}) shows that in the general
case it has two types of solutions
\bea
W^{(+)}(\zeta;R^2)
&=& \sum_{k=0}^{\infty}
a_k(R^2)(\cos\zeta)^k,
\label{ECO.8a}
\\[2mm]
W^{(-)}(\zeta;R^2)
&=&
\sin\zeta\sum_{k=0}^{\infty}
b_k(R^2)(\cos\zeta)^k,
\label{ECO.8b}
\eea
the first being even and the second being odd with respect to the change $\zeta
\rightarrow - \zeta$. One can easily be convinced that solutions to
(\ref{ECO.8a}) and (\ref{ECO.9b}) satisfy the periodicity condition
(\ref{ECO.6}). Substitution of series (\ref{ECO.8a}), (\ref{ECO.8b}) in
equation (\ref{ECO.7}) and supplementing of the definition of the coefficients
$a_k$ and $b_k$ by the equalities $a_{-1}=a_{-2}=b_{-1}=b_{-2}=0$ lead to the
trinomial recurrence relations
\bea
(k+1)(k+2)a_{k+2}+\beta_ka_k+\frac{R^2}{2}(k-\epsilon-1)a_{k-2}=0,
\label{ECO.9a}
\\[2mm]
(k+1)(k+2)b_{k+2}+\tilde\beta_{k}b_k+
\frac{R^2}{2}(k-\epsilon)b_{k-2}=0,
\label{ECO.9b}
\eea
in which
\bea
\beta_k
&=&
-k^2+\frac{R^{2}}{4}(\epsilon-2k-1)+\frac{R^4}{64}-A(R^2),
\label{ECO.10a}
\\[2mm]
\tilde\beta_k
&=&
-(k+1)^2 + \frac{R^{2}}{4}(\epsilon-2k-1) + \frac{R^4}{64}-A(R^2).
\label{ECO.10b}
\eea
From (\ref{ECO.9a}) and (\ref{ECO.9b}) it follows that to determine the
coefficients $a_k$ and $b_k$ one needs four initial conditions. Let us choose
these conditions in the form $a_0=a_1=b_0=b_1=1$ and split each of the
relations (\ref{ECO.9a}) and (\ref{ECO.9b}) into two classes - with even and
odd $k$.

Let us trace the way in which the condition of finiteness singles out the
discrete energy spectrum. Let for definiteness $k=2s$ and the coefficients in
point are $a_{2s}$. At large $s$ equations (\ref{ECO.9a}) and (\ref{ECO.10a})
result in
\begin{eqnarray*}
\left(s^2+\frac{3}{2}s\right)\frac{a_{2s+2}}{a_{2s}}
\frac{a_{2s}}{a_{2s-2}}-
\left(s^2+\frac{R^2}{4}s\right)\frac{a_{2s}}{a_{2s-2}}+
\frac{R^2}{4}s=0.
\end{eqnarray*}
As $s^{-1} \ll 1$, the following expansions are valid:
\begin{eqnarray*}
\frac{a_{2s+2}}{a_{2s}}\sim c_0 + \frac{c_1}{s}+o\left(\frac{1}{s^2}\right),
\qquad \frac{a_{2s}}{a_{2s-2}}\sim c_0 +
\frac{c_1}{s-1}+o\left(\frac{1}{s^2}\right) =
c_0+\frac{c_1}{s}+o\left(\frac{1}{s^2}\right).
\end{eqnarray*}
Together with the previous relation they lead to two cases:
\bea
&& a) \,\,\,\, c_0=0, \qquad
c_1 = \frac{R^2}{4}; \qquad
a_{2s}\sim \left(\frac{R^2}{4}\right)^s\frac{1}{s!};
\nonumber
\\[2mm]
&& b) \,\,\,\, c_0=1, \qquad c_1=-\frac{3}{2};
\qquad
a_{2s}\sim s^{-\frac{3}{2}}.
\nonumber
\eea
One can easily be convinced that in both the cases
$Z(\zeta;R^2)\rightarrow\infty$ as $\zeta \rightarrow i\infty$. Hence it
follows that series (\ref{ECO.8a}) should be cut. Analogous reasoning can
easily be repeated for the rest three recurrence relations and gives the same
result. The cutoff conditions for series (\ref{ECO.8a}) and  (\ref{ECO.8b})
lead to a spectrum of energies
\bea
\epsilon_{N} = N+1, \qquad N = 0,1,2, \dots.
\label{ECO.11}
\eea
Depending on the number $N$ being even the following four Ince polynomials
appear as solutions admissible by the condition of finiteness:
\bea
C_{2n}^{2q}(\zeta;R^2) &=& \sum_{s=0}^{n}a_{2s}(R^2)(\cos\zeta)^{2s},
\qquad N=2n,
\label{ECO.12a}
\\[2mm]
C_{2n+1}^{2q+1}(\zeta;R^2) &=&
\sum_{s=0}^{n}a_{2s+1}(R^2)(\cos\zeta)^{2s+1}, \qquad N=2n+1,
\label{ECO.12b}
\\
S_{2n+1}^{2q+1}(\zeta;R^2)
&=&
\sin\zeta\sum_{s=0}^{n}b_{2s}(R^2)(\cos\zeta)^{2s}, \qquad N=2n+1,
\label{ECO.12c}
\\[2mm]
S_{2n+2}^{2q+2}(\zeta;R^2)
&=&
\sin\zeta\sum_{s=0}^{n}b_{2s+1}(R^2)(\cos\zeta)^{2s+1},
\qquad N=2n+2.
\label{ECO.12d}
\eea
As series (\ref{ECO.8a}) and (\ref{ECO.8b}) are cut off each of the recurrence
relations ($k=2s; k=2s+1$) (\ref{ECO.9a}), (\ref{ECO.9b}) turns into the system
of $n+1$ linear homogeneous equations with respect to the coefficients
$a_{2s},a_{2s+1}, b_{2s}, b_{2s+1}$. Equating the corresponding determinants to
zero leads to four algebraic equations of the $(n+1)$th degree which are used
to determine eigenvalues of the separation constant $A_{N}^t(R^2)$. For
solutions \ref{ECO.12a}) - (\ref{ECO.12d}) the index $t$ takes values
$2q,2q+1,2q+1,2q+2,$ with $0\leq q\leq n$ in all the cases.

So the elliptic basis of the circular oscillator (\ref{ECO.4}) is divided into
the following four subbases:
\bea
\Psi^{(c,c)}
= C^{(1,1)}hc_{2n}^{2q}(i\xi;R^2)
hc_{2n}^{2q}(\eta;R^2), \quad N=2n,
\quad
D=n+1=\frac{N+2}{2},
\label{ECO.13a}
\eea
\bea
\Psi^{(s,c)}
= C^{(1,2)}hc_{2n+1}^{2q+1}(i\xi;R^2)
hc_{2n+1}^{2q+1}(\eta;R^2), \quad N=2n+1, \quad
D=n+1=\frac{N+1}{2}, \label{ECO.13b}
\eea
\bea
\Psi^{(c,s)} = C^{(2,1)}hs_{2n+1}^{2q+1}(i\xi;R^2)
hs_{2n+1}^{2q+1}(\eta;R^2), \quad N=2n+1, \quad
D=n+1=\frac{N+1}{2},
\label{ECO.13c}
\eea
\bea
\Psi^{(s,s)} = C^{(2,2)}hs_{2n+2}^{2q+2}(i\xi;R^2)
hs_{2n+2}^{2q+2}(\eta;R^2), \quad N=2n+2, \quad
D=n+1=\frac{N}{2}.
\label{ECO.13d}
\eea
Here $D$ is the number of states at given $N$, and $hc$ and $hs$ denote the
polynomials (\ref{ECO.12a})-(\ref{ECO.12d}) multiplied by the factor
$\exp\left(-\frac{R^2}{16}\cos 2\zeta\right)$.

The wave functions (\ref{ECO.13a}) - (\ref{ECO.13d}) as eigenfunctions of the
Hamiltonian are orthogonal to $N \ne N'$:
\begin{eqnarray*}
\int \Psi^{(i,j)*}_{N,t} \Psi_{N',t}^{(i,j)}dV = 0.
\end{eqnarray*}
From the Sturm-Liouville theory follow the equalities
\begin{eqnarray*}
\int_{0}^{\infty}hc_{j}^{i*}(i\xi;R^2)hc^{i'}_{j}(i\xi;R^2)d\xi
&=& \int_{0}^{\infty}hs_{j}^{i*}(i\xi;R^2)
hs_{j}^{i'}(i\xi;R^2)d\xi=0,
\\[2mm]
\int_{0}^{2\pi}hc_{j}^{i*}(\eta;R^2)hc^{i'}_{j}(\eta;R^2)d\eta
&=& \int_{0}^{2\pi}hs_{j}^{i*}(\eta;R^2)
hs_{j'}^{i'}(\eta;R^2)d\eta=0
\end{eqnarray*}
with the help of which one can easily prove that at $t\ne t$'
\begin{eqnarray*}
\int \psi_{N,t}^{*(i,j)}\psi_{N,t'}^{i,j}dV=0.
\end{eqnarray*}

Equation (\ref{ECO.5}) forms the eigenvalue problem for the separation constant
$A(R^2)$ and the relevant eigenfunctions. Since the operator in the left-hand
side of (\ref{ECO.5}) is invariant under the transformation
\bea
\zeta\rightarrow \zeta+\frac{\pi}{2}, \qquad
R^2\rightarrow -R^2,
\label{ECO.14}
\eea
then under the change $R^2\rightarrow -R^2$ the set of eigenvalues of the
separation constant $A(R^2)$ remains the same: in fact, only the numeration
changes
$$
A_{N}^t(R^2)=A_{N}^{t'}(-R^2).
$$
This fact leads to that transformation (\ref{ECO.14}) turns solutions
(\ref{ECO.13a}) - (\ref{ECO.13d}) into each other. There exist the following
rules of correspondence (see also subsect. 1.2)
\bea
hc_{2n}^{2q}\left(\zeta+\frac{\pi}{2};-R^2 \right)
\rightarrow
hc_{2n}^{2q'}(\zeta;R^2),
\quad
hc_{2n+1}^{2q+1}\left(\zeta+\frac{\pi}{2};-R^2\right)
\rightarrow hs_{2n+1}^{2q'+1}(\zeta;R^2),
\label{ECO.15a}
\\[2mm]
\label{ECO.15b}
hs_{2n+1}^{2q+1}\left(\zeta+\frac{\pi}{2};-R^2 \right)
\rightarrow
hc_{2n+1}^{2q'+1}(\zeta;R^2),
\quad
hs_{2n+2}^{2q+2}\left(\zeta+\frac{\pi}{2};-R^2\right)
\rightarrow hs_{2n+2}^{2q'+2}(\zeta;R).
\eea
From (\ref{ECO.15a}) and (\ref{ECO.15b}) we conclude that the elliptic basis
(\ref{ECO.13a}) - (\ref{ECO.13d}) can be represented as
\bea
\Psi^{(c,c)}_{2n,2q,2q'}
&=& C^{(1,1)}hc_{2n}^{2q}\left(i\xi+\frac{\pi}{2};-R^2\right)
hc_{2n}^{2q'}(\eta;R^2),
\label{ECO.16a}
\\[2mm]
\Psi^{(s,c)}_{2n+1,2q+1,2q'+1}
&=& C^{(1,2)}hc_{2n+1}^{2q+1}\left(i\xi+\frac{\pi}{2};-R^2\right)
hc_{2n+1}^{2q'+1}(\eta;R^2),
\label{ECO.16b}
\\[2mm]
\Psi^{(c,s)}_{2n+1,2q+1,2q'+1}
&=& C^{(2,1)}hc_{2n+1}^{2q+1}\left(i\xi+\frac{\pi}{2};-R^2\right)
hs_{2n+1}^{2q'+1}(\eta;R^2),
\label{ECO.16c}
\\[2mm]
\Psi^{(s,s)}_{2n+2,2q+2,2q'+2}
&=& C^{(2,2)}hc_{2n+2}^{2q+2}\left(i\xi+\frac{\pi}{2};-R^2\right)
hs_{2n+2}^{2q'+2}(\eta;R^2).
\label{ECO.16d}
\eea
As $\eta\to -\eta$ and $\eta \to \eta+\pi$  subbases
(\ref{ECO.16a})-(\ref{ECO.16d}) transform as
\begin{eqnarray*}
\Psi^{(c,c)}_{2n,2q,2q'}(\xi,\eta;R^2)
&=&
\Psi^{(c,c)}_{2n,2q,2q'}(\xi,- \eta;R^2) =
\Psi^{(c,c)}_{2n,2q,2q'}(\xi, \eta+\pi;R^2),
\\[2mm]
\Psi^{(s,c)}_{2n+1,2q+1,2q'+1}(\xi,\eta;R^2)
&=&
\Psi^{(s,c)}_{2n+1,2q+1,2q'+1}(\xi,-\eta;R^2) =
- \, \Psi^{(s,c)}_{2n+1,2q+1,2q'+1}(\xi,\eta+\pi;R^2),
\\[2mm]
\Psi^{(c,s)}_{2n+1,2q+1,2q'+1}(\xi,\eta;R^2)
&=&
- \, \Psi^{(c,s)}_{2n+1,2q+1,2q'+1}(\xi,-\eta;R^2) =
- \, \Psi^{(c,s)}_{2n+1,2q+1,2q'+1}(\xi,\eta+\pi;R^2),
\\[2mm]
\Psi^{(s,s)}_{2n+2,2q+2,2q'+2}(\xi,\eta;R^2)
&=&
- \Psi^{(s,s)}_{2n+2,2q+2,2q'+2}(\xi,-\eta;R^2) =
\Psi^{(s,s)}_{2n+2,2q+2,2q'+2}(\xi,\eta+\pi;R^2).
\end{eqnarray*}
Below Tables \ref{t31}--\ref{t34} give information on the orthonormalized factorized
elliptic basis for the lowest quantum numbers $(A' = A(R^2)-\frac{R^4}{64})$.

Now let us trace how formulae (\ref{ECO.16a}) - (\ref{ECO.16d}) within the
limits $R \to 0$ and $R\to \infty$ result in polar and Cartesian wave functions
of the circular oscillator. Relation (\ref{ECO.14}) implies that as $R \to 0$
numbers $t$ and $t'$ coincide, i.e., $q=q'$. From (\ref{ECO.9a}) -
(\ref{ECO.10b}) it follows that as $R \to 0$ each of the four trinomial
recurrence relations (\ref{ECO.9a}) - (\ref{ECO.9b}) $(k=2s;\, k=2s+1)$ splits
into two binomial ones:
\bea
(2s+1)(2s+2)a_{2s+2}+4(q+s)(q-s)a_{2s}=0,
\nonumber
\\[2mm]
(2s+2)(2s+3)a_{2s+3}+4(q+s+1)(q-s)a_{2s+1}=0,\nonumber
\\
\label{ECO.17a}
\\
(2s+1)(2s+2)b_{2s+2}+4(q+s+1)(q-s)b_{2s}=0,
\nonumber
\\[2mm]
(2s+2)(2s+3)b_{2s+3}+4(q+s+2)(q-s)b_{2s+1}=0,
\nonumber
\eea
if $0 \leq s \leq q-1$, and
\bea
4(q+s)(q-s)a_{2s}+R^2(s-n-1)a_{2s-2}=0,
\nonumber
\\[2mm]
4(q+s+1)(q-s)a_{2s+1}+R^2(s-n-1)a_{2s-1}=0,\nonumber
\\
\label{ECO.17b}
\\
4(q+s+1)(q-s)b_{2s}+R^2(s-n-1)b_{2s-2}=0,
\nonumber
\\[2mm]
4(q+s+2)(q-s)b_{2s+1}+R^2(s-n-1)b_{2s-1}=0,
\nonumber
\eea
if $q+1 \leq s \leq n$. Disentangling (\ref{ECO.17a}) and (\ref{ECO.17b}) we have
\bea
a_{2s}(0) &=& \frac{(q)_s(-q)_s}{s!\left(\frac{1}{2}\right)_s},
\qquad
a_{2s+1}(0)=\frac{(q+1)_s(-q)_s}{s!\left(\frac{3}{2}\right)_s},
\nonumber   \\
\label{ECO.18a}
\\
b_{2s}(0) &=& \frac{(q+1)_s(-q)_s}{s!\left(\frac{1}{2}\right)_s},
\qquad
b_{2s+1}(0)=\frac{(q+2)_s(-q)_s}{s!\left(\frac{3}{2}\right)_s},
\nonumber
\eea
at $0 \leq s \leq q$, and
\bea
\label{k7.2b}
a_{2s}(-R^2) &\stackrel{R\rightarrow 0}\rightarrow&
(-1)^q\frac{2^{2q-1}(q-n)_{s-q}}{(s-q)!(2q+1)_{s-q}}
\left(-\frac{R^2}{4}\right)^{s-q},
\nonumber \\[2mm]
a_{2s+1}(-R^2) &\stackrel{R\rightarrow 0}\rightarrow&
(-1)^q\frac{2^{2q}}{2q+1}\frac{(q-n)_{s-q}}{(s-q)!(2q+2)_{s-q}}
\left(-\frac{R^2}{4}\right)^{s-q}, \nonumber \\
\label{ECO.18c}  \\
b_{2s}(-R^2) &\stackrel{R\rightarrow 0}\rightarrow&
(-1)^q2^{2q}\frac{(q-n)_{s-q}}{(s-q)!(2q+2)_{s-q}}
\left(-\frac{R^2}{4}\right)^{s-q},
\nonumber \\[2mm]
b_{2s+1}(-R^2) &\stackrel{R\rightarrow 0}\rightarrow&
\frac{(-1)^q2^{2q}}{q+1}\frac{(q-n)_{s-q}}{(s-q)!(2q+3)_{s-q}}
\left(-\frac{R^2}{4}\right)^{s-q},
\nonumber
\eea
at $q+1\leq s\leq n$. Using formulae (\ref{ECO.18a}) - (\ref{ECO.18c}) and
taking into account the relations
\begin{eqnarray*}
\cos az
&=& F\left(\frac{a}{2},-\frac{a}{2};\frac{1}{2};\sin^2 z\right)=
\cos z F\left(\frac{1}{2}+\frac{a}{2},\frac{1}{2}-\frac{a}{2};\frac{1}{2};
\sin^2 z\right),
\\[2mm]
\sin az &=& a\sin z F\left(\frac{1}{2}+\frac{a}{2},
\frac{1}{2}-\frac{a}{2};
\frac{3}{2};\sin^2 z\right)=a\sin z\cos zF\left(1+\frac{a}{2},1-
\frac{a}{2};\frac{3}{2};\sin^2 z\right)
\end{eqnarray*}
we arrive at the following behavior of the functions $hc$ and $hs$ as $R \to
0$:
\begin{eqnarray}
hc_{2n}^{2q}(\eta;0) &=& (-1)^q\cos 2q\varphi,
\qquad
hc_{2n+1}^{2q+1}(\eta;0) = \frac{(-1)^q}{2q+1}\cos(2q+1)\varphi,
\label{ECO.1-18c}
\\[2mm]
hs_{2n+1}^{2q+1}(\eta;0) &=& (-1)^q\sin(2q+1)\varphi,
\qquad
hs_{2n+2}^{2q+2}(\eta;0) = \frac{(-1)^q}{2q+2}\sin(2q+2)\varphi,
\label{ECO.2-18c}
\end{eqnarray}
\begin{eqnarray}
hc_{2n}^{2q}\left(i\xi+\frac{\pi}{2};-R^2\right)
&\stackrel{R\to 0}\longrightarrow&
(-1)^{n-q}\,
\frac{(2q)! 2^{4q-1}}{\sqrt{2} R^{2q}}\,
\sqrt{\frac{(n-q)!}{(n+q)!}}\,
R_{2n,2q}(r),
\nonumber
\\[2mm]
hc_{2n+1}^{2q+1}\left(i\xi+\frac{\pi}{2};-R^2\right)
&\stackrel{R\to 0}\longrightarrow&
-i(-1)^{n-q}\,
\frac{(2q)! 2^{4q+1}}{\sqrt{2} R^{2q+1}}\,
\sqrt{\frac{(n-q)!}{(n+q+1)!}}\,
R_{2n+1,2q+1}(r), \nonumber
\\
\label{ECO.3-18c}
\\
hs_{2n+1}^{2q+1}\left(i\xi+\frac{\pi}{2};-R^2\right)
&\stackrel{R\to 0}\longrightarrow&
(-1)^{n-q}\,
\frac{(2q+1)! 2^{4q+1}}{\sqrt{2} R^{2q+1}}\,
\sqrt{\frac{(n-q)!}{(n+q+1)!}}\,
R_{2n+1,2q+1}(r),
\nonumber
\\[2mm]
hs_{2n+2}^{2q+2}\left(i\xi+\frac{\pi}{2};-R^2\right)
&\stackrel{R\to 0}\longrightarrow&
-i(-1)^{n-q}\,
\frac{(2q+2)!2^{4q+3}}{R^{2q+2}}
\sqrt{\frac{(n-q)!}{(n+q+1)!}}\,
R_{2n+2,2q+2}(r).
\nonumber
\end{eqnarray}
The behavior of the normalization constants $C^{(i,j)}(R^2)$ $(i,j=1,2)$,
within the limits $R\to 0$ can be established from (\ref{ECO.1-18c}),
(\ref{ECO.2-18c}) and (\ref{ECO.3-18c}), and the normalization condition
(\ref{ECO.01-18}):
\begin{eqnarray}
|C^{(1,1)}(R^2)| &\stackrel{R\to 0}\longrightarrow&
\sqrt{\frac{2}{\pi}}\,
\frac{R^{2q}}{(2q)!2^{4q+1}}\,
\sqrt{\frac{(n+q)!}{(n-q)!}}\,,
\nonumber
\\[2mm]
|C^{(1,2)}(R^2)| &\stackrel{R\to 0}\longrightarrow&
\sqrt{\frac{2}{\pi}}\,
\frac{R^{2q+1}}{(2q)!2^{4q+1}}\,
\sqrt{\frac{(n+q+1)!}{(n-q)!}}\,,\nonumber
\\
\label{ECO.3-019}
\\
|C^{(2,1)}(R^2)| &\stackrel{R\to 0}\longrightarrow&
\sqrt{\frac{2}{\pi}}\,
\frac{R^{2q+1}}{(2q)!2^{4q+1}}\,
\sqrt{\frac{(n+q+1)!}{(n-q)!}}\,,
\nonumber
\\[2mm]
|C^{(2,2)}(R^2)| &\stackrel{R\to 0}\longrightarrow&
\sqrt{\frac{2}{\pi}}\,
\frac{R^{2q+2}}{(2q+1)!2^{4q+3}}\,
\sqrt{\frac{(n+q+2)!}{(n-q)!}}\,.
\nonumber
\end{eqnarray}
Let now $R$ tend to infinity in formulae (\ref{ECO.10a}) and (\ref{ECO.10b}).
Then instead of (\ref{ECO.9a}) and (\ref{ECO.9b}) we have
\bea
(2s+1)(2s+2)a_{2s+2}+R^2(q-s)a_{2s}=0, \qquad
(q-s)a_{2s}+(s-n-1)a_{2s-2}=0, \nonumber \\[2mm]
(2s+2)(2s+3)a_{2s+3}+R^2(q-s)a_{2s+1}=0, \qquad
(q-s)a_{2s}+(s-n-1)a_{2s-2}=0,  \nonumber\\[2mm]
(2s+1)(2s+2)b_{2s+2}+R^2(q-s)b_{2s}=0, \qquad
(q-s)b_{2s}+(s-n-1)b_{2s-2}=0,  \nonumber  \\[2mm]
(2s+2)(2s+3)b_{2s+3}+R^2(q-s)b_{2s+1}=0,
\qquad
(q-s)b_{2s+1}+(s-n-1)b_{2s-1}=0, \nonumber
\eea
and eight more recurrence relations resulting from the latter under the change
$q \to q', R^2 \to -R^2$. Uncovering these recurrence relations and taking into
account (\ref{ECO.2b}) we arrive at formulae $(R\to \infty)$
\begin{eqnarray}
\lim\limits_{R\to\infty}
C_{2n}^{2q}(\eta;R^2)
&=&
\lim\limits_{R\to\infty}
S_{2n+1}^{2q+1}(\eta;R^2)
= F\left(-q,\frac{1}{2};x^2\right),
\nonumber
\\[2mm]
\lim\limits_{R\to\infty}
C_{2n+1}^{2q+1}(\eta;R^2)
&=&
\lim\limits_{R\to\infty}
S_{2n+2}^{2q+2}(\eta;R^2)
= \frac{2x}{R} F\left(-q,\frac{3}{2};x^2\right),\nonumber
\\
\label{1-ECO.3-019}
\\
\lim\limits_{R\to\infty}
C_{2n}^{2q'}(i\xi+\frac{\pi}{2};-R^2)
&=&
\lim\limits_{R\to\infty}
S_{2n+1}^{2q'+1}(\eta;R^2)
\left(i\xi+\frac{\pi}{2};-R^2\right)
= F\left(-q',\frac{1}{2};y^2\right),
\nonumber
\\[2mm]
\lim\limits_{R\to\infty}
C_{2n+1}^{2q'+1}(i\xi+\frac{\pi}{2};-R^2)
&=&
\lim\limits_{R\to\infty}
S_{2n+2}^{2q'+2}(\eta;R^2)
\left(i\xi+\frac{\pi}{2};-R^2\right)
= -i \frac{2y}{R}F\left(-q',\frac{3}{2};y^2\right).
\nonumber
\end{eqnarray}
It follows from the latter formulae that in the limit $R\to \infty$\, $q'=n-q$,
$q'=n-q-1/2$, $q'=n-q-1/2$ и $q'=n-q-1/2$ for (\ref{ECO.16a}) -
(\ref{ECO.16d}), respectively. As in the case of limiting relations
(\ref{ECO.3-019}) the normalization condition (\ref{ECO.01-18}) gives
\begin{eqnarray}
|C^{(1,1)}(R^2)| &\stackrel{R\to \infty}\longrightarrow&
\frac{1}{\sqrt{\pi}}\,
\frac{\sqrt{(2q)!(2n-2q)!}}{2^n q! (n-q)!},
\nonumber
\\[2mm]
|C^{(1,2)}(R^2)| &\stackrel{R\to \infty}\longrightarrow&
\frac{R}{\sqrt{\pi}}\,
\frac{\sqrt{(2q)!(2n-2q+1)!}}{2^{n+1/2} q! (n-q)!},\nonumber
\\
\label{ECO.3-119}
\\
|C^{(2,1)}(R^2)| &\stackrel{R\to \infty}\longrightarrow&
\frac{R}{\sqrt{\pi}}\,
\frac{\sqrt{(2q+1)!(2n-2q+1)!}}{2^{n+1/2} q! (n-q)!},
\nonumber
\\[2mm]
|C^{(2,2)}(R^2)| &\stackrel{R\to \infty}\longrightarrow&
\frac{R^2}{\sqrt{\pi}}\,
\frac{\sqrt{(2q+1)!(2n-2q+1)!}}{2^{n+1} q! (n-q)!}.
\nonumber
\end{eqnarray}
So as  $R\to 0$ and $R\to \infty$ the elliptic basis of the circular oscillator
turns into the polar and Cartesian bases, respectively.

\section{Expansion over the polar and Cartesian bases of the circular oscillator}
\markboth{CHAPTER 3. CIRCULAR OSCILLATOR}{3.3. EXPANSION OVER THE POLAR AND CARTESIAN BASES}
\subsection{Polar basis expansion}
We start with the expansion of the elliptic subbases with given parity
over the corresponding polar subbases:
\begin{eqnarray}
\Psi_{2n, 2q,2q'}^{(c,c)}(\xi,\eta; R^2) &=&
\sum_{p=0}^n\,W^{(+)}_{2p}(R^2) \,
\Psi_{2n,2p}^{(+)}(r, \varphi),
\label{CO-INT-6}
\\ [2mm]
\Psi_{2n+1, 2q+1,2q'+1}^{(s,c)}(\xi,\eta; R^2) &=&
\sum_{p=0}^n\,W^{(+)}_{2p+1}(R^2)\,
\Psi_{2n+1,2p+1}^{(+)}(r,\varphi),
\label{CO-INT-7}
\\ [2mm]
\Psi_{2n+1, 2q+1,2q'+1}^{(c,s)}(\xi,\eta; R^2) &=&
\sum_{p=0}^n\,W^{(-)}_{2p+1}(R^2)\,
\Psi_{2n+1,2p+1}^{(-)}(r,\varphi),
\label{CO-INT-8}
\\ [2mm]
\Psi_{2n+2, 2q+2,2q'+2}^{(s,s)}(\xi,\eta; R^2) &=&
\sum_{p=0}^n\,W^{(-)}_{2p+2}(R^2)\,
\Psi_{2n+2,2p+2}^{(-)}(r,\varphi).
\label{CO-INT-9}
\end{eqnarray}
In (\ref{CO-INT-6}) tending $r$ to zero with $\cosh\to 1$, $\sin^2\eta \to 1$
taken into account we get
\begin{eqnarray*}
W_{0}^{(+)} = (-1)^n \sqrt{\frac{\pi}{2}} C_{2n,2q,2q'}^{(c,c)}(R^2).
\end{eqnarray*}
Express the polar coordinates in terms of the elliptic ones via the formulae
\begin{eqnarray*}
r = \frac{R}{2}\sqrt{\cosh^2\xi - \cos^2\eta},
\qquad
\cos\varphi = \frac{\cosh\xi\cos\eta}
{\sqrt{\cosh^2\xi - \cos^2\eta}},
\qquad
\sin\varphi = \frac{\sinh\xi\sin\eta}
{\sqrt{\cosh^2\xi - \cos^2\eta}}.
\end{eqnarray*}
To calculate the rest of the coefficients, we pass in
(\ref{CO-INT-6})-(\ref{CO-INT-9}) from the polar coordinates to the elliptic
ones, tend $\eta \to \pi/2$ (correspondingly $\varphi\to \pi/2$), and use the
property of orthogonality(\ref{CO-INT-5}). This procedure leads to the
following results:
\begin{eqnarray}
W^{(+)}_{2p}
&=&
A^{(c,c)} \, e^{\frac{R^2}{16}} \,
\int\limits_{0}^{\infty}
hc_{2n}^{2q}\left(i\xi+\frac{\pi}{2};-R^2\right)\,
R_{2n,2p}\left(\frac{R}{2}\sinh\xi\right)
\frac{d\sinh\xi}{\sinh\xi},
\label{CO-INT0-10}
\\[3mm]
W^{(+)}_{2p+1}
&=&
A^{(s,c)} \, e^{\frac{R^2}{16}} \,
\int\limits_{0}^{\infty}
hc_{2n+1}^{2q+1}\left(i\xi+\frac{\pi}{2};-R^2\right)\,
R_{2n+1,2p+1}\left(\frac{R}{2}\sinh\xi\right)
d\sinh\xi,
\label{CO-INT1-10}
\\[3mm]
W^{(-)}_{2p+1}
&=&
A^{(c,s)} \,
e^{\frac{R^2}{16}}
\,
\int\limits_{0}^{\infty}
hc_{2n+1}^{2q+1}\left(i\xi+\frac{\pi}{2};-R^2\right)\,
R_{2n+1,2p+1}\left(\frac{R}{2}\sinh\xi\right)
\frac{d\sinh\xi}{\sinh\xi},
\label{CO-INT2-10}
\\[3mm]
W^{(-)}_{2p+2}
&=&
A^{(s,s)} \, e^{\frac{R^2}{16}}\,
\int\limits_{0}^{\infty}
hc_{2n+2}^{2q+2}\left(i\xi+\frac{\pi}{2};-R^2\right)\,
R_{2n+2,2p+2}\left(\frac{R}{2}\sinh\xi\right)
d\sinh\xi,
\label{CO-INT3-10}
\end{eqnarray}
where the following notation is introduced:
\begin{eqnarray*}
A^{(c,c)}  =
(-1)^p 2p \sqrt{\pi} \,
\,
C^{(c,c)}_{2n,2q,2q'}(R^2),
\qquad
A^{(s,c)}  =
(-1)^p \sqrt{\pi} \, C^{(s,c)}_{2n+1,2q+1,2q'+1}(R^2),
\\
A^{(c,s)} \,
=
(-1)^{p-\frac12} (2p+1) \sqrt{\pi} \,
C^{(c,s)}_{2n+1,2q+1,2q'+1}(R^2),
\quad
A^{(s,s)} =
(-1)^{p+\frac32} \sqrt{\pi} \,
C^{(s,s)}_{2n+2,2q+2,2q'+2}(R^2).
\end{eqnarray*}
By formulae (\ref{ECO.12a}) - (\ref{ECO.12d}) one can easily relate these
coefficients with $a_{2s}$, $a_{2s+1}$, $b_{2s}$, and $b_{2s+1}$ determining
the factorized elliptic basis
\begin{eqnarray*}
W^{(+)}_{2p} &=&
\frac{(-1)^n\sqrt{2\pi}n!}{\sqrt{(n+p)!(n-p)!}}\,
C^{(c,c)}(R^2)\,
\sum_{s=0}^p\, \left(-\frac{4}{R^2}\right)^s \,
a_{2s}\, \frac{4^s(p)_s(-p)_s}{(-n)_s},  \,\,\, p\ne 0,
\\ [2mm]
W^{(+)}_{2p+1} &=& \frac{(-1)^n\sqrt{2\pi}C^{(s,c)}(R^2)}
{R\sqrt{(n+p+1)!(n-p)!}}
\sum_{s=0}^p\,\frac{b_{2s}(-R^2)}{(-R^2)^s}
\frac{4^s(p+1)_s(-p)_s}{(-n)_s},
\\ [2mm]
W^{(-)}_{2p+1} &=&
\frac{(-1)^{n+1}\sqrt{2\pi}(2p+1)C^{(s,s)}(R^2)}
{R\sqrt{(n+p+1)!(n-p)!}}
\sum_{s=0}^p\,\frac{a_{2s+1}(-R^2)}{(-R^2)^s}
\frac{4^s(p+1)_s(-p)_s}{(-n)_s},
\\ [2mm]
W^{(-)}_{2p+2} &=&
\frac{(-1)^n\sqrt{2\pi}(2p+2)n!C^{(c,s)}(R^2)}
{\sqrt{(n+p+2)!(n-p)!}}
\sum_{s=0}^p\,
\frac{b_{2s+1}(-R^2)}{(-R^2)^{s+1}}
\frac{4^s(p+2)_s(-p)_s}{(-n)_s}.
\end{eqnarray*}
Now let us study the limits $R\to 0$ and $R\to\infty$ in the expressions
derived. From the integral representations (\ref{CO-INT0-10}) -
(\ref{CO-INT3-10}), according to(\ref{ECO.3-18c}), (\ref{ECO.3-019})
 and (\ref{CO-INT-5}), we have
\begin{eqnarray*}
W_{2p}^{(+)}(R^2) &\stackrel{R\to 0}\longrightarrow
(-1)^{n} e^{i\Phi^{(c,c)}(0)}\, \delta_{pq},
\qquad
W_{2p+1}^{(+)}(R^2) &\stackrel{R\to 0}\longrightarrow
(-1)^{n+1} e^{i\Phi^{(s,c)}(0)}\, \delta_{pq},
\\
W_{2p+1}^{(+)}(R^2) &\stackrel{R\to 0}\longrightarrow
(-1)^{n} e^{i\Phi^{(c,s)}(0)}\, \delta_{pq},
\qquad
W_{2p+2}^{(+)}(R^2)
&\stackrel{R\to 0}\longrightarrow
(-1)^{n+1} e^{i\Phi^{(s,s)}(0)}\, \delta_{pq},
\end{eqnarray*}
where the following notation is accepted:
\begin{eqnarray}
\label{CO-INT4-10}
C^{(i,j)}(R^2) = |C^{(i,j)}(R^2)|\, e^{i\Phi^{(i,j)}(R^2)}, \qquad
i,j=s,c.
\end{eqnarray}
Thus, the elliptic subbases turn into the polar ones if
\begin{eqnarray*}
\Phi^{(c,c)}(0) = \Phi^{(c,s)}(0) = \pi n,
\qquad
\Phi^{(s,c)}(0) = \Phi^{(s,s)}(0) = \pi (n+1).
\end{eqnarray*}
Now let us consider the limit $R\to\infty$. According to (\ref{1-ECO.3-019}),
(\ref{ECO.3-119}), and (\ref{d4})
\begin{eqnarray*}
W^{(+)}_{2p} &\stackrel{R\to 0}\longrightarrow&
\frac{(-1)^{n-p} n! e^{i\Phi^{(c,c)}(\infty)}}
{2^{n-1/2}q!(n-q)!}\,
\sqrt{\frac{(2q)!(2n-2q)!}{(n+p)!(n-p)!}}
{_3F}_2 \left\{
\begin{array}{l}
-q, -p, p \cr \\
\frac{1}{2},  -n
\end{array}
\Biggr| 1 \right\},
\\ [2mm]
W^{(+)}_{2p+1} &\stackrel{R\to 0}\longrightarrow&
\frac{(-1)^{n-p-1} n! e^{i\Phi^{(s,c)}(\infty)}}
{2^{n}q!(n-q)!}\,
\sqrt{\frac{(2q)!(2n-2q+1)!}{(n+p+1)!(n-p)!}}
{_3F}_2 \left\{
\begin{array}{l}
-q, -p, p+1 \cr \\
\frac{1}{2},  -n
\end{array}
\Biggr| 1 \right\},
\\ [2mm]
W^{(-)}_{2p+1} &\stackrel{R\to 0}\longrightarrow&
\frac{(-1)^{n-p} n! (2p+1)e^{i\Phi^{(c,s)}(\infty)}}
{2^{n}q!(n-q)!}\,
\sqrt{\frac{(2q+1)!(2n-2q)!}{(n+p+1)!(n-p)!}}
{_3F}_2 \left\{
\begin{array}{l}
-q, -p, p+1 \cr \\
\frac{3}{2},  -n
\end{array}
\Biggr| 1 \right\},
\\ [2mm]
W^{(-)}_{2p+2} &\stackrel{R\to 0}\longrightarrow&
\frac{(-1)^{n-p-1} n! (2p+2)e^{i\Phi^{(s,s)}(\infty)}}
{2^{n+\frac12}q!(n-q)!}\,
\sqrt{\frac{(2q+1)!(2n-2q+1)!}{(n+p+2)!(n-p)!}}
{_3F}_2 \left\{
\begin{array}{l}
-q, -p, p+2 \cr \\
\frac{3}{2},  -n
\end{array}
\Biggr| 1 \right\}.
\end{eqnarray*}
From the above-derived relations and formulae
(\ref{CO-INT4-10}) and (\ref{OV-CO-INT-1}) it follows that the elliptic
subbases transform into the Cartesian ones as $R\to\infty$ if

\begin{eqnarray}
\label{FAZA-1}
\Phi^{(c,c)}(\infty) = \Phi^{(s,s)}(0) = \pi q,
\qquad
\Phi^{(s,c)}(0) = \Phi^{(c,s)}(0) = \pi (q+1),
\end{eqnarray}

\subsection{Cartesian basis expansion}

Let us pass to expansions of the elliptic
subbases of the circular oscillator over the Cartesian ones
\begin{eqnarray}
\Psi_{2n,2q,2q'}^{(c,c)}(\xi,\eta;R^2) &=&
\sum_{k=0}^n\, U^{(+)}_{2k}(R^2)\, \Psi_{2k,2n-2k}^{(+)}(x,y),
\label{CO-INT-10}
\\ [2mm]
\Psi_{2n+1,2q+1,2q'+1}^{(s,c)}(\xi,\eta;R^2)
&=&
\sum_{k=0}^n\,U^{(+)}_{2k+1}
\Psi_{2k+1,2n-2k}^{(+)}(x,y),
\label{CO-INT-11}
\\ [2mm]
\Psi_{2n+1,2q+1,2q'+1}^{(c,s)}(\xi,\eta;R^2)
&=&
\sum_{k=0}^n\,U^{(-)}_{2k}
\Psi_{2k,2n-2k+1}^{(-)}(x, y),
\label{CO-INT-12}
\\ [2mm]
\Psi_{2n+2,2q+2,2q'+2}^{(s,s)}(\xi,\eta;R^2)
&=&
\sum_{k=0}^n\,U^{(-)}_{2k+1}
\Psi_{2k+1,2n-2k+1}^{(-)}(x, y).
\label{CO-INT-13}
\end{eqnarray}
Passing in these expansions to the limit $\eta \to \pi/2$ and using the
orthogonality of the Hermitian polynomials we get
\begin{eqnarray}
U^{(+)}_{2k}
&=&
B^{(c,c)} \,
\int\limits_{0}^{\infty}
e^{-\frac{R^2}{8}\cosh\xi} \,
hc_{2n}^{2q}\left(i\xi+\frac{\pi}{2};-R^2\right)\,
H_{2n-2k}\left(\frac{R}{2}\cosh\xi\right)\,
d(\frac{R}{2}\sinh\xi),
\nonumber
\\[2mm]
U^{(+)}_{2k+1}
&=&
B^{(s,c)} \,
\int\limits_{0}^{\infty}
e^{-\frac{R^2}{8}\cosh\xi} \,
hc_{2n+1}^{2q+1}\left(i\xi+\frac{\pi}{2};-R^2\right)\,
H_{2n-2k}\left(\frac{R}{2}\cosh\xi\right)\,
d\xi, \nonumber
\\
\label{DEC-CO-INT-1}
\\
U^{(-)}_{2k+1}
&=&
B^{(c,s)} \,
\int\limits_{0}^{\infty}
e^{-\frac{R^2}{8}\cosh\xi} \,
hc_{2n+1}^{2q+1}\left(i\xi+\frac{\pi}{2};-R^2\right)\,
H_{2n-2k+1}\left(\frac{R}{2}\cosh\xi\right)\,
d(\xi\sinh\xi),
\nonumber
\\[2mm]
U^{(-)}_{2k+2}
&=&
B^{(s,s)} \,
\int\limits_{0}^{\infty}
e^{-\frac{R^2}{8}\cosh\xi} \,
hc_{2n+2}^{2q+2}\left(i\xi+\frac{\pi}{2};-R^2\right)\,
H_{2n-2k+2}\left(\frac{R}{2}\cosh\xi\right)\,
d\xi,
\nonumber
\end{eqnarray}
where the following notation is used:
\begin{eqnarray}
B^{(c,c)}
&=&
\frac{(-1)^n 2^{2k-n+1} k!}{\sqrt{(2k)!(2n-2k)!}} \,
e^{\frac{R^2}{16}}
\,
C^{(c,c)}_{2n,2q,2q'}(R^2),
\nonumber
\\[2mm]
B^{(s,c)}
&=&
\frac{(-1)^n 2^{2k-n+\frac12} k!}{\sqrt{(2k+1)!(2n-2k)!}} \,
e^{\frac{R^2}{16}}
\,
C^{(c,c)}_{2n+1,2q+1,2q'+1}(R^2),\nonumber
\\
\label{DEC-CO-INT-2}
\\
B^{(c,s)}
&=&
\frac{(-1)^{n+\frac12} 2^{2k-n+\frac12} k!}{\sqrt{(2k)!(2n-2k+1)!}} \,
e^{\frac{R^2}{16}}
\,
C^{(c,s)}_{2n+1,2q+1,2q'+1}(R^2),
\nonumber
\\[2mm]
B^{(s,s)}
&=&
\frac{(-1)^{n+\frac12} 2^{2k-n} k!}{\sqrt{(2k+1)!(2n-2k+1)!}} \,
e^{\frac{R^2}{16}}
\,
C^{(c,s)}_{2n+2,2q+2,2q'+2}(R^2).
\nonumber
\end{eqnarray}
For a concrete calculation of the coefficients $U$ it is convenient to express
them in terms of the coefficients determining the elliptic basis. With the help
of expansions (\ref{ECO.12a}) - (\ref{ECO.12d}), after simple calculations we
have
\begin{eqnarray}
U^{(+)}_{2k}
&=&
2^{2n-2k-1}\,
e^{-\frac{R^2}{16}}\,
B^{(c,c)} \,
\sum_{s=0}^k\, \left(-\frac{4}{R^2}\right)^s
a_{2s}(-R^2)
\frac{\Gamma(s+1)\Gamma(s+1/2)}{\Gamma(s+k+1-n)},
\nonumber
\\ [2mm]
U^{(+)}_{2k+1}
&=&
2^{2n-2k}\,
e^{-\frac{R^2}{16}}\,
B^{(s,c)} \,
\sum_{s=0}^k\, \left(-\frac{4}{R^2}\right)^s
b_{2s}(-R^2)
\frac{\Gamma(s+1)\Gamma(s+1/2)}{\Gamma(s+k+1-n)},\nonumber
\\
\label{DEC-CO-INT-3}
\\
U^{(-)}_{2k+1}
&=&
i2^{2n-2k+1}\,
\frac{e^{-\frac{R^2}{16}}}{R^2}\,
B^{(c,s)} \,
\sum_{s=0}^k\, \left(-\frac{4}{R^2}\right)^s
a_{2s+1}(-R^2)
\frac{\Gamma(s+1)\Gamma(s+1/2)}{\Gamma(s+k+1-n)},
\nonumber
\\[2mm]
U^{(-)}_{2k+2}
&=&
-i2^{2n-2k+2}\,
\frac{e^{-\frac{R^2}{16}}}{R^2}\,
B^{(s,s)} \,
\sum_{s=0}^k\, \left(-\frac{4}{R^2}\right)^s
b_{2s+1}(-R^2)
\frac{\Gamma(s+1)\Gamma(s+1/2)}{\Gamma(s+k+1-n)}.
\nonumber
\end{eqnarray}
Formulae (\ref{ECO.9a}) - (\ref{ECO.9b}) are convenient in studying the limit
$R\to 0$. Passing in them to small $R$, using relations (\ref{ECO.3-18c}),
(\ref{ECO.3-019}), and (\ref{d4}), and comparing the obtained results with
(\ref{OV-CO-INT-1}) and (\ref{CO-INT4-10}) one can determine that
\begin{eqnarray*}
U^{(+)}_{2k}(R^2) \stackrel{R\to 0}\longrightarrow \omega^{(+)}_{2q},
\qquad\qquad
U^{(+)}_{2k+1}(R^2) \stackrel{R\to 0}\longrightarrow \omega^{(+)}_{2q+1},
\\[2mm]
U^{(-)}_{2k+1}(R^2) \stackrel{R\to 0}\longrightarrow \omega^{(-)}_{2q+1},
\qquad\qquad
U^{(-)}_{2k+2}(R^2) \stackrel{R\to 0}\longrightarrow \omega^{(-)}_{2q+2}.
\end{eqnarray*}
Formulae (\ref{DEC-CO-INT-1}), (\ref{1-ECO.3-019}), (\ref{ECO.3-119}),
(\ref{DEC-CO-INT-2}), (\ref{FAZA-1}) imply that
\begin{eqnarray*}
U^{(+)}_{2k}(R^2) \stackrel{R\to \infty}\longrightarrow
\delta_{rq},
\qquad\qquad
U^{(+)}_{2k+1}(R^2) \stackrel{R\to \infty}\longrightarrow
\delta_{rq},
\\[2mm]
U^{(-)}_{2k+1}(R^2) \stackrel{R\to \infty}\longrightarrow
\delta_{rq},
\qquad\qquad
U^{(-)}_{2k+2}(R^2) \stackrel{R\to \infty}\longrightarrow
\delta_{rq}.
\end{eqnarray*}

\section{Recurrence relations and elliptic corrections}
\markboth{CHAPTER 3. CIRCULAR OSCILLATOR}{3.4. RECURRENCE RELATIONS AND ELLIPTIC CORRECTIONS}

The process of variable separation in the elliptic coordinates generates the
separation constant $A(R^2)$ and leads to a pair of differential equations
(\ref{ECO.3a}) and (\ref{ECO.3b}). Eliminating the energy term from these
equations we get the operator
\begin{eqnarray}
\label{EL-OPERAT-1}
{\hat A}
= \frac{1}{\cosh2\xi - \cos2\eta}
\left(\cos2\eta\frac{\partial^2}{\partial \xi^2}
+ \cosh2\xi\frac{\partial^2}{\partial \eta^2}\right) - \frac{R^4}{64}
\cosh2\xi\cos2\eta,
\end{eqnarray}
whose eigenvalues are the constants $A(R^2)$ and eigenfunctions are the
solutions
\begin{eqnarray}
\label{EL-OPERAT-2}
{\hat A} \Psi = A(R^2) \Psi.
\end{eqnarray}
Passing in the operator (\ref{EL-OPERAT-1}) to the Cartesian coordinates one
can prove that
\begin{eqnarray}
\label{EL-OPERAT-3}
{\hat A} = - L^2 - \frac{R^2}{2} {\cal P} + \frac{R^4}{64}
\end{eqnarray}
and rewrite equation (\ref{EL-OPERAT-2}) in the form
\begin{eqnarray}
\label{EL-OPERAT-4}
\Lambda \Psi = \lambda \Psi,
\end{eqnarray}
where
\begin{eqnarray}
\label{EL-OPERAT-5}
\Lambda = {\hat A} - \frac{R^4}{64} =
- L^2 - \frac{R^2}{2} {\cal P}.
\end{eqnarray}
The elliptic bases(\ref{ECO.16a})-(\ref{ECO.16d}) are common eigenfunctions of
the operators ${\cal H}$, $\Lambda$, $P_{y}$, and $P_{xy}$. Therefore,
substituting expansion (\ref{CO-INT-6})-(\ref{CO-INT-9}) into
(\ref{EL-OPERAT-4}) we arrive at the following system of linear equations:
\begin{eqnarray}
\sum_{p=0}^n\,\
\left\{
{\cal P}_{2p',2p}^{(+)} + \frac{2}{R^2}
(\lambda^{(c,c)}+4p^2) (1+\delta_{p0}) \delta_{pp'}
\right\}\, W_{2p}^{(+)} &=& 0,
\label{CO-INT-17}
\\[2mm]
\sum_{p=0}^n\,
\left\{
{\cal P}_{2p'+1,2p+1}^{(+)} + \frac{2}{R^2}
[\lambda^{(s,c)}+(2p+1)^2] \delta_{pp'}\right\}\,
W_{2p+1}^{(+)} &=& 0,
\label{CO-INT-18}
\\[2mm]
\sum_{p=0}^n\,
\left\{
{\cal P}_{2p'+1,2p+1}^{(-)} + \frac{2}{R^2}
[\lambda^{(c,s)}+(2p+1)^2]
\delta_{pp'}\right\}\, W_{2p+1}^{(-)}
&=& 0,
\label{CO-INT-19}
\\[2mm]
\sum_{p=0}^n\,
\left\{
{\cal P}_{2p'+2,2p+2}^{(-)} + \frac{2}{R^2}
[\lambda^{(s,s)}+(2p+2)^2] \delta_{pp'}\right\}\,
W_{2p+2}^{(-)} &=& 0,
\label{CO-INT-20}
\end{eqnarray}
where $0\leq p' \leq n$. Let us calculate the matrix elements of the operator
${\cal P}$:
\begin{eqnarray}
\label{EL-OPERAT-6}
{\cal P}_{m',m}^{(\pm)} =  \int\, \Psi_{nm'}^{(\pm)*}(r, \varphi)\,
{\cal P}\, \Psi_{nm}^{(\pm)}(r, \varphi)\, dV.
\end{eqnarray}
Using the fact that
\begin{eqnarray}
\label{EL-OPERAT-7}
\Psi_{nm}^{(\pm)}(r, \varphi) = \frac{1}{\sqrt{2}}\,
\left\{\Psi_{n,m}(r, \varphi) \pm
\Psi_{n, -m}(r, \varphi)\right\},
\end{eqnarray}
one can easily derive the equality
\begin{eqnarray}
\label{EL-OPERAT-8}
{\cal P}_{m',m}^{(\pm)} =  \frac{1}{2}
\left\{{\cal P}_{m',m} \pm {\cal P}_{m',-m} \pm {\cal P}_{-m',m} + {\cal P}_{-m',-m}\right\},
\end{eqnarray}
in which
\begin{eqnarray}
\label{EL-OPERAT-9}
{\cal P}_{m',m} =  \int\, \Psi_{n,m'}^{*}(r, \varphi)\,
{\cal P}\, \Psi_{n,m} (r, \varphi)\, dV.
\end{eqnarray}
From (\ref{2D-OSC-8}) it follows that
\begin{eqnarray}
\label{EL-OPERAT-10}
\Psi_{n,m}(r,\varphi)
=
{\sum_{p = -n}^{n}}'\, d^{n}_{\frac{m}{2}, \frac{p}{2}}
\left(\frac{\pi}{2}\right)\, \Psi_{n,p}(x,y),
\end{eqnarray}
where the prime of the sum implies that numbers $n$ and $p$ have the same
parity. Substitute (\ref{EL-OPERAT-10}) into (\ref{EL-OPERAT-9}) and take into
account that
\begin{eqnarray}
\label{EL-OPERAT-11}
{\cal P} \Psi_{np} (x,y) = \frac{p}{2} \Psi_{np}(x,y).
\end{eqnarray}
Then by using the trinomial recurrence relation \cite{VAR}
\begin{eqnarray*}
- Md_{M,M'}^J\left(\frac{\pi}{2}\right)
&=&
\frac{1}{2}
\sqrt{(J+M')(J-M'+1)}d_{M,M'-1}^J\left(\frac{\pi}{2}\right)+
\\ [2mm]
&+&
\frac{1}{2}\sqrt{(J-M')(J+M'+1)}d_{M,M'+1}^J
\left(\frac{\pi}{2}\right),
\end{eqnarray*}
we get
\begin{eqnarray}
\label{EL-OPERAT-12}
{\cal P}_{m',m} =
\frac{1}{4} \sqrt{(n-m'+2)(n+m')} \delta_{m, m'-2}
+
\frac{1}{4} \sqrt{(n+m'+2)(n-m')} \delta_{m, m'+2}.
\end{eqnarray}
From the latter equation one can conclude that ${\cal P}_{m,'-m} = {\cal P}_{-m,'m}$ and
${\cal P}_{-m,'-m} = {\cal P}_{m,'m}$ and, therefore, from (\ref{EL-OPERAT-8}) we determine
\begin{eqnarray}
\label{EL-OPERAT-13}
{\cal P}_{m',m}^{(\pm)} =  {\cal P}_{m',m} \pm P_{m',-m}.
\end{eqnarray}
Further
\begin{eqnarray*}
{\cal P}_{m',-m} =
\frac{1}{4} \sqrt{(n-m'+2)(n+m')}\delta_{-m, m'-2}
+
\frac{1}{4} \sqrt{(n+m'+2)(n-m')}\delta_{-m, m'+2}.
\end{eqnarray*}
As $m\geq 0$ and $m'\geq 0$, the second term is identically equal to
zero and
the first term is nonzero only at $(m,m')= (0,2); (1,1);(2,0)$. Hence
\begin{eqnarray}
\label{EL-OPERAT-14}
{\cal P}_{m',m}^{(\pm)} =  \frac{1}{4} \sqrt{(n-m'+2)(n+m')}
(\delta_{m0} \delta_{m'2} + \delta_{m1} \delta_{m'1} +
\delta_{m2} \delta_{m'0}).
\end{eqnarray}
With (\ref{EL-OPERAT-12})-(\ref{EL-OPERAT-14}) taken into account we
get
\begin{eqnarray*}
\sum_{p=0}^{n} \,
{\cal P}_{2p',2p}^{(+)} W_{2p}^{(+)}
&=&
\frac{1}{2}\sqrt{(n+p')(n-p'+1)}\,W_{2p'-2}^{(+)}
+ \frac{1}{2}\sqrt{(n+p'+1)(n-p')}\, W_{2p'+2}^{(+)}+
\\ [2mm]
&+&\frac{1}{2}\sqrt{n(n+1)}\left(W_{0}^{(+)}\delta_{p',1}
+ W_{2}^{(+)}\delta_{p',0}\right),
\\[4mm]
\sum_{p=0}^{n} \,
{\cal P}_{2p'+1,2p+1}^{(+)} W_{2p+1}^{(+)}
&=&
\frac{1}{2}\sqrt{(n+p'+1)(n-p'+1)}\, W_{2p'-2}^{(+)}
+\frac{1}{2}\sqrt{(n+p'+2)(n-p')}\, W_{2p'+3}^{(+)}+
\nonumber \\ [2mm]
&+& \frac{1}{2}(n+1)\, W_{1}^{(+)} \delta_{p',0},
\\[4mm]
\sum_{p=0}^{n} \,
{\cal P}_{2p'+1,2p+1}^{(-)} W_{2p+1}^{(-)}
&=&
\frac{1}{2}\sqrt{(n+p'+1)(n-p'+1)}\, W_{2p'-2}^{(-)}
+\frac{1}{2}\sqrt{(n+p'+2)(n-p')}\, W_{2p'+3}^{(-)}+
\\ [2mm]
&+& + \frac{1}{2}(n+1)\, W_{1}^{(-)} \delta_{p',0},
\\[4mm]
\sum_{p=0}^{n} \,
{\cal P}_{2p'+2,2p+2}^{(-)}W_{2p+2}^{(-)}
&=&
\frac{1}{2}\sqrt{(n+p'+1)(n-p'+2)}\, W_{2p'}^{(-)}
+\frac{1}{2}\sqrt{(n+p'+3)(n-p')}\, W_{2p'+4}^{(-)},
\end{eqnarray*}
whence instead of (\ref{CO-INT-17}) - (\ref{CO-INT-20}) we have
\begin{eqnarray}
\label{EL-OPERAT-15}
&&\sqrt{(n+p')(n-p'+1)}\left(1-\delta_{p',0}\right)W_{2p'-2}^{(+)}
+ \sqrt{(n+p'+1)(n-p')}W_{2p'+2}^{(+)}+
\nonumber \\ [2mm]
&+&\sqrt{n(n+1)}\left(W_0^{(+)}\delta_{p',1}+W_2^{(+)}
\delta_{p',0}\right) + \frac{4}{R^2}\left(\lambda^{(c,c)} +
4{p'}^2\right) W_{2p'}^{(+)} = 0,
\end{eqnarray}
\begin{eqnarray}
\label{EL-OPERAT-16}
&&\sqrt{(n+p'+1)(n-p'+1)}\left(1-\delta_{p',0}\right)W_{2p'-1}^{(+)}
+ \sqrt{(n+p'+2)(n-p')}W_{2p'+3}^{(+)}+
\nonumber \\ [2mm]
&+&(n+1)W_1^{(+)}\delta_{p',0} + \frac{4}{R^2}
\left[\lambda^{(s,c)} + (2p'+1)^2\right]W_{2p'+1}^{(+)} = 0,
\end{eqnarray}
\begin{eqnarray}
\label{EL-OPERAT-17}
&&\sqrt{(n+p'+1)(n-p'+1)}\left(1-\delta_{p',0}\right)
W_{2p'-1}^{(-)}
+ \sqrt{(n+p'+2)(n-p')}W_{2p'+3}^{(-)}-
\nonumber
\\ [2mm]
&-&(n+1)W_1^{(-)}\delta_{p',0} + \frac{4}{R^2}
\left[\lambda^{(c,s)}+(2p'+1)^2\right]\, W_{2p'+1}^{(-)} = 0,
\end{eqnarray}
\begin{eqnarray}
\label{EL-OPERAT-18}
&&\sqrt{(n+p'+2)(n-p'+1)}\left(1-\delta_{p',0}\right)
W_{2p'}^{(-)}
+ \sqrt{(n+p'+3)(n-p')}W_{2p'+4}^{(-)}+
\nonumber
\\ [2mm]
&+& \frac{4}{R^2}\left[\lambda^{(s,s)}
+ (2p'+2)^2\right]\, W_{2p'+2}^{(-)} = 0.
\end{eqnarray}
Recurrence relations (\ref{EL-OPERAT-15})-(\ref{EL-OPERAT-18}) form the basis
for constructing expansions of the elliptic basis over the polar one and should
be solved together with the normalization conditions
\begin{eqnarray*}
2\left|W_{0}^{(+)}\right|^2 +
\sum_{p=1}^n\,\left|W_{2p}^{(+)}\right|^2 = 1, \qquad
\sum_{p=0}^n\,\left|W_{2p+1}^{(+)}\right|^2 =
\sum_{p=0}^n\,\left|W_{2p+1}^{(-)}\right|^2 =
\sum_{p=0}^n\,\left|W_{2p+2}^{(-)}\right|^2 =1.
\end{eqnarray*}
The above-described method is also applicable for determination of expansions
of the elliptic subbases as a superposition of the Cartesian bases. We give
here only the final results. The recurrence relations for the coefficients $U$
in expansions (\ref{CO-INT-10}) - (\ref{CO-INT-13}) have the form
\begin{eqnarray*}
\sqrt{(k+1)(2k+1)(n-k)(2n-2k-1)}U_{2k+2}^{(+)} +
\sqrt{k(2k-1)(n-k+1)(2n-2k+1)}U_{2k-2}^{(+)}+
\\ [2mm]
+
\frac{1}{2}
\left[\lambda^{(с,c)}+8k(n-k)+2n+\frac{R^2}{2}(2k-n)\right]
U_{2k}^{(+)} = 0,
\end{eqnarray*}
\begin{eqnarray*}
\sqrt{(k+1)(2k+3)(n-k)(2n-2k-1)}U_{2k+3}^{(+)} +
\sqrt{k(2k+1)(n-k+1)(2n-2k+1)}U_{2k-1}^{(+)} +
\\ [2mm]
+
\frac{1}{2}\left[\lambda^{(s,c)}+4(n-k)(2k+1)+2n+1+
\frac{R^2}{4}(4k-2n+1)\right]
U_{2k+1}^{(+)} = 0,
\end{eqnarray*}
\begin{eqnarray*}
\sqrt{(k+1)(2k+1)(n-k)(2n-2k+1)}U_{2k+2}^{(-)} +
\sqrt{k(2k-1)(n-k+1)(2n-2k+3)}U_{2k-2}^{(-)}+
\\ [2mm]
+
\frac{1}{2}\left[\lambda^{(s,s)}+4k(2n-2k+1)+2n+1+
\frac{R^2}{4}(4k-2n-1)\right]
U_{2k}^{(-)} = 0,
\end{eqnarray*}
\begin{eqnarray*}
\sqrt{(k+1)(2k+3)(n-k)(2n-2k+1)}U_{2k+3}^{(-)} +
\sqrt{k(2k+1)(n-k+1)(2n-2k+3)}U_{2k-1}^{(-)}+
\\ [2mm]
+
\frac{1}{2}\left[\lambda^{(c,s)}+2(2k+1)(2n-2k+1)+2n+2+
\frac{R^2}{2}(2k-n)\right]
U_{2k+1}^{(-)} = 0.
\end{eqnarray*}
The following normalization conditions are valid:
\begin{eqnarray*}
\sum_{k=0}^n\,\left|U_{2k}^{(+)}\right|^2 =
\sum_{k=0}^n\,\left|U_{2k+1}^{(s,c)}\right|^2 =
\sum_{k=0}^n\,\left|U_{2k}^{(s,s)}\right|^2 =
\sum_{k=0}^n\,\left|U_{2k+1}^{(c,s)}\right|^2 =1.
\end{eqnarray*}
Investigation of the above-derived trinomial recurrence relations is connected
with solutions of higher-order algebraic equations and, obviously, in the
general case cannot be carried out analytically. At $R \ll 1$ and $R \gg 1$ the
properties of the elliptic integral of motion $\Lambda$ are mainly determined
by $L^2$ or ${\cal P}$ and there is a possibility to use perturbation theory.
At $R \ll 1$ the second term in the operator $\Lambda$ is taken as
perturbation. If we apply eigenvalues of the quantity $\lambda$ with index $q$
changing in the range $0 \leq q \leq n$, divide them into four groups -
$\lambda_{2q}^{(+)}$, $\lambda_{2q+1}^{(+)}$, $\lambda_{2q+1}^{(-)}$, and
$\lambda_{2q+2}^{(-)}$, and make use of the standard procedure of perturbation
theory \cite{LL} and explicit expressions for the matrix elements of the
operator ${\cal P}$ over the polar bases with given $P_y$- and $P_{xy}$ parity,
we arrive at the following results:
\begin{eqnarray*}
\lambda_{2q}^{(+)} &=& -(2q)^2 - \frac{R^4}{32}\frac{n^2+q^2+n} {4q^2-1} +
\frac{R^4}{64}n(n+1)\left(\delta_{q,0}-\delta_{q,1}\right),
\\ [2mm]
\lambda_{2q+1}^{(+)} &=& -(2q+1)^2 - \frac{R^2}{4}(n+1)\delta_{q,0}+
\\ [2mm]
&+&\frac{R^4}{128}\left[\frac{(n-q)(n+q+2)}{q+1} -
\frac{(n-q+1)(n+q+1)}{q}\left(1-\delta_{q,0}\right)\right],
\\ [2mm]
\lambda_{2q+1}^{(-)} &=& -(2q+1)^2 + \frac{R^2}{4}(n+1)\delta_{q,0}+
\\ [2mm]
&+&\frac{R^4}{128}\left[\frac{(n-q)(n+q+2)}{q+1} -
\frac{(n-q+1)(n+q+1)}{q}\left(1-\delta_{q,0}\right)\right],
\\ [2mm]
\lambda_{2q+2}^{(-)} &=& -(2q+2)^2 -
\frac{R^4}{32}\frac{n^2+q^2+2q-3n+3}{(2q+1)(2q+3)},
\end{eqnarray*}
\begin{eqnarray*}
&&\Psi_{2n,2q}^{(c,c)} =
\Psi_{2n,2q}^{(+)}(r,\varphi) +
\frac{R^2}{16}\Biggl[\frac{\sqrt{(n-q)(n+q+1)}}{2q+1}
\Psi_{2n,2q+2}^{(+)}(r,\varphi)-
\\ [2mm]
&-& \frac{\sqrt{(n-q+1)(n+q)}}{2q-1}
\Psi_{2n,2q-2}^{(+)}(r,\varphi) + \sqrt{n(n+1)}
\left(\delta_{q,0}\, \Psi_{2n,2}^{(+)}(r,\varphi)-
\delta_{q,1}\, \Psi_{2n,0}^{(+)}(r,\varphi)\right)\Biggr],
\\[2mm]
&&\Psi_{2n+1, 2q+1}^{(s,c)} = \Psi_{2n+1, 2q+1}^{(c,s)} =
\Psi_{2n+1,2q+1}^{(\pm)}(r,\varphi) +
\frac{R^2}{16}\Biggl[\frac{\sqrt{(n-q)(n+q+2)}}{q+1}
\Psi_{2n+1,2q+3}^{(\pm)}(r,\varphi)-
\\ [2mm]
&-& \frac{\sqrt{(n-q+1)(n+q+1)}}{q}
\Psi_{2n+1,2q-1}^{(+)}(r,\varphi)\left(1-\delta_{q,0}\right)
\Biggr],
\\[2mm]
&&\Psi_{2n+2, 2q+2}^{(s,s)} =
\Psi_{2n+2,2q+2}^{(-)}(r,\varphi) +
\frac{R^2}{16}\Biggl[\frac{\sqrt{(n-q)(n+q+3)}}{2q+3}
\Psi_{2n+2,2q+4}^{(-)}(r,\varphi)-
\\ [2mm]
&-& \frac{\sqrt{(n-q+1)(n+q+2)}}{2q+1}
\Psi_{2n+2,2q}^{(-)}(r,\varphi)\Biggr].
\end{eqnarray*}
In the above formulae the factor $(1-\delta_{q,0})/q$ is by definition put to
zero at $q=0$.

At $R \gg 1$ by dividing both sides of equation (\ref{EL-OPERAT-4}) by $R^2/2$
and considering the term $2L^2/R^2$ to be perturbation we get
\begin{eqnarray*}
\lambda_{2q}^{(+)} &=& -\frac{R^2}{2}(q-n) - 2q(2n-q)-2n,
\\ [2mm]
\lambda_{2q+1}^{(+)} &=& \lambda_{2q+1}^{(-)} =
-\frac{R^2}{4}(2q-2n-1)- \left[2q(2n-2q+1) +2n+1\right],
\\ [2mm]
\lambda_{2q+2}^{(-)} &=& -\frac{R^2}{2}(q-n-1) -
2\left[q(2n-q+2)+n+1\right],
\end{eqnarray*}
\begin{eqnarray*}
&&\Psi_{2n,2q}^{(c,c)} = \Psi_{2q,2n-2q}^{(+)}(x,y) +
\frac{2}{R^2}\Biggl[\sqrt{q(2q-1)(n-q+1)(2n-2q+1)}\times
\\ [2mm]
&\times& \Psi_{2q-2,2n-2q+2}^{(+)}(x,y)
- \sqrt{(q+1)(2q+1)(n-q)(2n-2q-1)}
\Psi_{2q+2,2n-2q-2}^{(+)}(x,y)\Biggr],
\\[2mm]
&&\Psi_{2n+1,2q+1}^{(s,c)} =
\Psi_{2q+1,2n-2q}^{(-+)}(x,y) +
\frac{2}{R^2}\Biggl[\sqrt{q(2q+1)(n-q+1)(2n-2q+1)}\times
\\ [2mm]
&\times&\Psi_{2q-1,2n-2q+2}^{(+)}(x,y)
- \sqrt{(q+1)(2q+3)(n-q)(2n-2q-1)}
\Psi_{2q+3,2n-2q-2}^{(+)}(x,y)\Biggr],
\\[2mm]
&&\Psi_{2n+1,2q+1}^{(c,s)} =
\Psi_{2q+1,2n-2q+1}^{(+)}(x,y) +
\frac{2}{R^2}\Biggl[\sqrt{q(2q+1)(n-q+1)(2n-2q+3)}\times
\\ [2mm]
&\times&\Psi_{2q-1,2n-2q+3}^{(+)}(x,y)
- \sqrt{(q+1)(2q+3)(n-q)(2n-2q+1)}
\Psi_{2q+3,2n-2q-1}^{(+)}(x,y)\Biggr],
\\[2mm]
&&\Psi_{2n+2,2q+2}^{(s,s)} =
\Psi_{2q+1,2n-2q+1}^{(-)}(x,y) +
\frac{2}{R^2}\Biggl[\sqrt{q(2q-1)(n-q+1)(2n-2q+3)}\times
\\ [2mm]
&\times&\Psi_{2q-2,2n-2q+3}^{(-)}(x,y)
- \sqrt{(q+1)(2q+1)(n-q)(2n-2q+1)}
\Psi_{2q+2,2n-2q-1}^{(-)}(x,y)\Biggr.
\end{eqnarray*}
We have derived these expressions using the limiting relations (see Sect.2)
\begin{eqnarray*}
\lim_{R \to \infty} \frac{2\lambda_{2q}^{(+)}}{R^2} &=& - (q-n),
\qquad
\lim_{R \to \infty} \frac{2\lambda_{2q+2}^{(-)}}{R^2} = - (q-n-1),
\\ [2mm]
\lim_{R \to \infty} \frac{2\lambda_{2q+1}^{(+)}}{R^2} &=&
\lim_{R \to \infty} \frac{2\lambda_{2q+1}^{(-)}}{R^2} =
-(q-n-1/2).
\end{eqnarray*}
Higher perturbation orders can be calculated in an analogous way.

\section{Expansion in the bases describing a motion of a non-relativistic
charged particle in the homogeneous magnetic field}
\markboth{CHAPTER 3. CIRCULAR OSCILLATOR}{3.5. A
CHARGED PARTICLE IN THE HOMOGENEOUS MAGNETIC FIELD}

The behavior of a non-relativistic charged spinless particle in the constant and
homogeneous magnetic field is determined by the equation
\bea
\frac{1}{2\mu}\left({\hat\textbf{p}}
- \frac{e}{c}\textbf{A}\right)^2\Psi = E\Psi,
\label{M.1}
\eea
in which the vector potential has the form
\bea
\textbf{A}\left(\alpha; \textbf{r}\right) = \left\{-\alpha y,
(H-\alpha)x, 0\right\},
\label{M.2}
\eea
if the magnetic field is directed along the axis $z$, i.e. $\textbf{H}= (0, 0,
H )$. The parameter $\alpha$ specifies gauge and runs real values. Substitution
of (\ref{M.2}) into (\ref{M.1}) and the study of the resultant equation lead to
the conclusion that at arbitrary $\alpha$ variables in it are not separated in
any of the known two-dimensional orthogonal systems of coordinates. The
exclusion is represented by two cases: a) $\alpha = H$; \, b) $\alpha = H/2$.
In the first, the separation of variables is possible in the Cartesian
coordinates; and in the second, in the polar ones. The analysis of
equation(\ref{M.1}) for $\alpha = H$ was carried out by Landau \cite{LL}, who
showed that the Cartesian basis has the form
\bea
\Psi_{n y_0}(x, y) = \frac{1}{a \sqrt{2 \pi}} e^{ix
\frac{y_0}{a^2}} \,
\overline{H}_n \left(\frac{y-y_0}{a}\right),
\label{M.3}
\eea
where
\begin{eqnarray*}
y_0 = \frac{a^2}{\hbar}p_x,
\qquad
a = \sqrt{\frac{\hbar}{\mu\omega_H}},
\qquad \omega_H = \frac{|e|H}{\mu c},
\end{eqnarray*}
and $\overline{H}_n$ are the Hermitian functions normalized to unity that are
expressed in terms of the Hermite polynomials ${H}_n(x)$ by the formula
\begin{eqnarray*}
\overline{H}_n(z) = \frac{1}{\sqrt{2^n n!}}\,e^{-z^2}\,H_n(z).
\end{eqnarray*}

A motion of a particle that is in the stationary state (\ref{M.3}) is infinite
along the axis $x$ and finite along the axis $y$. The discrete energy spectrum
generated by finiteness of a motion along the axis $y$ has the form
\bea
E_n = \hbar \omega_H \left(n+\frac{1}{2}\right),
\label{M.4}
\eea
where $n=0,1,2,\dots$.

The case b) was systematically studied by Jonson and Lippmann
\cite{Lippmann}\footnote{References to earlier papers are given in \cite{Lippmann}.}.
According to \cite{LL}, the polar basis is determined by the expression
\bea
\psi_{N m}(r, \varphi) = \frac{(-i)^m}{\sqrt{2 \pi}}
\frac{(-1)^N}{a^{|m|+1}} \sqrt{\frac{(N+|m|)!}{2^{|m|} N!}}
\frac{r^{|m|}}{|m|!}\, e^{-\frac{r^2}{4a^2}} \,
F\left(-N, |m|+1, \frac{r^2}{2 a^2}\right)\,
e^{im\varphi},
\label{M.5}
\eea
in which $N=0,1,2,\dots$, and the phase factor $(-i)^m(-1)^N$ is added for
convenience. In terms of polar quantum numbers the spectrum (\ref{M.4}) has the
form\footnote{States (\ref{M.4}) describe a motion that is finite in the plane
$(x,y)$.}
\bea
E_N= \hbar \omega_H \left(N + \frac{m+|m|}{2} + \frac{1}{2}\right).
\label{M.6}
\eea
One can easily show (see, for example, \cite{KOGAN}) that the quantity $m$ at a
given principal quantum number $n$ in the case of a negatively charged particle
takes integer values from the domain $(-\infty, n]$. We will need this
information below when deriving interbasis expansions.

The bases (\ref{M.3}) and (\ref{M.5})are normalized as follows:
\bea
\int_{-\infty}^{\infty} \int_{-\infty}^{\infty}
{\psi_{n' y_0'}}^*(x, y) \psi _{n y_0}(x, y)dx dy &=& \delta_{n n'}
\delta \left(\frac{y_0'-y_0}{a}\right),
\label{M.7}
\\[2mm]
\int_0^{\infty} \int_0^{2\pi} {\psi_{N' m'}}^* (r, \varphi)
\psi_{N m}(r, \varphi) r dr d\varphi &=& \delta_{N N'}\delta_{m m'}.
\label{M.8}
\eea

The problem of interbasis expansions has the following specific feature in the
case of a magnetic field. Bases (\ref{M.3}) and (\ref{M.5}) are specified by
different Hamiltonians the form of which depends on a concrete choice of gauge
for the vector potential though these bases describe the same physical system.
As a result, the expansion of the Cartesian basis (\ref{M.3}) over the polar
one (\ref{M.5}) should have the summation over $N$ in addition to that over
$m$. This complication can be avoided by using gauge invariance. Under the
gauge transformation
\bea
\textbf{A} \to {\textbf{A}'} = \textbf{A} + \nabla f(x,y,z)
\label{M.9}
\eea
equation (\ref{M.1}) does not change if simultaneously with the change of
(\ref{M.9}) in the Hamiltonian one also substitutes the wave function,
according to
\bea
\Psi \to \Psi' \exp\left(\frac{ie}{\hbar c}f\right).
\label{M.10}
\eea
Passing from the Cartesian gauge $\textbf{A} = (-yH,0,0)$ to the polar one
$\textbf{A}=(-\frac{yH}{2}$, $\frac{xH}{2},0)$ corresponds to transformation
with $f=-\frac{xy}{2}H$. All this considered, for the Cartesian basis in the
polar gauge we have
\bea
\Psi'_{ny_0}(x,y) = \frac{1}{a\sqrt{2\pi}}\,
e^{-ix\frac{y-2y_0}{2a^2}}\, \overline{H}_n \,
\left(\frac{y-y_0}{2}\right).
\label{M.11}
\eea
Expansion of the basis (\ref{M.11}) over the polar one (\ref{M.5}) contains
only the summation over $m$ and we write it as
\bea
\Psi_{n y_0}(x, y) = \sum _{m=-{\infty}}^n C_{nm}(y_0) \Psi_{n-
\frac{m+|m|}{2}, m}(r, \varphi).
\label{M.12}
\eea
Both the bases in expansion (\ref{M.12}) correspond to the same energy value of
a transverse motion and, therefore, $N=n-\frac{|m|+m}{2}$. According to the
orthonormalization condition (\ref{M.8}), the expansion coefficients
(\ref{M.12}) can be written as an integral
\bea
C_{m n} (y_0) = \int {\Psi^*_{n-\frac{m+|m|}{2},m}}(r, \varphi)
\Psi_{n y_0}(x, y) dx dy.
\label{M.13}
\eea
The main difficulty in calculating this integral consists in that
the integrand is factorized neither in the Cartesian nor polar
coordinates. To overcome this difficulty, we should like to note
that the wave function satisfies the Schr\"{o}dinger equation
(\ref{M.5})
\bea
\Delta \Psi + \frac{2\mu}{\hbar^2} \left(\varepsilon - \frac{\mu \omega_H^2
r^2}{8}\right) \Psi=0,
\label{M.14}
\eea
if energy $\varepsilon$ equals
\begin{eqnarray*}
\varepsilon = \frac{\hbar \omega_H}{2} \left(2N+ |m|+1\right).
\end{eqnarray*}
Hence it follows that the polar basis (\ref{M.3}) may be interpreted as a wave
function of the circular oscillator with mass $\mu$ and frequency
${\omega_H}/{2}$. This conclusion allows us to make use of the above-derived
expansion for the polar basis of the circular oscillator over the Cartesian
one, i.e., the relation inverse to (\ref{2D-OSC-8}). The Cartesian basis
involved in this sort of expansion has the form
\begin{eqnarray*}
\Psi_{n_1 n_2}(x, y) = \frac{(-i)^{n_2}}{a \sqrt 2} \,
\overline{H}_{n_1}\left(\frac{x}{a \sqrt2}\right)\,
\overline{H}_{n_2}\left(\frac{y}{a \sqrt 2}\right),
\end{eqnarray*}
and its relevant energy level is equal to
\begin{eqnarray*}
\varepsilon = \frac{\hbar \omega_H}{2} (n_1+n_2+1).
\end{eqnarray*}
Substituting the above-mentioned expansion in integral (\ref{M.13}) and taking
formula \cite{GRAD} into account
\begin{eqnarray*}
\int_{-\infty}^{\infty} e^{i\alpha x} \overline{H}_n (x) dx
= (i)^n \sqrt{2\pi}\, \overline{H}_n (\alpha)
\end{eqnarray*}
we arrive at the relation
\bea
C_{nm}(y_0) = (-1)^{n-m} \sum_{k=0}^{2n-m}\, (-1)^k  \,
d_{\frac{m}{2}, n-\frac{m}{2}-k}^{n-\frac{m}{2}}
\left(\frac{\pi}{2}\right)\, R_{nm}^k,
\label{M.15}
\eea
in which the following notation is used:
\bea
R_{nm}^k = \int_{-\infty}^{\infty} dy \overline{H}_n
\left(\frac{y-y_0}{a}\right) \overline{H}_k
\left(\frac{y_0}{a\sqrt 2}\right) \overline{H}_{2n-m-k}
\left(\frac{y-2y_0}{a\sqrt 2}\right).
\label{M.16}
\eea
To calculate this integral, we apply theorem (\ref{ERMIT-1}) to the product
$\overline{H}_k \overline{H}_{2n-m-k}$ and use the orthonormality of these
polynomials
\bea
R_{nm}^k = (-1)^{n_m} \bar H_{n-m}\left(\frac{y_0}{a}\right)
d_{-\frac{m}{2}, n - \frac{m}{2}-k}^{n-\frac{m}{2}}
\left(\frac{\pi}{2}\right).
\label{M.17}
\eea
Substituting this expression into (\ref{M.15}) and taking into account the
symmetry property \cite{VAR}
\begin{eqnarray}
d_{M, M'}^j (\beta)= d_{-M, -M'}^j (\beta),
\label{M.18}
\end{eqnarray}
as well as substituting the summation index in (\ref{M.15}) by
$r=n-\frac{m}{2}-k$ and using the identity \cite{VAR}
\bea
\sum_{M''=-j}^j (-1)^{M''-M'} d_{M,M''}^j (\beta)d_{M'',M'}^j (\beta)
= \delta_{M M'}
\label{M.19}
\eea
we calculate the coefficient $C_{nm}(y_0)$ and prove that expansion
(\ref{M.12}) has the form
\bea
\Psi'_{n y_0}(x, y) = \sum_{m = -\infty}^n
\overline{H}_{n-m}\left(\frac{y_0}{a}\right)
\Psi_{n-\frac{m+|m|}{2}, m}(r, \varphi).
\label{M.20}
\eea
Hence we immediately derive the inverse expansion
\bea
\Psi_{N m}(r, \varphi)  = \int\overline{H}_{N-\frac{m-|m|}{2}}
\left(\frac{y_0}{a}\right)
\Psi'_{N-\frac{m+|m|}{2}, y_0}(x, y)dy_0.
\label{M.21}
\eea
In the relativistic case, the Klein-Gordon equation for a constant and
homogeneous magnetic field is also solved within the method of separation of
variables only in the Cartesian and polar coordinates. The corresponding bases
exactly coincide with their nonrelativistic partners. The only difference
arises in the energy spectrum (of course, as $c \to \infty$ it turns into a
nonrelativistic one). Nothing new, as concerns bases, can be obtained from the
Dirac equation in which the variables in the problem of homogeneous constant
magnetic field are also separated in the Cartesian and polar coordinates.

\begin{table}[h!]
\caption
{ Elliptic basis of the circular oscillator (\ref{ECO.16a})
for $N=0,2,4$.}\label{t31}
\begin{center}
\begin{tabular}{|c|c|c|}
\hline
&&\\$N$&$\Psi^{(c,c)}(\xi,\eta;R^2)$&$A^{'}$
\\
&&\\
\hline
&&\\
$0$& $\frac{1}{\sqrt \pi}e^{-\frac{R^2}{16}(\cosh 2\xi+\cos 2\eta)}$&
$A^{'}=0$\\
&&\\
\hline
&&\\
$2$&$-\frac{1}{2\pi}\left\{\frac{A'+4}{A'+2}\right\}
e^{-\frac{R^2}{16}(\cosh2\xi+\cos2\eta)}$&
$A^{'}(A^{'}+4)=\frac{R^4}{4}$\\
&&\\
&$\left\{1+\frac{1}{2}(A'-\frac{R^2}{2})\cos^2\eta\right\}
\left\{1-\frac{1}{2}(A'+\frac{R^2}{2})\sinh^2\xi\right\}$&\\
&&\\
\hline
&&\\
$4$&
$\frac{1}{\sqrt\pi}
\left\{1+\frac{4}{3}\left(\frac{A^{'}}{R}\right)^2 +
\frac{1}{3}\left(\frac{A^{'}}{A^{'}+16}\right)^2\right\}$&
$A^{'}(A^{'}+4)(A'+16)$\\
&&\\
&$\left\{1+\frac{1}{2}(A'-R^2)\cos^2\eta+\frac{A}{24}
(A+4-R^2)\cos^{4}\eta\right\}$&
$=R^4(A'+12)$\\
&&\\
&$\left\{1-\frac{1}{2}(A'+R^2)\sinh^2\xi+\frac{A^{'}}{24}
(A^{'}+4+R^2)\sinh^4\xi\right\}$&\\
&&\\
\hline
\end{tabular}
\end{center}
\end{table}

\begin{table}[h!]
\caption{ Elliptic basis of the circular oscillator (\ref{ECO.16b})
for $N=1,3,5$.}
\label{t32}

\begin{center}
\begin{tabular}{|c|c|c|}
\hline
&&\\
$N$&$\Psi^{(s,c)}(\xi,\eta;R^2) $&$ A^{'}$\\
&&\\
\hline
&&\\
$1$& $\frac{R}{\sqrt {2\pi}}\cos\eta \cosh\xi \,
e^{- \frac{R^2}{16}(\cosh 2\xi+\cos 2\eta)}$& $A^{'}=-1-\frac{R^2}{4}$\\
&&\\
\hline
&&\\
$3$&$\frac{4}{3\pi}\frac{A'+9}{R}\left\{1+\frac{16}{3}
\left(\frac{A'+9}{R^2}\right)^2\right\}^{-\frac{1}{2}}
\cosh \xi\cos\eta \, e^{-\frac{R^2}{16}(\cosh2\xi+\cos2\eta)}$&
$(A^{'}+1)(A^{'}+9)=$
\\
&&\\
&$\left\{1+\frac{1}{6}(A'+1-\frac{R^2}{4})\right\}
\left\{1-\frac{1}{2}(A'+1+\frac{3R^2}{4})\sinh^2\xi\right\}$&
$-\frac{R^2}{2}(A'+9)+\frac{3R^4}{16}$\\
&&\\
\hline
&&\\
$5$&$\sqrt{\frac{3}{2\pi}}R
\left\{1+2\left(\frac{A'+1}{R^2}+\frac{3}{4}\right)^2+
\frac{5}{8}\left(\frac{R^2}{A^{(-)'}+25}\right)^2
\left(\frac{A'+1}{R^2}+\frac{3}{4}\right)^2\right\}^{-\frac{1}{2}}$&
$(A'+1)(A'+9)(A'+25)$\\
&&\\
&$\times \cos\eta \cosh\xi \,
e^{-\frac{R^2}{16}(ch 2\xi+\cos 2\eta)}
\biggl\{1+\frac{1}{6}\left(A'+1-\frac{3R^2}{4}\right)\cos^2\eta$&
$=-\frac{3R^2}{4}(A'+9)(A'+25)$\\
&&\\
&$+\frac{1}{120}\left[\left(A'+1-\frac{3R^2}{4}\right)
\left(A'+9+\frac{R^2}{4}\right)+ 12R^2\right]\cos^4\eta\biggr\}$
&$+\frac{R^4}{16}(13A'+205)+\frac{15R^6}{64}$\\
&&\\
&$\times\biggl\{1-\frac{1}{2}\left(A'+1+\frac{5R^2}{4}\right)
\sinh^2\xi$&\\
&& \\
&$+\frac{1}{24}\left[\left(A'+1+\frac{5R^2}{4}\right)
\left(A'+9+\frac{R^2}{4}\right)-4R^2\right]\sinh^2\xi\biggr\}$&\\
&&\\
\hline
\end{tabular}
\end{center}
\end{table}

\begin{table}[h!]
\caption{ Elliptic basis of the circular oscillator (\ref{ECO.16c})
for $N=1,3,5$.}
\label{t33}

\begin{center}
\begin{tabular}{|c|c|c|}\hline
&&\\$N$&$\Psi^{(c,s)}(\xi,\eta;R^2)$&$A^{'}$
\\
&&\\ \hline
&&\\
$1$&$i\frac{R}{\sqrt {2\pi}}\sin\eta \sinh\xi
e^{-\frac{R^2}{16}(\cosh 2\xi+\cos 2\eta)}$&
$A^{'}=-1-\frac{R^2}{4}$\\
&&\\
\hline
&&\\
$3$&$-i\frac{R}{\sqrt\pi}\left\{1+\frac{3}{16}
\left(\frac{R^2}{A'+9}\right)^2\right\}^{-\frac{1}{2}}
\sin\eta \cosh\xi e^{-\frac{R^2}{16}(ch2\xi+\cos2\eta)}$&
$(A^{'}+1)(A^{'}+9)=$\\
&&\\
&$\left\{1+\frac{1}{2}(A'+1-\frac{3R^2}{4})\cos^2\eta\right\}
\left\{1-\frac{1}{6}(A'+1+\frac{R^2}{4})\sinh^2\xi\right\}$&
$\frac{R^2}{2}(A'+9)+\frac{3R^4}{16}$ \\
&&\\
\hline
&&\\
$5$&
$i\sqrt{\frac{3}{2\pi}}R
\left\{1+2\left(\frac{A'+1}{R^2}-\frac{3}{4}\right)^2+
\frac{5}{2}\left(\frac{R^2}{A^{(-)'}+25}\right)^2
\left(\frac{A'+1}{R^2}-\frac{3}{4}\right)^2\right\}^{-\frac{1}{2}}
$&
$(A'+1)(A'+9)(A'+25)$\\
&&\\
&$\sin\eta \sinh\xi\, e^{-\frac{R^2}{16}(\cosh 2\xi+\cos 2\eta)}\,
\biggl\{1+\frac{1}{2}\left(A'+1-\frac{5R^2}{4}\right)\cos^2\eta$&
$=\frac{3R^2}{4}(A'+9)(A'+25)+$\\
&&\\
&$+\frac{1}{24}\left[\left(A'+1-\frac{5R^2}{4}\right)
\left(A'+9-\frac{R^2}{4}\right)+4R^2\right]
\cos^4\eta\biggr\}$&$\frac{R^4}{16}(13A'+205)-$\\
&&\\
&$\biggl\{1-\frac{1}{6}\left(A'+1+\frac{3R^2}{4}\right)
\sinh^2\xi+\frac{1}{120}\biggl[\left(A'+1-\frac{3R^2}{4}\right)$
&$-\frac{15R^6}{64}$\\
&&\\
&$\left(A'+9-\frac{R^2}{4}\right)-12R^2\biggr]\sinh^4\xi\biggr\}$&\\
&&\\
\hline
\end{tabular}
\end{center}
\end{table}

\begin{table}[h!]
\caption{Elliptic basis of the circular oscillator (\ref{ECO.16d})
for $N=2,4,6$.}
\label{t34}

\begin{center}
\begin{tabular}{|c|c|c|}\hline
&&\\
$N$&$\Psi^{(s,s)}(\xi,\eta;R^2)$&$A^{'}$\\
&&\\
\hline
&&\\
$2$&$\frac{i}{8}\sqrt{\frac{2}{\pi}}R^2\sin 2\eta \sinh2\xi
e^{-\frac{R^2}{16}(\cosh2\xi+\cos 2\eta)}$&
$A'=-4$\\
&&\\
\hline
&&\\
$4$&$-i\frac{6}{\sqrt\pi}\frac{R}{16}\left\{\frac{A'+16}{A'+10}
\right\}^{\frac{1}{2}}
\sin 2\eta \sinh 2\xi e^{-\frac{R^2}{16}(\cosh2\xi+\cos2\eta)}$&
$(A^{'}+4)(A^{'}+16)$\\
&&\\
&$\left\{1+\frac{1}{6}\left(A'+4-\frac{R^2}{2}\right)\cos^2\eta\right\}
\left\{1-\frac{1}{6}(A'+4+\frac{R^2}{2})sh^2\xi\right\}$&
$=\frac{R^4}{4}$ \\
&&\\
\hline
&&\\
$6$&
$i\sqrt{\frac{3}{2\pi}}\frac{R}{4}
\left\{1+\frac{8}{5}\left(\frac{A'+4}{R}\right)^2+
\frac{3}{5}\left(\frac{A'+4}{A'+36}\right)^2\right\}^
{-\frac{1}{2}}\sin 2\eta \sinh2\xi$&
$(A'+4)(A'+16)(A'+36)$\\
&&\\
&$e^{-\frac{R^2}{16}(\cosh 2\xi+\cos 2\eta)}
\biggl\{1+\frac{1}{6}\left(A'+4-R^2\right)\cos^2\eta$
&$=R^4(A'+24)$\\
&&\\
&$+ \frac{A'+4}{120}(A'+16-R^2)\cos^4\eta\biggr\}
\biggl\{1-\frac{1}{6}\left(A'+4+R\right)\sinh^2\xi$&\\
&&\\
&$+\frac{A'+4}{120}\left(A'+16+R^2\right)\sinh^4\xi\biggr\}$&\\
\hline
\end{tabular}
\end{center}
\end{table}

\newpage

\chapter{Expansions in the multidimensional isotropic oscillator}
\markboth{CHAPTER 4. MULTIDIMENSIONAL ISOTROPIC OSCILLATOR}{}

This chapter is devoted to the consideration of interbasis expansions in the
multidimensional isotropic oscillator. Here, relation of hidden symmetry with
separation of variables has a twofold manifestation. First, variables are
separated in all so-called hyperspherical systems of coordinates \cite{VKS}.
Second, they are separated in all Cartesian coordinates related with each other
by the rotation transformation. That is the reason that we can speak of three
types of interbasis expansions: a hyperspherical basis over another
hyperspherical basis, a Cartesian basis over a hyperspherical one, and a
Cartesian basis over another Cartesian one. Interbasis expansions of the first
type are related to pure geometric symmetry of the isotropic oscillator and are
independent of a concrete form of the centrosymmetrical potential. This type of
interbasis expansions was calculated in \cite{KIL}. Expansions of the Cartesian
basis over hyperspherical ones and between the Cartesian bases connected with
each other by an arbitrary rotation were calculated, respectively, in
\cite{PSMT,POGOSYAN2}. From the mathematical standpoint, the
knowledge of matrices realizing the last two transitions opens up the way for
elucidation of transformation properties of hyperspherical harmonics at finite
rotations.

This chapter is organized as follows. The first section is the derivation of
the general integral representation of the transition matrix from the Cartesian
basis to an arbitrary hyperspherical one in terms of a hyperspherical harmonics
related to a given tree. In the second section, contributions of individual
cells to the general transition matrix are calculated. In the third section,
the diagrammatic method of constructing a transition matrix is developed. The
fourth section is devoted to the calculation of a manifest dependence of the
Wigner oscillator function on the Euler angle. In the fifth section, an
illustrative diagrammatic interpretation is given to the results obtained.

\section{Expansion of the Cartesian oscillator basis over hyperspherical ones}
\markboth{CHAPTER 4. MULTIDIMENSIONAL ISOTROPIC OSCILLATOR}{4.1. EXPANSION OF THE CARTESIAN BASIS OVER HYPERSPHERICAL}

Consider p-dimensional isotropic oscillator with mass $m$ and cyclic frequency
$\omega$. Denote the wave functions of this type of oscillator in the Cartesian
basis by $\Psi_{\vec n}(\vec x)$, where $\vec x= (x_1 x_2....x_p)$, $\vec
n=(n_1 n_2... n_p)$. It is obvious that $\Psi_{\vec n}(\vec x)$ equals the
product of the wave functions $p$ of linear oscillators. In an arbitrary
hyperspherical basis
\begin{eqnarray*}
\Psi_{n, \vec l}(r, \vec \theta)
= R_{nl_1}(r)Y_{\vec l}(\vec\theta)\,,
\end{eqnarray*}
where $N=n_1+n_2+....+n_p$, $\vec\theta=(\theta_1,.....,\theta_{p-1})$, $\vec
l=(l_1 l_2.....l_{p-1})$, $l_1$ is the global moment, and $l_2,\dots,l_{p-1}$
are hypermoments. The radial wave function has the form
\begin{eqnarray}
\label{NDIM-OSC-1}
R_{Nl_1}(r)=
\left\{\frac{2\alpha\Gamma\left(\frac{N+l_1+p}{2}\right)}
{\Gamma\left(\frac{N-l_1}{2}+1\right)}\right\}^{\frac{1}{2}}
\frac{(\alpha r)^{l_1}}{\Gamma\left(l_1+\frac{p}{2}\right)}
e^{-\frac{\alpha^2 r^2}{2}} F\left(-\frac{N-l_1}{2};
l_1+\frac{p}{2};\alpha^2 r^2 \right)\,,
\end{eqnarray}
where $\alpha= \sqrt{m\omega/\hbar}$.

Let us represent the sought expansion of the Cartesian oscillator basis over
the hyperspherical one by the following equality:
\bea
\Psi_{\vec n}(\vec x) =
\sum_{\vec l} W_{\vec n}^{N\vec l}
\Psi_{N\vec l}(r, \vec\theta)\,.
\label{4.1O}
\eea
The variation limits of $\vec l$ are determined by the condition $N= const$ and
the structure of a hyperspherical tree. Expansion (\ref{4.1O}) undergoes the
following operations: multiply it by $r^{-N}$ , tend all $x_i$ to infinity,
then multiply the derived limit equality by $Y_{\vec l'}^*(\vec \theta)$,
integrate over the solid angle, and use the orthonormalization condition of
hyperspherical harmonics
\begin{eqnarray*}
\int Y_{\vec l'}^*(\vec \theta) Y_{\vec l}(\vec \theta)d\Omega
= \delta_{\vec l\vec l'}\,.
\end{eqnarray*}
This results in the following formula that is common for all hyperspherical
systems of coordinates:
\bea
W_{\vec n}^{N\vec l} =  K \int d\Omega Y_{\vec l}^* (\vec \theta) \,
\prod_{i=1}^p (\hat x_i)^{n_i}\,,
\label{4.2O}
\eea
in which we use the notation $\hat x_i = {x_i}/{r}$ and the multiplier  ${\cal
K}$ is equal to
\bea
{\cal K} = \frac{(-1)^{\frac{N-l_1}{2}}}{\sqrt 2
\pi^{\frac{p}{4}}} \left\{\frac{2^N
\Gamma\left(\frac{N-l_1}{2}+1\right)
\Gamma\left(\frac{N+l_1+p}{2}\right)}{(n_1)!(n_2)!.....(n_p)!}
\right\}\,.
\label{4.3O}
\eea
To calculate the transition matrix (\ref{4.2O}), one needs an explicit form of
the functions $Y_{\vec l}(\vec \theta)$ and $\hat x_i (\vec\theta)$, and also
the formula expressing an element of the solid angle $d\Omega$ as the product
of differentials $d\theta_1..........d\theta_{p-1}$.

According to \cite{VKS}, it is convenient to put in correspondence to any
hyperspherical system of coordinates and its relevant hyperspherical harmonics
a certain graph called the hyperspherical tree. In the $ð$-dimensional space
every tree has  $ð$ vertices and each of these vertices is connected with its
cell. In the general case cells can be of four types. The contribution of each
cell to the functions $Y_{\vec l}(\vec\theta)$, $\hat x_{i}(\vec\theta)$ and
$d\Omega(\vec \theta)$ is given in Table \ref{t41}.

\begin{table}[t]
\caption{Contribution of each type of cells to the functions $Y_{\vec
l}(\vec\theta)$, $\hat x_{i}(\vec\theta)$ and  $d\Omega(\vec \theta)$.}
\label{t41}
\begin{center}
\begin{tabular}{|c|c|c|c|c|}\hline
&
\unitlength=1.50mm \special{em:linewidth 0.4pt}\linethickness{0.4pt}
\begin{picture}(20.00,20.00)
\put(10.00,00.00){\line(0,1){5.00}}
\put(10.00,5.00){\line(1,1){10.00}}
\put(10.00,5.00){\line(-1,1){10.00}}
\put(10.00,5.00){\circle*{1}}
\put(12.00,3.00){$l$}
\put(9.50,7.00){$\theta$}
\end{picture}
&
\unitlength=1.50mm \special{em:linewidth 0.4pt}\linethickness{0.4pt}
\begin{picture}(20.00,20.00)
\put(10.00,00.00){\line(0,1){5.00}}
\put(10.00,5.00){\line(1,1){10.00}}
\put(10.00,5.00){\line(-1,1){10.00}}
\put(10.00,5.00){\circle*{1}}
\put(12.00,3.00){$l$}
\put(9.50,7.00){$\theta$}
\put(13.00,15.00){$l_s,\nu_s$}
\put(20.00,15.00){\circle*{1}}
\end{picture}
&
\unitlength=1.50mm \special{em:linewidth 0.4pt}\linethickness{0.4pt}
\begin{picture}(20.00,20.00)
\put(10.00,00.00){\line(0,1){5.00}}
\put(10.00,5.00){\line(1,1){10.00}}
\put(10.00,5.00){\line(-1,1){10.00}}
\put(10.00,5.00){\circle*{1}}
\put(12.00,3.00){$l$}
\put(9.50,7.00){$\theta$}
\put(1.00,15.00){$l_c,\nu_c$}
\put(0.00,15.00){\circle*{1}}
\end{picture}
&
\unitlength=1.50mm \special{em:linewidth 0.4pt}\linethickness{0.4pt}
\begin{picture}(20.00,20.00)
\put(10.00,00.00){\line(0,1){5.00}}
\put(10.00,5.00){\line(1,1){10.00}}
\put(10.00,5.00){\line(-1,1){10.00}}
\put(10.00,5.00){\circle*{1}}
\put(12.00,3.00){$l$}
\put(9.50,7.00){$\theta$}
\put(1.00,15.00){$l_c,\nu_c$}
\put(0.00,15.00){\circle*{1}}
\put(13.00,15.00){$l_s,\nu_s$}
\put(20.00,15.00){\circle*{1}}
\end{picture}
\\
\hline
&&&&\\
$\theta$&$ 0\leq\theta\leq 2\pi$&$0\leq\theta\leq\pi$
&$-\frac{\pi}{2} \leq\theta\leq\frac{\pi}{2}$
&$0\leq\theta\leq\frac{\pi}{2}$\\
&&&&\\
\hline
&&&&\\
$d\Omega$&$d\theta$&$
(\sin\theta)^{v_s}d\theta$&$(\cos\theta)^{v_c} d\theta$&$
(\sin\theta)^{v_s}(\cos\theta)^{v_c}d\theta$\\
&&&&\\
\hline
&&&&\\
$Y_{\vec l}(\vec \theta)$&$\frac{e^{il\theta}}{\sqrt {2\pi}}$&$
f_{l_s,v_s}^{s;l}(\theta)$&$f_{l_c,v_c}^{c;l}(\theta)$
&$f_{l_c,v_c,l_s,v_s}^{c,v;l}(\theta)$\\
&&&&\\
\hline
&&&&\\
$\prod_{i=1}^{p-1}\hat x_i$&$(\sin\theta)^{n_s}(\cos\theta)^{n_c}$
&$(\sin\theta)^{N_s}(\cos\theta)^{n_c}$
&$(\sin\theta)^{n_s}(\cos\theta)^{N_c}$
&$(\sin\theta)^{N_s}(\cos\theta)^{N_c}$\\
&&&&\\
\hline
\end{tabular}
\end{center}
\end{table}

The first row of the table is the types of cells, the second - limits of
variation of the angles related to each of them. The parameters $v_c$ and $v_s$
mean the number of vertices lying above the cell's base, i.å. vertices with
hypermoment $l$ to the left and right of it. Numbers $N_c$ and $N_s$ from the
last row of the table equal the sums of the Cartesian quantum numbers $n_i$
which are assigned to free ends "growing" to the left and right of the cell's
base; in this case, for free ends $N_c=n_c$, $N_s=n_s$. The functions in the
fourth row of the table have the form
\bea
f_{l_s,v_s}^{s;l}(\theta)
&=&
(N_{l-l_s}^{\alpha_s})^{-\frac{1}{2}}
(\sin\theta)^{l_s}P_{l-l_s}^{(\alpha_s,\alpha_s)}(\cos\theta)\,,
\label{4.4O}
\\[3mm]
f_{l_c,v_c}^{c;l}(\theta)
&=&
(N_{l-l_c}^{\alpha_c})^{-\frac{1}{2}}
(\cos\theta)^{l_c}P_{l-l_c}^{(\alpha_c,\alpha_c)}(\sin\theta)\,,
\label{4.5O}
\\[3mm]
f_{l_c,v_c;l_s,v_s}^{c,s;l}(\theta)
&=&
2^{\frac{\alpha_c+\alpha_s}{2}+1}
\left(N_{\frac{l-l_s-l_c}{2}}^{\alpha_s,\alpha_c}\right)^{-\frac{1}{2}}
(\cos\theta)^{l_c}(\sin\theta)^{l_s}
P_{\frac{l-l_s-l_c}{2}}^{(\alpha_s,\alpha_c)}(\cos 2\theta)\,,
\label{4.6O} \eea
where $P_n ^{(a,b)}(x)$ is the Jacobi polynomial and the remaining notation is:
$$
\alpha_s=l_s+\frac{v_s-1}{2},
\qquad
\alpha_c=l_c+\frac{v_c-1}{2}\,,
$$
\begin{eqnarray*}
N_n^{\alpha,\beta} = \frac{2^{\alpha+\beta+1}}{2n+\alpha+\beta+1}
\frac{\Gamma(n+\alpha+1)\Gamma(n+\beta+1)}{\Gamma(n+1)\Gamma(n+\alpha+
\beta+1)},
\qquad
N_n^{\alpha}\equiv N_n^{\alpha,\alpha}\,.
\end{eqnarray*}
In formula (\ref{4.6O}) index $l-l_s-l_c$ should be even, otherwise the
left-hand side of (\ref{4.6O}) is assumed to vanish.

\section{Contributions of individual cells to the transition matrix}
\markboth{CHAPTER 4. MULTIDIMENSIONAL ISOTROPIC OSCILLATOR}{4.2. CONTRIBUTIONS OF INDIVIDUAL CELLS TO THE TRANSITION MATRIX}

According to Table \ref{t41}, each cell gives the following contribution to the
transition matrix (\ref{4.2O}):
\bea
T_{n_c,n_s}^l
&=&
\frac{1}{\sqrt{2\pi}} \int\limits_{0}^{2\pi}\,
(\cos\theta)^{n_c}(\sin\theta)^{n_s}e^{-il\theta}d\theta\,,
\label{4.7O}
\\[3mm]
T_{N_c,n_s}^{s;l}(l_s,v_s)
&=&
\int\limits_0^\pi \, (\cos\theta)^{n_c}(\sin\theta)^{N_s+v_s}
f_{l_s,v_s}^{s;l}(\theta)d\theta\,,
\label{4.8O}
\\[3mm]
T_{N_c,n_c}^{c;l}(l_c,v_c)
&=&
\int\limits_{-\frac{\pi}{2}}^{\frac{\pi}{2}}\,
(\cos\theta)^{N_c+v_c}(\sin\theta)^{n_s}
f_{l_c,v_c}^{c;l}(\theta)d\theta\,,
\label{4.9O}
\\[3mm]
T_{N_c,N_s}^{c,s;l}(l_c,v_c;l_s,v_s)
&=&
\frac{1+(-1)^{l-l_s-l_c}}{2}
R_{N_c,N_s}^{c,s;l}(l_c,v_c;l_s,v_s)\,,
\label{4.10O}
\eea
where
\bea
R_{N_c,N_s}^{c,s;l}(l_c,v_c;l_s,v_s)
=
\int\limits_0^{\frac{\pi}{2}}
(\cos\theta)^{N_c+v_c}(\sin\theta)^{N_s+v_s}
f_{l_c,v_c;l_s,v_s}^{c,s;l}(\theta)d\theta\,.
\label{4.11O}
\eea
Let us show that integrals (\ref{4.7O})-(\ref{4.9O}) are reduced to
(\ref{4.11O}). We begin with integral (\ref{4.7O}) corresponding to the cell
with free ends. Let us divide in it the integration interval into the intervals
$(0,\pi)$ è $(\pi,2\pi)$ and make in the second integral a replacement
$\varphi= \theta-2\pi$. Then taking into account that
\begin{eqnarray*}
\cos l\varphi
=
\sqrt{\frac{\pi}{2}}(N_l^{-\frac{1}{2}})^{-\frac{1}{2}}
P_l^{(-\frac{1}{2},-\frac{1}{2})}(\cos\varphi)\,,
\quad
\sin l\varphi
=
\sqrt{\frac{\pi}{2}}(N_{l-1}^{\frac{1}{2}})^{-\frac{1}{2}}
(\sin\varphi)P_{l-1}^{(\frac{1}{2},\frac{1}{2})}(\cos\varphi)
\end{eqnarray*}
and turning back to (\ref{4.8O}), we get
\bea
T_{n_c,n_s}^l = \frac{1+(-1)^{n_s}}{2}
T_{n_c,n_s}^{s;l}(0,0) - i \frac{1-(-1)^{n_s}}{2}
T_{n_c,n_s}^{s,l}(1,0)\,.
\label{4.12O}
\eea
Then integrals (\ref{4.8O}) and (\ref{4.9O}), after analogous operations and
the use of the formula
\begin{eqnarray*}
(N_k^\alpha)^{-\frac{1}{2}}P_k^{(\alpha,\alpha)}(x)
&=&
\frac{1+(-1)^k}{2}2^{\frac{\alpha}{2}-\frac{1}{4}}
(N_{\frac{k}{2}}^{\alpha,-\frac{1}{2}})^{-\frac{1}{2}}
P_{\frac{k}{2}}^{(\alpha,-\frac{1}{2})}(2x^2-1) +
\\[3mm]
&+& \frac{1-(-1)^k}{2}2^{\frac{\alpha}{2}+\frac{3}{4}}
(N_{\frac{k-1}{2}}^{\alpha,\frac{1}{2}})^{-\frac{1}{2}}x
P_{\frac{k-1}{2}}^{(\alpha,\frac{1}{2})}(2x^2-1),
\end{eqnarray*}
take the form
\bea
T_{n_c,N_s}^{s;l}(l_s,v_s)
&=& \frac{1}{\sqrt 2}\,\left[1+(-1)^{l-l_s+n_c}\right]
\, \biggl[ \frac{1+(-1)^{l-l_s}}{2}
R_{n_c,N_s}^{c,s;l}(0,0;l_s,v_s) +
\nonumber \\
\label{4.13O}
\\
&+& \frac{1-(-1)^{l-l_s}}{2}
R_{n_c,N_s}^{c,s;l}(1,0;l_s,v_s)\biggr]\,,
\nonumber
\\ [3mm]
T_{N_c,n_s}^{c;l}(l_c,v_c)
&=& \frac{(-1)^{l-l_c}}{\sqrt 2}
\left[1+(-1)^{l-l_c+n_s}\right]
\biggl[\frac{1+(-1)^{l-l_c}}{2}
R_{N_c,n_s}^{c,s;l}(l_c,l_c;0,0) +
\nonumber\\
\label{4.14O}
\\
&+& \frac{1-(-1)^{l-l_c}}{2} \,
R_{N_c,n_s}^{c,s;l}(l_c,v_c;1,0)\biggr]\,.
\nonumber
\eea
Now consider integral (\ref{4.11O}). Turning to the variable $x = \cos
2\theta$, writing down the Jacobi polynomial in accordance with the Rodrigues
formula
\begin{eqnarray*}
P_n^{(\alpha,\beta)}(x)= \frac{(-1)^n}{n!2^n}(1-x)^{-\alpha}
(1+x)^{-\beta} \frac{d^n}{dx^n}
\left[(1-x)^{\alpha+n}(1+x)^{\beta+n}\right]
\end{eqnarray*}
then substituting in the integral $x$ by $-x$ and keeping in mind the integral
representation for the Clebsch--Gordan coefficients (\ref{KG.1.1}) we have
\bea
R_{N_c,N_s}^{c,s;l}(l_c,v_c;l_s,v_s)=
\frac{(-1)^{\frac{N_c+l_s-l}{2}}}{\sqrt 2} J_{a\alpha;b\beta}^c
C_{a\alpha;b\beta}^{c\gamma}\,,
\label{4.15O}
\eea
where
\bea
J_{a\alpha;b\beta}^c =
\left\{\frac{\Gamma(a-\alpha+1)\Gamma(b+\beta+1)
\Gamma(b-\beta+1)\Gamma(a+\alpha+1)}{\Gamma(a+b-c+1)
\Gamma(a+b+c+2)}\right\}^{\frac{1}{2}}\,,
\label{4.16O}
\eea
and the indices $c,\gamma $ and so on have the form
\bea
c&=&\frac{l}{2}+\frac{v_c+v_s}{4}-\frac{1}{4},
\qquad\qquad\qquad
\gamma=\frac{l_c+l_s}{2}-\frac{v_c+v_s}{4} -\frac{1}{2}\,,
\nonumber
\\[3mm]
\label{4.17O}
a&=&\frac{N_c+N_s+l_s+l_c}{4}+\frac{v_c-1}{4},
\qquad
\alpha = \frac{l_c+l_s}{4}+\frac{N_s-N_c}{4}+\frac{v_c-1}{4}\,,
\\[3mm]
b&=&\frac{N_c+N_s+l_c-l_s}{4}+\frac{v_c-1}{4}, \qquad
\beta=\frac{l_c+l_s}{4}+\frac{N_c-N_s}{4}+\frac{v_c-1}{4}\,.
\nonumber
\eea
So integral (\ref{4.11O}) is expressed in terms of the Clebsch--Gordan
coefficients of the $SU(2)$ group, if they are formally extended to one-quarter
values of the moment. The reason for the occurrence of one-quarter moments and
the relation of their respective $3j$- symbols with analogous objects of the
group $Sp(2,R)$ were elucidated in \cite{KPS}.

It follows from formulae (\ref{4.13O}) and (\ref{4.14O}) that the cells with
one free end are contributed by both $l_c = 0$ and $l_c = 1$ values
(analogously, $l_s = 0$ and $l_s = 1$). Let us study these contributions in
more detail. According to (\ref{ecsf.1.13}), the Clebsch--Gordan coefficients
are expressed in terms of generalized hypergeometric functions ${_3F_2}$ of the
unit argument. Substitute into this formula instead of the parameters
$c,\gamma$, etc. their values from (\ref{4.17O}). Then instead of (\ref{4.15O})
we have
\begin{eqnarray*}
&&R_{N_c,N_s}^{c,s;l}(l_c,v_c;l_s,v_s)
=
(-1)^{\frac{N_c+l_s-l}{2}}
\frac{\sqrt{l+\frac{v_s+v_c}{2}}
\Gamma\left(\frac{N_s-l_s}{2}+1\right)
\Gamma\left(\frac{N_s+l_s}{2}+\frac{v_s+1}{2}\right)}
{\Gamma\left(\frac{N_s+N-l}{2}+1\right)\Gamma\left(\frac{N_c+N_s+l}{2}+
\frac{v_c+v_s}{2}+1\right)} \times
\\[3mm]
&\times&\frac{\sqrt{\Gamma\left(\frac{l+l_s+l_c}{2}+
\frac{v_c+v_s}{2}\right)\Gamma\left(\frac{l+l_s-l_c}{2}+
\frac{v_s+1}{2}\right)\Gamma\left(\frac{l-l_s-l_c}{2}+1\right)
\Gamma\left(\frac{l+l_c-l_s}{2}+
\frac{v_c+1}{2}\right)}}
{\Gamma\left(\frac{l-N+l_s}{2}+\frac{v_s+1}{2}\right)
\Gamma\left(\frac{l-N_c-l_s}{2}+1 \right)} \times
\\[3mm]
&\times& {_3F_2} \left\{\matrix{
-\frac{N_s+N_c-l}{2},\,\,\frac{l_c-N_c}{2},\,\,
-\frac{N_c+l_c}{2}-\frac{v_c-1}{2},\,\, \cr \cr
\frac{l-N_c-l_s}{2}+1,\,\, \frac{l-N_c+l_s}{2}+\frac{v_s+1}{2}
\,\, \cr}\Biggr |1\right\}.
\end{eqnarray*}
By a direct substitution we verify the validity of the identities
\begin{eqnarray*}
R_{n_c,N_s}^{c,s;l}(0,0;l_s,v_s)
=
R_{n_c,N_s}^{c,s;l}(1,0;l_s,v_s)\,,
\qquad
R_{N_c,n_s}^{c,s;l}(l_c,v_c;0,0)= i
R_{N_c,n_s}^{c,s;l}(l_c,v_c;1,0)\,.
\end{eqnarray*}
It follows from these identities that formulae (\ref{4.12O})-(\ref{4.14O}) can
be written down in a more compact form
\bea
T_{n_c,n_s}^l
&=& \frac{1+(-1)^{l+n_c+n_s}}{\sqrt 2}
R_{n_c,n_s}^{c,s;l}(0,0;0,0)\,,
\label{4.18O}
\\[2mm]
T_{n_c,N_s}^{s;l}(l_s,v_s)
&=&\frac{1+(-1)^{l-l_s+n_c}}{\sqrt 2}
R_{n_c,N_s}^{c,s;l}(0,0;l_s,v_s)\,,
\label{4.19O}
\\[2mm]
T_{N_c,n_c}^{c;l}(l_c,v_c)
&=& (-1)^{\frac{l-l_c}{2}}
\frac{1+(-1)^{l-l_c+n_s}}{\sqrt 2}
R_{N_c,n_s}^{c,s;l}(l_c,v_c;0,0)\,.
\label{4.20O}
\eea
Thus, the contributions of all four types of cells to the transition matrix are
calculated. They are given by formulae (\ref{4.15O}) and (\ref{4.18O})
-(\ref{4.20O}). Therefore, let us now discuss how these results can be used to
calculate the transition matrix.

\section{Transition Tree and Correspondence Rules}
\markboth{CHAPTER 4. MULTIDIMENSIONAL ISOTROPIC OSCILLATOR}{4.3. TRANSITION TREE AND CORRESPONDENCE RULES}

Let us correlate the coefficient $W_{\vec n}^{\vec l N}$ with the so-called
transition tree which is identical in structure to the relevant hyperspherical
tree, with the only difference that now we ascribe to the vertex not the
hypermoment and number $v$, but the hypermoment, number $v$ and quantum number
$N$ equal to the sum of Cartesian quantum numbers $n_i$ with which free ends
"growing" from this vertex are endowed with.

For example, in the six-dimensional space there can be a transition tree
depicted in Fig. \ref{f41}. We associated each vertex with an ordinal number that
stands for three numbers if written in detail; thus $1=(l,v,N)$, $7=(0,n_7,0)$
è ò.ä.

\begin{figure}[t]\unitlength=1mm \special{em:linewidth 0.4pt} \linethickness{0.4pt}
\begin{picture}(114.00,68.00)
\put(80.00,00.00){\line(0,1){30.00}}
\put(80.00,30.00){\line(1,1){34.00}}
\put(80.00,30.00){\line(-1,1){34.00}}
\put(58.00,52.00){\line(1,1){11.00}}
\put(102.00,52.00){\line(-1,1){12.00}}
\put(96.00,58.00){\line(1,1){6.00}}
\put(63.00,57.00){\line(-1,1){7.00}}
\put(69.00,63.00){\line(1,1){1.00}}
\put(80.00,30.00){\circle*{2.00}}
\put(102.00,52.00){\circle*{2.00}}
\put(96.00,58.00){\circle*{2.00}}
\put(58.00,52.00){\circle*{2.00}}
\put(63.00,57.00){\circle*{2.00}}
\put(85.00,28.00){\makebox(0,0)[cc]{$1$}}
\put(57.00,47.00){\makebox(0,0)[cc]{$2$}}
\put(65.00,53.00){\makebox(0,0)[cc]{$3$}}
\put(105.00,47.00){\makebox(0,0)[cc]{$4$}}
\put(100.00,58.00){\makebox(0,0)[cc]{$5$}}
\put(45.00,68.00){\makebox(0,0)[cc]{$6$}}
\put(55.00,68.00){\makebox(0,0)[cc]{$7$}}
\put(68.00,68.00){\makebox(0,0)[cc]{$8$}}
\put(89.00,68.00){\makebox(0,0)[cc]{$9$}}
\put(101.00,68.00){\makebox(0,0)[cc]{$10$}}
\put(114.00,68.00){\makebox(0,0)[cc]{$11$}}
\end{picture}\caption{}
\label{f41}
\end{figure}

Let us trace the contribution of the coefficients (\ref{4.16O}) to a given
transition tree.For simplicity, denote this coefficient by $J(m,q,r)$, where
$m$, $q$ and $r$ are ordinal numbers of a given cell with $m$ relating to the
base, and $q$ and $r$ to the above left and right vertices of it. It is obvious
that $v_m=v_q+v_r+1$ and, therefore, according to (\ref{4.16O}),
\begin{eqnarray*}
J(m,q,r) = \frac{f(q)f(r)}{f(m)}\,,
\end{eqnarray*}
where the function $f$ has the form
\begin{eqnarray*}
f(i) = \left\{\Gamma\left(\frac{N_i-l_i}{2}+1\right)
\Gamma\left(\frac{N_i+l_i}{2}
+ \frac{v_i+1}{2}\right)\right\}^{\frac{1}{2}}\,.
\end{eqnarray*}
The total contribution of the coefficients (\ref{4.16O}) to a given tree equals
the product of the factors $J(m,q,r)$, and one can easily verify that
\begin{eqnarray*}
\prod J(m,q,r) = \frac{1}{f(1)} \prod_{k=6}^{11}f(k)\,,
\end{eqnarray*}
i.e., the final result contains only the information about free ends and the
base of the transition tree. An analogous conclusion is also valid for any
$p$-dimensional transition tree. Hence, it follows that for an arbitrary tree
\begin{eqnarray*}
\prod J(m,q,r) = \frac{\pi^{\frac{p}{4}}}{2^{\frac{N}{2}}}
\left\{\frac{(n_1)!(n_2)!....(n_p)!}
{\Gamma\left(\frac{N-l_1}{2}+1\right)
\Gamma\left(\frac{N+l_1}{2}
+\frac{p}{2}\right)}\right\}^{\frac{1}{2}}\,.
\end{eqnarray*}
Comparing this formula with (\ref{4.3O}) we come to the conclusion that the
total contribution of the coefficients (\ref{4.16O}) to any transition tree
reduces to a phase factor and factor $\frac{1}{\sqrt 2}$ with the coefficient
${\cal K}$ inâ (\ref {4.3O}). That is the reason why they should not be taken
into account at intermediate stages as well, i.e., within a given cell.

Let us denote by $j$, $t$ and $q$ the number of free, partially free and closed
cells on the transition tree. Obviously $2j+t=q, j+t+q=p-1$, whence $j=q+1$;
therefore, taking into account formulae (\ref{4.10O}), (\ref{4.15O}) and
(\ref{4.18O})- (\ref{4.20O}) and the rules of parity embedded in them one can
formulate the following rule of construction of transition matrices. First, one
draws a transition tree with number triples $(l_i, N_i, v_i)$ on it. Then each
cell, depending on which type it belongs to, is put into correspondence with
its contribution, according to Table \ref{t42}. The last step is multiplication of
the contributions from Table \ref{t42} with the allowance made for the construction
of the transition tree. The final result should also include the factor
$(-1)^{\frac{N-l_1}{2}}$ as a multiplier, where $l_1$ is the global moment, and
$N=n_1+n_2+....n_p$.

\begin{table}[t]
\caption{Contribution of each type of cells to the transition tree.}
\label{t42}

\begin{tabular}{|c|c|c|c|}
\hline
\unitlength=1.50mm \special{em:linewidth 0.4pt}\linethickness{0.4pt}
\begin{picture}(24.00,20.00)
\put(10.00,00.00){\line(0,1){5.00}}
\put(10.00,5.00){\line(1,1){10.00}}
\put(10.00,5.00){\line(-1,1){10.00}}
\put(10.00,5.00){\circle*{1}}
\put(12.00,3.00){$l$}
\put(11.00,0.00){$n_c{+}n_s$}
\put(1.00,15.00){$n_c$}
\put(16.00,15.00){$n_s$}
\end{picture}
&
\unitlength=1.50mm \special{em:linewidth 0.4pt}\linethickness{0.4pt}
\begin{picture}(24.00,20.00)
\put(10.00,00.00){\line(0,1){5.00}}
\put(10.00,5.00){\line(1,1){10.00}}
\put(10.00,5.00){\line(-1,1){10.00}}
\put(10.00,5.00){\circle*{1}}
\put(12.00,3.00){$l,\nu_s{+}1$}
\put(11.00,0.00){$N_s{+}n_c$}
\put(1.00,15.00){$n_c$}
\put(8.00,15.00){$l_s,\nu_s,N_s$}
\put(20.00,15.00){\circle*{1}}
\end{picture}
&
\unitlength=1.50mm \special{em:linewidth 0.4pt}\linethickness{0.4pt}
\begin{picture}(24.00,20.00)
\put(10.00,00.00){\line(0,1){5.00}}
\put(10.00,5.00){\line(1,1){10.00}}
\put(10.00,5.00){\line(-1,1){10.00}}
\put(10.00,5.00){\circle*{1}}
\put(12.00,3.00){$l,\nu_c{+}1$}
\put(11.00,0.00){$N_c{+}n_s$}
\put(1.00,15.00){$l_c,\nu_c,N_c$}
\put(0.00,15.00){\circle*{1}}
\put(16.00,15.00){$n_s$}
\end{picture}
&
\unitlength=1.50mm \special{em:linewidth 0.4pt}\linethickness{0.4pt}
\begin{picture}(24.00,20.00)
\put(10.00,00.00){\line(0,1){5.00}}
\put(10.00,5.00){\line(1,1){10.00}}
\put(10.00,5.00){\line(-1,1){10.00}}
\put(10.00,5.00){\circle*{1}}
\put(12.00,3.00){$l$}
\put(11.00,0.00){$\nu_c{+}\nu_s{+}1$}
\put(0.00,0.00){$N_c{+}N_s$}
\put(1.00,15.00){$l_c,\nu_c,N_c$}
\put(0.00,15.00){\circle*{1}}
\put(13.00,15.00){$l_s,\nu_s$}
\put(20.00,15.00){\circle*{1}}
\end{picture}
\\
\hline
&&&\\
$\frac{(-1)^{\frac{n_c-l}{2}}}{\sqrt 2}C_{a,\alpha;b,\beta}^{c,\gamma}$&
$(-1)^{\frac{n_c+l_s-l}{2}}C_{a,\alpha;b,\beta}^{c,\gamma}$&
$(-1)^{\frac{N_c-l_c}{2}}C_{a,\alpha,b,\beta}^{c,\gamma}$&
$(-1)^{\frac{N_c+l_s-l}{2}}C_{a,\alpha,b,\beta}^{c,\gamma}$
\\
&&&\\
$l_c=l_s=v_c=v_s=0$&$l_c=v_c=0$&$l_s=v_s=0$&\\
&&&\\
\hline
\end{tabular}
\end{table}

It follows from the above-listed rules that formally the transition matrix for
any tree can be written down as
\bea
W_{\vec n}^{N,\vec l}=
\frac{(-1)^{\frac{N-l_1}{2}}}{2^{\frac{f}{2}}}
\prod_{i=1}^{p-1}(-1)^{Q_i}
C_{a_i,\alpha_i;b_i,\beta_i}^{c_i,\gamma_i}\,,
\label{4.21O}
\eea
where $f$ is the number of cells with free ends, $Q_i$ for each cell depends on
the power of the Clebsch-Gordan coefficients in Table \ref{t42} to which $(-1)$ is
raised, and the product is over all vertices. In the case of the canonical
tree, $j=1, v_{c_i}=0$ and, as is easy to show, $v_{s_i}=p-i-1,\sum Q_i =
\frac{(N-l_1-n_p)}{2}$ where $i$ is the ordinal number of the vertex as counted
from the tree basis.

The reasoning behind the fact that an additional factor equal to
$\frac{1}{\sqrt 2}$ corresponds to the cells with free ends is as follows. In
partially free and closed cells the hypermoments and their respective numbers
$N_i$ have the same parity and, therefore, the projections of the moments
$\alpha$ and $\beta$ in (\ref{4.17O}) change with the step equal to unity and
run all admissible values.There is no this sort of rule of parity for cells
with free ends and, therefore, the step in $\alpha$ and $\beta$ equals
$\frac{1}{2}$. This leads to the fact that in the latter case the values of the
projections $\alpha$ and $\beta$ are groupped into two subsets, each having the
step equal to unity and being complete in the sense that it covers all
permitted values of the projections. That is the reason why for the
Clebsch--Gordan coefficients corresponding to the cells with free ends the
orthonormalization condition takes the form
\begin{eqnarray*}
\sum_{\alpha+\beta=\gamma} C_{a\alpha;b\beta}^{c;\gamma}
C_{a\alpha;b\beta}^{c,\gamma}=2\delta_{cc'}\,.
\end{eqnarray*}
It follows from this formula that the additional factor $\frac{1}{\sqrt 2}$ in
the first column of Table \ref{t42} is urged "to restore" standard orthonormality of
the contributions of the cells with free ends.

On this basis, it is easy to show that for any tree the transition matrix obeys
the orthonormality condition
\begin{eqnarray*}
\sum_{n_1+...+n_p=N} W_{\vec n}^{N,\vec l}W_{\vec n}^{* N,l'}
= \delta_{\vec l,\vec l'}
\end{eqnarray*}
as it must with the general reasoning in mind. Hence it follows that the
inverse transition matrix, i.e., a transition from the hyperspherical basis to
the Cartesian one equals $W_{\vec n}^{N,\vec l *}$, and the inverse
transformation itself has the form
\begin{eqnarray*}
\Psi_{N,\vec l}(r,\vec \theta) =
\sum_{n_1+n_2...+n_p=N} W_{\vec n}^{N,\vec l*}
\Psi_{\vec n}(\vec x)\,.
\end{eqnarray*}
In \cite{KSMOR1}, it was found out that in some particular cases the
Clebsch--Gordan coefficients can degenerate into the Wigner $d$-function. This
is just the case when we have to do with the cells with free ends. Indeed,
according to (\ref{inter1.2}) and formula (\ref{2.2}),
\begin{eqnarray*}
T_{n_c,n_s}^l= (-1)^{\frac{n_s}{2}}\sqrt 2
\left\{\frac{(n_c)!(n_s)!}{\pi
2^{n_c+n_s}\left(\frac{n_c+n_s+l}{2}
\right)!\left(\frac{n_c+n_s-l}{2}\right)!}\right\}^{\frac{1}{2}}
d_{\frac{l}{2},\frac{n_c-n_s}{2}} \left(\frac{\pi}{2}\right)\,.
\end{eqnarray*}
On the other hand, it follows from (\ref{4.15O}) and (\ref{4.18O}) that
\begin{eqnarray*}
T_{n_c,n_s}^l=
\left\{\frac{(-1)^{n_c-l}\, (n_c)!(n_s)!}{\pi
2^{n_c+n_s}\left(\frac{n_c+n_s+l}{2}
\right)!\left(\frac{n_c+n_s-l}{2}\right)!}\right\}^{\frac{1}{2}}
d_{\frac{l}{2},\frac{n_c-n_s}{2}} \left(\frac{\pi}{2}\right)
C_{\frac{n_c+n_s}{4}-\frac{1}{4},\frac{n_s-n_c-1}{4};
\frac{n_c+n_s}{4}-\frac{1}{4},\frac{n_c-n_s-1}
{4}}^{\frac{l}{2}-\frac{1}{2}, -\frac{1}{2}}\,.
\end{eqnarray*}
Comparing the last two formulae we get
\begin{eqnarray*}
C_{a\alpha;a\beta}^{c,-\frac{1}{2}}=
(-1)^{\alpha-\beta+\gamma-c}\sqrt 2
d_{c-\gamma,\beta-\alpha}^{2a-\gamma}
\left(\frac{\pi}{2}\right)\,,
\end{eqnarray*}
i.e., the result analogous to that given in \cite{KSMOR1}.

As mentioned, the fact that the transition matrix $W_{\vec n}^{N,\vec l}$ is
expressed in terms of the Clebsch--Gordan coefficients of the group $Sp(2,R)$
containing moments divisible by $1/4$ has group-theoretical justification. In
\cite{KPS}, it was shown that the construction of an arbitrary hyperspherical
oscillator basis from the Cartesian one is equivalent to the scheme of addition
of moments divisible by $-3/4$ and $-1/4$ that is specific to this basis. The
scheme of addition depends on the structure of the corresponding hyperspherical
tree. In this regard, some information on the structure of the transition
matrix $W_{\vec n}^{N,\vec l}$ can be found in \cite{KPS}. However, an explicit
form of $W_{\vec n}^{N,\vec l}$ can be obtained by analytical calculations, we
made in this section following \cite{POGOSYAN2}.

\section{Oscillator Wigner Functions}
\markboth{CHAPTER 4. MULTIDIMENSIONAL ISOTROPIC OSCILLATOR}{4.4. OSCILLATOR WIGNER FUNCTIONS}

In the multidimensional isotropic oscillator the operator of finite rotations
can be put in correspondence with the matrix elements in different bases. The
matrix elements constructed with respect to the hyperspherical basis are
universal for all centrosymmetrical fields and are generalizations of ordinary
Wigner functions. Following \cite{PSMT} we will pass on to the study
of the matrix elements of the operator of finite rotations over the Cartesian
basis. We call these objects the oscillator Wigner functions, as they carry
information on hidden symmetry of the multidimensional oscillator. Let us begin
with the circular oscillator.

{\bf 4.4.1.}
In passing from a system of coordinates
$S(x,y)$ to another $S(x',y')$ rotated with respect to the first one by an
angle $\alpha$ the Cartesian wave functions of a circular oscillator are
transformed by the orthogonal matrix \footnote{In what follows, we choose the
system of units: $\hbar=m=\omega=1$; $\varphi$ and $\varphi'$ in systems $S$
and $S'$ are related by $\varphi'=\varphi+\alpha$.}
\bea
|j+m',j-m';x',y'\rangle
&=&
\sum_{m=-j}^j \, \langle j+m,j-m;x,y|j+m',j-m';x',y'\rangle \times
\nonumber \\
\label{4.22}
\\
&\times&
|j+m,j-m;x,y \rangle\,.
\nonumber
\eea
Let us expand coefficients of this transformation over polar wave functions
with the same energy. As $E^{\rm car}=2j+1$, $E^{\rm pol} = 2p+|M|+1 p\geq 0$,
where ${p}$ is integer and M is the azimuthal quantum number, then at equal
energies $p=j-\frac{M}{2}$, so that $-2j\leq M\leq 2j$ and, consequently,
\begin{eqnarray*}
\langle j+m,j-m;x,y|j+m',j-m';x',y'\rangle
&=&
\sum_{k=-j}^i
\langle j+m,j-m;x,y|j-|k|,2k;r,\varphi\rangle \times
\\[3mm]
&\times& e^{2ik\alpha}\,
\langle j-|k|,2k;r,\varphi|j+m',j-m';x',y'\rangle\,.
\end{eqnarray*}
In this formula the coefficients of transition from Cartesian to polar wave
functions are determined by (\ref{2D-OSC-8}) and have the form
\begin{eqnarray*}
\langle j+m,j-m;x,y|j-|k|,2k;r,\varphi\rangle
=
(-i)^{j-m}(-1)^{j-|k|}\, d_{k,m}^j \left(\frac{\pi}{2}\right)\,.
\end{eqnarray*}
Using this result and the addition theorem for the Wigner  $d$-functions
\cite{VAR}
\bea
\sum_{M^{''}= - J}^{J}\,
d^{J}_{M^{''} M^{'}}\left(\frac{\pi}{2}\right)\,
d^{J}_{M^{''} M}\left(\frac{\pi}{2}\right)\,
e^{-iM^{''}\beta} =
(-1)^{-M^{'}} (-i)^{M+M^{'}}\,
d^{J}_{M^{'} M}(\beta)
\label{SUMMER-1}
\eea
we obtain the transformation law
\bea
|j+m',j-m';x',y'\rangle =
\sum_{m=-j}^j \, d_{m',m}^j (2\alpha)|j+m,j-m;x,y\rangle\,.
\label{4.23}
\eea
Formula (\ref{4.23}) results in an interesting integral representation of the
Wigner $d$-functions in terms of the normalized Hermite function
\begin{eqnarray*}
d_{m',m}^j (\varphi)
&=&
\int\limits_{-\infty}^\infty dx
\int\limits_{-\infty}^\infty dy\,
{\bar H}_{j-m}(y) {\bar H}_{j+m}(x) \times
\\[3mm]
&\times& {\bar H}_{j+m'}
\left(x\cos\frac{\varphi}{2}+y\sin\frac{\varphi}{2}\right) {\bar
H}_{j-m'}
\left(-x\sin\frac{\varphi}{2}+y\cos\frac{\varphi}{2}\right)\,.
\end{eqnarray*}
The explicit form of transformation (\ref{4.23}) can be established without
passing to the polar wave functions but by considering relation (\ref{4.22}) at
large $x$ and $y$ and substituting the Hermite polynomials by the corresponding
highest powers of arguments.

{\bf 4.4.2}
The form of an $n$- dimensional
finite-rotation operator naturally depends on the number of angles used for
transition from one point on the sphere to another. In the Cartesian basis the
simplest rotations are in separate coordinate planes, as each rotation like
that transforms only a relevant pair of one-dimensional oscillator functions by
the law obtained above. An arbitrary $n$-dimensional rotation is known to be
equivalent to a sequence of such "elementary" rotations by the
$\frac{n(n-1)}{2}$ Euler angles, chosen in a definite manner.

In the three-dimensional case the finite rotation operator is expressed in
terms of the Euler angles $\theta_1^2,\theta_2^2$ and $\theta_1^1$ as follows:
\bea
\label{4.24}
\hat G(3) = \hat G_{21}(\theta_1^2)
\hat G_{32}(\theta_2^2) \hat G_{21}(\theta_1^1).
\eea
Indices of the operators $\hat G_{j+1,j}$  label the coordinate plane in which
a rotation occurs and the rotation from the axis $j+1$ to the  j-th axis is
taken to be positive. Since the wave functions in the systems $S(x,y,z)$ and
$S(x',y',z')$ are related to the same energy, the nonzero matrix elements are
$\langle j+m,j-m,n-2j|\hat G(3)|j'+m',j'-m',n-2j'\rangle$. Expressing,
according to (\ref{4.24}),  the operator $\hat G(3)$ in terms of the operators
$\hat G_{j+1,j}$ and separating the rotation in the plane $(x,y)$ by the angle
$\theta_1^1$ we get
\begin{eqnarray*}
&&\langle j+m,j-m,n-2j|\hat G(3)|j'+m',j'-m',n-2j'\rangle =
\\[3mm]
&=& \sum_{j''=0}^{\frac{n}{2}} \sum_{m''=-j''}^{j''}
\langle j+m,j-m,n-2j|\hat G_{21}(\theta_1^2) \hat
G_{32}(\theta_2^2)|j''+m'',j''-m'',n-2j''\rangle \times
\\[3mm]
&\times&
\langle i''+m'',i''-m'',n-2j''|\hat G_{21}(\theta_1^1)|
l'+m',j'-m',n-2j'\rangle\,.
\end{eqnarray*}

The rotation $\hat G_{21}(\theta_1^1)$ does not affect the coordinates $z$ and,
therefore, the latter matrix element differs from zero only at $j''=j'$, so
that there is no longer summation over $j''$. The coefficient containing the
product of two rotation operators is, in its turn, reduced to a linear
combination of products of matrix elements of the operators $\hat
G_{21}(\theta_1^2)$ and $\hat G_{32}(\theta_2^2)$, and with analogous selection
rules we have:
\begin{eqnarray*}
&&\langle j+m,j-m,n-2j|\hat G(3)|j'+m',j'-m',n-2j'\rangle =
\\[3mm]
&=& \sum_{m''=-j'}^{j' } \sum_{\tilde m=-j}^j \langle j+m,j-m|\hat
G_{21}(\theta_1^2)|j+\tilde m,j-\tilde m\rangle \,
\delta_{j+\tilde m,j'+m''} \times
\\[3mm]
&\times& \langle j-\tilde m,n-2j|\hat
G_{32}(\theta_2^2)|j'-m'',n-2j'\rangle \langle j'+m'',j'-m''|\hat
G_{21}(\theta_1^1)|j'+m',j'-m'\rangle\,.
\end{eqnarray*}
Then passing from summation over $\tilde m $ and $m''$ to summation over $k=j+
\tilde m$ and $t=j'+m''$,and using the fact that by (\ref{4.23})
\bea
&&
\langle j+m,j-m;x_k,x_{k+1}|j'+m',j'-m';x'_k,x'_{k+1}\rangle =
\nonumber \\
\label{4.25}
\\
&=& \langle j+m,j-m|\hat G_{k+1,k}(\alpha)|j'+m',j'-m'\rangle
= d_{m',m}^j (2\alpha)
\nonumber
\eea
we arrive at the explicit expression for the three-dimensional oscillator
Wigner function:
\bea
&&\langle j+m,j-m,n-2j|\hat G(3)|j'+m',j'-m',n-2j'\rangle =
\nonumber\\[3mm]
\label{4.26}
\\
&=& \sum_{t=o}^{\min(j,j')} d_{t-j,m}^j (2\theta_1^2)
d_{2j-\frac{n+t}{2},2j-\frac{n+t}{2}} (2\theta_2^2)
d_{m',t-j'}^{j'}(2\theta_1^1)\,.
\nonumber
\eea
The limits of summation over $t$ may not be written explicitly if $t$ is
assumed to run only those values at which the absolute value of lower
$t$-dependent indices of the Wigner $d$ functions does not exceed that of the
relevant upper index.

As is clear from (\ref{4.24}), at $\theta_2^2=0$ there is a pure rotation in
the plane $(x.y)$ by an angle $\theta_1^1+\theta_1^2$ and at $\theta_1^1=
\theta_1^2=0$ the same rotation by an angle $\theta_2^2$ in the plane $(z,y)$,
so in these cases expression (\ref{4.26}) should turn into the Wigner $d$ -
function of the arguments $2(\theta_1^1+\theta_1^2)$ and $2\theta_2^2$,
respectively. The validity of this transition is easily verified with the use
of $d_{m',m}^j (0) = \delta_{mm'}$ and the addition theorem for the Wigner $d$
-function (\ref{SUMMER-1}).

The four-dimensional rotation is given by six Euler angles and the operator
$\hat G(4)$ is expressed in terms of the rotation operators in the coordinate
planes by the formula \cite{VILEN1}
\bea
\hat G(4) =
\hat G_{21}(\theta_1^3) \hat G_{32}(\theta_2^3)
\hat G_{43}(\theta_3^3) \hat G_{21}(\theta_1^2)
\hat G_{32}(\theta_2^2) \hat G_{21}(\theta_1^1).
\label{4.27}
\eea
The Wigner oscillator function is calculated analogously
\bea
&&\langle j+m, j-m, k-2j,n-k|\hat
G(4)|j'+m',j'-m',k-2j',n-k'\rangle =
\nonumber
\\[3mm]
&=&
\sum_{\mu,\lambda,\sigma} d_{m',\sigma-j'}^{j'} (2\theta_1^1)
d_{\sigma-\mu,\lambda-\mu}^{\mu} (2\theta_1^2) d_{\lambda-j,m}^j
(2\theta_1^3) \times
\label{4.28}
\\[3mm]
&\times&
d_{2j'-\frac{k'+\sigma}{2},2j-\frac{k'+\sigma}{2}}
^{\frac{k'-\sigma}{2}}(2\theta_2^2)
d_{2\mu-\frac{k+\lambda}{2},2j-\frac{k+\lambda}{2}}^{\frac{k-\lambda}{2}}
(2\theta_2^3)d_{k'-\mu-\frac{n}{2},k-\mu-\frac{n}{2}}^{\frac{n}{2}-\mu}
(2\theta_1^3)\,.
\nonumber
\eea
As for the limits of summation, see the comments after formula (\ref{4.26}). At
$\theta_1^3=\theta_2^3=\theta_3^3=0$ the rotation (\ref{4.27}) becomes
three-dimensional and formula (\ref{4.28}) reduces to (\ref{4.26}).

{\bf 4.4.3}
Consider now the  $n$-dimensional case. The
operator of $n$-dimensional rotations $\hat G(n)$ can be represented by the
following product of rotation operators in the coordinate planes \cite{VILEN1}
\bea
\hat G(n)=\hat
G^{(n-1)}(\theta_1^{n-1}.....\theta_{n-1}^{n-1})... \hat
G^{(1)}(\theta_1^1)\,,
\label{4.29}
\\[3mm]
\hat G^{(k)}(\theta_1^k,....\theta_k^k) =
\hat G_{21}(\theta_1^k)......\hat G_{k+1,k}(\theta_k^k)\,.
\label{4.30}
\eea
It follows from (\ref{4.29}) that the Wigner $n$-dimensional oscillator
function can be expressed in terms of the matrix elements of the operators
(\ref{4.30}) as follows:
\bea
&&\langle N_1,\dots, N_n|\hat G(n)|N'_1,\dots, N'_n \rangle =
\nonumber\\[3mm]
&=& \sum_{p_1^{(n-1)}\dots p_{n-1}^{(n-1)}}\cdots \sum_{p_1^{(2)}
p_2^{(2)}} \langle N_1,\dots, N_n|\hat
G^{(n-1)}|p_1^{(n-1)},\dots, p_{n-1}^{(n-1)},N'_n \rangle \times
\label{4.31}
\\[3mm]
&\times& \langle p_1^{(n-1)},\dots, p_{n-1}^{(n-1)}|\hat
G(n-2)|p_1^{(n-2)},\dots, p_{n-2}^{(n-2)},N'_{n-1} \rangle \cdots
\langle p_1^{(2)}p_2^{(2)}|\hat G^{(1)}|N'_1,N'_2\rangle
\nonumber
\eea
with $p_1^{(i)}+\dots+p_i^{(i)}=N'_1+\dots+N'_i$. In the notation
\begin{eqnarray*}
p_1^{(1)}=N_1 ; \qquad p_i^{(n)} = n_i ,
\qquad i=1,2,...n
\end{eqnarray*}
expression (\ref{4.31}) can be written in a more compact form:
\bea
&& \langle N_1,...N_n|\hat G(4)|N'_1,...N'_n\rangle =
\nonumber\\
\label{4.32}
&=& \sum_{p_1^{(i)},...p_i^{(i)}} \prod_{j=1}^{n-1}
\langle p_1^{(j+1)},...p_{j+1}^{(j+1)}|\hat G^{(j)}(\theta_1^j,...
\theta_j^j)|p_1^{(j)},...p_j^{(j)},N'_{j+1}\rangle\,.
\nonumber
\eea
With the help of (\ref{4.30}) we single out the rightmost factor of the product
of the rotation operators in the coordinate planes and make use of the
expression (resulting from \ref{4.23})
\begin{eqnarray*}
\left\langle
\hat G_{j+1,j} (\theta_j^j)\biggl|p_j^{(j)},N'_{j+1}\right\rangle
&=&
\sum_m \left\langle \frac{p_j^{(j)}+N'_{j+1}}{2}+m,
\frac{p_j^{(j)}+n'_{j+1}}{2}
-m\left|\hat G_{j+1,j}(\theta_j^j)\right| p_j,^{(j),N'_{j+1}}
\right\rangle \times
\\[3mm]
&\times& \left|\frac{p_j^{(j)}+N'_{j+1}}{2} - m ,
\frac{p_j^{(j)}+N'_{j+1}}{2}+ m\right\rangle\,,
\end{eqnarray*}
where summation over $m$  is in the limits $2|m|\leq p_j^{(j)}+N'_{j+1}$.
Carrying the state vectors $\langle p_{j+1}^{(j+1)}|$ out of the product of the
first $j-1$ operators of rotation in the coordinate planes and using the
condition of orthonormalization of one-dimensional oscillator wave functions we
arrive at the relation
\begin{eqnarray*}
\langle p_1^{(j+1)}....p_{j+1}^{(j+1)}|\prod_{l=1}^j
\hat G_{l+1,l}(\theta_l^j)|p_1^{(j)},...,p_j^{(j)},N'_{j+1}\rangle
=
\langle p_j^{(j)}+N'_{j+1}-p_{j+1}^{(j+1)}| \hat
G_{j+1,j(\theta_j^j)}|p_j^{(j)},N'_{j+1}\rangle \times
\\[3mm]
\times \langle p_1^{(j+1)},...,p_j^{(j+1)}|\prod_{l=1} ^{j-1}
\hat G_{l+1,l}(\theta_l^j)|
p_1^{(j)},....p_{j-1}^j,p_j^{(j)}+N'_{j+1}-p_{j+1}^{(j+1)}\rangle
\,.
\end{eqnarray*}
Further decreasing the number of multipliers in the product of rotation
operators in the coordinate planes, after $j-1$ steps we find:
\bea
&&\langle p_1^{(j+1)}, \dots,
p_{j+1}^{(j+1)}\left|\prod_{l=1}^{j}
{\hat G}_{l+1,l}(\theta_l^j)\right| p_1^{(j)}, \dots,
p_j^{(j)},{N'}_{j+1}\rangle =
\nonumber \\
\label{4.33}
\\
&=&
\prod_{k=1}^j \langle J_{jk}+m_{jk},J_{jk}-m_{jk}|\hat
G_{j-k+2}(\theta_{j-k+1}^j)| J_{jk}+m'_{jk},J_{jk}-m'_{jk}\rangle\,.
\nonumber
\eea
The quantities $J_{jk}, m_{jk}$ and $m'_{jk}$ are determined in the following
way:
\begin{eqnarray*}
2J_{jk}
&=&
N'_{j+1}+\sum_{\nu=j-k+1}^j p_{\nu}^{(j)} -
\sum_{\nu=j-k+2}^{j+1} p_{\nu}^{j+1} + p_{j-k+2}^{(j+1)}\,,
\\[3mm]
2m_{ik}
&=&
N'_{j+1}+\sum_{\nu=j-k+1}^j p_{\nu}^{(j)}-
\sum_{\nu=j-k+2}^{j+1}p_{\nu}^{(j+1)}- p_{j-k+2}^{j+1}\,,
\\[3mm]
2m'_{jk}
&=&
\sum_{\nu=1}^{j-k+1} p_{\nu}^{(j)} -
\sum_{\nu=1}^{j-k+2}p_{\nu}^{(j+1)} + p_{j-k+1}^{(j)}\,.
\end{eqnarray*}
Now with formula (\ref{4.25}) we arrive at the final result:
\bea
\langle N_1,...,N_n|\hat G(n)|N'_1,...,N'_n\rangle =
\sum_{p_1^{(i)}....p_i^{(i)}} \prod_{j=1}^{n-1} \prod_{k=1}^j
d_{m_{jk},m'_{jk}}^{J_{jk}}(2\theta_{j-k+1}^j)\,.
\label{4.34}
\eea
As can be verified, at $n=2,3$ and $4$ formula (\ref{4.34}) turns into
\ref{4.25}), (\ref{4.26}) and (\ref{4.28}).

\section{Diagrammatic method of calculating the Wigner oscillator wave functions}
\markboth{CHAPTER 4. MULTIDIMENSIONAL ISOTROPIC OSCILLATOR}{4.5. DIAGRAMMATIC METHOD}

The results obtained can be made much more transparent with the use of the
following analogy. The operator $\hat G(n)$ relates states 
$| N_1,...,N_n;x_1,...,x_n\rangle$ and $|n'_1,...,N'_n;x'_1,...,x'_n\rangle$; therefore, the
matrix element $\langle N_1,...,N_n|\hat G(n)|N'_1,...,N'_n\rangle$ can be
interpreted as an "amplitude" of transition of a system from state $N'$ to
state $N$, and this amplitude can be associated with the diagram:
%
%
 \begin{eqnarray*} \langle N_1...N_n|\hat G(n)|N'_1...N'_n\rangle =  \parbox{0.2\textwidth} {\includegraphics[width=0.2\textwidth]{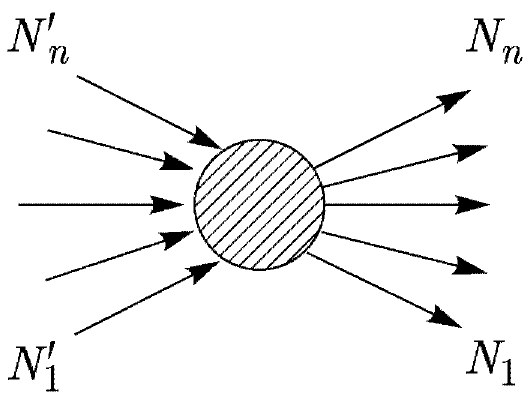}}
\end{eqnarray*}
\noindent Under this approach expansion (\ref{4.31}) of the matrix
element of the operator $\hat G(n)$ over the matrix elements of the operators
$\hat G^{(j)}$  is graphically represented as a sequence of "transitions".
\begin{figure}[t]

\begin{center}
\includegraphics[width=0.8\textwidth]{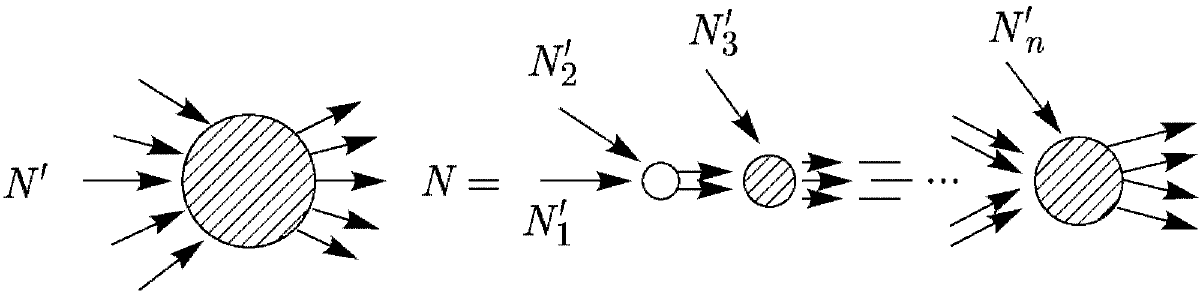}


\end{center}
\caption{}
\label{f42}
\end{figure}

\begin{figure}[h!]
\includegraphics[width=0.8\textwidth]{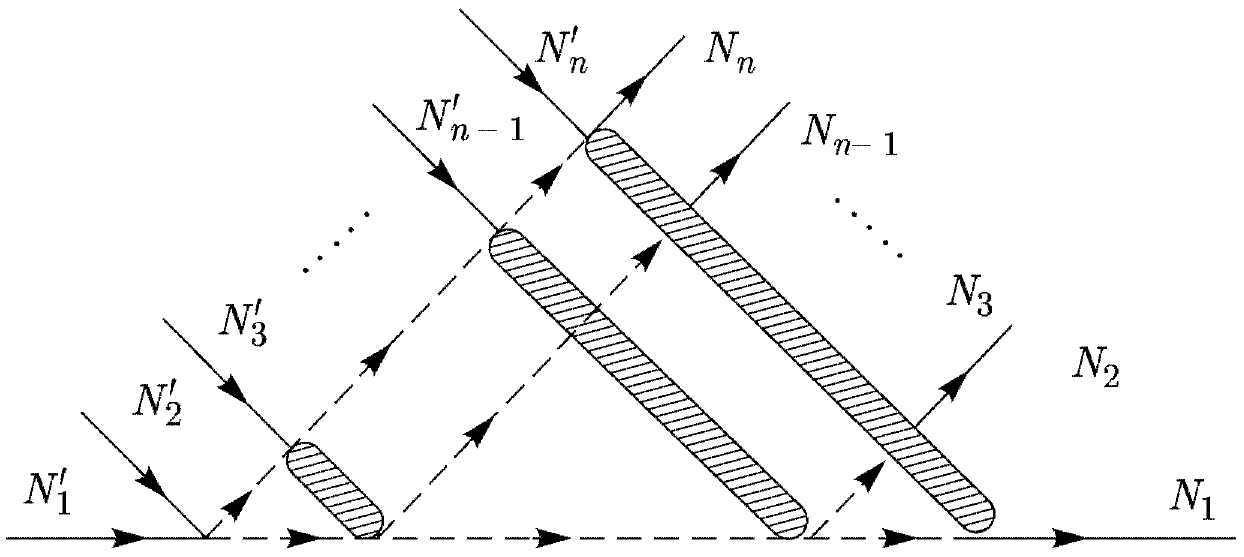}
\caption{}
\label{f43}
\end{figure}

Summation is implied over broken lines, and
intermediate blocks with one full line correspond to the matrix elements of the
operators $\hat G^{(j)}$, namely:
\bea
\langle p_1^{(j+1)}....p_{j+1}^{(j+1)}|\hat
G^j(\theta_1^j...\theta_j^j)|
p_1^{(j)}...p_j^{(j),N'_{j+1}}\rangle = p^{(j)}\parbox{0.18\textwidth} {\includegraphics[width=0.18\textwidth]{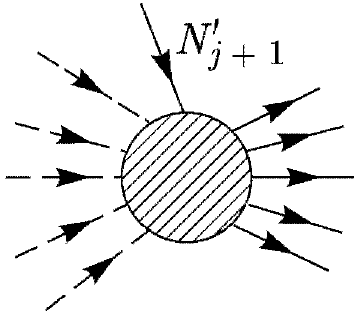}} p^{(j+1)}\,.
\label{4.36}
\eea
According to the condition preceding formula (\ref{4.32}), at $j=1$ and $j=n$
these blocks turn into the first and last blocks of expansion Fig. \ref{f42}.
Each block obeys the conservation law
\begin{eqnarray*}
p_1^{(j+1)}+...+p_{j+1}^{(j+1)}=
p_1^{(j)}+....+p_j^{(j)}+N'_{j+1}\,.
\end{eqnarray*}
The graphical expansion Fig. \ref{f42} is topologically equivalent to the diagram
in Fig. \ref{f43}.
The block (\ref{4.36}) in Fig. \ref{f42} looks like (see Fig. 4.4).

\begin{figure}[h!]
\includegraphics[width=0.8\textwidth]{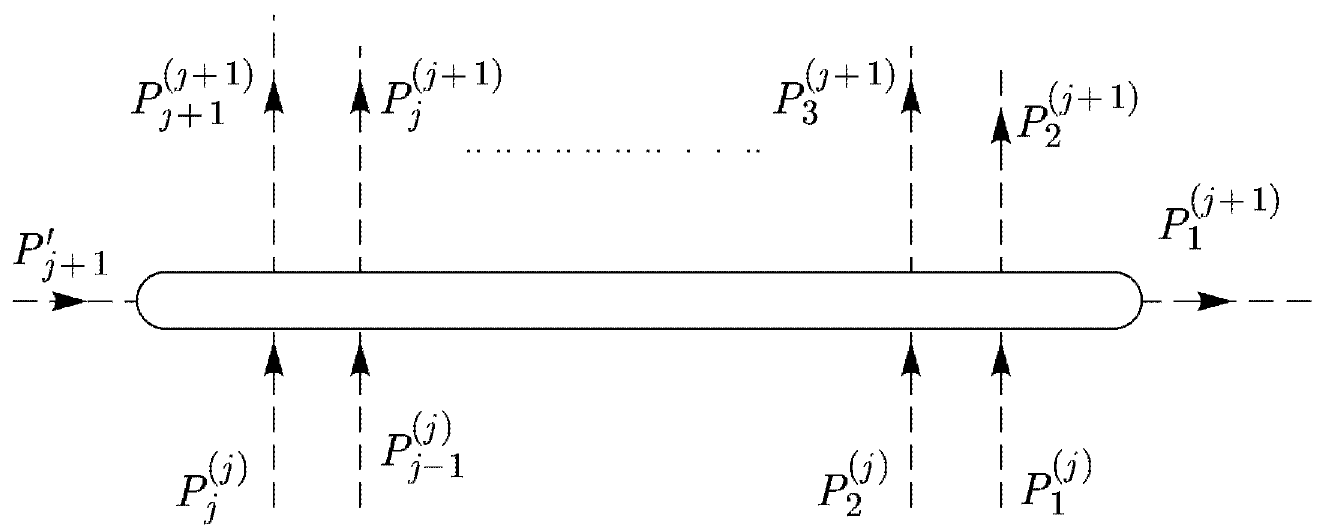}
\caption{}
\label{f44}
\end{figure}
According to formula (\ref{4.33}), the structure of this block has the form (see Fig.4.5).

\begin{figure}[h!]
\includegraphics[width=0.8\textwidth]{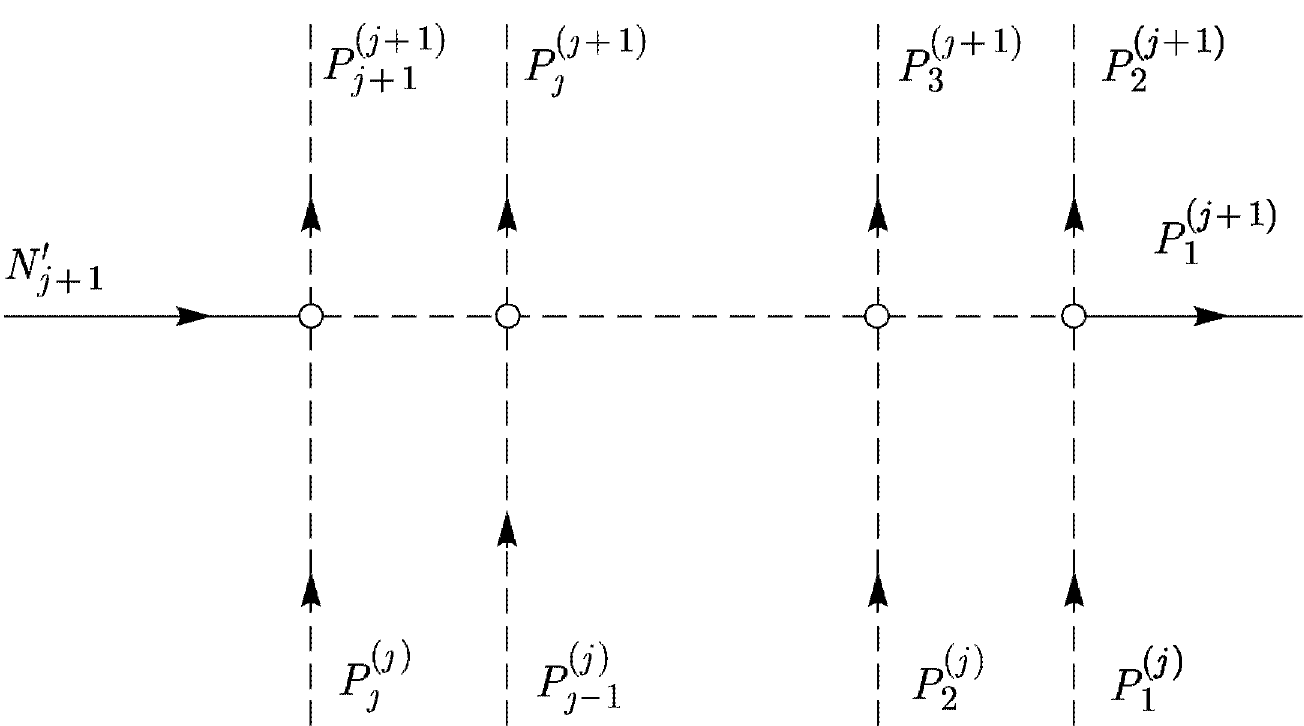}
\caption{}
\label{f45}
\end{figure}

\begin{figure}[h!]
\includegraphics[width=0.8\textwidth]{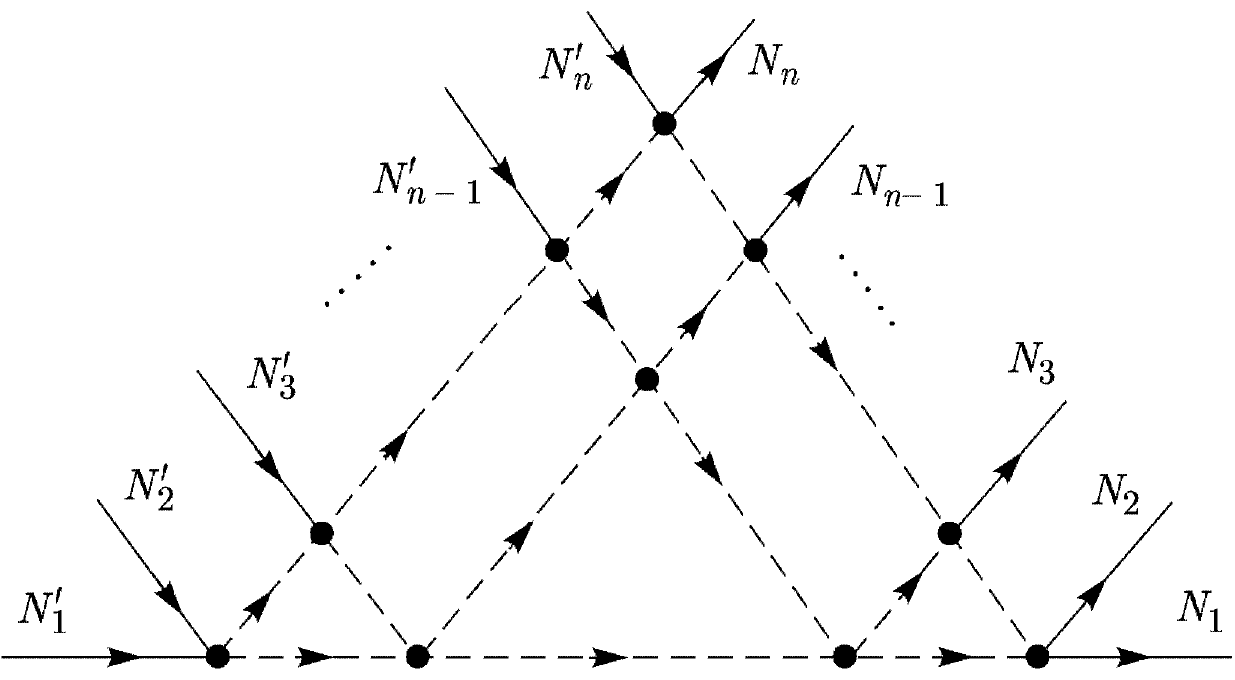}
\caption{}
\label{f46}
\end{figure}

\begin{figure}[h!]
\begin{center}
\includegraphics[width=0.8\textwidth]{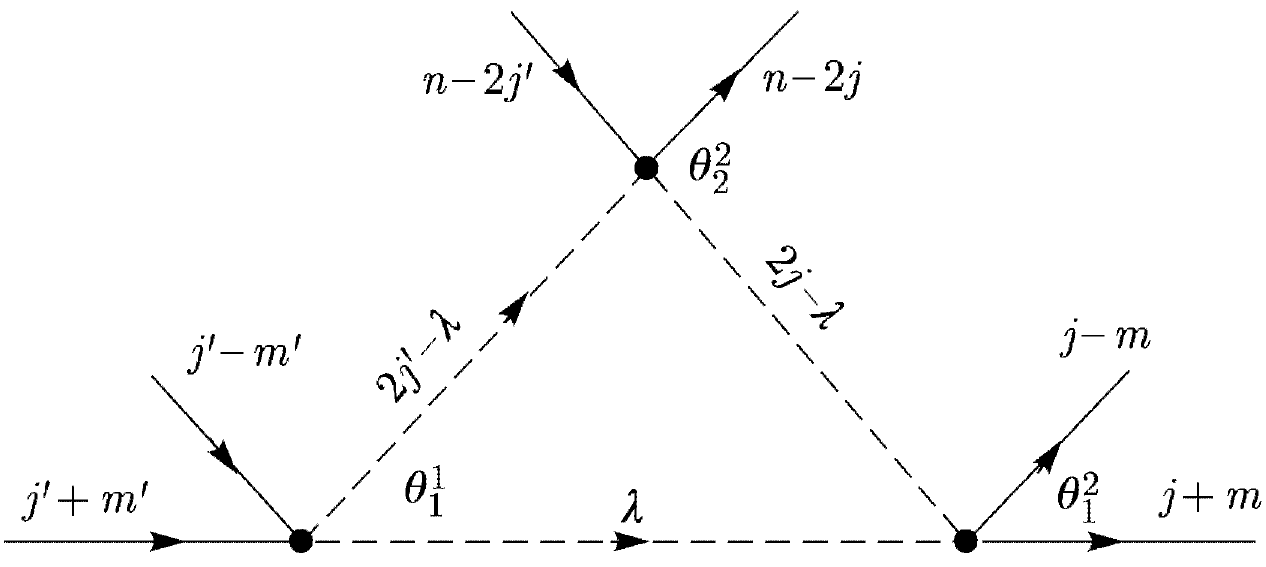}
\end{center}

\caption{Diagram for the three-dimensional oscillator Wigner
function}
\label{f47}

\begin{center}
\end{center}
\includegraphics[width=0.8\textwidth]{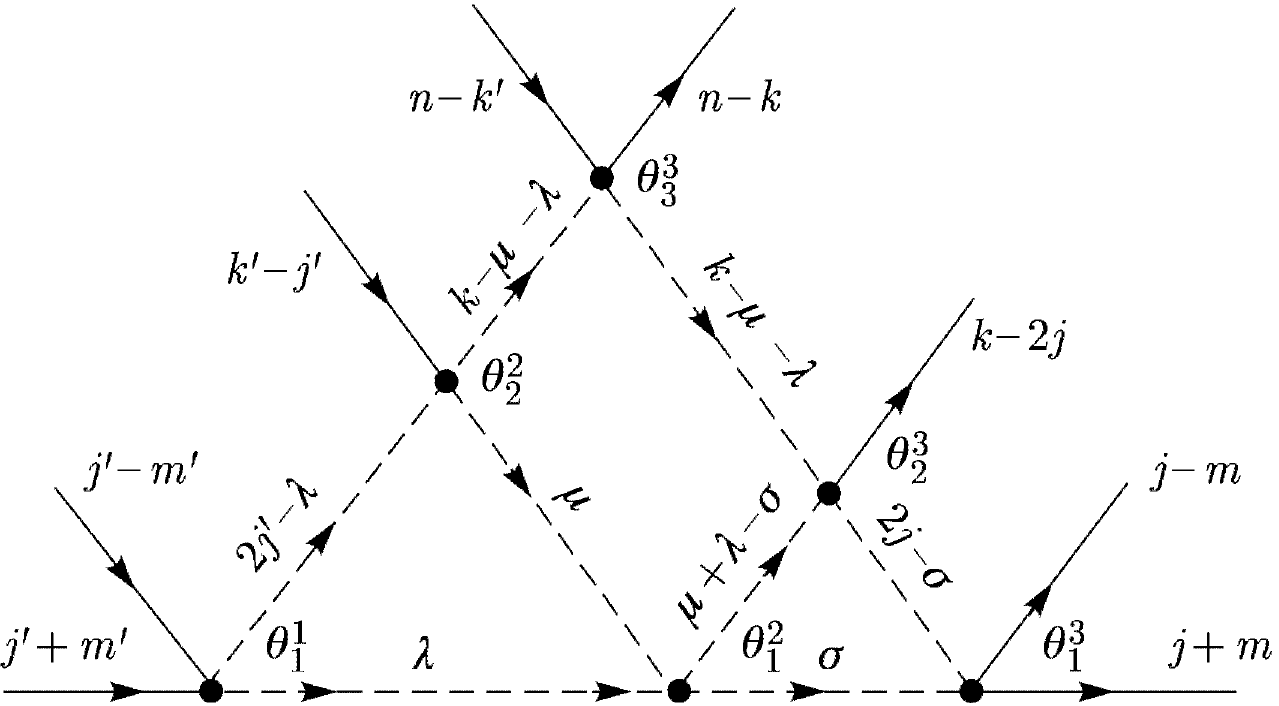}
\caption{}
\label{f48}
\end{figure}

In contrast with Fig. \ref{f42} on the Fig. \ref{f45}
intermediate broken lines at given outer lines have a definite index, so that a
transition from Fig. \ref{f44} to Fig. \ref{f45}  is not related to additional
summation. The diagrams in Fig. \ref{f43} and Fig. \ref{f45} make it possible
to represent the amplitudes of elementary transitions $\langle N|{\hat
G}(n)|N'\rangle$ by the amplitudes of elementary transitions of two lines into
two lines (see Fig. \ref{f46}).

In this diagram the number of intersections, i.e., elementary blocks, equals
$\frac{n(n-1)}{2}$. This means that rotation in the coordinate plane by a
certain Euler angle is associated with each of these blocks. The number of
summations in formula (\ref{4.34}) equals $\frac{(n-1)(n-2)}{2}$ and coincides
with the number of elementary loops in the diagram in Fig. \ref{f43}. According to
(\ref{4.25}), to each intersection there corresponds an amplitude of an
elementary "transition".
\bea
\parbox{0.18\textwidth} {\includegraphics[width=0.18\textwidth]{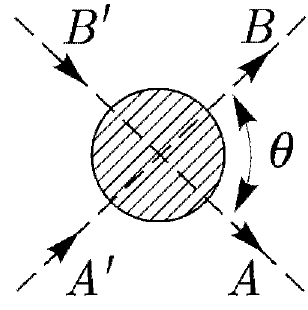}} = \langle A,B|{\hat G}^j(\theta)| A',B'\rangle
=
d^{\frac{A+B}{2}}_{\frac{A'-B'}{2}\,,\frac{A-B}{2}}\,(\theta)\,.
\label{4.39}
\eea
The operator $\hat G(\theta)$ in the last relation should be supplied with
indices of the coordinate axes defining the plane in which a rotation occurs.
We omitted these indices, as they are unambiguously defined by the very diagram
of Fig. \ref{f46}

The procedure of calculation is as follows. For each $n$ we draw the
corresponding diagram, mark the Euler angles, and place indices for all lines.
The number of nonfixed indices should be equal to the number of elementary
loops of the diagram. Further, with each intersection we associate the Wigner
$d$-function of the angle  "attached" to this intersection. Indices of the
Wigner $d$-function obey the rule (\ref{4.39}) and the summation is carried out
over nonfixed indices, over those values for which the $d$-function is
meaningful. The latter condition has been discussed earlier in connection with
formula (\ref{4.26}).

Let us exemplify this procedure for $n=3,4$. Choosing indices of wave functions
as in subsection 4, we arrive at the diagrams \ref{f47} and \ref{f48}.
The formulated rules for these diagrams give rise to formulae
(\ref{4.26}) and (\ref{4.28}).

Thus, the diagrammatic method reduces the problem of calculating the Wigner
oscillator function to simple geometric constructions.

\newpage

\chapter{Nonbijective transformations and Coulomb-oscillator analogy}
\markboth{CHAPTER 5. NONBIJECTIVE TRANSFORMATIONS}{}

The objective of the present chapter is to illustrate the property of the
Schr\"{o}dingier equation which is called here the dyon-oscillator duality
\cite{TERANT} (remind that a dyon is a hypothetic particle introduced by
Schwinger \cite{schwinger} which, unlike a monopole, is endowed with not just a
magnetic but also an electric charge). The property consists in the following.
The Schr\"{o}dinger equation for an oscillator possesses two parameters -- the
energy $E$ and the cyclic frequency $\omega$. The quantization leads to the
constraint $E=\hbar \omega(N+D/2)$ where $N=0,1,2\dots$ and $D$ is the
dimension of the configuration space of the oscillator. If $\omega$ is fixed,
then energy is quantized, and this is a standard situation. Imagine for a
moment that now $E$ is fixed. Then $\omega$ is necessarily quantized and we are
in a nonstandard situation. The question is whether the nonstandard situation
corresponds to any physics, i.e., whether it impossible to find such a
transformation that converts the oscillator into some physical system with the
coupling constant $\alpha$ as a function of $E$ and the energy $\varepsilon$
depending on $\omega$. If this transformation exists, we can confirm that the
"nonstandard oscillator" is identical to that physical system, and the initial
oscillator and the final system are dual to each other. That explains the
relevancy of the term "dyon-oscillator duality". It should be noted that in the
initial system the spectrum is discrete only, i.e., the particle has a finite
motion (for such cases it is used to say that we have a model with
confinement). Generally speaking, the final system has both the discrete and
continuous spectra \cite{Kibler-Negadi,MARDOYAN,MARDOYAN1}, i.e., in that model
there is no confinement. However, unlike the first model,the second model has
monopoles. There is an analogy between the dyon-oscillator and the
Seiberg--Witten duality, according to which the gauge theories with strong
interactions are equivalent to those having a weak interaction, on one hand,
and topological nontrivial objects as monopoles and dyons, on the other hand
\cite{SW}.

Let us consider the equation
\bea
\frac{d^2 R}{d u^2}+\frac{D-1}{u}\frac{d R}{d u} -
\frac{L(L+D-2)}{u^2}R +\frac{2\mu_0}{\hbar^2} \left(
E-\frac{\mu_0\omega^2 u^2}{2} \right)R=0.
\label{1.1.1}
\eea
Here $R$ is the radial part of the wave function for the $D$-dimensional
oscillator ($D>2$),and $L=0,1,2,\dots$ are the eigenvalues of the global
angular momentum.

After substitution $r=u^2$ equation (\ref{1.1.1}) turns into
\bea \frac{d^2 R}{d r^2}+\frac{d-1}{r}\frac{d R}{d
r}-\frac{l(l+d-2)}{r^2}R +\frac{2\mu_0}{\hbar^2}
\left(\varepsilon+\frac{\alpha}{r}\right)R=0, \label{1.1.2} \eea
where $d=D/2+1$, $l=L/2$,
\bea \varepsilon=-\mu_0\omega^2/8, \qquad \alpha=E/4.
\label{1.1.3} \eea
This is quite an unexpected result. If $D=4,6,8,10,\dots$, then
$d=3,4,5,6,\dots$, and equation (\ref{1.1.2}) is formally identical to the
radial equation for the  $d$-dimensional Kepler--Coulomb problem (for odd $D>2$
the value of $d$ is half-integer and, therefore, cannot have the meaning of the
dimension of the space in the usual sense). Then $l$ takes not only integer but
half-integer values as well; hence, it has the meaning of the total momentum
and there arises a question about the origin of the fermion degree of freedom.
The answer to the question will be given later. Finally, as mentioned above,
equations (\ref{1.1.1}) and (\ref{1.1.2}) are dual to each other and $r=u^2$ is
the duality transformation.

Up to now, only the radial part of the wave function of the oscillator has been
considered. In passing to the Schr\"{o}dinger equation we should take into
account the equation related to angular variables along with the radial
equation. Thus, the duality transformation must also include the transformation
of angular variables. If the substitution of variable $r=u^2$ is treated as a
mechanism of generation of an electric charge, then (as will be shown later)
the transformation of some angular variables is responsible for the generation
of magnetic charges.

The condition $r=u^2$ in the Cartesian coordinates has the form
\bea x_0^2+x_1^2 +\dots +x_{d-1}^2=(u_0^2+u_1^2+\dots
+u_{D-1}^2)^2, \label{1.1.4} \eea
which is called the Euler identity. According to the Hurwitz theorem
\cite{HUR}, if $x_i$ $(i=0,1,\dots,d-1)$ is a bilinear combination $u_\mu$
$(\mu=0,1,\dots,D-1)$, then identity (\ref{1.1.4}) is valid only for the
following pairs of numbers:
\begin{eqnarray*}
(D,\,d) = (1,\,1),\,(2,\,2),\,(4,\,3),\,(8,\,5).
\end{eqnarray*}
The "magic" numbers $D=1,2,4,8$ have a direct relation to the fact of existence
of four basic algebraic structures: real numbers, complex numbers and quaternions. 
Moreover, transformation (\ref{1.1.4}) establishes a connection
between two fundamental problems of mechanics, the oscillator and Kepler
problems.

Transformation $(D,d)=(1,1)$ relates the problem of linear oscillator with that
of one-dimensional Coulomb anyon \cite{TERANT}.

Transformation $(D,d)=(2,2)$ is the Levi--Civita transformation known from the
celestial mechanics \cite{LC}. It is the duality transformation reducing the
problem of circular oscillator to the problem of two-dimensional anyon \cite
{ARTUR,NTAT,NTA}.

Transformation corresponding to a pair of numbers $(D,d)=(4,3)$ is called in
the celestial mechanics the Kustaanheimo--Stiefel transformation 
(KS-transformation) \cite{KS}. The KS-transformation reduces the problem of
four-dimensional isotropic oscillator to the well-known MIC--Kepler system
\cite{Iwai-Uwano,MSTA,NTA1,TN}. This system was constructed by Zwanziger
\cite{ZWANZIG} and then revived by McIntosh and Cisneros \cite{MIC} it is a
generalization of the Coulomb problem in the presence of the Dirac monopole
Дирака \cite{DIRAC}. There are generalizations of the MIC--Kepler system to the
three-dimensional sphere \cite{GRITKUROT} and hyperboloid \cite{NERPOG}. The
MIC--Kepler system was treated from different points of view in
\cite{DRAGAN,INOMATA,IWAI1,IWAI2,IWAI3,MARDOYAN3,MARDOYAN2,MARDOYAN4,MLADENOV}.

Finally, the Hurwitz transformation corresponding to the case $(D,d)=(8,5)$
reduces the problem of eight-dimensional oscillator to the five-dimensional
Coulomb problem \cite{DMPST,KIB2}. Supplementing the Hurwitz transformation by
three specially chosen angles $(\alpha_T, \beta_T, \gamma_T)$ \cite{HVT}, we
find a transformation converting the spaces $\rm I\!R^8$ to a direct product
$\rm I\!R^8 = \rm I\!R^5 \otimes S^3$ of the space $\rm I\!R^5({\bf x})$ and
the three-dimensional sphere $S^3(\alpha_T, \beta_T,\gamma_T)$ \cite{MSTA1}. As
a result of such a separation of the space $\rm I\!R^8$, the eight-dimensional
isotropic oscillator can be used to construct a bound system of a charged
particle and the five-dimensional $SU(2)$ Yang monopole with a topological
charge $\pm 1$ \cite{MST1,MARDOYAN5} which we call the  $SU(2)$ Yang--Coulomb
monopole. In conclusion, it should be emphasized that a group of
six-dimensional rotations $SO(6)$ is a group of hidden symmetry $SU(2)$ of the
Yang--Coulomb monopole \cite{MSTA2} and this symmetry is the reason of
separation of variables in the Schr\"{o}dinger equation in the five-dimensional
hyperspherical, parabolic, and spheroidal coordinates \cite{MSTA3}.

\section{The MIC--Kepler problem}
\markboth{CHAPTER 5. NONBIJECTIVE TRANSFORMATIONS}{5.1. THE MIC--KEPLER PROBLEM}

The MIC--Kepler system, or the charge-dyon system, is described by the
Hamiltonian \cite{MIC,ZWANZIG}
\begin{eqnarray}
{\hat H} =\frac{1}{2\mu_0}\left(-i\hbar{\bf{\nabla}}+
\frac{e}{c}{\bf A}^{(\pm)}\right)^2 +\frac{\hbar^2{s}^2}{2\mu_0
r^2}-\frac{e^2}{r}, \label{3.1.1}
\end{eqnarray}
in which
\begin{eqnarray}
{\bf A}^{(\pm)} = \frac{g}{r(r \mp z)}\left(\pm y, \mp x.
0\right). \label{3.1.a}
\end{eqnarray}
The vector potentials ${\bf A}^{(\pm)}$ correspond to the Dirac monopole
\cite{DIRAC} with a magnetic charge $g=\frac{\hbar cs}{e}$ ($s=0,\pm
\frac{1}{2},\pm 1,\dots$) and the singularity axis at $z
> 0$ and $z < 0$, respectively.

One can easily see that the vector potentials $A^{(+)}_i$ and $A^{(-)}_i$ are
related to each other by the gauge transformation
\begin{eqnarray*}
A^{(-)}_i = A^{(+)}_i + \frac{\partial f}{\partial x_i},
\end{eqnarray*}
where $f=2g\arctan y/x$, and the intensity of a magnetic field created by the
dyon has the form
\begin{eqnarray*}
{\bf B} = {\bf \nabla} \times {\bf A}^{(\pm)} = g\frac{{\bf r}}
{r^3}.
\end{eqnarray*}
In what follows all calculations will be performed for the vector potential
$A^{(+)}_i$ only and, therefore, for brevity we will omit everywhere the sign
$(+)$.

A specific feature of the MIC--Kepler problem is hidden symmetry inherent in
the Coulomb problem. A direct calculation can really show that the following six
operators commute with Hamiltonian (\ref{3.1.1}):
\begin{eqnarray}
{\hat {\bf J}}&=& {\bf r}\times \left(-i{\bf\nabla}
-\frac{e}{\hbar c}{\bf A}\right)
+ s\frac{{\bf r}}{r},
\nonumber
\\
\label{3.1.2}
\\
{\hat {\bf I}}&=&\frac{1}{2\sqrt{\mu_0}}\left[{\hat {\bf J}}\times
\left(-i\hbar{\bf \nabla}+ \frac{e}{c}{\bf
A}\right)+\left(-i\hbar{\bf \nabla} + \frac{e}{c}{\bf
A}\right)\times{\hat{\bf J}}\right] +\frac{e^2}{\hbar
\sqrt{\mu_0}}\frac{\bf r}{r}.
\nonumber
\end{eqnarray}
The operators ${\hat {\bf J}}$ determine the rotational moment of the system
and the operator ${\hat{\bf I}}$ are an analog of the Runge-Lenz vector.

At fixed negative values of energy the integrals of motion (\ref{3.1.2}) form
an algebra isomorphic to the algebra $so(4)$; whereas at positive values of
energy, $so(3.1)$ \cite{ZWANZIG}. By virtue of hidden symmetry the MIC--Kepler
problem splits up not only in the spherical and parabolic but also prolate
spheroidal coordinates. Thus, the MIC--Kepler system can be considered a
natural generalization of the Coulomb problem in the presence of the Dirac
monopole.

\subsection{Spherical basis} 

In the spherical coordinates a solution of the spectral problem
\begin{eqnarray}
{\hat H}\Psi=\varepsilon\Psi,\qquad {\hat{\bf
J}}^2\Psi=j(j+1)\Psi, \qquad {\hat J}_z
\Psi=-\left(s+i\frac{\partial}{\partial \varphi}\right)\Psi=
m\Psi, \label{3.1.4}
\end{eqnarray}
which describes a bound MIC--Kepler system with energy $\varepsilon$,
rotational moment $j$ and $z$-component of the rotational moment $m$, can be
written as follows:
\begin{eqnarray}
\Psi_{njm}^{(s)}({\bf r}) =
\left(\frac{2j+1}{4\pi}\right)^{1/2}R_{nj}^{(s)}(r)
d^j_{ms}(\theta)e^{i(m+s)\varphi},
\label{3.1.5}
\end{eqnarray}
where $d^j_{ms}$ is the Wigner $d$-function \cite{VAR}, and $R_{nj}^{(s)}(r)$
is determined by the expression
\begin{eqnarray}
R_{nj}^{(s)}(r) = \frac{2e^{-\frac{r}{r_0n}}}{n^2r_0^{3/2}(2j+1)!}
\sqrt{\frac{(n+j)!}{(n-j-1)!}}\,\left(\frac{2r}{r_0n}\right)^j
\, F\left(j-n+1, 2j+2,\frac{2r}{r_0n}\right).
\label{3.1.6}
\end{eqnarray}
Now taking into account formula \cite{VAR}
\begin{eqnarray}
d^{J}_{MM'}(\beta) = (-1)^{\frac{M-M'+|M-M'|}{2}}\,
\left[\frac{k!(k+a+b)!}{(k+a)!(k+b)!}\right]^{\frac{1}{2}}\,
\left(\sin\frac{\beta}{2}\right)^{a}\left(\cos\frac{\beta}{2}\right)^{b}
P^{(a,b)}_{k}(\cos\beta),
\label{3.1.7}
\end{eqnarray}
where the indices of the Jacobi polynomial $k, a, b$ are related to $J, M, M'$
by
\begin{eqnarray*}
a= |M-M'|, \qquad b= |M+M'|,  \qquad k = J - \frac{1}{2}(a+b),
\end{eqnarray*}
one can write the angular wave function as
\begin{eqnarray}
&&Z_{jm}^{(s)}(\theta,
\varphi) = \left(\frac{2j+1}{4\pi}\right)^{1/2}d^j_{ms}(\theta)
e^{i(m+s)\varphi}= \left[\frac{(2j+1)(j-m_+)!(j+m_+)!}
{4\pi(j-m_-)!(j+m_-)!}\right]^{1/2}\times
\nonumber \\[1.5mm]
\label{3.1.8} \\[1.5mm]
&\times& (-1)^{\frac{m-s+|m-s|}{2}}\,
\left(\sin\frac{\theta}{2}\right)^{|m-s|}
\left(\cos\frac{\theta}{2}\right)^{|m+s|}\,
P^{(|m-s|,|m+s|)}_{j-m_+}(\cos\theta)\,e^{i(m+s)\varphi}.
\nonumber
\end{eqnarray}
Here we introduced the notation
\begin{eqnarray}
m_{\pm} = \frac{|m+s| \pm |m-s|}{2}.
\label{3.1.9}
\end{eqnarray}
The angular function $Z_{jm}^{(s)}(\theta, \varphi)$ is called in the
literature the Tamm monopole harmonics \cite{TAMM}. It follows from the
afore-said that the total wave function (\ref{3.1.5}) of the bound MIC--Kepler
system can be written as
\begin{eqnarray}
\Psi_{njm}^{(s)}({\bf r}) = R_{nj}^{(s)}(r)\,Z_{jm}^{(s)}(\theta,
\varphi).
\label{3.1.10}
\end{eqnarray}
The spectrum of the system is specified by the conditions:
\begin{eqnarray}
&\varepsilon_n& = -\frac{\mu_0 e^4}{2\hbar^2n^2},\qquad
n=|s|+1,|s|+2, \ldots ,
\label{3.1.11}
\\ [3mm]
&j&= |{s}|, |{s}|+1,\ldots, n-1;\quad  m =-j, -j+1,\ldots, j-1, j.
\nonumber
\end{eqnarray}
Quantum numbers $j$ and $m$ define the total moment of the system and its
projection onto the axis $z$. At integer $s$ the system has система has an
integer spin; and at half-integer, $s$ half-integer. At ${s}=0$ it turns into a
hydrogen-like system.

Under the identity transformation $\varphi\to\varphi+2\pi$ the wave function of
the system is unambiguous at integer ${s}$ and changes the sign at half-integer
${s}$. In the second case, the ambiguity of the wave function can be
interpreted as the presence of a magnetic field of an infinitely thin solenoid
(directed along the axis $z$) generating spin $1/2$ in the system.

\subsection{Parabolic basis} 

Let us now consider the MIC-Kepler system
in the parabolic basis. After substitution
\begin{eqnarray*}
\Psi(\mu,\nu,\varphi) = \Phi_1(\mu)
\Phi_2(\nu)\,\frac{e^{i(m+s)\varphi}}{\sqrt{2\pi}}
\end{eqnarray*}
the variables in the Schr\"{o}dinger equation are separated and we arrive at
the following system of equations:
\begin{eqnarray*}
\frac{d}{d \mu}\left(\mu \frac{d\Phi_1}{d \mu}\right) +
\left[\frac{\mu_0E^{(0)}}{2\hbar^2}\mu - \frac{(m+s)^2}{4\mu} +
\frac{\sqrt{\mu_0}}{2\hbar}\Omega + \frac{1}{2r_0}\right]\Phi_1
&=& 0, \\ [3mm] \frac{d}{d \nu}\left(\nu
\frac{d\Phi_2}{d \nu}\right) + \left[\frac{\mu_0
E^{(0)}}{2\hbar^2}\nu - \frac{(m-s)^2}{4\nu}-
\frac{\sqrt{\mu_0}}{2\hbar}\Omega + \frac{1}{2r_0}\right]\Phi_2
&=& 0,
\end{eqnarray*}
where $\Omega$ is the separation constant that is an eigenvalue of the
$z$-component of the Runge--Lenz vector ${\hat{\bf I}}$.

At $s=0$ these equations coincide with those for the hydrogen atom in the
parabolic coordinates \cite{LL}. Hence, we have
\begin{eqnarray}
\Psi_{n_1n_2 m}^{(s)}(\mu, \nu, \varphi) = \frac{1}{n^2r_0^{3/2}}
\Phi_{n_1 m+s}(\mu)\Phi_{n_2 m-s}(\nu)
\frac{e^{i(m+s)\varphi}}{\sqrt{\pi}},
\label{3.1.16}
\end{eqnarray}
where
\begin{eqnarray*}
\Phi_{p q}(x) = \frac{1}{|q|!} \sqrt{\frac{(p+|q|)!}{p!}}
e^{-x/2r_0n}\left(\frac{x}{r_0n}\right)^{|q|/2} F\left(-p;|q|+1;
\frac{x}{r_0n}\right).
\end{eqnarray*}
Here $n_1$ and $n_2$ are non-negative integers
\begin{eqnarray*}
n_1 = - \frac{|m+s|+1}{2} + \frac{\sqrt{\mu_0}}{2\kappa
\hbar}\Omega + \frac{1}{2r_0\kappa}, \qquad n_2 = -
\frac{|m-s|+1}{2} - \frac{\sqrt{\mu_0}}{2\kappa \hbar}\Omega +
\frac{1}{2r_0\kappa},
\end{eqnarray*}
and $\kappa=\sqrt{-2\mu_0 E^{(0)}}/\hbar$. The latter relations with the
allowance made for (\ref{3.1.11}) result in that the parabolic quantum numbers
$n_1$ and $n_2$ are related with the principal quantum number $n$ as
\begin{eqnarray}
n = n_1 + n_2 + \frac{|m-s|+|m+s|}{2} + 1.
\label{3.1.17}
\end{eqnarray}
Thus, we have solved the spectral problem
\begin{eqnarray}
{\hat H}\Psi=\varepsilon\Psi, \qquad {\hat I}_z\Psi=
\frac{e^2\sqrt{\mu_0}}{\hbar n}\left(n_1-n_2+m_-\right)\Psi,
\qquad {\hat J}_z\Psi= m\Psi,
\label{3.1.18}
\end{eqnarray}
where ${\hat H}, {\hat I}_z, {\hat J}_z$ are determined by expressions
(\ref{3.1.1}), (\ref{3.1.2}).

\section{Generalization of the Park--Tarter expansion}
\markboth{CHAPTER 5. NONBIJECTIVE TRANSFORMATIONS}{5.2. GENERALIZATION OF THE PARK--TARTER EXPANSION}

The sought expansion at a fixed energy value can be written as follows:
\begin{eqnarray}
\Psi_{n_1n_2m}^{{(s)}}(\mu, \nu, \varphi) =
\sum_{j=m_+}^{n-1}\,W^{j}_{nn_1ms}\Psi_{njm}^{(s)}(r, \theta,
\varphi).
\label{3.2.2}
\end{eqnarray}
Passing in the left-hand side of relation (\ref{3.2.2}) from the parabolic
coordinates to the spherical ones, according to formulae $\mu =
r(1+\cos\theta)$, $\nu = r(1-\cos\theta)$, assuming $\theta =0$, and taking
into account that
\bea P_{n}^{(a,b)}(1) = \frac{(a+1)_{n}}{n!},
\label{1.5.3}
\eea
we establish the equality which includes the variable $r$ only. Then using the
orthogonality condition for the radial wave function in the angular momentum
\begin{eqnarray*}
\int\limits_{0}^{\infty}\,R_{nj'}^{(s)}(r)R_{nj}^{(s)}(r)dr
=\frac{2}{r_0^2n^3} \frac{\delta_{jj'}}{2j+1},
\end{eqnarray*}
(see Appendix A), we get
\begin{eqnarray}
W_{nn_1ms}^j = (-1)^{\frac{m-s+|m-s|}{2}}\frac{\sqrt{(2j +1)\,
(n+j)!}} {|m+s|!\,(2j +1)!}\, E_{nn_1n_2}^{jms}\,K_{jms}^{nn_1}\,,
\label{3.2.4}
\end{eqnarray}
where
\begin{eqnarray}
E_{nn_1n_2}^{jms} = \left[\frac{(n_1+|m+s|)!(n_2+|m-s|)!
\left(j+m_-\right)! \left(j - m_+\right)!} {(n_1)!(n_2)!(n-j -1)!
\left(j + m_+\right)! \left(j-m_-\right)!}\right]^{1/2}, 
\nonumber \\ [1.5mm]
\label{3.2.5}
\\ [1.5mm]
K_{jms}^{nn_1}=\int \limits_{0}^{\infty} e^{-x} x^{j+m_+}F(-n_1,
|m+s|+1; x) F(-n+j+1, 2j +2; x)dx. \nonumber
\end{eqnarray}
Here $x=2r/r_0n$, and $r_0 = \hbar^2/\mu_0e^2$ is the Bohr radius. For
calculation of the integral $K_{jms}^{nn_1}$ we write down the degenerate
hypergeometric $F(-n_1, |m+s|+1; x)$ as a polynomial, perform integration by
formula (\ref{ort.1.4}) and with relation (\ref{ort.1.5}) taken into account we
get
\begin{eqnarray}
K_{jms}^{nn_1}= \frac{(2j +1)!\,(j+m_+)!\,\left(n-m_- -1\right)!}
{(j+m_+)!\,(n+j)!} {_3F}_2 \left\{
\begin{array}{l}
-n_1,  -j +m_+, j +m_+  +1
\\ [2mm] |m+s|+1,  -n+m_+ +1
\end{array}
\biggr| 1 \right\}.
\label{3.2.6}
\end{eqnarray}
Substitution of (\ref{3.2.5}) and (\ref{3.2.6}) into (\ref{3.2.4}) gives
\begin{eqnarray}
W_{nn_1ms}^j
&=& \sqrt{\frac{(2j+1)(n_1+|m+s|)!(n_2+|m-s|)! \left(j
+ m_+\right)! \left(j+m_-\right)!} {(n_1)!(n_2)!(n-j -1)!
(n+j)!\left(j -
m_+\right)! \left(j - m_- \right)!}}\,
\frac{\left(n - m_+ -1\right)!}{|m+s|!}\times
\nonumber \\ [1.5mm]
\label{3.2.7}
\\ [1.5mm]
&\times&
(-1)^{n_1+\frac{m-s+|m-s|}{2}}
{_3F}_2 \left\{
\begin{array}{l}
-n_1,  -j +m_+, j+m_+ +1
\\ [2mm] |m+s|+1,   -n+m_+ +1
\end{array}
\biggr| 1 \right\}.
\nonumber
\end{eqnarray}
Finally, comparing (\ref{3.2.7}) with
\bea
C_{a \alpha ; b \beta}^{c\gamma}
&=& \delta_{\gamma,\alpha+\beta} \,
\left[
\frac{(2c+1)(b-a+c)!(a+\alpha)!(b+\beta)!(c+\gamma)!}
{(b-\beta)!(c-\gamma) ! (a+b-c)!(a-b+c)!(a+b+c+1)!} \right] ^{1/2} \times
\nonumber \\ [1.5mm]
\label{1.5.14} 
\\ [1.5mm]
&\times&
\frac{(-1)^{a-\alpha}}{\sqrt{(a-\alpha)!}}
\frac{(a+b-\gamma)!}{(b-a+\gamma)!} {_3F}_2 \left\{
\begin{array}{l}
-a+\alpha, c+\gamma+1, -c+\gamma \\
\gamma-a-b,  b-a+\gamma+1  \\
\end{array}
\biggr| 1 \right\},
\nonumber
\eea
we arrive at the following representation:
\begin{eqnarray}
W_{nn_1ms}^j = (-1)^{n_1+\frac{m-s+|m-s|}{2}}
C_{\frac{n_1+n_2+|m-s|}{2},\frac{n_2-n_1+|m-s|}{2};
\frac{n_1+n_2+|m+s|}{2},\frac{n_1-n_2+|m+s|}{2}}^ {j,\,m_+}.
\label{3.2.8}
\end{eqnarray}
Inverse transformation, i.e., expansion of the spherical basis over the
parabolic one разложение сферического базиса по параболическому, has the form
\begin{eqnarray}
\Psi_{n j m}^{(s)}(r,\theta,\varphi) = \sum_{n_1=0}^{n-m_+ -1}
{\tilde W}_{n j ms}^{n_1} \Psi_{n_1n_2m}^{(s)}(\mu,\nu,\varphi).
\label{3.2.9}
\end{eqnarray}
Then using the orthonormalization condition of the Clebsch--Gordan coefficients
of the $SU(2)$ group
\begin{eqnarray}
\sum_{c=|\gamma|}^{a+b} C_{a\alpha ;b\beta }^{c\gamma}
C_{a\alpha';b\beta'}^{c\gamma} = \delta_{\alpha\alpha'}
\delta_{\beta\beta'},
\label{1.6.16}
\end{eqnarray}
we get
\begin{eqnarray}
{\tilde W}_{n j ms}^{n_1} = (-1)^{n_1+\frac{m-s+|m-s|}{2}}
C_{\frac{n-m_- -1}{2},\frac{n-m_- -1}{2}-n_1; \frac{n+m_-
-1}{2},\frac{m_+ +|m+s|-n+1}{2}+n_1}^ {j, m_+}.
\label{3.2.10}
\end{eqnarray}
Taking into account formula (\ref{1.5.14}) ${\tilde W}_{n j ms}^{n_1}$ can be
written in terms of the functions ${_3}F_2$. At $s=0$ formulae (\ref{3.2.7})
(\ref{3.2.8}) turn into the corresponding relations, we derived in the second
chapter.

\section{Spheroidal basis of the MIC--Kepler problem}
\markboth{CHAPTER 5. NONBIJECTIVE TRANSFORMATIONS}{5.3. SPHEROIDAL BASIS OF THE MIC--KEPLER PROBLEM}

Let us construct the spheroidal basis of the MIC--Kepler problem with the help
of formulae (\ref{3.2.2}) and (\ref{3.2.8}) - (\ref{3.2.10}) derived in the
previous section.

In the spheroidal system of coordinates, equation (\ref{3.1.1}) with the use of
the substitution
\begin{eqnarray}
\Psi^{(s)}(\xi, \eta, \varphi; R) = X^{(s)}(\xi;R)Y^{(s)}(\eta;R)
\frac{e^{i(m+s)\varphi}}{\sqrt{2\pi}}.  \label{3.3.2}
\end{eqnarray}
splits into two ordinary differential equations
\begin{eqnarray}
\left[\frac{d}{d\xi}(\xi^2-1)\frac{d}{d\xi} +
\frac{(m+s)^2}{2(\xi+1)}- \frac{(m-s)^2}{2(\xi -1)} +
\frac{\mu_0R^2\varepsilon} {2\hbar^2}(\xi^2-1) + \frac{R}{r_0}\xi
-\Lambda(R)\right]X^{(s)} = 0, \nonumber \\ \label{3.3.3}
\\
\left[\frac{d}{d\eta}(1-\eta^2)\frac{d}{d\eta} -
\frac{(m+s)^2}{2(1+\eta)}- \frac{(m-s)^2}{2(1-\eta)} +
\frac{\mu_0R^2\varepsilon}{2\hbar^2}(1-\eta^2) - \frac{R}{r_0}\eta +
\Lambda(R)\right]Y^{(s)} = 0, \nonumber
\end{eqnarray}
where $\Lambda(R)$ is the separation constant.

Excluding the energy from the system of equations (\ref{3.3.3}) we obtain the
spheroidal integral of motion
\begin{eqnarray*}
\hat \Lambda &=&
\frac{1}{\xi^2-\eta^2}\left[(1-\eta^2)\frac{\partial}{\partial
\xi} (\xi^2-1)\frac{\partial}{\partial \xi}- (\xi^2-1)
\frac{\partial}{\partial \eta}(1-\eta^2)\frac{\partial}{\partial
\eta} \right]-\frac{\xi^2+\eta^2-2}{(\xi^2-1)(1-\eta^2)}
\frac{\partial^2}{\partial \varphi^2}+  \\[2mm]
&+& \frac{2s}{(\xi - 1)(1-\eta)}
\left(\xi+\eta-1-\frac{\xi\eta+1}{\xi+\eta}
\right)\left(s+i\frac{\partial}{\partial \varphi}\right) +
\frac{R}{r_0}\frac{\xi\eta+1}{\xi+\eta}.
\end{eqnarray*}
In the last expression, passing to the Cartesian coordinates we find
\begin{eqnarray}
\hat \Lambda = {\hat {\bf J}}^2 + \frac{R\sqrt{\mu_0}}{\hbar}
{\hat I}_z.
\label{3.3.4}
\end{eqnarray}
So we have
\begin{eqnarray}
\hat \Lambda \Psi^{(s)}_{nqm}(\xi, \eta, \varphi; R)=
\Lambda_q(R)\Psi^{(s)}_{nqm}(\xi, \eta, \varphi; R). \label{3.3.5}
\end{eqnarray}
Index $q$ numbers eigenvalues for the operator ${\hat \Lambda}$ and takes the
following values: $1 \leq q \leq n-m_+$.

Let us now express the spheroidal basis as expansion over the spherical
(\ref{3.1.11}) and parabolic (\ref{3.1.16}) bases:
\begin{eqnarray}
\Psi_{nqm}^{(s)}(\xi,\eta,\varphi;R)
&=& \sum_{j=m_+}^{n-1}
V_{nqms}^j \Psi_{njm}^{(s)}(r,\theta,\varphi),
\label{3.3.6}
\\[3mm]
\Psi_{nqm}^{(s)}(\xi,\eta,\varphi;R)
&=& \sum_{n_1=0}^{n-m_+ -1}
U_{nqms}^{n_1} \Psi_{n_1n_2m}^{(s)}(\mu,\nu,\varphi).
\label{3.3.7}
\end{eqnarray}
Acting by operator (\ref{3.3.4}) on (\ref{3.3.6}) and (\ref{3.3.7}) and using
the spectral equations for the operators ${\hat J}^2$ (\ref{3.1.4}), ${\hat
I}_z$ (\ref{3.1.18}), and ${\hat \Lambda}$ (\ref{3.3.5}), we get
\begin{eqnarray}
\label{3.3.8}
[\Lambda_q(R)-j(j +1)]V_{nqms}^j(R)
&=& \frac{R\sqrt{\mu_0}}{\hbar}
\sum_{j'}\,V_{nqms}^{j'}(R) \left({\hat I}_z\right)_{j j'},
\\[3mm]
\label{3.3.8-1}
\left[\Lambda_q(R) - \frac{R}{nr_0}\left(n_1-n_2 +
m_-\right)\right]U_{nqms}^{n_1}(R)
&=&
\sum_{n_1'}\,U_{nqms}^{n_1'}(R)
\left({\hat {\bf J}}^2\right)_{n_1 n_1'},
\end{eqnarray}
where
\begin{eqnarray*}
\left({\hat I}_z\right)_{j j'}
&=& \int \Psi_{nj
m}^{(s)*}(r,\theta,\varphi){\hat I}_z \Psi_{nj'm}^{(s)}
(r,\theta,\varphi) dV,
\\ [3mm]
\left({\hat {\bf J}}^2\right)_{n_1 n_1'}
&=&
\int \Psi_{n_1n_2m}^{(s)*}(\mu,\nu,\varphi)
{\hat J}^2 \Psi_{n_1' n_2' m}^{(s)}(\mu,\nu,\varphi) dV.
\end{eqnarray*}
Then proceeding in the same way as in determining the spheroidal bases of the
hydrogen atom (see the second chapter), we find for the matrix elements
$\left({\hat I}_z\right)_{jj'}$, and $\left({\hat {\bf
J}}^2\right)_{n_1{n'}_1}$
\begin{eqnarray}
\label{3.3.9}
\left({\hat I}_z\right)_{j j'}
&=& \frac{e^2\sqrt{\mu_0}}{\hbar n}
\left(A_{j +1}\delta_{j',j+1} + B_j\delta_{j',j} +
A_j\delta_{j',j-1}\right),
\\[3mm]
\label{3.3.9-1}
\left({\hat {\bf J}}^2\right)_{n_1 n_1'}
&=&
C_{n_1+1}\delta_{n_1',n_1+1}+ D_{n_1}\delta_{n_1',n_1} +
C_{n_1}\delta_{n_1',n_1-1},
\end{eqnarray}
where
\begin{eqnarray*}
A_j &=& -\sqrt{\frac{\left(j^2-m_+^2\right)
\left(j^2-m_-^2\right)(n^2-j^2)} {j^2(2j-1)(2j+1)}}, \qquad B_j =
\frac{nm_+ m_-}{j(j+1)},\\
[3mm] C_{n_1} &=& - \sqrt{n_1(n_1+|m+s|) \left(n-n_1+m_-\right)
\left(n-n_1-m_+\right)}, \\ [3mm]
D_{n_1} &=& n_2(n_1+1)
+\left(n_1+|m+s|\right)\left(n_2+|m-s|+1\right)+m_-(m_- -1).
\end{eqnarray*}
Now substituting the matrix elements (\ref{3.3.9})-(\ref{3.3.9-1}) into
(\ref{3.3.8})-(\ref{3.3.8-1}) we derive the trinomial recurrence relations
\begin{eqnarray*}
A_{j+1}U_{nqms}^{j+1}(R)
&+& \left\{B_j -
\frac{nr_0}{R}\left[\Lambda_q(R)-j(j+1)
\right]\right\}U_{nqms}^j(R) + A_jU_{nqms}^{j-1}(R) = 0,
\\[3mm]
C_{n_1+1}V_{nqms}^{n_1+1}(R)
&+& \left[D_{n_1}-\frac{R}{nr_0}
\left(n_1-n_2+m_-\right)-\Lambda_q(R)\right]V_{nqms}^{n_1}(R) +
C_{n_1}V_{nqms}^{n_1-1}(R) = 0,
\end{eqnarray*}
to be solved simultaneously with the normalization conditions
\begin{eqnarray*}
\sum_{j}\,\left|U_{nqms}^j(R)\right|^2 = 1, \qquad
\sum_{n_1}\,\left|V_{nqms}^{n_1}(R)\right|^2 = 1.
\end{eqnarray*}

\section{Transformation of the MIC--Kepler oscillator}
\markboth{CHAPTER 5. NONBIJECTIVE TRANSFORMATIONS}{5.4. TRANSFORMATION OF THE MIC--KEPLER OSCILLATOR}

Let us show that the MIC-Kepler problem is dual to the four-dimensional
isotropic oscillator. In the spherical coordinates the Schr\"{o}dinger equation
(\ref{3.1.1}) has the form
\begin{eqnarray}
&&\frac{1}{r^2}\frac{\partial}{\partial r}\left( r^2\frac{\partial
\Psi^{(s)}}{\partial r}\right)
+ \frac{1}{r^2}\left[\frac{1}{\sin
\theta}\frac{\partial}{\partial \theta}\left( \sin \theta
\frac{\partial \Psi^{(s)}}{\partial \theta}\right) +
\frac{1}{\sin^2\theta}\frac{\partial^2\Psi^{(s)}} {\partial
\varphi^2}\right]-
\nonumber \\ [1.5mm]
\label{3.7.1}
\\ [1.5mm]
&-&
\frac{2is}{r^2(1-\cos\theta)}\frac{\partial \Psi^{(s)}}{\partial
\varphi}-
 \frac{2s^2}{r^2(1-\cos\theta)}\Psi +
\frac{2\mu_0}{\hbar^2}\left(\varepsilon +
\frac{e^2}{r}\right)\Psi^{(s)} = 0.
\nonumber
\end{eqnarray}
Making some changes in the last equation
\begin{eqnarray*}
\Psi^{(s)}({\bf r}) \to \Psi({\bf r},\gamma) = \Psi^{(s)}({\bf r})
\frac{e^{is(\gamma-\varphi)}}{\sqrt{4\pi}}, \qquad s \to
-i\frac{\partial}{\partial \gamma},\qquad {\rm где} \quad \gamma
\in [0, 4\pi),
\end{eqnarray*}
instead of (\ref{3.7.1}) we have
\begin{eqnarray}
\left[\frac{1}{r^2}\frac{\partial}{\partial r}\left(r^2
\frac{\partial}{\partial r}\right) - \frac{{\hat {\bf L}
}^2}{r^2}\right] \Psi + \frac{2\mu_0}{\hbar^2}\left(\varepsilon +
\frac{e^2}{r}\right) \Psi = 0, \label{3.7.2}
\end{eqnarray}
where
\begin{eqnarray*}
{\hat {\bf L}}^2 = -\left[\frac{1}{\sin
\beta}\frac{\partial}{\partial \beta} \left(\sin
\beta\frac{\partial}{\partial \beta}\right) +
\frac{1}{\sin^2\beta}\left(\frac{\partial^2}{\partial \alpha^2} -
2\cos\beta\frac{\partial^2}{\partial \alpha \partial \gamma}+
\frac{\partial^2}{\partial \gamma^2}\right)\right].
\end{eqnarray*}
Here we have introduced the following notation: $\beta =\theta$ and $\alpha
=\varphi$.

Passing from the spherical coordinates ($r, \alpha, \beta, \gamma$) to the new
ones ($u_0, u_1, u_2, u_3$), according to
\begin{eqnarray}
u_0 + iu_1 = u\cos{\frac{\beta}{2}} e^{i\frac{\alpha +
\gamma}{2}}, \qquad u_2 + iu_3 = u\sin{\frac{\beta}{2}}
e^{i\frac{\alpha - \gamma}{2}}
\label{3.7.3}
\end{eqnarray}
so that $u^2 = u_0^2+u_1^2+u_2^2+u_3^2 = r$, taking into account the fact that
the Laplace operator in the coordinates $u_{\mu}$ has the form
\begin{eqnarray}
\Delta_4 =
\frac{1}{u^3}\frac{\partial}{\partial u}
\left(u^3\frac{\partial}{\partial u}\right) - \frac{4}{u^2}{\hat
{\bf L}}^2, \quad  where \quad \mu =0,1,2,3
\label{3.7.4}
\end{eqnarray}
and introducing the following notation
\begin{eqnarray}
E = 4e^2, \qquad \varepsilon = -\frac{\mu_0 \omega^2}{8}
\label{3.7.5}
\end{eqnarray}
one can easily see that equation (\ref{3.7.2}) turns exactly into the
Schr\"{o}dinger equation for the four-dimensional isotropic oscillator
\begin{eqnarray}
\left[\Delta_4 +
\frac{2\mu_0}{\hbar}\left(E - \frac{\mu_0\omega^2
u^2}{2}\right)\right] \psi({\bf u}) = 0,
\label{3.7.6}
\end{eqnarray}
the energy spectrum of which is given by the well-known formula
\begin{eqnarray}
E_N = \hbar \omega (N+2), \quad where \quad
N=0,1,2,.... \label{3.7.7}
\end{eqnarray}
From a comparison of the energy spectra of the MIC--Kepler problem
(\ref{3.1.11}) and the $4D$ oscillator (\ref{3.7.7}), with taking into account
(\ref{3.7.5}) we get the following relation between the principal quantum
numbers $n$ and $N$:
\begin{eqnarray}
N = 2n-2.
\label{3.7.8}
\end{eqnarray}

Establish a relation between the Cartesian coordinates $x, y, z$ and the
coordinates $u_\mu$. From the definition of the coordinates $u_\mu$
(\ref{3.7.3}) with the allowance made for the notation
\begin{eqnarray}
r=u^2,\qquad \theta = \beta, \qquad \varphi = \alpha,
\label{3.7.8a}
\end{eqnarray}
we obtain the following relations:
\begin{eqnarray*}
\cos\theta &=&
\frac{u_0^2+u_1^2-u_2^2-u_3^2}{u_0^2+u_1^2+u_2^2+u_3^2}, \qquad
\sin\theta =
\frac{2\sqrt{(u_0^2+u_1^2)(u_2^2+u_3^2)}}{u_0^2+u_1^2+u_2^2+u_3^2},
\\ [3mm]
\cos\varphi &=&
\frac{u_0u_2-u_1u_3}{\sqrt{(u_0^2+u_1^2)(u_2^2+u_3^2)}}, \qquad
\sin\varphi
=\frac{u_0u_3+u_1u_2}{\sqrt{(u_0^2+u_1^2)(u_2^2+u_3^2)}}\,.
\end{eqnarray*}
Substituting the latter into the definition of the spherical coordinates
$$ x =r\sin\theta \cos\varphi, \qquad y = r\sin\theta \sin\varphi, \qquad z =
r\cos\theta $$
and recalling that $r=u^2=u_0^2+u_1^2+u_2^2+u_3^2$, we have

\begin{eqnarray}
x &=& 2(u_0u_2 - u_1u_3)\,,
\nonumber \\ [2mm]
y &=& 2(u_0u_3 + u_1u_2)\,,
\nonumber\\ 
\label{3.7.9}
\\
z &=& u_0^2 + u_1^2 - u_2^2 - u_3^2\, \nonumber
\\ [2mm]
\gamma &=& \frac{i}{2}\ln{\frac{(u_0-iu_1)(u_2+iu_3)}
{(u_0+iu_1)(u_2-iu_3)}}\,.
\nonumber
\end{eqnarray}
The first three equations in (\ref{3.7.9}) are the transformation ${\rm I
\!R}^4 \to {\rm I \!R}^3$ used by Kustaanheimo and Stiefel for regularization
of equations of the celestial mechanics \cite{KS}, and are called the
Kustaanheimo--Stiefel transformation (KS-transformation).

Thus, we have proved that the bound MIC--Kepler system is dual to the
four-dimensional isotropic oscillator and that the generalized version of the
KS-transformation (\ref{3.7.9}) together with the ansatz $\psi^{(s)}({\bf r})
\to \psi({\bf r}, \gamma)$ represent the duality transformation.

Let us discuss the problem of degeneracy multiplicity of energy levels
(\ref{3.1.11}) and (\ref{3.7.7}). It follows from formula (\ref{3.1.17}) that
at fixed quantum numbers $n, m$ and $s$ the degeneracy multiplicity of energy
levels (\ref{3.1.11}) equals
\begin{eqnarray*}
g_{nm}^s = n - \frac{|m-s|+|m+s|}{2}.
\end{eqnarray*}
For $s \geq 0$ the degeneracy multiplicity of energy levels(\ref{3.1.11}) at
fixed $n$ and $s$ will be
\begin{eqnarray*}
g_{n}^s = \sum_{|m|\geq s}g_{nm}^s + \sum_{|m|\leq s-1}g_{nm}^s\,,
\end{eqnarray*}
where the upper limit of summation is determined from the condition $g_{nm}^s
\geq 1$, i.е., $|m-s|+|m+s| \leq 2n - 2$. Consequently,
\begin{eqnarray}
g_{n}^s = \sum_{m=-s+1}^{s-1}(n - s) + 2\sum_{m=s}^{n-1}(n - m) =
(n-s)(n+s)\,.
\label{3.7.10}
\end{eqnarray}
An analogous result is obtained if $s < 0$. Quantum numbers $n$ and $s$ take
simultaneously either integer or half-integer values; therefore, summation of
(\ref{3.7.10}) over $s$ gives
\begin{eqnarray*}
g_n = \sum_{s=-n+1}^{n-1} g_{n}^s = \frac{1}{3}n(2n-1)(2n+1),
\end{eqnarray*}
where $g_n$ is the degeneracy multiplicity of energy levels $4D$ of the
oscillator (\ref{3.7.7}). Taking into consideration the fact that, according to
(\ref{3.7.8}), $N = 2n -2$ we arrive at the well-known result \cite{BAKER}:
\begin{eqnarray}
g_N = \frac{1}{6}(N+1)(N+2)(N+3).
\label{3.7.11}
\end{eqnarray}

\section{Four-dimensional oscillator}
\markboth{CHAPTER 5. NONBIJECTIVE TRANSFORMATIONS}{5.5. FOUR-DIMENSIONAL OSCILLATOR}

We consider the problem of four-dimensional isotropic oscillator only in the
following five systems of coordinates: 1) Cartesian coordinates $u_\mu \in
(-\infty, \infty)$ $(\mu =0,1,2,3)$; 2) Euler coordinates $u \in [0, \infty)$,
$\alpha \in [0, 2\pi)$, $\beta \in [0, \pi]$, $\gamma \in [0, 4\pi)$
(\ref{3.7.3}); 3) double polar coordinates $\rho_1,\rho_2 \in [0,\infty)$,
$\varphi_1,\varphi_2 \in [0,2\pi)$; 4) canonical spherical coordinates $u \in
[0, \infty)$, $\psi \in [0, \pi]$, $\theta \in [0, \pi]$, $\varphi \in [0,
2\pi)$; 5) four-dimensional spheroidal coordinates $\xi \in [1, \infty)$, $\eta
[-1, 1]$, $\alpha \in [0, 2\pi)$, $\gamma \in [0, 4\pi)$.

We give the definitions of the last three of the above-listed four-dimensional
coordinates:

\begin{itemize}
\item
Double polar coordinates
\begin{equation}
u_0+i u_1= \rho_1\,e^{i\varphi_1}, \qquad u_2+i u_3= \rho_2\,
e^{i\varphi_2},
\label{3.8.3}
\end{equation}
\begin{eqnarray*}
\Delta_4 = \frac{1}{\rho_1} \frac{\partial}{\partial
\rho_1}\left(\rho_1\frac{\partial} {\partial \rho_1}\right) +
\frac{1}{\rho_2} \frac{\partial}{\partial
\rho_2}\left(\rho_2\frac{\partial} {\partial \rho_2}\right) +
\frac{1}{\rho_1^2}\frac{\partial^2}{\partial \varphi_1^2}+
+\frac{1}{\rho_2^2}\frac{\partial^2}{\partial \varphi_2^2}.
\end{eqnarray*}
\item
Canonical spherical coordinates
\begin{eqnarray}
u_0+iu_1= u\sin\psi \sin\theta\, e^{i\phi}, \qquad
u_2= u\sin\psi \cos\theta, \qquad  u_3= u\cos\psi,
\label{3.8.4}
\end{eqnarray}
\begin{eqnarray*}
\Delta_4 = \frac{1}{u^3} \frac{\partial}{\partial
u}\left(u^3\frac{\partial} {\partial u}\right) + \frac{1}{u^2}
\left[\frac{1}{\sin^2\psi}\frac{\partial}{\partial
\psi}\left(\sin^2\psi\frac{\partial} {\partial \psi}\right)
-\frac{{\hat l}^2}{\sin^2\psi}\right],
\end{eqnarray*}
where
\begin{eqnarray*}
{\hat l}^2 = -\left[\frac{1}{\sin\theta}\frac{\partial}{\partial
\theta}\left(\sin\theta\frac{\partial}{\partial \theta}\right)
+\frac{1}{\sin^2\theta}\frac{\partial^2}{\partial \phi^2}\right].
\end{eqnarray*}
\item
Spheroidal coordinates
\begin{eqnarray}
u_0+iu_1 = \frac{d}{2}\sqrt{(\xi+1)(1+\eta)}\,\,
e^{i\frac{\alpha+\gamma}{2}}, \quad u_2+iu_3 =
\frac{d}{2}\sqrt{(\xi-1)(1-\eta)}\,\,
e^{i\frac{\alpha-\gamma}{2}}.
\label{3.11.1}
\end{eqnarray}
Here $d$ is the spheroidal parameter $(0\leq d < \infty)$, and the Laplace
operator has the form
\begin{eqnarray*}
\Delta_4 &=& \frac{8}{d^2(\xi - \eta)}
\Biggl\{\frac{\partial}{\partial \xi}\left[(\xi^2-1)
\frac{\partial}{\partial \xi}\right] + \frac{\partial}{\partial
\eta}\left[(1- \eta^2) \frac{\partial}{\partial \eta}\right]-   \\
[3mm]
&-&\frac{1}{2}\left(\frac{1}{\xi + 1} - \frac{1}{1 +
\eta}\right) \left(\frac{\partial}{\partial \alpha} +
\frac{\partial}{\partial \gamma} \right)^2 +
\frac{1}{2}\left(\frac{1}{\xi - 1} + \frac{1}{1 - \eta}\right)
\left(\frac{\partial}{\partial \alpha} - \frac{\partial}{\partial
\gamma} \right)^2\Biggr\}.
\end{eqnarray*}
\end{itemize}

Now we give explicit expressions for the fundamental bases of the
four-dimensional isotropic oscillator in the above-listed systems of
coordinates:

\begin{itemize}
\item
The Cartesian basis
\begin{eqnarray}
\left|\,{\cal C}\,\right\rangle =\left|N_0, N_1, N_2,
N_3\right\rangle = a^{2}\,\overline{H}_{N_0}\left(a u_0
\right)\,\overline{H}_{N_1}\left(a u_1\right)\,
\overline{H}_{N_2}\left(a u_2\right)\,\overline{H}_{N_3}\left(a u_3\right),
\label{3.8.5}
\end{eqnarray}
where $a= (\mu_0\omega/\hbar)^{1/2}$, and $\overline{H}_{n}\left(x\right)$ is
the orthonormalized Hermite polynomial.
\item
The Euler basis
\begin{eqnarray}
\left|\,{\cal E}\,\right\rangle \equiv \left|N, L, M, M'\right\rangle =
\sqrt{\frac{2L+1}{2\pi^2}}R_{NL}\left(u\right)
D_{M M'}^L(\alpha,\, \beta,\, \gamma),
\label{3.8.6}
\end{eqnarray}
where
\begin{eqnarray}
R_{NL}\left(u\right) =
\sqrt{\frac{2\left(\frac{N}{2}+L+1\right)!}
{\left(\frac{N}{2}-L\right)!}}\,
\frac{a^2(a u)^{2L}}{(2L+1)!}\,e^{-\frac{a^2 u^2}{2}}
F\left(-\frac{N}{2}+L; 2L+2; a^2 u^2\right).
\label{3.8.6a}
\end{eqnarray}
The Euler basis of the four-dimensional oscillator (\ref{3.8.6}) is
simultaneously proper for the operators ${\hat {\bf L}}^2$, ${\hat
L}_3=-i\partial/\partial \alpha$ and ${\hat L}_{3'}=-i\partial/\partial
\gamma$, with
\begin{eqnarray}
{\hat {\bf L}}^2\left|N, L, M, M'\right\rangle  &=& L(L+1)\left|N,
L, M, M'\right\rangle,  \nonumber \\ [3mm]
{\hat L}_3 \left|N,
L,M, M'\right\rangle &=& M \left|N, L, M, M'\right\rangle,
\label{3.8.7}
\\ [3mm]
{\hat L}_{3'}\left|N, L, M, M'\right\rangle
&=& M' \left|N, L, M, M'\right\rangle. \nonumber
\end{eqnarray}
\item
The double polar basis
\begin{eqnarray}
\left|\,2P\,\right\rangle \equiv \left|N_{\rho_1}, N_{\rho_2}, m_1,
m_2\right\rangle = \frac{1}{2\pi}f_{N_{\rho_1},m_1}(a^2\rho_1^2)\,
f_{N_{\rho_2},m_2}(a^2\rho_2^2)\, e^{im_1\varphi_1}\,e^{im_2\varphi_2},
\label{3.8.9}
\end{eqnarray}
where
\begin{eqnarray}
f_{pq}(x) = \frac{a}{|q|!} \sqrt{\frac{2(p+|q|)!}{p!}}
e^{-x/2}x^{|q|/2} F(-p,\, |q|+1,\, x).
\label{3.8.10}
\end{eqnarray}
Quantum numbers $N_{\rho_1}$, $N_{\rho_2}$, $m_1$ and $m_2$ are related to the
principal quantum number $N$ as follows:
\begin{eqnarray}
N = 2N_{\rho_1} + 2N_{\rho_2} + |m_1| + |m_2|.
\label{3.8.11}
\end{eqnarray}
The double polar basis is an eigenvalue of the following operators:
\begin{eqnarray}
{\hat {\cal P}} \left|N_{\rho_1}, N_{\rho_2}, m_1,
m_2\right\rangle &=& (2N_{\rho_1}-2N_{\rho_2}+|m_1|-|m_2|)
\left|N_{\rho_1}, N_{\rho_2}, m_1, m_2\right\rangle,
\nonumber
\\[3mm]
{\hat L}_3 \left|N_{\rho_1}, N_{\rho_2}, m_1, m_2\right\rangle &=&
\frac{m_1+m_2}{2} \left|N_{\rho_1}, N_{\rho_2}, m_1,
m_2\right\rangle, \\ [3mm] {\hat L}_{3'} \left|N_{\rho_1},
N_{\rho_2}, m_1, m_2\right\rangle &=& \frac{m_1-m_2}{2}
\left|N_{\rho_1}, N_{\rho_2}, m_1, m_2\right\rangle, \nonumber
\label{3.8.12}
\end{eqnarray}
where the operator ${\hat {\cal P}}$ is
\begin{eqnarray}
{\hat {\cal P}} = \frac{\hbar}{2\mu_0\omega}\left(
-\frac{\partial^2}{\partial u_0^2} - \frac{\partial^2}{\partial
u_1^2} + \frac{\partial^2}{\partial u_2^2} +
\frac{\partial^2}{\partial u_3^2} \right) + \frac{\mu_0
\omega}{2\hbar}\left(u_0^2+u_1^2-u_2^2-u_3^2\right).
\label{3.8.13}
\end{eqnarray}
\item
The canonical basis
\begin{eqnarray}
\left|\,{\cal K}\,\right\rangle \equiv \left|N, J, l, \overline{m}
\right\rangle = R_{NJ}(u)Y_{Jl\overline{m}}(\psi, \theta, \phi),
\label{3.8.14}
\end{eqnarray}
where
\begin{eqnarray}
R_{NJ}(u) = \frac{a^2(au)^{J}}{(J+1)!}
\sqrt{\frac{2\left(\frac{N+J}{2}+1\right)!}
{\left(\frac{N-J}{2}\right)!}}\, e^{-\frac{a^2 u^2}{2}}
F\left(-\frac{N-J}{2}; J+2; a^2u^2\right)
\label{3.8.14a}
\end{eqnarray}
and
\begin{eqnarray}
Y_{Jl\overline{m}}(\psi, \theta, \phi) =
2^l\,l!\sqrt{\frac{(2J+2)(J-l)!}{\pi
(J+l+1)!}}\,\left(\sin\psi\right)^l\,C_{J-l}^{l+1}\left(\cos\psi\right)\,
Y_{l\overline{m}}\left(\theta, \phi\right).
\label{3.8.15}
\end{eqnarray}
Here  $C_{n}^{\lambda}(x)$ is the Gegenbauer polynomial and
$Y_{l\overline{m}}\left(\theta, \phi\right)$ is an ordinary spherical function.

For the canonical basis the following formulae are valid:
\begin{eqnarray}
{\hat{\bf J}}^2\,\left|N, J, l, \overline{m}\right\rangle &=&
J(J+2)\,\left|N, J, l, \overline{m}\right\rangle,
\nonumber \\ [2mm]
{\hat{\bf l}}^2\,\left|N, J, l, \overline{m}\right\rangle &=&
l(l+1)\,\left|N, J, l, \overline{m}\right\rangle,
\label{3.8.16}
\\ [2mm]
{\hat l}_3\,\left|N, J, l, \overline{m}\right\rangle &=&
\overline{m}\,\left|N, J, l, \overline{m}\right\rangle.
\nonumber
\end{eqnarray}
The ranges of variation of quantum numbers are: $J=0,2,...,N$ or $J=1,3,...,N$
for even and odd $N$, respectively, and $0 \leq l \leq J$, $|\overline{m}| \leq
l$.
\end{itemize}

Consider the problem of interbasis expansions of the four-dimensional isotropic
oscillator. The general number of nontrivial interbasis expansions equals
twelve. By virtue of the unitarity condition the coefficients of "direct" and
"inverse" expansions are obtained from each other by complex conjugation. Thus,
it is sufficient to study the following six expansions:
\begin{eqnarray}
\left|\,{\cal C}\,\right\rangle
&=&
{\sum_{m_1=-N_0-N_1}^{N_0+N_1}}'\,{\sum_{m_2=
-N_2-N_3}^{N_2+N_3}}'\,\left\langle\, 2P\,|\,{\cal C}
\,\right\rangle\, \left|\,2P\, \right\rangle,
\label{3.9.1}
\\[3mm]
\left|\,{\cal C}\,\right\rangle
&=&
\sum_{L=\frac{|M|+|M'|}{2}}^{N/2}\,{\sum_{M=-N_0-N_1}^{N_0+N_1}}'\,
{\sum_{M'=-N_2-N_3}^{N_2+N_3}}'\,\left\langle\, {\cal E} \,|\,{\cal C}
\,\right\rangle\, \left|\,{\cal E}\, \right\rangle,
\label{3.9.2}
\\[3mm]
\left|\,{\cal C}\,\right\rangle
&=&
{\sum_{J=0,1}^{N}}\,'\,\sum_{l=0}^{J}\,\sum_{\overline{m}=
-l}^{l}\,\left\langle\, {\cal K} \,|\,{\cal C}\,\right\rangle\,
\left|\,{\cal K}\, \right\rangle,
\label{3.9.3}
\\[3mm]
\left|\,{\cal K}\,\right\rangle
&=&
\sum_{L=\frac{|M|+|M'|}{2}}^{N/2}\,\sum_{M,M'=-L}^{L}\,\left\langle\,
{\cal E} \,|\,{\cal K}\,\right\rangle\, \left|\,{\cal E}\,
\right\rangle,
\label{3.9.4}
\\[3mm]
\left|\,2P\,\right\rangle
&=&
\sum_{L=\frac{|M|+|M'|}{2}}^{N/2}\,\sum_{M,M'=-L}^{L}\,\left\langle\,
{\cal E} \,|\,2P\,\right\rangle\, \left|\,{\cal E}\,
\right\rangle,
\label{3.9.5}
\\[3mm]
\left|\,{\cal K}\,\right\rangle
&=&
{\sum_{J=0}^{N/2}}\,'\,\sum_{l=0}^{J}\,\sum_{\overline{m}=-l}^{l}\,
\left\langle\,2P\,|\,{\cal K}\,\right\rangle\, \left|\,2P\,\right\rangle.
\label{3.9.6}
\end{eqnarray}
Primes over the summation symbol means that the corresponding summation is only
over those values of indices whose parity coincides with that of a maximum
possible value indicated in the sum.

All six coefficients (\ref{3.9.1}) - (\ref{3.9.6}) are explicitly expressed in
terms of three structural elements: the Wigner $d$ function
$d_{m,m'}^j(\pi/2)$, Clebsch--Gordan coefficients
$C_{a,\alpha;b,\beta}^{c,\gamma}$ and the objects ${\cal
M}_{a,\alpha;b,\beta}^{c,\gamma}$ obtained by an analytic continuation of
ordinary Clebsch--Gordan coefficients $C_{a,\alpha;b,\beta}^{c,\gamma}$ to the
region of one-fourth values of some indices. According to \cite{VAR},
\begin{eqnarray*}
{\cal M}_{a,\alpha;b,\beta}^{c,\gamma} &=&
\frac{\delta_{\gamma,\alpha+\beta}\Delta(a,b,c)}
{\Gamma(a+b-c+1)\Gamma(c-b+\alpha+1)\Gamma(c-a-\beta+1)}\times
\\ [3mm] &\times&
\left[\frac{(2c+1)\Gamma(a+\alpha+1)\Gamma(b-\beta+1)
\Gamma(c+\gamma+1)\Gamma(c-\gamma+1)}{\Gamma(a-\alpha+1)
\Gamma(b+\beta+1)}\right]^{1/2}\times \\ [3mm]
&\times& {_3F}_2
\left\{
\begin{array}{l}
c-a-b, -a + \alpha, -b - \beta \\ [3mm]
c-a-\beta +1, c-b + \alpha +1\\
\end{array}
\biggr| 1 \right\}.
\end{eqnarray*}
In the last formula
\begin{eqnarray*}
\Delta(a,b,c) =
\left[\frac{\Gamma(a+b-c+1)\Gamma(a-b+c+1)\Gamma(d-a+c+1)}
{\Gamma(a+b+c+2)}\right]^{1/2}.
\end{eqnarray*}
It should be noted that the objects ${\cal M}_{a,\alpha;b,\beta}^{c,\gamma}$
have already appeared in the fourth chapter in the problem of interbasis
expansions in the multidimensional isotropic oscillator.

Let us write down the formulae determining the explicit form of the
coefficients of interbasis expansions (\ref{3.9.1}) - (\ref{3.9.6})
\begin{eqnarray}
\left\langle\, 2P\,|\,{\cal C}\,\right\rangle
&=& e^{i\pi\left(N_1+N_3-|m_1|-|m_2|\right)/2}\,
d_{\frac{m_1}{2},\,\frac{N_0-N_1}{2}}^{\frac{N_0+N_1}{2}}
\left(\frac{\pi}{2}\right)\,
d_{\frac{m_2}{2},\,\frac{N_2-N_3}{2}}^{\frac{N_2+N_3}{2}}
\left(\frac{\pi}{2}\right),
\label{3.9.9}
\\[3mm]
\left\langle\, {\cal E} \,|\,{\cal C}\,\right\rangle
&=& e^{i\pi
\left(N+N_1-N_3-2|M|-M'\right)/2}\,
d_{\frac{M+M'}{2},\,\frac{N_0-N_1}{2}}^{\frac{N_0+N_1}{2}}
\left(\frac{\pi}{2}\right)\,
d_{\frac{M-M'}{2},\,\frac{N_2-N_3}{2}}^{\frac{N_2+N_3}{2}}
\left(\frac{\pi}{2}\right)\times
\nonumber \\[1mm]
\label{3.9.10}
\\[1mm]
&\times&\, C^{L, \frac{|M|+|M'|}{2}}_{\frac{N-|M|+|M'|}{4},\,
\frac{N-2N_0-2N_1+|M|+|M'|}{4};
\,\frac{N+|M|-|M'|}{4},\,\frac{2N_0+2N_1-N+|M|+|M'|}{4}}\,,
\nonumber
\\[3mm]
\left\langle\, {\cal K} \,|\,{\cal C}\,\right\rangle\,
&=&
e^{i\pi \left(|M|-N_1\right)/2}\,
d_{\frac{M}{2},\,\frac{N_0-N_1}{2}}^{\frac{N_0+N_1}{2}}
\left(\frac{\pi}{2}\right)\,{\cal M}_{\frac{N+l+1}{4},
\frac{N-2N_3+l+1}{4};\,\frac{N-l-1}{4},\frac{2N_3-N+l-1}{4}}
^{\frac{J}{2},\frac{l}{2}}\times\,
\nonumber \\[1mm]
\label{3.9.11}
\\[1mm]
&\times&\,{\cal M}_{\frac{N-N_4+|M|}{4},\frac{N_0+N_1-N_2+|M|}{4};\,
\frac{N-N_4-|M|-1}{4},
\frac{N_2-N_1-N_0+|M|-1}{4}}^{\frac{2l-1}{4},\,\frac{2|M|-1}{4}\gamma}\,,
\nonumber
\\[3mm]
\left\langle\, {\cal E} \,|\,{\cal K}\,\right\rangle
&=& e^{i\pi
\left(J+l+2|M|-|M'|\right)/2}\,\delta_{L,\frac{J}{2}}\,
\delta_{\overline{m},M+M'}
C_{\frac{J}{2}, M'; \frac{J}{2}, M}^{l,\overline{m}},
\label{3.9.12}
\\[3mm]
\left\langle\, {\cal E} \,|\,2P\,\right\rangle
&=&
(-1)^{N_{\rho_1}+\frac{m_2+|m_2|}{2}}\,\delta_{M,\frac{m_1+m_2}{2}}\,
\delta_{M',\frac{m_1-m_2}{2}}\times\,
\nonumber \\[1mm]
\label{3.9.13}
\\[1mm]
&\times&\, C^{L,
\frac{|m_1|+|m_2|}{2}}_{\frac{N_{\rho_1}+N_{\rho_2}+|m_2|}{2},\,
\frac{N_{\rho_2}-N_{\rho_1}+|m_2|}{2}\,;
\,\frac{N_{\rho_1}+N_{\rho_2}+|m_1|}{2},\,
\frac{N_{\rho_1}-N_{\rho_2}+|m_1|}{2}}\,,
\nonumber
\\[3mm]
\left\langle\, 2P \,|\,{\cal K}\,\right\rangle
&=& e^{i\pi
\left(2N_{\rho_1}+J+l+2|m_1|-|m_2|\right)/2}\,
\delta_{\overline{m}, m_1}\, C_{\frac{J}{2}, \frac{m_1-m_2}{2};
\frac{J}{2}, \frac{m_1+m_2}{2}}^{l,\overline{m}}\times
\nonumber \\[1mm]
\label{3.9.14}
\\[1mm]
&\times& C^{\frac{J}{2},
\frac{|m_1|+|m_2|}{2}}_{\frac{N_{\rho_1}+N_{\rho_2}+|m_2|}{2},\,
\frac{N_{\rho_2}-N_{\rho_1}+|m_2|}{2}\,;
\,\frac{N_{\rho_1}+N_{\rho_2}+|m_1|}{2},\,
\frac{N_{\rho_1}-N_{\rho_2}+|m_1|}{2}}.
\nonumber
\end{eqnarray}
We now describe the methods used in calculating interbasis coefficients. In
calculating the coefficients (\ref{3.9.9}), (\ref{3.9.11}), (\ref{3.9.12}) and
(\ref{3.9.13}) we made use of the known recipes suggested in the fourth chapter
and paper \cite{KIL} which are useful for the isotropic oscillator of arbitrary
dimension. The results (\ref{3.9.10}) и (\ref{3.9.14}) are obtained by the
formulae
\begin{eqnarray*}
\left\langle\,{\cal E}\,|\,{\cal C}\,\right\rangle =
\sum_{\left|\,2P\,\right\rangle}\,\left\langle\, {\cal E}
\,|\,2P\,\right\rangle\,\left\langle\,2P\,|\,{\cal
C}\,\right\rangle, \qquad
\left\langle\,2P\,|\,{\cal K}\,\right\rangle =
\sum_{\left|\,2P\,\right\rangle}\,\left\langle\, 2P \,|\,{\cal
E}\,\right\rangle\,\left\langle\,{\cal E}\,|\,{\cal
K}\,\right\rangle.
\end{eqnarray*}
Now having all information on interbasis expansions for the fundamental bases
of the four-dimensional isotropic oscillator we can construct the spheroidal
basis. First of all, it is to establish that the spheroidal basis of the
four-dimensional isotropic oscillator is the solution of the following system
of equations:
\begin{eqnarray*}
{\hat H}\,\left|N, p, M, M'\right\rangle &=& E
\,\left|N, p, M, M'\right\rangle, \qquad
{\hat Q}\,\left|N, p, M, M'\right\rangle =
Q_p\,\left|N, p, M, M'\right\rangle\,, \\ [2mm]
{\hat L}_3 \left|N, p,M, M'\right\rangle &=&
M \left|N, p, M, M'\right\rangle, \qquad
{\hat L}_{3'}\left|N, L, M, M'\right\rangle
= M' \left|N, L, M, M'\right\rangle,
\end{eqnarray*}
where ${\hat H}$ is the Hamiltonian of the four-dimensional isotropic
oscillator and
\begin{eqnarray*}
{\hat Q} = {\hat{\bf L}}^2 + \frac{a^2 d^2}{4}{\hat {\cal P}}.
\end{eqnarray*}
The explicit form of the spheroidal integral of motion suggests that the
spheroidal basis of the four-dimensional isotropic oscillator can most
conveniently  be represented as expansion over the Euler and double polar
bases, i.e., in the form
\begin{eqnarray*}
\left|\,NpMM'\,\right\rangle
&=&
\sum_L\,
U_{NpMM'}^{L}(d)\left|\,NLMM'\,\right\rangle,
\\[3mm]
\left|\,NpMM'\,\right\rangle
&=&
\sum_{N_{\rho_1}}
V_{NpMM'}^{N_{\rho_1}}(d)\,
\left|\,N_{\rho_1}, N_{\rho_2}, (M+M')/2,
(M-M')/2\,\right\rangle,
\end{eqnarray*}
where summation over $L$ and over $N_{\rho_1}$ is carried out within the
limits
$$
\frac{|M+M'|+|M-M'|}{2} \leq L \leq \frac{N}{2}, \qquad
0 \leq N_{\rho_1} \leq \frac{N-|M+M'|+|M-M'|}{2}.
$$
The coefficients $U_{NpMM'}^{L}(d)$ and $V_{NpMM'}^{N_{\rho_1}}(d)$ are
determined from the following trinomial recurrence
relations:
\begin{eqnarray*}
A_{L+1} U_{NpMM'}^{L+1} + A_L U_{NpMM'}^{L-1}
&=&
\left\{\frac{4}{a^2 d^2}\left[Q_p-L(L+1)\right]-B_L\right\}
U_{NpMM'}^{L},
\\[3mm]
C_{N_{\rho_1}+1}V_{NpMM'}^{N_{\rho_1}+1} +
C_{N_{\rho_1}} V_{NpMM'}^{N_{\rho_1}-1}
&=&
\left[Q_p - a^2 d^2 \left(N_{\rho_1} - \frac{N -
|M+M'|}{4}\right) - D_{N_{\rho_1}}\right]V_{NpMM'}^{N_{\rho_1}},
\end{eqnarray*}
and the orthonormalization conditions
\begin{eqnarray*}
\sum_{L}\,U_{Np_1ms}^L (d)\,U_{Np_2ms}^{L*} (d) = \delta_{p_1p_2},
\qquad \sum_{N_{\rho_1}}\,{\tilde V}_{Np_1MM'}^{N_{\rho_1}} (d)
\,{\tilde V}_{Np_2MM'}^{N_{\rho_1}*} (d)= \delta_{p_1p_2}.
\end{eqnarray*}
The quantities $A_L$, $B_L$, $C_{N_{\rho_1}}$ and $D_{N_{\rho_1}}$ are given by
the relations
\begin{eqnarray*}
A_{L} &=& - \left\{\frac{[(N+2)^2-4L^2](L^2-M_+^2)
(L^2-M_-^2)}
{L^2(2L-1)(2L+1)}\right\}^{\frac{1}{2}},  \qquad
B_L = \frac{M_+M_-(N+2)}{L(L+1)},
\\ [3mm]
C_{N_{\rho_1}} &=& -\sqrt{\,N_{\rho_1}\,(N_{\rho_2}+1)\,
\left(N_{\rho_1}+\frac{|M+M'|}{2}\right)\,
\left(N_{\rho_2}+\frac{|M-M'|}{2}+1\right)},      \\ [3mm]
D_{N_{\rho_1}} &=&
\frac{N(N+4)}{8}+\frac{1}{8}M_-^2
+\frac{1}{2}\left(N_{\rho_1}-N_{\rho_2} + \frac{|M+M'|}{2}\right)
\left(N_{\rho_2}-N_{\rho_1}+\frac{|M-M'|}{2}\right),
\end{eqnarray*}
where for convenience we introduced the notation $M_{\pm} = (M \pm M')/2$.

\section{Interbasis expansions between the MIC--Kepler problem and
the four-dimensional isotropic oscillator}
\markboth{CHAPTER 5. NONBIJECTIVE TRANSFORMATIONS}{5.6. INTERBASIS EXPANSIONS}

In this section, we calculate coefficients relating the fundamental bases of
the bound MIC--Kepler problem and the four-dimensional isotropic oscillator.

Write down all eight expansions
\begin{eqnarray}
\left|\,n j m s\,\right\rangle &=& \sum\,\left\langle\,N L M
M'\,|\,n j m s\,\right\rangle\, \left|\,N L M M'\, \right\rangle,
\label{3.10.1}
\\ [3mm]
\left|\,n j m s\,\right\rangle &=&
\sum\,\left\langle\,N_{\rho_1} N_{\rho_2} m_1 m_2\,|\,n j m s
\,\right\rangle\, \left|\,N_{\rho_1} N_{\rho_2} m_1 m_2\,
\right\rangle,
\label{3.10.2}
\\ [3mm]
\left|\,n j m s\,\right\rangle &=& \sum\,\left\langle\,N J l
\overline{m}\,|\,n j m s \,\right\rangle\, \left|\,N J l
\overline{m}\, \right\rangle,
\label{3.10.3}
\\ [3mm]
\left|\,n j m s\,\right\rangle &=& \sum\,\left\langle\,N_0 N_1 N_2
N_3 \,|\,n j m s \,\right\rangle\, \left|\,N_0 N_1 N_2 N_3\,
\right\rangle,
\label{3.10.4}
\\ [3mm]
\left|\,n_1 n_2 m s\,\right\rangle &=&
\sum\,\left\langle\,N_{\rho_1} N_{\rho_2} m_1 m_2 \,|\,n_1 n_2 m s
\,\right\rangle\, \left|\,N_{\rho_1} N_{\rho_2} m_1 m_2\,
\right\rangle,
\label{3.10.5}
\\ [3mm]
\left|\,n_1 n_2 m s\,\right\rangle &=& \sum\,\left\langle\,N L M
M' \,|\,n_1 n_2 m s \,\right\rangle\, \left|\,N L M M'\,
\right\rangle,
\label{3.10.6}
\\ [3mm]
\left|\,n_1 n_2 m s\,\right\rangle &=& \sum\,\left\langle\,N J l
\overline{m}\,|\,n_1 n_2 m s \,\right\rangle\, \left|\,N J l
\overline{m}\, \right\rangle,
\label{3.10.7}
\\ [3mm]
\left|\,n_1 n_2 m s\,\right\rangle &=& \sum\,\left\langle\,N_0 N_1
N_2 N_3\,|\,n_1 n_2 m s \,\right\rangle\, \left|\,N_0 N_1 N_2
N_3\, \right\rangle.
\label{3.10.8}
\end{eqnarray}
Let us first consider expansion (\ref{3.10.1}). Taking account of the relation
connecting the three-dimensional spherical and four-dimensional Euler
coordinates (\ref{3.7.8a}), substituting into (\ref{3.10.1}) explicit
expressions for the spherical basis of the MIC--Kepler problem (\ref{3.1.10})
and the Euler basis of the four-dimensional isotropic oscillator (\ref{3.8.6}),
passing in the left-hand side from the spherical coordinates to the Euler ones,
and taking finally into account relation (\ref{3.1.7}), we get that
\begin{eqnarray}
\left\langle\,N L M M'\,|\,n j m s\,\right\rangle =
4n\,\sqrt{r_0}\,\delta_{n,\frac{N}{2}+1}\,\delta_{L,j}\,
\delta_{M,m}\,\delta_{M',s}.
\label{3.10.9}
\end{eqnarray}
In an analogous way we can show that the transition matrix in expansion
(\ref{3.10.5}) is also diagonal and has the form
\begin{eqnarray}
\left\langle\,N_{\rho_1} N_{\rho_2} m_1 m_2 \,|\,n_1 n_2 m s
\,\right\rangle = 4n\, \sqrt{r_0}\, \delta_{N_{\rho_1},
n_1}\,\delta_{N_{\rho_2}, n_2}\,
\delta_{m_1,m+s}\,\delta_{m_2,m-s}.
\label{3.10.10}
\end{eqnarray}
In deriving the last formula we proceeded from that the parabolic and double
polar coordinates were related with each other by: $\mu = 2\rho_1^2$, $\nu =
2\rho_2^2$, $\alpha = \varphi_1 + \varphi_2$, $\gamma = \varphi_1 - \varphi_2$.
The latter can easily be established with the use of the definitions of these
coordinates and the $KS$ transformation (\ref{3.7.9}).

Using now formulae (\ref{3.2.8}), (\ref{3.2.10}), (\ref{3.10.9}), and
(\ref{3.10.5}), and the identities
\begin{eqnarray*}
\left\langle\,N_{\rho_1} N_{\rho_2} m_1 m_2 \,|\,n j m s
\,\right\rangle
&=& \sum_{\left|\,n_1 n_2 m
s\,\right\rangle}\,\left\langle\,N_{\rho_1} N_{\rho_2} m_1 m_2
\,|\,n_1 n_2 m s \,\right\rangle \left\langle\,n_1 n_2 m s \,|\,n
j m s \,\right\rangle,
\\[3mm]
\left\langle\,N L M M' \,|\,n_1 n_2 m s \,\right\rangle
&=&
\sum_{\left|\,n j m s\,\right\rangle}\left\langle\,N L M M' \,|\,n
j m s \,\right\rangle \left\langle\,n j m s \,|\,n_1 n_2 m s
\,\right\rangle
\end{eqnarray*}
we find explicit expressions for the expansion coefficients(\ref{3.10.2}) and
(\ref{3.10.6}):
\begin{eqnarray}
\left\langle\,N_{\rho_1} N_{\rho_2} m_1 m_2 \,|\,n j m s
\,\right\rangle = (-1)^{N_{\rho_1}+\frac{m-s+|m-s|}{2}}\, 4n\,
\sqrt{r_0}\, \delta_{m_1,m+s}\,\delta_{m_2,m-s}\times
\nonumber \\ [1mm]
\label{3.10.11}
\\ [1mm]
\times \, C_{\frac{N_{\rho_1}+N_{\rho_2}+|m-s|}{2},
\frac{N_{\rho_2}-N_{\rho_1}+|m-s|}{2};
\frac{N_{\rho_1}+N_{\rho_2}+|m+s|}{2},
\frac{N_{\rho_1}-N_{\rho_2}+|m+s|}{2}}^
{j,\,m_+}\,,  
\nonumber
\end{eqnarray}
\begin{eqnarray}
\left\langle\,N L M M' \,|\,n_1 n_2 m s \,\right\rangle =
(-1)^{n_1+\frac{m-s+|m-s|}{2}}\, 4n\,
\sqrt{r_0}\,\delta_{n,\frac{N}{2}+1}\,\delta_{M,m+s}\,\delta_{M',s}\times
\nonumber \\ [1mm]
\label{3.10.12}
\\ [1mm]
\times \,
C_{\frac{n-m_- -1}{2},\frac{n-m_- -1}{2}-n_1; \frac{n+m_-
-1}{2},\frac{m_+ +|m+s|-n+1}{2}+n_1}^ {L, m_+}\,.
\nonumber
\end{eqnarray}
For complete solution of the problem we need to calculate the coefficients
$\left\langle\,N J l \overline{m}\,|\,n j m s \,\right\rangle$,\,\,\,
$\left\langle\,N_0 N_1 N_2 N_3 \,|\,n j m s \,\right\rangle$,\,\,\,
$\left\langle\,N J l \overline{m}\,|\,n_1 n_2 m s \,\right\rangle$\, и\,
$\left\langle\,N_0 N_1 N_2 N_3\,|\,n_1 n_2 m s \,\right\rangle$. The
coefficients can be obtained by the following formulae:
\begin{eqnarray*}
\left\langle\,N J l \overline{m} \,|\,n j m s \,\right\rangle
&=&
\sum\,\left\langle\,N J l \overline{m} \,|\,N L M M'
\,\right\rangle \left\langle\,N L M M' \,|\,n j m s
\,\right\rangle\,,
\\[3mm]
\left\langle\,N_0 N_1 N_2 N_3\,|\,n j m s \,\right\rangle
&=&
\sum\,\left\langle\,N_0 N_1 N_2 N_3 \,|\,N L M M' \,\right\rangle
\left\langle\,N L M M' \,|\,n j m s \,\right\rangle\,,
\\[3mm]
\left\langle\,N J l \overline{m} \,|\,n_1 n_2 m s \,\right\rangle
&=& \sum\,\left\langle\,N J l \overline{m} \,|\,N_{\rho_1}
N_{\rho_2} m_1 m_2\,\right\rangle \left\langle\,N_{\rho_1}
N_{\rho_2} m_1 m_2 \,|\,n_1 n_2 m s \,\right\rangle\,,
\\[3mm]
\left\langle\,N_0 N_1 N_2 N_3\,|\,n_1 n_2 m s \,\right\rangle
&=&
\sum\,\left\langle\,N_0 N_1 N_2 N_3 \,|\,N_{\rho_1} N_{\rho_2} m_1
m_2 \,\right\rangle \left\langle\,N_{\rho_1} N_{\rho_2} m_1 m_2
\,|\, n_1 n_2 m s\,\right\rangle\,.
\end{eqnarray*}
Here $\left\langle\,N J l \overline{m} \,|\,N L M M' \,\right\rangle\equiv
\left\langle\,{\cal K}\,|\,{\cal E}\,\right\rangle$, $\left\langle\,N_0 N_1 N_2
N_3 \,|\,N L M M' \,\right\rangle \equiv \left\langle\,{\cal C}\,|\,{\cal
E}\,\right\rangle$, \\ $\left\langle\,N J l \overline{m} \,|\,N_{\rho_1}
N_{\rho_2} m_1 m_2\,\right\rangle \equiv \left\langle\,{\cal
K}\,|\,2P\,\right\rangle$ и $\left\langle\,N_0 N_1 N_2 N_3 \,|\,N_{\rho_1}
N_{\rho_2} m_1 m_2 \,\right\rangle \equiv \left\langle\,{\cal
C}\,|\,2P\,\right\rangle$.
Then using explicit expressions for the coefficients (\ref{3.9.9}),
(\ref{3.9.10}), (\ref{3.9.12}), (\ref{3.9.14}), (\ref{3.10.9}), and
(\ref{3.10.10}), we have
\begin{eqnarray}
\left\langle\,N J l \overline{m} \,|\,n j m s \,\right\rangle =
4n\sqrt{r_0}\,e^{i\pi (2j+l+2|m|-|s|)/2}\,\delta_{n,\frac{N}{2}+1}
C_{j,s;j,m}^{l,m+s},
\label{3.10.13}
\end{eqnarray}
\begin{eqnarray}
\left\langle\,N_0 N_1 N_2 N_3\,|\,n j m s \,\right\rangle =
4n\sqrt{r_0}\,e^{i\pi
(N+N_1-N_3-2|m|-s)/2}\,\delta_{n,\frac{N}{2}+1} \,
d_{\frac{m+s}{2},\,\frac{N_0-N_1}{2}}^{\frac{N_0+N_1}{2}}
\left(\frac{\pi}{2}\right)\times \nonumber \\ [1mm]
\label{3.10.14}
\\ [1mm]
\times\,d_{\frac{m-s}{2},\,\frac{N_2-N_3}{2}}^{\frac{N_2+N_3}{2}}
\left(\frac{\pi}{2}\right)\,
C^{j, \frac{|m|+|s|}{2}}_{\frac{N-|m|+|s|}{4},\,
\frac{N-2N_0-2N_1+|m|+|s|}{4};
\,\frac{N+|m|-|s|}{4},\,\frac{2N_0+2N_1-N+|m|+|s|}{4}}\,,
\nonumber
\end{eqnarray}
\begin{eqnarray}
\left\langle\,N J l \overline{m} \,|\,n_1 n_2 m s \,\right\rangle
= 4n\sqrt{r_0}\,e^{i\pi (2n_1+ J +l +2|m+s|-|m-s_)/2}\,
\delta_{n,\frac{N}{2}+1}\, \delta_{\overline{m},m+s}\times
\nonumber \\ [1mm]
\label{3.10.15}
\\ [1mm]
\times\, C_{\frac{J}{2}, s; \frac{J}{2}, m}^{l, m+s}\,
C^{\frac{J}{2}, \frac{|m+s|+|m-s|}{2}}_{\frac{n_1+n_2+|m-s|}{2},\,
\frac{n_2-n_1-+|m-s|}{2};
\,\frac{n_1+n_2+|m+s|}{2},\,\frac{n_1-n_2-N+|m+s|}{2}}\,,
\nonumber
\end{eqnarray}
\begin{eqnarray}
\left\langle\,N_0 N_1 N_2 N_3\,|\,n_1 n_2 m s \,\right\rangle =
4n\sqrt{r_0}\,e^{i\pi (N_1+N_3-|m+s|-|m-s|)/2}\,
\delta_{n,\frac{N}{2}+1}\times \nonumber \\ [1mm]
\label{3.10.16}
\\ [1mm]
\times\,d_{\frac{m+s}{2},\,\frac{N_0-N_1}{2}}^{\frac{N_0+N_1}{2}}
\left(\frac{\pi}{2}\right)\,
d_{\frac{m-s}{2},\,\frac{N_2-N_3}{2}}^{\frac{N_2+N_3}{2}}
\left(\frac{\pi}{2}\right)\,.
\nonumber
\end{eqnarray}

Lastly, it is easily seen that the definition of prolate three- and
four-dimensional spheroidal coordinates and the $KS$ transformation
(\ref{3.7.9}) leads to that the spheroidal parameters $R$ and $d$ are related
by $R=d^2$ and that the coefficients relating the spheroidal bases of the
MIC--Kepler problem and the four-dimensional isotropic oscillator are diagonal
and have the form:
\begin{eqnarray*}
\left\langle\, NpMM';d\,|\,nqms;R\,\right\rangle = 4n\sqrt{r_0}
\delta_{n, \frac{N}{2}+1}\, \delta_{p,q}\, \delta_{M,m}\,
\delta_{M',s}.
\end{eqnarray*}
All interbasis expansions considered in this section are a generalization of
the results obtained in \cite{MPSTA2,MPSTA13}. In conclusion, we should like to
note that the coefficients of interbasis expansions (\ref{3.10.9}) and
(\ref{3.10.10}) in the case $s=0$ (for hydrogen atom) were first discussed in
Kibler, Ronveaux and Negadi's paper \cite{Kibler4} which in its turn stimulated
our papers \cite{MPSTA2,MPSTA13}.

\section{Hurwitz transformation}
\markboth{CHAPTER 5. NONBIJECTIVE TRANSFORMATIONS}{5.7. HURWITZ TRANSFORMATION}

Let us introduce a nonbijective quadratic transformation binding the Cartesian
coordinates $(x_{0}, \dots, x_{4})$ of the space $\rm I\!R^5$ with the
coordinates $(u_{0},\dots,u_{7})$ of the space $\rm I\!R^8$
\begin{eqnarray}
\left(\begin{array}{c}
x_0\\x_1\\x_2\\x_3\\x_4\\0\\0\\0
\end{array}
\right)=\left(
\begin{array}{cccccccc}
u_0&u_1&u_2&u_3&-u_4&-u_5&-u_6&-u_7\\
u_4&u_5&-u_6&-u_7&u_0&u_1&-u_2&-u_3\\
u_5&-u_4&u_7&-u_6&-u_1&u_0&-u_3&u_2\\
u_6&u_7&u_4&u_5&u_2&u_3&u_0&u_1\\
u_7&-u_6&-u_5&u_4&u_3&-u_2&-u_1&u_0\\
u_1&-u_0&u_3&-u_2&u_5&-u_4&u_7&-u_6\\
u_2&-u_3&-u_0&u_1&-u_6&u_7&u_4&-u_5\\
u_3&u_2&-u_1&-u_0&-u_7&-u_6&u_5&u_4\\
\end{array}
\right) \left(
\begin{array}{c}
u_0\\u_1\\u_2\\u_3\\u_4\\u_5\\u_6\\u_7
\end{array}
\right).
\label{1.2.1}
\end{eqnarray}
Hence, it follows that
\begin{eqnarray}
x_0 &=& u_0^2 + u_1^2 + u_2^2 + u_3^2 -u_4^2 - u_5^2- u_6^2 -
u_7^2, \nonumber \\ [2mm] x_1 &=& 2(u_0u_4 + u_1u_5 - u_2u_6 -
u_3u_7), \nonumber \\ [2mm] x_2 &=& 2(u_0u_5 - u_1u_4 + u_2u_7 -
u_3u_6), \label{1.2.2} \\ [2mm] x_3 &=& 2(u_0u_6 + u_1u_7 + u_2u_4
+ u_3u_5), \nonumber \\ [2mm] x_4 &=& 2(u_0u_7 - u_1u_6 - u_2u_5 +
u_3u_4). \nonumber
\end{eqnarray}
It should be noted that to each element in $\rm I\!R^5$ there corresponds not a
single element but a whole set of elements in $\rm I\!R^8$ called a fibre.
Therein lies the property of nonbijectivity of the transformation $\rm
I\!R^8(\bf{u})\to\rm I\!R^5(\bf{x})$.

The matrix $H(u;8)$ in (\ref{1.2.1}) differs from the well-known Kelly
matrix \cite{Zhevl} by a certain rearrangement of lines. It is easy to verify
that for the matrix $H(u;8)$ the following condition holds:
\bea
H_{\mu \lambda}H_{\lambda \nu}^T = u^2 \delta_{\mu \nu},
\label{1.2.3}
\eea
which guarantees the fulfillment of the Euler identity. Hereafter, unless
otherwise specified, Greek letters will take values $0, 1,...,7$; and Latin
letters, $i,j,k,...=0,1,...,4$.

Using an explicit form of the matrix $H(u;8)$ one can show that
\bea
\frac{\partial H_{\mu \nu}}{\partial u_\nu} &=& 0,
\label{1.2.4}
\\ [3mm]
\frac{\partial x_j}{\partial u_\mu} &=& 2 H_{j \mu}.
\label{1.2.5}
\eea

Now we turn to a transformation of derivatives. Taking into account
(\ref{1.2.5}), we have
\bea
\frac{\partial}{\partial u_\mu} = H_{i \mu}
\frac{\partial}{\partial x_i}.
\label{1.2.6}
\eea
Multiplying (\ref{1.2.6}) by $H_{j \mu}$, summing over $\mu$, and bearing in
mind the condition (\ref{1.2.3}), it is easy to show that
\bea \frac{\partial}{\partial x_i} = \frac{1}{2u^2}H_{i \mu}
\frac{\partial}{\partial u_\mu}.
\label{1.2.7}
\eea
We introduce the notation
\bea
\frac{\partial}{\partial q_\mu} = \frac{1}{2u^2}H_{\mu \nu}
\frac{\partial}{\partial u_\nu},
\label{1.2.8}
\eea
where $q_\mu=(x_j,0,0,0)$. Formula (\ref{1.2.8}) generalizes relation
(\ref{1.2.7}) to the case $\mu=5,6,7$.

Let us consider the second derivatives. From (\ref{1.2.8}) it follows that
\begin{eqnarray*}
\frac{\partial^2}{\partial q^2_\mu} = \frac{1}{2}
\frac{\partial}{\partial q_\mu}\left(\frac{1}{u^2}H_{\mu \nu}
\frac{\partial}{\partial u_\nu}\right).
\end{eqnarray*}
Applying formula (\ref{1.2.8}) once more with allowance for (\ref{1.2.3}) we
have
\begin{eqnarray*}
\frac{\partial^2}{\partial q^2_\mu} = \frac{1}{4u^2}H_{\mu
\lambda} \left(\frac{\partial}{\partial
u_\lambda}\frac{1}{u^2}H_{\mu \nu}\right) \frac{\partial}{\partial
u_\nu} + \frac{1}{4u^2} \frac{\partial^2}{\partial u^2_\mu}.
\end{eqnarray*}
Differentiating by parts in the first term, using the condition (\ref{1.2.3})
and then identity (\ref{1.2.4}), we as a result obtain
\bea
\frac{\partial^2}{\partial q^2_\mu} = \frac{1}{4u^2}
\frac{\partial^2}{\partial u^2_\mu}.
\label{1.2.9}
\eea
Relation (\ref{1.2.9}) and the Euler identity (\ref{1.1.4}) are the only
"mathematical tools" required for transformation of the Schr\"{o}dinger
equation of the eight-dimensional isotropic oscillator into the Schr\"{o}dinger
equation of the five-dimensional Coulomb problem. Before proceeding to this
transformation let us return to formula (\ref{1.2.7}) and introduce the
operators
\bea
\hat{\cal L}_w = iu^2\frac{\partial}{\partial q_w} =
\frac{i}{2}H_{w \mu}\frac{\partial}{\partial u_\mu},
\label{1.2.10}
\eea
where $w=5,6,7$. It follows from this definition and formula (\ref{1.2.5}) that
\begin{eqnarray*}
\hat{\cal L}_w x_j = iu^2{\delta}_{wj} = 0,
\end{eqnarray*}
as $w=5,6,7$ and $j=0,1,...,4$. Thus, the operators (\ref{1.2.10}) are
independent of the coordinates $x_j$ and, therefore, for an arbitrary function
$f(\bf x)$ the following identity holds:
\begin{eqnarray*}
\hat{\cal L}_w f(\bf x) = 0.
\end{eqnarray*}
With the help of formula(\ref{1.2.10}) and matrix $H(u,8)$ one can prove that
the operators (\ref{1.2.10}) have the following explicit form:
\begin{eqnarray*}
\hat{\cal L}_5 = \frac{i}{2}\left(u_1\frac{\partial}{\partial
u_0}- u_0\frac{\partial}{\partial u_1}+u_3\frac{\partial}{\partial
u_2}- u_2\frac{\partial}{\partial u_3}+u_5\frac{\partial}{\partial
u_4}- u_4\frac{\partial}{\partial u_5}+u_7\frac{\partial}{\partial
u_6}- u_6\frac{\partial}{\partial u_7}\right),
\\ [3mm]
\hat{\cal L}_6 = \frac{i}{2}\left(u_2\frac{\partial}{\partial u_0}-
u_3\frac{\partial}{\partial u_1}-u_0\frac{\partial}{\partial u_2}+
u_1\frac{\partial}{\partial u_3}-u_6\frac{\partial}{\partial u_4}+
u_7\frac{\partial}{\partial u_5}+u_4\frac{\partial}{\partial u_6}-
u_5\frac{\partial}{\partial u_7}\right),
\\ [3mm]
\hat{\cal L}_7 =
\frac{i}{2}\left(u_3\frac{\partial}{\partial u_0}+
u_2\frac{\partial}{\partial u_1}-u_1\frac{\partial}{\partial u_2}-
u_0\frac{\partial}{\partial u_3}-u_7\frac{\partial}{\partial u_4}-
u_6\frac{\partial}{\partial u_5}+u_5\frac{\partial}{\partial u_6}+
u_4\frac{\partial}{\partial u_7}\right).
\end{eqnarray*}
Now denoting these operators in a different way
\begin{eqnarray*}
{\hat {\cal J}}_1 =  \hat{\cal L}_5, \qquad {\hat {\cal J}}_2 =
\hat{\cal L}_6, \qquad {\hat {\cal J}}_3 = \hat{\cal L}_7
\end{eqnarray*}
and using the explicit form of the operators $\hat{\cal L}_5,$ $\hat{\cal
L}_5,$ and $\hat{\cal L}_5,$ one can prove by direct calculation that the
operators ${\hat {\cal J}}_1,$ ${\hat {\cal J}}_2,$ ${\hat {\cal J}}_3,$
satisfy the commutation relations
\begin{eqnarray*} \left[{\hat {\cal J}}_a, {\hat {\cal J}}_b\right] =
i\epsilon_{abc} {\hat {\cal J}}_c,
\end{eqnarray*}
where $a,b,c=1,2,3$. Formula (\ref{1.2.9}) with taking account of
(\ref{1.2.10}) results in that the Laplacians $\frac{\partial^2}{\partial
x^2_j}$ and $\frac{\partial^2}{\partial u^2_\mu}$ are related by
\bea \frac{\partial^2}{\partial u^2_\mu} =
4r\frac{\partial^2}{\partial x^2_j} -\frac{4}{r}{\hat{\cal J}}^2,
\label{1.2.11}
\eea
in which ${\hat{\cal J}}^2$ is determined as

\begin{eqnarray*}
{\hat{\cal J}}^2 = {\hat{\cal J}}^2_1 + {\hat{\cal J}}^2_2 +
{\hat{\cal J}}^2_3.
\end{eqnarray*}

\section{Coulomb--oscillator analogy}
\markboth{CHAPTER 5. NONBIJECTIVE TRANSFORMATIONS}{5.8. COULOMB--OSCILLATOR ANALOGY}

Let us relate the eight-dimensional problem of the isotropic oscillator
\bea
\left(-\frac{\hbar^2}{2\mu_0}\frac{{\partial}^2}{\partial u^2_\mu}
+ \frac{\mu_0\omega^2u^2}{2}\right)\psi({\bf u}) = E\psi({\bf u}),
\label{1.2.12} \\
[3mm] E = \hbar \omega \left(N + 4 \right), \qquad N = 0,1,2,...,
\label{1.2.13} \eea
where $N$ is the principal quantum number of the isotropic oscillator.

As the operators ${\hat {\cal J}}_a$  do not depend on the coordinates $x_j$,
we assume that the wave function of the eight-dimensional isotropic oscillator
$\psi(\bf u)$ can be represented in the following factorized form:
\bea
\psi({\bf u}) = \psi({\bf x})\Phi(\Omega_a),
\label{1.2.14}
\eea
where $a=1,2,3$, $\Omega_a$ denote the angles on which the operators ${\hat
{\cal J}}_a$ depend, and $\Phi(\Omega_a)$ is an eigenfunction of the operator
${\hat{\cal J}}^2$, i.e.,
\bea
{\hat{\cal J}}^2\,\Phi(\Omega_a) = {\cal J}({\cal
J}+1)\Phi(\Omega_a).
\label{1.2.15}
\eea
Here ${\cal J}({\cal J}+1)$ are eigenvalues of the operator ${\hat{\cal J}}^2$.
Now substituting (\ref{1.2.11}) in equation (\ref{1.2.12}), taking account of
relations (\ref{1.2.14}), (\ref{1.2.15}) and formula (\ref{1.1.3}) we arrive at
\bea
\left[-\frac{\hbar^2}{2\mu_0}\frac{{\partial}^2}{\partial
x^2_j} - \frac{e^2}{r} + \frac{\hbar^2}{2\mu_0r^2}{\cal J}({\cal
J}+1) \right]\psi(\bf x) = \epsilon \psi(\bf x).
\label{1.2.16}
\eea
Thus, we have obtained that the eight-dimensional isotropic oscillator is dual
to an infinite system of five-dimensional Coulomb systems with an additional
term $1/r^2$ and coupling constant $\kappa=\hbar^2{\cal J}({\cal J}+1)/2\mu_0$.
At $J \ne 0$ equation (\ref{1.2.16}) represents a system called in \cite{TRUNK}
the $SU(2)$ Kepler problem. In that paper, a group of hidden symmetry, $SO(6)$,
was also determined and an energy spectrum was calculated by a pure algebraic
method.

At ${\cal J}=0$ equation (\ref{1.2.16}) describes the five-dimensional Coulomb
problem
\bea
\left(-\frac{\hbar^2}{2\mu_0}\frac{{\partial}^2}{\partial x^2_j}
- \frac{e^2}{r}\right)\psi(\bf x) = \epsilon \psi(\bf x).
\label{1.2.17}
\eea
The condition ${\cal J}=0$ is equivalent to the requirement
\begin{eqnarray*}
{\hat{\cal J}}_a \psi({\bf x}) = 0.
\end{eqnarray*}
Moreover, it follows from (\ref{1.2.3}) that $\psi(\bf x)$ is an even function
of variables $u$:
\begin{eqnarray*} \psi\left(\bf x(-\bf u)\right) = \psi\left(\bf
x(\bf u)\right).
\end{eqnarray*}
Therefore, any solution of equation (\ref{1.2.17}) $\psi(\bf x)$ can be
expanded over a complete system of even solutions $\psi_{N\alpha}({\bf u})$
($\alpha$ are the remaining quantum numbers) of equation (\ref{1.2.12}), i.e.,
\begin{eqnarray*}
\psi_n({\bf x}) = \sum_{\alpha} C_{n \alpha}\psi_{N\alpha}({\bf u}),
\end{eqnarray*}
where
\bea
 N = 2n.
 \label{1.2.18}
 \eea
One can easily be convinced that $n$ coincides with the principal quantum
number of the five-dimensional Coulomb problem. Indeed, substituting in
(\ref{1.2.13}) relations $E=4e^2$ and (\ref{1.2.18}), we get
\bea
\omega_n = \frac{2e^2}{\hbar (n+2)}.
\label{1.2.19}
\eea
Thus,in our case the the oscillator energy is fixed and frequency $\omega$ is
quantized. Substituting now (\ref{1.2.19}) in the condition $\varepsilon =
-\mu_0\omega^2/8$, we arrive at the expression
\begin{eqnarray}
\varepsilon_n = - \frac{\mu_0e^4}{2\hbar^2 (n+2)^2},
\label{1.2.20}
\end{eqnarray}
which determines the energy spectrum of the five-dimensional Coulomb problem
\cite{ALLILUEV}.

\section{Euler parameterization}
\markboth{CHAPTER 5. NONBIJECTIVE TRANSFORMATIONS}{5.9. EULER PARAMETERIZATION}

Instead of the Cartesian coordinates $u_j$ we introduce the coordinates $u_T,\,
u_K,$ ${\alpha}_T,\, {\beta}_T,\, {\gamma}_T,$ ${\alpha}_K,$ ${\beta}_K,$
${\gamma}_K$ as follows:
\bea
u_0 + iu_1 &=& u_T \sin\frac{{\beta}_T}{2}
e^{-i\frac{{\alpha}_T-{\gamma}_T}{2}}, \qquad  u_2 + iu_3 = u_T
\cos\frac{{\beta}_T}{2} e^{i\frac{{\alpha}_T+{\gamma}_T}{2}},
\nonumber \\
\label{1.3.1} \\
u_4 + iu_5 &=& u_K \sin\frac{{\beta}_K}{2}
e^{i\frac{{\alpha}_K-{\gamma}_K}{2}}, \qquad  u_6 + iu_7 = u_K
\cos\frac{{\beta}_K}{2} e^{-i\frac{{\alpha}_K+{\gamma}_K}{2}}.
\nonumber
\eea
The new coordinates are determined in the ranges
\begin{eqnarray*}
0\leq u_T,u_K < \infty, \quad 0 \leq {\beta}_T, {\beta}_K \leq
\pi,\quad 0 \leq {\alpha}_T, {\alpha}_K < 2\pi, \quad 0 \leq
{\gamma}_T, {\gamma}_K < 4\pi.
\end{eqnarray*}
To determine types of coordinates (hyperspherical, cylindrical, parabolic,
spheroidal, etc.), $u_T$ and $u_K$ will be determined in addition. In this
case, $u_T$ and $u_K$ may be arbitrary  but relation to the coordinates $u_j$
is fixed
\begin{eqnarray*}
u_T = (u_0^2+u_1^2+u_2^2+u_3^2)^{1/2},  \qquad
u_K = (u_4^2+u_5^2+u_6^2+u_7^2)^{1/2}.
\end{eqnarray*}
Differential elements of length, volume, and the laplace operator in the
coordinates (\ref{1.3.1}) are
\begin{eqnarray*} dl_8^2 =
du_T^2+du_K^2+\frac{u_T^2}{4}dl_T^2+\frac{u_K^2}{4}dl_K^2, \qquad
dV_8   &=& u_T^3u_K^3du_Tdu_Kd{\Omega_T}d{\Omega_K},
\end{eqnarray*}
\bea \Delta_{8} = \frac{1}{u_T^3} \frac{\partial}{\partial u_T}
\left(u_T^3\frac{\partial}{\partial u_T} \right)+
\frac{1}{u_K^3}\frac{\partial}{\partial u_K}
\left(u_K^3\frac{\partial}{\partial u_K}\right)-
\frac{4}{u_T^2}{\hat {\bf T}}^2- \frac{4}{u_K^2}{\hat {\bf K}}^2,
\label{1.3.2}\eea
where
\bea
dl_a^2=d{\alpha_a}^2+d{\beta_a}^2+d{\gamma_a}^2+2\cos\beta_ad{\alpha_a}
d{\gamma_a}, \qquad d{\Omega_a} =
\frac{1}{8}\sin\beta_ad{\beta_a}d{\alpha_a}d{\gamma_a},
\label{1.3.3}
\eea
\bea
{\hat {\bf T}}^2 = -\left[\frac{\partial^2}{\partial
\beta_T^2}+ \cot \beta_T\frac{\partial}{\partial \beta_T}+
\frac{1}{\sin^2\beta_T}\left(\frac{\partial^2}{\partial
{\alpha_T}^2}- 2\cos \beta_T \frac{\partial^2}{\partial \alpha_T
\partial \gamma_T}+ \frac{\partial^2}{\partial
{\gamma_T}^2}\right)\right].
\label{1.3.4}
\eea
Index $a=T,\,K$, and the operator ${\hat {\bf K}}^2$ can be derived from the
operator ${\hat {\bf T}}^2$ by substituting $(\alpha_T, \beta_T, \gamma_T)$ for
$(\alpha_K, \beta_K, \gamma_K)$.

Let us find what coordinates the Hurwitz transformation (\ref{1.2.1}) converts
the coordinates (\ref{1.3.1}) into. After substitution of (\ref{1.3.1}) into
(\ref{1.2.2}) we have
\bea
x_0 = u_T^2 - u_K^2, \qquad x_j = 2u_Tu_K{\hat x}_j.
\label{1.3.5}
\eea
Here $j=1,2,3,4$, and ${\hat x}_j$ are given by the
expressions
\begin{eqnarray*}
{\hat x}_1 &=& \sin\frac{\beta_T}{2}\sin\frac{\beta_K}{2}
\cos\frac{\alpha_T+\alpha_K-\gamma_T-\gamma_K}{2}-
\cos\frac{\beta_T}{2}\cos\frac{\beta_K}{2}
\cos\frac{\alpha_T+\alpha_K+\gamma_T+\gamma_K}{2},  \\
[3mm] {\hat x}_2 &=& \sin\frac{\beta_T}{2}\sin\frac{\beta_K}{2}
\sin\frac{\alpha_T+\alpha_K-\gamma_T-\gamma_K}{2} -
\cos\frac{\beta_T}{2}\cos\frac{\beta_K}{2}
\sin\frac{\alpha_T+\alpha_K+\gamma_T+\gamma_K}{2}, \\
[3mm] {\hat x}_3 &=& \sin\frac{\beta_T}{2}\cos\frac{\beta_K}{2}
\cos\frac{\alpha_T-\alpha_K-\gamma_T-\gamma_K}{2}+
\cos\frac{\beta_T}{2}\sin\frac{\beta_K}{2}
\cos\frac{\alpha_T-\alpha_K+\gamma_T+\gamma_K}{2}, \\
[3mm] {\hat x}_4 &=& \sin\frac{\beta_T}{2}\cos\frac{\beta_K}{2}
\sin\frac{\alpha_T-\alpha_K-\gamma_T-\gamma_K}{2}+
\cos\frac{\beta_T}{2}\sin\frac{\beta_K}{2}
\sin\frac{\alpha_T-\alpha_K+\gamma_T+\gamma_K}{2}.
\end{eqnarray*}
Determine the following coordinates:
\bea
x_0 = u_T^2 - u_K^2, \quad x_2 + ix_1 = 2u_T u_K \sin
\frac{\beta}{2} {\rm e}^{i \frac{\alpha-\gamma}{2}}, \quad x_4 +
ix_3 = 2 u_T u_K \cos \frac{\beta}{2}{\rm e}^{i
\frac{\alpha+\gamma}{2}}.
\label{1.3.6}
\eea
In these coordinates
\begin{eqnarray*}
dl_5^2 = \frac{\mu+\nu}{4\mu}d{\mu}^2+
\frac{\mu+\nu}{4\nu}d{\nu}^2+\frac{\mu \nu}{4}dl^2, \qquad dV_5 =
\frac{\mu \nu}{4}(\mu+\nu)d{\mu}d{\nu}d{\Omega},
\end{eqnarray*}
\bea
\Delta_{5} = \frac{4}{\mu+\nu}
\left[\frac{1}{\mu}\frac{\partial}{\partial \mu}
\left({\mu}^2\frac{\partial}{\partial \mu}\right)+
\frac{1}{\nu}\frac{\partial}{\partial \nu}
\left({\nu}^2\frac{\partial}{\partial \nu}\right)\right]-
\frac{4}{\mu \nu}{\hat {\bf J}}^2,
\label{1.3.7}
\eea
where $\mu=2u_T^2,\, \nu=2u_K^2$, and $dl^2, d{\Omega}$, and ${\bf J}^2$ can be
found from relations (\ref{1.3.3}) и (\ref{1.3.4}) with the help of the
substitution $(\alpha_a,\beta_a,\gamma_a) \rightarrow (\alpha,\beta,\gamma)$.
By identifying the coordinates (\ref{1.3.6}) and (\ref{1.3.5}) we arrive at the
following system of trigonometric equations
\begin{eqnarray*}
\cos \frac{\beta}{2} e^{\frac{i}{2}(\alpha+\gamma)} &=&
\sin\frac{\beta_T}{2}\cos\frac{\beta_K}{2}
e^{-\frac{i}{2}(\alpha_T-\alpha_K-\gamma_T-\gamma_K-\pi)}+
\cos\frac{\beta_T}{2}\sin\frac{\beta_K}{2}
e^{-\frac{i}{2}(\alpha_T-\alpha_K+\gamma_T+\gamma_K-\pi)}, \\
[2mm] \sin \frac{\beta}{2}{\rm e}^{\frac{i}{2}(\alpha-\gamma)} &=&
\sin\frac{\beta_T}{2}\sin\frac{\beta_K}{2}
e^{-\frac{i}{2}(\alpha_T+\alpha_K-\gamma_T-\gamma_K-\pi)}-
\cos\frac{\beta_T}{2}\cos\frac{\beta_K}{2}
e^{-\frac{i}{2}(\alpha_T+\alpha_K+\gamma_T+\gamma_K-\pi)}.
\end{eqnarray*}
Having solved these equations we have
\bea
\cot (\alpha + \alpha_T - \pi) = \cos (\beta_T - \pi) \cot
(\gamma_T+\gamma_K)+ \cot \beta_K \frac{\sin(\beta_T - \pi)}{\sin
(\gamma_T+\gamma_K)},
\nonumber\\[3mm]
\cos \beta = \cos (\beta_T - \pi)\cos \beta_K- \sin(\beta_T -
\pi)\sin \beta_K \cos (\gamma_T+\gamma_K), \label{1.3.8}
\\ [3mm]
\cot (\gamma-\alpha_K) = \cos \beta_K \cot (\gamma_T+\gamma_K)+
\cot(\beta_T - \pi)\frac{\sin \beta_K}{\sin
(\gamma_T+\gamma_K)}.\nonumber
\eea
Thus, the Hurwitz transformation consists of the conformal Levi--Cevita
transformation $(u_T+iu_K) \to (u_T+iu_K)^2$ and the Euler addition of angles
\begin{eqnarray*}
(\pi - \alpha_T, \beta_T - \pi, \gamma_T) \oplus (\gamma_K,
\beta_K, \alpha_K) = (\alpha, \beta, \gamma).
\end{eqnarray*}
Therefore, parameterization (\ref{1.3.1}) is called the Euler parameterization.

It is seen from (\ref{1.3.2}) and (\ref{1.3.7}) that the Laplace operators
$\Delta_{8}$ and $\Delta_{5}$ contain the squares of the angular momentum
operators. If in the Schr\"{o}dinger equation the form of the potential admits
separation of variable in the systems of coordinates related to the operators
${\hat {\bf T}}^2$, ${\hat {\bf K}}^2$ and ${\hat {\bf J}}^2$, then solutions
have a universal dependence on the angles entering into these operators.

Fixing the coordinates $u_T$ and $u_K$ we can consider the hyperspheres $S^6$
and $S^3$ in the spaces $\rm I\!R^8(\bf{u})$ and $\rm I\!R^5(\bf{x})$,
respectively. The solutions are of the form
\begin{eqnarray*}
 \psi_6 = \left(\frac{2T+1}{2\pi^2}\right)^{1/2}
\left(\frac{2K+1}{2\pi^2}\right)^{1/2} D_{t,
t'}^{T}(\alpha_T,\beta_T,\gamma_T) D_{k,
k'}^{K}(\alpha_K,\beta_K,\gamma_K),
\end{eqnarray*}
\begin{eqnarray*}
\psi_3 = \left(\frac{2L+1}{2\pi^2}\right)^{1/2}
D_{m,m'}^L(\alpha,\beta,\gamma).
\end{eqnarray*}
For the angles in relation (\ref{1.3.8}) the following addition theory is
valid \cite{VAR}:
\begin{eqnarray}
D_{m,m'}^L(\alpha,\beta,\gamma)= {\rm e}^{im\pi} \sum_{k=-L}^L\,
(-1)^{m+m'-L-k}\, D_{m,k}^L(\alpha_T,\beta_T,\gamma_T)\,
D_{k,m'}^L(\alpha_K,\beta_K,\gamma_K).
\label{1.3.9}
\end{eqnarray}

Thus, in the Euler parameterization the correspondence between the solutions in
the spaces $\rm I\!R^8(\bf{u})$ and $\rm I\!R^5(\bf{x})$  is fulfilled
according to the addition theorem of the Wigner $D$-function (see also
\cite{SISTER1,SISTER2}).

\newpage

\addcontentsline{toc}{section}{{\bf Appendix A.} \\
Hidden symmetry and orthogonality in orbital moment}
\section*{Appendix A  \\
Hidden symmetry and orthogonality in orbital moment}
\markboth{APPENDIX A}{HIDDEN SYMMETRY AND ORTHOGONALITY IN ORBITAL MOMENT}
It is well known that radial wave functions of a particle moving in the
centrosymmetric field at a given orbital moment $l$ are orthogonal in the
principal quantum number $n$:
\begin{eqnarray}
\int\limits_0^\infty r^2 R_{nl}(r) R_{n'l}(r) dr = \delta_{nn'}.
\label{ort.1.1}
\end{eqnarray}

Let us prove that for a hydrogen atom in parallel with (\ref{ort.1.1}) there
holds an "additional" condition of orthogonality in orbital moment $l$:
\begin{eqnarray}
J_{ll'} = \int\limits_0^\infty  R_{nl'}(r) R_{nl}(r) dr
= \frac{2}{n^3}\frac{\delta_{ll'}}{2l+1}.
\label{ort.1.2}
\end{eqnarray}
Substituting into (\ref{ort.1.2}) the normalized radial wave function of the
hydrogen atom (\ref{rh.1.1}), writing down the degenerate hypergeometric
function $R_{nl'}(r)$ as a polynomial
\begin{eqnarray}
F\left(-n+l'+1; 2l'+2; \frac{2r}{n}\right) = \sum_{s=0}^{n-l'-1}\,
\frac{(-n+l'+1)_s}{s!(2l'+2)_s}\,\left(\frac{2r}{n}\right)^s,
\label{ort.1.3}
\end{eqnarray}
integrating by formula \cite{LL}
\bea
\int \limits_{0}^{\infty} e^{-\lambda x} x^\nu F(\alpha,
\gamma; kx) \, dx = \frac{\Gamma(\nu+1)}{\lambda^{\nu+1}} \,
{_2F}_1 \left( \alpha, \nu+1, \gamma ; \frac{k}{\lambda} \right)
\label{ort.1.4}
\eea
and taking account of that \cite{BE1}
\bea
{_2F}_1 \left(a, b; c; 1\right) =
\frac{\Gamma(c)\Gamma(c-a-b)} {\Gamma(c-a)\Gamma(c-b)},
\label{ort.1.5}
\eea
we have
\bea
J_{ll'} = \frac{2}{n^3}\,\frac{\Gamma\left(l+l'+1\right)}{(2l+1)!}
\sqrt{\frac{(n+l')!}{(n+l)!(n-l-1)!(n-l'-1)!}}\times \nonumber
\\
\label{ort.1.6}\\
\times \sum_{s=0}^{n-l'-1}\,
\frac{(-n+l'+1)_s}{s!}\frac{\left(l+l'+1\right)_s}{
(2l'+2)_s}\,\frac{\Gamma\left(n-l'-s\right)}
{\Gamma\left(l-l'-s+1\right)}.
\nonumber
\eea
Applying to gamma-functions under the summation sign the formula \cite{BE1}
\bea
\frac{\Gamma(z)}{\Gamma(z-n)} = (-1)^n\frac{\Gamma(-z+n+1)}{\Gamma(-z+1)},
\label{ort.1.7}
\eea
we note that the sum over $s$ in (\ref{ort.1.6}) turns into a hypergeometric
function of type (\ref{ort.1.5}) and, therefore,
\bea
J_{ll'} = \frac{2}{n^3}\,\frac{\left(l+l'\right)!}{\left(l+l'+1\right)!}
\sqrt{\frac{(n+l')!(n-l'-1)!}{(n+l)!(n-l-1)!}}\frac{1}
{\Gamma\left(l-l'+1\right)\Gamma\left(l'-l+1\right)}.
\label{ort.1.8}
\eea
The last relation vanishes at $l'\neq l$ due to the product of gamma-functions
of $(l-l'+1)$ è $(l'-l+1)$, which results in the orthogonality condition
(\ref{ort.1.2}).

Following the above-described scheme it is easy to verify that orthogonality in
orbital moment of number $l$ also holds for the radial wave function of the
isotropic oscillator:
\begin{eqnarray}
\int\limits_0^\infty  R_{nl'}(r) R_{nl}(r) dr= 2\alpha^2\,
\frac{2\delta_{ll'}}{2l+1},
\label{ort.1.9}
\end{eqnarray}
where $\alpha= \sqrt{M\omega/\hbar}$, and the explicit form  $R_{nl}(r)$ is
given by formula (\ref{NDIM-OSC-1}). Let us prove that the orthogonality
conditions (\ref{ort.1.2}) and (\ref{ort.1.9}) are a consequence of accidental
degeneracy of the energy spectrum.

The radial Schr\"{o}dinger equation in the field  $U(r)$ has the form
\begin{eqnarray}
\hat {\cal H} R_{EL}(r) = ER_{EL}(r),
\label{ort.1.10}
\end{eqnarray}
where $\hat {\cal H}$ denotes the Hermite operator
\begin{eqnarray}
\hat {\cal H} = - \frac{\hbar^2}{2M}\, \frac{1}{r^2}\frac{d}{dr}
\left(r^2\frac{d}{dr}\right) - \frac{\hbar^2}{2M}\,
\frac{l(l+1)}{r^2} + U(r).
\label{ort.1.11}
\end{eqnarray}
Relations (\ref{ort.1.10}) and (\ref{ort.1.11}) for the discrete spectrum
result in
\begin{eqnarray}
(l-l')(l+l'+1)\int \limits_{0}^{\infty}\,R_{E'L'}(r)R_{EL}(r)
\,dr = \frac{2M}{\hbar^2}\,(E-E')\,\int \limits_{0}^{\infty}\,
R_{E'l'}(r)R_{El}(r)\,r^2\,dr.
\label{ort.1.12}
\end{eqnarray}
If the spectrum is degenerate in $l$, then at $E'=E$ and $l'\neq l$ we have
\begin{eqnarray}
\int \limits_{0}^{\infty}\,R_{EL'}(r)R_{EL}(r)\,dr = 0.
\label{ort.1.13}
\end{eqnarray}
It is well known \cite{LL} that the Hermite operator ${\hat F}$ depending on
some parameter $\lambda$ satisfies the identity
\begin{eqnarray}
\left(\frac{\partial \hat F(\lambda)}{\partial \lambda}\right)_{nn}
= \frac{\partial F_n}{\partial \lambda},
\label{ort.1.14}
\end{eqnarray}
where averaging is over eigenfunctions of the operator ${\hat F}$. Applying
this identity to the operator (\ref{ort.1.11}) and using as a parameter
$\lambda$ the orbital moment$l$ we obtain
\begin{eqnarray}
\int \limits_{0}^{\infty}\,R_{nl}(r)R_{nl}(r)\,dr =
\frac{2M}{\hbar^2}\,
\left(\frac{\partial E_n}{\partial l}\right)_{n_r} \frac{1}{2l+1},
\label{ort.1.15}
\end{eqnarray}
where a derivative with respect to $l$ is taken at a fixed radial number $n_r$.
Combining (\ref{ort.1.13}) with (\ref{ort.1.15}) we finally get
\begin{eqnarray}
\int \limits_{0}^{\infty}\,R_{nl}(r)R_{nl'}(r)\,dr =
\left(\frac{\partial E_n}{\partial l}\right)_{n_r}
\frac{\delta_{ll'}}{2l+1}.
\label{ort.1.16}
\end{eqnarray}
It is obvious that this formula involves the conditions (\ref{ort.1.2}) and
(\ref{ort.1.9}).

For a multidimensional analog of the hydrogen atom \cite{ALLILUEV} and the
isotropic oscillator \cite{BAKER} the role of the operator (\ref{ort.1.11}) is
played by the Hermite operator
\begin{eqnarray}
\hat {\cal H} = - \frac{\hbar^2}{2M}\,\frac{1}{r^{D-1}}
\frac{d}{dr}\left(r^{D-1}\frac{d}{dr}\right) + \frac{\hbar^2}{2M}\,
\frac{L(L+D-2)}{r^2} + U(r),
\label{ort.1.17}
\end{eqnarray}
in which $D \geq 3$ is integer determining the dimension of space, and $L$ is
the so-called global moment \cite{VKS} which takes both integer and
half-integer nonnegative values.

The reasoning analogous to that given above shows that in the multi-dimensional
case the orthogonality condition holds
\begin{eqnarray}
\int \limits_{0}^{\infty}\,R_{nL}(r)R_{nL'}(r)\,r^{D-3}dr =
\frac{2M}{\hbar^2}\,
\left(\frac{\partial E_n}{\partial L}\right)_{n_r}
\frac{\delta_{LL'}}{2L+D-2}.
\label{ort.1.18}
\end{eqnarray}
The validity of this condition is proved by direct calculation with the use of
the explicit form of radial wave functions of the multi-dimensional hydrogen
atom \cite{ALLILUEV} and the isotropic oscillator (\ref{NDIM-OSC-1}).

\newpage
\addcontentsline{toc}{section}{{\bf Appendix B.} \\
Kinematic aspects of interbasis expansions}
\section*{Appendix B. \\
Kinematic aspects of interbasis expansions}
\markboth{APPENDIX B}{KINEMATIC ASPECTS OF INTERBASIS EXPANSIONS}

As was noted in the first chapter, the generators $L$, ${\cal P}$ and ${\cal
K}$ of the hidden symmetry $SO(3)$ group of the two-dimensional hydrogen atom
with the Hamiltonian  ${\cal H}$ form three total sets $\left({\cal H},
{L}\right)$, $\left({\cal H}, {\cal P}\right)$ and $\left({\cal H}, {\cal
K}\right)$ of operators the eigenfunctions of which (fundamental bases of the
two-dimensional hydrogen atom) are solutions of the Schr\"{o}dinger equation
resulting after separation of variables in polar and parabolic, turned relative
to each other by a right angle, systems of coordinates. In this connection
there arises an interesting question if there exists such a general basis of
the two-dimensional hydrogen atom that: 1) has information on all three
generators $L$, ${\cal P}$ è ${\cal K}$; 2) in some limiting cases turns into
the above-pointed fundamental bases; 3) has a factorized form, i.e., is
solution in separated variables. In the second chapter we showed that by
separating variables in the elliptic coordinates we managed to "unify" only two
fundamental bases, polar $\left({\cal H}, {L}\right)$ and parabolic
$\left({\cal H}, {\cal P}\right)$. Below we will show that  further
generalization satisfying all three requirements mentioned above can be
achieved by separation of variables in the Schr\"{o}dinger equation for an
arbitrarily directed, with respect to the fixed Cartesian system of
coordinates, elliptic system of coordinates.

Introduce elliptic coordinates $\xi^{(\alpha)}$ $(0\leq \xi^{(\alpha)} <
\infty)$ è $\eta^{(\alpha)}$ $(0\leq \eta^{(\alpha)} < 2\pi)$ in the following
way:
\begin{eqnarray}
x = \frac{R}{2}\cos\alpha\left(\cosh\xi^{(\alpha)}
\cos\eta^{(\alpha)}+1\right) - \frac{R}{2}\sin\alpha
\sinh\xi^{(\alpha)}\sinh\eta^{(\alpha)}, \nonumber \\
\label{KIN1} \\
y = \frac{R}{2}\sin\alpha\left(\cosh\xi^{(\alpha)}
\cos\eta^{(\alpha)}+1\right) + \frac{R}{2}\cos\alpha
\sinh\xi^{(\alpha)}\sinh\eta^{(\alpha)}. \nonumber
\end{eqnarray}
At $\alpha=0$ from (\ref{KIN1}) one obtains ordinary elliptic coordinates
$\xi^{(0)}\equiv\xi$, $\eta^{(0)}\equiv\eta$ (see formula (\ref{EL-COOR-2})).

In the Coulomb units $(\hbar=m=e=1)$ the Hamiltonian of the two-dimensional
hydrogen atom has the form
\begin{eqnarray}
{\hat H} = - \frac{2}{R^2\left(\cosh^{2}\xi^{(\alpha)} -
\cos^{2}\eta^{(\alpha)}\right)}\left(\frac{\partial^2} {\partial
{\xi^{(\alpha)}}^2} + \frac{\partial^2} {\partial
{\eta^{(\alpha)}}^2}\right) - \frac{2}
{R\left(\cosh\xi^{(\alpha)}+\cos\eta^{(\alpha)}\right)}.
\label{KIN2}
\end{eqnarray}
Separating in it variables and introducing the separation constant $Q$, we
arrive at the "elliptic" integral of motion
\begin{eqnarray}
{\hat Q} &=& \frac{1}{\cosh^{2}\xi^{(\alpha)} -\cos^{2}\eta^{(\alpha)}}
\left(\cos^{2}\eta^{(\alpha)}\frac{\partial^2}
{\partial {\xi^{(\alpha)}}^2} + \cosh^{2}\xi^{(\alpha)}
\frac{\partial^2} {\partial {\eta^{(\alpha)}}^2}\right)-\nonumber \\
\label{KIN3} \\
&-&
\frac{R\cosh\xi^{(\alpha)}\cos\eta^{(\alpha)}}
{\cosh\xi^{(\alpha)}+\cos\eta^{(\alpha)}}.
\nonumber
\end{eqnarray}
The constant $Q$ is an eigenvalue of the operator ${\hat Q}$. Passing in
(\ref{KIN3}) to the Cartesian coordinates and taking account of the explicit
form of the generators $L$, ${\cal P}$ and ${\cal K}$ (\ref{symdym1-2}) -
(\ref{symdym1-4}) we can prove the relation
\begin{eqnarray}
{\hat Q} = -{\hat L}^2 - \frac{2R\cos\alpha}{2N+1}{\hat P} -
\frac{2R\sin\alpha}{2N+1}{\hat K} - \frac{R^2}{2}{\hat H},
 \label{KIN4}
\end{eqnarray}
where $N$ is the principal quantum number. We call a basis in the elliptic
coordinates (\ref{KIN1}) an arbitrarily directed elliptic basis.  It follows
from (\ref{KIN4}) that this basis satisfies the above three requirements,
accumulating most completely information about the hidden symmetry group of the
two-dimensional hydrogen atom.

An arbitrarily directed elliptic basis of the two-dimensional hydrogen atom is
related to the elliptic basis (\ref{4.1}) by a rotation transformation. In the
second chapter we showed that the elliptic basis can be determined by the
expansion
\begin{eqnarray}
\Psi_{N}(\xi^{(0)}, \eta^{(0)}; R) =
\sum_{m=-N}^{N}\,W_m(R)\,\Psi_{Nm}(r, \varphi),
\label{KIN5}
\end{eqnarray}
in which $\Psi_{Nm}(r, \varphi)$ is the polar basis of the two-dimensional
hydrogen atom, and the coefficients $W_m(R)$ and eigenvalues of the elliptic
separation constant $Q$ are determined by the trinomial recurrence relation
\begin{eqnarray}
\sqrt{(N-m)(N+m+1)}W_{m+1}(R) &+& \sqrt{(N-m+1)(N+m)}W_{m-1}(R)=
\nonumber \\ [2mm]
 \label{KIN6} \\ [2mm]
 &=& \frac{(2N+1)^2Q-R^2}{R(2N+1)}W_m(R)
 \nonumber
\end{eqnarray}
and additional conditions
\begin{eqnarray*}
W_{N+1}(R) = W_{-N-1}(R) \equiv 0, \qquad \sum_{m=-N}^N\,
\left|W_m(R)\right|^2 =1.
\end{eqnarray*}
Thus, the sought arbitrarily directed elliptic basis has the form
\begin{eqnarray}
\Psi_{N}(\xi^{(\alpha)}, \eta^{(\alpha)}; R) =
\sum_{m=-N}^{N}\,W_m(R)\,e^{-im\alpha}\,\Psi_{Nm}(r, \varphi).
\label{KIN7}
\end{eqnarray}
It should be stressed that basis (\ref{KIN7}) is obtained in two
steps: 1) introduction of the elliptic basis (\ref{KIN5}) taking
into account the specific features of the Coulomb field; 2)
rotation involving  a "deficient" generator of the hidden symmetry
group and being of a pure kinematic nature. The recurrence
relation (\ref{KIN6}) does not include the angle $\alpha$ and,
therefore,  eigenvalues of the operator ${\hat Q}$ are invariant
with respect to rotations in the plane $(x,y)$. On the other hand,
the operator (\ref{KIN4}) has the angle $\alpha$. The above
discrepancy can be removed if in passing in (\ref{KIN4}) to
"accompanying" coordinates $(x',y')$, rotated relative to $(x,y)$
by the angle $\alpha$, the dependence on $\alpha$ disappears at
all. The mechanism explaining independence of eigenfunctions of
the operator (\ref{KIN4}) of the angle $\alpha$ can also be
established on a more formal basis. The equations
\begin{eqnarray*}
{\hat H}\Psi_{N}(\xi^{(\alpha)}, \eta^{(\alpha)}; R)
&=&
- \frac{2}{\left(2N+1\right)^2}\Psi_{N}(\xi^{(\alpha)},
\eta^{(\alpha)}; R),
\\[2mm]
{\hat Q}\Psi_{N}(\xi^{(\alpha)},
\eta^{(\alpha)}; R)
&=& Q\Psi_{N}(\xi^{(\alpha)}, \eta^{(\alpha)};
R),
\end{eqnarray*}
expansion
\begin{eqnarray*}
\Psi_{N}(\xi^{(\alpha)}, \eta^{(\alpha)}; R) =
\sum_{m=-N}^{N}\,T(\alpha)\,\Psi_{Nm}(r, \varphi)
\end{eqnarray*}
and formulae
\begin{eqnarray*}
&& \int\,\Psi_{Nm'}^*(r, \varphi){\cal P}\,\Psi_{Nm}(r, \varphi)dV
=i^{m-m'} \int\,\Psi_{Nm'}^*(r, \varphi){\cal K}\,\Psi_{Nm}(r,
\varphi)dV = \\ [3mm]
&=&
-\frac{1}{2}\sqrt{(N-m)(N+m+1)}\,\delta_{m,m'+1} -
\frac{1}{2}\sqrt{(N+m)(N-m+1)}\,\delta_{m,m'-1}
\end{eqnarray*}
lead to the recurrence relation
\begin{eqnarray}
\sqrt{(N-m)(N+m+1)}e^{i\alpha}\,T_{m+1}(\alpha) &+&
\sqrt{(N-m+1)(N+m)}e^{-i\alpha}\,T_{m-1}(\alpha)=
\nonumber \\[2mm]
\label{KIN8} \\
&=& \frac{(2N+1)^2Q-R^2}{R(2N+1)}T_m(\alpha).
\nonumber
\end{eqnarray}
It is obvious now that (\ref{KIN8}) turns into (\ref{KIN6}) under the
substitution $T_m(\alpha)= W_me^{-im\alpha}.$

Generalization of the elliptic basis of the circular oscillator, the spheroidal
bases of the hydrogen atom, and the three-dimensional isotropic oscillator to
the case of arbitrary direction along which a dimensional parameter $R$ is
introduced can be performed in the same way as above. In this connection we
give here only formulae determining additional integrals of motion:

\vspace{0.2cm} a) circular oscillator
\begin{eqnarray*}
{\hat Q} = -{\hat L}^2 - \frac{R^2}{2}{\hat P}\cos 2\alpha -
\frac{R^2}{2}{\hat K}\sin 2\alpha + \frac{R^4}{64},
\end{eqnarray*}
where ${L} = -i\frac{\partial}{\partial\varphi}$, and  ${\cal P}$ and ${\cal K}$
are determined by formulae (\ref{2D-OSC-3}) è (\ref{2D-OSC-03});

\vspace{0.2cm} b) hydrogen atom
\begin{eqnarray*}
{\hat Q}= -{\hat L}^2 - \frac{R}{n}\left({\hat
A}_x\cos\alpha\sin\beta + {\hat A}_y\sin\alpha\sin\beta + {\hat
A}_z\cos\beta\right),
\end{eqnarray*}
where ${\hat L}^2$ is the total momentum square, ${\hat A}_i$ is the
Runge--Lentz vector, and $n$ is the principal quantum number;

\vspace{0.2cm} c) three-dimensional isotropic oscillator
\begin{eqnarray*}
{\hat Q} &=& {\hat L}^2 + \frac{R^2}{4}\Bigl({\hat A}_{11}\sin^2\alpha\cos^2\beta +
{\hat A}_{22}\sin^2\alpha\sin^2\beta + {\hat A}_{33}\cos^2\alpha -\\
[2mm]
&-& {\hat A}_{21}\sin^2\alpha\sin 2\beta - {\hat A}_{31}\sin
2\alpha\cos \beta + {\hat A}_{32}\sin^2\alpha\sin \beta\Bigr), \\
[2mm]
{\hat A}_{ik} &=& x_ix_k -\frac{\partial}{\partial x_i
\partial x_k},
\end{eqnarray*}
${\hat A}_{ik}$ is Demkov's tensor.

It is easy to understand that the coefficients entering into the expansion of
bases generated by the above operators in the relevant polar and spherical
bases contain exponents and the Wigner $D$ functions depending on the angles
defining orientation of a system of coordinates.

\newpage
\pagestyle{headings}


\addcontentsline{toc}{section}{\bf References}
\begin{thebibliography}{99}
\bibitem{ALLILUEV}
Alliluev, {\it On the relation between" accidental" degeneracy and" hidden" symmetry of a system}.
Sov. Phys. JETP, {\bf 6}, 156, 1958.
\bibitem{ARSS}
F.M.Arscott. {\it Periodic Differential Equations}. Macmillan,
New York. 1964.
\bibitem{ARUT1}
M.H.Arutyunyan, G.S.Pogosyan and V.M.Ter-Antonyan.
{\it To the Connection between Parabolic and Polar
Wave Functions of the Hydrogen Atom.}
Izv. AN Arm.SSR, Fizika, {\bf XIII(2)}, 152-154, 1978,
(in Russian).
\bibitem{ARUT2}
G.M.Arutyunyan, M.H.Arutyunyan, G.S.Pogosyan and
V.M. Ter-Antonyan.
{\it The Connection between Wave Functions of simple Quantum
Systems with Hidden Symmetry.} Preprint PLRF-77-10, Yerevan State University,
Yerevan, 1977, (in Russian).
\bibitem{ARUT3}
G.M.Arutyunyan, L.S.Davtyan, L.G.Mardoyan, G.S.Pogosyan and
V.M.Ter-Antonyan.
{\it Connection between Wave Functions of the Two Dimensional
Quantum Systems with Hidden Symmetry.}
Preprint PLRF-79-17, Yerevan State University, Yerevan, 1979, (in Russian).
\bibitem{BACRY}
H.Bacry. {\it The de Sitter group $L_{4,1}$ and the bound states
of hydrogen atom}. Nuovo Cimento, {\bf A41}, 222-234, 1966.
\bibitem{BA}
W.N.Bailey. {\it Generalized Hypergeometric Series}. Cambridge
Tracts No.32 (Cambridge University Press), Cambridge, 1935.
\bibitem{BAKER}
G.A.Baker. {\it Degeneracy of the n-dimensional isotropic harmonic
oscillator}. Phys. Rev., {\bf 103}, 1119-1120, 1956.
\bibitem{BANDER}
M.Bander and C.Itzykson. {\it Group Theory and the Hydrogen Atom.
I,II.} Rev. Mod.Phys. \ {\bf 38}, 330--345; 346--358, 1968.
\bibitem{BARGMAN}
V.Bargman. {\it Zur Theorie des Wasserstoffatoms}. Zs. Phys., {\bf
B39}, 576-582, 1936.
\bibitem{BARUT1}
A.O.Barut and H.Kleinert. {\it Transition probabilities of the
hydrogen atom from noncompact dynamical groups}. Phys. Rev., {\bf
156}, 1541-1545, 1967.
\bibitem{BIJ1}
A.O.Barut, A.Inomata and G.Junker. {\it Path Integral Treatment of
the Hydrogen Atom in a Curved Space of Constant Curvature}.
J.Phys., {\bf A20}, 6271-6280, 1987.
\bibitem{BIJ2}
A.O.Barut, A.Inomata and G.Junker. {\it Path Integral Treatment of
the Hydrogen Atom in a Curved Space of Constant Curvature: II. Hyperbolic Space}.
J.Phys., {\bf A23}, 1179-1190, 1990.
\bibitem{BARUT-RON}
A.O.Barut and R.Raczka. {\it Theory of group representations and applications}.
World Scientific Press, 1986.
\bibitem{BATEMAN}
H.Bateman. {\it A Generalization of the Legandre Polynomial}.
Proc. London Math. Soc., {\bf 3}, 111-123, 1905.
\bibitem{BE1}
H.Bateman and A.Erdelyi. {\it Higher Transcendental Functions}. vol. 1,
McGraw-Hill New York, Toronto, London, 1953.
\bibitem{BE2}
H.Bateman and A. Erdelyi. {\it Higher transcendetal functions}. vol. 2 Graw-Hill.
New York, Toronto, London, 1953.
\bibitem{BAZ}
A.I.Baz', Ya.B.Zeldovich and A.M.Perelomov. {\it Scattering, Reactions and
Decompositions in Nonrelativistic Quantum Mechanics}. Moscow, Nauka,
1971, (in Russian).
\bibitem{BELL}
S.Bell. {\it Matrix elements in harmonic representations}.
J.Phys., {\bf B9}, 507-514, 1976.
\bibitem{BERTRAN}
J.Bertrand. {\it Th\'eorime relatif au mouvement d\'un point
atlire vers un centre fixe}. Comptes Rendus, {\bf 77}, 849-853, 1873.
\bibitem{BB}
N.Bessis and G.Bessis. {\it Electronic wavefunctions in a space of
constant curvature}. J.Phys., {\bf A12}, 1991-1997, 1979.
\bibitem{SHIR}
N.N.Bogoliubov and D.V.Shirkov. {\it Introduction to the theory of quantized fields}.
John Wiley and Sons, 1980.
\bibitem{BOT}
A.A.Bogush and V.S.Otchik.  {\it Problem of two Coulomb centres at
large intercentre separation: asymptotic expansions from
analytical solutions of the Heun eqution}. J.Phys., {\bf A30},
559-571, 1997.
\bibitem{BKO}
A.A.Bogush, Yu.A.Kurochkin and V.S.Otchik. {\it Quatum Mechanical Kepler
Problem in the 3-Dimensional Lobachevsky Space}.
Doklady Akademii Nauk Belarusi, {\bf 24}, 19-22, 1980, (in Russian).
\bibitem{BOR}
A.A.Bogush, V.S.Otchik and V.M.Red'kov. {\it Separation of variables in Schrodinger
equation and normed wave functions for
the Kepler problem in tree-dimensional spaces of constant curvature}.
Proceedings of the National Academy of Sciences of Belarus.
Ser. fiz.-mat, {\bf 3}, 56-62, 1983 (in Russian).
\bibitem{DAS1}
D.Bonatos, C.Daskaloyannis and K.Kokkotas. {\it Deformed
Oscillator Algebras for Two-Dimensional Quantum Superintegrable
Systems}. Phys. Rev., {\bf A50}, 3700-3709, 1994.
\bibitem{CHACON}
E.Chacon and M. de Llano. {\it Transformation Brakets Between
Cartesian and Angular Momentum Harmonic Oscillator Basis Functions
with and without Spin-Orbit Coupling. Tables for the 2s-1d Nuclear
Shell}. Rev. Mex. de Fisica, {\bf 12}, 57-68, 1963.
\bibitem{CHEN}
A.C.Chen.  {\it The zero-energy Coulomb problem}. J.Math.Phys.,
{\bf 19}, 1037-1040, 1978.
\bibitem{MAC}
A.Cisneros and H.V.McIntosh. {\it Symmetry of the Two-Dimensional
Hydrogen Atom}. J.Math.Phys., {\bf 10}, 277-286, 1969.
\bibitem{KHARE}
F.Cooper, A.Khare and U.Sukhatme. {\it Supersymmetry and quantum
mechanics}. Phys. Report, {\bf 251}, 267-385, 1995.
\bibitem{COJO}
C.A.Coulson and A.Joseph. {\it Spheroidal Wave Functions for the
Hydrogen Atom}. Proc. Phys. Soc. London, {\bf 90}, 887-893, 1967.
\bibitem{COULSON1}
C.A.Coulson and P.D.Robertson. {\it Wave Functions for the
Hydrogen Atom in Spheroidal Coordinates. I. The Derivation and
Properties of the Functions}. Proc. Phys. Soc. London, {\bf 71},
815-827, 1958.
\bibitem{DAS2}
C.Daskaloyannis. {\it Quadratic Poisson algebras of
two-dimensional classical superintegrable systems and quadratic
associative algebras of quantum superintegrable systems}. J.Math.
Phys., {\bf 42}, 1100-1119, 2001.
\bibitem{DPST1}
L.S.Davtyan, G.S.Pogosyan, A.N.Sissakian and V.M.Ter-Antonyan.
{\it On the Hidden Symmetry of a One-Dimensional Hydrogen Atom}.
J.Phys. {\bf A20}, 2765-2772, 1987.
\bibitem{DMPST}
L.S.Davtyan, L.G.Mardoyan, G.S.Pogosyan, A.N.Sissakian and
V.M.Ter-Antonyan. {\it Generalized KS transformation: from
five-dimensional hydrogen atom to eight-dimensional oscillator}.
J.Phys., {\bf A20}, 6121-6125, 1987.
\bibitem{DPST}
L.S.Davtyan, G.S.Pogosyan, A.N.Sisakyan and  V.M.Ter-Antonyan.
{\it Two - dimensional Hydrogen Atom.
Reciprocal Expansions of the Polar and Parabolic Bases in the
Continuous Spectrum}.
Theor. and Math. Phys., {\bf 66}, 146-153, 1986.
\bibitem{T-5}
L.S.Davtyan, G.S.Pogosyan, A.N.Sisakyan and V.M.Ter-Antonyan.
{\it Transformations between Parabolic Bases of a
Two-Di\-men\-sio\-nal Hydrogen Atom in the Continuous Spectrum}.
Theor. and Math. Phys., {\bf 74}, 157-161, 1988.
\bibitem{T-7}
L.S.Davtyan, G.S.Pogosyan, A.N.Sissakyan and V.M.Ter-Antonyan.
{\it Kinematic Aspects of Interbasis Expansions in Systems with Hidden
Symmetry}. Communication JINR, P2-85-503, Dubna, 1985,
(in Russian).
\bibitem{POG-10}
L.S.Davtyan, G.S.Pogosyan, A.N.Sissakyan and V.M.Ter-Antonyan.
{\it Transformations connecting the parabolic bases of the two-dimensional
hydrogen atom in continuous spectrum}. Theor. and Math. Phys., {\bf 74}, 240-246,
1988, (in Russian).
\bibitem{T-9}
L.S.Davtyan, L.G.Mardoyan, G.S.Pogosyan, A.N.Sissakian and
V.M.Ter-Antonyan. {\it KS-Transformation of the Three Dimensional Prolate Spheroidal
Coordinates}. Preprint JINR, P2-87-323, Dubna, 1987, (in Russian).
\bibitem{T-10}
L.S.Davtyan, L.G.Mardoyan, G.S.Pogosyan, A.N.Sissakian and
V.M.Ter-Antonyan. {\it Spheroidal Basis of the Four-Dimensional Isotropic Oscillator}.
Preprint JINR, P2-87-453, Dubna, 1987, (in Russian).
\bibitem{T-11}
L.S.Davtyan, G.S.Pogosyan, A.N.Sissakian and V.M.Ter-Antonyan.
{\it Algebraization of elliptic Coulomb wave functions in
continuous spectrum}. Preprint JINR, P2-87-454, Dubna, 1987.
\bibitem{DEMKOV1}
Yu.N.Demkov. {\it Symmetry group of the isotropic oscillator}. Sov. Phys. JETP.
{\bf 26}, 757, 1954, (in Russian); {\bf 36}, 63-66, 1959.
\bibitem{DEMKOV2}
Yu.N.Demkov. {\it The definition of the symmetry group of a quantum system.
The anisotropic oscillator}. Soviet Physics JETP, {\bf 17}, 1349-1351, 1963.
\bibitem{DICKE}
R.Dicke. {\it Coherence in spontaneous radiation processes}. Phys.
Rev., {\bf 93}, 99-110, 1954.
\bibitem{DIRAC}
P.A.M. Dirac. {\it Quantised singularities in the Electromagnetic
Field}. Proc. Roy. Soc., A, V. 133, 60-72, 1931.
\bibitem{DOTAN}
Y.Dothan, M.Gell-Mann and Y.Ne'eman. {\it Series of hadron energy
levels as representations of non-compact groups}. Phys. Lett.,
{\bf 17}, 148-151, 1965.
\bibitem{DRAGAN}
Gh.E. Dr{$\breve a$}g{$\breve a$}nescu, C. Campigotto and M.
Kibler. {\it On a generalized Aharonov-Bohm plus Coulomb system}.
Phys. Lett., {\bf A170}, 339-343, 1992.
\bibitem{DULOCK}
V.A.Dulock and H.V.McIntosh. {\it On the degeneracy of the
two-dimensional harmonic oscillator}. Amer. J. Phys., {\bf 33},
109-118, 1969.
\bibitem{EDEN}
R.J.Eden, P.V.Lnadshoff, D.I.Olive, J.C.Polkinghorne. {\it
Analytic S-matrix}. Cambridge: University Press, 1966.
\bibitem{EISENHART}
L.P.Eisenhart. {\it Enumeration of potentials for which
one-particle Shr\"odinger equations are separable}. Phys. Rev.,
{\bf 74}, 87-89, 1948.
\bibitem{ELLIOT2}
J.P.Elliot. {\it Collective motion in the nuclear shell model}.
Proc. Roy. Soc., Part I,II. {\bf 245}, 128-145, 562-581, 1958.
\bibitem{ELLIOT1}
J.P.Elliott, A.M.Lane. {\it Encyclopedia of Physics}, {\bf 39}, Ed. S.Fluge,
Berlin-Go ttingen-Heidelberg, 1957; {\it The structure of the atomic nucleus}.
Moscow, IL, 1958, (in Russian).
\bibitem{ENGLEF}
M.J.Englefield. {\it Group theory and the Coulomb problem}.
Wiley-Interscience, New York, London, Sydney, Toronto, 1972.
\bibitem{ESSLER}
F.H.L.Essler and V.Rittenberg. {\it Representations of the
quadratic algebra and partially asymmetric diffusion with open
boundaries}. J.Phys., {\bf A29}, 3375-3407, 1996.
\bibitem{EVANS}
N.W.Evans.  {\it Superintegrability in Classical Mechanics}.
Phys.Rev., {\bf A 41}, 5666-5676, 1990; {\it Super-Integrability of
the Winternitz System}. Phys.Lett., {\bf A 147}, 483-486, 1990; {\it
Group Theory of the Smorodinsky-Winternitz System} J.Math.Phys.,
{\bf 32}, 3369-3375, 1991.
\bibitem{FAIFMAN}
M.P.Faifman, L.I.Ponomarev, S.I.Vinitsky. {\it Asymptotic form
of effective potentials of the Coulomb three-body problem in the
adiabatic representations}. J. Phys., {\bf B9}, 2255-2268, 1976.
\bibitem{FISHER}
J.Fisher. {\it Beitr\"age zur Theorie der Absorption von
R\"ontgenstrahlen }. Ann. Phys., {\bf 8}, 821-850, 1931.
\bibitem{FOCK1}
V.A.Fock. {\it Zur Theorie des Wasserstoffatoms}. Zs. Phys., {\bf
98}, 145-154, 1935.
\bibitem{FOCK2}
V.A.Fock.
{\it Foundations of quantum mechanics}. Nauka, Moscow. 1976, (in Russian).
\bibitem{FMSUW}
J.Fri\v s, V.Mandrosov, Ya.A.Smorodinsky, M.Uhlir and
P.Winternitz. {\it On Higher Symmetries in Quantum Mechanics}.
Phys. Lett., {\bf 16}, 354-356, 1965.
\bibitem{FRONC}
C.Fronsdal. {\it Infinite multiplets and the hydrogen atom}. Phys.
Rev., {\bf 156}, 1665-1667, 1967.
\bibitem{FUJIK}
K.Fujikawa. {\it Path integral of the hydrogen atom, Jacobi's
principle of least action and one-dimensional quantum gravity}.
Nucl. Phys., {\bf B484}, 495-520, 1997.
\bibitem{GRAN3}
O.F.Gal'bert, Yu.A.Granovskii and A.S.Zhedanov {\it Dynamical
symmetry of anisotropic singular oscillator}. Phys. Lett., {\bf
A153}, 177-180, 1991.
\bibitem{KOGAN}
V.M.Galitsky, B.M.Karnakov and V.I.Kogan. {\it Problems in Quantum Mechanics}.
Textbook, 1992.
\bibitem{GELMAN}
M.Gell-Mann. {\it The symmetry group of vector and axial vector
currents}. Physics, {\bf 1}, 63-75, 1964.
\bibitem{GIBBONS}
G.W.Gibbons and N.S.Manton.  {\it Classical and quantum dynamics of
BPS monopole}. Nucl. Phys., {\bf B274}, 183-224,
1986.
\bibitem{GRAD}
I.S.Gradshteyn and I.M.Ryzhik. {\it Table of integrals, series, and products}.
Academic press, 2014.
\bibitem{GRAN1}
Yu.A.Granovskii, I.M.Lutsenko and A.S.Zhedanov. {\it Mutual
integrability, quadratic algebras and dynamical symmetry}. Ann.
Phys., {\bf 217}, 1-20, 1992.
\bibitem{GRAN2}
Yu.A.Granovskii, A.S.Zhedanov and I.M.Lutsenko. {\it Quadratic
algebra as a hidden symmetry of a Hartmann potential}. J.Phys.,
{\bf A24}, 3887-3894, 1991.
\bibitem{GZLa}
Yu.I.Granovskii, A.S.Zhedanov and I.M.Lutsenko. {\it Quadratic algebras
and dynamics in curved spaces. I. Oscillator}.
Theor. and Math. Phys., {\bf 91}, 474-480, 1992.
\bibitem{GZLb}
Yu.I.Granovskii, A.S.Zhedanov and I.M.Lutsenko. {\it Quadratic algebras
and dynamics in curved spaces. II. The Kepler problem}.
Theor. and Math. Phys., {\bf 91}, 604-612, 1992.
\bibitem{GRITKUR}
V.V.Gritsev and Yu.A.Kurochkin. {\it Model of excitations in
quantum dots based on quantum mechanics in spaces of quantum
curvature}. Phys. Rev., {\bf B64}, 035308, 2001.
\bibitem{GRITKUROT}
V.V. Gritsev, Yu.A. Kurochkin and V.S. Otchik. {\it Nonlinear
symmetry of the MIC-Kepler problem on the sphere $S^2$}. J. Phys., {\bf
A33}, 4903-4910, 2000.
\bibitem{GRO1}
C.Grosche. {\it The Path Integral for the Kepler Problem on the
Pseudosphere}. Ann. Phys., {\bf 204}, 208-222, 1990.
\bibitem{GRO2}
C.Grosche. {\it On the Path Integral in Imaginary Lobachevsky
Space}. J. Phys., {\bf A27}, 3475-3489, 1994.
\bibitem{GROP5}
C.Grosche, G.S.Pogosyan and A.N.Sissakian. {\it Path Integral
discussion for Smorodinsky - Winternitz Potentials: II. Two - and
Three Dimensional Sphere.} Fortschritte der Physik, {\bf 43(6)},
523-563, 1995.
\bibitem{GROP4}
C.Grosche, G.S.Pogosyan and A.N.Sissakian. {\it Path Integral
Approach to Superintegrable Potentials. Two - Dimensional
Hyperboloid.} Phys. Part. Nucl., {\bf 27}, 593-674, 1996.
\bibitem{GROP1}
C.Grosche, G.S.Pogosyan and A.N.Sissakian. {\it Path Integral
discussion for Superintegrable Potentials: IV. Three Dimensional
Pseudosphere}. Phys. Part. Nucl., {\bf 28}, 1230-1294, 1997.
\bibitem{GYORGYI}
G.Gy\"orgyi. {\it Kepler's equation, Fock variables, Bacry's
generators and Dirac brackets}. Nuovo Cimento, {\bf A53}, 717-736,
1968.
\bibitem{revai}
G.Gy\"orgyi and J.Revai. {\it Hidden Symmetry of the Kepler Problem}.
Soviet Physics JETP, {\bf 21}, 967-968, 1965.
\bibitem{HKPS1}
Ye.M.Hakobyan, M.Kibler, G.Pogosyan and A.N.Sissakian. {\it On a
Generalized Oscillator: Invariance Algebra and Interbasis
Expansions.} Phys. At. Nucl., {\bf 61}, 1782-1788, 1998.
\bibitem{HPSV1}
Ye.M.Hakobyan, S.I.Vinitsky, G.S.Pogosyan and A.N.Sissakian.
{\it Isotropic Oscillator in the space of Constant
Positive Curvature. Interbasis expansions}.
Phys. At. Nucl., {\bf 62}, 623-637, 1999.
\bibitem{HAN}
M.Y.Han. {\it Quantum-mechanical generators of the group for
hydrogen atom bound states}. Nuovo Cimento, {\bf B42},
367-370,1966.
\bibitem{HARTMAN}
H.Hartmann. {\it Bewegung eines K\"orpers in einem ringf\"ormigen
Potentialfield}. Theor. Chim. Acta {\bf 24}, 201-206, 1972.
\bibitem{HIG}
P.W.Higgs. {\it Dynamical Symmetries in a Spherical Geometry}.
J.Phys., {\bf A12}, 309-323, 1979.
\bibitem{HUR}
A.Hurwitz. {\it \"{U}ber die Zahlenthorie der Quaternionen}.
Narch. Ges. Wiss. G\" ottingen Math.-Phys. Kl, 313-340, 1896.
{\it Mathematische Werke}, Band II, 641,
(Birkh\"auser), Basel, 1933.
\bibitem{HYLLERAAS}
E.A.Hylleraas. {\it Die Wellengleichung des Keplerproblems im Impulsraume}. Z. Phys.,
{\bf 74}, 216-224, 1932.
\bibitem{IK}
M.Ikeda and N.Katayama. {\it On Generalization of Bertrand's
theorem to spaces of constant curvature}. Tensor, N.S., {\bf 38},
37-40, 1982.
\bibitem{ILKAL}
L.A.Il'kaeva. {\it Symmetry group of the anisotropic oscillator}. Vestnik Leningrad. Univ.,
{\bf 22}, 56-62, 1963.
\bibitem{INF}
L.Infeld. {\it On a New Treatment of Some Eigenvalue Problems}.
Phys.Rev., {\bf 59}, 737-747, 1941.
\bibitem{IS}
L.Infeld and A.Schild. {\it A Note on the Kepler Problem in a
Space of Constant Negative Curvature}. Phys.Rev.,  {\bf 67},
121-122, 1945.
\bibitem{IW}
E.In\"on\"u and E.P.Wigner. {\it On the contractions of groups and
their representations}. Proc. Nat. Acad. Sci. (US), {\bf 39}
510-524, 1953.
\bibitem{INOMATA}
A. Inomata, G. Junker, R.Wilson. {\it Topological charge
quantization via path integration: An application of the
Kustaanheimo-Stiefel transformation}. Found. Phys. {\bf 23},
1073-1091, 1993.
\bibitem{IWAI1}
T.Iwai. {\it Quantization of the conformal Kepler problem and its
application to the hydrogen atom} J.Math.Phys., {\bf 23}, 1093-1099,
1982.
\bibitem{IWAI2}
T.Iwai and Y. Uwano. {\it The four-dimensional conformal Kepler problem
reduces to the three-dimensional Kepler problem with a centrifugal
potential and the Dirac's monopole field. Classical theory}.
J.Math.Phys., {\bf 27}, 1523-1529, 1986.
\bibitem{Iwai-Uwano}
T.Iwai and Y.Uwano. {\it The quantized MIC-Kepler problem and its
symmetry group for negative energies}. J. Phys., {\bf A21},
4083-4104, 1988.
\bibitem{IWAI3}
T. Iwai. {\it The symmetry group of the
harmonic oscillator and its reduction}. J. Math. Phys., {\bf 23},
1088-1092, 1982.
\bibitem{IPSW4}
A.A.Izmest'ev, G.S.Pogosyan, A.N.Sissakian and P.Winternitz. {\it
Contraction of Lie Algebras and Separation of Variables.} J.Phys.,
{\bf A29}, 5940-5962, 1996.
\bibitem{IPSW2}
A.A.Izmest'ev, G.S.Pogosyan, A.N.Sissakian and P.Winternitz. {\it
Contraction of Lie Algebras and Separation of Variables.
N-dimensional sphere}, J.Math.Phys., {\bf 40}, 1549-1573, 1999.
\bibitem{IZ1}
A.A.Izmest'ev. {\it Exactly solvable potential model for quarkonia. Gluon propagator}.
Sov. J. Nucl. Phys., {\bf 53}, 1402-1409, 1991.
\bibitem{JAUCH1}
J.M.Jauch. {\it Groups of quantum-mechanical contact
transformations and the degeneracy of energy-levels}. Phys. Rev.,
{\bf 55}, 1132(A), 1939.
\bibitem{JAUCH2}
J.M.Jauch and E.L.Hill. {\it On the problem of degeneracy in
quantum mechanics}. Phys. Rev., {\bf 57}, 641-645, 1940.
\bibitem{Lippmann}
M.H.Jonson, B.A.Lippmann. {\it Motion in a Constant Magnetic
Field.} Phys. Rev. {\bf 76}, 828-832, 1949.
\bibitem{KADYSH}
V.G.Kadyshevskii, R.M.Mir-Kasimov and N.B.Skachkov.
{\it Three-dimensional formulation of the relativistic two-body problem}.
Particles and Nuclei: {\bf 2}, 69-109, 1973.
\bibitem{KMJ}
E.G.Kalnins and W.Miller, Jr. {\it Separation of variables on
n-dimensional Riemannian  manifolds 1. The n-sphere  $S_n$  and
Euclidean n-space  $R_n$}. J. Math. Phys., {\bf 27}, 1721-1736, 1986.
\bibitem{KAL1}
E.G.Kalnins. {\it Separation of Variables for Riemannian spaces
of constant curvature}. Pitman, Monographs and Surveys in Pure and
Applied Mathematics 28, Longman, Essex, England, 1986.
\bibitem{KMW}
E.G.Kalnins, W.Miller, Jr. and P.Winternitz. {\it The Group
$O(4)$, Separation of Variables and the Hydrogen Atom}. SIAM
J.Appl.Math., {\bf 30}, 630-664, 1976.
\bibitem{IPSW1}
E.G.Kalnins, W.Miller, Jr. and G.S.Pogosyan. {\it  Completeness of
multiseparable superintegrability in two dimensions}. Phys. At. Nucl., {\bf
65(6)}, 1066-1068, 2002.
\bibitem{KMJP2}  E.G.Kalnins, W.Miller, Jr. and G.S.Pogosyan.
{\it Completeness of multiseparable superintegrability in
$E_{2,C}$}. J.Phys., {\bf A33}, 4105-4120, 2000.
\bibitem{KMJP3}
E.G.Kalnins, W.Miller, Jr. and G.S.Pogosyan. {\it Completeness of
multiseparable superintegrability on the complex 2-sphere}.
J.Phys. {\bf A33}, 6791-6806, 2000.
\bibitem{KKMP1}
E.G.Kalnins, J.M.Kress, W.Miller, Jr. and G.S.Pogosyan. {\it
Completeness of superintegrability in two dimensional constant
curvature spaces}. J.Phys. {\bf A34}, 4705-4720, 2001.
\bibitem{KMP7}
E.G.Kalnins, W.Miller, Jr. and G.S.Pogosyan. {\it
Coulomb-oscillator duality in spaces of constant curvature}.
J.Math.Phys. {\bf 41}, 2629-2657, 2000.
\bibitem{KMP9}
E.G.Kalnins, W.Miller, Jr. and G.S.Pogosyan. {\it The
Coulomb-Oscillator Relation on n-Dimensional Spheres and
Hyperboloids}. Phys. At. Nucl., {\bf 65}, 1086-1094, 2002.
\bibitem{KAMPE}
Appel D.Kampe, J.de Feriet.
{\it Functions hypergeometriques et hypersferiques, Polinomes d'Hermite}.
Paris, 1926.
\bibitem{KATA}
N.Katayama. {\it A Note on a Quantum-Mechanical Harmonic
Oscillator in a Space of Constant Curvature}. Nuovo Cimento, {\bf
B107}, 763-768, 1992.
\bibitem{KIBLER}
M.Kibler and G.Grenet {\it On the $SU_2$ Unit Tensor},
J.Math.Phys., {\bf 21}, 422-439, 1980.
\bibitem{Kibler-Negadi}
M.Kibler and T.Negadi. {\it On the connection between the hydrogen atom
and the harmonic oscillator: the continuum case}.
J. Phys., {\bf A16}, 4265-4268 1983.
\bibitem{Kibler4}
M.Kibler, A.Ronveaux and T. Negadi.{\it On the hydrogen-oscillator
connection: Passage formulas between wave functions}. J. Math.
Phys., {\bf 27}, 1541-1548, (1986).
\bibitem{KIL}
M.S.Kildyushov. {\it Hyperspherical "tree" type functions in
problem of $n$ particles}. Sov. J. Nucl. Phys., {\bf 15}, 113-123, 1972.
\bibitem{KPS}
V.A.Knyr, P.P.Pepiraite and Yu.F.Smirnov. {\it Canonical transformations "trees" and
moments multiple of 1/4}. Sov. J. Nucl. Phys., {\bf 22}, 554, 1975.
\bibitem{KOMPOCLA}
I.V.Komarov, L.I.Ponomarev and S.Yu.Slavyanov.
{\it Spheroidal and Coulombian Spheroidal functions}.
 "Nauka", Moscow, 1971, (in Russian).
\bibitem{KO}
Yu.A.Kurochkin and V.S.Otchik. {\it Runge-Lenz Vector Analog and the Energy-Spectrum in the
Kepler-Problem on a 3-Dimensional Sphere}. DOKLADY AKADEMII NAUK BELARUSI,
{\bf 23}, 987-990, 1979.
\bibitem{KSMOR1}
G.I.Kuznetsov and Y.A.Smorodinsky. {\it Degenerate T-coefficients and their connection with 3j-symbols
and Wigner D-functions}.  Sov. J. Nucl. Phys., {\bf 25}, 447-449, 1977.
\bibitem{KUZNECOV}
G.I.Kuznetsov. {\it A Remark on the Multi-dimensional Coulomb Problem}.
Soviet Physics JETP, {\bf 24}, No.1, 1967.
\bibitem{KS}
P.Kustaanheimo and E.Stiefel. {\it Perturbation Theory of Kepler
Motion Based on Spinor Regularization}. J.Rein.Angew.Math., {\bf
218}, 204-219, 1965.
\bibitem{KIB2}
D.Lambert and M.Kibler: {\it An Algebraic and geometric approach
to non-bijective quadratic transformation}. J.Phys., {\bf A21},
307-343, 1988.
\bibitem{LANDAU-30}
L.D.Landau. {\it Diamagnetismus der Metalle}. Zs. Phys., {\bf
64}, 629-637, 1930.
\bibitem{LL}
L.D.Landau and E.M.Lifshits. {\it Quantum Mechanics}.
Pergamon Press, 1981.
\bibitem{LAPLACE}
M.Laplace. {\it Traite de macanique celeste}. V.I. Ch. 3, Paris,
Bachelier, 1889.
\bibitem{LAPORTE1}
O.Laporte. {\it Note on Kowalewski's Top in Quantum Mechanics}.
Phys. Rev. {\bf 43}, 548-552, 1933.
\bibitem{LAPORTE2}
O.Laporte and G.Y.Rainich. {\it Stereographic parameters and
pseudo-minimal hypersurfaces}. Trans. Am. Math. Soc.  {\bf 39},
154-182, 1936.
\bibitem{LEEMON}
H.I.Leemon. {\it Dynamical Symmetries in a Spherical Geometry}.
J.Phys., {\bf A12}, 489-501, 1979.
\bibitem{LENZ}
W.Lenz.  {\it \"Uber den Bewegungsverlauf und die
Quantenzust\"ande der gest\"orten Keplerbewegung}. Zeitschr.Phys.,
{\bf 24}, 197-207, 1924.
\bibitem{VILE}
P.Letourneau and L.Vinet. {\it Superintegrable systems: Polynomial
Algebras and quasi-exactly-solvable Hamiltonians}. Ann. Phys.,
{\bf 243}, 144-168, 1995.
\bibitem{HVT}
Le Van Hoang, Tony J.Viloria and Le anh Thu: {\it On the
hydrogen-like atom in five dimensional space}. J.Phys., {\bf A24},
3021-3030, 1991.
\bibitem{LC}
T.Levi-Civita. {\it Sur la Resolution Qualitative di Probleme
Restreint des Trois Corps}. {\it Opere Mathematiche}, {\bf 2},
411-417, 1956.
\bibitem{LEVY}
M.Levy. {\it Wave equations in momentum space}. Prog. Roy. Soc.,
{\bf 204}, 145-169, 1950.
\bibitem{LMPST1}
I.V.Lutsenko, L.G.Mardoyan, G.S.Pogosyan, A.N.Sissakian and
V.M.Ter-Antonyan. {\it Non-Relativistic Coulomb Problem in a
One-Dimensional Quantum Mechanics}. J.Phys. {\bf A22}, 2739-2749,
1989.
\bibitem{LUC-POG-1}
I.V.Lutsenko, G.S.Pogosyan, A.N.Sissakian and V.M.Ter-Antonyan.
{\it Supersymmetry of a One-Dimensional Hedrogen Atom}. Phys.
Lett., {\bf A143}, 247-249, 1990.
\bibitem{ARTUR}
A. Maghakian, A. Sissakian and V. Ter-Antonyan.
{\it Electromagnetic duality for anyons}. Phys. Lett., {\bf A236},
5-7, 1997.
\bibitem{MAIELLA}
G.Maiella and G.Vilasi. {\it Reducible representations of the
symmetry group of the anisotropic harmonic oscillator}. Lett.
Nuovo Cimento, {\bf 1}, 57-64, 1969.
\bibitem{BESU}
S.D.Majundar and D.Basu. {\it $O(3,1)$ Symmetry of the Hydrogen
Atom}. J.Phys., {\bf A7}, 787-793, 1974.
\bibitem{MSVW}
A.A.Makarov, Ya.A.Smorodinsky, Kh.Valiev and P.Winternitz: {\it A
Systematic Search for Nonrelativistic Systems with Dynamical
Symmetries}. Nuovo Cimento,  {\bf A52}, 1061-1084, 1967.
\bibitem{MANKO1}
I.A.Malkin and V. I.Manko.
{\it Dynamic symmetry and coherent states of quantum systems}.
"Nauka", Moscow, 1979, (in Russian).
\bibitem{MANKO2}
I.A.Malkin and V. I.Manko. {\it Symmetry of the hydrogen atom}.
Sov. J. Nucl. Phys., {\bf 3}, 372-382, 1966.
\bibitem{T-1}
L.G.Mardoyan, G.S.Pogosyan, A.N.Sissakian and V.M.Ter-Antonyan. {\it
Spheroidal analysis of hydrogen atom}. J.Phys., {\bf A16},
711-728, 1983.
\bibitem{T-2}
L.G.Mardoyan, G.S.Pogosyan, A.N.Sissakyan and V.M.Ter-Antonyan.
{\it Interbasis Expansions in a Circular Oscillator}. Nuovo
Cimento, {\bf A86}, 324-336, 1985.
\bibitem{T-4}
L.G.Mardoyan, G.S.Pogosyan, A.N.Sissakian and V.M.Ter-Antonyan.
{\it Hidden symmetry, Separation of Variables and Interbasis
Expansions in the Two-Di\-men\-sio\-nal Hydrogen Atom}. J.Phys.,
{\bf A18}, 455-466, 1985.
\bibitem{MPS1}
L.G.Mardoyan, G.S.Pogosyan, A.N.Sissakyan and V.M.Ter-Antonyan.
{\it Elliptic Basis of Circular Oscillator}. Nuovo Cimento {\bf
B88}, 43-56, 1985.
\bibitem{MSTA}
L.G. Mardoyan, A.N. Sissakian and V.M. Ter-Antonyan.
{\it Park-Tarter matrix for a dyon-dyon system}. Int. J. Mod.
Phys., {\bf A12}, 237-242, 1997.
\bibitem{MSTA1}
L.G. Mardoyan, A.N. Sissakian and V.M. Ter-Antonyan. {\it The
Eulerian Parametrization of the Hurwitz Transformation.}  In
Proceedings: International Workshop Finite-Dimensional Integrable
Systems, Eds. A.N. Sissakian, G.S. Pogosyan, JINR Publishing
Department, Dubna, E2-95-525, 137-142, 1995.
\bibitem{MST1}
L.G.Mardoyan, A.N.Sissakian, and V.M.Ter-Antonyan.
{\it Oscillator as a hidden non-Abelian monopole.} Preprint JINR,
E2-96-24, Dubna, 1996; hep-th/9601093.
{\it 8D Oscillator as a hidden SU(2)-monopole}. Phys. At. Nucl., {\bf 61}, 1746-1750,
1998.
\bibitem{MARDOYAN5}
L.G.Mardoyan. {\it Dyon-Oscillator Duality. Hidden Symmetry of
the Yang-Coulomb Monopole}. CRM Proceedings and Lecture Notes
"Superintegrability in Classical and Quantum Systems", {\bf 37},
99-108, 2004; quant-ph/0302162.
\bibitem{MSTA2}
L.G.Mardoyan, A.N.Sissakian and V.M.Ter-Antonyan. {\it Hidden
symmetry of the Yang-Coulomb monopole}. Mod. Phys. Lett. {\bf
A14}, 1303-1307, 1999.
\bibitem{MARDOYAN}
L.G.Mardoyan. {\it The Five--Dimensional SU(2) -- Monopole:
Continuous Spectrum.} Phys. At. Nucl., {\bf 65}, 1096-1102, 2002.
\bibitem{MARDOYAN3}
L.G. Mardoyan, L.S. Petrosyan and H.A. Sarkisyan. {\it Charge-dyon bound
system in the spherical quantum well}. Phys. Rev., {\bf A68}, 014103(R),
2003.
\bibitem{MPSTA2}
L.G. Mardoyan, G.S. Pogosyan, A.N. Sissakian and V.M. Ter-Antonyan.
{\it KS-Trans\-formation and Coulomb-Oscillator Interbasis
Expansions}. In Proceedings: Schrodinger Operators: Standard and
Non-Standard, 333-349, World Scientific Publishing, 1989.
\bibitem{MPSTA1}
L.G.Mardoyan, G.S.Pogosyan, A.N.Sisakyan and V.M.Ter-Antonyan.
{\it Spheroidal corrections to the spherical and
parabolic bases of the hydrogen atom}. Theor. and Math. Phys.,
{\bf 64} 762-764, 1985.
\bibitem{T-3}
L.G.Mardoyan, G.S.Pogosyan, A.N.Sisakyan and V.M.Ter-Antonyan.
{\it Interbasis expansions in the two-dimensional hydrogen atom}.
Theor. and Math. Phys., {\bf 63}, 597-603, 1985.
\bibitem{MPS}
L.G.Mardoyan, G.S.Pogosyan, A.N.Sisakyan and V.M.Ter-Antonyan.
{\it Two-dimensional hydrogen atom. I. Elliptic basis}.
Theor. and Math. Phys., {\bf 61}, 1021-1034, 1984.
\bibitem{MARDOYAN1}
L.G.Mardoyan. {\it Coulomb-oscillator duality and scattering problem
in the five-dimensional coulomb field}.
Izvestiya NAN Armenii, Fizika, {\bf 39}, 99-106, 2004, (in Russian).
\bibitem{MARDOYAN2}
L.G.Mardoyan. {\it Scattering of Electrons on the Dyon}.
Theor. and Math. Phys., {\bf 136}, 1110-1118, 2003.
\bibitem{MARDOYAN4}
L.G.Mardoyan, A.P.Nersesyan and M.G.Petrosyan. {\it Stark Effect in Charge–Dyon System}.
Theor. and Math. Phys., {\bf 140}, 958-964, 2004.
\bibitem{MSTA3}
L.G.Mardoyan, A.N.Sisakyan and V.M.Ter-Antonyan. {\it Bases and interbasis transformations for the SU(2) monopole}.
Theor. and Math. Phys., {\bf 123}, 451-462, 2000.
\bibitem{MPSTA4}
L.G.Mardoyan, G.S.Pogosyan, A.N.Sisakyan and V.M.Ter-Antonyan.
{\it Elliptic basis for a circular oscillator}. Theor. and Math. Phys.,
{\bf 65},  1113-1122, 1985.
\bibitem{MPTA}
L.G.Mardoyan, G.S.Pogosyan and V.M.Ter-Antonyan.
{\it On the expansion of hydrogen spheroidal wave functions in spherical ones}.
Izvestiya NAN Armenii, Fizika, {\bf 19}, 3-9, 1984, (in Russian).
\bibitem{MPSTA5}
L.G.Mardoyan, G.S.Pogosyan, A.N.Sisakyan and V.M.Ter-Antonyan.
{\it To the elliptic basis of the two-dimensional hydrogen atom}.
Preprint JINR P2-84-110, Dubna, 1984, (in Russian).
\bibitem{MPSTA8}
L.G.Mardoyan, G.S.Pogosyan, A.N.Sisakyan and V.M.Ter-Antonyan.
{\it Isotropic oscillator: three-term recurrence relations for the coefficients of expansions
of a spheroidal basis in terms of spherical and cylindrical}.
Preprint JINR P2-85-139, Dubna, 1985, (in Russian).
\bibitem{MPSTA9}
L.G.Mardoyan, G.S.Pogosyan, A.N.Sisakyan and V.M.Ter-Antonyan.
{\it Isotropic oscillator: expansion of a spheroidal basis in spherical and cylindrical}.
Preprint JINR P2-85-140, Dubna, 1985, (in Russian).
\bibitem{MPSTA10}
L.G.Mardoyan, G.S.Pogosyan, A.N.Sisakyan and V.M.Ter-Antonyan.
{\it Isotropic Oscillator: Spheroidal Wave Functions}.
Preprint JINR P2-85-141, Dubna, 1985, (in Russian).
\bibitem{MPSTA11}
L.G.Mardoyan, G.S.Pogosyan, A.N.Sisakyan and V.M.Ter-Antonyan.
{\it Exact solution of the Kibler-Ronveaux-Negadi problem}.
Preprint JINR P2-86-431, Dubna, 1986, (in Russian).
\bibitem{MPSTA12}
L.G.Mardoyan, G.S.Pogosyan, A.N.Sisakyan and V.M.Ter-Antonyan.
{\it Some interbasis expansions in a four-dimensional isotropic oscillator}.
Preprint JINR P2-86-436, Dubna, 1986, (in Russian).
\bibitem{MPSTA13}
L.G.Mardoyan, G.S.Pogosyan, A.N.Sisakyan and V.M.Ter-Antonyan.
{\it On the problem of interbasis expansions in quantum mechanics}.
Proceedings of the Interuniversity Conference "Computational Physics and Mathematical Modeling"
65-73, Moscow, 1989, (in Russian).
\bibitem{MARIWALLA}
K.Mariwalla. {\it Dynamical symmetries in mechanics}. Phys.
Report, {\bf C20}, 289-362, 1975.
\bibitem{MIC}
H.McIntosh and A. Cisneros. {\it Degeneracy in the presence of a
magnetic monopole}. J. Math. Phys., {\bf 11}, 896-916, 1970.
\bibitem{MEIXNER}
J.Meixner. {\it Die Greensche Function des Wellenmechanischen
Keplerproblem}. Math. Z., {\bf 36}, 677-707, 1933.
\bibitem{MILLER-1974}
W.Miller, Jr. {\it Lie Theory and Separation of Variables.
Parabolic Coordinates.} SIAM J. Math. Anal. {\bf 5}, 822-836,
1974.
\bibitem{MIL1}
W. Miller, Jr. {\it Symmetry and Separation of Variables}.
Addison-Wesley Publishing Company, 1977.
\bibitem{MLADENOV}
I.M.Mladenov and V.V.Tsanov. {\it Geometric quantization of the
MIC-Kepler problem}. J. Phys., {\bf A20}, 5865-5871, 1987; {\it
Reduction in stages and complete quantization of the MIC-Kepler
problem}. J. Phys., {\bf A32}, 3779-3791, 1999.
\bibitem{MORS-FESHBACH}
P.M.Morse and H.Feshbach. {\it Methods of theoretical physics}.
Mc Graw-Hill book company, New York-Toronto-London. 1953.
\bibitem{MUKUNDA}
N.Mukunda, L.O'Raifeartaigh and E.C.G.Sudarshan. {\it Characteristic
noninvariance groups of dynamical systems}. Phys. Rev. Lett., {\bf
15}, 1041-1044, 1965.
\bibitem{MUSTO}
R.Musto. {\it Generators of $O(4,1)$ for the quantum-mechanical
hydrogen atom}.  Phys. Rev., {\bf 148}, 1274-1275, 1966.
\bibitem{NTAT}
A.Nersessian, V.M.Ter-Antonyan and M.M. Tsulaia. {\it A note on
quantum Bohlin transformation}. Mod.Phys.Lett., {\bf A11},
1605-1610, 1996.
\bibitem{NTA}
A. Nersessian and V. Ter-Antonyan. {\it Anyons, monopole and Coulomb problem}.
Phys. At. Nucl., {\bf 61}, 1868-1872, 1998.
\bibitem{NTA1}
A. Nersessian and V. Ter-Antonyan.
{\it Charge Dyon System As The Reduced Oscillator}. Mod. Phys.
Lett., {\bf A9}, 2431-2435, 1994.
\bibitem{NERPOG}
A. Nersessian and G.Pogosyan. {\it On the relation of the
oscillator and Coulomb systems on (pseudo)spheres}. Phys. Rev.,
{\bf A63}, 020103(R) 2001.
\bibitem{OLEV}
M.N.Olevskii. {\it Triorthogonal systems in spaces of constant curvature in which the equation
$\Delta_2u+\lambda u=0$ allows a complete separation of variables}.
Matematicheskii Sbornik. {\bf 69(27)}, 379-426, 1950.
\bibitem{OR}
V.S.Otchik and V.M.Red'kov. {\it Quantum mechanical Kepler problem in spaces of constant curvature}.
Preprint, 298, 1986.
\bibitem{PARK}
D.Park. {\it Relation Between the Parabolic and Spherical
Eigenfunctions of Hydrogen}. Zs. Phys., {\bf 159}, 155-157, 1960.
\bibitem{PAULI}
W.Pauli. {\it \"Uber das Wasserstoffspektrum vom Standpunkt der
neuen Quantenmechanik}. Zs. Phys., {\bf 36}, 336-363, 1926.
\bibitem{PERELOMOV}
A.M.Perelomov. {\it Integrable Systems of Classical Mechanics and Lie Algebras}.
Birkhauser Basel, 1990.
\bibitem{PER-POP}
A.M.Perelomov and V.S.Popovþ {\it The Lorentz group as a dynamic symmetry group of the hydrogen atom}.
SOVIET PHYSICS JETP, {\bf 23}, 118-129, 1966.
\bibitem{PER-POP1}
A. M.Perelomov and V.S.Popov. {\it Casimir operators for the orthogonal and symplectic groups}.
ZhETF Pisma Redaktsiiu, {\bf 2}, 34-37, 1965.
\bibitem{PER-POP2}
A.M.Perelomov and V.S.Popov. {\it On the" Hidden" Symmetry of the Hydrogen Atom}.
SOVIET PHYSICS JETP, {\bf 27}, 967-970, 1968.
\bibitem{TOLAR}
Z.Pluhar and J.Tolar. {\it Transformation Matrix for the Isotropic
Harmonic Oscillator $SU(3)$ and Representations}. Czech. J.Phys.,
{\bf B14}, 287-293, 1964.
\bibitem{PODOLSKY}
D.Podolsky and L.Pauling. {\it The Momentum Distribution in
Hydrogen-Like Atoms}. Phys. Rev., {\bf 34}, 109-116, 1929.
\bibitem{POGSIS}
G.S.Pogosyan, A.N.Sissakian and P.Winternitz. {\it Separation of
Variables and Lie Algebra Contractions. Applicationa to Special
Functions }. Phys. Part. Nucl., {\bf 33}, 123-144, 2002.
\bibitem{PSMT}
G.S.Pogosyan, Ya.A.Smorodinsky and V.M.Ter-Antonyan.
{\it Oscillator Wigner Functions}. J. Phys., {\bf A14}, 769-776, 1981.
\bibitem{POGOSYAN1}
G.S.Pogosyan and V. M.Ter-Antonyan. {\it Connection between spheric and parabolic
Coulomb wave functions in a continuous spectrum}.
Preprint JINR, R2-80-318, 1980, (in Russian).
\bibitem{POGOSYAN2}
G.S.Pogosyan, Y.A.Smorodinskij and V. M.Ter-Antonyan. {\it Multidimensional isotropic oscillator:
transitions between the cartesian and hypersherical bases}.
Preprint JINR, R2-82-118, 1982, (in Russian).
\bibitem{POGOSYAN3}
G.S.Pogosyan, V. M.Ter-Antonyan and H.T.Torosyan. {\it Relationship between Cartesian and polar
wave functions of a circular oscillator}. Preprint PLRF-77-04, Yerevan, 1977, (in Russian).
\bibitem{POGOSYAN4}
G.S.Pogosyan and V. M.Ter-Antonyan.
{\it The Connection between Cartesian and Polar Wavefunctions of a Circular Oscillator and the
Dynamical $O(3)$ Symmetry}. Izv. AN Arm. SSR. Fizika. {\bf 13}, 235-237, 1978, (in Russian).
\bibitem{POGOSYAN5}
G. S.Pogosyan and V.M.Ter-Antonyan.
{\it Connection between Cartesian and polar wave functions of a nonrelativistic
charged particle in a homogeneous magnetic field}. Theor. and Math. Phys.,
{\bf 40}, 651-653, 1979.
\bibitem{POGOSYAN6}
G.S.Pogosyan and V.M.Ter-Antonyan. {\it Conversion factors for the Cartesian, polar and cylindrical wave
functions of isotropic oscillator}. Preprint JINR, R2-11962, 1978, (in Russian).
\bibitem{POPOV}
V.S.Popov. {\it On a "hidden" symmetry of a hydrogen atom}.
High Energy Physics and Theory of Elementary Particles, Naukova Dumka, Kiev,
1967, (in Russian).
\bibitem{PRAT1}
R.H.Pratt and T.F.Jordan. {\it Generators of the de Sitter group
for the hydrogen atom}. Phys. Rev., {\bf 148}, 1276-1279, 1966.
\bibitem{PRAT2}
R.H.Pratt and T.F.Jordan. {\it Coulomb group theory for any spin}.
Phys. Rev., {\bf 188}, 2534-2535, 1969.
\bibitem{PRONKO}
G.P.Pron'Ko and Y.G.Stroganov. {\it A new example of the quantum mechanical
problem involving latent symmetry}.
SOVIET PHYSICS JETP, {\bf 72}, 2048-2054, 1977.
\bibitem{RUNGE}
C.Runge. {\it Vektoranalysis}, {\bf 1}.  Hirzel, Leipzig, 1919.
\bibitem{SCHR}
E.Schr\"{o}dinger. {\it A Method of Determining Quantum Mechanical
Eigenvalues and Eigenfunctions}. Proc.Roy.Irish Soc.,  {\bf 46},
9-10, 1941;
{\it Further Studies on Solving Eigenvalue Problems by
Factorization}. Proc.Roy.Irish Soc.,  {\bf 46}, 183-206, 1941;
{\it The Factorization of the Hypergeometric Equation}.
Proc.Roy.Irish Soc.,  {\bf 47}, 53-54, 1941.
\bibitem{schwinger}
J.Schwinger. {\it A magnetic model of matter}. Science, {\bf
165}, 757-761, 1969.
\bibitem{SEGO}
 G.Sege. {\it Orthogonal Polynomials}. Fizmatgiz, Moscow, 1962,
 (in Russian).
\bibitem{SW}
N. Seiberg and E. Witten. {\it Monopoles, duality and chiral symmetry
breaking in $N=2$ supersymmetric QCD}. Nucl.Phys. B, {\bf 431},
484-550, 1994.
\bibitem{SHIBUYA}
T.Shibuya and C.E.Wufman. Am. J. Phys., {\bf 33}, 570-574, 1965.
\bibitem{SISTER1}
A.N.Sisakyan and V.M.Ter-Antonyan. {\it The Structure of the Hurwitz Transformation}.
Preprint JINR, P5-94-219, 1994, (in Russian).
\bibitem{SISTER2}
A.N.Sisakyan and V.M.Ter-Antonyan. {\it The Hurwitz-Eulerian Reduction}.
Preprint JINR, P5-94-220, 1994, (in Russian).
\bibitem{SKLYAN1}
E.K.Sklyanin. {\it Some algebraic structures connected with the Yang—Baxter equation}.
Functional Analysis and its Applications, {\bf 16}, 263-270, 1982.
\bibitem{SKLYANIN2}
E.K.Sklyanin. {\it Some algebraic structures connected with the Yang—Baxter equation.
Representations of quantum algebras}. Functional Analysis and Its Applications,
{\bf 17}, 273-284, 1983.
\bibitem{SMIRNOV}
Yu.F.Smirnov and K.V.Shitikova. {\it Method of K harmonics and the shell model}.
Sov. J. Particles Nucl., {\bf 8}, 847-910, 1977.
\bibitem{SMOR-SHEL}
Ya.A Smorodinskii and L.A. Shelepin.
{\it Clebsch-Gordan Coefficients, Viewed From Different Sides}.
Sov. Phys. Uspekhi, {\bf 15}, 3-24, (1972).
\bibitem{SMOR-SHEP}
Ya.A. Smorodinskii and G.I. Shepelev.
{\it Matrix Elements of the Boost for the Reduction O(3, 1), O(2, 1), O(2)}.
Sov. J. Nucl. Phys. 18, 483 (1974).
\bibitem{STEV}
A.F.Stevenson. {\it Note on the ``Kepler Problem'' in a Spherical
Space, and the Factorization Method of Solving Eigenvalue
Problems}. Phys.Rev.,\ {\bf 59}, 842-843, 1941.
\bibitem{STONE}
A.P.Stone {\it Some Properties of Wigner Coefficients and
Hyperspherical Harmonics}, Proc. Camb.Phil.Soc., {\bf 52},
424-430, 1956.
\bibitem{MUKUNDA1}
E.C.G.Sudarshan, N.Mukunda and L.O'Raifeartaigh. {\it Group theore
of the Kepler problem}. Phys. Lett., {\bf 19}, 322-326, 1965.
\bibitem{TAMM}
I.E.Tamm. {\it Selected papers}. Springer Science and Business Media,
2012.
\bibitem{TARTER}
C.B.Tarter. {\it Coefficients Connecting the Stark and Field-Free
Wavefunctions of Hydrogen}. J.Math.Phys., {\bf 11}, 3192-3195,
1970.
\bibitem{TERANT}
V.M.Ter-Antonyan.  {\it Dyon-Oscillator Duality}.
quant-ph/0003106.
\bibitem{TN}
V.M.Ter-Antonyan and A.Nersessian: {\it Quantum oscillator and a
bound system of two dyons}. Modern Phys.Lett., {\bf A10},
2633-2638, 1995.
\bibitem{TODOR}
I.T.Todorov. {\it Non-compact groups and dynamical symmetries}.
High Energy Physics and Theory of Elementary Particles, Naukova Dumka, Kiev,
1967, (in Russian).
\bibitem{TRUNK}
M. Trunk. {\it The five-dimensional Kepler problem as an
$SU(2)$ gauge system: Algebraic constraint quantization} Int. J.
Mod. Phys., {\bf A11}, 2329-2355, 1996.
\bibitem{VAR}
D.A.Varshalovich, A.N.Moskalev and V.K.Khersonskii. {\it Quantum theory of
angular momentum}. World scientific, 1988.
\bibitem{VENDRAMIN}
I.Vendramin. {\it On the dynamical symmetry of the nonisotropic
oscillators}. Nuovo Cimento, {\bf 54A}, 190-192, 1968.
\bibitem{VILEN2}
N.Ya.Vilenkin. {\it Polyspherical and oryspherical functions}. Mathematicheskii Sbornik,
{\bf 68}, 432-443, 1965 (in Russian).
\bibitem{VKS}
N.Ya.Vilenkin, G.I.Kuznetsov and Ya.A. Smorodinsky. {\it Eigenfunctions of the Laplace Operator
Providing Representations of the U(2), SU(2), SO(3), U(3), and SU(3) Groups and the Symbolic Method}.
Sov. J. Nucl. Phys., {\bf 2}, 906-917, 1965.
\bibitem{VILEN1}
 N.Ya.Vilenkin. {\it Special functions and group representation theory}.
 Transl. of Math. Monographs Amer. Math. Soc. 22, 1968.
\bibitem{VMPSS}
S.I.Vinitsky, L.G.Mardoyan, G.S.Pogosyan, A.N.Sissakian and
T.A.Strizh.\\
{\it A Hydrogen Atom in the Curved Space. Expansion over Free Solution
on the Three-Di\-men\-sio\-nal sphere}.
Yader. Fiz. {\bf 56(3)}, 61-73, 1993, (in Russian).
\bibitem{Watson}
G.N.Watson. {\it Note on Generalized Hypergeometrical Series}.
Proc. London Math. Soc. \textbf{23}, XIII-XV, 1925.
\bibitem{WENTZER}
G.Wentzer.  {\it Uber den R\"{u}ckstossbeim Comptoneffekt am
Wasserstoffatom}. Zs. Phys., {\bf A 58}, 348-367, 1929.
\bibitem{WITWAT}
E.T.Whittaker and G.N.Watson. {\it A Course of Modern Analysis}.
Cambridge Univerity Press, 1927.
\bibitem{FSUW}
P.Winternitz, Ya.A.Smorodinsky, M.Uhlir and I.Frish. {\it Symmetry groups in quantum mechanics}.
Sov. J. Nucl. Phys., {\bf 4}, 625-635, 1967.
\bibitem{WITTEN}
E.Witten. {\it Anti-de-Sitter space and holography}. Adv. Theor.
Math. Phys., {\bf 2}, 253-291, 1998; [hep-th/9802150].
\bibitem{Zaslow}
B.Zaslow, M.E.Zandler. {\it Two-dimensional Analog to the
Hydrogen Atom}. Am. J. Phys. {\bf 35}, 1118-1119, 1967.
\bibitem{Zhevl}
K.A.Zhevlakov, A.M. Slin'ko, I.P. Shestakov, and A.I. Shirshov.
{\it Rings that are nearly associative}. Moscow, Nauka, 1978.
\bibitem{ZWANZIG}
D. Zwanziger. {\it Exactly soluble nonrelativistic model of
particles with both electric and magnetic charges}. Phys. Rev.,
{\bf 176}, 1480-1488, 1968.
\bibitem{ELCHAS}
High Energy Physics and Theory of Elementary Particles, Naukova Dumka, Kiev,
1967, (in Russian).
\end{thebibliography}
\end{document}